# Physics at Super $B$ Factory

February 11, 2010

Possible updates of this document will be in the future available at
http://belle2.kek.jp/physics.html

A. G. Akeroyd,[25] T. Aushev,[10,17] W. Bartel,[3] A. Bondar,[1] J. Brodzicka,[15] T. E. Browder,[5]
P. Chang,[25] Y. Chao,[25] K. F. Chen,[25] J. Dalseno,[21] A. Drutskoy,[2] Y. Enari,[18] T. Gershon,[34]
B. Golob[‡],[16] T. Goto,[13] F. Handa,[31] K. Hara,[23] S. Hashimoto[‡],[13] H. Hayashii,[24] M. Hazumi[‡],[13]
T. Higuchi,[13] J. Hisano,[8] W. S. Hou,[25] T. Iijima,[23] K. Ikado,[23] K. Inami,[23] H. Itoh,[8]
R. Itoh,[13] H. Ishino,[26] N. Katayama,[13] Y. Y. Keum,[14] K. Kinoshita,[2] E. Kou,[27] P. Križan,[16]
P. Krokovny,[6] T. Kurimoto,[32] Y. Kwon,[35] A. Limosani,[20] T. Matsumoto,[11] T. Morozumi,[7]
Y. Nakahama,[28] M. Nakao,[13] S. Nishida,[13] T. Ohshima,[23] Y. Okada,[13] K. Okumura[19]
S. L. Olsen,[30] T. Onogi,[36] G. Pakhlova,[10] H. Palka,[15] P. Pakhlov,[10] A. Poluektov,[1]
S. Recksiegel,[22] H. Sagawa,[8] M. Saigo,[31] Y. Sakai,[13] A. I. Sanda,[12] C. Schwanda,[33]
A. Schwartz,[2] K. Senyo,[23] Y. Shimizu,[31] T. Shindou,[13] R. Sinha,[9] M. Starič,[16] K. Sumisawa,[13]
M. Tanaka,[29] K. Trabelsi,[13] P. Urquijo,[4] Y. Ushiroda,[13] E. Won,[14] H. Yamamoto,[31]
M. Yamauchi,[13] T. Yoshikawa,[23] J. Zupan [16]

[‡] Editor

[1] *Budker Institute of Nuclear Physics, Novosibirsk, Russia*
[2] *University of Cincinnati, Cincinnati, Ohio 45221, USA*
[3] *DESY, Hamburg, Germany*
[4] *University of Geneva, Geneva, Switzerland*
[5] *University of Hawaii, Honolulu, Hawaii 96822, USA*
[6] *Heidelberg University, Heidelberg, Germany*
[7] *Graduate School of Science, Hiroshima University, Higashi-Hiroshima, 739-8526, Japan*
[8] *ICRR, University of Tokyo, Kashiwa 277-8582, Japan*
[9] *Institute of Mathematical Sciences, C.I.T. Campus Taramani Chennai 600 113, India*
[10] *Institute of Theoretical and Experimental Physics, Moscow, Russia*
[11] *Japan Synchrotron Radiation Research Institute, Harima, Japan*
[12] *Kanagawa University, Kanagawa, Japan*
[13] *High Energy Accelerator Research Organization (KEK), Tsukuba, Japan*
[14] *Korea University, Seoul, Korea*
[15] *Henryk Niewodniczanski Institute of Nuclear Physics, Krakow, Poland*
[16] *J. Stefan Institute, Ljubljana, and Faculty of Mathematics and Physics, University of Ljubljana, Slovenia*
[17] *École Polytechnique Fédérale de Lausanne (EPFL), Lausanne, Switzerland*
[18] *LPNHE, IN2P3/CNRS, Universities Paris VI and VII, Paris, France*
[19] *Kyushu University, Fukuoka 812-8581 Japan*
[20] *University of Melbourne, Melbourne, Australia*
[21] *Max-Planck-Institut für Physik, München, Germany*

[22] *Technische Universität München, D-85748 Garching, Germany*
[23] *Nagoya University, Nagoya, Japan*
[24] *Nara Women's University, Nara, Japan*
[25] *National Taiwan University, Taipei, Taiwan*
[26] *Okayama University, Okayama, Japan*
[27] *Laboratoire de Physique Theorique, Orsay, France*
[28] *LAL Orsay, Orsay, France*
[29] *Osaka University, Osaka, Japan*
[30] *Seoul National University, Seoul, Korea*
[31] *Tohoku University, Sendai, Japan*
[32] *Toyama University, Toyama, Japan*
[33] *Institute of High Energy Physics, Vienna, Austria*
[34] *Department of Physics, University of Warwick, Coventry CV4 7AL, U.K.*
[35] *Yonsei University, Seoul, Korea*
[36] *YITP, Kyoto University, Kyoto 606-8502, Japan*

**Abstract**

This report presents the results of studies that investigate the physics reach at a Super $B$ factory, an asymmetric-energy $e^+e^-$ collider with a design luminosity of $8 \times 10^{35}$ cm$^{-2}$s$^{-1}$, which is around 50 times as large as the peak luminosity achieved by the KEKB collider. The studies focus on flavor physics and $CP$ violation measurements that could be carried out in the LHC era. The physics motivation, key observables, measurement methods and expected precisions are presented.

# Contents

# Chapter 1

# Introduction

## 1.1 Motivation for the Higher Luminosity $B$ Factory

Since the end of the last century, two asymmetric-energy $e^+e^-$ $B$ factories, the KEKB collider for the Belle experiment at KEK and the PEPII collider for the BaBar experiment at SLAC, have been achieving a tremendous success that lead to the confirmation of the Standard Model (SM) in the quark flavor sector. In the summer of 2001, the presence of $CP$ violation in the $B$ meson system was established by the Belle collaboration [1–4] and simultaneously by the BaBar collaboration [5–8] through the measurement of the time dependent asymmetry in the decay process $B^0(\bar{B}^0) \to J/\psi K_S^0$. This measurement was the main target of the present $B$ factories, and it was achieved as originally planned. The experimental data indicated that the Kobayashi-Maskawa mechanism, which is now a part of the SM of elementary particles, is indeed the dominant source of the observed $CP$ violation in Nature. Following the experimental confirmation, M. Kobayashi and T. Maskawa were awarded the 2008 Nobel Prize for physics.

The present paper, which is a thorough update of a document published in 2004 [9], aims to provide the motivation, experimental methods and at least part of the scientific output that could be expected with an upgraded $B$ factory, based on the successful KEKB collider and Belle detector, and planned to start operation in few years.

The Belle experiment proved its ability to measure a number of decay modes of the $B$ meson and to extract Cabibbo-Kobayashi-Maskawa (CKM) matrix elements and other interesting observables: the precision of the measurement of the angle $\phi_1$ of the unitarity triangle through the $B^0 \to J/\psi K_S^0$ time-dependent asymmetry has improved to better than 5% [10, 11]; direct $CP$ violation was observed in $B^0 \to \pi^+\pi^-$ [12] and $K^+\pi^-$ [13] decays; the angle $\phi_2$ [12, 14–16] has been measured with $B \to \pi\pi$, $\rho\pi$ and $\rho\rho$ systems using isospin symmetries; the angle $\phi_3$ has also been measured through the processes $B \to D^{(*)}K^{(*)}$ and the evidence of direct $CP$ violation in $B \to DK$ decays was obtained [17]; the magnitudes of the CKM matrix elements have been measured much more precisely than before; rare $B$ decays such as $B \to K^{(*)}\ell\ell$ [18], $\rho\gamma$ [19] and $\tau\nu$ [20] have been observed for the first time; the first evidence of $D^0 - \bar{D}^0$ mixing surfaced [21]; the quantum entanglement of neutral $B$ meson pairs was directly confirmed [22]. Through these precise measurements and new observations we have succeeded to overconstrain the quark flavor sector of the SM. The latter proves to be self-consistent within the current accuracy of the experimental results.

In spite of the tremendous success mentioned above, several fundamental questions remain in the flavor sector of quarks and leptons. First of all, the SM includes too many parameters - the masses and mixing parameters of the quarks and leptons - all of which are apriori unknown and should be determined experimentally. This is due to the fact that there is no principle to govern the Yukawa terms in the SM Lagrangian. Any Yukawa coupling between two fermions, irrespective of the generation they belong to, is allowed, as far as it is gauge-invariant and renormalizable. In spite of this fact, the measured CKM matrix elements show a clear pattern as shown in Fig. 1.1. The origin of this hierarchy is a mystery at the moment; it may indicate that some hidden mechanism, e.g. some new flavor symmetry, exists at a higher energy scale. Secondly, from a cosmological viewpoint, there is a serious problem with the matter-antimatter asymmetry in the universe. While the $CP$ violation is one of the conditions for the evolution of a matter dominated universe [23] the magnitude of the asymmetry cannot be explained solely by the $CP$ violation within the Standard Model, which originates from the quark mixing.

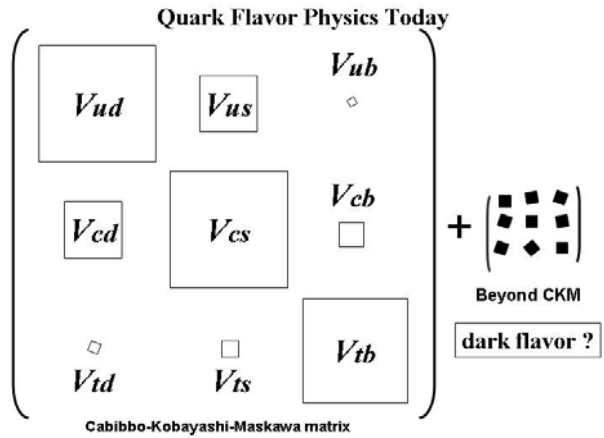

Figure 1.1: Current knowledge on the CKM unitary matrix.

There are also other fundamental questions in the SM, which have a deep impact on the studies of flavor mixing. Due to quadratically divergent radiative corrections, the Higgs mass is naturally of the same order as its cutoff scale. This implies that some new physics exists not far above the electroweak scale, most likely at the TeV energy scale. The mechanism to suppress the Flavor-Changing Neutral Current (FCNC) processes should be present in new physics models if the new physics lies at the TeV energy scale, because otherwise such FCNC processes would violate current experimental limits. Even with such a mechanism, many TeV new physics scenarios almost inevitably introduce new flavor mixing that can be detected with precision measurements at the energy scale of $B$ factories. Information obtained from flavor physics experiments is thus essential to uncover the details of the physics beyond the Standard Model, even after energy frontier machines discover new particles.

The history of particle physics tells us that the flavor physics experiments often provided a breakthrough in their own period. In fact, before its discovery, the existence of the charm quark was postulated to explain the smallness of strangeness-changing neutral currents (the Glashow-Illiopolous-Maiani (GIM) mechanism [24]). The third family of quarks and leptons was predicted by Kobayashi and Maskawa to explain the small $CP$ violation seen in kaon mixing [25]. The top quark mass was predicted based on the $B^0 - \bar{B}^0$ mixing measurements, before it was directly measured at the Tevatron. These are all examples of FCNC processes, with which one can investigate the effect of heavier particles appearing only in quantum loop corrections. Moreover, there have been many unexpected discoveries in the past and Belle has followed the tradition: several new states, which according to the so far known properties cannot be placed in the conventional quark model of hadrons, were discovered [26–28]. Therefore, if the history can be any guide, a long-term step-by-step strategy for flavor experiments should be part of the grand challenge of particle physics. It will play a complementary role to the TeV energy-frontier colliders such as the Large Hadron Collider (LHC) and the International Linear Collider (ILC), which are expected to produce new heavy elementary particles.

A natural place to investigate a wide range of FCNC processes is in $B$ meson decays. This is because the bottom quark belongs to the third generation of fermions and hence the processes with $b$ quarks involve all existing generations of quarks. In addition to $B^0 - \bar{B}^0$ mixing, which is an analog of the traditional $K^0 - \bar{K}^0$ mixing, there are many FCNC decay processes induced by so-called penguin diagrams, such as the radiative decay $b \to s\gamma$, the semileptonic decay $b \to s\ell^+\ell^-$, and the hadronic decays $b \to dq\bar{q}$ and $b \to sq\bar{q}$. All of these processes are suppressed in the Standard Model by the GIM mechanism, and, therefore, the effect of new physics may be relatively enhanced.

The present two $B$ factories have not only confirmed the CKM picture of the quark mixing but also discovered many new processes. Examples include the semileptonic FCNC processes $B \to K\ell^+\ell^-$ [29], $B \to K^*\ell^+\ell^-$ [18], $B \to X_s\ell^+\ell^-$ [30] with inclusive reconstructions of the $X_s$ system, the only FCNC process involving uplike quarks so far - mixing in the $D^0$ system, and a purely leptonic decay $B \to \tau\nu$.

Recent results include some which prove at least to be difficult to explain within the SM, if not some hints of deviations from its predictions. The values of the angle $\phi_1$ measured in some penguin process $b \to sq\bar{q}$ and the precisely measured value in $B \to J/\psi K_S^0$ differ by two to three standard deviations ($B^0 \to \pi^0\pi^0 K_S^0$, $B^0 \to K^+K^-K^0$, [11, 31, 32]) and may suggest the existence of a new $CP$ phase in this penguin process; the lepton forward-backward asymmetry in $B \to K^*\ell^+\ell^-$ is around two standard deviations higher than the SM prediction [33]; direct $CPV$ asymmetries in $B^0 \to K^+\pi^-$ and $B^+ \to K^+\pi^0$ differ significantly, although naively one would expect them to be the same [34]; the branching fraction for $B^+ \to \tau\nu$ is up to two standard deviations higher than the expectation, depending on the theoretical input chosen [11, 35]; in purely leptonic $D_s^+ \to \mu\nu$ and $D_s^+ \to \tau\nu$ decays the deviation is even larger [36, 37] if one uses the recent lattice QCD calculations of the meson decay constant; the measured production cross-section for pairs of $c\bar{c}$ states is higher than the calculations predict [38]. By collecting many such observations we may probe new physics, and, once its existence is established, these measurements will determine the properties of the new physics. This is possible, however, only if the luminosity at present $B$ factories is upgraded by a substantial amount. As we discuss in the following chapters, a factor of 50 improvement will greatly enhance the possibility to discover possible new physics and identify its nature. The Higher Luminosity B Factory SuperKEKB is a machine designed to explore such interesting $B$ decay processes.

There are several important questions in flavor physics where SuperKEKB plays an essential role to answer: Is there any new $CP$-violating phase?; Is there any new right-handed current?; Is there any effect from new Higgs fields?; Is there any new flavor violation such as lepton-flavor violation?; Is there any new flavor symmetry that explains the enigmatic CKM hierarchy? SuperKEKB offers unique measurements to answer these questions. Examples include time-dependent $CP$ asymmetries in $B^0 \to \phi K^0$, $K_S^0 \pi^0 \gamma$, branching fractions for $B \to \tau\nu$, $D\tau\nu$, $\tau \to \mu\gamma$, $CPV$ for charmed mesons, and of course angles and sides of the CKM unitarity triangle. These are "clean" measurements in a sense that both experimental and theoretical systematic uncertainties are small. There are many other "clean" measurements possible at SuperKEKB as will be described in this report. Indeed the true strength of SuperKEKB lies in the fact that correlations among the large number of such measurements identify the nature of flavor structure beyond the Standard Model.

$B$ physics programs are also being pursued at hadron machines, including the ongoing Teva-

tron experiments [39] and the $B$ physics programs at the Large Hadron Collider (LHC) [40], which started operation in 2009. Because of the very large production cross section for $b$-quarks in the hadron environment, some of the measurements such as $B_s \to \mu\mu$ can be performed better at the hadron colliders. However, an $e^+e^-$ machine provides a much cleaner environment, which is essential for important observables that involve $\gamma$'s, $\pi^0$'s, $K_L^0$'s or neutrinos in the final states. At the $\Upsilon(4S)$ resonance, the $B\bar{B}$ pair is produced near the energy threshold and there are no associated particles. This means that by reconstructing the full energy-momentum vector of a $B$ ($\bar{B}$) meson from its daughter particles (the full reconstruction technique), one can infer the missing momentum in the decay of the other $\bar{B}$ ($B$) meson. This technique is essential for the measurement of channels including neutrino(s) in the final state. The measurement of the CKM element $|V_{ub}|$ through the semi-leptonic decay $b \to ul\bar{\nu}$, the search for a charged Higgs effect in $B \to D\tau\bar{\nu}$, and measurements of $B \to K\nu\bar{\nu}$, $B \to \tau\nu$ fall in this class. Therefore, there is no doubt that SuperKEKB is complementary to the future experiments at the hadron machines from both, the motivational and the experimental aspect.

In the following two chapters some theoretical grounds of the flavor physics within and beyond the SM are addressed. In Chapter 4 we briefly overview the current status and the main results of the Belle experiment. Experimental methods and expected results of SuperKEKB are detailed in the main part of the paper, Chapter 5. Some of possible New Physics scenarios and their identification through measurements at the planned Super $B$ factory are given in Chapter 6, and the last chapter summarizes the paper.

# References

[1] A. Abashian *et al.* [Belle Collaboration], "Measurement of the CP violation parameter $\sin(2\phi_1)$ in $B_d^0$ meson decays," Phys. Rev. Lett. **86**, 2509 (2001).

[2] K. Abe *et al.* [Belle Collaboration], "Observation of large CP violation in the neutral B meson system," Phys. Rev. Lett. **87**, 091802 (2002).

[3] K. Abe *et al.* [Belle Collaboration], "Observation of mixing-induced CP violation in the neutral B meson system," Phys. Rev. D **66**, 032007 (2002).

[4] K. Abe *et al.* [Belle Collaboration], "An improved measurement of mixing-induced CP violation in the neutral B meson system. ((B))," Phys. Rev. D **66**, 071102 (2002).

[5] B. Aubert *et al.* [BABAR Collaboration], "Measurement of CP violating asymmetries in $B^0$ decays to CP eigenstates," Phys. Rev. Lett. **86**, 2515 (2001).

[6] B. Aubert *et al.* [BABAR Collaboration], "A study of time dependent CP-violating asymmetries and flavor oscillations in neutral B decays at the $\Upsilon(4S)$," Phys. Rev. D **66**, 032003 (2002).

[7] B. Aubert *et al.* [BABAR Collaboration], "Observation of CP violation in the $B^0$ meson system," Phys. Rev. Lett. **87**, 091801 (2001).

[8] B. Aubert *et al.* [BABAR Collaboration], "Measurement of the CP-violating asymmetry amplitude $\sin 2\beta$. ((B))," Phys. Rev. Lett. **89**, 201802 (2002).

[9] A.G. Akeroyd *et al.*, "Physics at Super B Factory," arXiv:hep-ex/0406071.

[10] K. F. Chen *et al.* [Belle Collaboration], "Observation of time-dependent $CP$ violation in $B^0 \to \eta' K^0$ decays and improved measurements of $CP$ asymmetries in $B^0 \to \phi K^0$, $K_S^0 K_S^0 K_S^0$ and $B^0 \to J/\psi K^0$ decays," Phys. Rev. Lett. **98**, 031802 (2007).

[11] E. Barberio *et al.* [Heavy Flavor Averaging Group (HFAG)], "Averages of *b*-hadron and *c*-hadron Properties at the End of 2007," arXiv:0808.1297, and online update for 2008 Summer at http://www.slac.stanford.edu/xorg/hfag/.

[12] K. Abe *et al.* [Belle Collaboration], "Observation of large CP violation and evidence for direct CP violation in $B^0 \to \pi^+\pi^-$ decays," Phys. Rev. Lett. **93**, 021601 (2004).

[13] Y. Chao *et al.* [Belle Collaboration], "Evidence for Direct $CP$ Violation in $B^0 \to K^+\pi^-$ Decays," Phys. Rev. Lett. **93**, 191802 (2004).

[14] K. Abe *et al.* [Belle Collaboration], "Measurement of CP-violating asymmetries in $B0 \to \pi^+\pi^-$ decays," Phys. Rev. Lett. **89**, 071801 (2002).

[15] K. Abe *et al.* [Belle Collaboration], "Evidence for CP-violating asymmetries $B^0 \to \pi^+\pi^-$ decays and constraints on the CKM angle $\phi_2$," Phys. Rev. D **68**, 012001 (2003).

[16] C.-C. Chiang *et al.* [Belle Collaboration], "Measurement of $B^0 \to \pi^+\pi^-\pi^+\pi^-$ decays and search for $B^0 \to \rho^0\rho^0$," Phys. Rev. D **78**, 111102 (2008).

[17] K. Abe *et al.* [Belle Collaboration], "Updated Measurement of $\phi_3$ with a Dalitz Plot Analysis of $B^+ \to D^{(*)}K^+$ Decay," arXiv:0803.3375 [hep-ex].

[18] A. Ishikawa *et al.* [Belle Collaboration], "Observation of $B \to K^* l^+ l^-$," Phys. Rev. Lett. **91**, 261601 (2003).

[19] D. Mohapatra *et al.* [Belle Collaboration], "Observation of $b \to d\gamma$ and Determination of $|V_{td}/V_{ts}|$," Phys. Rev. Lett. **96**, 221601 (2006).

[20] K. Ikado *et al.*, "Evidence of the purely leptonic decay $B^- \to \tau^- \bar{\nu}_\tau$," Phys. Rev. Lett. **97**, 251802 (2006).

[21] M. Starič *et al.* [Belle Collaboration], "Evidence for $D^0 - \bar{D}^0$ Mixing," Phys. Rev. Lett. **98**, 211803 (2007).

[22] A. Go *et al.* [Belle Collaboration], "Measurement of Einstein-Podolsky-Rosen-Type Flavor Entanglement in $\Upsilon(4S) \to B^0\bar{B}^0$ Decays," Phys. Rev. Lett. **99**, 131802 (2007).

[23] A. D. Sakharov, "Violation of CP invariance, C asymmetry, and baryon asymmetry of the universe," JETP Lett. **5**, 24 (1967).

[24] S. L. Glashow, J. Iliopoulos and L. Maiani, "Weak Interactions With Lepton – Hadron Symmetry," Phys. Rev. D **2**, 1285 (1970).

[25] M. Kobayashi and T. Maskawa, "CP Violation In The Renormalizable Theory Of Weak Interaction," Prog. Theor. Phys. **49**, 652 (1973).

[26] S.-K. Choi *et al.* [BELLE Collaboration], "Observation of a Narrow Charmoniumlike State in Exclusive $B^\pm \to K^\pm \pi^+\pi^- J/\psi$ Decays," Phys. Rev. Lett. **91**, 262001 (2003).

[27] S.-K. Choi *et al.* [BELLE Collaboration], "Observation of a Resonancelike Structure in the $\pi^\pm \psi'$ Mass Distribution in Exclusive $B \to K\pi^\pm\psi'$ Decays," Phys. Rev. Lett. **100**, 142001 (2008).

[28] R. Mizuk *et al.* [BELLE Collaboration], "Observation of two resonancelike structures in the $\pi^+\chi_{c1}$ mass distribution in exclusive $\bar{B}^0 \to K^-\pi^+\chi_{c1}$ decays," Phys. Rev. D **78**, 072004 (2008).

[29] K. Abe *et al.* [BELLE Collaboration], "Observation of the decay $B \to K\mu^+\mu^-$," Phys. Rev. Lett. **88**, 021801 (2002).

[30] J. Kaneko *et al.* [Belle Collaboration], "Measurement of the electroweak penguin process $B \to X_s l^+ l^-$. ((B))," Phys. Rev. Lett. **90**, 021801 (2003).

[31] K. Abe *et al.* [Belle Collaboration], "Measurements of $CP$ Violation Parameters in $B^0 \to K^0_S\pi^0\pi^0$ and $B^0 \to K^0_S K^0_S$ Decays," arXiv:0708.1845 [hep-ex].

[32] Y. Chao *et al.* [Belle Collaboration], "Measurements of time-dependent $CP$ violation in $B^0 \to \omega K^0_S$, $f_0(980)K^0_S$, $K^0_S\pi^0$ and $K^+K^-K^0_S$ decays," Phys. Rev. D **76**, 091103(R) (2007).

[33] I. Adachi *et al.* [Belle Collaboration], "Measurement of the Differential Branching Fraction and Forward-Backward Asymmetry for $B \to K^{(*)}\ell^+\ell^-$," arXiv:0810.0335 [hep-ex].

[34] S.-W. Lin *et al.* [Belle Collaboration], "Difference in direct charge-parity violation between charged and neutral $B$ meson decays," Nature **542**, 332 (2008).

[35] I. Adachi [The Belle Collaboration], "Measurement of $B^- \to \tau^- \bar{\nu}_\tau$ Decay With a Semileptonic Tagging Method," arXiv:0809.3834 [hep-ex].

[36] L. Widhalm *et al.* [Belle Collaboration], "Measurement of $\mathcal{B}(D_s^+ \to \mu^+\nu_\mu)$," Phys. Rev. Lett. **100**, 241801 (2008).

[37] J. Rosner, S. Stone, and C. Amsler *et al.* "Decay constants of charged pseudo-scalar mesons," Phys. Lett. B **667**, 1 (2008).

[38] K. Abe *et al.* [Belle Collaboration], "Observation of double $c\bar{c}$ production in $e^+e^-$ annihilation at $\sqrt{s} \approx 10.6$ GeV," Phys. Rev. Lett. **89**, 142001 (2002).

[39] K. Anikeev *et al.*, "B physics at the Tevatron: Run II and beyond," arXiv:hep-ph/0201071.

[40] P. Ball *et al.*, "B decays at the LHC," arXiv:hep-ph/0003238.

## Chapter 2

# Flavor Structure of the Standard Model

The Standard Model has a rich structure in its flavor sector, mainly because it contains three generations of quarks and leptons. In the quark sector, it is well established that the misalignment between the weak interaction eigenstates and mass eigenstates leads to the Cabibbo-Kobayashi-Maskawa (CKM) matrix, which is the source of the transitions between different generations. Even more importantly it offers the source of the $CP$ violation. This flavor structure has been confirmed by many experimental measurements to a good precision. In this chapter we summarize the basic flavor structure of the Standard Model and discuss the principles of various measurements to probe the flavor structure through the $B$ meson decays.

Among experimental tests of the $CP$ violation, measurements of the mixing induced $CP$ violation in the neutral $B$ meson system played a central role at the present $B$ factories. The angle $\phi_1$ of the unitarity triangle has been measured very precisely, and precision measurement of the angle $\phi_2$ is also possible by accumulating more statistics.

The present $B$ factories have also demonstrated the sensitivity in other measurements investigating the flavor structure. The direct $CP$ violation in $B$ meson decays can be measured in the decay modes such as $B \to \pi\pi$ and $B \to K\pi$. The angle $\phi_3$ can be measured through the interference of decay amplitudes involving intermediate $D^{(*)}$ mesons. Several other Flavor Changing Neutral Current (FCNC) processes that are sensitive to possible new heavy particles exchanged in the loops of Feynman diagrams have also been investigated.

Since the start of the measurements with the present $B$ factories, the knowledge of the CKM matrix elements has been greatly improved. We will briefly summarize the current status in Sec. 2.4.

## 2.1 Flavor Structure of the Standard Model

In the Standard Model of elementary particles there are three generations of leptons and quarks

$$\begin{pmatrix} \nu_e \\ e \end{pmatrix} \begin{pmatrix} \nu_\mu \\ \mu \end{pmatrix} \begin{pmatrix} \nu_\tau \\ \tau \end{pmatrix}, \tag{2.1}$$

$$\begin{pmatrix} u \\ d \end{pmatrix} \begin{pmatrix} c \\ s \end{pmatrix} \begin{pmatrix} t \\ b \end{pmatrix}, \tag{2.2}$$

and their interactions are described by a gauge field theory with the gauge group $SU(3)_C \times SU(2)_L \times U(1)_Y$. The strong interaction among quarks (QCD) is described by $SU(3)_C$. The

left- and right-handed fermion fields transform differently under the electroweak gauge group $SU(2)_L \times U(1)_Y$. The right-handed components of the leptons and quarks are singlets under the weak $SU(2)_L$, The weak hypercharge $Y$ has values $Y = -1$ for $(\nu_\ell, \ell)_L$, $Y = 1/3$ for $(q_u, q_d)_L$, $Y = -2$ for $\ell_R$, $Y = 4/3$ for $q_{uR}$ and $Y = -2/3$ for $q_{dR}$, such that the lepton and quark charges are obtained. The left-handed leptons are doublets of the weak $SU(2)_L$, while the weak doublets of quarks differ slightly from (2.2) and are given by

$$\begin{pmatrix} u \\ d' \end{pmatrix}_L \quad \begin{pmatrix} c \\ s' \end{pmatrix}_L \quad \begin{pmatrix} t \\ b' \end{pmatrix}_L. \tag{2.3}$$

The weak eigenstates $(d', s', b')$ are linear combinations of the mass eigenstates $(d, s, b)$. They are rotated by a $3 \times 3$ unitary matrix, referred to as the CKM matrix $\hat{V}_{\text{CKM}}$ [1,2],

$$\begin{pmatrix} d' \\ s' \\ b' \end{pmatrix} = \hat{V}_{\text{CKM}} \begin{pmatrix} d \\ s \\ b \end{pmatrix} \equiv \begin{pmatrix} V_{ud} & V_{us} & V_{ub} \\ V_{cd} & V_{cs} & V_{cb} \\ V_{td} & V_{ts} & V_{tb} \end{pmatrix} \begin{pmatrix} d \\ s \\ b \end{pmatrix}. \tag{2.4}$$

The charged current interactions of quarks are mediated by the $W$ bosons and are described by the interaction Lagrangian

$$\begin{aligned} \mathcal{L}^{\text{CC}} &= -\frac{g_2}{\sqrt{2}} \begin{pmatrix} \bar{u} & \bar{c} & \bar{t} \end{pmatrix}_L \gamma^\mu \begin{pmatrix} d' \\ s' \\ b' \end{pmatrix}_L W_\mu^\dagger + \text{h.c.} \\ &= -\frac{g_2}{\sqrt{2}} \begin{pmatrix} \bar{u} & \bar{c} & \bar{t} \end{pmatrix}_L \gamma^\mu \hat{V}_{\text{CKM}} \begin{pmatrix} d \\ s \\ b \end{pmatrix}_L W_\mu^\dagger + \text{h.c.}, \end{aligned} \tag{2.5}$$

where $W_\mu$ denotes the $W$ boson field, and $g_2$ is the gauge coupling of $SU(2)_L$. In the low energy effective Hamiltonian, obtained by integrating out the boson fields, the strength of the charged weak interaction is given by the Fermi constant $G_F/\sqrt{2} = g_2^2/8M_W^2$. Due to the misalignment between the up-type and down-type quark fields, the charged current induces transitions among different generations.

In contrast, the neutral current is flavor-conserving, which is ensured by the unitarity of the CKM matrix, and, thus, Flavor Changing Neutral Currents (FCNC) are absent at the tree level in the Standard Model. This is the Glashow-Iliopoulos-Maiani (GIM) mechanism [3]. Even including loop corrections, the FCNC interactions vanish in the limit of degenerate (up- or down-type) quark masses, due to the unitarity of the CKM matrix.

The CKM matrix is a unitary $N \times N$ matrix with $N (= 3)$ number of generations, and thus contains $N^2$ parameters in general. However, $2N - 1$ phases may be absorbed by re-phasing the $2N$ quark fields (one overall phase is related to the total baryon number conservation and is irrelevant for the quark mixing), and $(N - 1)^2$ independent parameters remain. Of these, $\frac{1}{2}(N-1)N$ are real parameters, which correspond to rotation angles among different generations, while $\frac{1}{2}(N - 2)(N - 1)$ are imaginary parameters, which are sources of $CP$-violation. In the three-generation Standard Model, there are 3 mixing angles and 1 $CP$-phase.

The standard parametrization of the CKM matrix is the following [4]:

$$V_{\text{CKM}} = \begin{pmatrix} c_{12}c_{13} & s_{12}c_{13} & s_{13}e^{-i\delta} \\ -s_{12}c_{23} - c_{12}s_{23}s_{13}e^{i\delta} & c_{12}c_{23} - s_{12}s_{23}s_{13}e^{i\delta} & s_{23}c_{13} \\ s_{12}s_{23} - c_{12}c_{23}s_{13}e^{i\delta} & -s_{23}c_{12} - s_{12}c_{23}s_{13}e^{i\delta} & c_{23}c_{13} \end{pmatrix}, \tag{2.6}$$

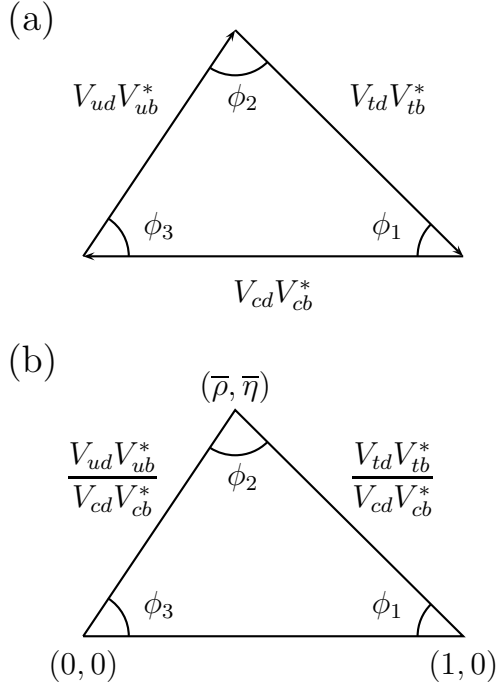

Figure 2.1: Unitarity triangle

where $c_{ij} = \cos\theta_{ij}$ and $s_{ij} = \sin\theta_{ij}$ with $\theta_{ij}$ ($ij = 12, 13$ and $23$) the mixing angles, and $\delta$ is the complex phase. It is known experimentally that the angles are small and exhibit the hierarchy $1 \gg s_{12} \gg s_{23} \gg s_{13}$. To make this structure manifest, the Wolfenstein parametrization [5] is often used, in which one sets $\lambda = |V_{us}| \simeq 0.22$ and

$$V_{\text{CKM}} = \begin{pmatrix} 1 - \lambda^2/2 & \lambda & A\lambda^3(\rho - i\eta) \\ -\lambda & 1 - \lambda^2/2 & A\lambda^2 \\ A\lambda^3(1 - \rho - i\eta) & -A\lambda^2 & 1 \end{pmatrix} + O(\lambda^4), \tag{2.7}$$

with $A$, $\rho$ and $\eta$ being real parameters of order unity. In this parametrization the source of $CP$-violation is carried by the most off-diagonal elements $V_{ub}$ and $V_{td}$.

Among these four parameters, $\lambda$ and $A$ are relatively well known from corresponding semileptonic decays: $|V_{us}| = 0.2255 \pm 0.0019$ from $K_{l3}$ decays and $|V_{cb}| = (41.2 \pm 1.1) \times 10^{-3}$ from inclusive and exclusive $b \to cl\bar{\nu}_l$ decays [6]. The determination of the other two parameters $\rho$ and $\eta$ is conveniently depicted as a contour in the $(\rho, \eta)$ plane. It corresponds to the unitarity relation of the CKM matrix applied to the first and third columns

$$V_{ud}V_{ub}^* + V_{cd}V_{cb}^* + V_{td}V_{tb}^* = 0. \tag{2.8}$$

This relation may be presented in the complex plane as in Fig. 2.1 (a), which is called the "unitarity triangle". Since $V_{cd}V_{cb}^*$ is real to a good approximation (up to $O(\lambda^7)$), it is convenient to normalize the triangle by $|V_{cd}V_{cb}^*| = A\lambda^3$ so that the apex has the coordinate $(\bar{\rho}, \bar{\eta})$ where

$$\bar{\rho} = \rho(1 - \lambda^2/2), \quad \bar{\eta} = \eta(1 - \lambda^2/2), \tag{2.9}$$

(Fig. 2.1 (b)). The three angles of the unitarity triangle represent the complex phase of the combinations

$$\phi_1 = \arg\left[-\frac{V_{cd}V_{cb}^*}{V_{td}V_{tb}^*}\right], \quad \phi_2 = \arg\left[-\frac{V_{td}V_{tb}^*}{V_{ud}V_{ub}^*}\right], \quad \phi_3 = \arg\left[-\frac{V_{ud}V_{ub}^*}{V_{cd}V_{cb}^*}\right]. \tag{2.10}$$

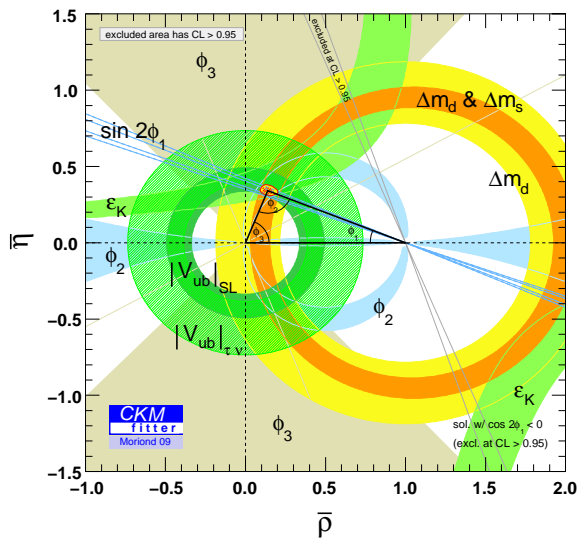

Figure 2.2: A fit of the parameters $(\overline{\rho}, \overline{\eta})$ using several experimental constraints as of Winter 2009. The plot is taken from [7].

The notation $\alpha \equiv \phi_2$, $\beta \equiv \phi_1$, $\gamma \equiv \phi_3$ is also used in the literature.

The present constraints on the parameter $(\overline{\rho}, \overline{\eta})$ are summarized in Fig. 2.2 [7].

## 2.2 Low Energy Effective Hamiltonians

In $B$ decays the exchange of a $W$ boson and virtual loops involving the top quark are effectively point-like interactions, since the relevant length scale of $B$ meson decays is of the order of $O(1/m_b)$ while the $W$ exchange takes place at a much shorter distance scale $O(1/M_W)$. It is theoretically inefficient to calculate the physical amplitudes using the entire $W$ and top quark propagators, and one may instead introduce a low energy effective Hamiltonian. This framework is based on the Operator Product Expansion (OPE) [8], which allows one to separate the long distance physics from the short distance interactions, occurring at a length scale of $1/M_W$, up to the corrections of order $m_b/M_W$, which can be safely neglected in many cases.

For instance, $B^0 - \overline{B}^0$ mixing occurs through the box diagrams shown in Fig. 2.3. The interaction can be described in terms of the $\Delta B = 2$ effective Hamiltonian [9, 10]

$$\mathcal{H}_{eff}^{\Delta B=2} = \frac{G_F^2}{16\pi^2}(V_{tb}V_{td}^*)^2 M_W^2 C^{\Delta B=2}(\mu_b) Q^{\Delta B=2}(\mu_b), \qquad (2.11)$$

where the $\Delta B = 2$ effective four-quark operator is

$$Q^{\Delta B=2} = \overline{d}\gamma_\mu(1-\gamma_5)b\,\overline{d}\gamma_\mu(1-\gamma_5)b. \qquad (2.12)$$

The operator is defined at the renormalization scale $\mu_b \sim m_b$, with the corresponding Wilson coefficient given at leading order by $C^{\Delta B=2}(\mu_b) = [\alpha_s(\mu_W)/\alpha_s(\mu_b)]^{6/23} S_0(m_t^2/M_W^2)$. The function $S_0(m_t^2/M_W^2)$ is the Inami-Lim function [11] and describes the leading order loop effect in the box diagrams at the weak scale.

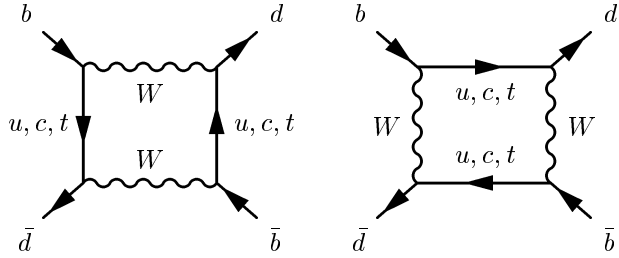

Figure 2.3: Box diagrams to produce the $\Delta B = 2$ four-quark operator

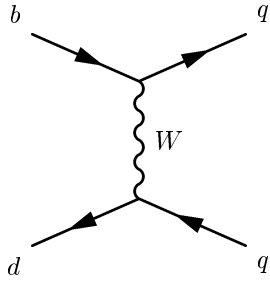

Figure 2.4: Tree-type $W$ boson exchange diagram.

The $\Delta B = 1$ transitions are described by the following effective Hamiltonian

$$\mathcal{H}_{\text{eff}}^{\Delta B=1} = \frac{4G_F}{\sqrt{2}} V_{\text{CKM}} \sum_i C_i(\mu_b) O_i(\mu_b) + \text{h.c.}, \qquad (2.13)$$

where $V_{\text{CKM}}$ is the corresponding CKM matrix element. The operators $O_i(\mu_b)$ defined at the scale $\mu_b$ are listed below, and the couplings $C_i(\mu_b)$ are the Wilson coefficients.

The effective operators representing the tree-level $W$ exchange diagram depicted in Fig. 2.4 are

$$O_1^q = \overline{d}^\alpha \gamma_\mu (1-\gamma_5) q^\beta \, \overline{q}^\beta \gamma^\mu (1-\gamma_5) b^\alpha, \quad q = u, c, \qquad (2.14)$$

$$O_2^q = \overline{d}^\alpha \gamma_\mu (1-\gamma_5) q^\alpha \, \overline{q}^\beta \gamma^\mu (1-\gamma_5) b^\beta, \quad q = u, c. \qquad (2.15)$$

The summation over the color indices $\alpha$ and $\beta$ is implied. There is another set of operators obtained by replacing $d$ by $s$.

Many important FCNC decays occur through the so-called penguin diagrams. Some examples are shown in Fig. 2.5. The gluon penguin diagram (Fig. 2.5 (left)) generates the following operators

$$O_3 = \sum_{q=u,d,s,c} \overline{d}^\alpha \gamma_\mu (1-\gamma_5) b^\alpha \, \overline{q}^\beta \gamma^\mu (1-\gamma_5) q^\beta, \qquad (2.16)$$

$$O_4 = \sum_{q=u,d,s,c} \overline{d}^\alpha \gamma_\mu (1-\gamma_5) b^\beta \, \overline{q}^\beta \gamma^\mu (1-\gamma_5) q^\alpha, \qquad (2.17)$$

$$O_5 = \sum_{q=u,d,s,c} \overline{d}^\alpha \gamma_\mu (1-\gamma_5) b^\alpha \, \overline{q}^\beta \gamma^\mu (1+\gamma_5) q^\beta, \qquad (2.18)$$

$$O_6 = \sum_{q=u,d,s,c} \overline{d}^\alpha \gamma_\mu (1-\gamma_5) b^\beta \, \overline{q}^\beta \gamma^\mu (1+\gamma_5) q^\alpha. \qquad (2.19)$$

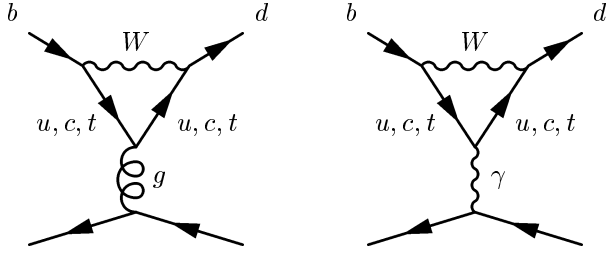

Figure 2.5: Penguin diagrams. The fermion line on the bottom are quarks for the gluon penguin diagram (left), while either quarks or leptons can be involved in the electro-weak penguin (right).

There are also diagrams in which the gluon or photon is not attached to the fermion line and directly appears in the final state. Such diagrams produce

$$O_{7\gamma} = \frac{e}{8\pi^2} m_b \overline{d}\sigma^{\mu\nu}(1+\gamma_5)F_{\mu\nu}b, \tag{2.20}$$

$$O_{8g} = \frac{g}{8\pi^2} m_b \overline{d}\sigma^{\mu\nu}(1+\gamma_5)G^a_{\mu\nu}T^a b, \tag{2.21}$$

where $F_{\mu\nu}$ and $G_{\mu\nu}$ are electromagnetic and QCD field strength tensors, respectively. These operators are responsible for the $b \to d\gamma$ and $b \to dg$ transitions. The operators relevant for the $b \to s\gamma$ and $b \to sg$ transitions are obtained by replacing $d$ by $s$ in (2.20) and (2.21).

The electroweak penguin diagram (Fig. 2.5 (right)) gives a higher order contribution in the electromagnetic coupling constant $\alpha$. However, its contribution is not negligible due to the heavy top quark. It is also important since heavy new particles could enhance its contribution. Moreover, the electroweak penguin process leads to isospin symmetry breaking and in some cases this aspect can be relevant. The corresponding operators are

$$O_7 = \frac{3}{2}\sum_{q=u,d,s,c} e_q \overline{d}^\alpha \gamma_\mu(1-\gamma_5)b^\alpha \overline{q}^\beta \gamma^\mu(1+\gamma_5)q^\beta, \tag{2.22}$$

$$O_8 = \frac{3}{2}\sum_{q=u,d,s,c} e_q \overline{d}^\alpha \gamma_\mu(1-\gamma_5)b^\beta \overline{q}^\beta \gamma^\mu(1+\gamma_5)q^\alpha, \tag{2.23}$$

$$O_9 = \frac{3}{2}\sum_{q=u,d,s,c} e_q \overline{d}^\alpha \gamma_\mu(1-\gamma_5)b^\alpha \overline{q}^\beta \gamma^\mu(1-\gamma_5)q^\beta, \tag{2.24}$$

$$O_{10} = \frac{3}{2}\sum_{q=u,d,s,c} e_q \overline{d}^\alpha \gamma_\mu(1-\gamma_5)b^\beta \overline{q}^\beta \gamma^\mu(1-\gamma_5)q^\alpha. \tag{2.25}$$

$e_q$ is the electromagnetic charge of quarks; $2/3$ for up-type quarks and $-1/3$ for down-type quarks. When the fermion line in the bottom of Fig. 2.5 (right) is a lepton ($e$, $\mu$ or $\tau$), the diagram also generates

$$O_{9V} = \frac{e^2}{16\pi^2}\overline{d}\gamma_\mu(1-\gamma_5)b\,\overline{l}\gamma_\mu l, \tag{2.26}$$

$$O_{10A} = \frac{e^2}{16\pi^2}\overline{d}\gamma_\mu(1-\gamma_5)b\,\overline{l}\gamma_\mu\gamma_5 l. \tag{2.27}$$

These operators give rise to the $b \to d(s)l^+l^-$ transitions.

The Wilson coefficients in Eq.(2.13) are calculated using a perturbation theory to next-to-leading order (NLO) [12–15]. Next-to-next-to-leading order (NNLO) values are also known [16]. The remaining task for the calculation of the mixing and decay processes is to evaluate the matrix elements $\langle f|O_i|i\rangle$ of the operator $O_i$ for a given initial state $|i\rangle$ and final state $|f\rangle$.

## 2.3 Theoretical Methods

In order to extract fundamental parameters, such as the quark masses and CKM matrix elements, from $B$ decay experiments, one needs model independent calculations of the decay amplitudes. However, since $B$ meson decays involve complicated QCD interactions, which are highly non-perturbative in general, the theoretical calculation of physical amplitudes is a non-trivial task. Some of theoretical methods are briefly described in the following.

### 2.3.1 Heavy Quark Symmetry

One useful theoretical method for avoiding hadronic uncertainties is to use symmetries. Using isospin or $SU(3)$ flavor symmetries different decay amplitudes can be related to each other. This approach is widely used in $B$ decay analyses, *e.g.* the isospin analysis of $\pi\pi$ decays to extract $\sin 2\phi_2$ discussed in Section 2.4.4.

Another symmetry which is especially important in $B$ physics is the heavy quark symmetry [17, 18]. In the limit of an infinitely heavy quark mass, the heavy quark behaves as a static color source and the QCD interaction cannot distinguish different flavors, *i.e.* charm or bottom. Consequently the decay amplitudes (or form factors) of $b$ and $c$ hadrons are related to each other (heavy quark flavor symmetry). Moreover, since the spin-dependent interaction decouples in the infinitely heavy quark mass limit, some form factors become redundant (heavy quark spin symmetry). The most famous example are the heavy-to-heavy semi-leptonic decay $\overline{B} \to D^{(*)}\ell\overline{\nu}_\ell$ form factors. In general there are 6 independent form factors for these exclusive decay modes, but in the heavy quark limit they reduce to one; the Isgur-Wise function, and its normalization in the zero-recoil limit is determined.

A more general formalism has also been developed in the language of effective field theory, *i.e.* Heavy Quark Effective Theory (HQET) [19–21]. It provides a systematic expansion in terms of $\Lambda_{\rm QCD}/m_Q$.

### 2.3.2 Heavy Quark Expansion

The inclusive decay rate of a $B$ meson to the final state $X$ can be written as

$$\Gamma(B \to X) = \frac{1}{2m_B} \sum_X (2\pi)^4 \delta^4(p_B - p_X) |\langle X | \mathcal{H}_{eff}^{\Delta B=1} | B \rangle|^2, \tag{2.28}$$

where the sum runs over all possible final states and momentum configurations. The effective Hamiltonian $\mathcal{H}_{eff}^{\Delta B=1}$ (see Sect. 2.2) is proportional to $(\bar{c}\gamma^\mu P_L b)(\bar{l}\gamma_\mu P_L \nu_l)$ when the $b \to c$ semileptonic decay is considered, or to $(\bar{u}\gamma^\mu P_L b)(\bar{l}\gamma_\mu P_L \nu_l)$ if we are interested in using the $b \to u$ semileptonic decay process to determine $|V_{ub}|$. It could also describe non-leptonic decay by considering a four-quark operator $(\bar{c}\gamma^\mu P_L b)(\bar{q}\gamma_\mu P_L q')$. Using the optical theorem, Eq.(2.28) can be rewritten in terms of an absorptive part of a $B$ meson matrix element

$$\Gamma(B \to X) = \frac{1}{m_B} \text{Im} \, \langle B | \mathcal{T} | B \rangle, \tag{2.29}$$

where the operator $\mathcal{T}$ is

$$\mathcal{T} = i \int d^4x T \left( \mathcal{H}_{eff}^{\Delta B=1}(x) \mathcal{H}_{eff}^{\Delta B=1}(0) \right). \tag{2.30}$$

and $T$ denotes the time ordered product. Since the momentum flowing into the final state quark propagator is large ($\sim m_b$), one can expand the time-ordered product of operators in terms of local operators, using the Operator Product Expansion (OPE) technique [8]. It gives an expansion in terms of the inverse heavy quark mass and is thus called the Heavy Quark Expansion (HQE) [22–26].

The lowest dimensional operator is $\bar{b}b$, whose matrix element is unity at LO. The first non-trivial higher order correction appears at $O(1/m_b^2)$ with the chromomagnetic operator $\bar{b}\sigma_{\mu\nu}g_s G^{\mu\nu}b$. Therefore, at $O((\Lambda_{\rm QCD}/m_b)^2)$ the heavy quark expansion can be expressed in terms of two non-perturbative parameters

$$\lambda_1 = \frac{1}{2m_B}\langle B(v)|\bar{h}_v(i\vec{D})^2 h_v|B(v)\rangle, \qquad (2.31)$$

$$3\lambda_2 = \frac{g_s}{2m_B}\langle B(v)|\frac{1}{2}\bar{h}_v\sigma_{\mu\nu}G^{\mu\nu}h_v|B(v)\rangle, \qquad (2.32)$$

where the operators are defined with the HQET field $h_v$ and the $B$ meson state is also defined in the heavy quark limit. $\lambda_2$ is known from the hyper-fine splitting of the $B$ meson ($B$-$B^*$ splitting) to be $\lambda_2 \simeq 0.12$ GeV$^2$, while $\lambda_1$ has to be calculated using non-perturbative methods, such as QCD sum rules [27, 28] or lattice QCD [29–32], or to be fitted with experimental data of inclusive $B$ decays [33–37].

### 2.3.3 Perturbative Methods

Model independent theoretical calculations of exclusive non-leptonic decay amplitudes are known to be very challenging, as they involve both soft and hard gluon exchanges and clear separation of the perturbative (hard) and non-perturbative (soft) parts is intractable. In the heavy quark limit, however, a formulation to realize such separation of short and long distance physics has been developed in the last decade, ameliorating the perturbative calculation of decay amplitudes.

The intuitive idea is the color transparency argument due to Bjorken [38]. An energetic light meson emitted from $B$ decay resembles a color dipole, and its soft interaction with the remaining decay products is suppressed by $\Lambda_{\rm QCD}/m_b$. A systematic formulation of such an idea is provided by QCD factorization [39–43]. In this formalism, the decay matrix elements of $B \to \pi\pi$ can be factorized in the form

$$\langle \pi(p')\pi(q)|Q_i|\overline{B}(p)\rangle = f^{B\to\pi}(q^2)\int_0^1 du\, T_i^I(u)\Phi_\pi(u)$$
$$+ \int_0^1 d\xi du dv\, T_i^{II}(\xi, u, v)\Phi_B(\xi)\Phi_\pi(u)\Phi_\pi(v), \qquad (2.33)$$

for a four-quark operator $Q_i$. The first term represents a factorization of the amplitude into the $B \to \pi$ form factor and an out-going pion wave function $\Phi_\pi(u)$ convoluted with the hard scattering kernel $T_i^I(u)$. Here, the term "factorization" is used for two meanings: one is the factorization of the diagram to $\langle\pi|V|\overline{B}\rangle\langle\pi|A|0\rangle$, while the other is the separation of hard, collinear and soft interactions. The kernel $T_i^I(u)$ describes the hard interaction only and, thus, is calculable using perturbation theory. The second term in Eq.(2.33) describes a factorization of the amplitude into three pieces: $\overline{B} \to 0$, $0 \to \pi$, and $0 \to \pi$ convoluted with a hard interaction kernel $T_i^{II}(\xi, u, v)$. To make the factorization of hard, collinear and soft degrees of freedom more explicit, an effective theory has also been developed, which is called the Soft Collinear Effective Theory (SCET) [44–47]. Using SCET, the factorization can be generalized to all orders of the perturbative expansion. Recently factorization has been shown at NNLO for vertex corrections to the tree amplitudes [48, 49]. For NLO results in spectator scattering including penguins see [50, 51].

The form factor $f^{B\to\pi}(q^2)$ and the light-cone distribution function (or wave function) $\Phi_B(\xi)$ and $\Phi_\pi(u)$ contain long-distance dynamics, which has to be treated with non-perturbative methods.

Another method of factorization, Perturbative QCD (PQCD) [52–58], which is based on the $k_T$ factorization theorem, has also been proposed and used in the analysis of various $B$ decay modes. It relies on the Sudakov suppression, which ensures the absence of end-point singularities. The form factor $f^{B\to\pi}(q^2)$ is then made factorizable, and the entire matrix element can be factorized into convolutions of the hard kernel $H_i$, the jet function $J$, the Sudakov factor $S$, and the wave functions as

$$\langle\pi\pi|Q_i|\overline{B}\rangle = H_i \otimes J \otimes S \otimes \Phi_B \otimes \Phi_\pi \otimes \Phi_\pi, \tag{2.34}$$

where the arguments are suppressed for brevity and $\otimes$ stands for the convolution integrals over momentum fractions and transverse intervals of the constituent quarks inside the hadrons.

### 2.3.4 Lattice QCD

Lattice QCD provides a method to calculate non-perturbative hadronic matrix elements from the first principles of QCD [59]. It is a regularization of QCD on a four-dimensional hypercubic lattice, which enables numerical simulation on the computer. Since the calculation is numerically so demanding, one has to introduce several approximations in the calculation and these lead to systematic uncertainties.

For more than a decade, lattice QCD has been applied to the calculation of matrix elements relevant to $B$ physics. The best-known quantity is the $B$ meson leptonic decay constant $f_B$, for which the systematic uncertainty is now under control at the level of 10% accuracy. The important tools to achieve this goal are the following.

- *Effective theories for heavy quarks.* Since the Compton wave-length of the $b$ quark is shorter than the lattice spacing $a$, the discretization error is out of control with the usual lattice fermion action for relativistic particles. Instead, Heavy Quark Effective Theory (HQET) [19] or Non-relativistic QCD (NRQCD) [60–62] is formulated on the lattice and used to simulate the $b$ quark. Another related effective theory is the so-called Fermilab action [63, 64], which covers the entire (light to heavy) mass regime with the same lattice action. For the $B$ meson, the next-to-leading $(1/m_Q)$ order calculation provides $\lesssim 5\%$ accuracy [65].

- *Effective theories to describe discretization errors.* The discretization effect can be expressed in terms of the Lagrangian language, *i.e.* Symanzik effective theory [66, 67]. It also provides a method to eliminate the error by adding irrelevant operators to the lattice action. The $O(a)$ error existing in the Wilson fermion action can be removed by adding a dimension-five operator [68]. The cancellation of the $O(a)$ error can also be done non-perturbatively [69, 70], so that the remaining discretization error is $O(a^2)$ and not $O(\alpha_s^n a)$.

- *Renormalized perturbation theory.* To relate the lattice operators to their continuum counterparts, one has to rely on perturbation theory. For a long time lattice perturbation had bad convergence behavior and the perturbative error was too large if one calculated only the one-loop terms. This problem was cured by Lepage and Mackenzie by taking a renormalized coupling constant as an expansion parameter [71]. Even better method is the non-perturbative renormalization, which has been developed in some limited cases, *e.g.* the static quark self-energy [72] and the static-light current [73], and will become more important in the future to improve the precision of the lattice calculation.

Lattice QCD calculations can be applied to several other important quantities:

- $B_B$. The $B$ parameter in the $B^0 - \overline{B}^0$ mixing has been calculated in the unquenched QCD.

- *Heavy-to-heavy semileptonic decay.* The zero recoil form factor of the semileptonic decay $B \to D^{(*)} l \nu$ has been calculated rather precisely using a double-ratio technique.

- *Heavy-to-light semileptonic decay.* The $B \to \pi \ell \nu$ form factor has been calculated in the quenched and unquenched QCD.

The relevant references can be found in the recent review [74].

## 2.4 CP violating processes

### 2.4.1 $B - \bar{B}$ mixing and time-dependent CP asymmetry

Neutral $B$ meson mixings are one of the most important FCNC processes in $B$ physics. In the Standard Model, the $B_d - \bar{B}_d$ mixing involves the CKM matrix element $V_{td}$ and thus gives a $CP$-violating amplitude, which induces a variety of $CP$-violating observables through its quantum mechanical interference with other amplitudes. The recently observed $B_s - \bar{B}_s$ mixing is related to $|V_{ts}|$. Using this observation, the ratio $|V_{td}/V_{ts}|$ can be precisely extracted, since the hadronic uncertainty largely cancels between $B_s$ and $B_d$ mesons.

A $B^0$ meson produced as an initial state may evolve into its antiparticle $\overline{B}^0$ through the interaction given by the $\Delta B = 2$ effective Hamiltonian (2.11). In quantum mechanics the state $|B^0(t)\rangle$ at time $t$ is a superposition of two states $|B^0\rangle$ and $|\overline{B}^0\rangle$. The time evolution is described by a Shrödinger equation

$$i\frac{d}{dt}|B(t)\rangle = \left(M - i\frac{\Gamma}{2}\right)|B(t)\rangle, \quad (2.35)$$

where the two-by-two Hermitian matrices $M$ and $\Gamma$ denote mass and decay matrices, respectively. The diagonal parts are constrained from $CPT$ invariance implying $M_{11} = M_{22}$ and $\Gamma_{11} = \Gamma_{22}$, and the off-diagonal parts $M_{12}$ ($M_{21}$) and $\Gamma_{12}/2$ ($\Gamma_{21}/2$) are dispersive and absorptive parts of the $\Delta B = 2$ transition. The eigenstates of the matrix $M - i\Gamma/2$ are given by

$$|B_1\rangle = p|B^0\rangle + q|\overline{B}^0\rangle, \quad (2.36)$$
$$|B_2\rangle = p|B^0\rangle - q|\overline{B}^0\rangle, \quad (2.37)$$

and their coefficients $p$ and $q$ are obtained by solving

$$\frac{q}{p} = +\sqrt{\frac{M_{12}^* - i\Gamma_{12}^*/2}{M_{12} - i\Gamma_{12}/2}} \quad (2.38)$$

together with the normalization condition $|p|^2 + |q|^2 = 1$. The eigenvalues $M_{1,2} - i\Gamma_{1,2}/2$ are related to observables as follows: the $B$ meson mass $M = (M_1 + M_2)/2$, the $B$ meson width $\Gamma = (\Gamma_1 + \Gamma_2)/2$, the $B^0 - \overline{B}^0$ mixing frequency $\Delta M \equiv M_2 - M_1$, and the width difference $\Delta \Gamma \equiv \Gamma_1 - \Gamma_2$.

In the $B$ meson system there is a relation $\Delta\Gamma \ll \Delta M$, which follows from $\Gamma_{12} \ll M_{12}$. We may then approximately obtain

$$\Delta M = -2|M_{12}|\left[1 + O\left(\left|\frac{\Gamma_{12}}{M_{12}}\right|^2\right)\right], \tag{2.39}$$

$$\Delta\Gamma = 2|\Gamma_{12}|\cos\zeta\left[1 + O\left(\left|\frac{\Gamma_{12}}{M_{12}}\right|^2\right)\right], \tag{2.40}$$

and

$$\frac{q}{p} = +\sqrt{\frac{M_{12}^*}{M_{12}}}\left[1 - \frac{1}{2}\left|\frac{\Gamma_{12}}{M_{12}}\right|\sin\zeta + O\left(\left|\frac{\Gamma_{12}}{M_{12}}\right|^2\right)\right]. \tag{2.41}$$

The angle $\zeta$ is the $CP$ violating phase difference between $M_{12}$ and $\Gamma_{12}$

$$\frac{\Gamma_{12}}{M_{12}} = \left|\frac{\Gamma_{12}}{M_{12}}\right|e^{i\zeta}. \tag{2.42}$$

The time evolution of the state $|B^0\rangle$ and $|\overline{B}^0\rangle$ produced at time $t = 0$ is then given by

$$|B^0(t)\rangle = g_+(t)|B^0\rangle + \frac{q}{p}g_-(t)|\overline{B}^0\rangle, \tag{2.43}$$

$$|\overline{B}^0(t)\rangle = g_+(t)|\overline{B}^0\rangle + \frac{p}{q}g_-(t)|B^0\rangle, \tag{2.44}$$

with

$$g_+(t) = e^{-iMt-\Gamma t/2}\left[\cosh\frac{\Delta\Gamma t}{4}\cos\frac{\Delta Mt}{2} - i\sinh\frac{\Delta\Gamma t}{4}\sin\frac{\Delta Mt}{2}\right], \tag{2.45}$$

$$g_-(t) = e^{-iMt-\Gamma t/2}\left[-\sinh\frac{\Delta\Gamma t}{4}\cos\frac{\Delta Mt}{2} + i\cosh\frac{\Delta\Gamma t}{4}\sin\frac{\Delta Mt}{2}\right]. \tag{2.46}$$

As the initial $B^0$ (or $\overline{B}^0$) state evolves, it oscillates between $B^0$ and $\overline{B}^0$ states with the frequency $\Delta M$. The $CP$-violating phase arises in the mixing parameter $q/p$, which carries the phase of $M_{12}$ as shown in (2.41).

Now, let us consider a decay of the $B$ meson to a final state $f$. The decay rate $\Gamma(B^0(t) \to f)$ is time dependent, since the decaying state is a time-dependent superposition of $|B^0\rangle$ and $|\overline{B}^0\rangle$, as discussed in the previous section. We write the decay amplitude of flavor eigenstates as $A_f = \langle f|B^0\rangle$ and $\overline{A}_f = \langle f|\overline{B}^0\rangle$, and define a parameter

$$\lambda_f = \frac{q}{p}\frac{\overline{A}_f}{A_f}. \tag{2.47}$$

If the final state is a $CP$-eigenstate $CP|f\rangle = \xi_f|f\rangle$ with an eigenvalue $\xi_f = \pm 1$, then the time dependent asymmetry

$$a_f(t) = \frac{\Gamma(\overline{B}^0(t) \to f) - \Gamma(B^0(t) \to f)}{\Gamma(\overline{B}^0(t) \to f) + \Gamma(B^0(t) \to f)} \tag{2.48}$$

becomes

$$a_f(t) = \mathcal{A}_f\cos(\Delta Mt) + \mathcal{S}_f\sin(\Delta Mt), \tag{2.49}$$

neglecting the small width difference of the $B$ meson. Here, the direct and indirect (or mixing-induced) $CP$ asymmetries are written as

$$\mathcal{A}_f = \frac{|\lambda_f|^2 - 1}{|\lambda_f|^2 + 1},$$
$$\mathcal{S}_f = \frac{2\,\text{Im}\lambda_f}{1 + |\lambda_f|^2}. \quad (2.50)$$

Since the absolute value of $q/p$ is approximately 1, direct $CP$-violation $|\mathcal{A}_f| \neq 0$ requires $|A_f| \neq |\overline{A}_f|$, which could happen if $A_f$ is a sum of (more than one) decay amplitudes having different $CP$-phases. Indirect $CP$-violation, on the other hand, arises from the quantum mechanical interference between the mixing and decay amplitudes.

### 2.4.2 Measurement of $\sin 2\phi_1$

The mixing-induced asymmetry provides a variety of methods to measure the angles of the Unitarity Triangle (Fig. 2.1). It was first proposed by Bigi, Carter, and Sanda in 1980–1981 [75–77], and gave strong motivation to construct the present KEK B Factory. The best known example is the case where the final state is $J/\psi K_s$, whose quark level process is $\bar{b} \to \bar{c}c\bar{s}$ followed by the $K^0 - \overline{K}^0$ mixing. In the case where the decay is dominated by a single amplitude, the ratio of decay amplitudes is given by

$$\frac{\overline{A}_{J/\psi K_s}}{A_{J/\psi K_s}} = -\left(\frac{V_{cb}V_{cs}^*}{V_{cb}^*V_{cs}}\right)\left(\frac{V_{cs}V_{cd}^*}{V_{cs}^*V_{cd}}\right), \quad (2.51)$$

where the minus sign appears due to the $CP$ odd final state $J/\psi K_s$. Together with the phase in the mixing

$$\frac{q}{p} \simeq \frac{V_{tb}^*V_{td}}{V_{tb}V_{td}^*}, \quad (2.52)$$

the entire ratio $\lambda_{J/\psi K_s}$ becomes

$$\lambda_{J/\psi K_s} = -\left(\frac{V_{tb}^*V_{td}}{V_{tb}V_{td}^*}\right)\left(\frac{V_{cb}V_{cs}^*}{V_{cb}^*V_{cs}}\right)\left(\frac{V_{cs}V_{cd}^*}{V_{cs}^*V_{cd}}\right) = -e^{-2i\phi_1}. \quad (2.53)$$

Thus, one can precisely measure the angle $\phi_1$ from the time-dependent asymmetry

$$a_{J/\psi K_s}(t) = \sin(2\phi_1)\sin\Delta Mt. \quad (2.54)$$

There exists an additional decay amplitude through the penguin diagram $\bar{b} \to \bar{s}c\bar{c}$, which involves the CKM factor $V_{ts}V_{tb}^*$. Using the unitarity of the CKM matrix $V_{ts}V_{tb}^* = -V_{cs}V_{cb}^* - V_{us}V_{ub}^*$, the weak phase of the penguin contribution is the same as that of the tree amplitude $V_{cs}V_{cb}^*$ up to a doubly Cabibbo-suppressed correction. Therefore, the relation (2.54) holds to an excellent approximation ($\sim 1\%$), and the mode $J/\psi K_s$ is called the "gold-plated" mode (for a review of corrections to the relation see [78]). Other $\bar{b} \to \bar{s}c\bar{c}$ decay mode, such as $B^0 \to J/\psi K_L$, $B^0 \to \psi(2S)K_S$, *etc.*, can also be used to measure $\sin 2\phi_1$. The precision expected at SuperKEKB is discussed in Section 5.6.

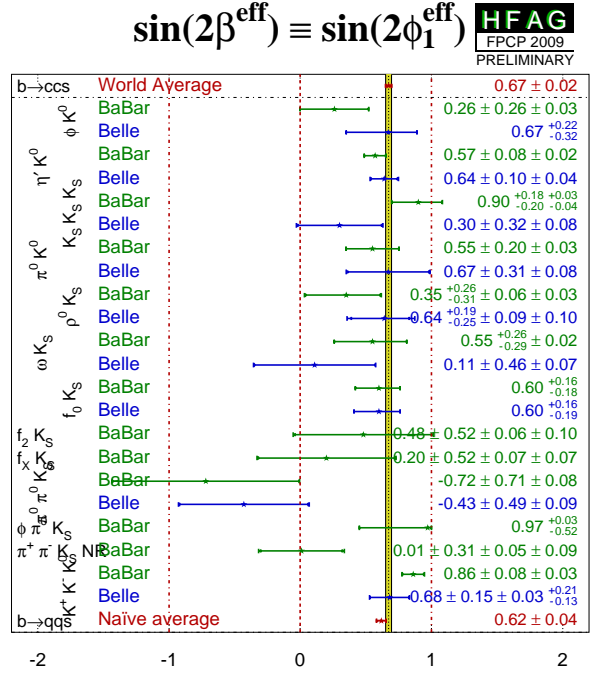

Figure 2.6: $\sin 2\phi_1$ as measured through the $\overline{b} \to \overline{s}q\overline{q}$ processes [80].

### 2.4.3 Other decay modes leading to $\sin 2\phi_1$

There are other decay modes which develop the same weak phase. Namely, the penguin decays $\overline{b} \to \overline{s}s\overline{s}$ are accompanied by the CKM factor $V_{ts}V_{tb}^*$ just as the penguin contribution to $\overline{b} \to \overline{s}c\overline{c}$. Since there is no direct tree diagram for $\overline{b} \to \overline{s}s\overline{s}$, measurement of the same angle $\sin 2\phi_1$ through its time dependent asymmetry can be a probe of any new physics phase in the penguin loop process [79]. At the hadron level, the corresponding modes are $B^0 \to \phi K_s$ and $B^0 \to \eta' K_s$, $B^0 \to K_S K_S K_S$, $B^0 \to \pi^0 K_S$, $B^0 \to \rho^0 K_S$, $B^0 \to \omega K_S$, etc. These asymmetries have already been measured as summarized in Figure 2.6 [80], whose difference from the $\sin 2\phi_1$ from $\overline{b} \to \overline{s}c\overline{c}$ may suggest a hint of new physics contribution in the penguin diagram. The expected sensitivity at SuperKEKB is studied in Section 5.2. Possible contaminations from the tree-level process $\overline{b} \to \overline{u}u\overline{s}$ with a rescattering of $u\overline{u}$ to $s\overline{s}$ may distort the measurement. However, these contributions are expected to be small ($O(\lambda^2) \sim 5\%$) [81], and a model independent bound can be obtained using $SU(3)$ relations provided that the related modes are observed at higher luminosity $B$ factories [82, 83]. A theoretically clean measurement of weak phase in the penguin modes can be obtained from three body $B$ decays [84, 85].

### 2.4.4 Measurement of $\phi_2$

If a decay can occur through more than one amplitude with different weak phases, the analysis is more involved. As an example, we consider $B^0 \to \pi^+\pi^-$, whose decay amplitude can be parametrized as

$$A(B^0 \to \pi^+\pi^-) = T_{\pi\pi} + P_{\pi\pi}. \tag{2.55}$$

The first term represents an amplitude for the tree level $W$ exchange process $\overline{b} \to \overline{u}u\overline{d}$, which picks up the CKM matrix elements $V_{ud}V_{ub}^*$, while the second term is a penguin diagram contribution $\overline{b} \to \overline{d}u\overline{u}$ containing the CKM factor $V_{td}V_{tb}^*$. If the penguin contribution can be neglected, the ratio $\lambda_{\pi^+\pi^-}$ in (2.47) reads as

$$\lambda_{\pi^+\pi^-} = \left(\frac{V_{tb}^*V_{td}}{V_{tb}V_{td}^*}\right)\left(\frac{V_{ud}^*V_{ub}}{V_{ud}V_{ub}^*}\right) = e^{2i\phi_2}, \tag{2.56}$$

and the time-dependent asymmetry could be used to determine the angle $\phi_2$. However, both amplitudes in (2.55) are of the same order in $\lambda$ ($\sim \lambda^3$), and the penguin contribution is not so suppressed compared to the tree level contribution; the ratio of amplitudes $|P_{\pi\pi}/T_{\pi\pi}|$ is roughly estimated to be around 0.3 from $B \to K\pi$ decays assuming the flavor $SU(3)$ symmetry.

One solution to the problem is to consider the isospin symmetry [86]. The $B \to \pi\pi$ decay amplitudes are written as

$$\begin{aligned}
A(B^0 \to \pi^+\pi^-) &= \sqrt{2}(A_2 - A_0), \\
A(B^0 \to \pi^0\pi^0) &= 2A_2 + A_0, \\
A(B^+ \to \pi^+\pi^0) &= 3A_2.
\end{aligned} \tag{2.57}$$

$A_0$ and $A_2$ are amplitudes for isospin 0 and 2 of two-pion final state, respectively. The tree diagram contributes to both $A_0$ and $A_2$, while the penguin diagram produces only isospin 0. Then, one obtains a relation

$$A(B^0 \to \pi^+\pi^-) + \sqrt{2}A(B^0 \to \pi^0\pi^0) = \sqrt{2}A(B^+ \to \pi^+\pi^0). \tag{2.58}$$

and its $CP$ conjugate

$$\overline{A}(B^0 \to \pi^+\pi^-) + \sqrt{2}\overline{A}(\overline{B}^0 \to \pi^0\pi^0) = \sqrt{2}\overline{A}(B^- \to \pi^-\pi^0). \tag{2.59}$$

By measuring the branching ratios of the three decay modes and the time dependent asymmetry of $\pi^+\pi^-$, one can determine the absolute values $|A_0|$ and $|A_2|$ and their relative phase difference $\arg(A_0 A_2^*)$, through a simple geometric reconstruction. This determines the angle $\phi_2$ up to a four-fold ambiguity. The details of the analysis will be discussed in Section 5.7.1.

Another solution is to consider the isospin relations among $B \to \rho\pi$ decays [87]. There are three possible decay chains $B^0 \to \{\rho^+\pi^-, \rho^0\pi^0, \rho^-\pi^+\} \to \pi^+\pi^-\pi^0$. Together with their $CP$-conjugate amplitudes, there exist six different amplitudes, each of which has contributions from both tree and penguin diagrams. By combining the time-dependent asymmetry of this process in the Dalitz plot one may extract the pure tree amplitude, and thus the angle $\phi_2$. For the sensitivity study at Super-KEKB, see Section 5.7.2.

The $B^0 \to \rho\rho$ channel can also be used to measure $\sin 2\phi_2$, provided that the final state $\rho\rho$ is almost purely longitudinally polarized and thus a pure $CP$-even state [88,89]. The sensitivity study is presented in Section 5.7.3.

If there is non-negligible contribution from the electroweak penguin diagram, the isospin relations are violated, since up and down quarks have different charges [90]. However, such a contribution is suppressed compared to the gluon penguin by a factor $\alpha_{\text{weak}}(m_t^2/m_Z^2)/\alpha_s \ln(m_t^2/m_c^2) \sim 0.1$ [91]. At SuperKEKB, the actual size of the electroweak penguin amplitude can be estimated from the analysis of $K\pi$ decays [91–95].

The prospects of measuring the angle $\phi_2$ at SuperKEKB using these various methods are discussed in Section 5.7.

### 2.4.5 Other indirect CP asymmetries

There are other opportunities to probe the CP asymmetry through the interference with the neutral $B$ meson mixing amplitude. An interesting example is the $B^0 \to K^{*0}\gamma$ decay, since it is sensitive to new physics contributions. In the Standard Model, this process occurs through the penguin diagram carrying the CKM factor $V_{ts}V_{tb}^*$, and thus leads to the measurement of $\sin 2\phi_1$ as the $\bar{b} \to \bar{s}s\bar{s}$ modes. Since the interference requires a helicity flip of the final state photon, the asymmetry is expected to be small (of order of $m_s/m_b$) in the Standard Model. (There is an argument that the non-perturbative contribution may give a larger contribution of order $\Lambda_{\text{QCD}}/m_b$, though [96].) Therefore, if a large asymmetry is observed in this process, it implies the effect of new physics. So far, the signal is consistent with zero within the error bar [97, 98]. The sensitivity study for this and other related modes is given in Section 5.3.

### 2.4.6 Direct CP asymmetry

The CP asymmetry could occur even without the interference with the mixing amplitude if there are more than one decay amplitudes that interfere among themselves. Such direct CP asymmetry has been observed in $B \to \pi\pi$ and $B \to K\pi$ decays, for instance. In order to extract the fundamental parameters of the underlying theory, one has to theoretically calculate several different amplitudes (tree, penguin, electroweak penguin, *etc.*), including their phase shift or to fit them to the experimental data. Having precise experimental data for various decay modes, it is possible to extract some information on new physics contributions. For some recent analysis, see, for example [93, 94, 99, 100].

### 2.4.7 CP phases through $D$ meson intermediate states

The angle $\phi_3$ can be extracted by measuring the interference among different decay amplitudes in $B \to D\pi$ and $DK$ decays.

Let us consider, for example, the interference between the decay modes $B^- \to D^0 K^-$ and $B^- \to \bar{D}^0 K^-$ [101–103]. The $D^0 K^-$ mode involves a quark level process $b \to c\bar{u}s$, which is of a tree topology, while $\bar{D}^0 K^-$ occurs through another tree-type process $b \to u\bar{c}s$. Since the CKM factor, especially its phase, is different among these two processes ($V_{cs}V_{us}^*$ and $V_{ub}V_{cs}^*$), the CP asymmetry may be observed if these two processes interfere. The angle $\phi_3$ can then be extracted. It could actually occur in the final state common to these two processes, such as $B^- \to D^0 K^-$ followed by $D^0 \to K^+\pi^-$ and $B^- \to \bar{D}^0 K^-$ followed by $\bar{D}^0 \to K^+\pi^-$ [104]. A detailed analysis of this and other methods to measure the angle $\phi_3$ is discussed in Section 5.8.1.

## 2.5 Other processes of interest

At the $B$ factory experiment, there are many decay modes that can be used as a probe of new physics, since the Standard Model contribution is suppressed due to the GIM mechanism. There are other interesting processes even at the tree level, as they are sensitive to the small CKM matrix elements ($V_{td}$ and $V_{ub}$) or to new physics contribution appearing at the tree level.

### 2.5.1 $b \to s\gamma$ and related processes

Radiative decays $B \to X_s\gamma$ and $X_d\gamma$ are induced by the penguin diagram with a dominant contribution from the top quark loop. The inclusive decay rate $B \to X_s\gamma$ has been measured precisely (to the level of 10%) [105, 106]. Theoretically, this rate has been calculated within the Standard Model to the $O(\alpha_s^2)$ level [107], with which the remaining uncertainty is about 7%. From the experimental data, even the photon energy spectrum is measured, which provides important information to extract the $B$ meson shape function.

The exclusive rate for the $K^*\gamma$ mode has been measured more precisely, but its theoretical calculation is much harder as it requires non-perturbative methods. Still, this mode plays an important role to probe the CP violation with an interference with the $B$ meson mixing.

The $B \to \rho\gamma$ contains a quark level process $b \to d\gamma$, which involves the CKM element $V_{td}$ in its dominant top quark loop. By considering a ratio $B(B \to \rho\gamma)/B(B \to K^*\gamma)$, the bulk of the hadronic uncertainty cancels up to the SU(3) breaking effect and the ratio $|V_{td}/V_{ts}|$ can be extracted. The $B \to \rho\gamma$ decay has already been measured experimentally [108, 109], and the sensitivity at the SuperKEKB is discussed in Sec. 5.3.3.

A related mode is $B \to K^{(*)}\ell^+\ell^-$. This process may contain different loop diagrams and thus may have different sensitivity to the new physics. Furthermore, the angular distribution of the final state lepton pair provides more detailed information on the structure of the low-energy effective Hamiltonian [110, 111]. In particular, the forward-backward asymmetry is sensitive to a certain class of new physics contribution [112, 113]. Although the decay rate is much lower than $b \to s\gamma$, the exclusive modes and their asymmetry have already been measured [114, 115]. The expected sensitivity at the SuperKEKB is discussed in Sec. 5.3.7.

With neutrinos in the final state, i.e. $B \to K^{(*)}\nu\bar{\nu}$, the possible new physics contribution may enter with a different pattern. Experimental measurement of this decay mode is much more difficult, since there are two neutrinos in the final state. So far, only the experimental upper limit has been obtained [116].

### 2.5.2 Leptonic decays

The purely leptonic decays $B^+ \to e^+\nu_e$ and $B^+ \to \mu^+\nu_\mu$ are highly suppressed in the Standard Model due to the wrong helicity leptons in the final state. Presently, only the tauonic mode $B^+ \to \tau^+\nu_\tau$ is measurable, though the measurement involves difficult experimental technique due to the two missing neutrinos [117, 118]. The leptonic decay sets a constraint on the combination $f_B^2|V_{ub}|^2$. Therefore, it serves as an experimental test of the lattice calculation, for which the decay constant $f_B$ is one of the easiest quantity to calculate, once the CKM element $|V_{ub}|$ can be determined separately.

In the two Higgs doublet model, or in general in the supersymmetric models, the leptonic decay contains a contribution from the tree-level charged Higgs exchange. It is enhanced when $\tan\beta$ is large as $\sim [1 - \tan^2\beta(M_B^2/M_{H^\pm}^2]$ [119, 120] with $\tan\beta$ the ratio of two Higgs expectation values and $M_{H^\pm}$ the charged Higgs mass. Therefore, one can obtain a bound on the charged Higgs mass from this decay mode.

### 2.5.3 $B \to D^{(*)}\tau\bar{\nu}$

The semi-leptonic decay $B \to D^{(*)}\tau\bar{\nu}$ also has a good sensitivity to the charged Higgs exchange at the tree level [121–124]. It may therefore set a constraint on the charged Higgs mass, which is a new physics effect. Although the branching ratio is large for this semi-leptonic decay, the final state contains more than one missing neutrino and the experimental measurement is challenging. This decay mode has been observed recently [125], and will become one of the most interesting modes at the SuperKEKB, as discussed in Sec. 5.5.

### 2.5.4 $B \to X_u \ell \bar\nu$

The CKM element $|V_{ub}|$ can be determined through the quark level decay $b \to u\ell\bar\nu$. The corresponding exclusive decay modes $B \to \pi\ell\bar\nu$, $\rho\ell\bar\nu$ have be measured using different tagging techniques. The branching fraction of the $B \to \pi\ell\bar\nu$ mode is now measured to a 10% level [126–129], and the momentum transfer ($q^2$) dependence is also obtained. Model independent theoretical calculation relies on lattice QCD and is thus limited to the high $q^2$ region. The error in the lattice calculation is still rather large ($\sim 20$–$30\%$) [130, 131], but improvements are expected in the next several years.

Inclusive $B \to X_u \ell\bar\nu$ decay is used after cutting out the kinematical regime of much larger $B \to X_c \ell\bar\nu$. Such reduction of the phase space introduces complications in the theoretical descritption as it is not completely inclusive. There are several strategies for kinematical cuts that minimize the theoretical uncertainty while enhancing the experimental signal (see, for example, [132–134]).

A detailed discussion of the $|V_{ub}|$ determination is given in Sec. 5.9.

# References

[1] N. Cabibbo, "Unitary Symmetry And Leptonic Decays," Phys. Rev. Lett. **10**, 531 (1963).

[2] M. Kobayashi and T. Maskawa, "CP Violation In The Renormalizable Theory Of Weak Interaction," Prog. Theor. Phys. **49**, 652 (1973).

[3] S. L. Glashow, J. Iliopoulos and L. Maiani, "Weak Interactions With Lepton – Hadron Symmetry," Phys. Rev. D **2**, 1285 (1970).

[4] K. Hagiwara *et al.* [Particle Data Group Collaboration], "Review Of Particle Physics," Phys. Rev. D **66**, 010001 (2002).

[5] L. Wolfenstein, "Parametrization Of The Kobayashi-Maskawa Matrix," Phys. Rev. Lett. **51**, 1945 (1983).

[6] C. Amsler *et al.* [Particle Data Group], "Review of particle physics," Phys. Lett. B **667**, 1 (2008).

[7] J. Charles *et al.* [CKMfitter Group], "CP violation and the CKM matrix: Assessing the impact of the asymmetric B factories" Eur. Phys. J. C **41**, 1 (2005); updates at http://ckmfitter.in2p3.fr.

[8] K. G. Wilson, "Nonlagrangian Models Of Current Algebra," Phys. Rev. **179**, 1499 (1969).

[9] A. Buras, M. Jamin, P. Weisz, "Leading and Next/to/leading QCD Corrections to $\epsilon$ parameter and $B^0 - \bar{B}^0$ Mixing in the Presence of a Heavy Top Quark," Nucl. Phys. B**347**, 491 (1990).

[10] G. Buchalla, A. Buras, M. Lautenbacher, "Weak Decays Beyond Leading Logarithms," Rev. Mod. Phys. **68**, 1125 (1996).

[11] T. Inami and C. S. Lim, "Effects Of Superheavy Quarks And Leptons In Low-Energy Weak Processes $K_L \to \mu\bar{\mu}$, $K^+ \to \pi^+\nu\bar{\nu}$ and $K^0 \leftrightarrow \bar{K}^0$," Prog. Theor. Phys. **65**, 297 (1981).

[12] G. Altarelli, G. Curci, G. Martinelli and S. Petrarca, "QCD Nonleading Corrections To Weak Decays As An Application Of Regularization By Dimensional Reduction," Nucl. Phys. B **187**, 461 (1981).

[13] A. J. Buras and P. H. Weisz, "QCD Nonleading Corrections To Weak Decays In Dimensional Regularization and 't Hooft-Veltman Schemes," Nucl. Phys. B **333**, 66 (1990).

[14] A. J. Buras, M. Jamin, M. E. Lautenbacher and P. H. Weisz, "Effective Hamiltonians for $\Delta S = 1$ and $\Delta B = 1$ nonleptonic decays beyond the leading logarithmic approximation," Nucl. Phys. B **370**, 69 (1992).

[15] A. J. Buras, M. Jamin, M. E. Lautenbacher and P. H. Weisz, "Two loop anomalous dimension matrix for $\Delta S = 1$ weak nonleptonic decays. 1. $O(\alpha_s^2)$," Nucl. Phys. B **400**, 37 (1993).

[16] See p. 8 of T. Browder *et al.*, "New Physics at Super Flavor Factory", arXiv:0802.3201 [hep-ph], and references therein.

[17] N. Isgur and M. B. Wise, "Weak Decays Of Heavy Mesons In The Static Quark Approximation," Phys. Lett. B **232**, 113 (1989).

[18] N. Isgur and M. B. Wise, "Weak Transition Form-Factors Between Heavy Mesons," Phys. Lett. B **237**, 527 (1990).

[19] E. Eichten and B. Hill, "An Effective Field Theory For The Calculation Of Matrix Elements Involving Heavy Quarks," Phys. Lett. B **234**, 511 (1990).

[20] H. Georgi, "An Effective Field Theory For Heavy Quarks At Low-Energies," Phys. Lett. B **240**, 447 (1990).

[21] B. Grinstein, "The Static Quark Effective Theory," Nucl. Phys. B **339**, 253 (1990).

[22] J. Chay, H. Georgi and B. Grinstein, "Lepton Energy Distributions In Heavy Meson Decays From QCD," Phys. Lett. B **247**, 399 (1990).

[23] I. I. Bigi, N. G. Uraltsev and A. I. Vainshtein, "Nonperturbative corrections to inclusive beauty and charm decays: QCD versus phenomenological models," Phys. Lett. B **293**, 430 (1992).

[24] I. I. Bigi, M. A. Shifman, N. G. Uraltsev and A. I. Vainshtein, "QCD predictions for lepton spectra in inclusive heavy flavor decays," Phys. Rev. Lett. **71**, 496 (1993).

[25] A. V. Manohar and M. B. Wise, "Inclusive semileptonic B and polarized Lambda(b) decays from QCD," Phys. Rev. D **49**, 1310 (1994).

[26] B. Blok, L. Koyrakh, M. A. Shifman and A. I. Vainshtein, "Differential distributions in semileptonic decays of the heavy flavors in QCD," Phys. Rev. D **49**, 3356 (1994).

[27] P. Ball and V. M. Braun, "Next-to-leading order corrections to meson masses in the heavy quark effective theory," Phys. Rev. D **49**, 2472 (1994).

[28] M. Neubert, "QCD sum-rule calculation of the kinetic energy and chromo-interaction of heavy quarks inside mesons," Phys. Lett. B **389**, 727 (1996).

[29] M. Crisafulli, V. Gimenez, G. Martinelli and C. T. Sachrajda, "First lattice calculation of the $B$ meson binding and kinetic energies," Nucl. Phys. B **457**, 594 (1995).

[30] V. Gimenez, G. Martinelli and C. T. Sachrajda, "A high-statistics lattice calculation of $\lambda_1$ and $\lambda_2$ in the $B$ meson," Nucl. Phys. B **486**, 227 (1997).

[31] A. S. Kronfeld and J. N. Simone, "Computation of $\bar{\Lambda}$ and $\lambda_1$ with lattice QCD," Phys. Lett. B **490**, 228 (2000).

[32] S. Aoki *et al.* [JLQCD Collaboration], "Heavy quark expansion parameters from lattice NRQCD," Phys. Rev. D **69**, 094512 (2004).

[33] R. D. Dikeman, M. A. Shifman and N. G. Uraltsev, "$b \to s+\gamma$: A QCD consistent analysis of the photon energy distribution," Int. J. Mod. Phys. A **11**, 571 (1996).

[34] A. Kapustin and Z. Ligeti, "Moments of the photon spectrum in the inclusive $B \to X_s\gamma$ decay," Phys. Lett. B **355**, 318 (1995).

[35] A. F. Falk, M. E. Luke and M. J. Savage, "Hadron spectra for semileptonic heavy quark decay," Phys. Rev. D **53**, 2491 (1996).

[36] A. F. Falk, M. E. Luke and M. J. Savage, "Phenomenology of the $1/m_Q$ Expansion in Inclusive $B$ and $D$ Meson Decays," Phys. Rev. D **53**, 6316 (1996).

[37] M. Gremm, A. Kapustin, Z. Ligeti and M. B. Wise, "Implications of the $B \to X\ell\bar{\nu}_\ell$ lepton spectrum for heavy quark theory," Phys. Rev. Lett. **77**, 20 (1996).

[38] J. D. Bjorken, "Topics In B Physics," Nucl. Phys. Proc. Suppl. **11**, 325 (1989).

[39] M. Beneke, G. Buchalla, M. Neubert and C. T. Sachrajda, "QCD factorization for $B \to \pi\pi$ decays: Strong phases and CP violation in the heavy quark limit," Phys. Rev. Lett. **83**, 1914 (1999).

[40] M. Beneke, G. Buchalla, M. Neubert and C. T. Sachrajda, "QCD factorization for exclusive, non-leptonic $B$ meson decays: General arguments and the case of heavy-light final states," Nucl. Phys. B **591**, 313 (2000).

[41] M. Beneke, G. Buchalla, M. Neubert and C. T. Sachrajda, "QCD factorization in $B \to \pi K$, $\pi\pi$ decays and extraction of Wolfenstein parameters," Nucl. Phys. B **606**, 245 (2001).

[42] M. Beneke, M. Neubert, "Flavor Singlet $B$ Decay Amplitudes in QCD Factorization," Nucl. Phys. B**651**, 225 (2003).

[43] M. Beneke, M. Neubert, "QCD Factorization for $B \to PP$ and $B \to PV$ Decays," Nucl. Phys. B**675**, 333 (2003).

[44] C. W. Bauer, S. Fleming and M. E. Luke, "Summing Sudakov logarithms in B → X/s gamma in effective field theory," Phys. Rev. D **63**, 014006 (2001).

[45] C. W. Bauer, S. Fleming, D. Pirjol and I. W. Stewart, "An effective field theory for collinear and soft gluons: Heavy to light decays," Phys. Rev. D **63**, 114020 (2001).

[46] C. W. Bauer and I. W. Stewart, "Invariant operators in collinear effective theory," Phys. Lett. B **516**, 134 (2001).

[47] C. W. Bauer, D. Pirjol and I. W. Stewart, "Soft-collinear factorization in effective field theory," Phys. Rev. D **65**, 054022 (2002).

[48] G. Bell, "NNLO Vertex Corrections in Charmless Hadronic $B$ Decays: Real Part," Nucl. Phys. B**822**, 172 (2009).

[49] M. Beneke, T. Huber, Xin-Qiang Li, "NNLO Vertex Corrections to Non-leptonic $B$ Decays," arXiv:0911.3655 [hep-ph].

[50] M. Beneke, S. Jager, "Spectator Scattering at NLO in Non-leptonic $B$ Decays: Tree Amplitudes," Nucl. Phys. B**751**, 160 (2006).

[51] M. Beneke, S. Jager, "Spectator Scattering at NLO in Non-leptonic $B$ Decays: Leading Penguin Amplitudes," Nucl. Phys. B**768**, 51 (2007).

[52] H-n. Li and H. L. Yu, "Extraction of $V_{ub}$ from decay $B \to \pi l \nu$," Phys. Rev. Lett. **74**, 4388 (1995).

[53] H-n. Li and H. L. Yu, "PQCD Analysis Of Exclusive Charmless B Meson Decay Spectra," Phys. Lett. B **353**, 301 (1995).

[54] H-n. Li and H. L. Yu, "Perturbative QCD analysis of B meson decays," Phys. Rev. D **53**, 2480 (1996).

[55] Y. Y. Keum, H-n. Li and A. I. Sanda, "Fat penguins and imaginary penguins in perturbative QCD," Phys. Lett. B **504**, 6 (2001).

[56] Y. Y. Keum, H-n. Li and A. I. Sanda, "Penguin enhancement and $B \to K\pi$ decays in perturbative QCD," Phys. Rev. D **63**, 054008 (2001).

[57] Y. Y. Keum and H-n. Li, "Nonleptonic charmless B decays: Factorization vs. perturbative QCD," Phys. Rev. D **63**, 074006 (2001).

[58] C. D. Lu, K. Ukai and M. Z. Yang, "Branching ratio and CP violation of $B \to \pi\pi$ decays in perturbative QCD approach," Phys. Rev. D **63**, 074009 (2001).

[59] K. G. Wilson, "Confinement Of Quarks," Phys. Rev. D **10**, 2445 (1974).

[60] W. E. Caswell and G. P. Lepage, "Effective Lagrangians For Bound State Problems In QED, QCD, And Other Field Theories," Phys. Lett. B **167**, 437 (1986).

[61] B. A. Thacker and G. P. Lepage, "Heavy Quark Bound States In Lattice QCD," Phys. Rev. D **43**, 196 (1991).

[62] G. P. Lepage, L. Magnea, C. Nakhleh, U. Magnea and K. Hornbostel, "Improved nonrelativistic QCD for heavy quark physics," Phys. Rev. D **46**, 4052 (1992).

[63] A. X. El-Khadra, A. S. Kronfeld and P. B. Mackenzie, "Massive Fermions in Lattice Gauge Theory," Phys. Rev. D **55**, 3933 (1997).

[64] A. S. Kronfeld, "Application of heavy-quark effective theory to lattice QCD. I: Power corrections," Phys. Rev. D **62**, 014505 (2000).

[65] K. I. Ishikawa, H. Matsufuru, T. Onogi, N. Yamada and S. Hashimoto, "$f_B$ with lattice NRQCD including $O(1/m_Q^2)$ corrections," Phys. Rev. D **56**, 7028 (1997).

[66] K. Symanzik, "Continuum Limit And Improved Action In Lattice Theories. 1. Principles And $\Phi^4$ Theory," Nucl. Phys. B **226**, 187 (1983).

[67] K. Symanzik, "Continuum Limit And Improved Action In Lattice Theories. 2. $O(N)$ Nonlinear Sigma Model In Perturbation Theory," Nucl. Phys. B **226**, 205 (1983).

[68] B. Sheikholeslami and R. Wohlert, "Improved Continuum Limit Lattice Action For QCD With Wilson Fermions," Nucl. Phys. B **259**, 572 (1985).

[69] K. Jansen *et al.*, "Non-perturbative renormalization of lattice QCD at all scales," Phys. Lett. B **372**, 275 (1996).

[70] M. Luscher, S. Sint, R. Sommer, P. Weisz and U. Wolff, "Non-perturbative $O(a)$ improvement of lattice QCD," Nucl. Phys. B **491**, 323 (1997).

[71] G. P. Lepage and P. B. Mackenzie, "On the viability of lattice perturbation theory," Phys. Rev. D **48**, 2250 (1993).

[72] M. Della Morte, N. Garron, M. Papinutto and R. Sommer, "Heavy quark effective theory computation of the mass of the bottom quark," JHEP **0701**, 007 (2007).

[73] M. Della Morte, P. Fritzsch and J. Heitger, "Non-perturbative renormalization of the static axial current in two-flavour QCD," JHEP **0702**, 079 (2007).

[74] V. Lubicz, C. Tarantino, "Flavour Physics and Lattice QCD: Averages of Lattice Inputs for the Unitarity Triangle Analysis," Nuovo Cim. **123B**, 674 (2008).

[75] A. B. Carter and A. I. Sanda, "CP Violation In Cascade Decays Of B Mesons," Phys. Rev. Lett. **45**, 952 (1980).

[76] A. B. Carter and A. I. Sanda, "CP Violation In B Meson Decays," Phys. Rev. D **23**, 1567 (1981).

[77] I. I. Y. Bigi and A. I. Sanda, "Notes On The Observability Of CP Violations In B Decays," Nucl. Phys. B **193**, 85 (1981).

[78] T. Browder *et al.*, "New Physics at Super Flavor Factory", arXiv:0802.3201 [hep-ph].

[79] D. London and A. Soni, "Measuring the CP angle beta in hadronic $b \to s$ penguin decays," Phys. Lett. B **407**, 61 (1997).

[80] E. Barberio *et al.* [HFAG Group], "Averages of $b-$hadron and $c-$hadron Properties at the End of 2007," arXiv:0808.1297 [hep-ex]; updates at http://www.slac.stanford.edu/xorg/hfag/charm/index.html.

[81] Y. Grossman, G. Isidori and M. P. Worah, "CP asymmetry in $B_d \to \phi K_S$: Standard model pollution," Phys. Rev. D **58**, 057504 (1998).

[82] Y. Grossman, Z. Ligeti, Y. Nir and H. Quinn, "SU(3) relations and the CP asymmetries in B decays to $\eta' K_S$, $\phi K_S$ and $K^+ K^- K_S$," Phys. Rev. D **68**, 015004 (2003).

[83] M. Gronau, J.L. Rosner, J. Zupan, "Updated Bounds on $CP$ Asymmetries in $B^0 \to \eta' K_S$ and $B^0 \to \pi^0 K_S$," Phys. Rev. D**74**, 093003 (2006).

[84] M. Chiuchini, M. Pierini, L. Silvestrini, "New Bounds on the CKM Matrix from $B \to K\pi\pi$ Dalitz Plot Analysis," Phys. Rev. D**74**, 051301 (2006).

[85] M. Gronau, D. Pirjol, A. Soni, J. Zupan, "Improved Method for CKM Constraints in Charmless Three-body B and $B_s$ Deacys," Phys. Rev. D**75**, 014002 (2007).

[86] M. Gronau and D. London, "Isospin Analysis Of CP Asymmetries In B Decays," Phys. Rev. Lett. **65**, 3381 (1990).

[87] A. E. Snyder and H. R. Quinn, "Measuring CP asymmetry in $B \to \rho\pi$ decays without ambiguities," Phys. Rev. D **48**, 2139 (1993).

[88] A. Somov *et al.*, "Measurement of the branching fraction, polarization, and CP asymmetry for $B^0 \to \rho^+\rho^-$ decays, and determination of the CKM phase $\phi_2$," Phys. Rev. Lett. **96**, 171801 (2006).

[89] K. Abe *et al.* [Belle Collaboration], "Improved measurement of CP-violating parameters in $\rho^+\rho^-$ decays," Phys. Rev. D **76**, 011104 (2007).

[90] N. G. Deshpande and X. G. He, "Isospin structure of penguins and their consequences in B physics," Phys. Rev. Lett. **74**, 26 (1995).

[91] M. Gronau, O. F. Hernandez, D. London and J. L. Rosner, "Electroweak penguins and two-body B decays," Phys. Rev. D **52**, 6374 (1995).

[92] M. Neubert, "Model-independent analysis of $B \to \pi K$ decays and bounds on the weak phase gamma," JHEP **9902**, 014 (1999).

[93] T. Yoshikawa, "A possibility of large electro-weak penguin contribution in $B \to K\pi$ modes," Phys. Rev. D **68**, 054023 (2003).

[94] M. Gronau and J. L. Rosner, "Rates and asymmetries in $B \to K\pi$ decays," Phys. Lett. B **572**, 43 (2003).

[95] A. J. Buras, R. Fleischer, S. Recksiegel and F. Schwab, "The $B \to \pi K$ puzzle and its relation to rare B and K decays," Eur. Phys. J. C **32**, 45 (2003).

[96] B. Grinstein, Y. Grossman, Z. Ligeti and D. Pirjol, "The photon polarization in $B \to X\gamma$ in the standard model," Phys. Rev. D **71**, 011504 (2005).

[97] Y. Ushiroda *et al.* [Belle Collaboration], "Time-dependent CP asymmetries in $B^0 \to K_S^0 \pi^0 \gamma$ transitions," Phys. Rev. D **74**, 111104 (2006).

[98] B. Aubert *et al.* [BABAR Collaboration], "Measurement of The Time-Dependent CP Asymmetry in $B^0 \to K^{*0}\gamma$ Decays," arXiv:0708.1614 [hep-ex].

[99] A. J. Buras, R. Fleischer, S. Recksiegel and F. Schwab, "The $B \to \pi\pi, \pi K$ puzzles in the light of new data: Implications for the standard model, new physics and rare decays," Acta Phys. Polon. B **36**, 2015 (2005).

[100] A. J. Buras, R. Fleischer, S. Recksiegel and F. Schwab, "New aspects of $B \to \pi\pi, \pi K$ and their implications for rare decays," Eur. Phys. J. C **45**, 701 (2006).

[101] M. Gronau and D. London., "How to determine all the angles of the unitarity triangle from $B_d^0 \to DK_s$ and $B_s^0 \to D^0$," Phys. Lett. B **253**, 483 (1991).

[102] M. Gronau and D. Wyler, "On determining a weak phase from CP asymmetries in charged B decays," Phys. Lett. B **265**, 172 (1991).

[103] I. Dunietz, "CP violation with selftagging $B_d$ modes," Phys. Lett. B **270**, 75 (1991).

[104] D. Atwood, I. Dunietz and A. Soni, "Enhanced CP violation with $B \to KD^0(\bar{D}^0)$ modes and extraction of the CKM angle $\gamma$," Phys. Rev. Lett. **78**, 3257 (1997).

[105] B. Aubert *et al.* [BABAR Collaboration], "Measurements of the $B \to X_s\gamma$ branching fraction and photon spectrum from a sum of exclusive final states," Phys. Rev. D **72**, 052004 (2005).

[106] P. Koppenburg *et al.* [Belle Collaboration], "An inclusive measurement of the photon energy spectrum in $b \to s\gamma$ decays," Phys. Rev. Lett. **93**, 061803 (2004).

[107] M. Misiak *et al.*, "The first estimate of $B(\bar{B} \to X_s\gamma)$ at $O(\alpha_s^2)$," Phys. Rev. Lett. **98**, 022002 (2007).

[108] K. Abe *et al.*, "Observation of $b \to d\gamma$ and determination of $|V_{td}/V_{ts}|$," Phys. Rev. Lett. **96**, 221601 (2006).

[109] B. Aubert *et al.* [BABAR Collaboration], "Branching fraction measurements of $B^+ \to \rho^+\gamma$, $B^0 \to \rho^0\gamma$, and $B^0 \to \omega\gamma$," Phys. Rev. Lett. **98**, 151802 (2007).

[110] A. Ali, T. Mannel, and T. Morozumi, "Forward backward asymmetry of dilepton angular distribution in the decay $b \to sl^+l^-$," Phys. Lett. B**273**, 505 (1991).

[111] A. Ali, G. Hiller, L.T. Handoko, T. Morozumi, "Power corrections in the decay rate and distributions in $b \to X_s l^+ l^-$ in the standard model," Phys. Rev. D**55**, 4105 (1997).

[112] E. Lunghi, A. Masiero, I. Scimemi and L. Silvestrini, "$B \to X_s \ell^+\ell^-$ decays in supersymmetry," Nucl. Phys. B **568**, 120 (2000).

[113] A. Ali, P. Ball, L. T. Handoko and G. Hiller, "A comparative study of the decays $B \to (K, K^*)\ell^+\ell^-$ in standard model and supersymmetric theories," Phys. Rev. D **61**, 074024 (2000).

[114] A. Ishikawa *et al.* [Belle Collaboration], "Observation of the electroweak penguin decay $B \to K^*\ell^+\ell^-$," Phys. Rev. Lett. **91**, 261601 (2003).

[115] B. Aubert *et al.* [BABAR Collaboration], "Measurements of branching fractions, rate asymmetries, and angular distributions in the rare decays $B \to K\ell^+\ell^-$ and $B \to K^*\ell^+\ell^-$," Phys. Rev. D **73**, 092001 (2006).

[116] K. F. Chen *et al.* [BELLE Collaboration], "Search for $B \to h(*)\nu\bar{\nu}$ Decays at Belle," Phys. Rev. Lett. **99**, 221802 (2007).

[117] K. Ikado *et al.*, "Evidence of the purely leptonic decay $B^- \to \tau^- \bar{\nu}_\tau$," Phys. Rev. Lett. **97**, 251802 (2006).

[118] B. Aubert *et al.* [BABAR Collaboration], "A search for $B^+ \to \tau^+\nu$ with Hadronic B tags," Phys. Rev. D **76**, 052002 (2007).

[119] W. S. Hou, "Enhanced charged Higgs boson effects in $B^- \to \tau\bar{\nu}$, $\mu\bar{\nu}$ and $b \to \tau\bar{\nu} + X$," Phys. Rev. D **48**, 2342 (1993).

[120] A. G. Akeroyd and S. Recksiegel, "The effect of $H^\pm$ on $B^\pm \to \tau^\pm \nu_t au$ and $B^\pm \to \mu^\pm \nu_\mu$," J. Phys. G **29**, 2311 (2003).

[121] M. Tanaka, "Charged Higgs effects on exclusive semitauonic B decays," Z. Phys. C **67**, 321 (1995).

[122] T. Miki, T. Miura and M. Tanaka, "Effects of charged Higgs boson and QCD corrections in $\bar{B} \to D\tau\bar{\nu}$," arXiv:hep-ph/0210051.

[123] H. Itoh, S. Komine and Y. Okada, "Tauonic B decays in the minimal supersymmetric standard model," Prog. Theor. Phys. **114**, 179 (2005).

[124] J.F. Kamenik, F. Mescia, "$B \to D\tau\nu$ Branching Ratios: Opportunity for Lattice QCD and Hadron Colliders," Phys. Rev. D **78**, 014003 (2008).

[125] A. Matyja *et al.* [Belle Collaboration], "Observation of $B^0 \to D^{*-}\tau^+\nu_\tau$ decay at Belle," Phys. Rev. Lett. **99**, 191807 (2007).

[126] B. Aubert *et al.* [BABAR Collaboration], "Measurement of the $B \to \pi\ell\nu$ branching fraction and determination of $|V_{ub}|$ with tagged B mesons," Phys. Rev. Lett. **97**, 211801 (2006).

[127] B. Aubert *et al.* [BABAR Collaboration], "Measurement of the $B^0 \to \pi^-\ell^+\nu$ form-factor shape and branching fraction, and determination of $|V_{ub}|$ with a loose neutrino reconstruction technique," Phys. Rev. Lett. **98**, 091801 (2007).

[128] T. Hokuue *et al.* [Belle Collaboration], "Measurements of branching fractions and $q^2$ distributions for $B \to \pi\ell\nu$ and $B \to \rho\ell\nu$ decays with $B \to D^{(*)}\ell\nu$ decay tagging," Phys. Lett. B **648**, 139 (2007).

[129] K. Abe *et al.* [BELLE Collaboration], "Measurement of exclusive $B \to X_u\ell\nu$ decays using a full-reconstruction tag at Belle," arXiv:hep-ex/0610054.

[130] E. Dalgic, A. Gray, M. Wingate, C. T. H. Davies, G. P. Lepage and J. Shigemitsu, "B Meson Semileptonic Form Factors from Unquenched Lattice QCD," Phys. Rev. D **73**, 074502 (2006).

[131] P. B. Mackenzie *et al.* [Fermilab Lattice, MILC and HPQCD Collaborations], "B and D meson semileptonic decays in three-flavor lattice QCD," PoS **LAT2005**, 207 (2006).

[132] C. W. Bauer, Z. Ligeti and M. E. Luke, "Precision determination of $|V_{ub}|$ from inclusive decays," Phys. Rev. D **64**, 113004 (2001).

[133] B. O. Lange, M. Neubert and G. Paz, "Theory of charmless inclusive B decays and the extraction of $V_{ub}$," Phys. Rev. D **72**, 073006 (2005).

[134] J. R. Andersen and E. Gardi, "Inclusive spectra in charmless semileptonic B decays by dressed gluon exponentiation," JHEP **0601**, 097 (2006).

# Chapter 3

# Flavor Structure of the Physics beyond the Standard Model

## 3.1 Motivation for New Physics

The Standard Model of elementary particles has been very successful in explaining a wide variety of existing experimental data. It covers a range of phenomena from low energy (less than a GeV) physics, such as kaon decays, to high energy (a few hundred GeV) processes involving real weak gauge bosons ($W$ and $Z$) and top quarks. There is, therefore, little doubt that the present Standard Model is the correct theory to describe the physics below the energy scale of several hundred GeV, which has been explored so far.

However, the Standard Model is not satisfactory as *the* theory of elementary particles beyond the TeV energy scale. First of all, it does not explain the characteristic pattern of the mass spectrum of quarks and leptons. The second generation quarks and leptons are several orders of magnitude heavier than the corresponding first generation particles, and the third generation is even heavier by another order of magnitude. The quark flavor mixing matrix — the CKM matrix — also has a striking hierarchical structure, *i.e.* the diagonal terms are close to unity and $1 \gg \theta_{12} \gg \theta_{23} \gg \theta_{13}$, where $\theta_{ij}$ denotes a mixing angle between the $i$-th and $j$-th generation. The observation of neutrino oscillations implies that there is also a rich flavor structure in the lepton sector. All of these masses and mixings are free parameters in the Standard Model, but ideally they should be explained by higher scale theories.

The particles in the Standard Model acquire masses from the Higgs mechanism. The Higgs potential itself is described by a scalar field theory, which contains a quadratic mass divergence. This means that a Higgs mass of order 100 GeV is realized only after a huge cancellation between the bare Higgs mass squared $\mu_0^2$ and the quadratically divergent mass renormalization, both of which are quantities of order $\Lambda^2$ where $\Lambda$ is the cutoff scale. If $\Lambda$ is of the order of the Planck scale, then a cancellation of more than 30 orders of magnitude is required. This is often called the hierarchy problem [1–3]. It would be highly unnatural if the Standard Model were *the* theory valid at a very high energy scale, such as the Planck scale. Instead, the Standard Model should be considered as an effective theory of some more fundamental theory, which most likely lies in the TeV energy region.

$CP$-violation is needed in order to produce the observed baryon number (or matter-antimatter) asymmetry in the universe. In the Standard Model, the complex phase of the CKM matrix provides the only source of $CP$-violation[1], but models of baryogenesis suggest that it is quanti-

---
[1]The $\theta$ parameter in the QCD Lagrangian is another possible source of the $CP$-violation, but its value has to

tatively insufficient (for a review, see [4]). This is another motivation to consider new physics models.

## 3.2 New physics scenarios

Several scenarios have been proposed for the physics beyond the Standard Model. They introduce new particles, dynamics, symmetries or even extra-dimensions at the TeV energy scale. In the supersymmetric (SUSY) scenarios, one introduces a new symmetry between bosons and fermions, and a number of new particles that form supersymmetric pairs with the existing Standard Model particles. The quadratic divergence of the Higgs mass term then cancels out among superpartners (for reviews, see [5,6]). Technicolor-type scenarios assume new strong dynamics (like QCD) at the TeV scale so that the Higgs field is realized as a composite state of more fundamental particles (for a recent review, see [7]). The large extra space-time dimension models [8,9] cure the problem by extending the number of spacetime dimensions beyond four (a recent review can be found in [10]). In Little Higgs models the Higgs is a pseudo-Nambu-Goldstone boson, and thus naturally light [11].

Flavor Changing Neutral Current (FCNC) processes, such as $B^0 - \overline{B}^0$ mixing and the $b \to s\gamma$ transition, provide strong constraints on new physics models. If there is no suppression mechanism for FCNC processes, such as the GIM mechanism in the Standard Model, the new physics contribution can easily become too large to be consistent with the experimental data. In fact, if one introduces a FCNC interaction through a higher dimensional operator, the associated energy scale is typically of order $10^3$ TeV, which is much higher than the expected scale of the new physics ($\sim$ TeV). Therefore, one has to introduce some flavor structure in new physics models.

## 3.3 Supersymmetric models

Supersymmetric (SUSY) models are an example of new physics models at the TeV scale. The SUSY models are attractive not only because they solve the Higgs mass hierarchy problem. They can also be consistent with Grand Unification [12,13], *i.e.* the renormalization group running of the three gauge couplings is modified by the supersymmetric partners, causing them to intersect at the same point at $M_{\text{GUT}} \simeq 10^{16}$ GeV.

General SUSY models have a number of free parameters corresponding to the masses and mixings of the superpartners for each Standard Model particle. Even in the minimal model — the Minimal Supersymmetric Standard Model (MSSM) — the number is more than a hundred. These mass and mixing parameters are, at least partially, governed by the soft supersymmetry breaking mechanism, which is necessary to make the superpartners heavy enough such that they are not detected at existing collider experiments. Therefore, to predict the mass spectrum and flavor mixing of the SUSY particles one has to specify the details of the SUSY breaking mechanism, which is tipically given at energy scales higher than the TeV scale.

The Minimal Supersymmetric Standard Model (MSSM) is a minimal supersymmetric extension of the Standard Model, containing a superpartner for each particle in the Standard Model and two Higgs doublets. Its matter content is organized in terms of chiral super fields as

$$Q_i(3, 2, 1/6), \quad \overline{U}_i(\overline{3}, 1, -2/3), \quad \overline{D}_i(\overline{3}, 1, 1/3) \tag{3.1}$$

---

be unnaturally small $\theta \leq 10^{-10}$ in order to be consistent with the neutron electron dipole moment experiment. This is another problem — the strong $CP$ problem.

for the left-handed ($Q$) and right-handed ($U$ and $D$) quark sector,

$$L_i(1, 2, -1/2), \quad \overline{E}_i(1, 1, 1) \tag{3.2}$$

for the left-handed ($L$) and right-handed ($E$) lepton sector, and

$$H_1(1, 2, -1/2), \quad H_2(1, 2, 1/2) \tag{3.3}$$

for the Higgs fields. The representation (or charge) for the gauge group $SU(3)_C \times SU(2)_L \times U(1)_Y$ is given in parentheses, and $i$ (= 1, 2, or 3) is a generation index. Under the assumption of $R$-parity conservation, which is required to avoid an unacceptably large proton decay rate, the superpotential is written as

$$\mathcal{W}_{\text{MSSM}} = Y_D^{ij} \overline{D}_i Q_j H_1 + Y_U^{ij} \overline{U}_i Q_j H_2 + Y_E^{ij} \overline{E}_i L_j H_1 + \mu H_1 H_2, \tag{3.4}$$

where $Y_U$ and $Y_D$ are the quark Yukawa couplings. The soft supersymmetry breaking terms are

$$\begin{aligned}
-\mathcal{L}_{\text{soft}} &= (m_Q^2)_{ij} \tilde{q}_i^\dagger \tilde{q}_j + (m_D^2)_{ij} \tilde{d}_i^\dagger \tilde{d}_j + (m_U^2)_{ij} \tilde{u}_i^\dagger \tilde{u}_j + (m_E^2)_{ij} \tilde{e}_i^\dagger \tilde{e}_j + (m_L^2)_{ij} \tilde{l}_i^\dagger \tilde{l}_j \\
&\quad + \Delta_1^2 h_1^\dagger h_1 + \Delta_2^2 h_2^\dagger h_2 - (B\mu h_1 h_2 + \text{h.c.}) \\
&\quad + \left( A_D^{ij} \tilde{d}_i^\dagger \tilde{q}_j h_1 + A_U^{ij} \tilde{u}_i^\dagger \tilde{q}_j h_2 + A_E^{ij} \tilde{e}_i^\dagger \tilde{l}_j h_2 + \text{h.c.} \right) \\
&\quad + \frac{M_1}{2} \overline{\tilde{B}} \tilde{B} + \frac{M_2}{2} \overline{\tilde{W}} \tilde{W} + \frac{M_3}{2} \overline{\tilde{g}} \tilde{g}
\end{aligned} \tag{3.5}$$

These consist of mass terms for scalar fields ($\tilde{q}_i$, $\tilde{u}_i$, $\tilde{d}_i$, $\tilde{l}_i$, $\tilde{e}_i$, $h_1$, and $h_2$), Higgs mixing terms, trilinear scalar couplings, and gaugino ($\tilde{B}$, $\tilde{W}$, and $\tilde{g}$) mass terms.

Flavor physics already places strong constraints on the possible structure of the SUSY breaking sector. Arbitrary terms would induce many flavor violating processes which are easily ruled out by present experimental data. In order to comply with the requirement of highly suppressed FCNC interactions, one has to introduce some structure in the soft SUSY breaking terms. Several scenarios have been proposed.

- *Universality (degeneracy).* SUSY breaking terms have a universal flavor structure at very high energy scale, such as the Planck scale ($\sim 10^{18}$ GeV) or the GUT scale ($\sim 10^{16}$ GeV). It could also be a lower scale ($\sim 10^{4-6}$ GeV). The universality comes from mediation of the SUSY breaking effects by flavor-blind interactions, such as gravity (for a review of gravity mediation see [5]), the Standard Model gauge interaction (gauge mediation [14–16], the gaugino mediation [17–19]), or the super-Weyl anomaly (anomaly mediation [20,21]). Since the soft SUSY breaking terms are flavor-blind, the squark masses are degenerate at the high energy scale where those terms are generated. Flavor violating effects appear through the renormalization group running of the squark masses to the low energy scale [22]. For the gauge mediation scenario the effect on FCNC processes is extremely suppressed, since the SUSY breaking scale is low and there is not enough room for the running.

- *Alignment.* Squark and slepton mass matrices could be diagonalized (no flavor changing interaction) in the same basis as quarks and leptons, if one assumes some symmetries involving different generations [23, 24]. Flavor violation is then suppressed and flavor violating processes are induced by incomplete alignment. However, the alignment among the left-handed squarks is inevitably violated due to the CKM mixing. Consequently, either one of the $s \to d$ transition or the $c \to u$ transition (or both) receive sizable SUSY contributions, unless the degeneracy (described above) or the decoupling (described below) cooperate for the first and the second generations. Since the SUSY effect in the $s \to d$ transition, such as $K - \overline{K}$ mixing, is already constrained experimentally, an alignment model generally predicts a signal in the $c \to u$ transition, namely the $D - \overline{D}$ mixing.

- *Decoupling.* The squarks and sleptons of the first and second generations are sufficiently heavy, 10–100 TeV, so that flavor violation in the first and second generation is suppressed [25–31]. In general such models predict large FCNC effects in the third generation, *i.e.* the $b$ quark and $\tau$ lepton decays.

Signals for FCNC processes and $CP$-violation largely depend on the structure of the soft SUSY breaking terms.

The Grand Unified Theory (GUT) [12,13] is one of the motivations for introducing supersymmetry. Besides the unification of couplings, GUTs also relate the Yukawa couplings of the quark and lepton sectors. Since the particle content and symmetry are modified above the GUT scale, they could generate different FCNC effects even if universal soft SUSY breaking is assumed. GUTs also predict some correlation between quark and lepton flavor violation processes. Such studies have been done by several authors [32–40].

## 3.4 Searching for SUSY effects in the $b \to s$ transitions

The $b \to s$ transitions are one of the most important channels to search for SUSY in B physics. The largest effect comes from the penguin diagrams, thus their precise measurements may allow us to reveal the effect of the SUSY particles propagating virtually in the loop. In this section, we focus particularly on the time-dependent CP asymmetries in $B \to \phi K_S$, $B \to \eta' K_S$ modes and also in $B \to K^* \gamma$ mode, for which the Super $B$ factory can improve the current experimental limits significantly. A few years ago, a possible large new physics effect has been announced by the Belle and Babar collaborations in the measurements of the time-dependent CP asymmetry of $B \to \phi K_S$ and $B \to \eta' K_S$ modes. Various theoretical models beyond SM have been tested in details through this phenomena [32–40]. The Super $B$ factory will allow further scrutiny of the possible SUSY contributions to these observables. In this section, we discuss the impact of the Super $B$ measurements on the SUSY models and also discuss the complementarity of the $B_s$ oscillation measurement, which has recently been done by Tevatron and will be improved by the LHCb in near future.

The time-dependent CP asymmetries in $B \to \phi K_S, \eta' K_S$ are expected to be very similar to the one in $B \to J/\psi K_S$ in SM: the $B - \overline{B}$ oscillation part is exactly the same and the decay parts (penguin diagram for the former and tree diagram for the later) do not contain any CP violating phase. Thus, in SM,

$$\mathcal{S}_{J/\psi K_S} \simeq \mathcal{S}_{\phi K_S, \eta' K_S} \ . \tag{3.6}$$

Once we introduce SUSY contributions, this relation can change. The decay amplitude of $B \to J/\psi K_S$ is mainly due to the tree level diagram which does not change. On the other hand, the decay of $B \to \phi K_S, \eta' K_S$ comes dominantly from the loop level diagram, so it may receive an extra CP violating contribution from SUSY particles running in the loop. In this section we discuss how large the departure from $S_{J/\psi K_S}$ can be in the SUSY models.

Another interesting process to search for SUSY effect is the time-dependent CP asymmetries in $B \to K^* \gamma$ (where $K^*$ decays to a CP eigenstate, e.g. $K_S \pi^0$ or $K_L \pi^0$). In this case, since the photon can have two chiralities ($L$ and $R$), there are four different amplitudes $\langle f_{L/R} | B^0 \rangle = A_{L/R}$ and $\langle f_{L/R} | \overline{B}^0 \rangle = \overline{A}_{L/R}$ where $f_{L,R} = K_S \pi^0 \gamma_{L,R}$. In SM, these decay amplitudes do not contain the CP violating phase and more importantly, there is a well-known chiral suppression $|A_L/\overline{A}_L| = |\overline{A}_R/A_R| \simeq m_s/m_b$. As a result, the mixing induced CP asymmetry is expected to be very small

$$\mathcal{S}_{B \to K_S \pi^0 \gamma} \simeq -2\frac{m_s}{m_b} \sin 2\phi_1 \quad \text{in SM} \tag{3.7}$$

The suppression factor $m_s/m_b$ comes from the fact that the quark level diagram for $b \to s\gamma$ originates from the left-handed coupling of $W$ and quarks[2]. However, if the loop contains a right-handed coupling SUSY contribution, this suppression can be avoided and $\mathcal{S}_{B \to K_S \pi^0 \gamma}$ can become large.

### 3.4.1 Hadronic uncertainties in the SM predictions

In order to search for a small SUSY contribution it is essential to have a good understanding of the theoretical uncertainties in the SM predictions. In this sub-section, we overview the SM prediction for $\Delta \mathcal{S}_f \equiv (-\xi_f)\mathcal{S}_f - \sin(2\phi_1)$ where $f$ represents the penguin-dominant final states, $\phi, \eta', \eta, \pi^0, \rho^0, \omega$ (more details can be found in [41–43]). Note that we take $\mathcal{S}_{J/\psi K_S} = \sin(2\phi_1)$ for simplicity since the discrepancy between the two are theoretically found to be minor [44,45]. We show in Figure 3.1 the theoretical calculations for $\Delta \mathcal{S}_f$ done by using several approaches. Both PQCD [46,47] and QCDF [48,49] predict positive deviations for most of the decay modes except for the $B^0 \to \rho^0 K_S$ decay. The sign and sizes of $\Delta \mathcal{S}_f$ can be understood from the following approximate relation,

$$\Delta \mathcal{S}_f \simeq 2\lambda^2 \sqrt{\rho^2 + \eta^2} \, \text{Re}\left(\frac{a_f^t - a_f^u}{a_f^t - a_f^c}\right) \cos(2\phi_1) \sin\phi_3, \tag{3.8}$$

where we parameterized the $B^0 \to f$ decay amplitude as $A_f = V_{ub}^* V_{us} a_f^u + V_{cb}^* V_{cs} a_f^c + V_{tb}^* V_{ts} a_f^t$. In the case of dominant penguin contribution in $a_f^t$, we have a positive and small ($\sim 0.02$) deviation. However, a sizable color-suppressed-tree amplitude $C$ contributing to $a_f^u$ could change this naive expectation. For the $B^0 \to \eta K_S$, $\pi^0 K_S$ and $\omega K_S$ modes, the effect of $C$ is constructive, whereas it is destructive for $\rho^0 K_S$. Therefore, the former decay modes exhibit positive and larger deviations compared to $\phi K_S$, while $\eta' K_S$ has a negative deviation. The model for long-distance (LD) rescattering suggests that the effects on $\mathcal{S}_f$ are minor except for $\rho^0 K_S$ and $\omega K_S$ [50]. Including the LD effects in QCDF approach, $\Delta \mathcal{S}_{\rho^0 K_S}$ turns out to be positive. We conclude that the $\phi K_S$ and $\eta' K_S$ modes are theoretically very clean in all of the available estimations, including SCET [51].

All the above approaches are based on factorization of decay amplitudes and perturbative expansions. There exists a somewhat complementary approach based on SU(3) flavor symmetry. Using the SU(3) relations, one can obtain a conservative bound on $\Delta \mathcal{S}_f$ from the measured decay rates of the related $b \to d$ modes [52]. In general, the bounds are much weaker than the factorization-based methods, although they may improve significantly with future data [52–57]. Note that the sign of $\Delta \mathcal{S}_f$ is not determined by the SU(3) method.

### 3.4.2 Mass insertion approximation for SUSY contributions

As discussed in the previous section, the soft SUSY breaking terms which provide a new source of flavour and CP violation contain a huge number of free parameters. In order to organize them, we employ the so-called mass insertion approximation (MIA) [22]. In the MIA, one adopts a basis where the fermion and sfermion mass matrices are rotated in the same way to diagonalize the fermion mass matrix (the super-CKM basis). In this basis, the couplings of fermions and sfermions to neutral gauginos are flavor diagonal, leaving all the sources of flavor violation in the off-diagonal terms of the sfermion mass matrix. These terms are denoted by $(\Delta^q_{AB})_{ij}$, where $A, B$ denote the chirality $(L, R)$ and $q$ indicates the $(u, d)$ type. The sfermion propagator can then be expanded as

$$\langle \tilde{q}_{Ai} \tilde{q}^*_{Bj} \rangle = i(k^2 \mathbf{1} - \tilde{m}^2 \mathbf{1} - \Delta^q_{AB})^{-1}_{ij} \simeq \frac{i\delta_{ij}}{k^2 - \tilde{m}^2} + \frac{i(\Delta^q_{AB})_{ij}}{(k^2 - \tilde{m}^2)^2} + \cdots, \quad (3.9)$$

where $\mathbf{1}$ is the unit matrix and $\tilde{m}$ is the average squark mass. In this way, the flavour violation in SUSY models, such as the SUSY GUT models, can be parameterized in a model independent way by the dimensionless parameters $(\delta^q_{AB})_{ij} = (\Delta^q_{AB})_{ij}/\tilde{m}^2$. In the following, we keep only the first term of this expansion.

---

[2]There are also QCD loop contributions leading to additional $\Lambda_{\text{QCD}}/m_b$ term.

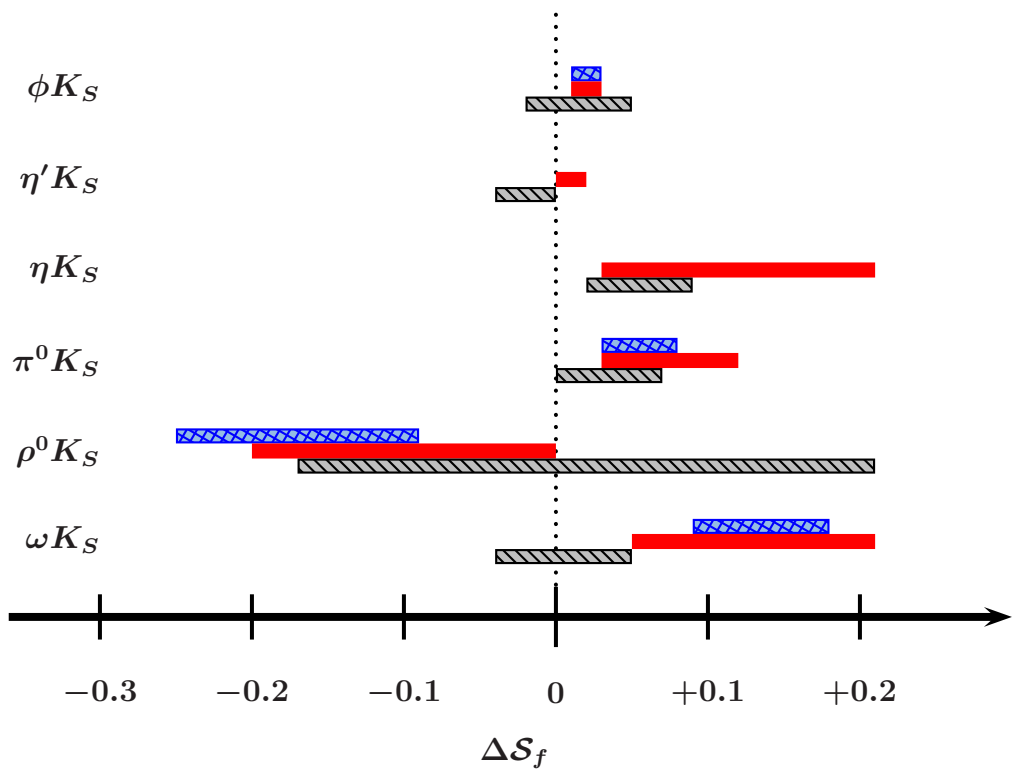

Figure 3.1: $\Delta\mathcal{S}_f$ in PQCD (blue and crosshatched), QCDF (red and solid) and QCDF+LD (black and hatched).

The mass insertion parameters $(\delta^q_{AB})_{ij}$ are constrained by various flavour experiments such as $K$ physics, electric-dipole-moment, as well as lepton flavour violation experiments which are to be discussed in Section 3.6. We summarize the current constraints in Table 3.1.

| $|(\delta^q_{AB})_{ij}|$ | | $AB = LL/RR$ | | $AB = LR/RL$ | |
|---|---|---|---|---|---|
| $q$ | $ij$ | Re. part | Im. part | Re. part | Im. part |
| $d$ | 11 | - | - | $< 10^{-3}$ | $< 10^{-6}$ |
| $u$ | 11 | - | - | - | $< 10^{-6}$ |
| $d$ | 12 | $< 10^{-2}$ | $< 10^{-3}$ | $< 10^{-3}$ | $< 10^{-4}$ |
| $u$ | 12 | $< 10^{-2}$ | – | $< 10^{-2}$ | – |
| $d$ | 13 | $< 10^{-1}$ | $< 10^{-1}$ | $< 10^{-2}$ | $< 10^{-1}$ |

Table 3.1: The constraints on the the mass insertions involving the first generation. Here, we consider only the gluino contributions with the SUSY masses $m_{\tilde{g}} = m_{\tilde{q}} = 500$ GeV. The imaginary part of $ij = 11$ element is constrained by the electric-dipole-moment [58–62]. The real and imaginary part of the down-type $ij = 12$ element is constrained by the $K - \overline{K}$ mixing parameter, $\Delta M_K$ and $\epsilon_K$, respectively [58, 63]. The real part of the up-type $ij = 12$ element is constrained by $\Delta M_D$ [58, 64]. Finally, the real and imaginary part of the down-type $ij = 13$ element is constrained by $\Delta M_{B_d}$ and $\sin 2\phi_1$, respectively [58, 65].

The mass insertion relevant to the subject of this section, $(\delta^q_{AB})_{23}$, is also constrained, in particular, by the branching ratio measurement of the $B \to X_s \gamma$ process and also the $B_s - \overline{B}_s$ oscillation measurement. In the following we present the allowed range of the CP violation in $B \to \phi K_S$, $B \to \eta' K_S$ and also in $B \to K^* \gamma$ mode using those constraints.

### 3.4.3 Effective Hamiltonian for the SUSY contributions

The effective Hamiltonian describing the SUSY contribution of $b \to s\gamma$ is given as:

$$\mathcal{H}^{\Delta B=1}_{\text{eff}} = -\frac{G_F}{\sqrt{2}} V_{tb} V^*_{ts} \left[ C_\gamma O_\gamma + \tilde{C}_\gamma \tilde{O}_\gamma \right] \quad (3.10)$$

with

$$O_\gamma = \frac{e}{8\pi^2} m_b \overline{s}_\alpha \sigma^{\mu\nu} R b_\beta F_{\mu\nu}, \quad \tilde{O}_\gamma = \frac{e}{8\pi^2} m_b \overline{s}_\alpha \sigma^{\mu\nu} L b_\beta F_{\mu\nu}, \quad (3.11)$$

where $L \equiv (1 - \gamma_5)$ and $R \equiv (1 + \gamma_5)$. Here we consider the dominant gluino contribution to the Wilson coefficient (chargino and Higgsino become important in the large $\tan \beta$ limit, which we discuss in Section 3.4.6). Then, the Wilson coefficient is written in terms of the down-type mass insertion as:

$$C_{7\gamma}^{\tilde{g}}(M_S) = -\frac{\sqrt{2}\alpha_s\pi}{6G_F V_{ts}^* V_{tb} m_{\tilde{q}}^2}$$
$$\left[(\delta_{LR}^d)_{23}\frac{m_{\tilde{g}}}{m_b}\frac{8}{3}M_1(x) + (\delta_{LL}^d)_{23}\left(\frac{8}{3}M_3(x) + (\delta_{LR}^d)_{33}\frac{m_{\tilde{g}}}{m_{\tilde{b}}}\frac{8}{3}M_a(x)\right)\right]. \quad (3.12)$$

where $m_{\tilde{g}}$ is the gluino mass, $m_{\tilde{q}}$ the squark mass, $x \equiv m_{\tilde{g}}^2/m_{\tilde{q}}^2$, and the loop functions $M_{1-4}$ and $M_{a,b}$ can be found in Ref. [58, 66][3]. In the SM, this Wilson coefficient is relatively suppressed because the chirality flip on the external $b$ and $s$ fields requires a suppression factor of $m_b$. In SUSY models with large $LR/RL$ mixing the factor $m_b$ is replaced by the internal gluino mass as can be seen in the first term in the square brackets. The last term comes from the double-mass-insertion diagrams with $(\delta_{LR}^d)_{33} = (A_D^{33}\langle h_1^0\rangle - m_b\mu\tan\beta)/m_{\tilde{q}}^2$, where the first trilinear term is sub dominant and can be neglected. In general, the double-mass insertion term becomes important for large $\tan\beta$. However, for $b \to s\gamma$ transition, it becomes significant even for a moderate value of $\tan\beta$ (starting from $\mu\tan\beta \sim 1$ TeV) due to the factor $m_{\tilde{g}}/m_b$. The current experimental data on this process [67]

$$\mathcal{B}(B \to X_s\gamma) = (3.55 \pm 0.26) \times 10^{-4}, \quad (3.13)$$

can be compared to the recent NLLO SM prediction $\mathcal{B}(\bar{B} \to X_s\gamma) = (3.15 \pm 0.23) \times 10^{-4}$. Since the difference between these two values is not very large, this measurement leads to a severe constraint on the mass insertion parameter $(\delta_{AB}^d)_{23}$ as we will show in the following.

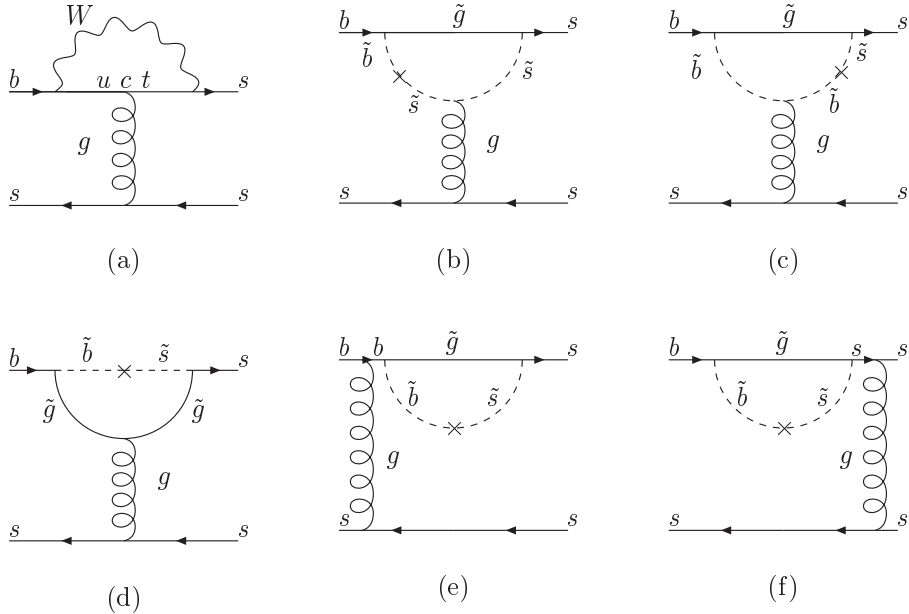

Figure 3.2: The SM contribution (a) and the gluino–down squark contributions (b)–(f) to the $b \to s\bar{s}s$ transition. The cross represents the mass insertions $(\delta_{AB}^d)_{23}$.

---

[3]$M_a(x)$ and $M_b(x)$ correspond to $M_1(x)$ and $M_2(x)$ in Ref. [66], respectively.

Let us now switch to $b \to s\bar{s}s$ transitions; the effective Hamiltonian can be expressed as

$$\mathcal{H}_{\text{eff}}^{\Delta B=1} = -\frac{G_F}{\sqrt{2}} V_{tb} V_{ts}^* \left[ \sum_{i=3}^{6} C_i O_i + C_g O_g + \sum_{i=3}^{6} \tilde{C}_i \tilde{O}_i + \tilde{C}_g \tilde{O}_g \right], \quad (3.14)$$

with

$$O_3 = \bar{s}_\alpha \gamma^\mu L b_\alpha \bar{s}_\beta \gamma^\mu L s_\beta, \quad (3.15)$$

$$O_4 = \bar{s}_\alpha \gamma^\mu L b_\beta \bar{s}_\beta \gamma^\mu L s_\alpha, \quad (3.16)$$

$$O_5 = \bar{s}_\alpha \gamma^\mu L b_\alpha \bar{s}_\beta \gamma^\mu R s_\beta, \quad (3.17)$$

$$O_6 = \bar{s}_\alpha \gamma^\mu L b_\beta \bar{s}_\beta \gamma^\mu R s_\alpha, \quad (3.18)$$

$$O_g = \frac{g_s}{8\pi^2} m_b \bar{s}_\alpha \sigma^{\mu\nu} R \frac{\lambda_{\alpha\beta}^A}{2} b_\beta G_{\mu\nu}^A, \quad (3.19)$$

where $L \equiv (1 - \gamma_5)$ and $R \equiv (1 + \gamma_5)$. The terms with a tilde are obtained from $C_{i,g}$ and $O_{i,g}$ by exchanging $L \leftrightarrow R$. We consider the dominant gluino contribution to the Wilson coefficient and postpone the discussion of large $\tan \beta$ where chargino and Higgsino become important to Section 3.4.6. The Wilson coefficients for the gluino contributions that come from the penguin diagrams (see Figure 3.2) were computed in [58]. The dominant contribution to $b \to s\bar{s}s$ decays turns out to arise from the Wilson coefficient of the chromo-magnetic operator which is very similar to the $C_\gamma$:

$$C_g^{\tilde{g}}(M_S) = -\frac{\sqrt{2}\alpha_s \pi}{2 G_F V_{tb} V_{ts}^* m_{\tilde{q}}^2} \Bigg\{ (\delta_{LR}^d)_{23} \frac{m_{\tilde{g}}}{m_b} \left( \frac{1}{3} M_1(x) + 3 M_2(x) \right) \quad (3.20)$$
$$+ (\delta_{LL}^d)_{23} \left[ \left( \frac{1}{3} M_3(x) + 3 M_4(x) \right) + (\delta_{LR}^d)_{33} \frac{m_{\tilde{g}}}{m_b} \left( \frac{1}{3} M_a(x) + 3 M_b(x) \right) \right] \Bigg\},$$

### 3.4.4 SUSY contributions to $\mathcal{S}_{\phi K_S, \eta' K_S}$

In order to see how the SUSY contribution can cause $\mathcal{S}_{\phi K_S, \eta' K_S}$ to deviate from $\mathcal{S}_{J/\psi K_S}$, let us first derive the general formulae for the time-dependent CP asymmetry including the SUSY effect. By parameterizing the decay amplitudes of these channels $\langle f | B^0 \rangle = A$ and $\langle f | \overline{B}^0 \rangle = \overline{A}$ ($f = \phi K_S, \eta' K_S$) for SM and SUSY as:

$$A^{\text{SM}} = |A^{\text{SM}}| e^{i\delta^{\text{SM}}}, \qquad A^{\text{SUSY}} = |A^{\text{SUSY}}| e^{i\theta^{\text{SUSY}}} e^{i\delta^{\text{SUSY}}}, \quad (3.21)$$

$$\overline{A}^{\text{SM}} = |\overline{A}^{\text{SM}}| e^{i\delta^{\text{SM}}}, \qquad \overline{A}^{\text{SUSY}} = |\overline{A}^{\text{SUSY}}| e^{-i\theta^{\text{SUSY}}} e^{i\delta^{\text{SUSY}}}, \quad (3.22)$$

where $\delta^{\text{SM(SUSY)}}$ is the strong ($CP$-conserving) phase and $\theta^{\text{SUSY}}$ is the weak ($CP$-violating) phase, the mixing induced CP asymmetry for $B \to \phi K_S, \eta' K_S$ can be written as

$$\mathcal{S}_f = \frac{\sin 2\phi_1 + 2 \left( \frac{|A^{\text{SUSY}}|}{|A^{\text{SM}}|} \right) \cos \delta_{12} \sin(\theta^{\text{SUSY}} + 2\phi_1) + \left( \frac{|A^{\text{SUSY}}|}{|A^{\text{SM}}|} \right)^2 \sin(2\theta^{\text{SUSY}} + 2\phi_1)}{1 + 2 \left( \frac{|A^{\text{SUSY}}|}{|A^{\text{SM}}|} \right) \cos \delta_{12} \cos \theta^{\text{SUSY}} + \left( \frac{|A^{\text{SUSY}}|}{|A^{\text{SM}}|} \right)^2}, \quad (3.23)$$

where $\delta_{12} \equiv \delta_{\text{SM}} - \delta_{\text{SUSY}}$. Recalling $\mathcal{S}_{J/\psi K_S} \simeq \sin 2\phi_1$, one can see that $\mathcal{S}_f \neq \mathcal{S}_{J/\psi K_S}$ requires not only non-zero $|A^{\text{SUSY}}|/|A^{\text{SM}}|$ but also non-zero CP violating phase $\theta^{\text{SUSY}} \neq 0$ as we naively expect.

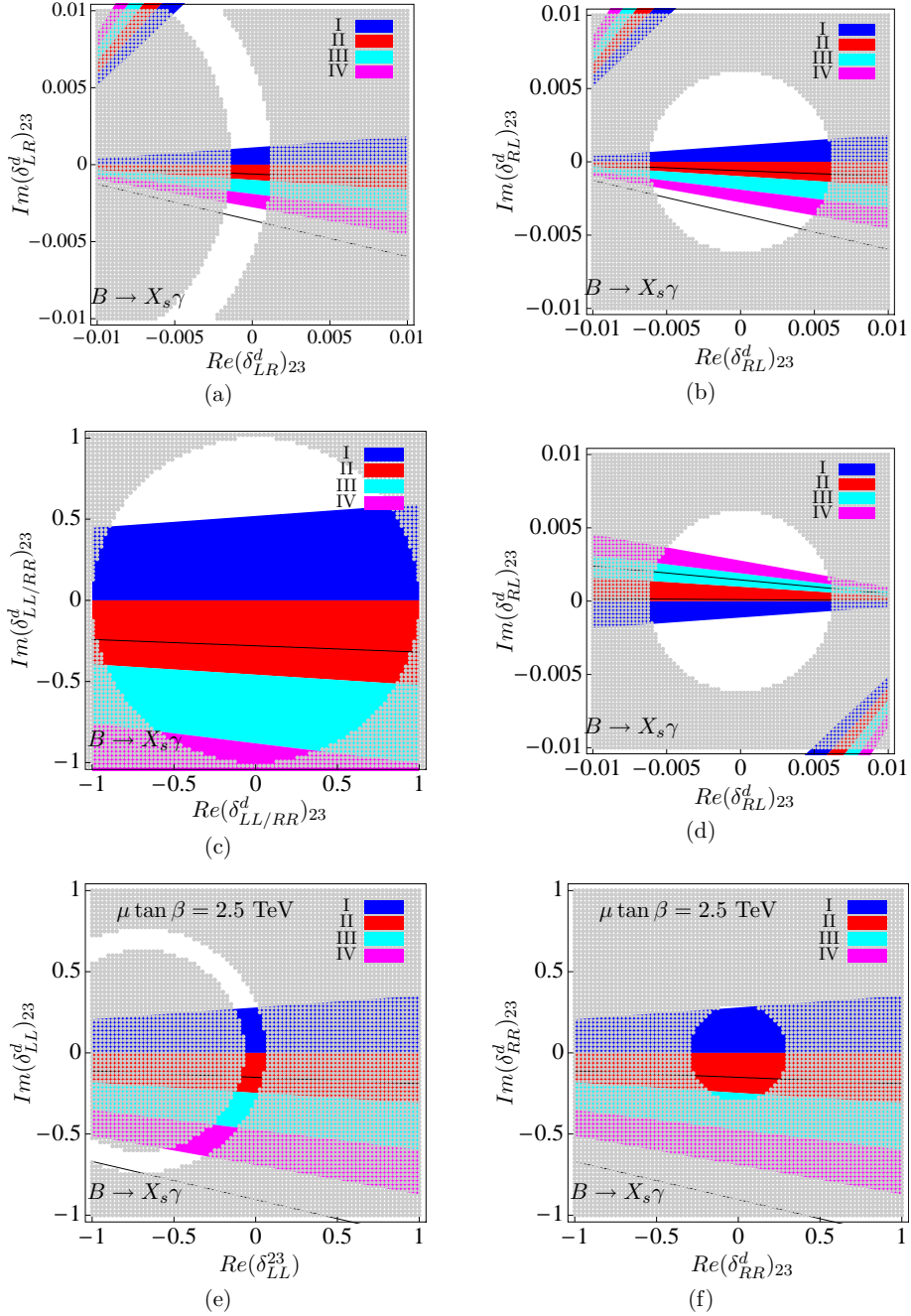

Figure 3.3: Constraints on the down-type $ij = 23$ elements of the mass insertion expected from $\Delta\mathcal{S}_{\phi K_S, \eta' K_S}$ measurements. Each coloured band represents a $\pm 0.1$ error which is roughly the Super $B$ factory sensitivity with 5 ab$^{-1}$. The colour indicates (I) $0 < \Delta\mathcal{S}_f < 0.1$, (II) $-0.1 < \Delta\mathcal{S}_f < 0$, (III) $-0.2 < \Delta\mathcal{S}_f < -0.1$, (IV) $-0.3 < \Delta\mathcal{S}_f < -0.2$. The current experimental bounds, $\Delta\mathcal{S}_{\phi K_S} = -0.23 \pm 0.18$ and $\Delta\mathcal{S}_{\eta' K_S} = -0.08 \pm 0.07$, are shown in the black lines. The patched area is excluded by the $B_s \to X_s\gamma$ branching ratio measurements. Figure (a) and (b) show the constraints from $\Delta\mathcal{S}_{\phi K_S}$ together with $B \to X_s\gamma$ for the $LR$ and $RL$ mass insertions. The asymmetric constraint from $B \to X_s\gamma$ for the $LR$ case occurs due to the interference between SM and SUSY. The constraint on the $LR$ mass insertion from $\Delta\mathcal{S}_{\eta' K_S}$ is approximately the same as (a) while for the $RL$ mass insertion, the chirality difference results in the 180° rotation as shown in Figure (d). In general, the $LL/RR$ mass insertions are not strongly constrained neither from $B_s \to X_s\gamma$ nor $\Delta\mathcal{S}_{\phi K_S, \eta' K_S}$ as shown in Figure (c). However, for the case of relatively large $\mu \tan\beta$, the double-mass-insertion effect enhances the $LL/RR$ term and leads to a very strong constraint (see (e) and (f)).

Now let us focus on the $B \to \phi K_S$. We first derive the SUSY to SM amplitude ratio in terms of the mass insertion parameters. The numerical result for the ratio of the SM to the SUSY amplitude for $m_{\tilde{g}} \simeq m_{\tilde{q}} = 500$ GeV is as follows [68]

$$\frac{A^{\text{SUSY}}(\phi K_S)}{A^{\text{SM}}(\phi K_S)} \simeq (0.14 + 0.02i)[(\delta^d_{LL})_{23} + (\delta^d_{RR})_{23}] + (65 + 11i)[(\delta^d_{LR})_{23} + (\delta^d_{RL})_{23}]$$
$$- (26.5 - 4.4i)[(\delta^d_{LL})_{23}(\delta^d_{LR})_{33} + (\delta^d_{RR})_{23}(\delta^d_{RL})_{33}].. \quad (3.24)$$

For computing the matrix element, $< \phi K_S | O_{i,g} | B >$, we applied the naive factorization approximation[4]. The tiny imaginary parts in (3.24) are the strong phases coming from the QCD correction terms in the effective Wilson coefficient in [71]. Note that in the following, we consider a class of models in which only one mass insertion, $LL, RR, LR$ or $RL$, dominates and the others are all negligible.

The resulting constraints on the real and imaginary parts of the mass insertions $(\delta^d_{AB})_{ij}$ that we will obtain from the future measurement of $\Delta S_f$ (coloured bands) together with the constraint from $B \to X_s \gamma$ (patched area) are shown in Figure 3.3. Let us first look at the constraint from the $B \to X_s \gamma$. The results for single mass insertions (and small $\tan\beta$) are shown in Figure 3.3 (a) - (c). We can see that the $LR/RL$ mass insertions are strongly constrained (see Figure 3.3 (a) and (b)) whereas $LL/RR$ are not (see Figure 3.3 (c)). This tendency comes from the chiral enhancement factor $m_{\tilde{g}}/m_b$ for the $LR/RL$ mass insertions term discussed previously. The interference between the SM and SUSY contributions for the $LR$ mass insertion results in the narrow asymmetric bound. Now let us analyse the constraint from $\Delta S$. Each coloured band represents a 0.1 error which is the sensitivity that can be surpassed at the Super $B$ factory already with 5 ab$^{-1}$ (for sensitivity estimates see Sect. 5.2). The colours indicate

$$\begin{aligned} \text{I}: & \quad 0 < \Delta S_f < 0.1, \\ \text{II}: & \quad -0.1 < \Delta S_f < 0, \\ \text{III}: & \quad -0.2 < \Delta S_f < -0.1, \\ \text{IV}: & \quad -0.3 < \Delta S_f < -0.2. \end{aligned} \quad (3.25)$$

We can see from Figure 3.3 (a) and (b) that the future experiments which will measure $\Delta S_{\phi K_S}$ with the accuracy at a few % level will provide further constraints on the $LR/RL$ mass insertions, in particular on its imaginary part. On the contrary, Figure 3.3 (c) shows that the $LL/RR$ mass insertions will be constrained only mildly by the $\Delta S_{\phi K_S}$ measurement.

Now let us consider the double-mass-insertion term. Resulting constraints on $LL/RR$ mass insertions with $\mu \tan\beta = 2.5$ TeV are shown in Figure 3.3 (e) and (f), respectively. In this case we obtained much stricter constraints compared to Figure 3.3 (c) both from $B \to X_s \gamma$ and $\Delta S_{\phi K_S}$.

Next we discuss $B \to \eta' K_S$. The $\eta'$ meson is known to be composed of $u\bar{u}$, $d\bar{d}$ and $s\bar{s}$ accompanied by a small amount of other particles such as gluonium and $c\bar{c}$ etc. [5]. Apart from such exotic components, the $B \to \eta' K_S$ process comes from the two penguin diagrams $b \to s\bar{s}s$ and $b \to d\bar{d}d$, and the Cabibbo suppressed tree diagram. The tree contribution is estimated to be less than 1%. Thus the $\mathcal{S}_f$ in $B \to \eta' K_S$ and $B \to \phi K_S$ are approximately the same apart from the parity of the final states. The numerical result for the ratio of the SM to the SUSY amplitude is obtained as [72]:

$$\frac{A^{\text{SUSY}}(\eta' K_S)}{A^{\text{SM}}(\eta' K_S)} \simeq (0.15 + 0.03i)[(\delta^d_{LL})_{23} - (\delta^d_{RR})_{23}] + (69 + 12i)[(\delta^d_{LR})_{23} - (\delta^d_{RL})_{23}], \tag{3.26}$$

for $m_{\tilde{g}} \simeq m_{\tilde{q}} = 500$ GeV. As expected, the result is very similar to the one for $B \to \phi K_S$ in (3.24) except for the overall signs of the $RR$ and $RL$ mass insertions. Therefore, the constraints on the $LR/LL$ mass insertions are approximately the same as for $\Delta \mathcal{S}_{\phi K_S}$ given in Figure 3.3 (a) and (c) while the constraints on the $RL$ and $RR$ mass insertions get rotated by $180°$ (see Figure 3.3 (d) for an example of the $RL$ constraint). Therefore, if the future experiment confirms $S_{\phi K_S} \neq S_{\eta' K_S}$, that would require a SUSY contribution with right-handed mass insertion $(\delta^d_{RR/LL})_{23}$.

---

[4]We show our numerical result with the SM input parameters fixed to have a simple illustration of the SUSY contribution effects. The matrix element of $B \to \phi K_S$ can also be computed using more recent methods: the QCD factorization and the pQCD approach method. Application of these to analyse the SUSY effect on $S_{\phi K_S}$ can be found in [38, 69] for the former and in [70] for the latter.

[5]We do not consider contributions from exotic components here. However, since an unexpectedly large branching ratio is observed in the $B \to \eta' K$ processes, possible large contributions cannot be excluded completely. The gluonium contributions to $B \to \eta' K_S$ including the possibility that SUSY effects also enhance the branching ratio of $B \to \eta' K$ are discussed in [72].

Before closing this section, it is interesting to discuss the complementarity with the other important $b \to s$ transition, the $B_s - \overline{B}_s$ mixing. In 2006, CDF has obtained the first evidence for the $B_s$ oscillation [73]:

$$\Delta M_{B_s} = 17.77 \pm 0.10(\text{stat.}) \pm 0.07(\text{syst.}) \text{ ps}^{-1} \tag{3.27}$$

More recently, CDF and D0 have announced their first results on the CP violating phase $\beta_s$ through the $B \to J/\psi \phi$ channel [74, 75]. The HFAG average is[6]:

$$\beta_s = 22°\,{}^{+10°}_{-8°}. \tag{3.28}$$

The theoretical prediction for the first result $\Delta M_{B_s}$ suffers from a large theoretical uncertainty mainly coming from the bag parameter and the decay constant. The measured value can be explained by the SM within 20% theoretical uncertainty. On the other hand, the measurement of the CP violating phase $\beta_s$ came as a surprise. In the SM, the $B_s$ box diagrams contain only a small CP violating phase, $\beta_s \simeq 2°$, so that the observed value is about $2\sigma$ away from the SM.

A discussion of the impact of this new measurement on the mass insertion parameter is in order. The gluino contribution to the ratio of the SUSY contribution to the SM one for $m_{\tilde{g}} \simeq m_{\tilde{q}} = 500$ GeV is given by [78]:

$$\begin{aligned}\frac{M_{12}^{\text{SUSY}}}{M_{12}^{\text{SM}}} &\simeq 1.44 \left[(\delta^d_{23})^2_{LL} + (\delta^d_{23})^2_{RR}\right] + 27.57 \left[(\delta^d_{23})^2_{LR} + (\delta^d_{23})^2_{RL}\right] \\ &\quad - 44.76 \left[(\delta^d_{23})_{LR}(\delta^d_{23})_{RL}\right] - 175.79 \left[(\delta^d_{23})_{LL}(\delta^d_{23})_{RR}\right].\end{aligned} \tag{3.29}$$

Note $\Delta M_{B_s} = 2|M_{12}|$. Taking into account the constraints from $B \to X_s \gamma$ (see Fig. 3.3), we can find that the maximum contribution from $LR$ or $RL$ dominated models leads to only a negligible effect of $\mathcal{O}(10^{-3})$, whereas $LL$ or $RR$ dominated models can give a $\mathcal{O}(1)$ contribution which leads to very strong constraints on these mass insertions. The resulting constraint on the real and imaginary parts of the $LL/RR$ mass insertion is shown in Figure 3.4. The patched region is excluded by the $\Delta M_{B_s}$ measurements of Tevatron. In Figure 3.4 (a) and (b), we include the previous result on $\Delta \mathcal{S}_f$ (coloured bounds), with and without the double-mass-insertion term, respectively (i.e. Figure 3.3 (c) and (e, f)). One can see that the real part of the $LL/RR$ mass

---

[6]Averaging the results from CDF and D0 is performed also in [76, 77] where resulting errors are slightly smaller.

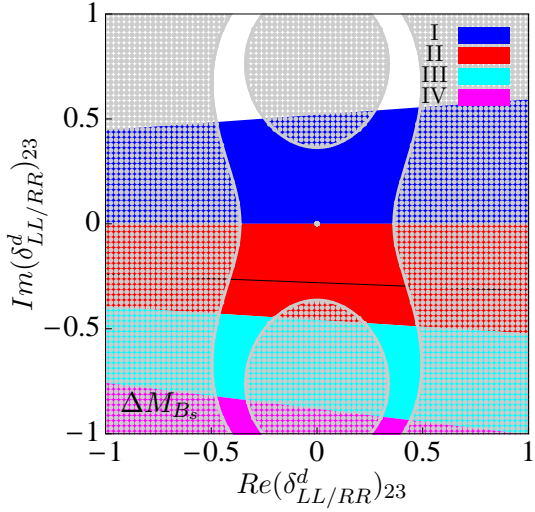
(a)

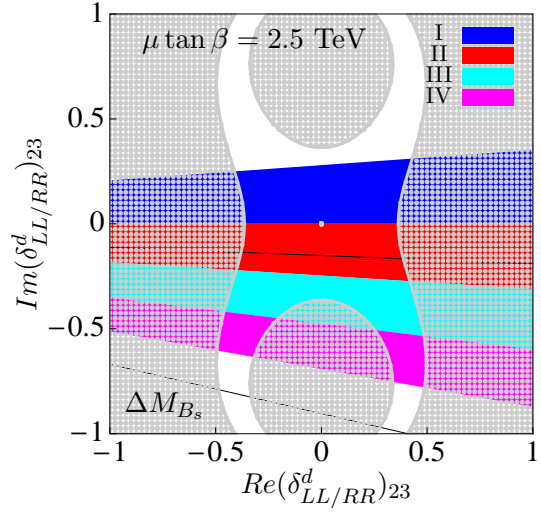
(b)

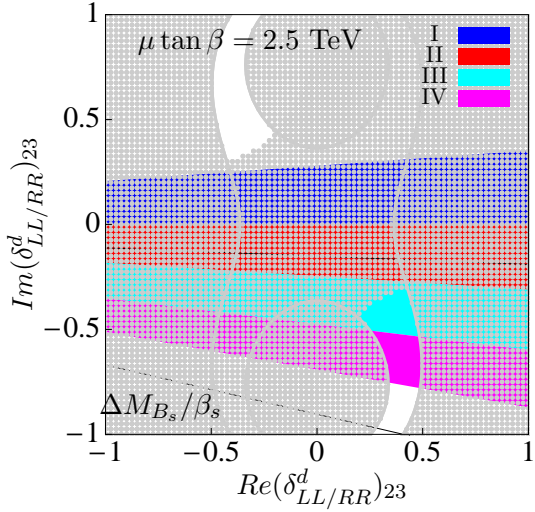
(c)

Figure 3.4: Figure (a) exhibits the constraints on the $LL/RR$ mass insertion obtained by the current experimental data on $\Delta M_{B_s}$ (patched area) together with the previously obtained $\Delta\mathcal{S}$ constraint (see also the caption of Figure 3.3 (c)). Figure (b) is in the case of moderate $\tan\beta$ scenario (see the caption of Figure 3.3 (e, f) for details). In Figure (c), we further overlap the recent measurement of the CP violating phase of the $B_s$ oscillation. We can see that the current experimental value (see Eq. (3.28)) requires a large imaginary part in $(\delta^d_{23})_{LL/RR}$.

insertion is nicely constrained by $\Delta M_{B_s}$ while we need $\Delta \mathcal{S}_f$ in order to determine the imaginary part. In Figure 3.4 (c), we further overlay the constraint from the new CP phase measurement of Eq. (3.28). Although the experimental errors are still very large there is a tendency in the current experimental data to require a very large imaginary part of $(\delta_{23}^d)_{LL/RR}$. If the experimental data on $B_s - \bar{B}_s$ mixing is confirmed at the future LHCb experiment, SUSY models with $(\delta_{23}^d)_{LL/RR}$ will lead to a 10-20% deviation between $S_{\phi K_S, \eta' K_S}$ and $S_{J/\psi K_S}$, as implied by Fig. 3.4.

### 3.4.5 SUSY contributions to $\mathcal{S}_{B \to K_S \pi^0 \gamma}$

The time-dependent CP asymmetry of $B \to K^* \gamma$, $S_{B \to K^* \gamma}$, is an interesting observable a precise measurement of which will become available at the Super $B$ factory. The SM predicts a very small $S_{B \to K^* \gamma}$ as given in Eq. (3.7). Let us first derive general formulae for the CP asymmetry including the SUSY contributions. A crucial point is to separate the two photon polarizations and to write the decay amplitude in terms of $\langle f_{L/R} | B^0 \rangle = A_{L/R}$ and $\langle f_{L/R} | \overline{B}^0 \rangle = \overline{A}_{L/R}$ where $f_L$ and $f_R$ represent $f_L = K_S \pi^0 \gamma_L, f_R = K_S \pi^0 \gamma_R$. Then, the time-dependent CP asymmetry for $B \to K_S \pi^0 \gamma$ can be written as:

$$\mathcal{S}_{K_S \pi^0 \gamma} = \frac{2 Im \left[ \frac{q}{p} (A_L^* \overline{A}_L + A_R^* \overline{A}_R) \right]}{|A_L|^2 + |A_R|^2 + |\overline{A}_L|^2 + |\overline{A}_R|^2} \quad (3.30)$$

As can be seen from this expression, a non-zero CP asymmetry in this mode occurs when there is an interference between left- and right-handed amplitude[7]. In order to see the source of the right-handed contribution, let us write the SM and SUSY contributions separately:

$$\overline{A}_L = A_L^{\rm SM} + A_L^{\rm SUSY} e^{i\theta_L^{\rm SUSY}} e^{i\delta^{\rm SUSY}} \quad (3.31)$$
$$A_L = A_R^{\rm SM} + A_R^{\rm SUSY} e^{-i\theta_R^{\rm SUSY}} e^{i\delta^{\rm SUSY}} \quad (3.32)$$
$$\overline{A}_R = A_R^{\rm SM} + A_R^{\rm SUSY} e^{i\theta_R^{\rm SUSY}} e^{i\delta^{\rm SUSY}} \quad (3.33)$$
$$A_R = A_L^{\rm SM} + A_L^{\rm SUSY} e^{-i\theta_L^{\rm SUSY}} e^{i\delta^{\rm SUSY}} \quad (3.34)$$

where we factored out the CP violating phase of the SUSY amplitudes, $\theta_{L/R}^{\rm SUSY}$. In SM, the right-handed part is suppressed by the factor $m_s/m_b$ compared to the left-handed one, thus we neglect the right-handed amplitude $A_R^{\rm SM}/A_L^{\rm SM} \simeq 0$ (the theoretical uncertainty in this argument is discussed for example in [79–82]). For the SUSY amplitude, the $A_L^{\rm SUSY}$ comes mainly from $O_\gamma^{\tilde{g}}$ and $A_R^{\rm SUSY}$ from $\tilde{O}_\gamma^{\tilde{g}}$. The former contribution (which comes from non-zero $(\delta_{23}^d)_{LL}, (\delta_{23}^d)_{LR}$ mass insertion) itself can not yield an extra contribution to $S_{K_S \pi^0 \gamma}$, and $(\delta_{23}^d)_{RR}$ and/or $(\delta_{23}^d)_{RL}$ mass insertion is necessary to have a deviation from the SM relation in Eq. (3.7). By assuming SUSY models in which the mass insertions $(\delta_{23}^d)_{RR}$ and/or $(\delta_{23}^d)_{RL}$ are non-zero, we have $A_L^{\rm SUSY} \simeq A_R^{\rm SUSY} \simeq 0$ and Eq. (3.30) can be simplified as:

$$\mathcal{S}_{K_S \pi^0 \gamma} \simeq -2 \frac{|A_L^{\rm SUSY}/A_L^{\rm SM}|}{1 + |A_L^{\rm SUSY}/A_L^{\rm SM}|^2} \sin(2\phi_1 - \theta_R^{\rm SUSY} - \delta_{12}) \quad (3.35)$$

where $\delta_{12}$ is the strong phase difference between SM and SUSY amplitude.

---

[7]Here, the left-handed amplitudes correspond to the quark level process $\bar{b}_L \to \bar{s}_R \gamma_R$ and $b_R \to s_L \gamma_L$ while the right-handed amplitudes correspond to $\bar{b}_R \to \bar{s}_L \gamma_L$ and $b_L \to s_R \gamma_R$.

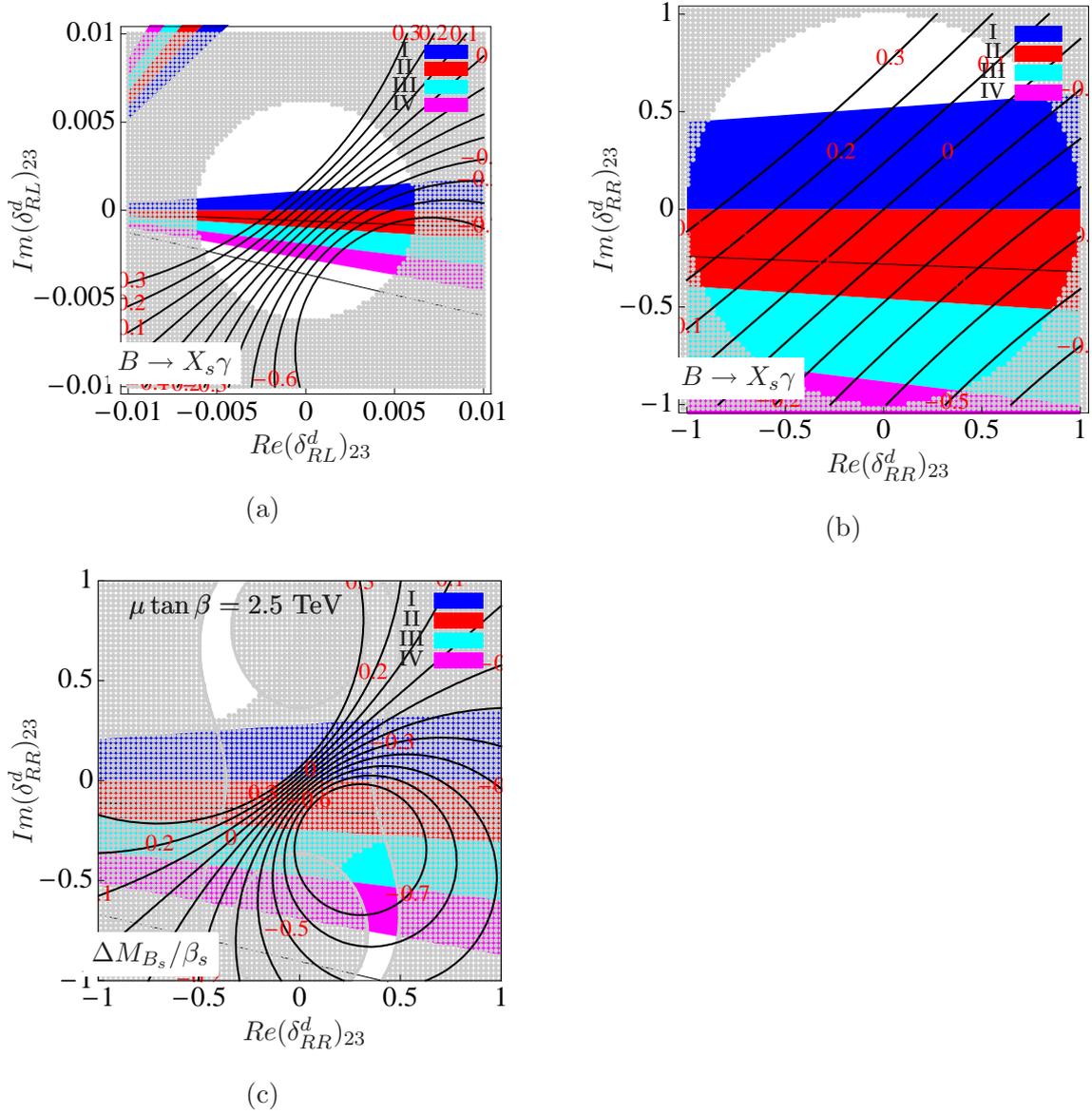

Figure 3.5: The SUSY contribution to the time-dependent CP asymmetry of $B \to K_S \pi^0 \gamma$ is shown as black lines labeled with the value of $\mathcal{S}_{K_S\pi^0\gamma}$ in red. The interval of the lines represents a $\pm 0.1$ uncertainty which can be easily achieved by the Super $B$ factory. Note that the current experimental bound is $\mathcal{S}_{K_S\pi^0\gamma} = -0.15 \pm 0.20$ (we show our result in a range allowed by including $3\sigma$ error). In Figure (a) and (b), we overlap the results for $\Delta\mathcal{S}_f$ (colored band) and $B \to X_s\gamma$ (patched region) (i.e. Figure 3.3 (b) and (c), respectively). We can see that the future precise measurement of $\mathcal{S}_{K_S\pi^0\gamma}$ together with $\mathcal{S}_{\phi K_S, \eta' K_S}$ will allow us to pin down both the real and imaginary parts of the mass insertions $(\delta^d_{RL})_{23}$ and also $(\delta^d_{RR})_{23}$. In Figure (c), we show the result for a moderate $\tan\beta$ scenario on top of the previous result on the $RR$ constraint from $\Delta\mathcal{S}_f$ (colored band) and $\Delta M_{B_s}\beta_s$ (patched region) (see also the caption of Figure 3.4 (c)). If the recent results on the CP violating phase in $B_s$ oscillation are confirmed (see Eq. 3.28), we can see that $\mathcal{S}_{B\to K_S\pi^0\gamma}$ can reach -0.5, which will be observed as a clear signal of new physics at Super $B$ factory.

The dominant gluino contribution comes from the electro-magnetic operator $\tilde{O}_\gamma$ whose Wilson coefficient is given in Eq. 3.12. With $m_{\tilde{g}} \simeq m_{\tilde{q}} = 500$ GeV, we find

$$\frac{A^{\text{SUSY}}(K_S\pi^0\gamma)}{A^{\text{SM}}(K_S\pi^0\gamma)} \simeq (0.28 + 0.0057i)(\delta^d_{RR})_{23} + (134 + 2.7i)(\delta^d_{RL})_{23} + (80 + 1.6i)(\delta^d_{RL})_{33}(\delta^d_{RR})_{23}. \tag{3.36}$$

In Figure 3.5 (a) and (b), we show the SUSY contribution to the time-dependent CP asymmetry (black lines labeled with the value of $\mathcal{S}_{B\to K_S\pi^0\gamma}$ in red). The line intervals represent a 0.1 accuracy which can be achieved by the Super $B$ factory with 5 ab$^{-1}$. Detailed estimates of the sensitivities are given in Sect. 5.3. Note that the current experimental bound is $\mathcal{S}_{K_S\pi^0\gamma} = -0.15 \pm 0.20$ (we show a range allowed by including 3$\sigma$ error). We can see that the future precise measurement of $\mathcal{S}_{K_S\pi^0\gamma}$ will constrain $(\delta^d_{RR})_{23}$ and $(\delta^d_{RL})_{23}$ mass insertions in a very narrow band going roughly in the diagonal direction (this direction comes from the $\phi_1$ term in Eq. (3.35)). Both the real and imaginary parts of the mass insertions can be constrained by combining the bounds from $\Delta \mathcal{S}_{\phi K_S, \eta' K_S}$.

In Figure 3.5 (c), we show the result of $\mathcal{S}_{K_S\pi^0\gamma}$ for the moderate $\tan\beta$ scenario on top of the previous result on the $RR$ constraint from $\Delta \mathcal{S}_f$ (colored band) and $\Delta M_{B_s}/\beta_s$ (patched region) (see also the caption of Figure 3.4 (c)). It is interesting to note that the current experimental bound on $\mathcal{S}_{B\to K_S\pi^0\gamma}$ excludes one of the allowed region for $\Delta M_{B_s}$ in the positive $Im(\delta^d_{RR})_{23}$. In the other allowed region $\mathcal{S}_{B\to K_S\pi^0\gamma}$ can reach -0.5, which would be a clear signal of new physics at Super $B$ factory.

### 3.4.6 SUSY models beyond leading logarithm (large $\tan\beta$ scenarios)

We have so far discussed the MIA parameters at tree level, however, subtlety arises once we include higher order corrections in the quark mass. The superpotential in exact supersymmetry is not renormalized by radiative correction. Renormalization of the canonical Yukawa coupling is purely a consequence of the wave function renormalization due to the corrections in the Kähler potential. However, this is not true once we include the soft SUSY breaking terms. Corrections to the original Yukawa coupling which are not multiplicative arise due to the SUSY breaking and they distort original flavor structure in the superpotential. This leads to corrections in the quark mass matrices as,

$$m_d = m_d^{(0)} + \delta m_d = \text{diag}(m_d, m_s, m_b), \tag{3.37}$$
$$m_u = m_u^{(0)} + \delta m_u = \text{diag}(m_u, m_c, m_t). \tag{3.38}$$

Here, $m_{d,u}^{(0)}$ are the tree level mass and $\delta m_{d,u}$ are correction from the SUSY breaking effect. Note that $\delta m_{d,u}$ are in general not diagonal in the physical super-CKM basis[8] and therefore $m_{d,u}^{(0)}$ are also not diagonal. The $F$-term contribution related by supersymmetry to the Yukawa coupling in the superpotential and the quark mass should be replaced by $m_{d,u}^{(0)}$. Yukawa coupling which appears in the Higgs or higgsino vertices also need to be estimated using $m_{d,u}^{(0)}$ and not the diagonal masses in the physical super-CKM basis. This introduces new sources of flavor mixing which do not appear at tree level. Formally these corrections are classified as part of a correction at the next-to-leading logarithm order (NLLO) in estimates of purely loop induced processes like $b \to s\gamma$, whose completion requires to include various two loop self energy and vertex corrections. However, these corrections beyond leading logarithmic order (BLLO) are enhanced by $\tan\beta$ and the SUSY breaking trilinear coupling which is not the case for the higher loop corrections.

---
[8]The basis using here is not the same as the ones applied in the previous sections. The detailed definition of the so-called physical super-CKM basis can be found in [83–87].

In the following, we discuss the constraints including the BLLO correction to the flavour observables. As described above these contributions are particularly important for large $\tan\beta$, conversely, the value of $\tan\beta$ can be constrained by these flavour observables. Figure 3.6 shows $\tan\beta$ dependence of the FCNC constraints on single insertion of $(\delta_{RL}^d)_{23}$ and $(\delta_{RR}^d)_{23}$. For the SUSY mass parameters, we choose $m_{\tilde{g}} = m_{\tilde{q}} = 1$ TeV and $\mu = -A_t = 500$ GeV and $m_A = 500$ GeV. For the other gaugino masses, the GUT relation is assumed.

In Fig. 3.6, the yellow regions are excluded due to the upper bound on $\mathcal{B}(\bar{B} \to X_s\gamma)$. Different from the analyses in the previous sections, the MFV contributions to $C_{\gamma,g}$ involving charged Higgs or chargino are significant in large $\tan\beta$ scenarios. The former always add up to the SM contribution, while the latter interfere either constructively ($\mu < 0$) or destructively ($\mu > 0$) with it, which determines the base line of $\mathcal{B}(\bar{B} \to X_s\gamma)$ at $\delta^d$'s =0 for given SUSY mass parameters. In Fig 3.6 we choose $\mu > 0$ which is favored by the anomaly in muon $g-2$ experiment at present. The pattern of the gluino contributions which comes from the $\delta_{AB}^d$ mass insertions was already discussed in the previous sections. It should be noted that for the $AB = LL$ mass insertion, $\delta_{LL}^d$ and $\delta_{LL}^u$ are not independent variables due to the $SU(2)$ symmetry and $\delta_{LL}^d$ also generates the chargino contribution on top of the MFV effect. Most important feature of having the BLLO correction is the partial cancellation between the gluino (LLO and BLLO) and chargino (BLLO) contribution for large $\tan\beta$ and $\mu > 0$. This *focusing effect* [83–86] is responsible for the unusual relaxation of the constraint at large $\tan\beta$ in Fig. 3.6. For $\delta_{LL}^d$ this effect is undermined by the LLO chargino contribution induced by $\delta_{LL}^u$.

Next, we discuss the constraint on the intermediate $\tan\beta$ region, where as can be seen in Figure 3.6, the $B_s \to \mu^+\mu^-$ process (orange) and the $B_s$ oscillation parameter $\Delta M_{B_s}$ (red) provide strong constraints. The current upper bound for $B_s \to \mu^+\mu^-$ is given by,

$$\mathcal{B}(B_s \to \mu^+\mu^-) < 8 \times 10^{-8}, \tag{3.39}$$

at 90% confidence level. The SM prediction is extremely small due to the chiral suppression factor $(m_\mu/m_{B_s})^2$:

$$\mathcal{B}(B_s \to \mu^+\mu^-) = (3.46 \pm 1.5) \times 10^{-9}. \tag{3.40}$$

In the SUSY scenario, this suppression can be compensated by a large $\tan\beta$. In particular, the Higgs mediated contribution can generate Wilson coefficients of scalar and pseudo scalar operators in which chirality flip is enhanced by $\tan^3\beta$. Their contributions to $\mathcal{B}(B_s \to \mu^+\mu^-)$ are thus proportional to $\tan^6\beta$. The Higgs mediated contributions to $B_s \to \mu^+\mu^-$ are quite interesting because they do not decouple in the heavy superpartner limit as long as the Higgs sector remains light. This MFV contribution determines the base line of $\mathcal{B}(B_s \to \mu^+\mu^-)$ at $\delta^d$'s =0 in Figure 3.6. The present bound does not impose any constraint on $\tan\beta$ for the given mass parameters. Once we turn on the MIA parameters, gluino loop diagram also starts to contribute to the non holomorphic Yukawa coupling. The current bound on $\mathcal{B}(B_s \to \mu^+\mu^-)$ excludes considerable part of intermediate $\tan\beta$ region with $\delta_{RR,RL}^d$ for which the BLLO $\mathcal{B}(\bar{B} \to X_s\gamma)$ is not sensitive. As future projection, we also plot a curve for $\mathcal{B}(B_s \to \mu^+\mu^-) = 5 \times 10^{-9}$ which is just above the SM prediction. The process is quite sensitive to $\tan\beta$, while it has relatively mild dependence on $\delta^d$'s. Therefore it is rather suitable to use for determining $\tan\beta$ by combining with other processes sensitive to $\delta^d$'s.

The SUSY contribution to $\Delta M_{B_s}$ emerges from the box diagrams and the double Higgs penguin diagram. The MFV contribution, which determines the base line at $\delta^d$'s = 0, comes from $W$ boson, charged Higgs and chargino diagrams. The gluino diagram gives dominant contribution for finite $\delta^d$'s. The double Higgs penguin diagram exchanges neutral Higgs bosons between the two Higgs FCNC vertices. Thus it is enhanced by $\tan^4\beta$ and dominates the process at large $\tan\beta$. Among three choices for external quark chirality, only $(\bar{s}_L b_R)(\bar{s}_R b_L)$ is allowed. Thus in the MFV, the process is suppressed by $m_s/m_b$. The MIA parameters $\delta^d_{LL,LR}$ do not change the situation. However once we include $\delta^d_{RR,RL}$, we can avoid the suppression factor $m_s/m_b$ in $\bar{s}_R b_L$ vertex. Combined with the MFV contribution in $\bar{s}_L b_R$ vertex, we can obtain rather strong bounds on $\delta^d_{RR,RL}$ from $\Delta M_{B_s}$ as shown in Figure 3.6, despite theoretical ambiguity in the SM prediction. The structure at moderate $\tan\beta$ emerges as a result of interference between the gluino box diagrams and the double Higgs penguin diagram. The contours at large $|\delta^d_{RR}|$ have mixing amplitude of opposite sign compared to the SM, which is caused by the gluino box diagram with a very light bottom squark.

Finally, we discuss the impact of the future measurements on the time-dependent CP asymmetry in $B \to K^* \gamma$. The MIA parameters $\delta^d_{RR,RL}$ give new contributions to $\tilde{C}_{\gamma,g}$ making the asymmetry larger as discussed in the previous sections. It has a relatively mild $\tan\beta$ dependence and can play a complementary role to the Higgs mediated FCNC's. We can see in Figure 3.6 that the remaining window at relatively small $\tan\beta$ can be nicely constrained by the time-dependent CP asymmetry in $B \to K^* \gamma$. The contours of $S_{B \to K^* \gamma}$ are shown as dashed lines in intervals of 0.1. Note that the expected sensitivity of Super $B$ factory is expected to be around 0.1 with 5 ab$^{-1}$.

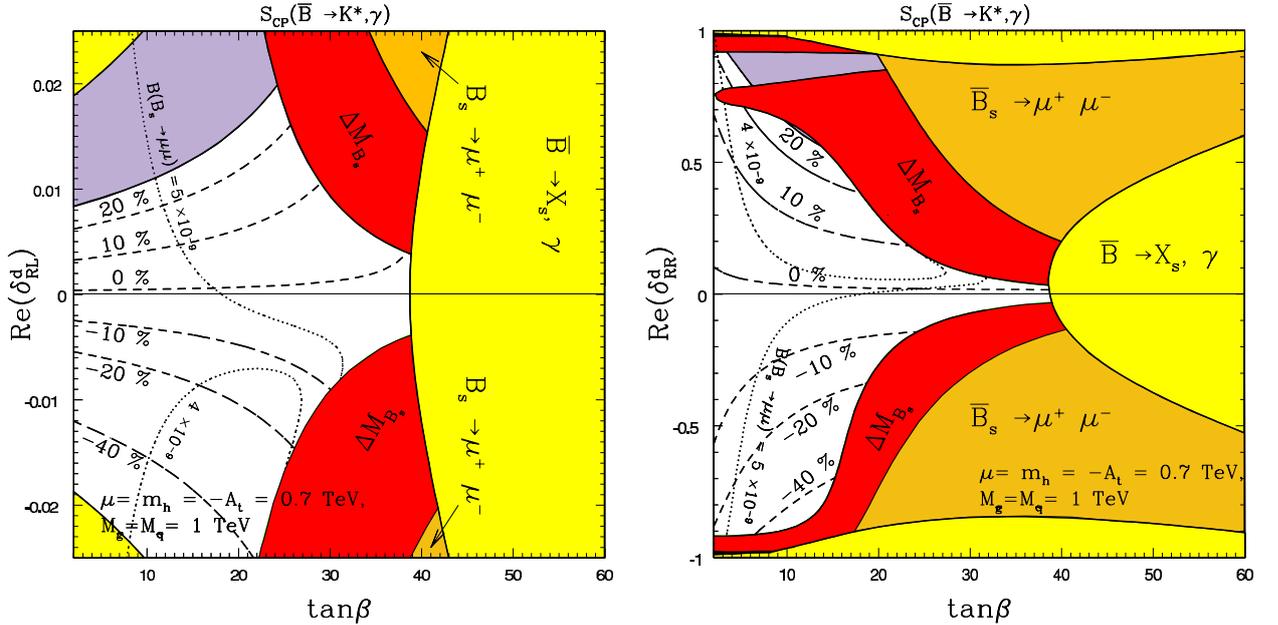

Figure 3.6: Constraints on the mass insertion parameters. Measurements of $\mathcal{B}(\bar{B} \to X_s \gamma)$ (light shaded/yellow), $\mathcal{B}(B_s \to \mu^+ \mu^-)$ (medium shaded/orange) and $\Delta M_{B_s}$ (dark shaded/red) are considered. The contours of $S_{B \to K^* \gamma}$ are shown in the intervals of 0.1.

## 3.5 Model independent study of $b \to s\gamma$ and $b \to sl^+l^-$ processes

In previous sections we considered the supersymmetric extension of the Standard Model and discussed its possible effects on the $b \to s$ transition processes. In contrast, in this section we explain what we can learn from the experimental data in a model independent way. To be specific, we consider the $b \to s\gamma$ and $b \to sl^+l^-$ decays [88–91].

As we discuss below, there are many types of interactions which may contribute to $b \to s\gamma$ and $b \to sl^+l^-$. The new physics effects originate from the energy scale higher than the electro-weak scale, and below that new physics scale, the effects evolve down to the electro-weak scale. Then, at the electro-weak scale, they are matched onto the low energy effective theory, which is valid below the electro-weak scale. The new physics effects can thus be expressed in terms of higher dimensional operators such as four-fermion interactions and dimension-five interactions. For $b \to sl^+l^-$ and $b \to s\gamma$ processes, $\bar{s}b\bar{l}l$ ($O_9$ and $O_{10}$), $\bar{s}\sigma_{\mu\nu}bF^{\mu\nu}$ ($O_7$) and $\bar{s}\sigma_{\mu\nu}T^a bG^a_{\mu\nu}$ ($O_8$) are such operators (there are several of these operators with different chiral structures). For each operator there is the Wilson coefficient $C_i^{\text{eff}}$, in which the new physics effects are encoded. The four-quark operators can contribute to $b \to sl^+l^-$ through the one-loop matrix elements and the operator mixing [91]. We assume that such effects are small compared to the contributions from the tree level matrix elements of $\bar{s}b\bar{l}l$ operators.

At the leading logarithmic approximation of QCD corrections, we obtain the following amplitude for $b \to s\gamma$.

$$M(b \to s\gamma) = \frac{4G_F}{\sqrt{2}} \frac{e}{16\pi^2} V_{ts}^* V_{tb} m_b \left[ C_{7L}^{\text{eff}}(\bar{s}\sigma_{\mu\nu}b_R) + C_{7R}^{\text{eff}}(\bar{s}\sigma_{\mu\nu}b_L) \right] F^{\mu\nu}, \quad (3.41)$$

where $F_{\mu\nu}$ stands for $-i(q_\mu \epsilon_\nu^* - q_\nu \epsilon_\mu^*)$. In the Standard Model $C_{7R}^{\text{eff}} = \frac{m_s}{m_b} C_{7L}^{\text{eff}}$. In the left-right symmetric models [92], $C_{7R}^{\text{eff}}$ can be as large as $C_{7L}^{\text{eff}}$ [93–95]. Using the branching ratio of $B \to X_s\gamma$, we can constrain $|C_{7L}^{\text{eff}}|^2 + |C_{7R}^{\text{eff}}|^2$.

Since the branching fraction does not tell us about the ratio $C_{7R}^{\text{eff}}/C_{7L}^{\text{eff}}$, the observables which are sensitive to the ratio are needed. Three methods using $B$ meson decays have been proposed. One can extract the ratio from the time dependent CP asymmetry of $b \to s\gamma$ [96]. Another measurement which is sensitive to the ratio is the transverse polarization of $B \to K^*\gamma$ [97]. This can be measured by using the decay chains $B \to K^*\gamma^* \to K\pi l^+l^-$ [98, 99]. The azimuthal angle distribution is sensitive to the ratio. The distribution at low invariant dilepton mass region must be measured so that the decay amplitude from a $Z$ exchange and box diagram contribution is suppressed. The other method uses $B \to K_{\text{res}}\gamma \to K\pi\pi\gamma$ and triple momentum correlation $p_\gamma \cdot (p_K \times p_\pi)$ [100, 101].

We next consider the possible new physics effects on $b \to sl^+l^-$. In addition to $C_7^{\text{eff}}$, there are ten local four Fermi interactions which contribute to $b \to sl^+l^-$. One can write down the amplitude including all the contributions [89, 90]:

$$\begin{aligned}
\mathcal{M}(b \to sl^+l^-) &= \frac{G_F \alpha}{\sqrt{2}\pi} V_{ts}^* V_{tb} \bigg\{ (C_{LL} + C_9^{\text{eff}} - C_{10})(\bar{s}_L \gamma_\mu b_L)(\bar{l}_L \gamma^\mu l_L) \\
&\quad + (C_{LR} + C_9^{\text{eff}} + C_{10})(\bar{s}_L \gamma_\mu b_L)(\bar{l}_R \gamma^\mu l_R) \\
&\quad + C_{RL}(\bar{s}_R \gamma_\mu b_R)(\bar{l}_L \gamma^\mu l_L) + C_{RR}(\bar{s}_R \gamma_\mu b_R)(\bar{l}_R \gamma^\mu l_R) \\
&\quad + C_{LRLR}(\bar{s}_L b_R)(\bar{l}_L l_R) + C_{RLLR}(\bar{s}_R b_L)(\bar{l}_L l_R) \\
&\quad + C_{LRRL}(\bar{s}_L b_R)(\bar{l}_R l_L) + C_{RLRL}(\bar{s}_R b_L)(\bar{l}_R l_L) \\
&\quad + C_T(\bar{s}\sigma_{\mu\nu}b)(\bar{l}\sigma^{\mu\nu}l) + iC_{TE}(\bar{s}\sigma_{\mu\nu}b)(\bar{l}\sigma_{\alpha\beta}l)\epsilon^{\mu\nu\alpha\beta}
\end{aligned}$$

$$- i2m_b \left[ C_{7L}^{\text{eff}}(\bar{s}\sigma_{\mu\nu}b_R) + C_{7R}^{\text{eff}}(\bar{s}\sigma_{\mu\nu}b_L) \right] (\bar{l}\gamma^\mu l)\frac{q^\nu}{q^2} \Bigg\}. \qquad (3.42)$$

The Standard Model predicts

$$(C_{LL}, C_{LR}, C_{RR}, C_{RL}, C_{RLRL}, C_{LRLR}, C_{LRRL}, C_{RLLR}) = 0, \qquad (3.43)$$

$$(C_{7R}^{\text{eff}}, C_{7L}^{\text{eff}}) = (\frac{m_s}{m_b}, 1)C_{7\text{SM}}^{\text{eff}}, \qquad (3.44)$$

$$(C_T, C_{TE}) = 0, \qquad (3.45)$$

where $(C_9^{\text{eff}}, C_{10}, C_{7SM}^{\text{eff}})$ are the Standard Model coefficients. The numerical values of the corresponding Wilson coefficients are $C_9^{\text{NDR}} = 4.153$, $C_{10} = -4.546$, $C_{7\text{SM}}^{\text{eff}} = -0.311$. $C_9^{\text{eff}}$ is close to $-C_{10}$. Two-loop calculation has also been completed recently [102, 103]. Beyond the Standard Model, one-loop calculation of these Wilson coefficients is available for the Minimal Supersymmetric Standard Model (see, for example, [104] for a recent paper).

The branching ratio of $b \to sl^+l^-$ is most sensitive to the coefficient $C_{LL}$, since the interference of the $C_{LL}$ and $C_9^{\text{eff}} - C_{10}$ is large. Depending on the sign of $C_{LL}$, the interference with the Standard Model contribution can be constructive or destructive.

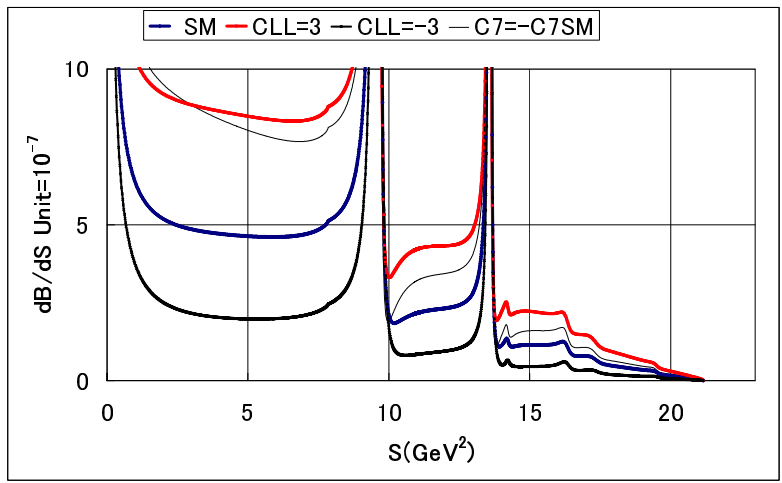

Figure 3.7: Differential branching ratio $\frac{dB}{ds}$ for $b \to sl^+l^-$.

In Figures 3.7 and 3.8, we show the differential branching ratio $\frac{dB}{ds}$ and the forward-backward asymmetry $\frac{dA}{ds}$, respectively. We choose three sets of the coefficients and show dilepton mass squared ($s$) distribution. The three cases correspond to: (1) $C_{LL} = 3$, (2) $C_{LL} = -3$, and (3) $(C_{7R}^{\text{eff}}, C_{7L}^{\text{eff}}) = -(\frac{m_s}{m_b}, 1) C_{7SM}$. When $C_{LL}$ is positive, the branching ratio is larger than the Standard Model value. If $C_{LL}$ is negative, it decreases the branching ratio. If we change the sign of $C_7$ compared with the Standard Model, the branching ratio also increases as in the case of $C_{LL} > 0$. The case (3) can be clearly distinguished from the Standard Model by studying the forward-backward asymmetry.

As shown in Figure 3.8, in the Standard Model there is a zero crossing point of the forward-backward asymmetry in the low invariant mass squared region. If the sign of $C_{7L}$ is different from the Standard Model, the zero crossing point may disappear from the forward-backward asymmetry. Such a scenario is suggested in a supergravity model [105]. In another new physics model scenario including the down type SU(2) singlet quark, $Z$ FCNC current may contribute to $b \to sl^+l^-$ at the tree level. The tree level $Z$ FCNC contribution may give a large contribution to $C_{10}$ and can change the sign of the forward-backward asymmetry. Besides the observables mentioned above, the forward-backward CP asymmetry of $b \to sl^+l^-$ is useful under the presence of new source of CP violation [106]. To conclude, with various observables and combination of them, we can test the new physics scenarios if they contribute to $b \to sl^+l^-$ transitions. The estimates of the sensitivity at Super KEKB are discussed in Sects. 5.3.7 and 5.3.8.

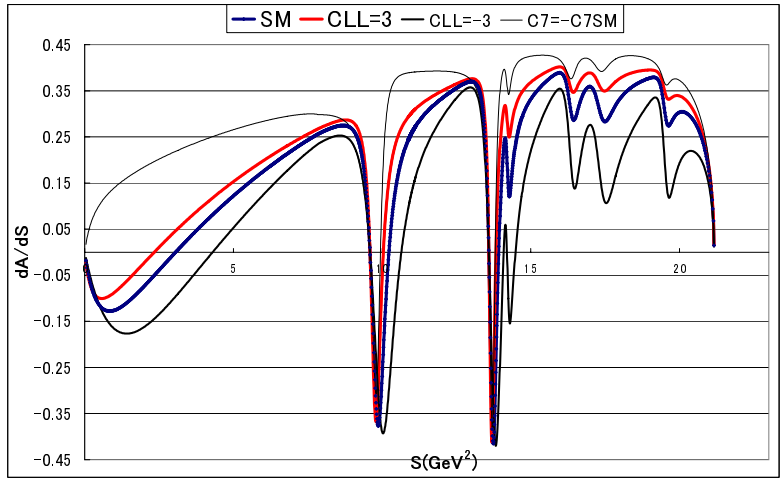

Figure 3.8: Forward-backward asymmetry $\frac{dA}{ds}$ for $b \to sl^+l^-$.

## 3.6 Lepton Flavor Violation

Strictly speaking, the Standard Model already has to be modified by introducing tiny neutrino masses, in order to incorporate the Lepton Flavor Violating (LFV) phenomena observed in the neutrino sector. Neutrino mixing $\nu_\mu$-$\nu_\tau$ was first discovered in atmospheric neutrino measurements at Super-Kamiokande [107], and was further confirmed by the K2K experiment [108]. The neutrino oscillation was also confirmed in solar neutrinos, which come from $\nu_e$-$\nu_\mu$ mixing, by both the Super-Kamiokande [109] and SNO [110, 111] experiments. More recently, the Kamland experiment pinned down the explanation of the solar neutrino problem to the large mixing angle solution [112]. It is very interesting that the $\nu_e$-$\nu_\mu$ and $\nu_\mu$-$\nu_\tau$ mixing angles are found to be almost maximal and the neutrino mass structure is quite different from that of the quark sector.

Now it is known that lepton-flavor symmetries are not exact in Nature. However, the magnitude of LFV processes in the charged lepton sector is not obvious. The tiny neutrino masses do not lead to sizable LFV processes in the charged lepton sector, since the event rates are suppressed by the fourth power of $(m_\nu/m_W)$. Thus, searches for LFV in the charged lepton sector will probe physics beyond the Standard Model and the origin of the neutrino masses.

The $\tau$ lepton is a member of the third generation and is the heaviest charged lepton. It can decay into quarks and leptons of the first and second generation. This may imply that $\tau$ lepton physics could provide some clues to puzzles in the family structure. In fact, one naively expects the heavier quarks and leptons to be more sensitive to the dynamics responsible for fermion mass generation.

In the following we discuss LFV $\tau$ lepton decays in SUSY models, especially in the supersym-metric seesaw mechanism and SUSY GUTs, and other models such as extra-dimension models and R-parity violating SUSY models.

### 3.6.1 LFV in the Supersymmetric Models

Lepton flavour violation in the SUSY extension of the Standard Model comes from the soft SUSY breaking terms, since the supersymmetric interactions have the same flavor structure as in the Standard Model. The soft SUSY breaking terms in the lepton sector are

$$-\mathcal{L} = (m_E^2)_{ij}\tilde{e}_{Ri}^\dagger \tilde{e}_{Rj} + (m_L^2)_{ij}\tilde{l}_{Li}^\dagger \tilde{l}_{Lj} + \left[(A_E)_{ij}H_1\tilde{e}_{Ri}^\dagger \tilde{l}_{Lj} + \text{h.c.}\right], \tag{3.46}$$

where $(m_E^2)_{ij}$ and $(m_L^2)_{ij}$ are mass matrices for the right-handed sleptons $\tilde{e}_{Rj}$ and the left-handed sleptons $\tilde{l}_{Lj}(\equiv (\tilde{\nu}_{Lj}, \tilde{e}_{Lj}))$, respectively. $(A_E)_{ij}$ is the trilinear scalar coupling matrix.

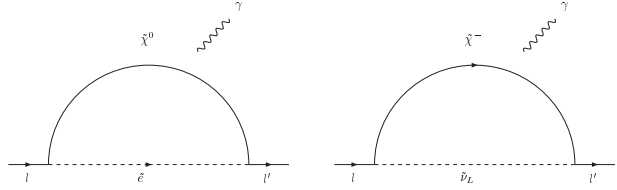

Figure 3.9: Feynman diagrams to generate $l^- \to l'^-\gamma$ in the SUSY models with conserved R parity. $\tilde{e}$, $\tilde{\nu}$, $\tilde{\chi}^0$ and $\tilde{\chi}^-$ represent charged slepton, sneutrino, chargino, and neutralino, respectively.

LFV processes for charged leptons are radiative if R parity is conserved, since the SUSY interactions must be bilinears of the SUSY fields. Thus, $\tau^- \to \mu^-(e^-)\gamma$ and $\mu^- \to e^-\gamma$ are the most sensitive to the flavor structure of the soft SUSY breaking terms except for some exceptional cases. These processes are generated by diagrams in Figure 3.9. The effective operators relevant to $l^- \to l'^-\gamma$ are flavor-violating magnetic moment operators,

$$-\mathcal{L}_{\text{eff}} = \sum_{l>l'} \frac{4G_F}{\sqrt{2}} \left[ m_l A_R^{ll'} \bar{l}\sigma^{\mu\nu}P_R l' + m_l A_L^{ll'} \bar{l}\sigma^{\mu\nu}P_L l' \right] F_{\mu\nu} + h.c., \tag{3.47}$$

where $P_{L/R} = (1 \mp \gamma_5)/2$. The branching ratios are then

$$\mathcal{B}(l^- \to l'^-\gamma) = 384\pi^2(|A_R^{ll'}|^2 + |A_L^{ll'}|^2)\,\mathcal{B}(l^- \to l'^-\nu_l\overline{\nu}_{l'}). \tag{3.48}$$

Here, $\mathcal{B}(\tau^- \to \mu^-(e^-)\nu_\tau\overline{\nu}_{\mu(e)}) \simeq 0.17$ and $\mathcal{B}(\mu^- \to e^-\nu_\mu\overline{\nu}_e) = 1$.

The coefficients in (3.47) are approximately given as

$$A_R^{\tau l'} = \frac{\sqrt{2}e}{4G_F}\frac{\alpha_2}{4\pi}\frac{\tan\beta}{m_{\text{SUSY}}^2}\left[-\frac{t_W^2}{120}\delta_{\tau l'}^R\right], \tag{3.49}$$

$$A_L^{\tau l'} = \frac{\sqrt{2}e}{4G_F}\frac{\alpha_2}{4\pi}\frac{\tan\beta}{m_{\text{SUSY}}^2}\left[(\frac{1}{30}+\frac{t_W^2}{24})\delta_{\tau l'}^L\right], \tag{3.50}$$

$$A_R^{\mu e} = \frac{\sqrt{2}e}{4G_F}\frac{\alpha_2}{4\pi}\frac{\tan\beta}{m_{\text{SUSY}}^2}\left[-\frac{t_W^2}{120}\delta_{\mu e}^R + \frac{t_W^2}{120}\delta_{\mu\tau}^R\delta_{\tau e}^R - \frac{t_W^2}{60}\frac{m_\tau}{m_\mu}\delta_{\mu\tau}^L\delta_{\tau e}^R\right], \tag{3.51}$$

$$A_L^{\mu e} = \frac{\sqrt{2}e}{4G_F}\frac{\alpha_2}{4\pi}\frac{\tan\beta}{m_{\text{SUSY}}^2}\left[(\frac{1}{30}+\frac{t_W^2}{24})\delta_{\mu e}^L - (\frac{1}{80}+\frac{7t_W^2}{240})\delta_{\mu\tau}^L\delta_{\tau e}^L - \frac{t_W^2}{60}\frac{m_\tau}{m_\mu}\delta_{\mu\tau}^R\delta_{\tau e}^L\right], \tag{3.52}$$

assuming for simplicity that all SUSY particle masses are equal to $m_{\text{SUSY}}$ and $\tan\beta \gg 1$. Here, $t_W \equiv \tan\theta_W$, where $\theta_W$ is the Weinberg angle, and the mass insertion parameters are given as

$$\delta_{ll'}^R = \frac{(m_E^2)_{ll'}}{m_{\text{SUSY}}^2}, \qquad \delta_{ll'}^L = \frac{(m_L^2)_{ll'}}{m_{\text{SUSY}}^2}. \tag{3.53}$$

When both the 13 and 23 components in the slepton mass matrices are non-vanishing, $\mu^- \to e^-\gamma$ is generated via a scalar $\tau$ lepton exchange. In particular, if both the left-handed and right-handed mixings are sizable, the branching ratio is enhanced by $(m_\tau/m_\mu)^2$ compared to the case where only left-handed or right-handed mixing angles are non-vanishing. The off-diagonal components in $(A_E)_{ij}$ are sub-dominant in these processes since the contribution is not proportional to $\tan\beta$.

We list constraints on $\delta_{ll'}^R$ and $\delta_{ll'}^L$ from current experimental bounds on $\mathcal{B}(\tau^- \to \mu^-(e^-)\gamma)$, which are derived by the Belle experiment [113], and $\mathcal{B}(\mu^- \to e^-\gamma)$ [114] in Table 3.2. In this table, we take $\tan\beta = 10$ and $m_{\text{SUSY}} = 100$ GeV and 300 GeV.

The constraints from $\mu^- \to e^-\gamma$ on the slepton mixings are quite stringent. Independently, the current bounds on the LFV $\tau$ lepton decay modes independently give serious constraints on $|\delta_{\tau\mu}^L|$ and $|\delta_{\tau e}^L|$. Furthermore, the current constraint on $|\delta_{\mu\tau}^L \delta_{\tau e}^L|$ from the LFV $\tau$ lepton decay is now comparable to that from the LFV muon decay. The MEG experiment to search for $\mu^+ \to e^+\gamma$ published preliminary results in [115], and the branching ratio sensitivity may be improved by two orders of magnitude in the experiment [116]. It is also desirable for the LFV $\tau$ lepton decay searches to be furthermore improved.

| $m_{\text{SUSY}}$ | $|\delta_{\tau\mu}^L|$ | $|\delta_{\tau e}^L|$ | $|\delta_{\mu e}^L|$ | $|\delta_{\mu\tau}^L \delta_{\tau e}^L|$ | $|\delta_{\mu\tau}^R \delta_{\tau e}^L|$ |
|---|---|---|---|---|---|
| 100 GeV | $7 \times 10^{-3}$ | $1 \times 10^{-2}$ | $4 \times 10^{-5}$ | $1 \times 10^{-4}$ $(8 \times 10^{-5})$ | $2 \times 10^{-5}$ $(2 \times 10^{-3})$ |
| 300 GeV | $6 \times 10^{-2}$ | $1 \times 10^{-1}$ | $4 \times 10^{-4}$ | $9 \times 10^{-4}$ $(7 \times 10^{-3})$ | $2 \times 10^{-4}$ $(1 \times 10^{-1})$ |
| $m_{\text{SUSY}}$ | $|\delta_{\tau\mu}^R|$ | $|\delta_{\tau e}^R|$ | $|\delta_{\mu e}^R|$ | $|\delta_{\mu\tau}^R \delta_{\tau e}^R|$ | $|\delta_{\mu\tau}^L \delta_{\tau e}^R|$ |
| 100 GeV | $1 \times 10^{-1}$ | $2 \times 10^{-1}$ | $9 \times 10^{-4}$ | $9 \times 10^{-4}$ $(3 \times 10^{-2})$ | $2 \times 10^{-5}$ $(2 \times 10^{-3})$ |
| 300 GeV | $1$ | $2$ | $8 \times 10^{-3}$ | $8 \times 10^{-3}$ $(2)$ | $2 \times 10^{-4}$ $(1 \times 10^{-1})$ |

Table 3.2: Constraints on $\delta_{ll'}^R$ and $\delta_{ll'}^L$ from experimental bounds on $\mathcal{B}(l^- \to l'^-\gamma)$. Here, we use $\mathcal{B}(\tau^- \to \mu^-\gamma) < 4.5 \times 10^{-8}$, $\mathcal{B}(\tau^- \to e^-\gamma) < 1.2 \times 10^{-7}$ [113] and $\mathcal{B}(\mu^- \to e^-\gamma) < 1.2 \times 10^{-11}$ [114]. We take $\tan\beta = 10$ and $m_{\text{SUSY}} = 100$ GeV and 300 GeV. The numbers in parentheses are derived from constraints on $|\delta_{\tau\mu}^{(L/R)}|$ and $|\delta_{\tau e}^{(L/R)}|$.

The flavor structure of the soft SUSY breaking terms depends on the origin of the SUSY breaking and the physics beyond the MSSM, as discussed in Section 3.3. Even in the Universal scalar mass scenario, the LFV Yukawa interaction may induce LFV slepton mass terms radiatively. If the heavier leptons have larger LFV Yukawa interactions, the $\tau$ lepton is the most sensitive to them. In the decoupling scenario, stau may be much lighter than other sleptons and have LFV interactions. Thus, the search for LFV $\tau$ lepton decay is important to probe such new physics. In the next section we present predictions for the LFV $\tau$ lepton decay in the SUSY seesaw model and SUSY GUTs.

Next, we discuss other tau LFV processes in SUSY models. $\tau^- \to \mu^-(e^-)\gamma$ are the largest tau LFV processes, unless they are suppressed by some accidental cancellation or much heavier SUSY particle masses. The LFV $\tau$ lepton decay modes to three leptons are dominantly induced by the photon-penguin contributions, and are correlated with $\tau^- \to \mu^-(e^-)\gamma$,

$$\mathcal{B}(\tau^- \to \mu^- e^+ e^-)/\mathcal{B}(\tau^- \to \mu^- \gamma) \simeq 1/94, \quad (3.54)$$

$$\mathcal{B}(\tau^- \to \mu^- \mu^+ \mu^-)/\mathcal{B}(\tau^- \to \mu^- \gamma) \simeq 1/440, \quad (3.55)$$

$$\mathcal{B}(\tau^- \to e^- e^+ e^-)/\mathcal{B}(\tau^- \to e^- \gamma) \simeq 1/94, \quad (3.56)$$

$$\mathcal{B}(\tau^- \to e^- \mu^+ \mu^-)/\mathcal{B}(\tau^- \to e^- \gamma) \simeq 1/440. \quad (3.57)$$

The LFV $\tau$ decays into pseudoscalar mesons tend to be smaller than those to three leptons since they do not have photon-penguin contributions.

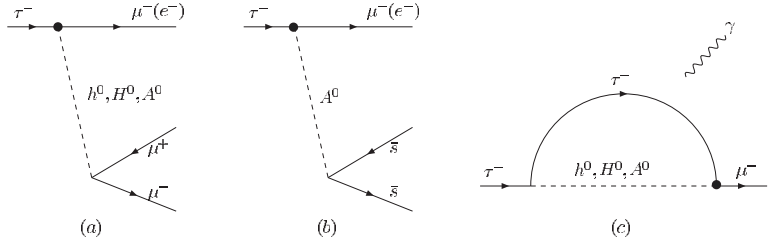

Figure 3.10: Feynman diagrams to generate (a) $\tau^- \to \mu^-(e^-)\mu^+\mu^-$, (b) $\tau^- \to \mu^-(e^-)\eta$, (c) $\tau^- \to \mu^-(e^-)\mu^-\gamma$, induced by anomalous Higgs boson couplings. Dots denote the LFV anomalous Yukawa couplings.

The Yukawa couplings of the Higgs bosons in the MSSM are lepton-flavor conserving at tree level. If sleptons have LFV mass terms, the LFV anomalous Yukawa couplings are generated by radiative correction. The Higgs-mediation contributions to the LFV processes, which are induced by the anomalous couplings, become important when the SUSY particle masses are much heavier than the weak scale. This is because the anomalous couplings are not suppressed by the SUSY breaking scale [117]. In this case the branching ratios of the LFV $\tau$ lepton decay modes have a different pattern from the above.

The anomalous Yukawa couplings for the $\tau$ lepton are given by

$$-\mathcal{L}_{\text{eff}} = \sum_{l=e,\mu} (2G_F^2)^{1/4} \frac{m_\tau}{\cos^2\beta} (\Delta_L^{\tau l} \bar{\tau} P_L l + \Delta_R^{\tau l} \bar{l} P_L \tau) \left(\cos(\beta-\alpha)h^0 - \sin(\beta-\alpha)H^0 - iA^0\right)$$

$$+ \sum_{l=e,\mu} (8G_F^2)^{1/4} \frac{m_\tau}{\cos^2\beta} (\Delta_L^{\tau l} \bar{\tau} P_L \nu_l + \Delta_R^{\tau l} \bar{l} P_L \nu_\tau) H^- + h.c., \quad (3.58)$$

where $\alpha$ is the mixing angle between the CP-even Higgs bosons $h^0$ and $H^0$. $A^0$ and $H^-$ are CP-odd and charged Higgs bosons, respectively. When the 13 and 23 components in the left-handed slepton mass matrix are non-vanishing, $\Delta_L^{\tau l}$ ($l=e,\mu$) are given by

$$\Delta_L^{\tau l} \simeq -\frac{\alpha_2 - \alpha_Y}{16\pi} \delta_{\tau l}^L. \quad (3.59)$$

Here we assumed for simplicity that all SUSY particle masses are equal to $m_{\text{SUSY}}$, though $\Delta_L^{\tau l}$ is not explicitly suppressed by $m_{\text{SUSY}}$ as mentioned above. If the right-handed sleptons have the LFV mass terms, $\Delta_R^{\tau l}$ ($l=e,\mu$) could be non-vanishing. However, it is very sensitive to the SUSY particle mass spectrum, and it is zero due to an accidental cancellation when all SUSY particle masses are taken to be equal. The detailed formulae for $\Delta^{\tau l}_{L/R}$ are provided in Ref. [118].

Diagrams (a-c) in Figure 3.10 induce $\tau^- \to \mu^-(e^-)\mu^+\mu^-$, $\tau^- \to \mu^-(e^-)\eta$, and $\tau^- \to \mu^-(e^-)\gamma$, respectively, due to the LFV anomalous Yukawa couplings in (3.58). The branching ratios are proportional to $\tan^6\beta/m_{A^0}^4$. For example, the approximate formula for $\mathcal{B}(\tau^- \to l^-l'^+l'^-)$ is given as

$$\begin{aligned}\mathcal{B}(\tau^- \to l^-l'^+l'^-) &\simeq \frac{m_\tau^2 m_{l'}^2}{32 m_{A^0}^4}\left(|\Delta_L^{\tau l}|^2 + |\Delta_R^{\tau l}|^2\right)\tan^6\beta\,(3+5\delta_{ll'})\\ &\quad \times \mathcal{B}(\tau^- \to l^-\nu_l\overline{\nu}_\tau),\end{aligned} \qquad (3.60)$$

where $m_{A^0}$ is the CP-odd Higgs boson mass. When $\tan\beta(m_{\rm SUSY}/m_{A^0}) \gtrsim O(10^3)$, the branching ratios of these processes are larger than those of $\tau^- \to \mu^-(e^-)\gamma$ induced by diagrams in Figure 3.9 [117, 119–121].

The ratios of the branching ratios for the LFV processes depend on the Higgs boson masses and also on the MSSM parameters in the Higgs-mediation case. The branching ratio for $\tau^- \to \mu^-\eta$ is about 5 times larger than that for $\tau^- \to \mu^-\mu^+\mu^-$ at the leading order in the perturbation theory. It can be a few times larger or smaller than that, depending on the radiative correction to the bottom and strange quark masses [118]. The branching ratio for $\tau^- \to \mu^-\gamma$ could be the largest in the Higgs-mediation case [121]. However, it is sensitive to the mass splitting between $H^0$ and $A^0$, which is dominated by the radiative corrections in a limit of $m_Z/m_{A^0} \ll 1$. Also, the two-loop diagram induced by the Higgs bosons is not negligible in the process.

### 3.6.2 SUSY Seesaw Mechanism and SUSY GUTs

In general, in seesaw and GUT models, LFV Yukawa interactions are introduced. If the SUSY breaking mediation scale is higher than the GUT [122–124] or the right-handed neutrino mass scale [125–128], sizable LFV processes are predicted as mentioned in the previous section.

The most economical way to generate the tiny neutrino masses is the seesaw mechanism. In the seesaw model, a neutrino Yukawa coupling $Y_\nu$ is introduced, which is lepton-flavor violating. In the supersymmetric extension, the off-diagonal components of the left-handed slepton mass matrix are radiatively induced, and they are approximately given by

$$(\delta m_{\tilde{L}}^2)_{ij} \simeq -\frac{1}{8\pi^2}(3m_0^2 + A_0^2)\sum_k (Y_\nu^\dagger)_{ki}(Y_\nu)_{kj}\log\frac{M_G}{M_{N_k}}, \qquad (3.61)$$

where $M_{N_i}$ is the $i$-th right-handed neutrino mass, and $M_G$ is the Planck scale. Here we assume the gravity mediation scenario, and the parameters $m_0$ and $A_0$ are the universal scalar mass and the universal trilinear scalar coupling. The predicted small neutrino mass matrix is

$$(m_\nu)_{ij} = \sum_k \frac{(Y_\nu)_{ki}(Y_\nu)_{kj}\langle H_2\rangle^2}{M_{N_k}}. \qquad (3.62)$$

Eq. (3.61) has a different structure than (3.62). Thus, we may obtain independent information about the seesaw mechanism from the charged LFV searches and the neutrino oscillation experiments.

In Figure 3.11 we show $\mathcal{B}(\tau^- \to \mu^-\gamma)$ and $\mathcal{B}(\tau^- \to e^-\gamma)$ in the SUSY seesaw mechanism, assuming the gravity mediation scenario for the SUSY breaking. We fix the neutrino Yukawa coupling using the neutrino oscillation data under assumptions for the neutrino Yukawa coupling $Y_\nu$, which suppresses $\mathcal{B}(\mu^- \to e^-\gamma)$. The experimental bounds on these $\tau$ LFV processes have already excluded some of the parameter space. However, the bulk of the parameter space will be covered by the future searches which will reach branching fractions of few times $10^{-9}$ (see Sect. 5.10). While a natural candidate for the largest LFV $\tau$ lepton decay mode is $\tau^- \to \mu^-\gamma$ from the atmospheric neutrino result, some model-parameters in the seesaw model predict larger $\mathcal{B}(\tau^- \to e^-\gamma)$ [129]. This is because (3.61) and (3.62) have different dependences on $Y_\nu$ and $M_N$ as mentioned above.

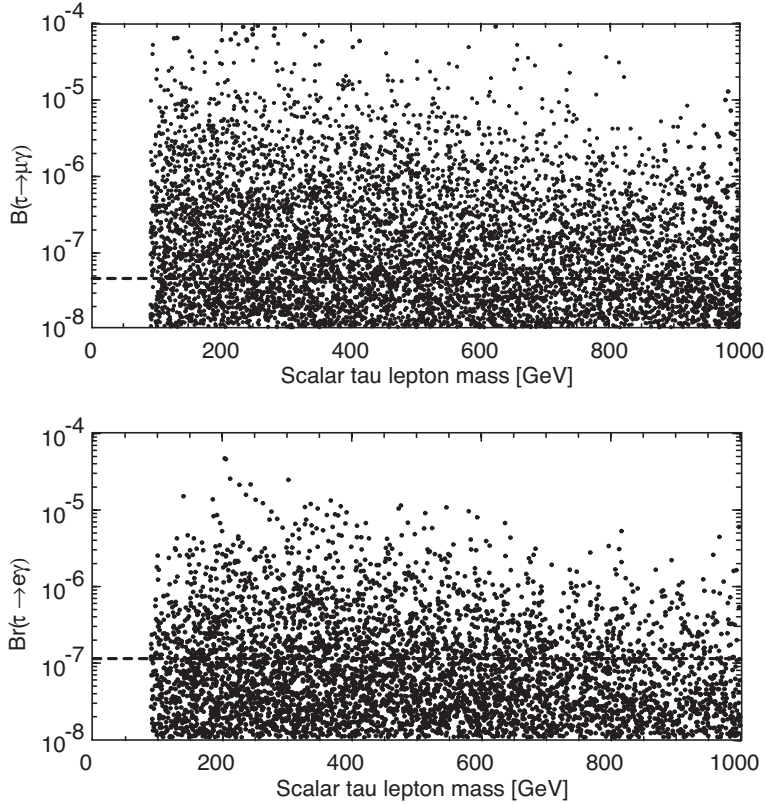

Figure 3.11: $\mathcal{B}(\tau^- \to \mu^-\gamma)$ and $\mathcal{B}(\tau^- \to e^-\gamma)$ in the SUSY seesaw mechanism, assuming the gravity mediation scenario for the SUSY breaking. Dashed lines show the current experimental bounds from the Belle experiment [113]. Here, we take $\tan\beta = 30$, the $SU(2)_L$ gaugino mass 200 GeV, $A_0 = 0$, and positive Higgsino mass. We fix the neutrino Yukawa coupling by using the neutrino oscillation data under assumptions of the neutrino Yukawa coupling.

In GUT models, even if the neutrino Yukawa contribution is negligible, LFV processes are predicted. In the $SU(5)$ SUSY GUT, the right-handed charged leptons are embedded in the **10**-dimensional multiplets with the left-handed quarks and the right-handed up-type quarks. The LFV SUSY breaking terms for the right-handed sleptons are generated by the top-quark Yukawa coupling $Y_t$ above the GUT scale. The off-diagonal components in the right-handed slepton mass matrix are given as

$$(\delta m^2_{\tilde{E}})_{ij} \simeq -\frac{3}{8\pi^2}(3m_0^2 + A_0^2)V_{i3}V_{j3}^*|Y_t|^2 \log\frac{M_G}{M_{\text{GUT}}}, \qquad (3.63)$$

where $V_{ij}$ is the CKM matrix in the SUSY $SU(5)$ GUT and $M_{\text{GUT}}$ is the GUT scale.

In the minimal $SU(5)$ SUSY GUT, in which the neutrino Yukawa coupling is negligible, $\mathcal{B}(\tau^- \to \mu^- \gamma)$ is smaller than $10^{-10} - 10^{-9}$. While the process is enhanced by the top-quark Yukawa coupling, it is suppressed by the CKM matrix element and the $U(1)_Y$ gauge coupling constant. Furthermore, when the right-handed sleptons have LFV mass terms, an accidental cancellation among the diagrams tends to suppress the branching ratio. However, notice that the CKM matrix elements at the GUT scale may not be the same as the ones determined in the low energy data. This is because the quark and lepton mass ratios are not well explained in this model. If $V_{32}$ is larger, the processes are enhanced [130, 131].

### 3.6.3 Other Theoretical Models

In the previous subsections we discussed LFV $\tau$ lepton decays in SUSY models assuming that R parity is conserved. In this subsection we review some examples of other models which predict LFV $\tau$ lepton decays.

In extra-dimension models [8, 9] the "fundamental" scale is expected to be comparable to the weak scale, and the classical seesaw mechanism does not explain the tiny neutrino masses. Instead, singlet neutrinos in the bulk space are introduced [132–134]. The Yukawa couplings of the left-handed neutrinos with the bulk neutrinos are suppressed by an overlap of the wave functions, and small Dirac neutrino masses are predicted.

In this model the loop diagrams from the Kaluza-Klein states of the bulk neutrinos generate the LFV $\tau$ lepton and $\mu$ decay [135, 136]. On the other hand, this model is constrained by short baseline experiments and charged current universality, since the kinetic mixing term for neutrinos is not negligible. In the minimal model, in which the unique source of LFV is the Yukawa coupling of the left-handed neutrinos with the bulk neutrinos, $\mathcal{B}(\tau^- \to \mu^- \gamma) \lesssim 10^{-9}$ from the existing constraints, while $\mathcal{B}(\mu^- \to e^- \gamma) \lesssim 10^{-13} - 10^{-11}$ depending on $U_{e3}$ [136]. In the non-minimal case, the constraints may be looser.

In R-parity violating SUSY models, there exist lepton flavor and baryon number non-conserving interactions at the tree level. Proton stability gives stringent bounds on such models, as well as other processes, such as the decays of charged leptons and $B$ and $D$ mesons, neutral current processes, and FCNC processes. Since the $\tau$ lepton is the heaviest lepton, various $\tau$ LFV modes can be induced. These include $\tau^- \to \mu^- \mu^+ \mu^-$, $\mu^- e^+ e^-$, $\mu^+ e^- e^-$ and $\tau^- \to \mu^- M^0$, $\mu^- V^0$. Here, $M^0$ and $V^0$ are pseudoscalar and vector mesons, respectively. Comprehensive studies of LFV $\tau$ lepton decay have been performed in [137].

If the masses of the right-handed neutrinos are $O(1-10)$ TeV in the seesaw mechanism, sizable LFV $\tau$ lepton decay might be possible [138]. It was found in [139] that $\mathcal{B}(\tau^- \to \mu^- \gamma)$ and $\mathcal{B}(\tau^- \to \mu^- \mu^+ \mu^-)$ are smaller than $10^{-9}$ and $10^{-10}$, respectively. It was, however, also argued that $\mathcal{B}(\tau^- \to e^- \gamma)$ and $\mathcal{B}(\tau^- \to e^- e^+ e^-)$ can reach $10^{-8}$ and $10^{-9}$, respectively, in various models [139].

# References

[1] S. Weinberg, "Implications Of Dynamical Symmetry Breaking," Phys. Rev. D **13**, 974 (1976).

[2] S. Weinberg, "Implications Of Dynamical Symmetry Breaking: An Addendum," Phys. Rev. D **19** (1979) 1277.

[3] L. Susskind, "Dynamics Of Spontaneous Symmetry Breaking In The Weinberg-Salam Theory," Phys. Rev. D **20**, 2619 (1979).

[4] A. G. Cohen, D. B. Kaplan and A. E. Nelson, "Progress in electroweak baryogenesis," Ann. Rev. Nucl. Part. Sci. **43**, 27 (1993).

[5] H. P. Nilles, "Supersymmetry, Supergravity And Particle Physics," Phys. Rept. **110**, 1 (1984).

[6] H. E. Haber and G. L. Kane, "The Search For Supersymmetry: Probing Physics Beyond The Standard Model," Phys. Rept. **117**, 75 (1985).

[7] C. T. Hill and E. H. Simmons, "Strong dynamics and electroweak symmetry breaking," Phys. Rept. **381**, 235 (2003).

[8] N. Arkani-Hamed, S. Dimopoulos and G. R. Dvali, ' 'The hierarchy problem and new dimensions at a millimeter," Phys. Lett. B **429**, 263 (1998).

[9] L. Randall and R. Sundrum, "A large mass hierarchy from a small extra dimension," Phys. Rev. Lett. **83**, 3370 (1999).

[10] J. Hewett and M. Spiropulu, "Particle physics probes of extra spacetime dimensions," Ann. Rev. Nucl. Part. Sci. **52**, 397 (2002).

[11] N. Arkani-Hamed, A. G. Cohen and H. Georgi, "Electroweak symmetry breaking from dimensional deconstruction," Phys. Lett. B **513**, 232 (2001).

[12] H. Georgi and S. L. Glashow, "Unity Of All Elementary Particle Forces," Phys. Rev. Lett. **32**, 438 (1974).

[13] J. C. Pati and A. Salam, "Lepton Number As The Fourth Color," Phys. Rev. D **10**, 275 (1974).

[14] M. Dine, A. E. Nelson and Y. Shirman, "Low-energy dynamical supersymmetry breaking simplified," Phys. Rev. D **51**, 1362 (1995).

[15] M. Dine, A. E. Nelson, Y. Nir and Y. Shirman, "New tools for low-energy dynamical supersymmetry breaking," Phys. Rev. D **53**, 2658 (1996).

[16] M. Dine, Y. Nir and Y. Shirman, "Variations on minimal gauge mediated supersymmetry breaking," Phys. Rev. D **55**, 1501 (1997).

[17] D. E. Kaplan, G. D. Kribs and M. Schmaltz, "Supersymmetry breaking through transparent extra dimensions," Phys. Rev. D **62**, 035010 (2000).

[18] Z. Chacko, M. A. Luty, A. E. Nelson and E. Ponton, "Gaugino mediated supersymmetry breaking," JHEP **0001**, 003 (2000).

[19] M. Schmaltz and W. Skiba, "Minimal gaugino mediation," Phys. Rev. D **62**, 095005 (2000).

[20] L. Randall and R. Sundrum, "Out of this world supersymmetry breaking," Nucl. Phys. B **557**, 79 (1999).

[21] G. F. Giudice, M. A. Luty, H. Murayama and R. Rattazzi, "Gaugino mass without singlets," JHEP **9812**, 027 (1998).

[22] L. J. Hall, V. A. Kostelecky and S. Raby, "New Flavor Violations In Supergravity Models," Nucl. Phys. B **267**, 415 (1986).

[23] Y. Nir and N. Seiberg, "Should squarks be degenerate?," Phys. Lett. B **309**, 337 (1993).

[24] M. Leurer, Y. Nir and N. Seiberg, "Mass matrix models: The Sequel," Nucl. Phys. B **420**, 468 (1994).

[25] M. Dine, A. Kagan and S. Samuel, "Naturalness In Supersymmetry, Or Raising The Supersymmetry Breaking Scale," Phys. Lett. B **243**, 250 (1990).

[26] M. Dine, R. G. Leigh and A. Kagan, "Flavor symmetries and the problem of squark degeneracy," Phys. Rev. D **48**, 4269 (1993).

[27] S. Dimopoulos and G. F. Giudice, "Naturalness constraints in supersymmetric theories with nonuniversal soft terms," Phys. Lett. B **357**, 573 (1995).

[28] A. Pomarol and D. Tommasini, "Horizontal symmetries for the supersymmetric flavor problem," Nucl. Phys. B **466**, 3 (1996).

[29] A. G. Cohen, D. B. Kaplan and A. E. Nelson, "The more minimal supersymmetric standard model," Phys. Lett. B **388**, 588 (1996).

[30] J. Hisano, K. Kurosawa and Y. Nomura, "Large squark and slepton masses for the first-two generations in the anomalous U(1) SUSY breaking models," Phys. Lett. B **445**, 316 (1999).

[31] J. Hisano, K. Kurosawa and Y. Nomura, "Natural effective supersymmetry," Nucl. Phys. B **584**, 3 (2000).

[32] T. Moroi, "CP violation in $B_d \to \phi K_S$ in SUSY GUT with right-handed neutrinos," Phys. Lett. B **493**, 366 (2000).

[33] N. Akama, Y. Kiyo, S. Komine and T. Moroi, "CP violation in kaon system in supersymmetric SU(5) model with seesaw-induced neutrino masses," Phys. Rev. D **64**, 095012 (2001).

[34] S. Baek, T. Goto, Y. Okada and K. i. Okumura, "Muon anomalous magnetic moment, lepton flavor violation, and flavor changing neutral current processes in SUSY GUT with right-handed neutrino," Phys. Rev. D **64**, 095001 (2001).

[35] T. Goto, Y. Okada, Y. Shimizu, T. Shindou and M. Tanaka, "Exploring flavor structure of supersymmetry breaking at B factories," Phys. Rev. D **66**, 035009 (2002).

[36] T. Goto, Y. Okada, Y. Shimizu, T. Shindou and M. Tanaka, "Exploring flavor structure of supersymmetry breaking from rare B decays and unitarity triangle," Phys. Rev. D **70**, 035012 (2004).

[37] D. Chang, A. Masiero and H. Murayama, "Neutrino mixing and large CP violation in B physics," Phys. Rev. D **67**, 075013 (2003).

[38] M. Ciuchini, E. Franco, A. Masiero and L. Silvestrini, "b → s transitions: A new frontier for indirect SUSY searches," Phys. Rev. D **67**, 075016 (2003).

[39] J. Hisano and Y. Shimizu, "GUT relation in neutrino induced flavor physics in SUSY SU(5) GUT," Phys. Lett. B **565**, 183 (2003).

[40] M. Ciuchini, A. Masiero, L. Silvestrini, S. K. Vempati and O. Vives, "Grand unification of quark and lepton FCNCs," Phys. Rev. Lett. **92**, 071801 (2004).

[41] Y. Grossman and M. P. Worah, "CP asymmetries in $B$ decays with new physics in decay amplitudes," Phys. Lett. B **395**, 241 (1997).

[42] D. London and A. Soni, "Measuring the CP angle $\beta$ in hadronic $b \to s$ penguin decays," Phys. Lett. B **407**, 61 (1997).

[43] Y. Grossman, G. Isidori and M. P. Worah, "CP asymmetry in $B_d \to \phi K_S$: Standard model pollution," Phys. Rev. D **58**, 057504 (1998).

[44] H. Boos, T. Mannel and J. Reuter, "The gold-plated mode revisited: sin(2$\beta$) and $B^0 \to J/\psi K_S$ in the standard model," Phys. Rev. D **70**, 036006 (2004).

[45] H. N. Li and S. Mishima, "Penguin pollution in the $B^0 \to J/\psi K_S$ decay," JHEP **03**, 009 (2007).

[46] H. N. Li, S. Mishima and A. I. Sanda, "Resolution to the $B \to \pi K$ puzzle," Phys. Rev. D **72**, 114005 (2005).

[47] H. N. Li and S. Mishima, "Penguin-dominated $B \to PV$ decays in NLO perturbative QCD," Phys. Rev. D **74**, 094020 (2006).

[48] M. Beneke and M. Neubert, "QCD factorization for $B \to PP$ and $B \to PV$ decays," Nucl. Phys. B **675**, 333 (2003).

[49] M. Beneke, "Corrections to sin(2$\beta$) from CP asymmetries in $B^0 \to (\pi^0, \rho^0, \eta, \eta', \omega, \phi) K_S$ decays," Phys. Lett. B **620**, 143 (2005).

[50] H. Y. Cheng, C. K. Chua and A. Soni, "Effects of final-state interactions on mixing-induced CP violation in penguin-dominated $B$ decays," Phys. Rev. D **72**, 014006 (2005).

[51] A. R. Williamson and J. Zupan, "Two body $B$ decays with isosinglet final states in SCET," Phys. Rev. D **74**, 014003 (2006).

[52] Y. Grossman, Z. Ligeti, Y. Nir and H. Quinn, "SU(3) relations and the CP asymmetries in B decays to $\eta' K_S$, $\phi K_S$ and $K^+ K^- K_S$," Phys. Rev. D **68**, 015004 (2003).

[53] C. W. Chiang, M. Gronau, Z. Luo, J. L. Rosner and D. A. Suprun, "Charmless $B \to VP$ decays using flavor SU(3) symmetry," Phys. Rev. D **69**, 034001 (2004).

[54] M. Gronau, Y. Grossman and J. L. Rosner, "Interpreting the time-dependent CP asymmetry in $B^0 \to \pi^0 K_S$," Phys. Lett. B **579**, 331 (2004).

[55] M. Gronau, J. L. Rosner and J. Zupan, "Correlated bounds on CP asymmetries in $B^0 \to \eta' K_S$," Phys. Lett. B **596**, 107 (2004).

[56] G. Raz, "Using SU(3) flavor to constrain the CP asymmetries in $B \to PP, VP, VV$ decays involving $b \to s$ transitions," arXiv:hep-ph/0509125.

[57] M. Gronau, J. L. Rosner and J. Zupan, "Updated bounds on CP asymmetries in $B^0 \to \eta' K_s$ and $B^0 \to \pi^0 K_s$," Phys. Rev. D **74**, 093003 (2006).

[58] F. Gabbiani, E. Gabrielli, A. Masiero and L. Silvestrini, "A complete analysis of FCNC and CP constraints in general SUSY extensions of the standard model," Nucl. Phys. B **477**, 321 (1996).

[59] T. Falk, K. A. Olive, M. Pospelov and R. Roiban, "MSSM predictions for the electric dipole moment of the Hg-199 atom," Nucl. Phys. B **560**, 3 (1999).

[60] S. Abel, S. Khalil and O. Lebedev, "EDM constraints in supersymmetric theories," Nucl. Phys. B **606**, 151 (2001).

[61] P. H. Chankowski, O. Lebedev and S. Pokorski, "Flavour violation in general supergravity," Nucl. Phys. B **717**, 190 (2005).

[62] S. Khalil, "Probing flavor structure in supersymmetric theories," Phys. Rev. D **72**, 055020 (2005).

[63] M. Ciuchini *et al.*, "$\Delta M_K$ and $\epsilon_K$ in SUSY at the next-to-leading order," JHEP **9810**, 008 (1998).

[64] D. Chang, W. F. Chang, W. Y. Keung, N. Sinha and R. Sinha, "Squark mixing contributions to CP violating phase gamma," Phys. Rev. D **65**, 055010 (2002).

[65] D. Becirevic *et al.*, "$B_d - \bar{B}_d$ mixing and the $B_d \to J/\psi K_S$ asymmetry in general SUSY models," Nucl. Phys. B **634**, 105 (2002).

[66] J. Hisano and Y. Shimizu, "Hadronic EDMs induced by the strangeness and constraints on supersymmetric CP phases," Phys. Rev. D **70**, 093001 (2004).

[67] E. Barberio *et al.* [Heavy Flavor Averaging Group (HFAG)], "Averages of b-hadron properties at the end of 2005," arXiv:hep-ex/0603003.

[68] S. Khalil and E. Kou, "On supersymmetric contributions to the CP asymmetry of the $B \to \phi K_S$," Phys. Rev. D **67**, 055009 (2003).

[69] G.L. Kane, P. Ko, H.B. Wang, C. Kolda, J.H. Park and L.T. Wang, "$B_d \to \phi K_S$ CP asymmetries as an important probe of supersymmetry," Phys. Rev. Lett. **90**, 141803 (2003).

[70] S. Mishima and A. I. Sanda, "An analysis of supersymmetric effects on $B \to \phi K$ decays in PQCD approach," Phys. Rev. D **69**, 054005 (2004).

[71] A. Ali and C. Greub, "An analysis of two-body non-leptonic B decays involving light mesons in the standard model," Phys. Rev. D **57**, 2996 (1998).

[72] S. Khalil and E. Kou, "A possible supersymmetric solution to the discrepancy between $B \to \phi K_S$ and $B \to \eta' K_S$ CP asymmetries," Phys. Rev. Lett. **91**, 241602 (2003).

[73] A. Abulencia et al. [CDF Collaboration], "Observation of $B_s^0 - \bar{B}_s^0$ oscillations," Phys. Rev. Lett. **97**, 242003 (2006).

[74] T. Aaltonen et al. [CDF Collaboration], "First Flavor-Tagged Determination of Bounds on Mixing-Induced CP Violation in $B_s^0 \to J/\psi\phi$ Decays," Phys. Rev. Lett. **100**, 161802 (2008).

[75] V. M. Abazov et al. [D0 Collaboration], "Measurement of $B_s^0$ mixing parameters from the flavor-tagged decay $B_s^0 \to J/\psi\phi$," Phys. Rev. Lett. **101**, 241801 (2008).

[76] M. Bona et al. [UTfit Collaboration], "First Evidence of New Physics in $b \longleftrightarrow s$ Transitions," arXiv:0803.0659 [hep-ph].

[77] M. Bona et al., "New Physics from Flavour," arXiv:0906.0953 [hep-ph].

[78] P. Ball, S. Khalil and E. Kou, "$B_s^0 - \bar{B}_s^0$ mixing and the $B_s \to J/\psi\Phi$ asymmetry in supersymmetric models," Phys. Rev. D **69**, 115011 (2004).

[79] M. Matsumori and A. I. Sanda, "The mixing-induced CP asymmetry in $B \to K^*\gamma$ decays with perturbative QCD approach," Phys. Rev. D **73**, 114022 (2006).

[80] P. Ball and R. Zwicky, "Time-dependent CP asymmetry in $B \to K^*\gamma$ as a (quasi) null test of the standard model," Phys. Lett. B **642**, 478 (2006).

[81] B. Grinstein and D. Pirjol, "The CP asymmetry in $B_t^0 \to K_S \pi^0 \gamma$ in the standard model," Phys. Rev. D **73**, 014013 (2006).

[82] A. Khodjamirian, R. Ruckl, G. Stoll and D. Wyler, "QCD estimate of the long-distance effect in $B \to K^*\gamma$," Phys. Lett. B **402**, 167 (1997).

[83] K. i. Okumura and L. Roszkowski, "Weakened Constraints from $b \to s\gamma$ on Supersymmetry Flavor Mixing Due to Next-To-Leading-Order Corrections," Phys. Rev. Lett. **92**, 161801 (2004).

[84] K. i. Okumura and L. Roszkowski, "Large beyond-leading-order effects in $b \to s\gamma$ in supersymmetry with general flavor mixing," JHEP **0310**, 024 (2003).

[85] J. Foster, K. i. Okumura and L. Roszkowski, "New Higgs effects in B physics in supersymmetry with general flavour mixing," Phys. Lett. B **609**, 102 (2005).

[86] J. Foster, K. i. Okumura and L. Roszkowski, "Probing the flavour structure of supersymmetry breaking with rare B-processes: A beyond leading order analysis," JHEP **0508**, 094 (2005).

[87] J. Foster, K. i. Okumura and L. Roszkowski, "Current and future limits on general flavour violation in $b \to s$ transitions in minimal supersymmetry," JHEP **0603**, 044 (2006).

[88] A. Ali, G. F. Giudice and T. Mannel, "Towards a model independent analysis of rare $B$ decays," Z. Phys. C **67**, 417 (1995).

[89] S. Fukae, C. S. Kim, T. Morozumi and T. Yoshikawa, "A model independent analysis of the rare $B$ decay $B \to X_s l^+ l^-$," Phys. Rev. D **59**, 074013 (1999).

[90] S. Fukae, C. S. Kim and T. Yoshikawa, "The effects of non-local operators in rare $B$ decays, $B \to X_s l^+ l^-$," Int. J. Mod. Phys. A **16**, 1703 (2001).

[91] G. Hiller and F. Kruger, "More model-independent analysis of $b \to s$ processes," Phys. Rev. D **69**, 074020 (2004).

[92] G. Senjanovic and R. N. Mohapatra, "Exact Left-Right Symmetry And Spontaneous Violation Of Parity," Phys. Rev. D **12**, 1502 (1975).

[93] K. Fujikawa and A. Yamada, "Test of the chiral structure of the top – bottom charged current by the process $b \to s\gamma$," Phys. Rev. D **49**, 5890 (1994).

[94] K. S. Babu, K. Fujikawa and A. Yamada, "Constraints on left-right symmetric models from the process $b \to s\gamma$," Phys. Lett. B **333**, 196 (1994).

[95] P. L. Cho and M. Misiak, "$b \to s\gamma$ Decay In SU(2)-L X SU(2)-R X U(1) Extensions Of The Standard Model," Phys. Rev. D **49**, 5894 (1994).

[96] D. Atwood, M. Gronau and A. Soni, "Mixing-induced CP asymmetries in radiative $B$ decays in and beyond the standard model," Phys. Rev. Lett. **79**, 185 (1997).

[97] D. Melikhov, N. Nikitin and S. Simula, "Probing right-handed currents in $B \to K^* l^+ l^-$ transitions," Phys. Lett. B **442**, 381 (1998).

[98] F. Kruger, L. M. Sehgal, N. Sinha and R. Sinha, "Angular distribution and CP asymmetries in the decays $\bar{B} \to K^- \pi^+ e^- e^+$ and $\bar{B} \to \pi^- \pi^+ e^- e^+$," Phys. Rev. D **61**, 114028 (2000).

[99] C. S. Kim, Y. G. Kim, C. D. Lu and T. Morozumi, "Azimuthal angle distribution in $B \to K^*(\to K\pi) l^+ l^-$ at low invariant $m_{l^+ l^-}$ region," Phys. Rev. D **62**, 034013 (2000).

[100] M. Gronau, Y. Grossman, D. Pirjol and A. Ryd, "Measuring the photon helicity in radiative $B$ decays," Phys. Rev. Lett. **88**, 051802 (2002).

[101] M. Gronau and D. Pirjol, "Photon polarization in radiative $B$ decays," Phys. Rev. D **66**, 054008 (2002).

[102] C. Bobeth, M. Misiak and J. Urban, "Photonic penguins at two loops and $m_t$-dependence of $BR(B \to X_s l^+ l^-)$," Nucl. Phys. B **574**, 291 (2000).

[103] H. H. Asatrian, H. M. Asatrian, C. Greub and M. Walker, "Two-loop virtual corrections to $B \to X_s l^+ l^-$ in the standard model," Phys. Lett. B **507**, 162 (2001).

[104] Z. Xiong and J. M. Yang, "Rare $B$-Meson Dileptonic Decays In Minimal Supersymmetric Model," Nucl. Phys. B **628**, 193 (2002).

[105] T. Goto, Y. Okada, Y. Shimizu and M. Tanaka, "$b \to s l \bar{l}$ in the minimal supergravity model," Phys. Rev. D **55**, 4273 (1997).

[106] G. Buchalla, G. Hiller and G. Isidori, "Phenomenology of non-standard $Z$ couplings in exclusive semileptonic $b \to s$ transitions," Phys. Rev. D **63**, 014015 (2001).

[107] Y. Fukuda *et al.* [Super-Kamiokande Collaboration], "Evidence for oscillation of atmospheric neutrinos," Phys. Rev. Lett. **81**, 1562 (1998).

[108] M. H. Ahn *et al.* [K2K Collaboration], "Indications of neutrino oscillation in a 250-km long-baseline experiment," Phys. Rev. Lett. **90**, 041801 (2003).

[109] S. Fukuda *et al.* [Super-Kamiokande Collaboration], "Solar B-8 and he p neutrino measurements from 1258 days of Super-Kamiokande data," Phys. Rev. Lett. **86**, 5651 (2001).

[110] Q. R. Ahmad *et al.* [SNO Collaboration], "Direct evidence for neutrino flavor transformation from neutral-current interactions in the Sudbury Neutrino Observatory," Phys. Rev. Lett. **89**, 011301 (2002).

[111] Q. R. Ahmad *et al.* [SNO Collaboration], "Measurement of day and night neutrino energy spectra at SNO and constraints on neutrino mixing parameters," Phys. Rev. Lett. **89**, 011302 (2002).

[112] K. Eguchi *et al.* [KamLAND Collaboration], "First results from KamLAND: Evidence for reactor anti-neutrino disappearance," Phys. Rev. Lett. **90**, 021802 (2003).

[113] K. Abe *et al.* [Belle Collaboration], "A new search for $\tau \to \mu\gamma$ and $\tau \to e\gamma$ decays at Belle," arXiv:hep-ex/0609049.

[114] C. Amsler *et al.* [Particle Data Group], "The Review of Particle Physics," Phys. Lett. B**667**, 1 (2008).

[115] J. Adam *et al.* [MEG Collaboration], "A Limit for the $mu \to e\gamma$ Decay From the MEG Experiment," arXiv:0908.2594 [hep-ex].

[116] T. Mori [MEG Collaboration], "The detector for the new mu $\to$ e gamma experiment MEG," http://www.slac.stanford.edu/spires/find/hep/www?irn=5685834SPIRES entry *Prepared for 31st International Conference on High Energy Physics (ICHEP 2002), Amsterdam, The Netherlands, 24-31 Jul 2002.*

[117] K. S. Babu and C. Kolda, "Higgs-mediated $\tau \to 3\mu$ in the supersymmetric seesaw model," Phys. Rev. Lett. **89**, 241802 (2002).

[118] A. Brignole and A. Rossi, "Anatomy and phenomenology of mu tau lepton flavour violation in the MSSM," Nucl. Phys. B **701**, 3 (2004).

[119] A. Dedes, J. R. Ellis and M. Raidal, "Higgs mediated $B^0_{s,d} \to \mu\tau$, $e\tau$ and $\tau \to 3\mu$, $e\mu\mu$ decays in supersymmetric seesaw models," Phys. Lett. B **549**, 159 (2002).

[120] M. Sher, "$\tau \to \mu\eta$ in supersymmetric models," Phys. Rev. D **66**, 057301 (2002).

[121] P. Paradisi, "Higgs-mediated tau $\to$ mu and tau $\to$ e transitions in II Higgs doublet model and supersymmetry," JHEP **0602**, 050 (2006).

[122] R. Barbieri and L. J. Hall, "Signals for supersymmetric unification," Phys. Lett. B **338**, 212 (1994).

[123] R. Barbieri, L. J. Hall and A. Strumia, "Violations of lepton flavor and CP in supersymmetric unified theories," Nucl. Phys. B **445**, 219 (1995).

[124] J. Hisano, T. Moroi, K. Tobe and M. Yamaguchi, "Exact event rates of lepton flavor violating processes in supersymmetric SU(5) model," Phys. Lett. B **391**, 341 (1997).

[125] F. Borzumati and A. Masiero, "Large Muon And Electron Number Violations In Supergravity Theories," Phys. Rev. Lett. **57**, 961 (1986).

[126] J. Hisano, T. Moroi, K. Tobe, M. Yamaguchi and T. Yanagida, "Lepton flavor violation in the supersymmetric standard model with seesaw induced neutrino masses," Phys. Lett. B **357**, 579 (1995).

[127] J. Hisano, T. Moroi, K. Tobe and M. Yamaguchi, "Lepton-Flavor Violation via Right-Handed Neutrino Yukawa Couplings in Supersymmetric Standard Model," Phys. Rev. D **53**, 2442 (1996).

[128] J. A. Casas and A. Ibarra, "Oscillating neutrinos and $\mu \to e\gamma$," Nucl. Phys. B **618**, 171 (2001).

[129] J. R. Ellis, J. Hisano, M. Raidal and Y. Shimizu, "A new parametrization of the seesaw mechanism and applications in supersymmetric models," Phys. Rev. D **66**, 115013 (2002).

[130] N. Arkani-Hamed, H. C. Cheng and L. J. Hall, "Flavor mixing signals for realistic supersymmetric unification," Phys. Rev. D **53**, 413 (1996).

[131] J. Hisano, D. Nomura, Y. Okada, Y. Shimizu and M. Tanaka, "Enhancement of $\mu \to e\gamma$ in the supersymmetric SU(5) GUT at large $\tan\beta$," Phys. Rev. D **58**, 116010 (1998).

[132] N. Arkani-Hamed, S. Dimopoulos, G. R. Dvali and J. March-Russell, "Neutrino masses from large extra dimensions," Phys. Rev. D **65**, 024032 (2002).

[133] K. R. Dienes, E. Dudas and T. Gherghetta, "Light neutrinos without heavy mass scales: A higher-dimensional seesaw mechanism," Nucl. Phys. B **557**, 25 (1999).

[134] Y. Grossman and M. Neubert, "Neutrino masses and mixings in non-factorizable geometry," Phys. Lett. B **474**, 361 (2000).

[135] A. E. Faraggi and M. Pospelov, "Phenomenological issues in TeV scale gravity with light neutrino masses," Phys. Lett. B **458**, 237 (1999).

[136] A. De Gouvea, G. F. Giudice, A. Strumia and K. Tobe, "Phenomenological implications of neutrinos in extra dimensions," Nucl. Phys. B **623**, 395 (2002).

[137] J. P. Saha and A. Kundu, "Constraints on R-parity violating supersymmetry from leptonic and semileptonic $\tau$, $B_d$ and $B_s$ decays," Phys. Rev. D **66**, 054021 (2002).

[138] A. Ilakovac and A. Pilaftsis, "Flavor violating charged lepton decays in a GUT and superstring inspired standard model," Nucl. Phys. B **437**, 491 (1995).

[139] G. Cvetic, C. Dib, C. S. Kim and J. D. Kim, "On lepton flavor violation in tau decays," Phys. Rev. D **66**, 034008 (2002).

# Chapter 4

# Belle Status

By the 2009 summer shutdown, Belle had accumulated data corresponding to an integrated luminosity of ∼950 fb$^{-1}$, of which ∼710 fb$^{-1}$ are at the $\Upsilon(4S)$ resonance, corresponding to ∼772 million $B\overline{B}$ pairs. We expect a total integrated luminosity of ∼1000 fb$^{-1}$ (= 1 ab$^{-1}$) as a final Belle data set.

In this section, the current physics results are briefly reviewed for selected topics in order to give an overview of the present status. Note that the theoretical framework of quark flavor physics in the Standard Model and beyond is described in Chapter 2 and 3, respectively, while some details of the experimental methods are further explained in chapter 5.

## 4.1 $\sin 2\phi_1$ in $b \to c\bar{c}s$ Processes

Decays of $B^0$ meson to $CP$-eigenstates with an underlying $b \to c\bar{c}s$ process are reconstructed for $J/\psi K_S^0$ ($\xi_f = -1$) and $J/\psi K_L^0$ ($\xi_f = +1$) final states [1], where $\xi_f$ denotes $CP$ parity. The two classes ($\xi_f = \pm 1$) should have $CP$-asymmetries that are opposite in sign.

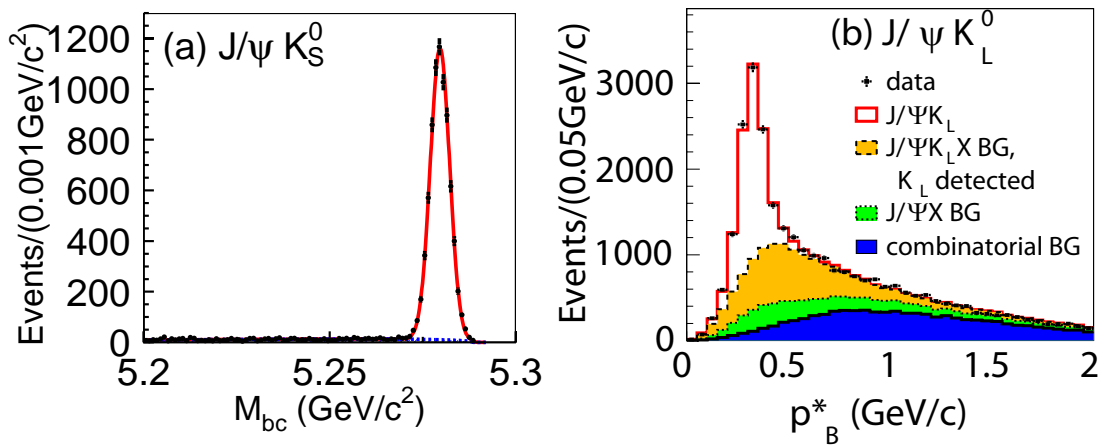

Figure 4.1: (a) $M_{\rm bc}$ distribution in the $\Delta E$ signal region for selected $B^0 \to J/\psi K_S^0$ candidates and (b) $p_B^*$ distribution for selected $B^0 \to J/\psi K_L^0$ candidates. The shaded portions show the contributions of different background components.

The reconstructed samples with 535 million $B\overline{B}$ pairs used for the $\sin 2\phi_1$ measurement [1]

---
[1]The inclusion of the charge conjugate decay mode is implied unless otherwise stated.

are shown in Figure 4.1. $B$ mesons produced at the $\Upsilon(4S)$ peak are identified using the energy difference $\Delta E = E_B^* - E_{\text{beam}}^*$ and the beam-energy constrained mass $M_{bc} = \sqrt{(E_{\text{beam}}^*)^2 - (p_B^*)^2}$, where $E_{\text{beam}}^*$ is the beam energy, and $E_B^*$ and $p_B^*$ are the energy and the momentum of the reconstructed $B$ meson, all measured in the $e^+e^-$ center-of-mass frame. The signal yields (purities) are $7484 \pm 87$ (97%) and $6521 \pm 123$ (59%) for $J/\psi K_S^0$ and $J/\psi K_L^0$, respectively. It is clear that the $CP$-eigenstate event samples used for the $CP$-violation measurements in $b \to c\bar{c}s$ are large and clean.

Figure 4.2 shows the $\Delta t$ distributions, where a clear shift between $B^0$ and $\overline{B}^0$ tags is visible, and the raw asymmetry plots (see Eq.(2.48)) for a sample with good tag quality. The final results are extracted from an unbinned maximum-likelihood fit to the $\Delta t$ distributions which takes into account resolution, mistagging and background dilution. The result is $\sin 2\phi_1 = 0.642 \pm 0.031 \pm 0.017$. The $CP$ asymmetries for $J/\psi K_S^0$ and $J/\psi K_L^0$ samples are $+0.643 \pm 0.038$ and $-0.641 \pm 0.057$ (statistical error only), respectively, and have opposite sign with the same magnitudes as expected. Combining a measurement of $\sin 2\phi_1$ with the $\psi(2S)K_S^0$ mode using 657 million $B\overline{B}$ pairs [2], we obtain

$$\sin 2\phi_1 = 0.650 \pm 0.029 \pm 0.018. \tag{4.1}$$

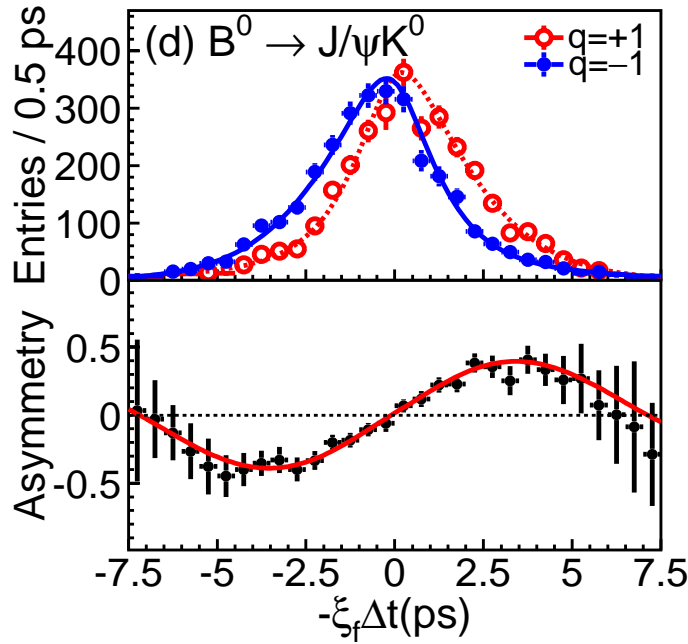

Figure 4.2: Background subtracted $\Delta t$ distributions and asymmetries for events with good tags for $B^0 \to J/\psi K^0$. The smooth curves are projections of the unbinned likelihood fit.

This result may be compared to the BaBar result with 465 million $B\overline{B}$ pairs of $\sin 2\phi_1 = 0.687 \pm 0.028 \pm 0.012$ [3]. Both experiments are in good agreement; the average of these results is [4]

$$\sin 2\phi_1 = 0.672 \pm 0.023. \tag{4.2}$$

The average can be interpreted as a constraint on the CKM angle $\phi_1$ and confronted with the indirect determinations of the unitarity triangle [5] (see Sec. 4.5). Both values are consistent with the hypothesis that the Kobayashi-Maskawa phase is the source of $CP$ violation. The measurement of $\sin 2\phi_1$ in $b \to c\bar{c}s$ modes has become a presision measurement with better than 4% accuracy. Still the systematic uncertainties are small and well understood.

The asymmetry with a cosine dependence, which indicates direct $CP$ violation, is measured to be $\mathcal{A} = 0.018 \pm 0.020 \pm 0.015$. The average with BaBar's result $\mathcal{A} = -0.024 \pm 0.020 \pm 0.016$ gives

$$\mathcal{A} = -0.004 \pm 0.019. \tag{4.3}$$

This result is consistent with the theoretical expectation $\mathcal{A} = 0$.

## 4.2 $CP$-Violation in $b \to sq\bar{q}$ Penguin Processes

One of the promising ways to probe additional $CP$-violating phases from new physics beyond the Standard Model is to measure the time-dependent $CP$-asymmetry in $b \to sq\bar{q}$ penguin-dominated modes such as $B^0 \to \phi K_S^0$ and $B^0 \to \eta' K_S^0$, where heavy new particles may contribute inside the loop, and to compare the results with the asymmetry in $B^0 \to J/\psi K_S^0$ and related $b \to c\bar{c}s$ charmonium modes. Belle has measured $CP$-violation in $B^0 \to \eta' K^0$, $K_S^0 K_S^0 K_S^0$, $K^+K^-K_S^0$, and $\omega K_S^0$ with 535 million $B\overline{B}$ pairs [1, 6]; and in $B^0 \to \pi^0\pi^0 K_S^0$, $\pi^0 K^0$, $\phi K_S^0$, $f_0(980)K_S^0$, and $\rho^0 K_S^0$ with 657 million $B\overline{B}$ pairs [7–9]

The decays $B^0 \to \phi K_S^0$, $\eta' K_S^0$ and $K_S^0 K_S^0 K_S^0$, which are dominated by the $b \to s\bar{s}s$ transition, are especially unambiguous and sensitive probes of new $CP$-violating phases from physics beyond the Standard Model [10]. The Standard Model predicts that measurements of $CP$-violation in this mode should yield $\sin 2\phi_1$ to a very good approximation [11–13]. A significant deviation in the time-dependent $CP$-asymmetry in this mode from what is observed in $b \to c\bar{c}s$ decays would be evidence for a new $CP$-violating phase.

Decay modes $B \to \eta' K_L^0$ and $\pi^0 K_L^0$ are used as well as $B \to \eta' K_S^0$ and $\pi^0 K_S^0$ to increase the signal samples. Figure 4.3 shows the $\Delta t$ distributions and asymmetries for good tag quality events for $B^0 \to \eta' K^0$ and $B^0 \to K_S^0 K_S^0 K_S^0$ decay modes. Note that these projections onto the

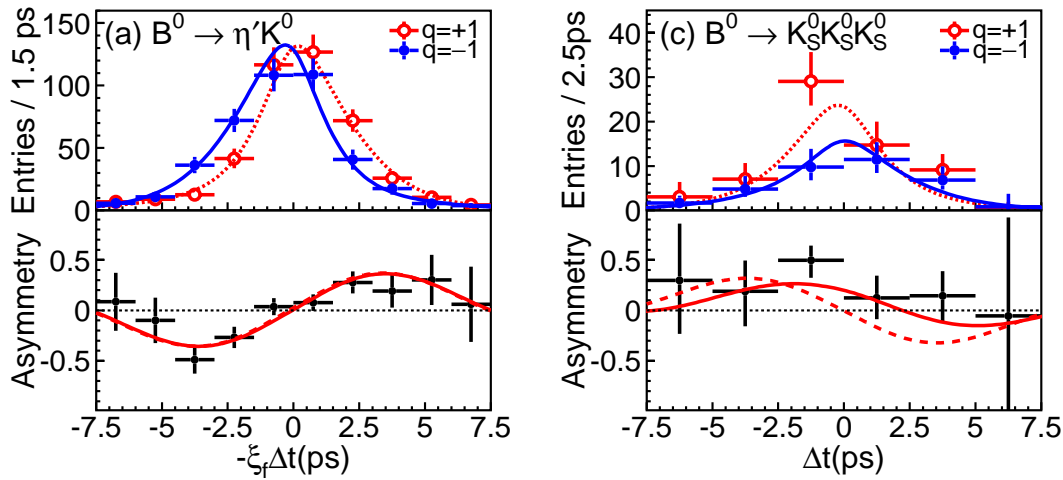

Figure 4.3: Background-subtracted $\Delta t$ distributions and asymmetries for events with good tags for (left) $B^0 \to \eta' K^0$, and (right) $B^0 \to K_S^0 K_S^0 K_S^0$. In the asymmetry plots, solid curves show the fit results, and dashed curves show the Standard Model expectation from the $B^0 \to J/\psi K^0$ measurement.

$\Delta t$ axis do not take into account event-by-event information (such as the signal fraction, the

wrong tag fraction and the vertex resolution) that is used in the unbinned maximum likelihood fit. For the $B^0 \to \eta' K^0$ mode, we obtain

$$\sin 2\phi_1^{\text{eff}} = +0.64 \pm 0.10 \pm 0.04, \qquad (4.4)$$

which is the first observation of $CP$ violation in the $b \to sq\bar{q}$ mode with a significance equivalent to $5.6\sigma$.

For $B^0 \to \phi K_S^0$, $f_0(980)K_S^0$, and $\rho^0 K_S^0$ modes, time-dependent Dalitz plot analysis is performed in $B^0 \to K_S^0 K^+ K^-$ and $K_S^0 \pi^+ \pi^-$ three-body decays in order to take into account effects of interferences among intermediate states.

BaBar has also made measurements for these decay modes with 227–467 million $B\bar{B}$ pairs [14]. The results of both experiments and averages are summarized in Fig. 4.4. The $\sin 2\phi_1^{\text{eff}}$ results of all individual modes are consistent with the world average of $b \to c\bar{c}s$ modes within their measurement errors, apart of $B^0 \to \pi^0\pi^0 K_S^0$ and $B^0 \to K^+K^- K^0$ where some discrepancies at the level of two to three standard deviations are currently observed. When all the $b \to s$ modes are naively averaged, the result is $0.7\sigma$ lower than the SM expectation. It should be noted that this average neglects both the theoretical uncertainty and the fact that experimental systematic uncertainties are correlated between the measurements of individual modes. Calculations within the SM predict that the $\sin 2\phi_1^{\text{eff}}$ in $b \to sq\bar{q}$ should be higher than the value measured in $b \to c\bar{c}s$ mode [15–18].

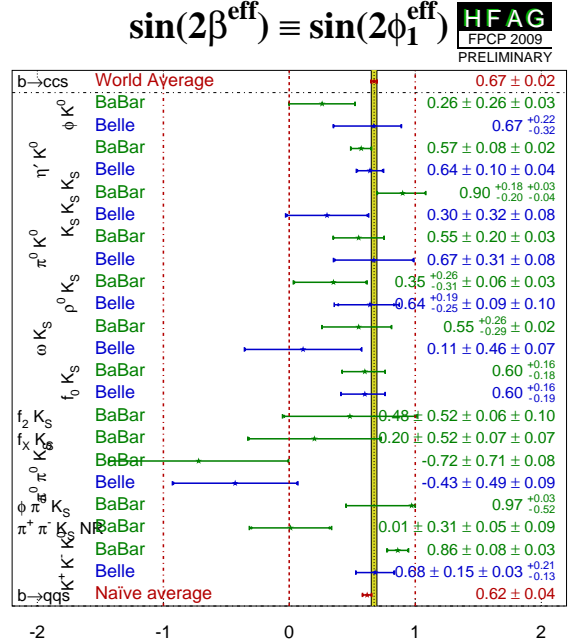

Figure 4.4: Summary of $\sin 2\phi_1^{\text{eff}}$ measurements for $b \to s\bar{q}q$ decay modes [4].

## 4.3 $\phi_2$ measurements

$\phi_2$ is an angle between $V_{ub}^* V_{ud}$ and $V_{tb}^* V_{td}$, and can be measured by a time-dependent $CP$ asymmetry analysis using $B^0$ decays to $CP$ eigenstates through a $b \to u$ tree diagram. For a pure $b \to u$ tree transition, a mixing-induced $CP$ asymmetry measurement yields $\sin 2\phi_2$ as the decay amplitude has the weak phase $V_{ub}^* V_{ud}$ (Eq.(2.56)). Unfortunately, in contrast with $B^0 \to J/\psi K^0$ decay case, the contribution of the $b \to d$ penguin amplitude could be comparable to that of the $b \to u$ tree amplitude. Because these amplitudes have different weak phases, $\mathcal{S}$ is no longer equal to $\sin 2\phi_2$ and is expressed as

$$\mathcal{S} = \sqrt{1 - \mathcal{A}^2} \sin(2\phi_2 + \kappa), \tag{4.5}$$

where the value of $\kappa$ can be extracted using the isospin relations among $B^0 \to \pi^+\pi^-$, $B^0 \to \pi^0\pi^0$, and $B^+ \to \pi^+\pi^0$ decays given in Eqs. (2.58),(2.59) [19].

### 4.3.1  $B^0 \to \pi^+\pi^-$

The penguin contribution to the decay may cause the $CP$ asymmetry $\mathcal{A}_{\pi\pi}$ to differ from the null value expected in the case of the tree contribution only. By defining the effective angle $2\phi_2^{\text{eff}} = 2\phi_2 + \kappa$ through $\lambda_{\pi\pi} = |\lambda_{\pi\pi}| e^{2i\phi_2^{\text{eff}}}$, and using $|\lambda_{\pi\pi}| = \sqrt{1 + \mathcal{A}_{\pi\pi}}/\sqrt{1 - \mathcal{A}_{\pi\pi}}$ (see Eq.(2.50)), one obtains the asymmetry $\mathcal{S}_{\pi\pi} = \sqrt{1 - \mathcal{A}_{\pi\pi}^2} \sin 2\phi_2^{\text{eff}}$. The phase shift $\kappa$ can be expressed using the isospin relations as [20]

$$\kappa = -\arccos(B^+) - \arccos(B^-)$$
$$B^- = \frac{\mathcal{B}^{+0} + \mathcal{B}^{+-}(1 - \mathcal{A}_{\pi\pi})/2 - \mathcal{B}^{00}(1 - \mathcal{A}_{\pi^0\pi^0})}{\sqrt{2\mathcal{B}^{+-}\mathcal{B}^{+0}(1 - \mathcal{A}_{\pi\pi})}}$$
$$B^+ = \frac{\mathcal{B}^{+0} + \mathcal{B}^{+-}(1 + \mathcal{A}_{\pi\pi})/2 - \mathcal{B}^{00}(1 + \mathcal{A}_{\pi^0\pi^0})}{\sqrt{2\mathcal{B}^{+-}\mathcal{B}^{+0}(1 + \mathcal{A}_{\pi\pi})}} \quad , \tag{4.6}$$

where $\mathcal{B}^{+0}$, $\mathcal{B}^{+-}$ and $\mathcal{B}^{00}$ denote the branching fractions for decays of a $B$ meson to a specific $\pi\pi$ charge combination.

Belle has measured $CP$ asymmetry parameters $\mathcal{S}_{\pi\pi}$ and $\mathcal{A}_{\pi\pi}$ using 535 million $B\bar{B}$ pairs [21]. The results are

$$\mathcal{S}_{\pi\pi} = -0.61 \pm 0.10 \pm 0.04 \quad \text{and} \quad \mathcal{A}_{\pi\pi} = +0.55 \pm 0.08 \pm 0.05. \tag{4.7}$$

Figure 4.5 shows the $\Delta t$ distributions and asymmetries together with fit results. The result shows the observation of the direct $CP$ violation with a 5.5 $\sigma$ significance, while BaBar result [22] with 467 million $B\bar{B}$ pairs shows somewhat smaller direct $CP$ violation ($\mathcal{A}_{\pi\pi} = +0.25 \pm 0.08 \pm 0.02$ and $\mathcal{S}_{\pi\pi} = -0.68 \pm 0.10 \pm 0.03$). The difference between two experiments is $\sim 1.9\sigma$.

The isospin analysis is performed using above results and the current world averages of $\mathcal{B}(B^+ \to \pi^+\pi^0)$, $\mathcal{B}(B^0 \to \pi^+\pi^-)$, $\mathcal{B}(B^0 \to \pi^0\pi^0)$, and $\mathcal{A}(B^+ \to \pi^0\pi^0)$ [4], resulting in the confidence level intervals of $\phi_2$ as shown in Fig. 4.9.

### 4.3.2  $B^0 \to \rho^+\rho^-$

The decay mode $B^0 \to \rho^+\rho^-$ contains two vector mesons in the final state. Therefore it is in general a mixture of $CP$-even and -odd final states. Since the $\rho$ meson is a wide resonance, it would also suffer from interference effects with other resonances and a non-resonant component. Because of these difficulties, it was thought that an analysis for this decay mode would be complicated and might not be useful for a $\phi_2$ determination. However, the fraction of longitudinal polarization component ($f_L$) was found to be close to 100% [23, 24] which means that the final state is an almost pure $CP$-even state. Furthermore, the branching fraction of $B^0 \to \rho^0\rho^0$ [25, 26] is found to be much smaller than those of $B^0 \to \rho^+\rho^-$ and $B^+ \to \rho^+\rho^0$ [4], which indicates that the effect of the penguin diagram ($\kappa$) is small enough. Also the branching fraction for $B^0 \to \rho^+\rho^-$ is relatively large and there are no other significant resonant contributions that could contribute to this final state.

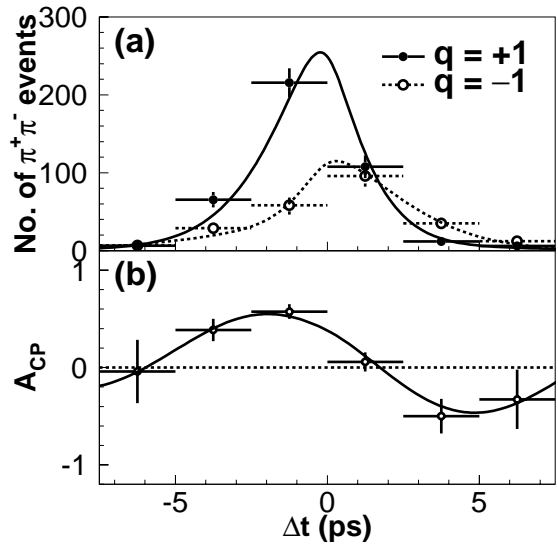

Figure 4.5: (a) $\Delta t$ distributions of $B^0 \to \pi^+\pi^-$ signal events with good tag quality after background subtraction for $B^0$-tagged (solid) and $\bar{B}^0$-tagged (dashed), and $\mathcal{A}_{CP}$ plot. The curves are projections of the fit result.

Belle has measured $CP$ asymmetry parameters using 535 million $B\bar{B}$ pairs [27]. The results are

$$\mathcal{S}_{\rho^+\rho^-} = +0.19 \pm 0.30 \pm 0.07 \quad \text{and} \quad \mathcal{A}_{\rho^+\rho^-} = 0.16 \pm 0.21 \pm 0.07. \tag{4.8}$$

Both $\mathcal{S}_{\rho\rho}$ and $\mathcal{A}_{\rho\rho}$ are consistent with zero within errors. Figure 4.6 shows the $\Delta t$ distributions and asymmetries together with fit results. BaBar has obtained a consistent result with 384 million $B\bar{B}$ pairs [23].

The isospin analysis similar to $B \to \pi\pi$ is performed using above results ($\mathcal{S}$, $\mathcal{A}$ and $f_L$) and the current world averages of $\mathcal{B}(B^+ \to \rho^+\rho^0)$, $\mathcal{B}(B^0 \to \rho^+\rho^-)$, and $\mathcal{B}(B^0 \to \rho^0\rho^0)$ [4] (Fig. 4.9).

### 4.3.3 $B^0 \to \rho^+\pi^-$

In the $B^0 \to \rho^+\pi^-$ decay, the final state is not a $CP$ eigenstate. This decay however proceeds via the same quark level diagram as that of $B^0 \to \pi^+\pi^-$ and both $B^0$ and $\overline{B}^0$ can decay to $\rho^+\pi^-$. Consequently, mixing induced $CP$ violations can occur in four decay amplitudes, $B^0 \to \rho^\pm\pi^\mp$ and $\overline{B}^0 \to \rho^\pm\pi^\mp$. The measurements of $CP$ violation parameters for these decays, where they are treated as quasi-two body decays, have been made both by BaBar [28] and the Belle [29]. However, an isospin analysis is rather complicated and no significant constraint on $\phi_2$ has been obtained [5].

Belle has performed a time-dependent Dalitz plot analysis to $B^0 \to \pi^+\pi^-\pi^0$ decays using 449 million $B\bar{B}$ pairs [30]. This method in principle allows to extract $\phi_2$ without multi-fold ambiguity utilizing the different behavior of time evolution in Dalitz plane due to interference of $\rho^+\pi^-$, $\rho^-\pi^+$, and $\rho^0\pi^0$ amplitudes [31]. "Square Dalitz plot" parameters ($m'$ and $\theta'$) are used rather than the usual squared masses of two particles. The maximum likelihood fit to the Dalitz plot and $\Delta t$ distributions extracts 26 coefficients which uniquely determine 6 complex amplitudes ($B^0$ and $\bar{B}^0 \to \rho^+\pi^-, \rho^-\pi^+, \rho^0\pi^0$). The mass and helicity distributions and $CP$ asymmetries (as a function of $\Delta t$) for candidate events containing $971\pm42$ $B^0 \to \pi^+\pi^-\pi^0$ signals are shown in Fig. 4.7 together with fit results. The constraint on $\phi_2$ from 26 coefficients is shown in Fig. 4.8 (dotted curve). Also, the constraint including isospin (pentagon) relations [32, 33] among $B^0 \to \rho\pi$ and $B^+ \to \rho\pi$ decays is obtained (solid curve). The similar analysis is also made by BaBar using 375 million $B\bar{B}$ pairs [34].

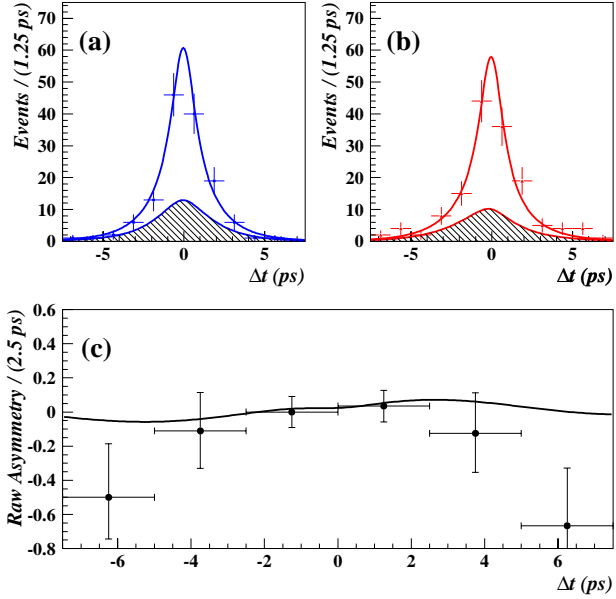

Figure 4.6: $\Delta t$ distributions of $B^0 \to \rho^+\rho^-$ candidate events with good tag quality for (a) $B^0$-tagged and (b) $\bar{B}^0$-tagged. The hatched region shows the background component. The raw $CP$ asymmetry is shown in (c). The plots are for the signal enhanced region with tightened continuum suppression requirement.

### 4.3.4 Combined $\phi_2$ Constraint

Figure 4.9 summarizes the constraints on $\phi_2$ from the $B \to \pi\pi$, $B \to \rho\rho$, $B \to \rho\pi$ decays using the isospin analysis, and combined results [5]. Averages of Belle and BaBar measurements are used. Selecting the solution closest to the value arising from the global fit to the Unitarity Triangle, $\phi_2 = (89.0^{+4.4}_{-4.2})°$ is obtained.

## 4.4 $\phi_3$ Measurement

$\phi_3$ is an angle between $V_{ub}^*V_{ud}$ and $V_{cb}^*V_{cd}$ and can not be measured using a mixing induced $CP$ violation in $B^0$ decays with a simple manner as done for other two angles. Naturally, decay modes useful to measure $\phi_3$ should involve an interference between an amplitude with $V_{ub}$ and that with any CKM element other than $V_{td}$. Also, some trick is required to get rid of the effect of strong phase in $CP$ violation. One of promising decay modes is $B$ decays to a neutral $D$ meson and a kaon. If both $D^0$ and $\bar{D}^0$ decay to a common final state ($f_{\text{com}}$), an amplitude of $B^- \to D^0K^- \to f_{\text{com}}K^-$ (via $b \to c\bar{u}s$ diagram) and that of $B^- \to \bar{D}^0K^- \to f_{\text{com}}K^-$ (via

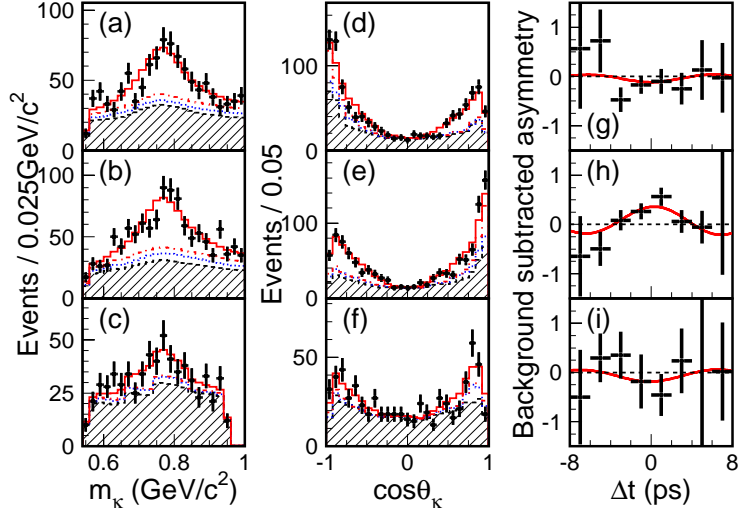

Figure 4.7: Mass (a)-(c) and helicity (d)-(f) distributions, and background subtracted $\Delta t$ asymmetry plots in the good tagging quality (g)-(i), corresponding to the $\rho^+\pi^-$ [(a),(d),(g)], $\rho^-\pi^+$ [(b),(e),(h)], and $\rho^0\pi^0$ [(c),(f),(i)] enhanced regions. Solid, dot-dashed, dotted, and dashed hatched histograms correspond to correctly reconstructed signal, self cross feed, $B\overline{B}$, and continuum PDF's, respectively.

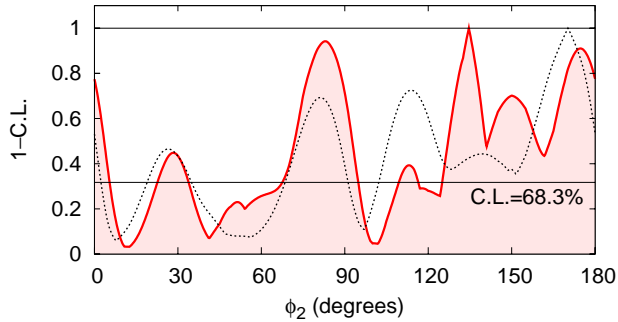

Figure 4.8: $1 - $ C.L. vs. $\phi_2$. Dotted and solid curves correspond to the result from the time-dependent Dalitz plot analysis only and that from the Dalitz and an isospin (pentagon) combined analysis, respectively.

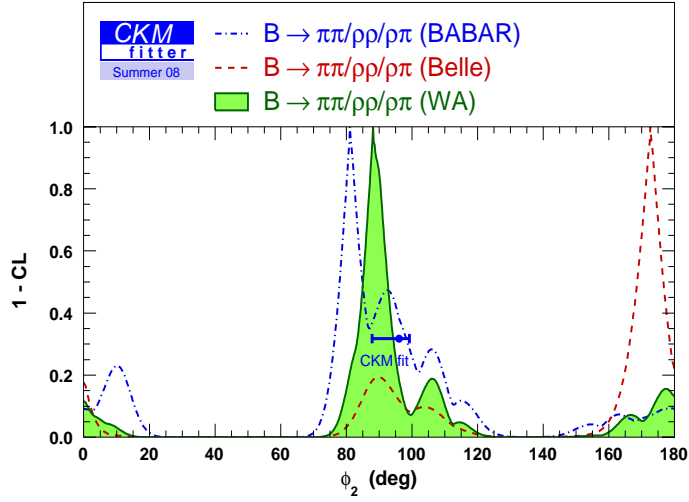

Figure 4.9: Summary of confidence level as a function of $\phi_2$ obtained from $B^0 \to \pi^+\pi^-\pi^0$ decay, $B \to \pi\pi$ isospin analysis, $B \to \rho\rho$ isospin analysis, and all three decay modes combined.

$b \to u\bar{c}s$ diagram) interfere, resulting in a $CP$ violation which is related to $\phi_3$. There are several types of $f_{\text{com}}$ considered to be useful [35–37]

The Dalitz method [37] uses $f_{\text{com}} = K_S^0\pi^+\pi^-$ and currently provides the most promising measurements. It was first tried by Belle with 152 million $B\bar{B}$ pairs [38]. Belle has updated the results for $B^+ \to DK^+$ and $B^+ \to D^*[D\pi^0]K^+$ with an additional mode $B^+ \to DK^{*+}$ using 657 million $B\bar{B}$ pairs [39].

The decay amplitude in Dalitz plane is given for $B^+$ as

$$M_+ = f(m_+^2, m_-^2) + re^{i\phi_3 + i\delta}f(m_-^2, m_+^2), \tag{4.9}$$

and the corresponding amplitude for $B^-$ decay is given as

$$M_- = f(m_-^2, m_+^2) + re^{-i\phi_3 + i\delta}f(m_+^2, m_-^2), \tag{4.10}$$

where $r$ and $\delta$ are the ratio of magnitudes and the strong phase difference of the two amplitudes, $b \to c\bar{u}s$ and $b \to u\bar{c}s$, respectively, $m_+^2 (m_-^2)$ is the invariant mass squared of $K_S^0\pi^+$ ($K_S^0\pi^-$), and $f(m_+^2, m_-^2)$ is the complex amplitude of $D^0 \to K_S^0\pi^+\pi^-$ decay. $f(m_+^2, m_-^2)$ is parametrized as a sum of intermediate quasi-two-body decays using $D^0$ from $D^{*+} \to D^0\pi^+$ whose flavor is tagged by the charge of the soft pion. Unbinned maximum likelihood fits are performed including the effects of background, efficiency, and resolutions. For statistical reasons the fitted parameters are defined as $x_\pm = r_\pm \cos(\pm\phi_3 + \delta)$ and $y_\pm = r_\pm \sin(\pm\phi_3 + \delta)$, where the indices $\pm$ correspond to $B^\pm$ decays. The results for the $DK$ final state are presented in Fig. 4.10 (left). Extraction of the angle $\phi_3$ from the measured $x_\pm$, $y_\pm$ results in

$$\phi_3 = (78.4^{+10.8}_{-11.6}(\text{stat}) \pm 3.6(\text{syst}) \pm 8.9(\text{model}))°. \tag{4.11}$$

BaBar has obtained $\phi_3 = (76 \pm 22(\text{stat}) \pm 5(\text{syst}) \pm 5(\text{model}))°$ using 383 million $B\bar{B}$ pairs [40]. The current world average of $\phi_3$ measurements is presented in Fig. 4.10 (right) and is found to be $(73^{+22}_{-25})°$ [5].

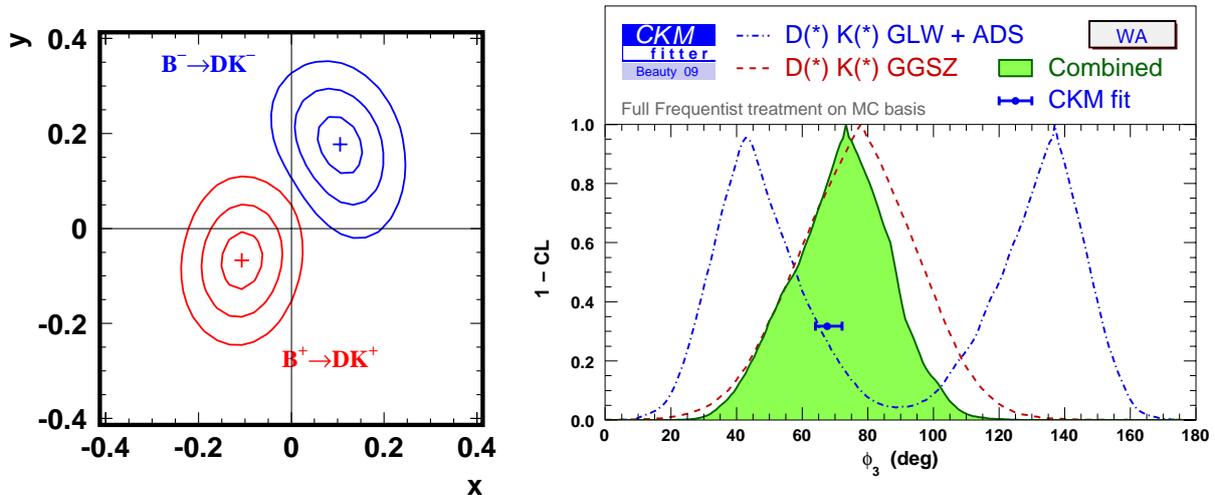

Figure 4.10: Left: $x_\pm$ and $y_\pm$ as obtained from the Dalitz analysis of $B^- \to D^0 K^- \to K_S^0 \pi^+ \pi^- K^-$ [39]. Solid lines correspond to one, two and three standard deviations contours. Right: The combined C.L. for measurements of $\phi_3$. Dashed (red) line shows results of Dalitz analyses, dashed-dotted (blue) line results of ADS and GLW methods, and shaded (green) area the combination of all [5]. For further description of the methods see Sect. 5.8.1.

## 4.5 Status of Unitarity Triangle Determination

Figure 2.2 summarizes the measurements and constraints on the CKM Unitarity Triangle (Eq. (2.8)) described in previous sections together with other available measurements/constraints on $|V_{ub}/V_{cb}|$, $|V_{td}|$ etc. All measurements presented in the $(\bar\rho, \bar\eta)$ plane (Eq. (2.9)) are consistent within the uncertainties that include both experimental and theoretical ones.

## 4.6 Radiative $B$ decays

### 4.6.1 Exclusive $B \to K^* \gamma$

Measurement of the $B \to K^* \gamma$ exclusive branching fraction is straightforward and made with good precision[2]. The results from CLEO [41], BaBar [42] and Belle [43] are in good agreement. The world averages are [4]

$$\begin{aligned}
\mathcal{B}(B^0 \to K^{*0}\gamma) &= (4.40 \pm 0.15) \times 10^{-5}, & (4.12)\\
\mathcal{B}(B^+ \to K^{*+}\gamma) &= (4.57 \pm 0.19) \times 10^{-5}. & (4.13)
\end{aligned}$$

The theoretical predictions [44–46] have considerably larger uncertainties than experimental measurements.

A better approach to exploit the effect of possible new physics contribution in $B \to K^* \gamma$ is to measure the time-dependent $CP$ asymmetry in $B^0$ decay into self-charge conjugate ($C$ eigenstate) final state, $B^0 \to K^{*0}[\to K_S^0 \pi^0]\gamma$. Within the Standard Model, the photon emitted from a $B^0$ ($\overline{B}^0$) meson is predominantly right handed (left handed). A flip of photon polarization is suppressed by the quark mass ratio $2m_s/m_b$ [47]. Hence, for $B^0 \to K^{*0}[\to K_S^0 \pi^0]\gamma$ the

---

[2]$K^*$ denotes $K^*(892)$ throughout this section.

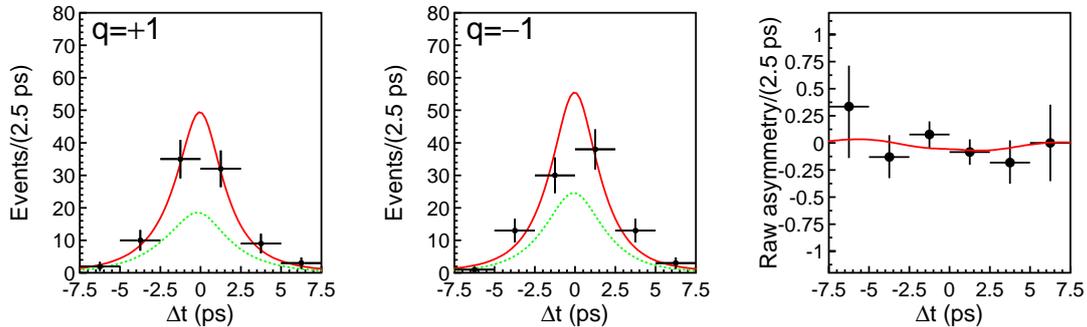

Figure 4.11: $\Delta t$ distributions for $B^0 \to K_S^0 \pi^0 \gamma$ for $B^0$-tagged (left) and $\overline{B}^0$-tagged (middle). The solid (dashed) curve shows the total (signal component) fit result. Asymmetry (right) in each $\Delta t$ bin for good tagging quality events together with fit result.

Standard Model predicts a small time-dependent $CP$ asymmetry, which arises from the interference between decay amplitudes with and without $B^0$-$\overline{B}^0$ mixing (up to 0.1 including a possible enhancement due to strong interactions [48]). The same suppression is expected for final states of the type $B^0 \to P^0 Q^0 \gamma$, where $P^0$ and $Q^0$ are spin-0 neutral particle [49]. A significant deviation from the small Standard Model prediction could indicate new physics. Belle has performed time-dependent $CP$ asymmetry measurement for $B^0 \to K_S^0 \pi^0 \gamma$ decay with a $K_S^0 \pi^0$ invariant mass up to 1.8 GeV/$c^2$ using 535 million $B\overline{B}$ pairs [50]. The result is $\mathcal{S} = -0.10 \pm 0.31 \pm 0.07$ and $\mathcal{A} = -0.20 \pm 0.20 \pm 0.06$. The $\Delta t$ distributions and asymmetry are shown in Fig. 4.11. The averages with BaBar measurement using 467 million $B\overline{B}$ pairs [51] are

$$\mathcal{S} = -0.15 \pm 0.20 \quad \text{and} \quad \mathcal{A} = +0.07 \pm 0.12, \tag{4.14}$$

which are consistent with the Standard Model expectation; however, considering the expected magnitude of the asymmetry, more data are needed to make any conclusion.

### 4.6.2 Other Exclusive Radiative Decays

The dominant radiative decay channel $B \to K^* \gamma$ covers only 12.5% of the total $B \to X_s \gamma$ branching fraction (world average $(3.52 \pm 0.25) \times 10^{-4}$ [4]). The remainder is due to decays with higher resonances or multi-body decays. Knowledge of these decay modes will eventually be useful to reduce the systematic error for the inclusive measurement.

Belle has extended the analysis into various multi-body decay channels, such as $B \to K\pi(\pi)\gamma$ [52, 53], $B \to K\phi\gamma$ [54], $B \to K\eta\gamma$ [55], and $B \to K\eta'\gamma$ [56]. Together with BaBar measurements [57, 58], the world average branching fractions for multi-body decay channels are summarized in Table 4.1.

Time-dependent $CP$-asymmetry measurements are also performed for $B^0 \to K_S^0 \rho^0[\to \pi^+\pi^-]\gamma$ by Belle [59] and for $B^0 \to K_S^0 \eta \gamma$ by BaBar [58], and are useful to exploit the new physics effect as described above.

### 4.6.3 $b \to d\gamma$ Decays

There are various interesting aspects to the $b \to d\gamma$ transition. Within the Standard Model, most of the diagrams are the same as those for $b \to s\gamma$, except for the replacement of the CKM matrix element $V_{ts}$ with $V_{td}$. A measurement of the $b \to d\gamma$ process will therefore provide the ratio $|V_{td}/V_{ts}|$ without large model-dependent uncertainties. This mode is also one where a large direct $CP$-asymmetry is predicted both within and beyond the Standard Model [60].

Table 4.1: Branching fractions for other exclusive $b \to s$ radiative decays.

| mode | $B^+$ ($\times 10^{-6}$) | $\overline{B}^0$ ($\times 10^{-6}$) |
|---|---|---|
| $B \to K\pi^+\pi^-\gamma$ | $27.7 \pm 1.8$ | $19.5 \pm 2.2$ |
| $B \to K\phi\gamma$ | $3.5 \pm 0.6$ | $< 2.7$ |
| $B \to K\eta\gamma$ | $9.4 \pm 1.1$ | $7.6^{+1.8}_{-1.7}$ |
| $B \to K\eta'\gamma$ | $3.2^{+1.2}_{-1.1}$ | $< 6.3$ |

The search for the exclusive decay modes $B \to \rho\gamma$ and $B \to \omega\gamma$ is the easiest and straightforward, and is quite similar to $B \to K^*\gamma$ mode except for its small branching fraction and the enormous combinatorial background from copious $\rho$ and $\omega$ mesons and random pions. The $B \to \rho\gamma$ mode suffers from the huge $B \to K^*\gamma$ background that overlaps with the $B \to \rho\gamma$ signal window, while $B \to \omega\gamma$ is not affected by the $B \to K^*\gamma$ background. In summer 2005, Belle has reported the first observation of $b \to d\gamma$ process with $B \to \rho\gamma$ and $\omega\gamma$ modes using 386 million $B\overline{B}$ pairs [61]. Besides the continuum background suppression, to suppress the $B \to K^*\gamma$ background in $B \to \rho\gamma$, the invariant mass information is used, where the kaon mass is assigned to one of charged pions. The measurements are updated using 657 million $B\overline{B}$ pairs [62]. The $\Delta E$ and $M_{\rm bc}$ distributions are shown in Fig. 4.12. Combining $B \to \rho\gamma$ and $\omega\gamma$ mode, Belle observes 139 signal yields with $6.2\sigma$ significance (including systematic uncertainties), which gives

$$\mathcal{B}(B \to (\rho,\omega)\gamma) = (1.14 \pm 0.20^{+0.10}_{-0.12}) \times 10^{-6}. \quad (4.15)$$

The ratio of this branching fraction to that of $B \to K^*\gamma$ gives $|V_{td}/V_{ts}| = 0.202 \pm 0.018 \pm 0.015$ using a prescription of Ref. [63]. BaBar has confirmed the $B \to \rho\gamma$ and $\omega\gamma$ signals using 347 million $B\overline{B}$ pairs [64] and updated the measutements using 465 million $B\overline{B}$ pairs [65] with a consistent branching fraction, $(1.63^{+0.30}_{-0.24} \pm 0.16) \times 10^{-6}$.

Belle has performed the first time-dependent $CP$ violation measurement for $B^0 \to \rho^0\gamma$ mode using 657 million $B\overline{B}$ pairs [66]. With the present statistics, the result is consistent with no $CP$ asymmetry and consistent with the Standard Model expectation.

## 4.7 Electroweak Rare $B$ Decays

A process $b \to s\ell^+\ell^-$ is suppressed compared to $b \to s\gamma$ by an additional electro-weak coupling. The first signal for $B \to K\ell^+\ell^-$ was observed by Belle [67] using 31 million $B\overline{B}$ pairs and later confirmed by BaBar [68]. Since then, inclusive ($B \to X_s\ell^+\ell^-$) [69–71] and $B \to K^*\ell^+\ell^-$ decays [68,72] were observed and measurements have been improved using larger data samples by Belle and BaBar. The branching fractions and their asymmetries are so far consistent with the Standard Model expectations within experimental and theoretical uncertainties [73–79]. This process has contributions from virtual photon and $Z^0$ (as well as from the box diagram). Since lepton pair enable us to measure additional quantities, dependence on momentum transfer squared $q^2 = M^2(\ell^+\ell^-)$ and forward-backward asymmetry $A_{\rm FB}$ of the lepton decay angle, these could be unique probes for new physics not seen in $b \to s\gamma$ decays.

Belle has performed the first measurement of the forward-backward asymmetry and extracted the ratios of Wilson coefficients, $A_9/A_7$ and $A_{10}/A_7$ in $B \to K^*\ell^+\ell^-$ using 386 million $B\bar{B}$ pairs [80]. The fit to $q^2$ and $\cos\theta_{\ell\ell}$ distributions assuming $A_7 = -0.330$ (the Standard Model expectation from the $\mathcal{B}(B \to X_s\gamma)$ measurements) gives $A_9/A_7 = -15.3\,^{+3.4}_{-4.8}\pm 1.1$ and $A_{10}/A_7 = 10.3\,^{+5.2}_{-3.5} \pm 1.8$. Here, $\theta_{\ell\ell}$ is the angle between the momenta of the negative (positive) lepton and the $B$ ($\bar{B}$) meson in the dilepton rest frame. These are consistent with the Standard Model expectation, $A_9/A_7 = -12.3$ and $A_{10}/A_7 = 12.8$. The result excludes the solutions with $A_9 A_{10} > 0$ with a C.L. of more than 95%.

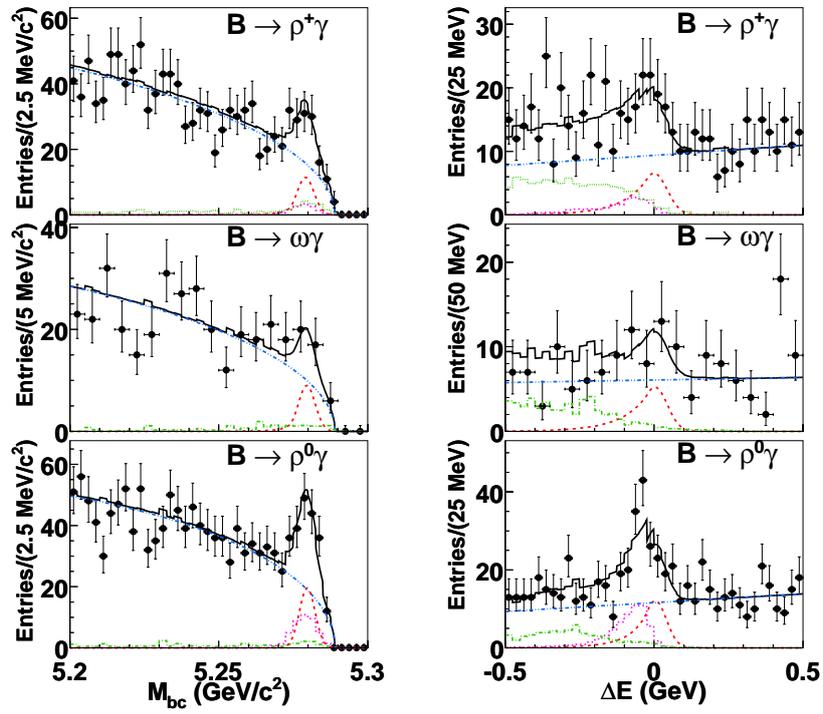

Figure 4.12: Projection of the fit results to $M_{\rm bc}$ in the $\Delta E$ signal region and $\Delta E$ in $M_{\rm bc}$ signal region for individual and simultaneous fits. Curves show the signal (red/dashed line), continuum (blue/dot-dot-dashed line), $B \to K^*\gamma$ (purple/dotted line), other $B$ decay background (green/dot-dashed line) components, and the total fit result (black/solid line).

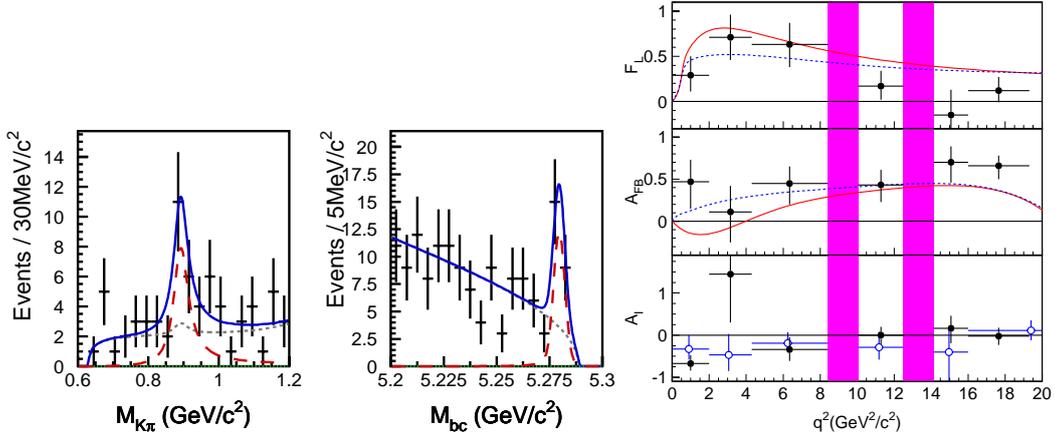

Figure 4.13: $K\pi$ invariant mass and $M_{\rm bc}$ distributions for $B \to K^*\ell^+\ell^-$ samples (left). The solid (blue), long-dashed (red), short-dashed (gray) and dotted (green) curves represent the combined fit result, fitted signal, combinatorial background, and $J/\psi(\psi')X$ background, respectively. Measured $A_{FB}$ (right) as a function of $q^2$. The solid (dashed) curve shows the Standard Model ($C_7 = -C_7^{SM}$) prediction.

The measurements have been updated using 657 million $B\bar{B}$ pairs [81]. Using $230^{+24}_{-23}$ signal yield, we measure the forward-backward asymmetry in six $q^2$ bins by fitting $\cos\theta_{\ell\ell}$ distributions. Figure 4.13 shows the $B \to K^*\ell^+\ell^-$ signal in $K\pi$ mass and $M_{\rm bc}$ distributions (left panel), and the extracted $A_{\rm FB}$ together with the Standard Model and $C_7 = -C_7^{SM}$ expectations (right panel, middle). Charge asymmetry, $K^*$ polarization (right panel, top) and isospin asymmetry (right panel, bottom) are also measured, and are consistent with Standard Model expectations within experimental errors. BaBar also performed a similar measurement of the forward-backward asymmetry and other properties of $B \to K^*\ell^+\ell^-$ using 384 million $B\bar{B}$ pairs [82,83] and obtained consistent results, except that they measure a large negative isosipn asymmetry in the low $q^2$ region while Belle observes no significant isospin asymmetry.

## 4.8 $B \to \tau\bar{\nu}_\tau$

In the Standard Model, the purely leptonic decay $B^- \to \tau^-\bar{\nu}_\tau$ proceeds via annihilation of $b$ and $\bar{u}$ quarks to a $W^-$ boson. It provides a direct determination of the product of the $B$ meson decay constant $f_B$ and the magnitude of the Cabibbo-Kobayashi-Maskawa (CKM) matrix element $|V_{ub}|$. Physics beyond the Standard Model, such as supersymmetry or two-Higgs doublet models, could modify $\mathcal{B}(B^- \to \tau^-\bar{\nu}_\tau)$ through the introduction of a charged Higgs boson (see Sect. 2.5.2 and [84]).

Belle has reported the first evidence of $B^- \to \tau^-\bar{\nu}_\tau$ decay using 449 million $B\bar{B}$ pairs [85]. Since the $B^- \to \tau^-\bar{\nu}_\tau$ decay involves at least two neutrinos which are not detected, it is not possible to reconstruct the decay in the same way as the other exclusive $B$ decays. Therefore, detecting $B^- \to \tau^-\bar{\nu}_\tau$ is experimentally quite challenging. To overcome this difficulty, one of the $B$ mesons (tag side, $B_{\text{tag}}$) in the event is fully (hadronic tagging) or partially (semileptonic tagging) reconstructed, and the properties of the remaining particles (signal side, $B_{\text{sig}}$) are compared to those expected for signal and background. The method allows one to suppress strongly the combinatorial background from both $B\overline{B}$ and continuum events.

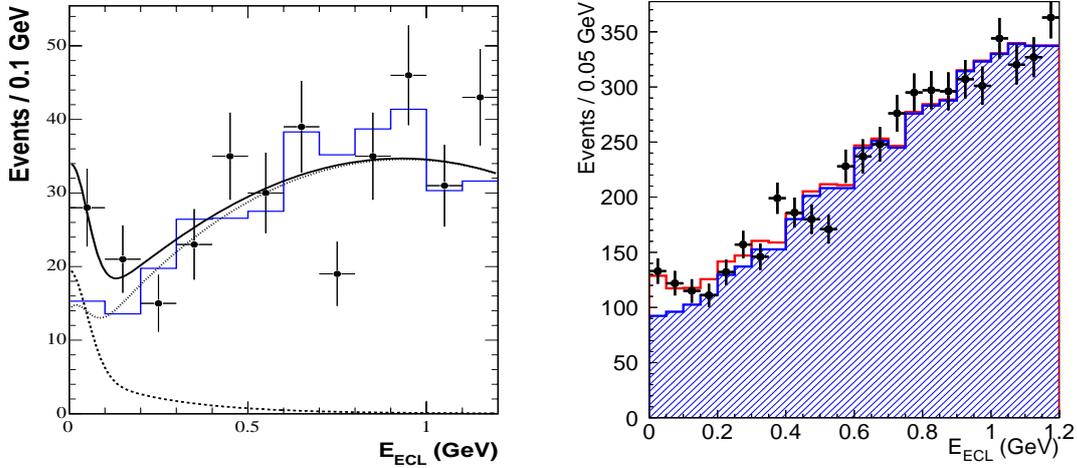

Figure 4.14: $E_{\text{ECL}}$ distributions in the data, using the hadronic tagging (left). The data and background MC samples are represented by the points with error bars and the solid histogram, respectively. The solid curve shows the result of the fit with the sum of the signal (dashed curve) and background (dotted curve) contributions. Same distribution for events with semileptonic tagging (right). The points with error bars are data. The hatched and open histograms are the background and the signal, respectively.

The $B_{\text{tag}}$ candidates are reconstructed in the hadronic decay modes, $B^+ \to \overline{D}^{(*)0}\pi^+$, $\overline{D}^{(*)0}\rho^+$, $\overline{D}^{(*)0}a_1^+$ and $\overline{D}^{(*)0}D_s^{(*)+}$ (hadronic tagging method) [85]. In the events where a $B_{\text{tag}}$ is reconstructed, we search for decays of $B_{\text{sig}}$ into a $\tau$ and a neutrino. The $\tau$ lepton is identified in the five decay modes, $\mu^-\bar{\nu}_\mu\nu_\tau$, $e^-\bar{\nu}_e\nu_\tau$, $\pi^-\nu_\tau$, $\pi^-\pi^0\nu_\tau$ and $\pi^-\pi^+\pi^-\nu_\tau$, which taken together correspond to 81% of all $\tau$ decays. The most powerful variable for separating signal and background is the remaining energy in the ECL ($E_{\text{ECL}}$), which is a sum of the energy of photons that are not associated with either the $B_{\text{tag}}$ or the $\pi^0$ candidate from the $\tau^- \to \pi^-\pi^0\nu_\tau$ decay. Figure 4.14 (left) shows the $E_{\text{ECL}}$ distribution when all $\tau$ decay modes are combined. One can see a significant excess of events in the $E_{\text{ECL}}$ signal region below $E_{\text{ECL}} < 0.25$ GeV. The maximum likelihood fit gives $17.2\,^{+5.3}_{-6.6}$ signal yield in the signal region with a significance of $3.5\sigma$ (including systematic errors).

Belle has also reported the result on $B^- \to \tau^-\bar{\nu}_\tau$ decay with a semileptonic tagging method using 657 million $B\bar{B}$ pairs [86], where $B^- \to D^{*0}\ell\nu$ and $B^- \to D^0\ell\nu$ decay modes are used as $B_{\text{tag}}$. Three $\tau$ decay modes ($\mu^-\bar{\nu}_\mu\nu_\tau$, $e^-\bar{\nu}_e\nu_\tau$, and $\pi^-\nu_\tau$) are used. The signal yield is extracted from the fit to the $E_{\text{ECL}}$ distribution (see Fig. 4.14 (right)) as done in hadronic tagging method. We obtain $154\,^{+36}_{-35}$ signal yield with a significance of $3.8\sigma$ and the branching fraction is

$$\mathcal{B}(B^- \to \tau^-\bar{\nu}_\tau) = (1.65\,^{+0.38}_{-0.37}\,^{+0.35}_{-0.37}) \times 10^{-4}. \tag{4.16}$$

From this value, $f_B \cdot |V_{ub}| = (9.7 \pm 1.1\,^{+1.1}_{-1.1}) \times 10^{-4}$ GeV is determined. Using the value of $|V_{ub}| = 3.99\,^{+0.35}_{-0.30}$ from Ref. [4], we obtain $f_B = 0.242\,^{+0.028}_{-0.027} \pm 0.033$ GeV.

Although the signals are not significant, BaBar reported consistent results, $\mathcal{B}(B^- \to \tau^- \bar{\nu}_\tau) = (1.8\,^{+0.9}_{-0.8} \pm 0.4) \times 10^{-4}$ [87] and $\mathcal{B}(B^- \to \tau^- \bar{\nu}_\tau) = (1.8 \pm 0.8 \pm 0.1) \times 10^{-4}$ [88].

Belle's results represent the first evidence of the pure leptonic $B$ decay and also of the $B$ decay mode with multiple neutrinos. Successful detection of signal yield demonstrates the advantage of the hadronic and semileptonic tagging techniques and encourages to explore other similar decay modes and improved measurement with larger data sample in the future.

## 4.9 Prospects

The above mentioned measurements as well as other measurements will be updated with improved accuracy using a full data sample accumulated by the Belle detector. The value of $\sin 2\phi_1$ will be measured with an accuracy with less than 4% using $b \to c\bar{c}s$ processes with Belle data alone ($\sim 3\%$ combined with BaBar measurement). Though the uncertainties are still large, $\phi_2$ and $\phi_3$ started to be measured using 400 to 657 million $B\bar{B}$ pairs. The accuracies will be improved using the final data set. Together with some improvement in $|V_{cb}|$ and $|V_{ub}|$ measurements, the tension between $\sin 2\phi_1$ and $|V_{ub}|$ (see Fig. 2.2), which is barely seen in the current measurements, may become more clear. More decay modes sensitive to new physics will be searched and established.

We have started to measure some of the observables that are sensitive to new physics and have not been measured in past, such as the forward-backward asymmetry of lepton-pairs in $B \to K^{(*)} \ell^+ \ell^-$, time-dependent $CP$-violation of radiative decays, and others. If the effect of new physics is large, it may be revealed; one example is the current indication of the $CP$-violation in $B^0 \to \phi K^0_S$ decays if the central value remains as currently measured. However, the confirmation of any new physics effect and the understanding of its nature will require much larger data samples, which will be the primary goal of the proposed SuperKEKB and upgraded Belle detector.

# References

[1] K. F. Chen *et al.* [Belle Collaboration], "Observation of time-dependent $CP$ violation in $B^0 \to \eta^{'}K^0$ decays and improved measurements of $CP$ asymmetries in $B^0 \to \phi K^0$, $K_S^0 K_S^0 K_S^0$ and $B^0 \to J/\psi K^0$ decays," Phys. Rev. Lett. **98**, 031802 (2007).

[2] H. Sahoo *et al.* [Belle Collaboration], "Measurements of time-dependent $CP$ violation in $B^0 \to \psi(2S)K_S^0$ decays," Phys. Rev. D **77**, 091103 (2008).

[3] B. Aubert *et al.* [BaBar Collaboration], "Measurement of Time-Dependent CP Asymmetry in $B0 \to c\bar{c}K^{*0}$ Decays," Phys. Rev. D **79**, 072009 (2009).

[4] E. Barberio *et al.*, [Heavy Flavor Averaging group (HFAG)] "Averages of *b*-hadron and *c*-hadron Properties at the End of 2007," arXiv:0808.1297, and online update for 2008 Summer at http://www.slac.stanford.edu/xorg/hfag/.

[5] J. Charles *et al.* [CKMFitter Group], "$CP$ violation and the CKM matrix: assessing the impact of the asymmetric $B$ factories," Eur. Phys. J. C **41**, 1 (2005); updates available at http://www.slac.stanford.edu/xorg/ckmfitter/.

[6] Y. Chao *et al.* [Belle Collaboration], "Measurements of time-dependent $CP$ violation in $B^0 \to \omega K_S^0$, $f_0(980)K_S^0$, $K_S^0\pi^0$ and $K^+K^-K_S^0$ decays," Phys. Rev. D **76**, 091103 (2007).

[7] K. Abe *et al.* [Belle Collaboration], "Measurements of $CP$ Violation Parameters in $B^0 \to K_S^0\pi^0\pi^0$ and $B^0 \to K_S^0 K_S^0$ Decays," arXiv:0708.1845.

[8] M. Fujikawa, Y. Yusa *et al.* [Belle Collaboration], "Measurement of $CP$ asymmetries in $B^0 \to K^0\pi^0$ decays," arXiv:0809.4366.

[9] J. Dalseno *et al.* [Belle Collaboration], "Time-dependent Dalitz Plot Measurement of CP Parameters in $B^0 \to K_s^0\pi^+\pi^-$ Decays," Phys. Rev. D **79**, 072004 (2009).

[10] Y. Grossman and M. P. Worah, "$CP$ asymmetries in $B$ decays with new physics in decay amplitudes," Phys. Lett. B **395**, 241 (1997).

[11] D. London and A. Soni, "Measuring the $CP$ angle $\beta$ in hadronic $b \to s$ penguin decays," Phys. Lett. B **407**, 61 (1997).

[12] Y. Grossman, G. Isidori and M. P. Worah, "$CP$ asymmetry in $B_d \to \phi K_S$: Standard model pollution," Phys. Rev. D **58**, 057504 (1998).

[13] Y. Grossman, Z. Ligeti, Y. Nir and H. Quinn, "SU(3) relations and the $CP$ asymmetries in $B$ decays to $\eta'K_S$, $\phi K_S$ and $K^+K^-K_S$," Phys. Rev. D **68**, 015004 (2003).

[14] B. Aubert *et al.* [BaBar Collaboration], "Measurement of $CP$-violating asymmetries in the $B^0 \to K^+K^-K^0$ Dalitz plot," arXiv:0808.0700 [hep-ex]; B. Aubert *et al.* [BaBar Collaboration], "Measurement of time dependent $CP$ asymmetry parameters in $B^0$ meson decays to $\omega K_S^0$, $\eta' K_S^0$, and $\pi^0 K_S^0$", Phys. Rev. D **79**, 052003 (2009); B. Aubert *et al.* [BaBar Collaboration], "Time-dependent amplitude analysis of $B0 \to K0_S\pi^+\pi^-$", arXiv:0905.3615; B. Aubert *et al.* [BaBar Collaboration], "Measurement of $CP$ asymmetry in $B^0 \to K_S^0 \pi^0 \pi^0$ decays," Phys. Rev. D **76**, 071101 (2007);

B. Aubert *et al.* [BaBar Collaboration], "Time-dependent and time-integrated angular analysis of $B \to \varphi K0_S\pi0$ and $\varphi K^\pm\pi^\mp$", Phys. Rev. D **78**, 092008 (2008).

[15] M. Beneke, "Corrections to sin(2β) from $CP$ asymmetries in $B^0 \to (\pi^0, \rho^0, \eta, \eta', \omega, \phi)K_S^0$ decays," Phys. Lett. B **620**, 143 (2005).

[16] H. Y. Cheng, C. K. Chua and K. C. Yang, "Charmless hadronic $B$ decays involving scalar mesons: Implications to the nature of light scalar mesons," Phys. Rev. D **73**, 014017 (2006).

[17] H. n. Li, S. Mishima and A. I. Sanda, "Resolution to the $B \to \pi K$ puzzle," Phys. Rev. D **72**, 114005 (2005).

[18] A. R. Williamson and J. Zupan, "Two body $B$ decays with isosinglet final states in SCET," Phys. Rev. D **74**, 014003 (2006); [Erratum-ibid. D **74**, 03901 (2006)].

[19] M. Gronau and D. London, "Isospin analysis of $CP$ asymmetries in $B$ decays," Phys. Rev. Lett. **65**, 3381 (1990).

[20] M. Pivk, F. R. Le Diberder, "Isospin constraints from/on $B \to \pi\pi$," Eur. Phys. J. C **39**, 397 (2005).

[21] H. Ishino [Belle Collaboration], "Observation of direct $CP$-violation in $B^0 \to \pi^+\pi^-$ decays with 535 million $B\bar{B}$ pairs," Phys. Rev. Lett. **98**, 211801 (2007).

[22] B. Aubert *et al.* [BaBar Collaboration], "Measurement of $CP$ Asymmetries and Branching Fractions in $B^0 \to \pi^+\pi^-$, $B^0 \to K^+\pi^-$, $B^0 \to \pi^0\pi^0$, $B^0 \to K^0\pi^0$ and Isospin Analysis of $B \to \pi\pi$ Decays," arXiv:0807.4226 [hep-ex].

[23] B. Aubert *et al.* [Babar Collaboration], "A Study of $B^0 \to \rho^+\rho^-$ Decays and Constraints on the CKM Angle $\alpha$," Phys. Rev. D **76**, 052007 (2007).

[24] A. Somov *et al.*, "Measurement of the branching fraction, polarization, and $CP$ asymmetry for $B^0 \to \rho^+\rho^-$ decays, and determination of the CKM phase $\phi_2$," Phys. Rev. Lett. **96**, 171801 (2006).

[25] B. Aubert *et al.* [BaBar Collaboration], "Measurement of the Branching Fraction, Polarization, and $CP$ Asymmetries in $B^0 \to \rho^0\rho^0$ Decay, and Implications for the CKM Angle $\alpha$," Phys. Rev. D **78**, 071104 (2008).

[26] C. C. Chiang *et al.* [Belle Collaboration], "Measurement of $B^0 \to \pi^+\pi^-\pi^+\pi^-$ Decays and Search for $B^0 \to \rho^0\rho^0$," Phys. Rev. D **78**, 111102 (2008).

[27] K. Abe *et al.* [Belle Collaboration], "Improved measurement of $CP$-violating parameters in $B \to \rho^+\rho^-$ decays," Phys. Rev. D **76**, 011104 (2007).

[28] B. Aubert *et al.* [BaBar Collaboration], "Measurements of branching fractions and $CP$-violating asymmetries in $B^0 \to \rho^\pm h^\mp$ decays," Phys. Rev. Lett. **91**, 201802 (2003).

[29] C. C. Wang *et al.* [Belle Collaboration], "Study of $B^0 \to \rho^\pm \pi^\mp$ time-dependent $CP$ violation at Belle," Phys. Rev. Lett. **94**, 121801 (2005).

[30] A. Kusaka, C. C. Wang and H. Ishino *et al.* [Belle Collaboration], "Measurement of $CP$ asymmetry in a time-dependent Dalitz analysis of $B^0 \to (\rho\pi)^0$ and a constraint on the CKM angle $\phi_2$," Phys. Rev. Lett. **98**, 221602 (2007); A. Kusaka *et al.* [Belle Collaboration], "Measurement of $CP$ Asymmetries and Branching Fractions in a Time-Dependent Dalitz Analysis of $B^0 \to (\rho\pi)^0$ and a Constraint on the Quark Mixing Angle $\phi_2$," Phys. Rev. D **77**, 072001 (2008).

[31] A. E. Snyder and H. R. Quinn, "Measuring $CP$ asymmetry in B $\to \rho\pi$ decays without ambiguities," Phys. Rev. D **48**, 2139 (1993).

[32] H. J. Lipkin, Y. Nir, H. R. Quinn and A. Snyder, "Penguin trapping with isospin analysis and $CP$ asymmetries in $B$ decays," Phys. Rev. D **44**, 1454 (1991).

[33] M. Gronau, "Elimination of penguin contributions to $CP$ asymmetries in $B$ decays through isospin analysis," Phys. Lett. B **265**, 389 (1991).

[34] B. Aubert *et al.* [BaBar Collaboration], "Measurement of $CP$-violating asymmetries in $B^0 \to (\rho\pi)^0$ using a time-dependent Dalitz plot analysis," Phys. Rev. D **76**, 012004 (2007).

[35] M. Gronau and D. London., "How to determine all the angles of the unitarity triangle from $B_d^0 \to DK_S$ and $B_s^0 \to D\phi$," Phys. Lett. B **253**, 483 (1991); M. Gronau and D. Wyler, "On determining a weak phase from $CP$ asymmetries in charged B decays," Phys. Lett. B **265**, 172 (1991).

[36] D. Atwood, I. Dunietz and A. Soni, "Enhanced $CP$ violation with $B \to KD^0(\bar{D}^0)$ modes and extraction of the CKM angle $\gamma$," Phys. Rev. Lett. **78**, 3257 (1997). D. Atwood, I. Dunietz and A. Soni, "Improved methods for observing $CP$ violation in $B^\pm \to KD$ and measuring the CKM phase $\gamma$," Phys. Rev. D **63**, 036005 (2001).

[37] A. Bondar, talk at the Belle analysis workshop, Novosibirsk, September 2002; A. Giri, Y. Grossman, A. Soffer and J. Zupan, "Determining $\gamma$ using $B^\pm \to DK^\pm$ with multibody $D$ decays," Phys. Rev. D **68**, 054018 (2003).

[38] A. Poluektov *et al.* [Belle Collaboration], "Measurement of $phi_3$ with Dalitz plot analysis of $B^\pm \to D^{(*)}K^\pm$ decay," Phys. Rev. D **70**, 072003 (2004).

[39] K. Abe *et al.* [Belle Collaboration], "Updated Measurement of $\phi_3$ with a Dalitz Plot Analysis of $B^+ \to D^{(*)}K$ decay," arXiv:0803.3375 [hep-ex];

A. Poluektov [for Belle Collaboration], talk at "The 2009 Europhysics Conference on High Energy Physics", Krakow, July 2009, http://www.ifj.edu.pl/hep2009/index.php

[40] B. Aubert *et al.* [BaBar Collaboration], "Improved measurement of the CKM angle gamma in $B^\pm \to D^{(*)}K^{(*)\pm}$ decays with a Dalitz plot analysis of $D$ decays to $K_S^0\pi^+\pi^-$ and $K_S^0K^+K^-$," Phys. Rev. D **78**, 034023 (2008).

[41] T. E. Coan *et al.* [CLEO Collaboration], "Study of exclusive radiative $B$ meson decays," Phys. Rev. Lett. **84**, 5283 (2000).

[42] B. Aubert *et al.* [BaBar Collaboration], "Measurement of Branching Fractions and $CP$ and Isospin Asymmetries in $B \to K^*\gamma$", Phys. Rev. Lett. **103**, 211802 (2009).

[43] M. Nakao *et al.* [Belle Collaboration], "Measurement of the $B \to K^*\gamma$ branching fractions and asymmetries," Phys. Rev. D **69**, 112001 (2004).

[44] A. Ali and A. Y. Parkhomenko, "Branching ratios for $B \to \rho\gamma$ decays in next-to-leading order in $\alpha_s$ including hard spectator corrections," Eur. Phys. J. C **23**, 89 (2002).

[45] S. W. Bosch and G. Buchalla, "The radiative decays $B \to V\gamma$ at next-to-leading order in QCD," Nucl. Phys. B **621**, 459 (2002).

[46] Y. Y. Keum, M. Matsumori, and A. I. Sanda, "$CP$ asymmetry, branching ratios and isospin breaking effects of $B \to K^*\gamma$ with perturbative QCD approach," Phys. Rev. D **72**, 014013 (2005).

[47] D. Atwood, M. Gronau and A. Soni, "Mixing-induced $CP$ asymmetries in radiative $B$ decays in and beyond the standard model," Phys. Rev. Lett. **79**, 185 (1997).

[48] B. Grinstein, Y. Grossman, Z. Ligeti and D. Pirjol, "The photon polarization in $B \to X\gamma$ in the standard model," Phys. Rev. D **71**, 011504 (2005).

[49] D. Atwood, T. Gershon, M. Hazumi and A. Soni, "Mixing-induced $CP$ violation in $B \to P_1 P_2 \gamma$ in search of clean new physics signals," Phys. Rev. D **71**, 076003 (2005).

[50] Y. Ushiroda *et al.* [Belle Collaboration], "Time-dependent $CP$ asymmetries in $B^0 \to K_S^0 \pi^0 \gamma$ transitions," Phys. Rev. D **74**, 111104 (2006).

[51] B. Aubert *et al.* [BaBar Collaboration], "Measurement of Time-Dependent $CP$ Asymmetry in $B^0 \to K_S^0 \pi^0 \gamma$ decays," Phys. Rev. D **78**, 071102 (2008).

[52] S. Nishida *et al.* [Belle Collaboration], "Radiative $B$ meson decays into $K\pi\gamma$ and $K\pi\pi\gamma$ final states," Phys. Rev. Lett. **89**, 231801 (2002).

[53] H. Yang *et al.*, "Observation of $B^+ \to K_1(1270)^+ \gamma$," Phys. Rev. Lett. **94**, 111802 (2005).

[54] A. Drutskoy *et al.* [Belle Collaboration], "Observation of radiative $B \to \phi K\gamma$ decays," Phys. Rev. Lett. **92**, 051801 (2004).

[55] S. Nishida *et al.* [Belle Collaboration], "Observation of $B^+ \to K^+ \eta \gamma$," Phys. Lett. B **610**, 23 (2005).

[56] R. Wedd *et al.* [Belle Collaboration], "Evidence for $B \to K\eta'\gamma$ Decays at Belle," arXiv:0810.0804.

[57] B. Aubert *et al.* [BaBar Collaboration], "Measurement of branching fractions and mass spectra of $B \to K\pi\pi\gamma$," Phys. Rev. Lett. **98**, 211804 (2007), [Erratum-ibid. **100**, 189903 (2008 ERRAT,100,199905.2008)]; B. Aubert *et al.* [BaBar Collaboration], "Measurement of $B$ decays to $\phi K \gamma$," Phys. Rev. D **75**, 051102 (2007); B. Aubert *et al.* [BaBar Collaboration], "Measurement of branching fractions in radiative $B$ decays to $\eta K \gamma$ and search for $B$ decays to $\eta' K \gamma$," Phys. Rev. D **74**, 031102 (2006).

[58] B. Aubert *et al.* [BaBar Collaboration], "Branching Fractions and $CP$-Violating Asymmetries in Radiative $B$ Decays to $\eta K \gamma$", Phys. Rev. D **79**, 011102 (2009).

[59] J. Li *et al.* [Belle Collaboration], "Time-dependent $CP$ Asymmetries in $B^0 \to K_S^0 \rho^0 \gamma$ Decays", Phys. Rev. Lett. **101**, 251601 (2008).

[60] T. Hurth, E. Lunghi, W. Porod, "Untagged $\bar{B} \to X_{s+d}\gamma$ $CP$ asymmetry as a probe for new physics," Nucl. Phys. B**704**, 56 (2005).

[61] D. Mohapatra *et al.* [Belle Collaboration], "Observation of $b \to d\gamma$ and determination of $|V_{td}/V_{ts}|$," Phys. Rev. Lett. **96**, 221601 (2006).

[62] N. Taniguchi *et al.* [Belle Collaboration], "Measurement of branching fractions, isospin and $CP$-violating asymmetries for exclusive $b \to d\gamma$ modes," Phys. Rev. Lett. **101**, 111801 (2008), [Erratum-ibid. **101**, 129904 (2008)].

[63] P. Ball, G. W. Jones and R. Zwicky, "$B \to V\gamma$ beyond QCD factorisation", Phys. Rev. D **75**, 054004 (2007).

[64] B. Aubert *et al.* [BaBar Collaboration], "Branching fraction measurements of $B^+ \to \rho^+\gamma$, $B^0 \to \rho^0\gamma$, and $B^0 \to \omega\gamma$," Phys. Rev. D **66**, 010001 (2002).

[65] B. Aubert *et al.* [BaBar Collaboration], "Measurements of Branching Fractions for $B^+ \to \rho^+\gamma$, $B^0 \to \rho^0\gamma$, and $B^0 \to \omega\gamma$," Phys. Rev. D **78**, 112001 (2008).

[66] Y. Ushiroda *et al.* [BELLE Collaboration], "Time-Dependent $CP$-Violating Asymmetry in $B^0 \to \rho^0\gamma$ Decays," Phys. Rev. Lett. **100**, 021602 (2008).

[67] K. Abe *et al.* [BELLE Collaboration], "Observation of the decay $B \to K\mu^+\mu^-$," Phys. Rev. Lett. **88**, 021801 (2002).

[68] B. Aubert *et al.* [BaBar Collaboration], "Evidence for the rare decay $B \to K^*\ell^+\ell^-$ and measurement of the $B \to K\ell^+\ell^-$ branching fraction," Phys. Rev. Lett. **91**, 221802 (2003).

[69] J. Kaneko *et al.* [Belle Collaboration], "Measurement of the electroweak penguin process $B \to X_s l^+ l^-$. ((B))," Phys. Rev. Lett. **90**, 021801 (2003).

[70] M. Iwasaki *et al.* [Belle Collaboration], "Improved measurement of the electroweak penguin process $B \to X_s \ell^+\ell^-$," Phys. Rev. D **72**, 092005 (2005).

[71] B. Aubert *et al.* [BaBar Collaboration], "Measurement of the $B \to X_s \ell^+\ell^-$ branching fraction with a sum over exclusive modes," Phys. Rev. Lett. **93**, 081802 (2004).

[72] A. Ishikawa *et al.* [Belle Collaboration], "Observation of $B \to K^* l^+ l^-$," Phys. Rev. Lett. **91**, 261601 (2003).

[73] A. Ali, E. Lunghi, C. Greub and G. Hiller, "Improved model-independent analysis of semileptonic and radiative rare $B$ decays," Phys. Rev. D **66**, 034002 (2002).

[74] E. Lunghi, "Improved model-independent analysis of semileptonic and radiative rare $B$ decays," arXiv:hep-ph/0210379.

[75] W. Jaus and D. Wyler, "The Rare Decays of $B \to K\ell\bar{\ell}$ And $B \to K * \ell\bar{\ell}$," Phys. Rev. D **41**, 3405 (1990).

[76] D. Melikhov, N. Nikitin and S. Simula, "Rare decays $B \to (K, K^*)(\ell^+\ell^-, \nu\bar{\nu})$ in the quark model," Phys. Lett. B **410**, 290 (1997).

[77] P. Colangelo, F. De Fazio, P. Santorelli and E. Scrimieri, "QCD Sum Rule Analysis of the Decays $B \to K\ell+\ell-$ and $B \to K*\ell+\ell-$," Phys. Rev. D **53**, 3672 (1996), [Erratum-ibid. D **57**, 3186 (1998)].

[78] T. M. Aliev, C. S. Kim and Y. G. Kim, "A systematic analysis of the exclusive $B \to K^*\ell^+\ell^-$ decay," Phys. Rev. D **62**, 014026 (2000).

[79] M. Zhong, Y. L. Wu and W. Y. Wang, "Exclusive B meson rare decays and new relations of form factors in effective field theory of heavy quarks," Int. J. Mod. Phys. A **18**, 1959 (2003).

[80] A. Ishikawa et al., "Measurement of forward-backward asymmetry and Wilson coefficients in $B \to K*\ell^+\ell^-$," Phys. Rev. Lett. **96**, 251801 (2006).

[81] J.T. Wei et al. [The Belle Collaboration], "Measurement of the Differential Branching Fraction and Forward-Backward Asymmetry for $B \to K^{(*)}\ell^+\ell^-$," Phys. Rev. Lett. **103**, 171801 (2009).

[82] B. Aubert et al. [BaBar Collaboration], "Angular Distributions in the Decays $B \to K^*\ell^+\ell^-$," Phys. Rev. D **79**, 031102 (2009).

[83] B. Aubert et al. [BaBar Collaboration], "Direct $CP$, Lepton Flavor and Isospin Asymmetries in the Decays $B \to K^{(*)}\ell^+\ell^-$," Phys. Rev. Lett. **102**, 091803 (2009).

[84] W. S. Hou, "Enhanced charged Higgs boson effects in $B- \to \tau\bar{\nu}$, $\mu\bar{\nu}$ and $b \to \tau\bar{\nu} + X$," Phys. Rev. D **48**, 2342 (1993).

[85] K. Ikado et al., "Evidence of the purely leptonic decay $B^- \to \tau^- \bar{\nu}_\tau$," Phys. Rev. Lett. **97**, 251802 (2006).

[86] I. Adachi [The Belle Collaboration], "Measurement of $B^- \to \tau^- \bar{\nu}_\tau$ Decay With a Semileptonic Tagging Method," arXiv:0809.3834.

[87] B. Aubert et al. [BaBar Collaboration], "A Search for $B^+ \to \tau^+\nu$ with Hadronic B tags," Phys. Rev. D **77**, 011107 (2008).

[88] B. Aubert et al. [BaBar Collaboration], "A Search for $B^+ \to \ell^+\nu_\ell$ Recoiling Against $B^- \to D^0\ell^-\bar{\nu}X$," arXiv:0809.4027.

## Chapter 5

# Sensitivity at SuperKEKB

## 5.1 Overview

### 5.1.1 Goals of sensitivity studies

As described in Chapter 1, the primary purpose of SuperKEKB is to perform a comprehensive study of $B$ decays governed by quark transitions induced by quantum loops, where sizable effects from physics beyond the Standard Model are expected. If the study leads to a conclusion that an observed pattern is inconsistent with the Standard Model expectation, we further proceed to identify underlying flavor structure.

One of major goals of sensitivity studies is to clarify the meaning of the statement above quantitatively. To this end, we estimate statistical, systematic and theoretical errors on key observables at SuperKEKB. The target luminosity of SuperKEKB is $8 \times 10^{35}$ cm$^{-2}$s$^{-1}$, which corresponds to an annual integrated luminosity of 8 ab$^{-1}$ assuming 100 days of operation (i.e. the Snowmass year definition). We estimate expected errors at 5 ab$^{-1}$, which is expected to be accumulated within a year, to describe what will be achieved at an early stage of the SuperKEKB experiment. We also provide errors at 50 ab$^{-1}$ as ultimate measurements that can be performed at SuperKEKB. Note that our experience at the Belle experiment tells that an integrated luminosity can be even larger if we achieve the design instantaneous luminosity.

There are a huge number of observables that can be measured at SuperKEKB. It is not the main purpose of our sensitivity studies to cover all of them. Instead, our strategy is to concentrate on observables that are indispensable to reach the primary goal mentioned above. The following points are considered to select such observables:

- A sizable deviation from a prediction of the Standard Model is expected.

- A hadronic uncertainty is negligible or very small.

- A measurement at SuperKEKB has a clear advantage to those at other facilities such as LHCb whose expected physics performance is regarded as the benchmark of next-generation $B$ physics programs at hadron colliders.

At SuperKEKB we expect a level of the beam-induced background that is about 20 times larger than what we currently observe at Belle. Our investigation leads to a conclusion that under such conditions there is a feasible detector design that guarantees performance equivalent to the present Belle detector [1]. Therefore, throughout our studies, we assume a detector that has the same performance as the present Belle detector unless otherwise noticed. Any possible gain from an improved detector performance is regarded as a bonus in this report, and our aim is to demonstrate that most physics goals at SuperKEKB are achieved even without an improvement in the detector performance. One important exception, however, is $B$ meson vertex reconstruction using the $K_S^0$. Here we clearly need to introduce a larger vertex detector to increase vertex reconstruction efficiencies. Therefore we incorporate the proposed detector design in this case.

One of the biggest advantages of the sensitivity studies for SuperKEKB compared with those for hadron collider experiments is that we can fully utilize information obtained by analyzing Belle data. In particular, many of studies described in this chapter rely on Monte Carlo pseudo-experiments (also called "toy Monte Carlo experiments") in which PDFs are constructed from data. A clear advantage of this approach over genuine Monte Carlo simulations (e.g. GEANT simulation) is that background fractions and detector resolutions are more reliable. For some topics for which pseudo-experiments are not available, however, we also use GEANT simulation and/or FSIM, a parametric Monte Carlo simulator that requires much less CPU power than GEANT.

In the rest of this section we overview two important analysis techniques that are used repeatedly in our studies. One is the procedure to fit a proper-time difference distribution for a time-dependent $CP$ asymmetry measurement. The other is to reconstruct one $B$ meson exclusively (or semi-inclusively) so as to study decays of an accompanying $B$ meson in the cleanest environment. This is called "full reconstruction $B$ tagging".

### 5.1.2 Time-dependent $CP$ asymmetries

In the decay chain $\Upsilon(4S) \to B^0\overline{B}^0 \to f_{CP}f_{\text{tag}}$, where one of the $B$ mesons decays at time $t_{CP}$ to a final state $f_{CP}$ and the other decays at time $t_{\text{tag}}$ to a final state $f_{\text{tag}}$ that distinguishes between $B^0$ and $\overline{B}^0$, the decay rate has a time dependence given by [2–4]

$$\mathcal{P}(\Delta t) = \frac{e^{-|\Delta t|/\tau_{B^0}}}{4\tau_{B^0}}\left\{1 + q \cdot \left[\mathcal{S}\sin(\Delta m_d \Delta t) + \mathcal{A}\cos(\Delta m_d \Delta t)\right]\right\}, \tag{5.1}$$

where $\tau_{B^0}$ is the $B^0$ lifetime, $\Delta m_d$ is the mass difference between the two $B^0$ mass eigenstates, $\Delta t = t_{CP} - t_{\text{tag}}$, and the $b$-flavor charge $q = +1$ $(-1)$ when the tagging $B$ meson is a $B^0$ ($\overline{B}^0$). $\mathcal{S}$ and $\mathcal{A}$ are $CP$-violation parameters. For example, to a good approximation, the Standard Model predicts $\mathcal{S} = -\xi_f \sin 2\phi_1$, where $\xi_f = +1(-1)$ corresponds to $CP$-even (-odd) final states, and $\mathcal{A} = 0$ for both $b \to c\bar{c}s$ and $b \to s\bar{s}s$ transitions.

We determine $q$ and $\Delta t$ for each event. Charged leptons, pions, kaons, and $\Lambda$ baryons that are not associated with a reconstructed $CP$ eigenstate decay are used to identify the $b$-flavor of the accompanying $B$ meson. The tagging algorithm is described in detail elsewhere [5]. We use two parameters, $q$ and $r$, to represent the tagging information. The first, $q$, is already defined above. The parameter $r$ is an event-by-event Monte Carlo-determined flavor-tagging dilution parameter that ranges from $r = 0$ for no flavor discrimination to $r = 1$ for an unambiguous flavor assignment. It is used only to sort data into six intervals of $r$, according to estimated flavor purity. We determine directly from data the average wrong-tag probabilities, $w_l \equiv (w_l^+ + w_l^-)/2$ ($l = 1, 6$), and differences between $B^0$ and $\overline{B}^0$ decays, $\Delta w_l \equiv w_l^+ - w_l^-$, where $w_l^{+(-)}$ is the wrong-tag probability for the $B^0(\overline{B}^0)$ decay in each $r$ interval. In some analyses, we remove events with very small $r$ values to ensure the quality of tagging. The event fractions and wrong-tag fractions are summarized in Table 5.1. The total effective tagging efficiency is determined to be $\epsilon_{\text{eff}} \equiv \sum_{l=1}^{6} \epsilon_l(1 - 2w_l)^2 \sim 0.30$, where $\epsilon_l$ is the event fraction for each $r$ interval. The error includes both statistical and systematic uncertainties.

| $l$ | $r$ interval | $\epsilon_l$ | $w_l$ | $\Delta w_l$ | $\epsilon^l_{\text{eff}}$ |
|---|---|---|---|---|---|
| 1 | 0.000 − 0.250 | 0.398 | 0.464 ± 0.006 | −0.011 ± 0.006 | 0.002 ± 0.001 |
| 2 | 0.250 − 0.500 | 0.146 | 0.331 ± 0.008 | +0.004 ± 0.010 | 0.017 ± 0.002 |
| 3 | 0.500 − 0.625 | 0.104 | 0.231 ± 0.009 | −0.011 ± 0.010 | 0.030 ± 0.002 |
| 4 | 0.625 − 0.750 | 0.122 | 0.163 ± 0.008 | −0.007 ± 0.009 | 0.055 ± 0.003 |
| 5 | 0.750 − 0.875 | 0.094 | 0.109 ± 0.007 | +0.016 ± 0.009 | 0.057 ± 0.002 |
| 6 | 0.875 − 1.000 | 0.136 | 0.020 ± 0.005 | +0.003 ± 0.006 | 0.126 ± 0.003 |

Table 5.1: A typical example of the event fractions $\epsilon_l$, wrong-tag fractions $w_l$, wrong-tag fraction differences $\Delta w_l$, and average effective tagging efficiencies $\epsilon^l_{\text{eff}} = \epsilon_l(1-2w_l)^2$ for each $r$ interval at Belle. The errors include both statistical and systematic uncertainties. The event fractions are obtained from the $J/\psi K^0_S$ simulation.

The vertex position for the $f_{CP}$ decay is reconstructed using leptons and charged hadrons, and that for $f_{\text{tag}}$ is obtained with well reconstructed tracks that are not assigned to $f_{CP}$. Tracks that are consistent with coming from a $K^0_S \to \pi^+\pi^-$ decay are not used. Each vertex position is required to be consistent with the interaction region profile, determined run-by-run, smeared in the $r$-$\phi$ plane to account for the $B$ meson decay length. With these requirements, we are able to determine a vertex even with a single track; the fraction of single-track vertices is about 10% for $z_{CP}$ and 22% for $z_{\text{tag}}$. The proper-time interval resolution function $R_{\text{sig}}(\Delta t)$ is formed by convolving four components: the detector resolutions for $z_{CP}$ and $z_{\text{tag}}$, the shift in the $z_{\text{tag}}$ vertex position due to secondary tracks originating from charmed particle decays, and the kinematic approximation that the $B$ mesons are at rest in the cms [6]. A small component of broad outliers in the $\Delta z$ distribution, caused by mis-reconstruction, is represented by a Gaussian function. We determine fourteen resolution parameters from the aforementioned fit to the control samples. We find that the average $\Delta t$ resolution is $\sim 1.43$ ps (rms). The width of the outlier component is determined to be $(39 \pm 2)$ ps; the fractions of the outlier components are $(2.1 \pm 0.6) \times 10^{-4}$ for events with both vertices reconstructed with more than one track, and $(3.1 \pm 0.1) \times 10^{-2}$ for events with at least one single-track vertex.

We determine $\mathcal{S}$ and $\mathcal{A}$ for each mode by performing an unbinned maximum-likelihood fit to the observed $\Delta t$ distribution. The probability density function (PDF) expected for the signal distribution, $\mathcal{P}_{\text{sig}}(\Delta t; \mathcal{S}, \mathcal{A}, q, w_l, \Delta w_l)$, is given by Eq. (5.1) incorporating the effect of incorrect flavor assignment. The distribution is also convolved with the proper-time interval resolution function $R_{\text{sig}}(\Delta t)$, which takes into account the finite vertex resolution. We determine the following likelihood value for each event:

$$\begin{aligned} P_i &= (1-f_{\text{ol}}) \int_{-\infty}^{\infty} \Big[ f_{\text{sig}} \mathcal{P}_{\text{sig}}(\Delta t') R_{\text{sig}}(\Delta t_i - \Delta t') \\ &+ (1-f_{\text{sig}}) \mathcal{P}_{\text{bkg}}(\Delta t') R_{\text{bkg}}(\Delta t_i - \Delta t') \Big] d(\Delta t') \\ &+ f_{\text{ol}} P_{\text{ol}}(\Delta t_i) \end{aligned} \quad (5.2)$$

where $P_{\text{ol}}(\Delta t)$ is a broad Gaussian function that represents an outlier component with a small fraction $f_{\text{ol}}$. The signal probability $f_{\text{sig}}$ depends on the $r$ region and is calculated on an event-by-event basis as a function of $\Delta E$ and $M_{\text{bc}}$. $\mathcal{P}_{\text{bkg}}(\Delta t)$ is a PDF for background events, which is modeled as a sum of exponential and prompt components, and is convolved with a sum of two Gaussians $R_{\text{bkg}}$. All parameters in $\mathcal{P}_{\text{bkg}}(\Delta t)$ and $R_{\text{bkg}}$ are determined by the fit to the $\Delta t$ distribution of a background-enhanced control sample; i.e. events away from the $\Delta E$-$M_{\rm bc}$ signal region. We fix $\tau_{B^0}$ and $\Delta m_d$ at their world-average values. The only free parameters in the final fit are $\mathcal{S}$ and $\mathcal{A}$, which are determined by maximizing the likelihood function $L = \prod_i P_i(\Delta t_i; \mathcal{S}, \mathcal{A})$ where the product is over all events.

| Decay mode | Motivation |
|---|---|
| $B \to X_u l \nu$ | Precise measurement of $|V_{ub}|$ |
| $B \to \tau \nu$ | Measurement of $f_B$ |
| $B \to K \nu \bar{\nu}$, $D \tau \nu$ | Search for new physics |
| Inclusive $B$ decays | Detailed study, model independent analysis etc. |

Table 5.2: Physics topics that will be studied with the fully-reconstructed $B$ sample.

### 5.1.3 $B$ tagging with full reconstruction

At SuperKEKB, $B\bar{B}$ meson pairs will be produced from $\Upsilon(4S)$ decays. To study $B$ meson decays that include neutrinos, photons, $\pi^0$ mesons in the final states, it is useful to tag one of the $B$ mesons through the full reconstruction. This method has the following attractive features:

- The momentum vector and flavor of the other $B$ meson can be identified, *i.e.* single $B$ meson beams can practically be obtained offline.

- Continuum and combinatoric $B\bar{B}$ backgrounds can be significantly reduced.

If we take advantage of these features, it will be possible to measure the $B$ decays listed in Table 5.2. Some of these decays have more than one neutrino in the final states. Therefore, it is very difficult to perform such studies even in the clean environment of $e^+e^-$ collider unless the full reconstruction $B$ tagging is applied.

Because of the modest full reconstruction efficiency [$\mathcal{O}(0.1\%)$], this method has not been extensively applied at the current $B$ factory experiments. However, a very large $B$ meson sample at SuperKEKB will make it possible to extract useful results.

#### Hadronic $B$ Tagging

Most $B$ meson decays are hadronic decays. However, the branching fraction of each mode is less than $\mathcal{O}(1\%)$. Therefore, we need to collect as many modes as possible to achieve a high efficiency. In Fig. 5.1, the beam-constrained mass distribution for the main decay modes, $B \to D^{(*)}(\pi, \rho, a_1)^-$, are shown. The yields including other decay modes are also shown in Table 5.3. With a 152 million $B\bar{B}$ sample, we already have a sample of more than $10^5$ fully-reconstructed $B$ meson tags. The tagging efficiency is 0.20% (0.09%) for charged (neutral) $B$ mesons with a purity of 78% (83%) [7].

The tagging efficiency can be improved largely by loosening selection criteria. In this case, however, the purity becomes lower. Therefore the best selection criteria should be searched for in each analysis. In the case of hadronic $B$ tagging mentioned above, we achieve a tagging efficiency of 0.33% (0.20%) for charged (neutral) $B$ mesons with a purity of 58% (52%).

If a $B$ meson is reconstructed semi-inclusively, we can further improve the efficiency. At BaBar, the $B \to D^{(*)}(n_1\pi^\pm n_2 K^\pm n_3 K_S n_4 \pi^0)^-$ process is reconstructed, where $n_1 + n_2 \leq 5$, $n_3 \leq 2$ and $n_4 \leq 2$. As a result, the higher efficiency, 0.5 (0.3) % for charged (neutral) $B$ mesons, is obtained. However, due to combinatoric backgrounds, the purity is only around

|  | decay mode | yield | eff. (%) | purity(%) |
|---|---|---|---|---|
| Charged B | $B^- \to D^{(*)0}(\pi, \rho, a_1)^-$ | 132723 | 0.17 | 79 |
|  | $B^- \to D^{(*)0}D_s^{(*)-}$ | 8700 | 0.01 | 60 |
|  | $B^- \to J/\psi K^-$ | 9373 | 0.01 | 96 |
|  | total | 150796 | 0.20 | 78 |
| Neutral B | $B^0 \to D^{(*)+}(\pi, \rho, a_1)^-$ | 56898 | 0.07 | 85 |
|  | $\bar{B}^0 \to D^{(*)+}D_s^{(*)-}$ | 4390 | 0.006 | 60 |
|  | $B^0 \to (J/\psi, \psi(2s), \chi_{c1})K_S, J/\psi K^{*0}$ | 7275 | 0.01 | 94 |
|  | total | 68563 | 0.09 | 83 |

Table 5.3: Full reconstructed $B$ events for hadronic modes with 152 million $B\bar{B}$ sample.

25 %. These samples are used for the $|V_{ub}|$ measurement using $B \to X_u l\nu$ decays, where the background can be reduced by requiring a lepton in the recoil side [8, 9].

**Semileptonic $B$ tagging**

If we use semileptonic $B$ decays to tag one $B$ meson, we cannot use information about the $B$ momentum vector due to the missing neutrino. However, we still have relatively clean tagged samples. Semileptonic $B$ decays are dominated by $B \to Dl\nu$ and $D^*l\nu$, with a total branching fraction of around 15%. At Belle, these samples were used for a $|V_{ub}|$ measurement with inclusive $B \to X_u l\nu$ decays. If we require a semileptonic decay on the recoil side, we can apply a kinematical constraint, and the $B$ momentum vector can be determined with a two-fold ambiguity. The background contribution is significantly suppressed in this case. This method yields a $B \to D^*l\nu$ efficiency of 0.3 (0.2)% for charged (neutral) $B$ decays [10].

It is also notable that we include other $B \to DXl\nu$ decays as $B$ tags in the $Dl\nu$ mode, because the signal in the missing mass spectrum is too broad for separation. The missing mass distribution for $B^- \to D^0 l\nu$ is shown in Fig. 5.2. The full reconstruction tagging efficiency is estimated to be 1.7% including the contribution from the $B^- \to D^{(*)0}l\nu$ decays of around 15% [7]. Other background is mostly combinatoric. Although the tagging quality is not good in this case, it is still useful for the study of rare $B$ decays with low track multiplicities in the final states. In BaBar, this method was used to search for the $B \to K\nu\bar{\nu}$ and $\tau\nu$ decays [11, 12] to provide improved upper limits.

**Conclusions**

In summary, the tagging efficiency will be around 0.2% (0.1%) for the clean hadronic tagging for charged (neutral) $B$ mesons, and around 1% for the semileptonic $B$ tagging that has a lower purity. The efficiency for the hadronic tagging can be around 0.3% (0.2%) if we loosen the selection criteria, with a tolerable decrease in the purity of the tagging.

With a 1 ab$^{-1}$ of data, there will be at least 2 million clean tags and around 10 million semileptonic tags. With these samples, the physics topics listed in Table 5.2 will be studied.

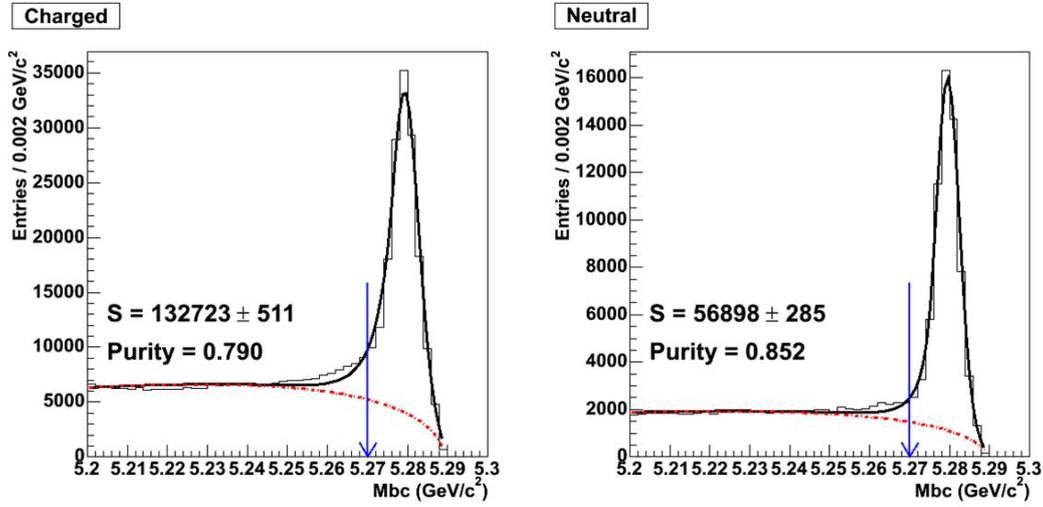

Figure 5.1: Beam-constrained mass distribution for $B \to D^{(*)}(\pi, \rho, a_1)^-$ with the 152 million $B\bar{B}$ sample recorded by Belle.

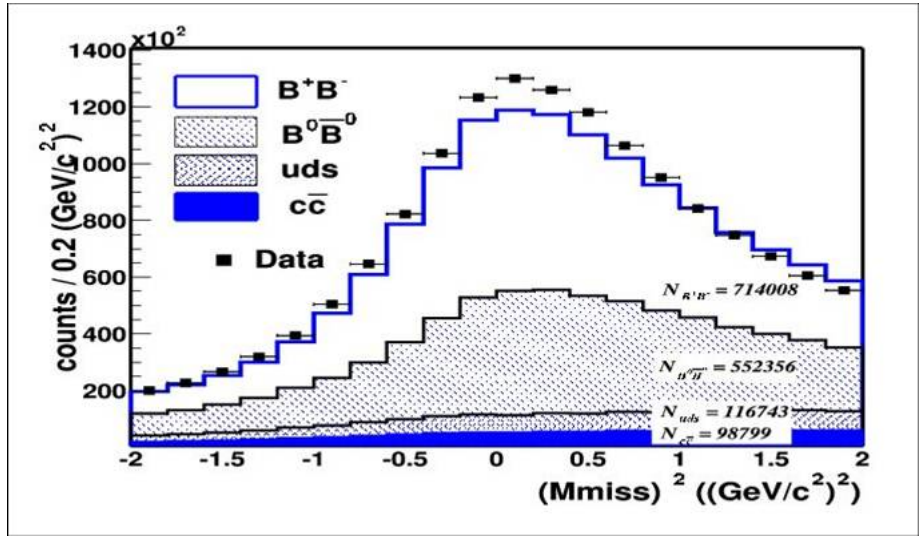

Figure 5.2: Missing mass squared distribution for $B^- \to D^0 l \nu$ with a sample of $85 \times 10^6$ $B\bar{B}$ pairs.

## 5.2 New $CP$-violating phase in $b \to s\bar{q}q$

### 5.2.1 Introduction

Despite the great success of the KM mechanism, additional $CP$-violating phases are inevitable in most theories involving new physics (NP) beyond the SM [13]. Some of them allow large deviations from the SM predictions for $B$ meson decays. Examples include supersymmetric grand-unified theories with the see-saw mechanism that can accommodate large neutrino mixing [14–16]. Therefore it is of fundamental importance to measure $CP$ asymmetries that are sensitive to the difference between the SM and NP. Additional sources of $CP$ violation are also highly desirable to understand the origin of the matter-antimatter asymmetry of the universe; detailed studies have found no way that $CP$ violation in the SM alone could explain baryogenesis [17]. Many methods to search for a new source of $CP$ violation in $B$ meson decays have been proposed up to now. One of the most promising ways is to compare the mixing-induced $CP$ asymmetries in the $B \to \phi K_S^0$ decay [18], which is dominated by the $b \to s\bar{s}s$ transition that is known to be sensitive to possible NP effects, with those in the $B^0 \to J/\psi K_S^0$ decay [3,4]. Ignoring a strong phase difference between the amplitude of NP ($A_{\rm NP}$) and SM ($A_{\rm SM}$)[1], we obtain

$$\mathcal{S}_{\phi K_S^0} = \frac{\sin 2\phi_1 + 2\rho \sin(2\phi_1 + \Theta_{\rm NP}) + \rho^2 \sin(2\phi_1 + 2\Theta_{\rm NP})}{1 + \rho^2 + 2\rho \cos \Theta_{\rm NP}}, \qquad (5.3)$$

where $\rho \equiv A_{\rm NP}/A_{\rm SM}$ is an amplitude ratio of NP to the SM. Since $\mathcal{S}_{J/\psi K_S^0} \simeq \sin 2\phi_1$ is expected in many extensions of the SM, the difference $\Delta \mathcal{S}_{\phi K_S^0} \equiv (-\xi_f)\mathcal{S}_{\phi K_S^0} - \mathcal{S}_{J/\psi K_S^0}$ is a gold-plated observable to search for a new $CP$-violating phase. The other gold-plated modes mediated by the $b \to s\bar{q}q$ are $B^0 \to \eta' K_S^0$ and $B^0 \to K_S^0 K_S^0 K_S^0$.

Recent measurements by Belle [19] and BaBar [20,21] collaborations show the smaller values than the SM expectation. The present $\mathcal{S}$ measurements in $b \to s\bar{q}q$ modes by Belle and BaBar and the world average of $\mathcal{S}$ in $b \to c\bar{c}s$ given by HFAG [22] are summarised in Figure 2.6. Though each $\mathcal{S}$ value for $b \to s\bar{q}q$ mode is not significantly different from the $\mathcal{S}$ in $b \to c\bar{c}s$ that is the SM expectation, the $b \to s\bar{q}q$ values tends to be lower than the SM expectation. Possible theoretical implications of these measurements are already discussed in Section 3.4.

In this section we describe the expected sensitivities for $\mathcal{S}(\phi K^0)$, $\mathcal{S}(\eta' K^0)$ and $\mathcal{S}(K_S^0 K_S^0 K_S^0)$ based on the measurements performed with the present Belle detector. While this section describes measurements of quasi two-body decays, the next section describes the more involved decay time dependent Dalitz plot analyses. It will be crucial to measure as many observables as possible and to check correlations among them. Therefore we also describe several other new methods to access a new $CP$-violating phase in the $b \to s$ transition.

### 5.2.2 $B^0 \to \phi K^0$, $\eta' K^0$ and $B^0 \to K_S^0 K_S^0 K_S^0$

As mentioned in the previous section, the $B^0 \to \phi K_S^0$ $B^0 \to \eta' K_S^0$ and $B^0 \to K_S^0 K_S^0 K_S^0$ decays are the most promising decays in which to search for a new $CP$-violating phase in $b \to s\bar{q}q$ transitions. As most of the experimental procedure for these modes is common, we discuss them here together. Note that the theoretical uncertainties for $\Delta \mathcal{S}$ within the SM depend on the decay mode. We discuss this issue in Section 5.2.5.

We estimate the expected sensitivities at SuperKEKB by extrapolating the present experimental results. Therefore we first explain Belle's analysis with a data sample of 492 fb$^{-1}$ [19].

---
[1] The formula with the strong phase is given in Section 3.4 [Eq. (3.23)].

We reconstruct $B^0$ decays to $\phi K_S^0$ and $\eta' K_S^0$ final states for $\xi_f = -1$ and $\phi K_L^0$, $\eta' K_L^0$ and $K_S^0 K_S^0 K_S^0$ final states for $\xi_f = +1$. The decays $B^0 \to \phi K_S^0$ and $\phi K_L^0$ are combined by redefining $\mathcal{S}$ as $-\xi_f \mathcal{S}$ to take the opposite $CP$ eigenvalues into account and are collectively called "$B^0 \to \phi K^0$". Similarly, $CP$ asymmetries for "$B^0 \to \eta' K^0$" is obtained by combining the decays $B^0 \to \eta' K_S^0$ and $\eta' K_L^0$.

The intermediate meson states are reconstructed from the following decays: $\pi^0 \to \gamma\gamma$, $K_S^0 \to \pi^+\pi^-$ (denoted by $K_S^{+-}$) or $\pi^0\pi^0$ (denoted by $K_S^{00}$), $\eta \to \gamma\gamma$ or $\pi^+\pi^-\pi^0$, $\rho^0 \to \pi^+\pi^-$, $\eta' \to \rho^0\gamma$ or $\eta\pi^+\pi^-$, and $\phi \to K^+K^-$. We use all combinations of the intermediate states with the exception of $\{\eta \to \pi^+\pi^-\pi^0, \eta' \to \rho\gamma\}$ candidates for $B^0 \to \{\eta' K_S^{00}, \eta' K_L^0\}$ decays, respectively. In addition, $\phi \to K_S^{+-} K_L^0$ decays are used for the $B^0 \to \phi K_S^{+-}$ sample. We reconstruct the $B^0 \to K_S^0 K_S^0 K_S^0$ decay in the $K_S^{+-} K_S^{+-} K_S^{+-}$ or $K_S^{+-} K_S^{+-} K_S^{00}$ final states.

Pairs of oppositely charged tracks are used to reconstruct $K_S^0 \to \pi^+\pi^-$ decays. The $\pi^+\pi^-$ vertex is required to be displaced from the IP by a minimum transverse distance of 0.22 cm for high momentum ($> 1.5$ GeV/$c$) candidates and 0.08 cm for those with momentum less than 1.5 GeV/$c$. The direction of the pion pair momentum must agree with the direction defined by the IP and the vertex displacement within 0.03 rad for high-momentum candidates, and within 0.1 rad for the remaining candidates.

To select $K_S^0 \to \pi^0\pi^0$ decays, we reconstruct $\pi^0$ candidates from pairs of photon candidates identified as isolated ECL clusters with $E_\gamma > 0.05$ GeV, where $E_\gamma$ is the photon energy measured with the ECL. Candidate $K_S^0 \to \pi^0\pi^0$ decays are required to have an invariant mass between 0.47 GeV/$c^2$ and 0.52 GeV/$c^2$, where we perform a fit with constraints on the $K_S^0$ vertex and the $\pi^0$ masses. We also require that the distance between the IP and the reconstructed $K_S^0$ decay vertex be larger than $-10$ cm, where the positive direction is defined by the $K_S^0$ momentum.

We select $K_L^0$ candidates from ECL and/or KLM hit patterns that are consistent with the shower induced by a $K_L^0$ meson. We also require that the cosine of the angle between the $K_L^0$ direction and the direction of the missing momentum of the event in the laboratory frame be greater than 0.6.

Candidate $\phi \to K^+K^-$ decays are found by selecting pairs of oppositely charged tracks that are not pion-like ($P(K/\pi) > 0.1$), where a kaon likelihood ratio, $P(K/\pi) = \mathcal{L}_K/(\mathcal{L}_K + \mathcal{L}_\pi)$, has values between 0 (likely to be a pion) and 1 (likely to be a kaon). The likelihood $\mathcal{L}_{K(\pi)}$ is derived from $dE/dx$, ACC and TOF measurements. The vertex of the candidate charged tracks is required to be consistent with the interaction point (IP) to suppress poorly measured tracks. In addition, candidates are required to have a $K^+K^-$ invariant mass that is less than 10 MeV/$c^2$ from the nominal $\phi$ meson mass.

To reconstruct $\eta'$ candidates, we first require that all of the tracks have radial impact parameters $|dr| < 0.1$ cm projected on the $r$-$\phi$ plane. Candidate photons from $\eta_{\gamma\gamma}$ ($\eta'_{\rho\gamma}$) decays are required have $E_\gamma > 50$ (100) MeV. The invariant mass of $\eta_{\gamma\gamma}$ candidates is required to be between 500 MeV/$c^2$ and 570 MeV/$c^2$. A kinematic fit with an $\eta$ mass constraint is performed using the fitted vertex of the $\pi^+\pi^-$ tracks from the $\eta'$ as the decay point. $\pi^0$ candidates are reconstructed from pairs of photons with $E_\gamma > 50$ MeV and required to have invariant mass with in 0.118 GeV/$c^2 < M_{\gamma\gamma} < 0.15$ GeV/$c^2$. The $\pi^+\pi^-\pi^0$ invariant mass is required to be between 0.535 and 0.558 GeV/$c^2$ for the $\eta \to \pi^+\pi^-\pi^0$ decay. For $\eta'_{\rho\gamma}$ decays, the candidate $\rho^0$ mesons are reconstructed from pairs of vertex-constrained $\pi^+\pi^-$ tracks with an invariant mass between 550 and 920 MeV/$c^2$. The $\eta'$ candidates are required to have a reconstructed mass from 940 to 970 MeV/$c^2$ for the $\eta'_{(\gamma\gamma)\pi\pi}$ mode, from 950 to 966 MeV/$c^2$ for the $\eta'_{\eta(\pi^+\pi^-\pi^0)\pi\pi}$ mode, and 935 to 975 MeV/$c^2$ for $\eta'_{\rho\gamma}$ mode.

For reconstructed $B \to f_{CP}$ candidates, we identify $B$ meson decays using the energy difference $\Delta E \equiv E_B^{\rm cms} - E_{\rm beam}^{\rm cms}$ and the beam-energy constrained mass $M_{\rm bc} \equiv \sqrt{(E_{\rm beam}^{\rm cms})^2 - (p_B^{\rm cms})^2}$, where $E_\text{beam}^\text{cms}$ is the beam energy in the cms, and $E_B^\text{cms}$ and $p_B^\text{cms}$ are the cms energy and momentum of the reconstructed $B$ candidate, respectively. The $B$ meson signal region is defined as $|\Delta E| < 0.06$ GeV for $B^0 \to \phi K_S^{+-}$, $|\Delta E| < 0.06$ GeV for $B^0 \to \eta'(\to \rho\gamma) K_S^{+-}$, $-0.10$ GeV $< \Delta E < 0.08$ GeV for $B^0 \to \eta'(\to \pi^+\pi^-\eta(\gamma\gamma)) K_S^{+-}$, $-0.08$ GeV $< \Delta E < 0.06$ GeV for $B^0 \to \eta'(\to \pi^+\pi^-\eta(\pi^+\pi^-\pi^0)) K_S^{+-}$, $|\Delta E| < 0.10$ GeV for $B^0 \to K_S^{+-} K_S^{+-} K_S^{+-}$, $-0.15$ GeV $< \Delta E < 0.1$ GeV for modes including $K_S^{00}$, and 5.27 GeV/$c^2 < M_\text{bc} < 5.29$ GeV/$c^2$ for all decays. In order to distiuguish background from the $e^+e^- \to u\bar{u}$, $d\bar{d}$, $s\bar{s}$, or $c\bar{c}$ continuum, we form signal and background likelihood functions, $\mathcal{L}_\text{S}$ and $\mathcal{L}_\text{BG}$, from a set of variables that characterize the event topology, and calcualte the likelihood ratio $\mathcal{L}_\text{S}/(\mathcal{L}_\text{S} + \mathcal{L}_\text{BG})$.

| Mode | $\xi_f$ | $N_\text{sig}$ | Statistical error | |
|---|---|---|---|---|
| | | | $\mathcal{S}$ | $\mathcal{A}$ |
| $\phi K_S^0$ | $-1$ | 3100 | 0.073 | 0.050 |
| $\phi K_L^0$ | $+1$ | 1200 | 0.178 | 0.121 |
| $\phi K^0$ Total | | 4300 | 0.067 | 0.046 |
| $\eta' K_S^0$ | $-1$ | 14400 | 0.033 | 0.023 |
| $\eta' K_L^0$ | $+1$ | 4600 | 0.075 | 0.050 |
| $\eta' K^0$ Total | | 19000 | 0.030 | 0.021 |
| $K_S^0 K_S^0 K_S^0$ | $+1$ | 1900 | 0.100 | 0.063 |

Table 5.4: Expected numbers of $B^0 \to f_{CP}$ signal events, $N_\text{sig}$, for each $f_{CP}$ mode at 5 ab$^{-1}$.

The vertex reconstruction, flavor tagging and the unbinned maximum likelihood fit to the $\Delta t$ distributions are the same as those described for the $\sin 2\phi_1$ measurement (Section 5.6) with $B^0 \to J/\psi K^0$ decays. By using the identical procedure for both cases, we can reduce the systematic uncertainties in the differences $\Delta \mathcal{S}_{\phi K^0} \equiv \mathcal{S}_{\phi K^0} - \mathcal{S}_{J/\psi K^0}$ etc.

After flavor tagging and vertex reconstruction, we expect the signal yields listed in Table 5.4.

Expected statistical errors on $\mathcal{S}$ and $\mathcal{A}$ extrapolated from the unbinned maximum likelihood fit results with the Belle 492 fb$^{-1}$ data are also shown in Table 5.4. It is seen that errors are rather small even for these rare $B$ decays. Hence systematic uncertainties become crucial.

Major sources of the systematic uncertainties on $\mathcal{S}$ and $\mathcal{A}$ are common to those for $B^0 \to J/\psi K^0$, which will be described in Section 5.6. An additional systematic error for $B^0 \to \phi K_S^0$ arises from the non-resonant $K^+K^- K_S^0$ and $B^0 \to f_0 K_S^0$ backgrounds. These background fractions are estimated with the current Belle data from the Dalitz plot for $B \to K^+K^-K$ to be $2.75 \pm 0.14\%$ and consistent with zero within error, respectively. We assume that the errors on these fractions will be reduced as the integrated luminosity increases. We also assume that the $CP$-violating parameters for these decays are also determined experimentally with errors that will also decrease as the integrated luminosity increases. Therefore, the systematic errors on $\mathcal{S}$ and $\mathcal{A}$ due to these backgrounds are also assumed to be *reducible*.

Some of the systematic errors cancel when we calculate $\Delta \mathcal{A}$ or $\Delta \mathcal{S}$. For example, the effect of the tag-side interference that dominates the systematic error for $\mathcal{A}$ cancels in $\Delta \mathcal{A}_{\phi K^0}$ since it causes a bias in the same direction for $\mathcal{S}_{\phi K^0}$ and $\mathcal{S}_{J/\psi K^0}$ measurements. Because the systematic bias from the tag-side interference does not cancel between $\mathcal{A}_{K_S^0 K_S^0 K_S^0}$ and $\mathcal{A}_{J/\psi K^0}$, which have different $CP$ contents, we use information from $B^0 \to J/\psi K_L^0$ decays to calculate $\Delta \mathcal{A}_{K_S^0 K_S^0 K_S^0}$ to reduce this uncertainty. Otherwise we use $B^0 \to J\psi K^0$ decays to estimate $\Delta \mathcal{S}$ and $\Delta \mathcal{A}$.

Figure 5.3 shows the resulting total errors on $\Delta \mathcal{S}$ as a function of integrated luminosity. Table 5.5 also shows the corresponding values at 5 ab$^{-1}$ and 50 ab$^{-1}$.

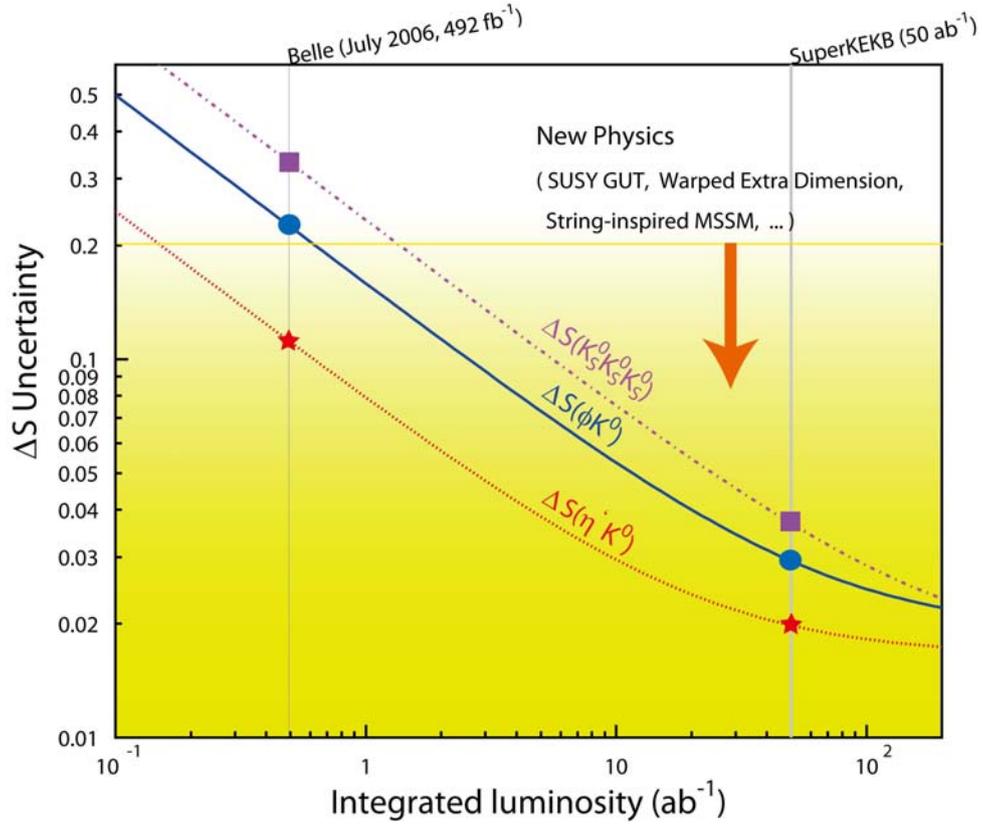

Figure 5.3: Expected total errors on $\Delta \mathcal{S}$ as a function of integrated luminosity.

| Mode | 5 ab$^{-1}$ | | 50 ab$^{-1}$ | |
| --- | --- | --- | --- | --- |
| | $\Delta \mathcal{S}$ | $\Delta \mathcal{A}$ | $\Delta \mathcal{S}$ | $\Delta \mathcal{A}$ |
| $\phi K^0$ | 0.073 | 0.049 | 0.029 | 0.018 |
| $\eta' K^0$ | 0.038 | 0.026 | 0.020 | 0.012 |
| $K_S^0 K_S^0 K_S^0$ | 0.105 | 0.067 | 0.037 | 0.024 |

Table 5.5: Expected total errors on $\Delta \mathcal{S}$ and $\Delta \mathcal{A}$ at 5 ab$^{-1}$ and 50 ab$^{-1}$.

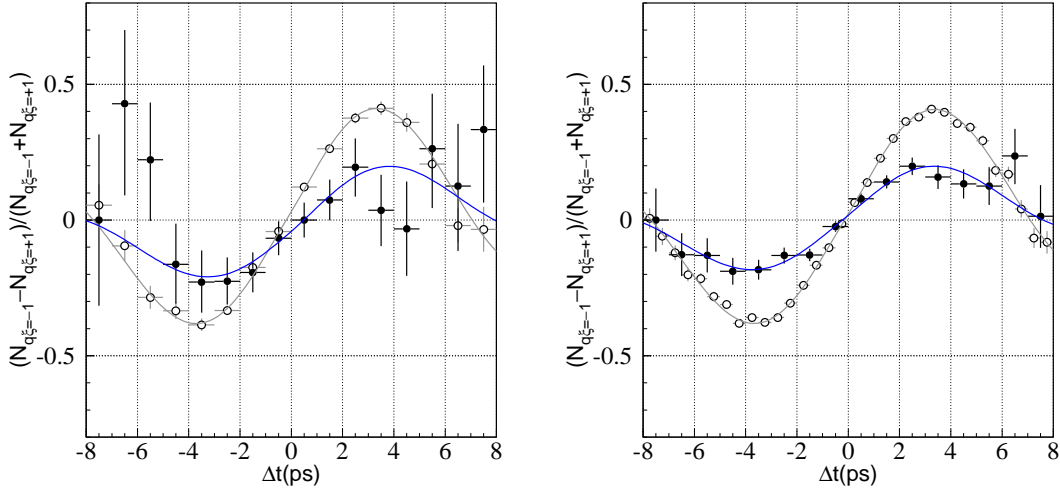

Figure 5.4: Raw asymmetries for $B^0 \to \phi K_S^0$ (closed circles) and $B^0 \to J/\psi K_S^0$ (open circles) at 5 ab$^{-1}$ (left) and at 50 ab$^{-1}$ (right). Input values are $\mathcal{S}_{\phi K_S^0} = +0.39$ and $\mathcal{A}_{\phi K_S^0} = 0$ for $B^0 \to \phi K^0$.

Based on the above estimates, we perform Feldman-Cousins analyses to obtain 5$\sigma$ discovery regions at 5 ab$^{-1}$ and at 50 ab$^{-1}$ in the 2-dimensional plane of $\mathcal{A}$ and $\mathcal{S}$. Results are shown in Figure 5.5. Figure 5.4 shows an example of a fit to events in a MC pseudo-experiment for the $B^0 \to \phi K_S^0$ and $J/\psi K_S^0$ decays at 5 ab$^{-1}$ and 50 ab$^{-1}$, where the input value of $\mathcal{S}(\phi K_S^0) = +0.39$ is chosen to be the world average value of the $\mathcal{S}$ term for $B^0 \to \phi K^0$. At SuperKEKB, even a small deviation of $\Delta \mathcal{S} \sim 0.1$ can be established with a 5$\sigma$ significance as far as the statistical and systematic errors are concerned. Therefore it is important to understand levels of theoretical uncertainties within the SM very well. This issue will be discussed in Section 5.2.5.

### 5.2.3 Time-dependent Dalitz plot analysis of $B^0 \to \phi K_S^0$

In the previous section quasi two-body time-dependent $CP$ analyses were performed on $B^0 \to \phi K^0$. However, a two-body approach to this mode is not ideal due to interference with other resonances as well as non-resonant decays into the same final state $K^+K^-K_S$. These effects can be taken into account with a time-dependent Dalitz plot analysis. In addition, the interference terms in some cases may be sensitive to the cosine of the effective weak phase difference ($\phi_1^{eff}$) in mixing, potentially resolving the two-fold ambiguity in $\phi_1^{eff}$, characteristic of quasi two-body analysis.

The two independent variables of the Dalitz plot are the invariant squared masses,

$$s_{\pm} \equiv (p_{\pm} + p_0)^2, \tag{5.4}$$

where $p_+$, $p_-$ and $p_0$ denote the four-momentum of $K^+$, $K^-$ and $K_S^0$, respectively. The third variable, $s_0 \equiv (p_+ + p_-)^2$, can be obtained from the 4-vector conservation:

$$s_0 = m_{B^0}^2 + 2m_{K^+}^2 + m_{K_S^0}^2 - s_+ - s_-. \tag{5.5}$$

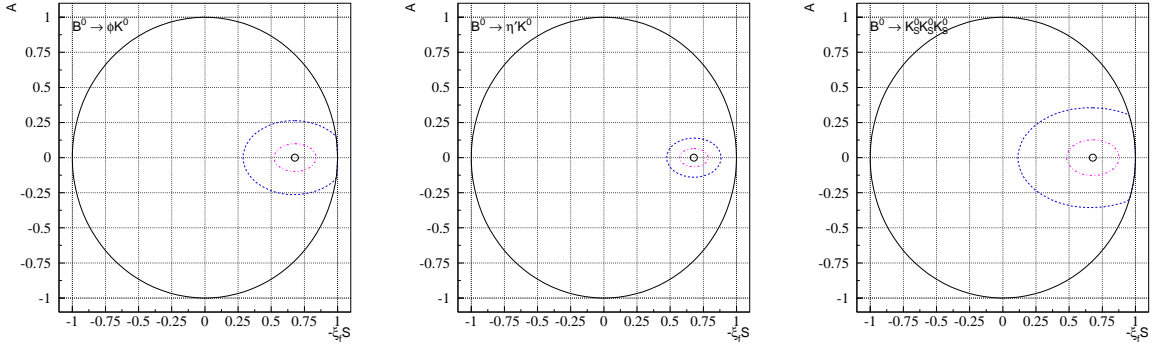

Figure 5.5: $5\sigma$ confidence regions for $\mathcal{A}$ and $\mathcal{S}$ in (left) $B^0 \to \phi K_S^0$, (middle) $B^0 \to K^+K^-K_S^0$ and (right) $B^0 \to \eta' K_S^0$ decays at 5 ab$^{-1}$ (larger dashed/blue ellipse) and 50 ab$^{-1}$ (smaller dashed/red ellipse). Input values are $\mathcal{S} = 0.73$ and $\mathcal{A} = 0$.

The differential $B^0$ decay width with respect to the Dalitz variables is

$$d\Gamma(B^0 \to K^+K^-K_S^0) = \frac{1}{(2\pi)^3} \frac{|A|^2}{32m_{B^0}^3} ds_+ ds_-, \tag{5.6}$$

where $A$ is the Lorentz-invariant amplitude of the decay. In the isobar approximation, the total amplitude of $B^0(\overline{B^0}) \to K^+K^-K_S^0$ is written as the sum of various intermediate Breit/Wigner resonances contributing to this final state,

$$A(s_+, s_-) = \sum_i a'_i F_i(s_+, s_-), \quad \bar{A}(s_-, s_+) = \sum_i \bar{a}'_i \bar{F}_i(s_-, s_+), \tag{5.7}$$

where $a'_i \equiv a_i e^{ib_i}$ are complex coefficients describing the relative magnitudes and phases between the decay channels and carry the weak phase dependence. The Dalitz-dependent amplitudes, $F_i(s_+, s_-)$ contain only strong dynamics and thus $F_i(s_+, s_-) = \bar{F}_i(s_-, s_+)$. They can be expanded in terms of invariant mass and angular dependence,

$$F_i^L(s_+, s_-) \equiv X_i^L(\vec{p}^*) \times X_i^L(\vec{q}) \times Z_i^L(\vec{p}, \vec{q}) \times R_i(s_+, s_-), \tag{5.8}$$

where $\vec{p}^*$ is the momentum of the bachelor particle in the $B^0$ rest frame, $\vec{p}$ and $\vec{q}$ are the momenta of the bachelor particle and one of the resonance daughters in the resonance frame, respectively, $L$ is the orbital angular momentum between the resonance and the bachelor particle, while $X_i^L$ are the Blatt-Weisskopf barrier factors [23]. The angular distribution, $Z_i^L(\vec{p}, \vec{q})$ depends on $L$,

$$\begin{aligned} Z_i^0(\vec{p}, \vec{q}) &= 1, \\ Z_i^1(\vec{p}, \vec{q}) &= -4\vec{p} \cdot \vec{q}. \end{aligned} \tag{5.9}$$

The mass shapes are denoted as $R_i(s_+, s_-)$ which differ depending on the decay channel. We use the Relativistic Breit-Wigner (RBW) [24] and Flatté [25] line shapes. The non-resonant component is empirically modeled by the sum of three exponential functions as described in the previous time-integrated Dalitz plot analysis of the $B^+ \to K^+K^-K^+$ decay from Belle [26].

Table 5.6 summarizes the resonances considered in the $B^0 \to K^+K^-K_S^0$ signal model, which was motivated by the time-integrated Dalitz plot analysis of the $B^+ \to K^+K^-K^+$ decay [26].

Table 5.6: Summary of the resonances considered in the $B^0 \to K^+ K^- K^0_S$ signal model.

| Resonances | Fixed parameters (GeV) | Form Factor, $R_i(s_+, s_-)$ | L |
|---|---|---|---|
| $f_0$ | $M = 0.965 \pm 0.010$ [28] | Flatté | 0 |
| | $g_\pi = 0.165 \pm 0.018$ | | |
| | $g_K = (4.21 \pm 0.09)g_\pi$ | | |
| $\phi$ | $M = 1.019455 \pm 0.020$ [24] | RBW | 1 |
| | $\Gamma = 0.00426 \pm 0.00004$ | | |
| $f_X$ | $M = 1.524 \pm 0.014$ [26] | RBW | 0 |
| | $\Gamma = 0.136 \pm 0.023$ | | |
| $\chi_{c0}$ | $M = 3.41475 \pm 0.00035$ [24] | RBW | 0 |
| | $\Gamma = 0.0104 \pm 0.0007$ | | |
| $(K^+ K^-)_{\rm NR}$ | | $e^{-\alpha s^0}$ | |
| $(K^0_S K^+)_{\rm NR}$ | | $e^{-\alpha s^+}$ | |
| $(K^0_S K^-)_{\rm NR}$ | | $e^{-\alpha s^-}$ | |

The $f_X$ resonance with unknown spin that appears in the table was first introduced in Ref. [26] and was also considered in the analysis by BaBar [27] in order to account for a wide enhancement of signal events observed at $M(K^+K^-) \sim 1.5$ GeV/$c^2$. Since it is best described as a scalar we assume the spin of $f_X$ to be 0.

The decay of the $\Upsilon(4S)$ produces a $B^0 \overline{B}^0$ pair of which one ($f_{CP}$) may be reconstructed as $K^+ K^- K^0_S$ while the other ($f_{\rm tag}$) may reveal its flavor. Since the $CP$ eigenvalue of $K^0_S$ is $\eta_{CP} = +1$, the time-dependent decay rate is given by

$$|A(\Delta t, q)|^2 = \frac{e^{-|\Delta t|/\tau_{B^0}}}{4\tau_{B^0}} \bigg[ (|A|^2 + |\bar{A}|^2) - q(|A|^2 - |\bar{A}|^2)\cos \Delta m_d \Delta t \quad (5.10)$$

$$+ 2q \mathrm{Im}(\bar{A} A^*) \sin \Delta m_d \Delta t \bigg], \quad (5.11)$$

where no $CP$ violation in mixing, $|q/p| = 1$, is assumed.

The Dalitz-dependent amplitudes, $A$, were previously defined in Eq. 5.7 and we choose a convention where the $B^0 \overline{B}^0$ phase of $q/p$ is absorbed into the $\overline{B}^0$ decay amplitude, $\bar{a}'_i$. These complex coefficients can be redefined in a way that parametrizes the direct and indirect $CP$ violation,

$$a'_i \equiv a_i (1 + c_i) e^{i(b_i + d_i)} \quad (5.12)$$

for $A$ and

$$\bar{a}'_i \equiv a_i (1 - c_i) e^{i(b_i - d_i)} \quad (5.13)$$

for $\bar{A}$. Thus an intermediate state $i$ exhibits a direct $CP$ violation asymmetry given by

$$\mathcal{A}(i) \equiv \frac{|\bar{a}'_i|^2 - |a'_i|^2}{|\bar{a}'_i|^2 + |a'_i|^2} = \frac{-2c_i}{1 + c_i^2}, \quad (5.14)$$

where $c_i$'s are restricted by its definition to lie between -1 and 1.

For an intermediate $CP$ eigenstate the CKM angle, $\phi_1^{eff}(i)$, is expressed by a fitted parameter

$$\phi_1^{eff}(i) \equiv \frac{\arg(a'_i \bar{a}'^*_i)}{2} = d_i \quad , \quad (5.15)$$

and its effective mixing-induced $CP$ violation asymmetry is calculated as

$$-\eta_i \mathcal{S}(i) \equiv \frac{-2\mathrm{Im}(\bar{a}'_i a'^*_i)}{|a'_i|^2 + |\bar{a}'_i|^2} = \frac{1-c_i^2}{1+c_i^2}\sin 2\phi_1^{eff}(i). \tag{5.16}$$

Here $\eta_i$ is the $CP$ eigenvalue of the final state. Note that $\mathcal{A}(i)$ and $\mathcal{S}(i)$ are restricted by these definitions to lie in the physical region.

We estimate the expected sensitivities at SuperKEKB by extrapolating the present experimental results [29].

The main contributions to the systematic uncertainties are originating from the background, the vertex reconstruction and the wrong tag fraction as well as the Dalitz model. With a larger integrated luminosity, some of the systematic uncertainties can be reduced. We estimate the uncertainties and treat the reducible and irreducible systematic sources separately. The uncertainties due to the vertex reconstruction, $\Delta t$ resolution function parametrization and tag-side interference are included in the former, while the remaining sources belong to the latter.

We calculate the experimental uncertainties as a function of integrated luminosity as shown in Fig. 5.6. At an integrated luminosity of 5 ab$^{-1}$, the uncertainty of $\phi_1^{eff}$ in the $B^0 \to \phi K_S^0$ decay is reduced to 3.3°. By converting $\delta\phi_1^{eff}$ to $\delta\mathcal{S}$, the experimental uncertainty of $\mathcal{S}$ in $B^0 \to \phi K_S^0$ is found to be $\sim 0.10$. At the integrated luminosity of 50 ab$^{-1}$ we could measure $\phi_1^{eff}$ in the $B^0 \to \phi K_S^0$ decays with an accuracy of 1.5°.

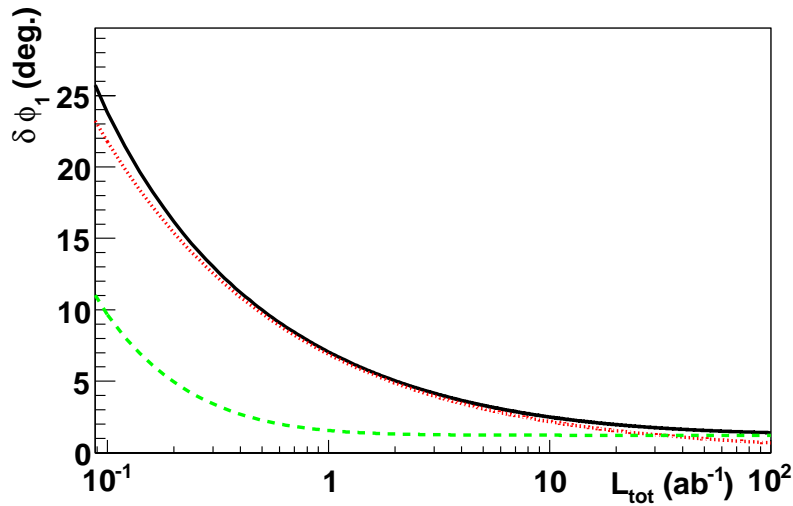

Figure 5.6: Estimated uncertainties as a function of the integrated luminosity in $B^0 \to \phi K_S^0$ decay. The dotted (red) and dashed (green) curves show the statistical and systematic errors, respectively. The solid (black) line shows the total error obtained as the quadratic sum of statistical and systematic errors.

### 5.2.4 $B^{\pm} \to \phi\phi X_s^{\pm}$

In this section we discuss a new method to study direct $CP$ violation that arises from a new $CP$-violating phase in $B^{\pm} \to \phi\phi X_s^{\pm}$ decays [30]. Here $X_s^{\pm}$ represents a final state with a specific strange flavor such as $K^{\pm}$ or $K^{*\pm}$. These non-resonant direct decay amplitudes are dominated by the $b \to s\bar{s}s\bar{s}s$ transition. A contribution from the $b \to u\bar{u}s$ transition followed by rescattering into $s\bar{s}s$ is expected to be below 1% because of CKM suppression and the OZI rule [30]. In these decays, when the invariant mass of the $\phi\phi$ system is within the $\eta_c$ resonance region, they interfere with the $B^{\pm} \to \eta_c(\to \phi\phi)X_s^{\pm}$ decay that is dominated by the $b \to c\bar{c}s$ transition. The decay width of $\eta_c$ is sufficiently large [4, 31] to provide a sizable interference. Within the SM, this interference does not cause sizable direct $CP$ violation because there is no weak phase difference between the $b \to s\bar{s}s\bar{s}s$ and the $b \to c\bar{c}s$ transitions. On the other hand, a NP contribution with a new $CP$-violating phase can create a large weak phase difference. Thus large $CP$ asymmetries can appear only from NP amplitudes, and an observation of direct $CP$ violation in these decays is an unambiguous manifestation of physics beyond the SM.

Although the same argument so far is applicable to the $B^{\pm} \to \phi X_s^{\pm}$ decays, there is no guaranteed strong phase difference that is calculable reliably for these decays. In contrast, the Breit-Wigner resonance provides the maximal strong phase difference in the case of $B^{\pm} \to (\phi\phi)_{m \sim m_{\eta_c}} X_s^{\pm}$ decays. Since present experimental knowledge of the decay rate for $b \to s\bar{s}s$ is still limited, a large $CP$ asymmetry up to 0.4 is allowed.

The Belle Collaboration recently announced evidence for $B \to \phi\phi K$ decays [32]. The signal purity is close to 100% when the $\phi\phi$ invariant mass is within the $\eta_c$ mass region. Belle [31] has also reported the first observation of the $B^0 \to \eta_c K^{*0}$ decay. This implies that other modes such as $B^+ \to \eta_c K^{*+}$ will also be seen with a similar branching fraction, so that we will be able to study semi-inclusive $B^{\pm} \to \eta_c X_s^{\pm}$ transitions experimentally. The semi-inclusive branching fraction of $B^{\pm} \to \eta_c X_s^{\pm}$ is not yet measured, but is expected to be comparable to the branching fraction of the semi-inclusive decay $B^{\pm} \to J/\psi X_s^{\pm}$ [33–35].

We have performed Monte Carlo simulation for the $B^{\pm} \to \phi\phi K^{\pm}$ decay and estimated statistical errors on the $CP$ asymmetry parameter. The procedure and the fit parameters are the same as those described in [30]. The reconstruction efficiency and the $\phi\phi$ mass resolution are estimated using a GEANT-based detector simulator for the present Belle detector [36]. We perform an unbinned maximum-likelihood fit to the differential decay rate distribution. Figure 5.7 shows the $5\sigma$ search regions at 5 ab$^{-1}$ (dotted line) and at 50 ab$^{-1}$ (solid line), where $r^2$ is the ratio between the NP amplitude and the SM amplitude, and $\Theta_{\rm NP}$ is the $CP$-violating phase from NP. Direct $CP$ violation can be observed in a large parameter space with significance above $5\sigma$.

Figure 5.8 shows the expected significance of the new phase $\Theta_{\rm NP}$ at 5 ab$^{-1}$ for $B^{\pm} \to \phi\phi K^{\pm}$ decay ($r^2 = 0.5$) and for time-dependent $CP$ violation in the $B^0 \to \phi K_S^0$ decay ($|A_{\rm NP}/A_{\rm SM}|^2 = 0.5$).

The significance for $\Delta S_{\phi K_S^0}$ depends on the sign of $\Theta_{\rm NP}$, which is not the case for the $B^{\pm} \to \phi\phi K^{\pm}$ decay. The sign dependence arises from an asymmetric range for $\Delta S_{\phi K_S^0}$; to a good approximation, we have $-1 - \sin 2\phi_1 \leq \Delta S \leq 1 - \sin 2\phi_1$ where $\sin 2\phi_1 = +0.650 \pm 0.029 \pm 0.018$ [22]. Therefore the $B^{\pm} \to \phi\phi K^{\pm}$ decay plays a unique role in searching for a new $CP$-violating phase.

Experimental sensitivities can be improved by adding more final states. The technique to reconstruct $X_s$, which has been successfully adopted for the measurements of semi-inclusive $B \to X_s \ell\ell$ transitions [37], can be used for this purpose. Flavor-specific neutral $B$ meson decays, such as $B^0 \to \phi\phi K^{*0}(\to K^+\pi^-)$, and other charmonia such as the $\chi_{c0} \to \phi\phi$ decay can also be included.

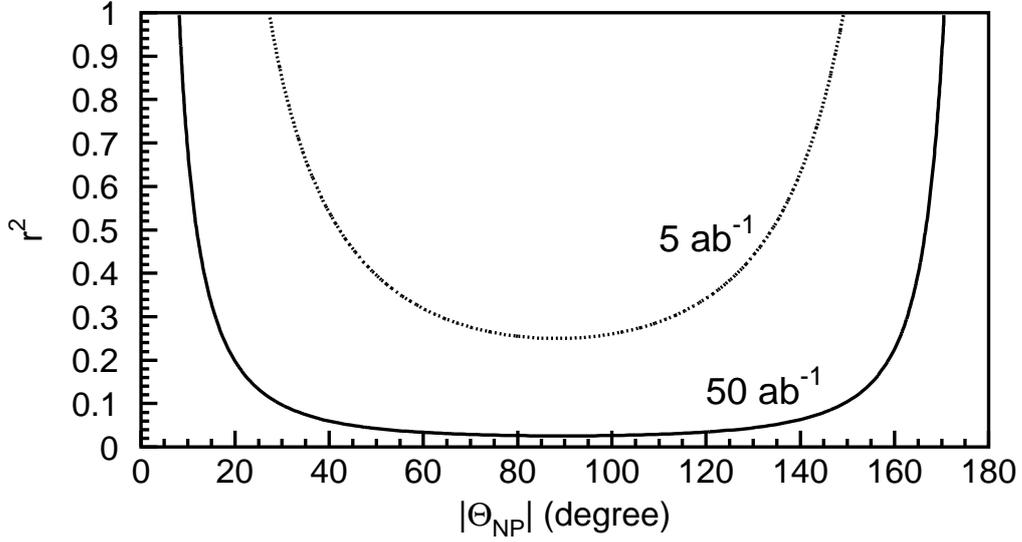

Figure 5.7: Expected sensitivities on direct $CP$ violation in the $B^\pm \to \phi\phi K^\pm$ decay at 5 ab$^{-1}$ (dotted line) and at 50 ab$^{-1}$ (solid line). In the regions above the curves, direct $CP$ violation can be measured with a $5\sigma$ significance or larger.

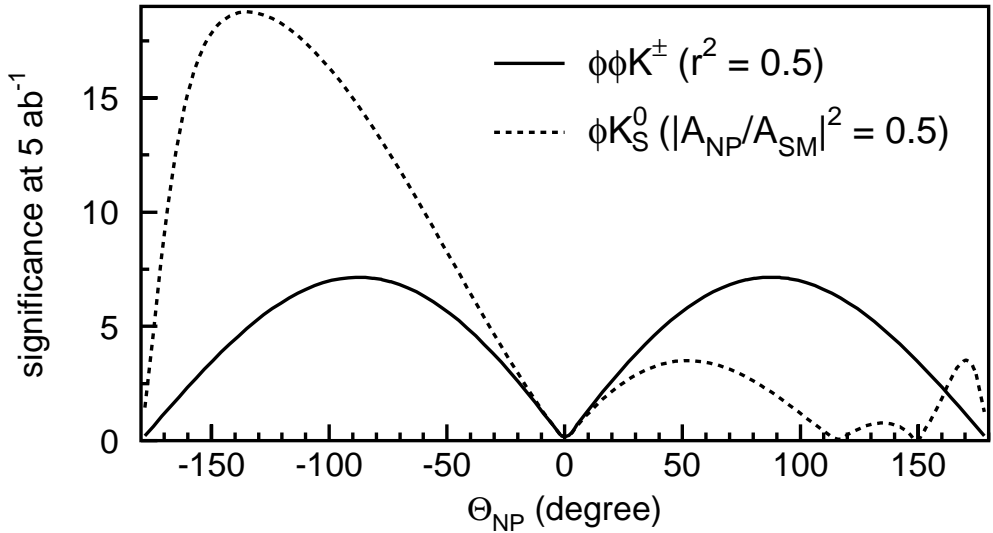

Figure 5.8: Expected statistical significance of deviations from the SM for direct $CP$ violation in the $B^\pm \to \phi\phi K^\pm$ decay with $r^2 = 0.5$ (solid line) and for time-dependent $CP$ violation in the $B^0 \to \phi K_S^0$ decay with $|A_{\rm NP}/A_{\rm SM}|^2 = 0.5$ (dashed line). For each case, significance is calculated at 5 ab$^{-1}$.

### 5.2.5 Discussion

As discussed in the previous sections, statistical errors in new phase measurements can be at a few percent level at SuperKEKB. This level of large deviation can easily be observed only with a single decay channel $B^0 \to \phi K_S^0$ at SuperKEKB. Combining all the available modes described in the previous sections allows us to measure a deviation of $\sim 0.1$. At this level, even the SM may be able to create non-zero values of $\Delta S$. Therefore it is important to evaluate $\Delta S$ within the SM. Grossman, Isidori and Worah [38] analyzed the possible pollution in the $B^0 \to \phi K_S^0$ decay, which comes from the $b \to u\bar{u}s$ transition that contains $V_{ub}$. They estimate that the pollution is at most $O(\lambda^2) \sim 5\%$. In addition, they claim that the upper limit of the pollution will be obtained experimentally from the ratios of branching fractions $\mathcal{B}(B^+ \to \phi\pi^+)/\mathcal{B}(B^0 \to \phi K_S)$ and $\mathcal{B}(B^+ \to K^*K^+)/\mathcal{B}(B^0 \to \phi K_S)$. This is due to the fact that enhancement of $b \to u\bar{u}s$ should also be detected in these modes. Therefore they conclude that new physics is guaranteed if $|\Delta S(\phi K_S^0)| > 0.05$ is established.

For the $B^0 \to \eta' K_S^0$ decay, London and Soni [39] discussed the tree $(b \to u\bar{u}s)$ pollution by evaluating $T(\eta' K_S)/P(\eta' K_S) = T(\eta' K_S)/T(\pi^+\pi^-) \times T(\pi^+\pi^-)/P(\eta' K_S)$, and concluded that manifestation of new physics is established if $|\Delta S(\eta' K_S^0)| > 0.1$ is observed.

The recent calculations based on the frame work of QCD factorization are also given by Beneke and Neubert [40] for $\phi K_S^0$ and $\eta' K_S^0$ and by Cheng, Chua and Soni [41] for $K_S^0 K_S^0 K_S^0$. Cheng, Chua and Soni also estimate the final-state rescattering effects in $b \to s\bar{q}q$ two-body decays [44]. Williamson and Zupan [42] give $\Delta S$ estimation based on the Soft Collinear Effective Theory at the leading order in $1/m_b$. They all show the $\Delta S$ in these modes are at the level of $O(0.01)$.

The above studies aim at providing estimates of the possible deviation within the SM based on some models of QCD. A different approach has been proposed by Y. Grossman, Z. Ligeti, Y. Nir and H. Quinn [43]. They do not rely on specific QCD models but instead use SU(3) relations to estimate or bound the contributions to these amplitudes proportional to $V_{ub}^* V_{us}$, which induce a non-zero $\mathcal{S}$ value within the SM. At present, the power of the method is limited by the uncertainties on branching fractions of charmless two-body decays. As a result, using measurements as of 2003 they conclude that $\Delta S(\phi K_S^0) < 0.25$ and $\Delta S(\eta' K_S^0) < 0.36$. As measurements improve, these bounds could become significantly stronger. M. Gronau, J. L. Rosner and J. Zupan [45] give the updated bound for $\eta' K_S^0$ based on the results at ICHEP06 to be $-0.133 < \Delta S(\eta' K_S^0) < 0.152$.

Taking these theoretical considerations into account, we conclude that SuperKEKB can provide precision measurements of $\Delta S$ up to the limit of hadronic uncertainties, which will be at a few percent level.

## 5.3 $b \to s\gamma$, $b \to d\gamma$ and $b \to s\ell^+\ell^-$

### 5.3.1 Introduction

In this section, we discuss the radiative and electroweak processes, $b \to s\gamma$, $b \to d\gamma$, and $b \to s\ell^+\ell^-$. The representative exclusive decay modes for these processes are $B \to K^*\gamma$, $B \to \rho\gamma$ and $\omega\gamma$, and $B \to K^{(*)}\ell^+\ell^-$, respectively. The radiative process $b \to s\gamma$ starts at one-loop order, but still has a relatively large branching fraction because of the non-decoupling effect of the top quark loop and the large CKM factor $V_{tb}V_{ts}^*$ [46]. The other processes, $b \to d\gamma$ and $b \to s\ell^+\ell^-$, are suppressed with respect to $b \to s\gamma$ in the SM by two orders of magnitude mainly due to additional $|V_{td}/V_{ts}|^2$ and $\alpha_{\rm em}$ factors, respectively [47]. One can also consider $b \to d\ell^+\ell^-$ which is suppressed by four orders of magnitude due to both factors. Since the loop diagrams can be accommodated by heavy virtual particles, these decay processes are sensitive to new physics effects that are predicted in extensions to the SM. Moreover, new physics effects could have different contributions to $b \to s\gamma$, $b \to d\gamma$ and $b \to s\ell^+\ell^-$ which may be experimentally distinguished. For example, new physics effects in the $b \to s(d)\gamma$ process always appear in the one-loop or higher diagrams, while there may be a tree-level new physics contribution to the $b \to s(d)\ell^+\ell^-$ process.

The $b \to s\gamma$ process was first observed by CLEO in the exclusive decay $B \to K^*\gamma$ in 1993 [48], inclusively measured in 1995, and has been extensively studied by Belle and BaBar. The $b \to d\gamma$ process was observed by Belle in the $B \to \rho\gamma$ and $B \to \omega\gamma$ exclusive modes [49] and confirmed by BaBar [50]. Measurements of the $b \to s\ell^+\ell^-$ process were also led by Belle and followed by BaBar; Belle first observed the $B \to K\ell^+\ell^-$ exclusive mode [51], then measured the inclusive $B \to X_s\ell^+\ell^-$ branching fraction [52], and observed the $B \to K^*\ell^+\ell^-$ decay [53]. Measurements are not limited to the branching fractions; observables such as ratios of branching fractions, $CP$ asymmetries, $q^2$ distribution and forward-backward asymmetries, isospin asymmetries, and angular distributions are useful probes for physics beyond the SM. In particular, the first trial was performed to fit the Wilson coefficients to the forward-backward asymmetry in $B \to K^*\ell^+\ell^-$ [54]. Despite the expected sensitivities to new physics in many modes, no measurement has so far uncovered a possible new physics effect as an inconsistency with SM predictions. These results already constrain physics beyond the Standard Model severely.

There are still more measurements that have not been statistically possible yet, and even for those which have been carried out, the measurement errors are still often limited by the experimental statistics. In many cases, a few orders of magnitude larger data sample is required to match the precision of SM predictions, and therefore the target integrated luminosity of SuperKEKB is ideal. These measurements will be essential in order to understand the parity, chirality and Lorentz structures that may differ from the SM, before, and especially after, the discovery of new physics elsewhere if not in these decays.

In order to search for or constrain so many predictions from extensions to the SM, a model independent study is a useful approach. New physics effects may appear as a modification to the short-distance couplings, which can be expressed as the Wilson coefficients. The $b \to s\gamma$ transition is sensitive to the coefficient $C_7$ for the $bs\gamma$-coupling and to a lesser extent to $C_8$ for the $bsg$-coupling through higher order corrections; the $b \to s\ell^+\ell^-$ transition is sensitive to $C_9$ and $C_{10}$ for the vector and axial-vector $bs\ell^+\ell^-$-couplings in addition [55]. In a further generalized approach [56, 57], new physics effects that may affect $b \to s\gamma$ can be parametrized by four types of interactions, which include two types of $b \to sg$ interactions and two types of $b \to s\gamma$ transitions. Similarly, there are four Fermi-interactions with the form of the bilinear products of $\bar{b}s$ and $\ell^+\ell^-$. They can be parametrized by 12 types of interactions in $B \to X_s\ell^+\ell^-$ [56].

Strong interaction also plays an important role in radiative and electroweak decays. In the past decade, perturbative QCD corrections to the inclusive $B \to X_s \gamma$ and $B \to X_s \ell^+ \ell^-$ decay rates have been computed for next-to-next-to-leading order (NNLO) to meet or exceed the experimental precision. The non-perturbative effects due to the Fermi-motion of $B$ meson also play an important role in the determination of the branching fraction for $B \to X_s \gamma$. The photon energy spectrum $d\Gamma/dE_\gamma$ in $b \to s\gamma$ decays has been an excellent tool together with $B \to X_c \ell \overline{\nu}$ measurements to decode the shape function and to reduce the uncertainties [58–60]. For $B \to X_s \ell^+ \ell^-$, the same technique is also applicable, and, in addition, there is a long-distance effect that comes from the decay chain $B \to (c\overline{c}) X_s \to \ell^+ \ell^- X_s$ [61] where $(c\overline{c})$ is a charmoniuum state such as $J/\psi$. There is no complete understanding of the long-distance effect which may be experimentally studied by measuring the dilepton mass distribution near the charmonium states.

There are many exclusive decays that have been observed. However, in terms of searches for deviations from SM predictions, exclusive decays suffer from large uncertainties in the form factors of the hadronic matrix elements. These form factors have usually been calculated using QCD sum rules, for which no clear idea on how to reduce their errors is available. New calculations based on lattice QCD have been attempted for $B \to K^* \gamma$ and $B \to \rho \gamma$, but the error is still large and unreliable since one has to make an extrapolation towards $q^2 = 0$ where the mesons escape from the lattice at the largest velocity. One way out is to cancel the large uncertainties by measuring a ratio or asymmetry of two decay rates. Predictions for exclusive rare $B$ decays by modified factorization techniques such as perturbative QCD (PQCD), QCD factorization (QCDF), soft-colinear effective theory (SCET), and other models can be compared with the experimental data.

An inexhaustible list of targets at SuperKEKB for radiative and electroweak decays may be represented by following measurements in which deviations from SM predictions can be searched for:

1. precise measurement of inclusive $B \to X_s \gamma$ branching fraction,

2. measurement of inclusive $B \to X_d \gamma$ branching fraction,

3. direct CP violation: $B \to X_s \gamma$, $B \to K^* \gamma$, $B \to X_s \ell^+ \ell^-$ and so on,

4. time-dependent CP violation in $B \to K^* \gamma$, $B \to \rho \gamma$ and related modes,

5. measurement of photon polarization with photon-conversion,

6. measurement of the forward-backward asymmetry and $q^2$ distribution of $B \to K^* \ell^+ \ell^-$ and $B \to X_s \ell^+ \ell^-$, and,

7. lepton flavor dependence of $b \to s \ell^+ \ell^-$,

We have performed experimental sensitivity studies for these items, which we discuss in the following sub-sections.

### 5.3.2 $B \to X_s \gamma$ branching fraction

The agreement between the measured $B \to X_s \gamma$ branching fraction [22] and the theoretical prediction [62,63] has constrained various new physics scenarios. One of the most popular examples is the lower bound on the type-II charged Higgs boson mass, since it always constructively interferes with the SM amplitude. The current limit is around 300 GeV if no other new physics destructively contributes. This limit is significantly higher than the direct search limit. The higher limit, which used to be around 500 GeV, has actually gone down with the latest measurements and calculations, but this is a good sign because it is a result of a better agreement between data and prediction than before.

Thanks to the enormous efforts, the NNLO calculations are very close to complete. The first estimate of the inclusive branching fraction has been recently reported, with a slightly lower branching fraction and a reduced error [62]. The full NNLO calculation with a $\sim 5\%$ total error is expected to be available soon. A model independent approach of parametrization the $Br(B \to X_s\gamma)$ using Wilson coefficients is described in Sect. 3.5.

The measurement error is also not so easy to reduce. Because the signal rate decreases and the background increases towards the lower photon energy, one has to require a minimum photon energy and make an extrapolation. Until recently this extrapolation is one of the irreducible systematic error source. There are two techniques to measure $B \to X_s\gamma$: a fully inclusive method by reconstruction of the photon only, in which the background subtraction is the main issue, and a method of summing up exclusive modes, in which the extrapolation due to the un-measured modes is the key issue. The former method has several options; to eliminate backgrounds one can either require a lepton-tag or a full-reconstruction tag for the other side $B$ meson, however this severely limits the statistical accuracy of the measurement. A huge background which has to be suppressed is shown in Fig.5.9(left). The latter method gives more information such as the more precise photon energy at the $B$ meson rest frame and flavor of the decay, but the hadronization model is hardly precise and reliable.

The spectrum can be described by heavy quark expansion (HQE) except for the non-perturbative effects that can be parametrized by a few parameters, e.g. the $b$ quark mass ($m_b$) and the Fermi momentum ($\mu_\pi^2$), that are considered to be universal in various inclusive $B$ decays: $B \to X_s\gamma$, $B \to X_s\ell^+\ell^-$, $B \to X_c\ell\overline{\nu}$ and $B \to X_u\ell\overline{\nu}$. Recently a significant effort was made to combine all available data for $B \to X_s\gamma$ and $B \to X_c\ell\overline{\nu}$ to determine the HQE parameters. Using the thus obtained parameters, $B \to X_s\gamma$ branching fraction results are combined together to provide a rather precise branching fraction [22],

$$\mathcal{B}(B \to X_s\gamma; E_\gamma > 1.6 \text{ GeV}) = (352 \pm 23 \pm 9) \times 10^{-6}, \tag{5.17}$$

where the first error is combined statistical and systematic uncertainty and the second the systematics due to the shape function. The size of the statistical and systematic errors are already about the same size for the fully inclusive measurement.

However, it has been argued that the energy scale $\Delta = M_b - 2E_\gamma^{\min}$ is significantly smaller than $m_b$ and therefore the above extrapolation error is underestimated. Therefore it is still very important to measure $b \to s\gamma$ spectrum as precise as possible. So far Belle has been able to lower the minimum photon energy to 1.7 GeV with a data set corresponding to the integrated luminosity of 605 fb$^{-1}$ [64]. In this measurement the main backgrounds (photons produced in continuum events, $\pi^0$ and $\eta$ decays) are subtracted using the off-resonance data and measured inclusive samples of $\pi^0 X$, $\eta X$. The main systematic uncertainty of the measurement arises from background photons in $B$ decays other than those from $\pi^0$ and $\eta$, for which the reliable control samples are scarce. Very conservatively assuming that this source of uncertainty would not scale with the increased size of the available data set one could expect a measurement of $Br(B \to X_s\gamma)$ with a relative accuracy of around 7% and 6% with the integrated luminosity of 5 ab$^{-1}$ and 50 ab$^{-1}$, respectively. Such a precision would match the anticipated precision of the theoretical predictions. The current situation is illustrated in Fig.5.9(right) where the comparison of the predicted branching fraction [63], of the current measurement and of conservatively anticipated experimental accuracy is shown.

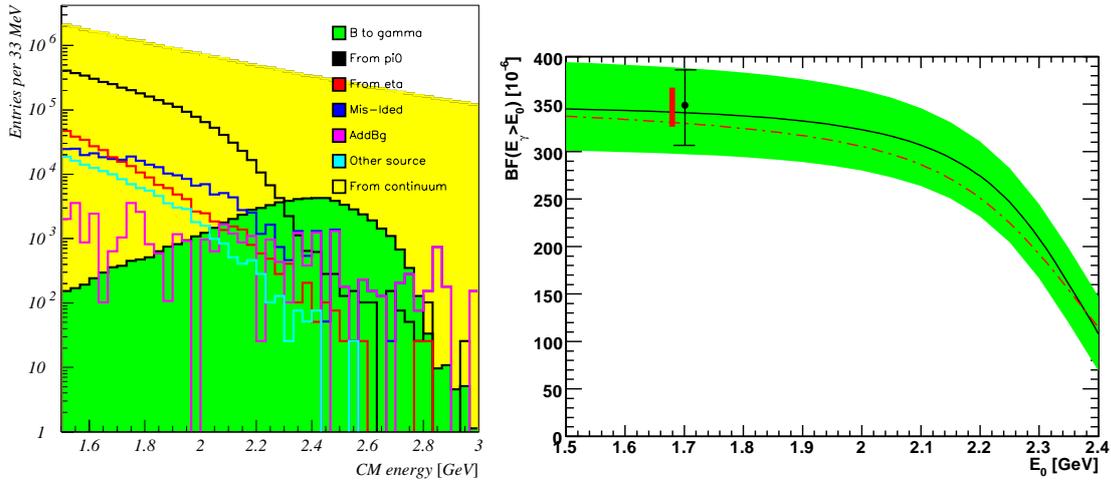

Figure 5.9: Left: Simulated photon spectrum of backgrounds and signal for the inclusive $B \to X_s\gamma$ measurement. Right: Predicted $Br(B \to X_s\gamma)$ for the photons of minimum energy $E_0$. The shaded (green) band reflects the sensitivity to various QCD parameters [63]. The accuracy of the recent Belle measurement is shown as a data point with error bars and the conservatively estimated accuracy with 50 ab$^{-1}$ as a dark (red) bar.

Further refinements of the measurement will be possible with a better knowledge of $B \to \omega$, $\eta'$, $J/\psi X$ decays and of the subsequent branchings to the photons. With more statistics, a problematic background source will be neutral hadrons such as $K_L^0$ or anti-neutron that may leave an energy shower in the calorimeter and mimic a photon. Since we cannot fully rely on the Monte Carlo simulation, a detailed study of control samples is needed. One of the possibilities is to use $\gamma \to e^+e^-$ conversion events that are free from the neutral hadron backgrounds. The conversion probability is typically 3% with the current Belle detector, which depends on the amount of material used for the vertex detector. A measurement of $B \to X_s\gamma$ with the photon conversion events could ideally be performed at SuperKEKB.

### 5.3.3 $B \to X_d\gamma$ branching fraction

Finding new physics effects in the $b \to d$ transition may be easier than in $b \to s$ because the SM amplitude is suppressed in $b \to d$. In the SM, $b \to s\gamma$ and $b \to d\gamma$ are both described by a common Wilson coefficient, $C_7$. This is also true in any model within a minimal flavor-violating framework in which the flavor changing interactions are determined by the CKM angles. However, in models with tree-level FCNCs, $C_7$ for $b \to d\gamma$ can differ from $C_7$ for $b \to s\gamma$. Examples include SUSY models with gluino mediated FCNCs [65] and models with a non-unitary CKM matrix [66].

Since SM predictions for exclusive modes such as $B \to \rho\gamma$ or $B \to \omega\gamma$ suffer from large model-dependent uncertainties, it is necessary to measure the inclusive rate for $B \to X_d\gamma$. The largest experimental challenge is the huge background due to $b \to s\gamma$. The situation is similar to $B \to X_u\ell\overline{\nu}$ measurement underneath the huge $B \to X_c\ell\overline{\nu}$ background, but much worse because of the larger suppression factor and very limited phase space in $b \to d\gamma$ where no $b \to s\gamma$ is allowed. The only possible way is probably to sum up exclusive $b \to d\gamma$ modes.

An extensive MC study was performed to reconstruct $B \to X_d\gamma$ with the following conditions:

- $B \to X_d\gamma$ modeled as the sum of $B \to (\rho, \omega)\gamma$ and an inclusive spectrum which is

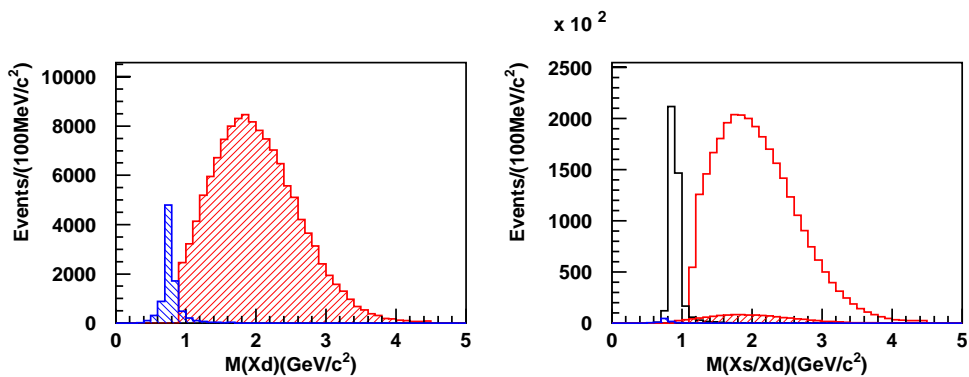

Figure 5.10: Mass spectra models for $b \to d\gamma$ (left) and $b \to s\gamma$ (right).

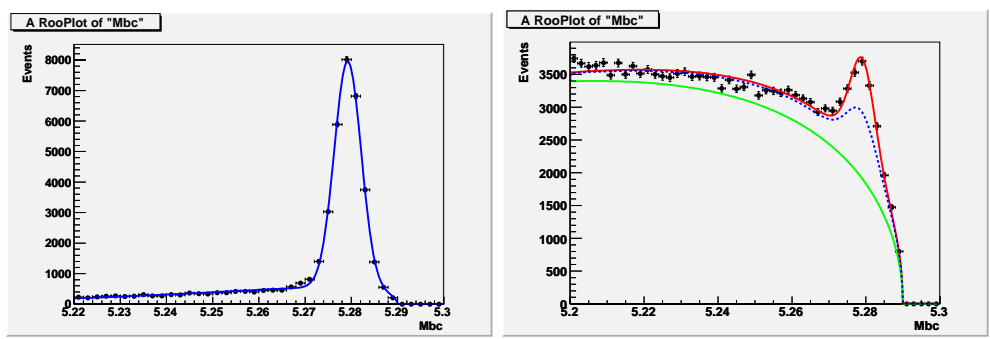

Figure 5.11: Beam constrained mass distribution for the $b \to d\gamma$ signal MC (left) and fit results for 5 ab$^{-1}$ equivalent MC sample (right). In the right plot the background including the peaking $b \to s\gamma$ component is shown as dotted (blue) line.

hadronized with JETSET (similar model is used in $B \to X_s\gamma$) as shown in Fig. 5.10,

- continuum background suppression using the Belle's $B \to \rho\gamma$ analysis [67],
- summing up 2 to 4 pions including up to 1 $\pi^0$ for $M(X_d) < 2.0$ GeV,
- $K_S^0$ veto to reduce peaking $b \to s\gamma$ background, and,
- $b \to c$ background taken into account.

The signal MC distribution and a typical fit result for a 5 ab$^{-1}$ MC sample are shown in Fig. 5.11. The signal is clearly seen on top of the peaking $b \to s\gamma$ background. Here, the evaluation of the $b \to s\gamma$ component is crucial; in the example we assume a 20% uncertainty. The fit yields $(4.2 \pm 0.2(\text{stat}) \pm 0.9(\text{fit})) \times 10^3$ events, where the second error is the fitting uncertainty. A 10% signal modeling error is assigned in addition. According to this study, inclusive $B \to X_d\gamma$ can be measured with a 2.9% efficiency and a 24% total error with 5 ab$^{-1}$ data. There are still systematic error sources that may not be properly taken into account, but certainly an accurate measurement of the inclusive $b \to d\gamma$ transitions is possible at SuperKEKB.

One can convert an inclusive $B \to X_d\gamma$ branching fraction result to $|V_{td}/V_{ts}|$ which is roughly proportional to the square root of $\mathcal{B}(B \to X_d\gamma)/\mathcal{B}(B \to X_s\gamma)$, except for higher order correction terms. Since there are no unknown form factors, this ratio should have about 5% theory error or less. This serves as a very good SM test, since $|V_{td}/V_{ts}|$ can be compared with the one obtained from the $B_s$ and $B_d$ mixing measurements and the one from a unitarity triangle fit.

### 5.3.4 $B \to X_s\gamma$ direct $CP$ asymmetry

The direct $CP$ asymmetry for $B \to X_s\gamma$ is one of the quantities with a substantially smaller theoretical uncertainty than the experimental error. The predicted SM $CP$ asymmetry is $A_{CP} = 0.0042^{+0.0017}_{-0.0012}$ [68], while its magnitude could be above 10% in many extensions of the SM.

The sensitivity at SuperKEKB can be obtained by extrapolating the latest results, $A_{CP}(B \to X_s\gamma) = +0.002 \pm 0.050(\text{stat}) \pm 0.030(\text{syst})$ from Belle with 140 fb$^{-1}$ [69], $A_{CP}(B \to X_s\gamma) = +0.025 \pm 0.050(\text{stat}) \pm 0.015(\text{syst})$ from BaBar with 82 fb$^{-1}$, and from the world average $A_{CP}(B \to X_s\gamma) = +0.005 \pm 0.036$ that also includes the CLEO result. The measurements were performed by summing up the exclusive modes. This method has an advantage of suppressing the $B \to X_d\gamma$ contribution to a negligible level. Another method that uses the inclusive photon and tags the charge by an additional lepton, cannot distinguish the $B \to X_d\gamma$ contribution; it represents, however, another interesting subject because of the partial sensitivity to the $B \to X_d\gamma$ channel.

The SM prediction for the combined asymmetry $A_{CP}^{s\gamma+d\gamma}$ is essentially zero, in contrast to $A_{CP}$ for $b \to s\gamma$ alone [68]. This is because $b \to s\gamma$ and $b \to d\gamma$ decays contribute to $A_{CP}^{s\gamma+d\gamma}$ with an opposite sign and practically an equal magnitude, which is a consequence of the unitarity of the CKM matrix, the small mass difference $m_s - m_d$ and the real Wilson coefficient $C_7$ [70]. In models beyond the SM, $A_{CP}^{s\gamma+d\gamma}$ can be non-zero, and is usually dominated by the $b \to s\gamma$ component. In addition, the contributions to $A_{CP}^{s\gamma+d\gamma}$ from $b \to s\gamma$ and $b \to d\gamma$ can have the same or opposite sign [66, 68, 71].

Although the current systematic uncertainty is not significantly smaller than the statistical error, most of the systematic errors arise due to the limited statistics of the control samples and can thus be reduced with a larger statistics. Note that systematic errors in the tracking efficiency and particle identification mostly cancel in the asymmetry. The breakdown of the uncertainty sources at 140 fb$^{-1}$ is as follows: the signal shape ($\sim 0.008$), partly due to the uncertainty in the $M(X_s)$ spectrum and partly due to the multiplicity distribution; possible $A_{CP}$ in the charmless $B$ decay background (0.02); a charge asymmetry in the background suppression requirements (0.029). The $M(X_s)$ shape and charmless contributions will be known better with more data, and other errors just scale with the statistics. The irreducible model part of the former error is about 0.003, giving the expected errors of

$$\begin{aligned} \delta A_{CP}(\text{at } 5 \text{ ab}^{-1}) &= \pm 0.009 \text{ (stat)} \pm 0.006 \text{ (syst)}, \\ \delta A_{CP}(\text{at } 50 \text{ ab}^{-1}) &= \pm 0.003 \text{ (stat)} \pm 0.002 \text{ (syst)} \pm 0.003 \text{ (model)}. \end{aligned} \quad (5.18)$$

Therefore, at 50 ab$^{-1}$, the measurement accuracy is limited by the systematic error. The precision will be at the level of the SM prediction and hence the measurement will enable tests of a large range of possible NP models. It will be possible to observe an asymmetry of $\sim 0.03$ with a $5\sigma$ significance. Furthermore, the measurement is complementary to the searches for the $CP$ violation in the $B_s$ system due to the correlations between the $A_{CP}$ and a possible NP amplitude contributing to the $B_s$ mixing in grand unification theories [72].

### 5.3.5 Mixing induced CP asymmetry in $b \to s\gamma$ and $b \to d\gamma$

Mixing-induced $CP$ asymmetry in an exclusive $b \to s\gamma$ decay into a $CP$ eigenstate such as $K^{*0}(\to K_S^0\pi^0)\gamma$ is an excellent probe for a particular class of new physics scenarios. In the SM, the expected asymmetry is $2(m_s/m_b)\sin 2\phi_1$ [73], because the two final states from $B^0$ and $\overline{B}^0$ decays have photons of different helicity (opposite helicity photon is suppressed by a $m_s/m_b$ factor) and do not mix.

The existence of the neutrino mass suggests that the left-right symmetry is restored at a higher energy, while parity is spontaneously broken at a low energy energy. In left-right symmetric models, the helicity of the photon from $b \to s\gamma$ can be a mixed state of two possible photon helicities, while the left-handed (right-handed) photon is dominant for $b \to s\gamma$ ($\bar{b} \to \bar{s}\gamma$) in the SM. The expected size of the $CP$ violation parameter is up to 4% in the SM (there is also an argument that it could be up to 10%), and therefore any large CP asymmetry is a sign of a sizable non-SM right-handed current in the $b \to s\gamma$ transition. In the left-right symmetric models the expected size of $CP$ asymmetry is $\sim 0.7 \sin 2\phi_1$ [73]. The same type of new physics can be searched for by using $b \to d\gamma$ modes, too, although the possible SM uncertainty may be larger. For predictions in SUSY models and constraints on the squark masses arising from the measurement of the time-dependent asymmetry in the mass insertion approximation see Sect. 3.4.5.

In order to measure the mixing-induced $CP$ asymmetry, a suitable final state is needed. It is recently found that this test can be applicable to a final state with any $P^0 Q^0 \gamma$ final state where $P^0$ is a neutral pseudoscalar meson and $Q^0$ is either another neutral pseudoscalar meson or a neutral vector meson. The candidate decay modes, for which at least the decay mode (or the charged partner) is observed, are $B^0 \to K_S^0 \pi^0 \gamma$ which includes $B^0 \to K^{*0}\gamma$ and $B^0 \to K_2^*(1430)\gamma$, $B \to K_1(1270)^0(\to K_S^0 \rho^0)\gamma$, $B^0 \to K_S^0 \eta\gamma$, $B^0 \to K_S^0 \phi\gamma$, $B^0 \to \rho^0\gamma$ and $B^0 \to \omega\gamma$.

Although the branching fraction for $B^0 \to K^{*0}\gamma$ is sizable ($\sim 4 \times 10^{-5}$), the number of events in the $K_S^0 \pi^0 \gamma$ final state is rather limited due to its small sub-decay branching fraction and lower reconstruction efficiency. In addition, the efficiency further reduces by requiring hits in the vertex detector associated with the charged pions from $K_S^0$ decay, in order to reconstruct the $B$ meson decay vertex from a displaced $K_S^0$ decay point and its momentum. Despite the experimental challenges, the mixing-induced $CP$ violation parameter in $B^0 \to K_S^0 \pi^0 \gamma$ is already measured using a 0.5 ab$^{-1}$ data with a sensitivity of around $\pm 0.3$ [74]. In the case of $B^0 \to K_S^0 \phi\gamma$, $B^0 \to K_S^0 \rho^0 \gamma$, $B^0 \to \rho^0\gamma$ and $B^0 \to \omega\gamma$, the vertex reconstruction is more straightforward.

Using the current analysis technique with an assumption of the current Belle detector, error of the $CP$ violation parameter is extrapolated as a function of the integrated luminosity. A similar exercise is also performed for $B \to \rho^0\gamma$, assuming the Belle's efficiency, backgrounds and measured branching fraction. In this case, the vertexing efficiency is taken from the $B^0 \to \phi K_S^0$ analysis. These results for both cases are shown in Fig. 5.12, or,

$$\begin{aligned} \delta S_{K_S^0 \pi^0 \gamma} &= 0.03 \text{ (at 50 ab}^{-1}), \\ \delta S_{\rho^0 \gamma} &= 0.15 \text{ (at 50 ab}^{-1}). \end{aligned} \quad (5.19)$$

Furthermore, the efficiency for the reconstruction of $K_S^0$ with associated charged pion hits will be improved by the upgrade of the silicon detector by around 30%. The size of the error with 50 ab$^{-1}$ is already smaller than the SM expectation in the case of $B^0 \to K_S^0 \pi^0 \gamma$.

### 5.3.6 $b \to s\gamma$ with $\gamma \to e^+e^-$ conversion

Another method to search for the non-SM right-handed photon is to use $B \to K^*\gamma$ events with $\gamma \to e^+e^-$ conversions that occur in the detector. If the photon helicity is mixed, the photon has an elliptical polarization (or linear polarization if the size of the left- and right-handed amplitudes are the same). The distribution of the angle $\phi$ between the plane of $K^* \to K\pi$ and the plane of $e^+e^-$ (see Fig. 5.13(left)) can be written as

$$1 + \xi(E_e, q^2) \frac{|A_R||A_L|}{|A_R|^2 + |A_L|^2}[\cos(2\phi + \delta)], \quad (5.20)$$

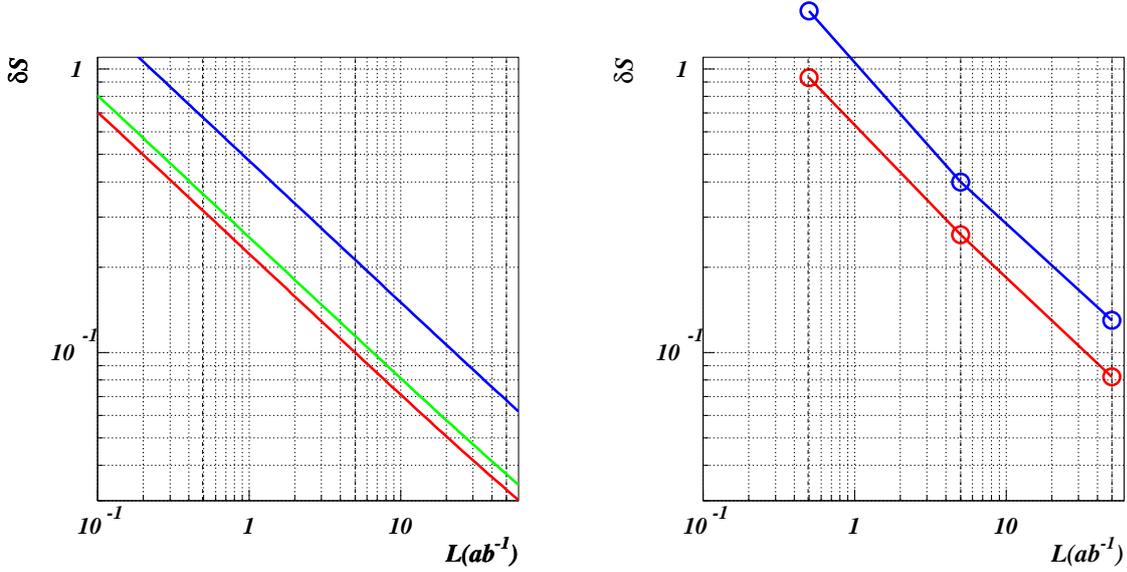

Figure 5.12: Expected errors on mixing-induced $CP$ violation parameters as a function of the integrated luminosity. Left plot for total $B^0 \to K^0_S \pi^0 \gamma$ (red), $B^0 \to K^{*0} \gamma$ (green) and other $B^0 \to K^0_S \pi^0 \gamma$ (blue); right plot for $B^0 \to \rho^0 \gamma$ with the Belle's measured branching fraction (red) and the SM branching fraction (blue).

where $A_{L,R}$ are the left- and right-handed amplitudes, $\xi$ is an efficiency factor as a function of the energy and the momentum transfer of the electron, and $\delta$ is a phase of the modulation that is determined from the relative phase between $A_L$ and $A_R$.

This is actually a small effect due to several reasons. First, $\xi$ is only about 0.1 when the $e^+e^-$ opening angle is not measured and one integrates over the energy and angle. Second, the factor $|A_R||A_L|/(|A_R|^2 + |A_L|^2)$ is at most 0.5. Third, the conversion probability is only 3% in the current Belle detector, and it is not preferable to increase the amount of material to gain more conversion events. Finally, the angle $\phi$ cannot be measured for a large fraction of events because the opening angle is too small and immediately distorted by the magnetic field.

A sensitivity study was performed assuming the Belle analysis techniques and the Belle detector. The efficiency is 0.36% for $B^0 \to K^{*0}(\to K^+ \pi^-)\gamma$ events, or 0.64% by summing up also $B^+ \to K^{*+} \gamma$ events, with very small background contaminations. It is not very likely to further increase the efficiency by improving the analysis, and it is possible only by increasing the amount of material in the vertex detector.

With the current track reconstruction code, there is no sensitivity to the opening angle at all. By assuming that the conversion point is correctly measured, only 35% of events can be effectively used for the $\phi$ measurement, with a $\phi$ resolution of $\delta\phi = 23°$ (Fig. 5.14(right)). This will give a $2\sigma$ modulation for the maximal right-handed current case with 50 ab$^{-1}$ data. However, the $\phi$ measurement performance strongly depends on the reconstruction code, and there is a large room for improvements. By assuming that all the events can be usable for $\phi$ measurement, the maximal right-handed current case will show up as a $4\sigma$ effect. In addition, if one can measure the opening angle of $e^+e^-$, the sensitivity significantly increases since $\xi$ can be larger than 0.5 for a large $\theta$.

We note that a similar study can be performed by using very low $q^2$ $B \to K^* e^+ e^-$ [75], and a combined analysis would be possible.

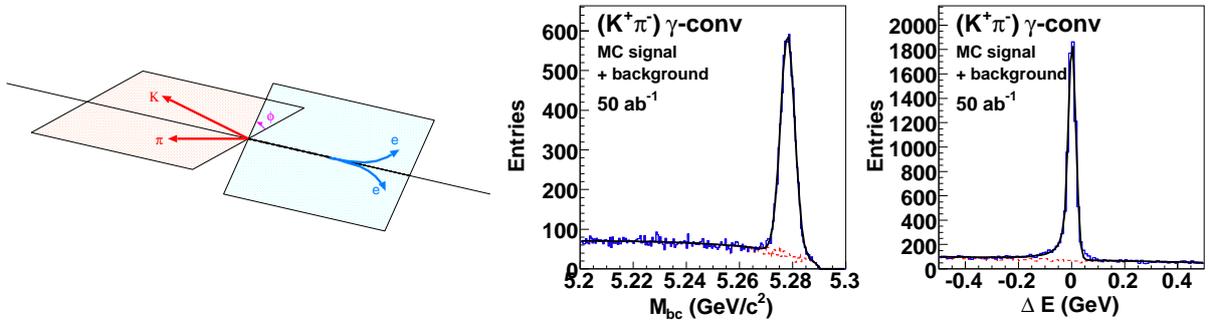

Figure 5.13: The definition of the angle $\phi$ (left), and the expected signal and background in $M_{\rm bc}$ (middle) and $\Delta E$ (right) in $B \to K^*\gamma(\to e^+e^-)$.

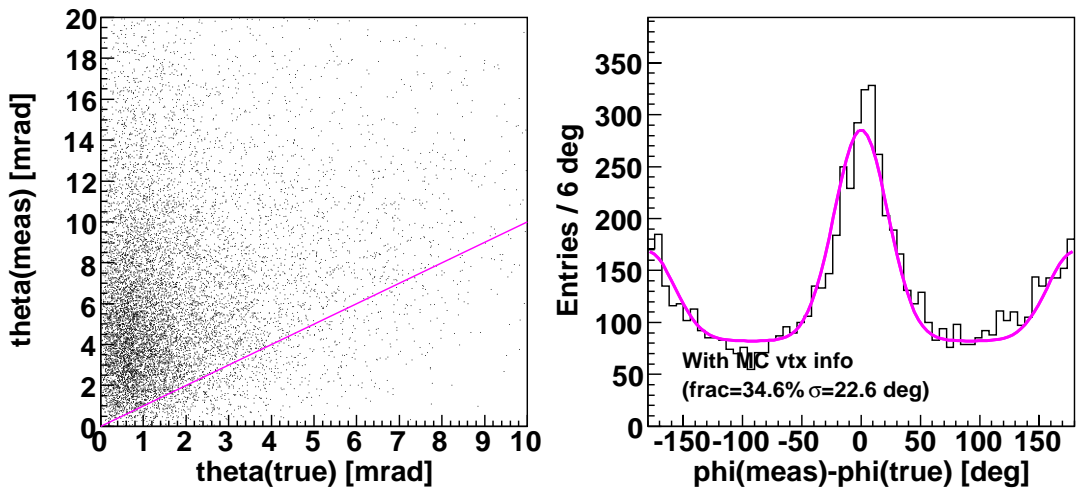

Figure 5.14: Measured versus true $\theta$ distribution (left) and measured minus true $\phi$ distribution (right) in $B \to K^*\gamma(\to e^+e^-)$.

### 5.3.7 $B \to K^*\ell^+\ell^-$ forward-backward asymmetry

The forward-backward asymmetry in $B \to K^*\ell^+\ell^-$, defined as

$$\frac{d\overline{A}_{\rm FB}(q^2)}{dq^2} = \frac{N(q^2; \theta_{B\ell^+} > \theta_{B\ell^-}) - N(q^2; \theta_{B\ell^+} < \theta_{B\ell^-})}{N(q^2; \theta_{B\ell^+} > \theta_{B\ell^-}) + N(q^2; \theta_{B\ell^+} < \theta_{B\ell^-})}, \quad (5.21)$$

as a function of $q^2 = m(\ell^+\ell^-)^2$, where $\theta_{B\ell^\pm}$ denotes the angle between the lepton and the $B$ meson direction in the rest frame of the meson (Fig. 5.15), is an ideal quantity to disentangle the Wilson coefficients $C_i$. The asymmetry of Eq. 5.21 can be expressed as [76]

$$-C_{10}\xi(q^2) \times \left[{\rm Re}(C_9)F_1 + \frac{1}{q^2}C_7 F_2\right] \quad (5.22)$$

where $\xi$ is a function of $q^2$, and $F_{1,2}$ are functions of form factors.

It is straightforward to determine $C_{10}$, ${\rm Re}(C_9)$ and the sign of $C_7$ from the $\overline{A}_{\rm FB}$ distribution as a function of $q^2$, using the value of $|C_7|$ from $B \to X_s\gamma$ and a few more assumptions: phases of $C_{10}$ and $C_7$ are neglected, and higher order corrections are known. These assumptions should be examined by comparing the results with the inclusive $B \to X_s\ell^+\ell^-$ differential branching fraction as a function of $q^2$, since it is also sensitive to $C_9$ and $C_{10}$ in a different way [77].

As discussed in Section 3.5, in a SUSY scenario the sign of the $b \to s\gamma$ amplitude ($C_7$) can be opposite to the SM prediction, while the transition rate may be the same as in the SM. Within the SM, there is a zero crossing point of the forward-backward asymmetry in the low $q^2$ region, while it disappears with the opposite sign $C_7$ if the sign of ${\rm Re}(C_9)$ is the same as in the SM. In another model with $SU(2)$ singlet down-type quarks, tree-level $Z$ flavor-changing-neutral-currents are induced. In this case, the larger effect is expected on the axial-vector coupling ($C_{10}$) to the dilepton than on the vector coupling ($C_9$). Because the forward-backward asymmetry is proportional to the axial-vector coupling, the sign of the asymmetry can be opposite to the SM. The same new physics effect is also effective for $B^0 \to \phi K_S^0$ where anomalous mixing-induced $CP$ violation can occur.

In the first trial to extract $C_9$ and $C_{10}$ from $\overline{A}_{\rm FB}$ in $B \to K^*\ell^+\ell^-$ with 357 fb$^{-1}$ data by Belle [78], higher order QCD correction terms are assumed to be the same as in the SM, and only the leading order terms $A_9$ and $A_{10}$ are allowed to float to fit the data, and it was possible to constrain the sign of the product $A_9 A_{10}$ to be negative as in the SM. Using the same framework, a sensitivity study is performed for a 5 ab$^{-1}$ data. One of the typical fit results is given in Fig. 5.15, in which the crossing pattern is clearly observed. From a 1000 pseudo-experiments assuming the SM values of Wilson coefficients, the expected statistical errors are

$$\begin{array}{ll} \delta A_9/A_9 = 0.11, & \delta A_{10}/A_{10} = 0.13 \quad (5~{\rm ab}^{-1}) \\ \delta A_9/A_9 = 0.04, & \delta A_{10}/A_{10} = 0.04 \quad (50~{\rm ab}^{-1}), \end{array} \quad (5.23)$$

where the 50 ab$^{-1}$ is a simple extrapolation. The zero crossing point, for the SM value of $\hat{s}_0 \sim -2\hat{m}_b \frac{C_7}{{\rm Re}(C_9)} \sim 0.16$ [76], can be measured with a 5% precision. Currently the most accurate measurement of $B \to K\ell^+\ell^-$ and $B \to K^*\ell^+\ell^-$ branching fractions, as well as of the $d\overline{A}_{\rm FB}(q^2)/dq^2$ is available from Belle in [79].

Since $B \to K^*\ell^+\ell^-$ is an all charged particle final state, LHCb will be able to measure the zero crossing point with a better precision than an $e^+e^-$ $B$ factory. However, in order to avoid the form factor uncertainty, especially when the new physics effect is only a small modification to the SM, one has to measure the forward-backward asymmetry in inclusive $B \to X_s\ell^+\ell^-$; it is experimentally more challenging, and is only possible in an $e^+e^-$ $B$ factory. In addition, combined analysis with the inclusive $B \to X_s\ell^+\ell^-$ differential branching fraction should improve the results.

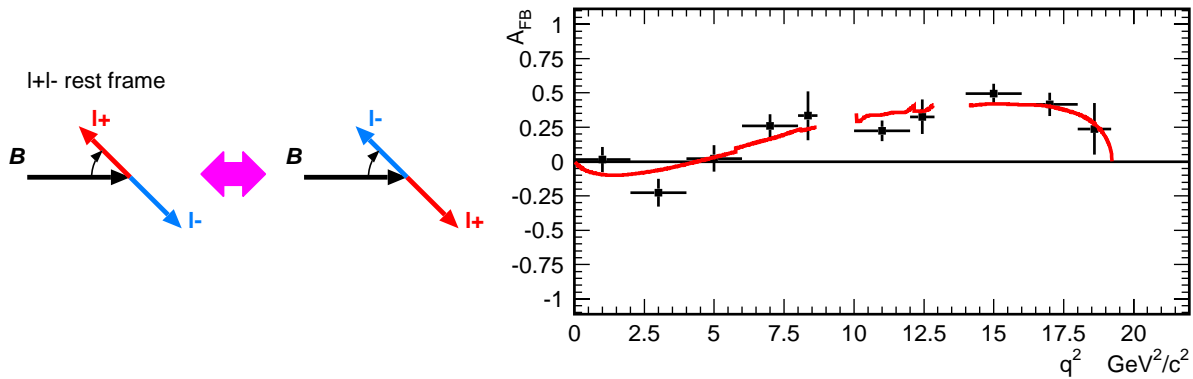

Figure 5.15: Forward-backward asymmetry in $B \to K^*\ell^+\ell^-$ at 5 ab$^{-1}$.

### 5.3.8 $B \to K\mu^+\mu^-$ versus $B \to Ke^+e^-$

Branching fractions for the exclusive decays $B \to K^{(*)}\ell^+\ell^-$ have already been measured to be consistent with SM predictions: the measurement error ($\sim 13\%$ [79]) is already smaller than the uncertainty of the theoretical prediction; the latter suffers from large model dependent and irreducible uncertainties in the form-factors of at least $\pm 30\%$ or even more if one takes into account the variations in the available predictions. However, one can still utilize the measurements in such a way that the theory uncertainties cancel.

In new physics models with a Higgs sector differing from that of the SM, scalar and pseudo-scalar types of interactions may arise in $b \to s\ell^+\ell^-$. Depending on the lepton flavor $\ell = e$ and $\ell = \mu$, the new physics effects can differ. By measuring $R_{K^{(*)}} = \mathcal{B}(B \to K^{(*)}\mu^+\mu^-)/\mathcal{B}(B \to K^{(*)}e^+e^-)$, such new physics effects can be searched for. A particular example can be found in a minimal supergravity model [57, 80].

In the SM, the branching fractions for $B \to Ke^+e^-$ and $B \to K\mu^+\mu^-$ are predicted to be equal except for a tiny phase space difference due to the lepton masses, $R_K = 0.95$. In $B \to K^*\ell^+\ell^-$ decays, the branching fraction for $B \to K^*e^+e^-$ is larger than $B \to K^*\mu^+\mu^-$ for small dilepton masses, $R_{K^*} = 0.75$, due to a larger interference contribution from $B \to K^*\gamma$ in $B \to K^*e^+e^-$. However, this situation may be modified in the models mentioned above, in which a neutral SUSY Higgs contribution can significantly enhance only the $B \to K^{(*)}\mu^+\mu^-$ channel if $\tan\beta$ is large. Therefore, the ratio $R_K = \mathcal{B}(B \to K\mu^+\mu^-)/\mathcal{B}(B \to Ke^+e^-)$ is an observable that is sensitive to new physics if it is larger than unity.

From the current Belle results on the branching fractions [79] we obtain $R_K = 1.03 \pm 0.19 \pm 0.06$ and $R_{K^*} = 0.83 \pm 0.17 \pm 0.05$. The uncertainties using 600 ab$^{-1}$ are still dominated by the statistical error. The systematic error in the ratio is dominated by the uncertainties of the lepton identification efficiency, which are obtained from the dilepton data samples and will also reduce with the increasing statistics. Scaling the total uncertainties to a larger luminosity the expected error is

$$\begin{aligned} \delta R_K &= 0.07 \quad \text{(at 5 ab}^{-1}\text{)}, \\ \delta R_K &= 0.02 \quad \text{(at 50 ab}^{-1}\text{)}, \end{aligned} \quad (5.24)$$

and similar for the $R_{K^*}$.

The $b \to s\ell^+\ell^-$ transition diagram is equivalent to $B_s \to \mu^+\mu^-$, which has been extensively searched for by CDF and D0. The current best limit is $\mathcal{B}(B_s^0 \to \mu^+\mu^-) < 4.7 \times 10^{-8}$ (90% C.L.) [81], which is still far from the SM prediction $(3.4 \pm 0.5) \times 10^{-9}$. Since $B_s \to e^+e^-$ is not expected to be observed at Tevatron or LHCb, a precision measurement on $R_K$ will be still useful even if the $B_s^0 \to \mu^+\mu^-$ branching fraction is measured to be near the SM predicted value.

| mode | 5 ab$^{-1}$ | 50 ab$^{-1}$ |
|---|---|---|
| $\mathcal{B}(B \to X_s\gamma)$ | 7% | 6% |
| $A_{CP}(B \to X_s\gamma)$ | $0.009 \oplus 0.006$ | $0.003 \oplus 0.002 \oplus 0.003$ |
| Mixing induced $S_{K_S^0\pi^0\gamma}$ | 0.1 | 0.03 |
| $\mathcal{B}(B \to X_d\gamma)$ | 24% | |
| Mixing induced $S_{\rho^0\gamma}$ | 0.4 | 0.15 |
| $R_K(B \to K\ell^+\ell^-)$ | 0.07 | 0.02 |
| $\overline{A}_{\rm FB}(B \to K^*\ell^+\ell^-)$ | | |
| $\quad C_9$ from $\overline{A}_{\rm FB}(B \to K^*\ell^+\ell^-)$ | 0.11 | 0.04 |
| $\quad C_{10}$ from $\overline{A}_{\rm FB}(B \to K^*\ell^+\ell^-)$ | 0.13 | 0.04 |
| $\quad \hat{s}_0$ from $\overline{A}_{\rm FB}(B \to K^*\ell^+\ell^-)$ | | 5% |

Table 5.7: Summary of the expected errors for some of the observables in $b \to s(d)\gamma$ and $b \to s\ell^+\ell^-$ decays at SuperKEKB.

### 5.3.9 Summary

We have discussed various decay channels of $b \to s(d)\gamma$ and $b \to s\ell^+\ell^-$ that with the increased sensitivity at SuperKEKB represent good probes of new physics effects. The expected sensitivities for some of the observables that can be measured are summarized in Table 5.7.

| Decay | SM Prediction |
|---|---|
| $B \to e\nu$ | $(1.67 \pm 0.42) \times 10^{-11}$ |
| $B \to \mu\nu$ | $(7.07 \pm 1.80) \times 10^{-7}$ |
| $B \to \tau\nu$ | $(1.59 \pm 0.40) \times 10^{-4}$ |

Table 5.8: The expected branching fraction for $B \to \ell\nu$.

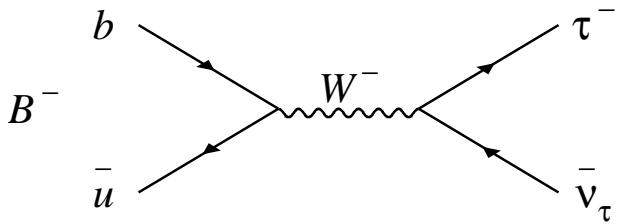

Figure 5.16: Purely leptonic $B$ decay proceeds via quark annihilation into a $W$ boson.

## 5.4 Leptonic Decays : $B \to \ell\nu$

### 5.4.1 Introduction

In the SM, the purely leptonic decay $B^- \to \tau^- \bar{\nu}_\tau$ proceeds via annihilation of $b$ and $\bar{u}$ quarks to a $W^-$ boson (Fig. 5.16). It provides a direct determination of the product of the $B$ meson decay constant $f_B$ and the magnitude of the Cabibbo-Kobayashi-Maskawa (CKM) matrix element $|V_{ub}|$. The branching fraction is given by

$$\mathcal{B}(B \to \ell\nu) = \frac{G_F^2 m_B m_\ell^2}{8\pi} \left(1 - \frac{m_\ell^2}{m_B^2}\right)^2 f_B^2 |V_{ub}|^2 \tau_B, \qquad (5.25)$$

where $G_F$ is the Fermi coupling constant, $m_B$ and $m_\ell$ are the $B$ and lepton masses, respectively, and $\tau_B$ is the $B^-$ lifetime. In models beyond the SM there can be additional tree-level contributions such as a $H^\pm$ (s-channel) [82, 83] or sfermions ($s, t$-channels) in $R$ parity violating SUSY models [84]. The SM predictions are shown in Table 5.8, where we take $|V_{ub}| = (4.39 \pm 0.33) \times 10^{-3}$, $\tau_B = 1.643 \pm 0.010$ ps, and $f_B = 0.216 \pm 0.022$ GeV. Observation of such decays would provide the direct measurement of $f_B$ or even evidence for new physics. In particular, the sensitivity of the $\tau^\pm \nu$ and $\mu^\pm \nu$ channels to any $H^\pm$ contribution is complementary and competitive with that of the exclusive semi-leptonic decay $B \to \overline{D}\tau^\pm\nu$ that is described in Sec. 5.5. The tree-level partial width (including only $W^\pm$ and $H^\pm$ contributions) is given as follows [82]:

$$\mathcal{B}(B \to \ell\nu) = \frac{G_F^2 m_B m_\ell^2}{8\pi} \left(1 - \frac{m_\ell^2}{m_B^2}\right)^2 f_B^2 |V_{ub}|^2 \tau_B \times r_H, \qquad (5.26)$$

where $r_H$ is independent of the lepton flavour and is given by

$$r_H = \left(1 - \frac{m_B^2}{m_{H^\pm}^2} \tan^2 \beta\right)^2. \qquad (5.27)$$

An additional observable in which $f_B$ cancels out is the ratio $R_{\tau\mu}$ defined by:

$$R_{\tau\mu} = \frac{\mathcal{B}(B^\pm \to \tau^\pm \nu)}{\mathcal{B}(B^\pm \to \mu^\pm \nu)}. \qquad (5.28)$$

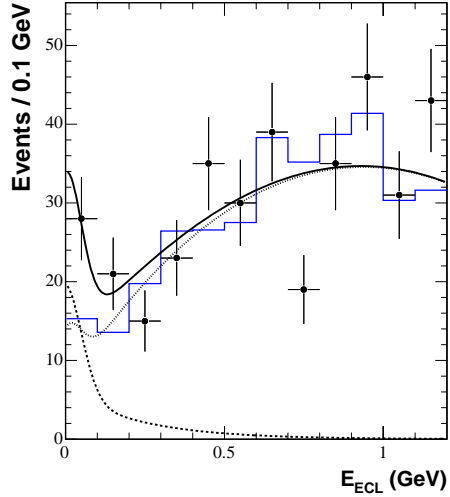

Figure 5.17: $E_{\text{ECL}}$ distributions in the data after all selection criteria except the one on $E_{\text{ECL}}$. The data and background MC samples are represented by the points and the solid histogram, respectively. The solid curve shows the result of the fit with the sum of the signal (dashed) and background (dotted) contributions.

Assuming only $W^{\pm}$ and $H^{\pm}$ contributions, $r_H$ would also cancel out, and thus $R_{\tau\mu} \sim m_\tau^2/m_\mu^2$. However, sizeable deviations of $R_{\tau\mu}$ from this value are possible in $R$ parity violating SUSY models, since $r_H$ is in general no longer independent of the lepton flavour. In addition, in such models the decay $B^{\pm} \to e^{\pm}\nu$ may be enhanced to experimental observability.

### 5.4.2 $B \to \tau\nu$

The measurement of $B \to \tau\nu$ is experimentally difficult due to multiple neutrinos in the final state, and is possible only at $B$ factories. We reconstruct one of the $B$ mesons produced in pair, referred hereafter as the tag side ($B_{\text{tag}}$), and compare properties of the remaining particle(s), referred to as the signal side ($B_{\text{sig}}$), to those expected for signal and background.

The first evidence for the decay $B^- \to \tau^-\bar{\nu}_\tau$ is found by using 414 fb$^{-1}$ of data collected at the $\Upsilon(4S)$ resonance with the Belle detector [87]. In this analysis, $B_{\text{tag}}$ is fully reconstructed in hadronic modes. We search for decays of $B_{\text{sig}}$ into a $\tau$ and a neutrino, where the $\tau$ lepton is identified in one of the five decay modes, $\mu^-\bar{\nu}_\mu\nu_\tau$, $e^-\bar{\nu}_e\nu_\tau$, $\pi^-\nu_\tau$, $\pi^-\pi^0\nu_\tau$ and $\pi^-\pi^+\pi^-\nu_\tau$, which in total correspond to 81% of all $\tau$ decays.

The most powerful variable for separating signal and background is the remaining energy in the ECL, denoted as $E_{\text{ECL}}$, which is a sum of the energies of neutral clusters that are not associated with either the $B_{\text{tag}}$ or the $\pi^0$ candidate from the $\tau^- \to \pi^-\pi^0\nu_\tau$ decay. $E_{\text{ECL}}$ is consistent with zero for signal events.

Figure 5.17 shows the $E_{\text{ECL}}$ distribution obtained when all $\tau$ decay modes are combined. Table 5.9 shows the number of events observed in the signal region ($N_{\text{obs}}$) for each $\tau$ decay mode, as well as the MC expectations of background.

The branching fractions are calculated as $\mathcal{B} = N_{\text{s}}/(2 \cdot \varepsilon \cdot N_{B^+B^-})$ where $N_{B^+B^-}$ is the number of $\Upsilon(4S) \to B^+B^-$ events, assuming $N_{B^+B^-} = N_{B^0\overline{B}^0}$. The efficiency is defined as $\varepsilon = \varepsilon^{\text{tag}} \times \varepsilon^{\text{sel}}$, where $\varepsilon^{\text{tag}}$ is the tag reconstruction efficiency for events with $B^- \to \tau^-\bar{\nu}_\tau$ decays on the signal side, determined by MC to be $0.136 \pm 0.001(\text{stat})\%$, and $\varepsilon^{\text{sel}}$ is the event selection efficiency listed in Table 5.9, as determined by the ratio of the number of events surviving all the selection criteria including the $\tau$ decay branching fractions to the number of fully reconstructed $B^\pm$.

| | $N_{\text{side}}^{\text{obs}}$ | $N_{\text{side}}^{\text{MC}}$ | $N_{\text{sig}}^{\text{MC}}$ | $N_{\text{obs}}$ | $N_{\text{s}}$ | $N_{\text{b}}$ | $\varepsilon^{\text{sel}}(\%)$ | $\mathcal{B}(10^{-4})$ | $\Sigma$ |
|---|---|---|---|---|---|---|---|---|---|
| $\mu^-\bar{\nu}_\mu\nu_\tau$ | 96 | $94.2 \pm 8.0$ | $9.4 \pm 2.6$ | 13 | $5.6^{+3.1}_{-2.8}$ | $8.8^{+1.1}_{-1.1}$ | $3.64 \pm 0.02$ | $2.57^{+1.38}_{-1.27}$ | $2.2\sigma$ |
| $e^-\bar{\nu}_e\nu_\tau$ | 93 | $89.6 \pm 8.0$ | $8.6 \pm 2.3$ | 12 | $4.1^{+3.3}_{-2.6}$ | $9.0^{+1.1}_{-1.1}$ | $4.57 \pm 0.03$ | $1.50^{+1.20}_{-0.95}$ | $1.4\sigma$ |
| $\pi^-\nu_\tau$ | 43 | $41.3 \pm 6.2$ | $4.7 \pm 1.7$ | 9 | $3.8^{+2.7}_{-2.1}$ | $3.9^{+0.8}_{-0.8}$ | $4.87 \pm 0.03$ | $1.30^{+0.89}_{-0.70}$ | $2.0\sigma$ |
| $\pi^-\pi^0\nu_\tau$ | 21 | $23.3 \pm 4.7$ | $5.9 \pm 1.9$ | 11 | $5.4^{+3.9}_{-3.3}$ | $5.4^{+1.6}_{-1.6}$ | $1.97 \pm 0.02$ | $4.54^{+3.26}_{-2.74}$ | $1.5\sigma$ |
| $\pi^-\pi^+\pi^-\nu_\tau$ | 21 | $18.5 \pm 4.1$ | $4.2 \pm 1.6$ | 9 | $3.0^{+3.5}_{-2.5}$ | $4.8^{+1.4}_{-1.4}$ | $0.77 \pm 0.02$ | $6.42^{+7.58}_{-5.42}$ | $1.0\sigma$ |

Table 5.9: The number of observed events in data in the sideband region ($N_{\text{side}}^{\text{obs}}$), number of background MC events in the sideband region ($N_{\text{side}}^{\text{MC}}$) and the signal region ($N_{\text{sig}}^{\text{MC}}$), number of observed events in data in the signal region ($N_{\text{obs}}$), number of signal ($N_{\text{s}}$) and background ($N_{\text{b}}$) in the signal region determined by the fit, signal selection efficiencies ($\varepsilon^{\text{sel}}$), extracted branching fraction ($\mathcal{B}$) for $B^- \to \tau^-\bar{\nu}_\tau$. The listed errors are statistical only. The last column gives the significance of the signal including the systematic uncertainty in the signal yield ($\Sigma$).

The combined fit gives $17.2^{+5.3}_{-4.7}$ signal events in the signal region ($N_{\text{s}}$) and $24.1^{+7.6}_{-6.6}$ in the entire region ($n_{\text{s}}$). The branching fraction is found to be $(1.79^{+0.56}_{-0.49}) \times 10^{-4}$.

Systematic errors for the measured branching fraction are associated with the uncertainties in the number of $B^+B^-$, signal yields and efficiencies. The total fractional systematic uncertainty of the combined measurement is $^{+26}_{-28}\%$, and the branching fraction is

$$\mathcal{B}(B^- \to \tau^-\bar{\nu}_\tau) = (1.79^{+0.56}_{-0.49}(\text{stat})^{+0.46}_{-0.51}(\text{syst})) \times 10^{-4}.$$

The significance is $3.5\sigma$ when all $\tau$ decay modes are combined,

In the two-Higgs doublet model, the branching fraction $\mathcal{B}(B^- \to \tau^-\bar{\nu}_\tau)$ is enhanced by a factor of $[1 - (m_B/m_H)^2 \tan^2\beta]^2$. With our experimental result and $\mathcal{B}(B^- \to \tau^-\bar{\nu}_\tau)_{\text{SM}} = (1.59 \pm 0.40) \times 10^{-4}$ from the SM prediction, we set a constraint

$$[1 - (m_B/m_H)^2 \tan^2\beta]^2 = 1.13 \pm 0.51. \tag{5.29}$$

Figure 5.18 shows the 95.5% C.L. exclusion boundaries in the $[M_{H^+}, \tan\beta]$ plane compared with other experimental searches at LEP and at the Tevatron. The direct search for charged Higgs bosons at LEP gives the constraint $M_{H^\pm} > 78.6$ GeV/$c^2$.

Further accumulation of luminosity helps to reduce both statistical and systematic errors on the measurement of the $B \to \tau\nu$ branching fraction. We estimate the sensitivity at Super $B$ Factory, assuming that the experimental error decreases as $\sim 1/\sqrt{\mathcal{L}}$, where $\mathcal{L}$ is the luminosity. This assumption is not unrealistic because some of the major systematic errors come from limited statistics of the control samples, and also because of possible improvements of the analysis in future. We also assume that the theoretical errors of both $|V_{ub}|$ and $f_B$ are reduced to 5.0% and 5.0% (2.5% and 2.5%) when Super $B$ accumulates 5 ab$^{-1}$ (50 ab$^{-1}$).

Figure 5.19 shows 95.5% C.L. exclusion region and $5\sigma$ discovery region for the charged Higgs boson with 5 ab$^{-1}$ (5 ab$^{-1}$) of data. Here, the exclusion regions are obtained assuming that $r_H$ coincides with the theoretical prediction.

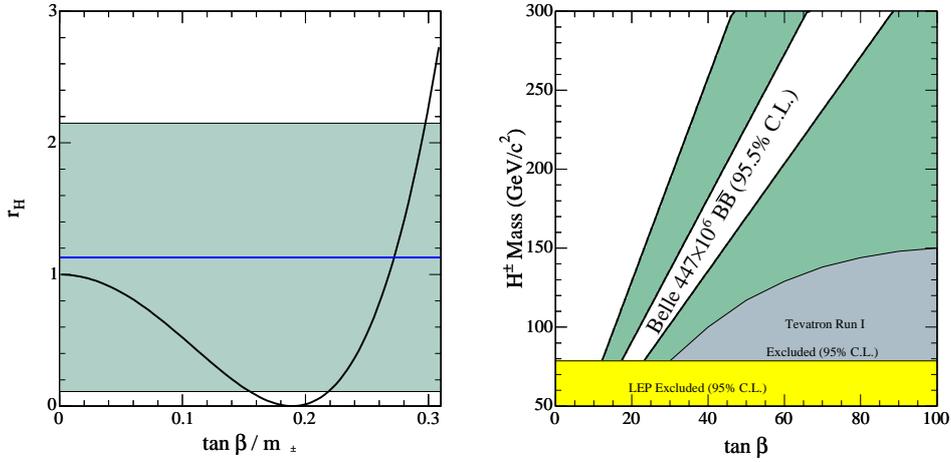

Figure 5.18: The 95.5% C.L. exclusion boundaries in the $[M_{H^+}, \tan\beta]$ plane obtained from the measured branching fraction of $B^- \to \tau^- \nu$.

The assumption on the error of $|V_{ub}|$ is reasonable because $|V_{ub}|$ can be constrained not only from its measurement from semi-leptonic $B$ decays but also from many other measurements on the unitarity triangle. On the other hand, the precision of $f_B$ depends on the development of the lattice calculation, and it is not clear whether this assumption is valid. It is essential to keep its error at the similar size as $|V_{ub}|$.

### 5.4.3  $B \to \ell\nu$

The expected branching fractions of $B \to \mu\nu_\mu$ and $B \to e\nu_e$ are by several orders lower than $B \to \tau\nu_\tau$ as listed in Table 5.8. On the other hand, these modes are experimentally easier to study than $B \to \tau\nu_\tau$ due to a single neutrino and a monochromatic lepton in the final state. In fact, $B \to \mu\nu_\mu$ is almost within a reach of present $B$ factories.

The strategy of the analysis is as follows. Because $B^+ \to \ell^+\nu_\ell$ is a two-body decay, the lepton has a fixed momentum in the signal $B$ meson ($B^{\text{sig}}$) rest frame, with $p_\ell^B$ equal to approximately half of the $B$ meson mass, $p_\ell^B \sim m_B/2$. The lepton momentum in the CM frame, $p_\ell^*$, is related to $p_\ell^B$ by

$$p_\ell^B \simeq p_\ell^* \left(1 - \frac{|\vec{p}_{B^{\text{sig}}}^*|}{m_B} \cos\theta_{\ell\text{-}B^{\text{sig}}}\right) \quad (5.30)$$

where $\vec{p}_{B^{\text{sig}}}^*$ is the momentum of the $B^{\text{sig}}$ in the CM frame and $\cos\theta_{\ell\text{-}B^{\text{sig}}}$ represents the cosine of the angle between the directions of the signal lepton and $B^{\text{sig}}$ in the CM frame.

Since the neutrino is not detected in the $B^{\text{sig}}$ decay, we need to obtain $\vec{p}_{B^{\text{sig}}}^*$ using the information of the companion $B$ meson ($B^{\text{comp}}$) recoiling against $B^{\text{sig}}$. The quantity $|\vec{p}_{B^{\text{sig}}}^*|$ is approximately given by $\sqrt{E_{\text{beam}}^2 - m_B^2} \simeq 0.32$ GeV/c using the beam energy in the CM frame ($E_{\text{beam}}$), while $\cos\theta_{\ell-B^{\text{sig}}}$ is related to the angle between the direction of the signal lepton and the momentum of $B^{\text{comp}}(\theta_{\ell-B^{\text{comp}}})$ by $\cos\theta_{\ell-B^{\text{sig}}} = -\cos\theta_{\ell-B^{\text{comp}}}$. Thus, equation (5.30) can be expressed in terms of two measurable quantities, $p_\ell^*$ and $\cos\theta_{\ell-B^{\text{comp}}}$:

$$p_\ell^B \simeq p_\ell^*(1 + 0.065 \cos\theta_{\ell-B^{\text{comp}}}). \quad (5.31)$$

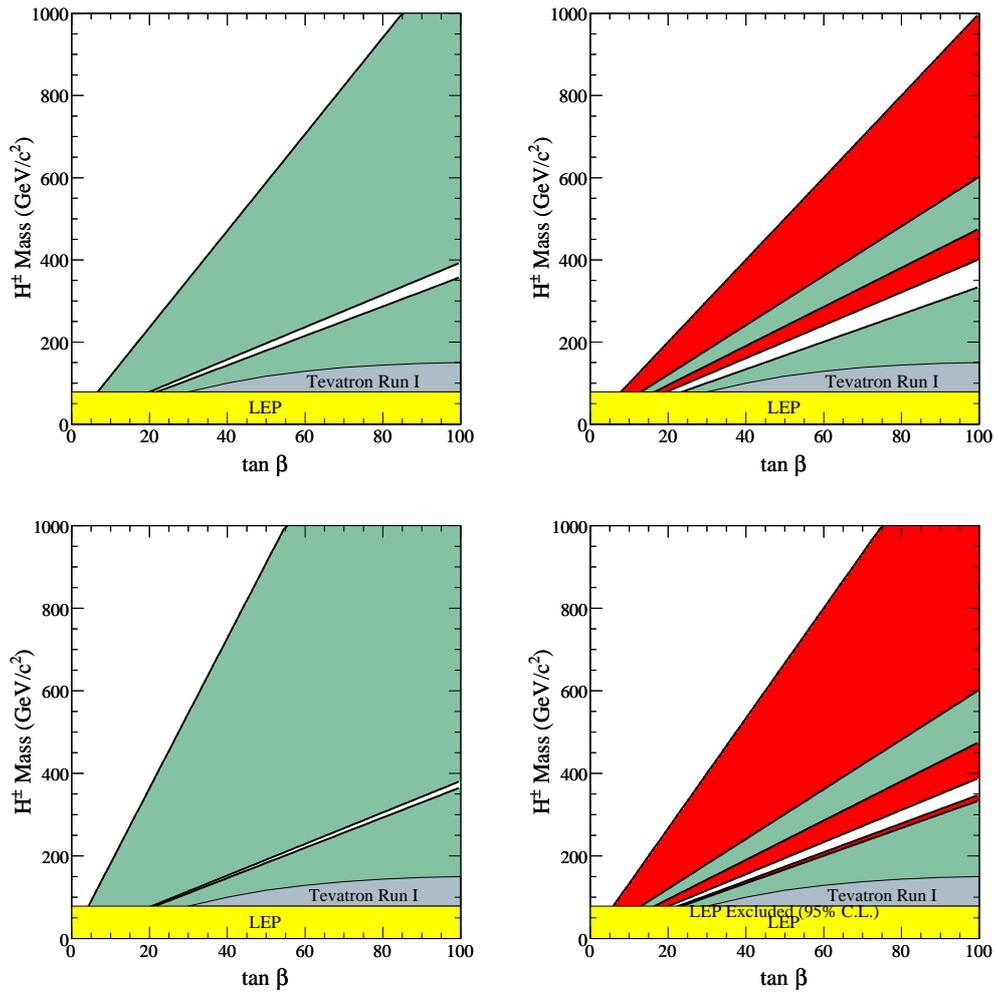

Figure 5.19: 95.5% C.L. exclusion region (left) and $5\sigma$ discovery region (right) of charged Higgs at 5 ab$^{-1}$ (top) and 50 ab$^{-1}$. In the left plots, the green area shows the exclusion region at Super $B$ Factory. In the right plots, the red area shows the discovery region, while the green area shows the present exclusion region.

The dominant background in this analysis arises from the continuum process $e^+e^- \to q\bar{q}$ ($q = u, d, s, c$) and semileptonic $B$ meson decays, mostly into charm final states ($B \to X_c \ell \nu$) with contributions from rare $b \to u$ modes ($B \to X_u \ell \nu$). These backgrounds can be suppressed by using the missing momentum information, because the contamination of the continuum events is often due to undetected particles that are outside the detector acceptance. Event shape variables that are used in various $B$ decay modes are also of use.

In the previous analysis by Belle with a $253\,\text{fb}^{-1}$ data sample, we have performed an inclusive reconstruction of the companion $B$ meson; i.e. we form $M_{\text{bc}}$ and $\Delta E$ from all the charged and neutral particles in an event except for the signal lepton. Figure 5.20 shows the $p_\ell^B$ distributions after all other selections have been applied. The background that remains in the signal region consists of approximately $76\,\%$ continuum and $24\,\%$ $B \to X_u \ell \nu$ according to the off-resonance data and MC studies in both modes. The signal efficiencies are estimated to be $3.15 \pm 0.07\,\%$ for the muon mode and $3.86 \pm 0.08\,\%$ for the electron mode.

Events that pass the $p_\ell^B$ selection (shown in Figure 5.20) are fitted with the $M_{\text{bc}}$ distribution of the companion $B$. The signal yield extracted from the fit is $4.1 \pm 3.1$ events for the muon mode and $-1.8 \pm 3.3$ events for the electron mode in the signal region. The significance of the signal in the muon mode is $1.3\sigma$, and no excess of events is observed in the electron mode.

The following upper limits on the branching fractions at the $90\,\%$ confidence level were obtained:

$$\mathcal{B}(B^+ \to \mu^+ \nu_\mu) < 1.7 \times 10^{-6} \quad (90\,\%\text{ C.L.}) \quad (5.32)$$
$$\mathcal{B}(B^+ \to e^+ \nu_e) < 9.8 \times 10^{-7} \quad (90\,\%\text{ C.L.}), \quad (5.33)$$

including the effect of the systematic uncertainties. The expected sensitivities on the upper limits with the present dataset, computed using toy MC studies with a null signal hypothesis, are $1.0 \times 10^{-6}$ for the muon mode and $1.1 \times 10^{-6}$ for the electron mode.

The upper limit measurement $\mathcal{B}(B^+ \to \mu^+ \nu_\mu)$ has a significance of $1.3\sigma$ with $250\,\text{fb}^{-1}$. By extrapolating this result into the future experiment, we need $1.6\,\text{ab}^{-1}$ for $3\sigma$ evidence and $4.3\,\text{ab}^{-1}$ for $5\sigma$ discovery for the mode $B^+ \to \mu^+ \nu_\mu$. assuming the Standard Model branching fraction. Figure 5.21 shows the standard deviation for the signal of $B^+ \to \mu^+ \nu_\mu$ as a function of luminosity.

Our concern in the Super B will be not only the establishment of the signal, but also precise measurement of the branching fraction. Further improvement of analysis may be possible by using other methods for the companion $B$ reconstruction. For example, full reconstruction technique presently used by BaBar (Ref.XXX) now seems to give higher upper limit, but will be also useful due to higher purity.

### 5.4.4 $B \to K \nu \bar{\nu}$

In the Standard Model the rare FCNC decay $b \to s \nu \bar{\nu}$ proceeds at the one-loop level through penguin and box diagrams.Additional new physics heavy particles may therefore contribute to this decay mode, leading to significant enhancements of the branching fraction. Since the final state leptons do not have electric charge, this mode is not affected by long distance effects from vector resonances ($\rho$, $J/\psi$, $\psi'$ etc.) like decays to charged leptons, $b \to s \ell^+ \ell^-$, and hence the theoretical predictions are more accurate. The inclusive branching fraction is estimated to be $4 \times 10^{-5}$ [88, 89] for the sum of three neutrino flavors, whereas the exclusive branching fractions are predicted to be $Br(B^- \to K^- \nu \bar{\nu}) \approx 4 \times 10^{-6}$ [90].

However, the experimental measurement of $b \to s \nu \bar{\nu}$ is quite challenging due to two missing neutrinos. The best inclusive limit to date is from ALEPH $Br(b \to s \nu \bar{\nu}) < 6.4 \times 10^{-4}$ [91], and a limit of $< 2.4 \times 10^{-4}$ at 90% confidence level on the exclusive branching fraction of $B \to K \nu \bar{\nu}$ was set by CLEO [92]. BABAR has recently reported a preliminary upper limit $Br(B \to K \nu \bar{\nu}) < 7.0 \times 10^{-5}$ [93]. At SuperKEKB, measurements of these decay branching fractions will become possible, as millions of fully reconstructed $B$ samples will be accumulated.

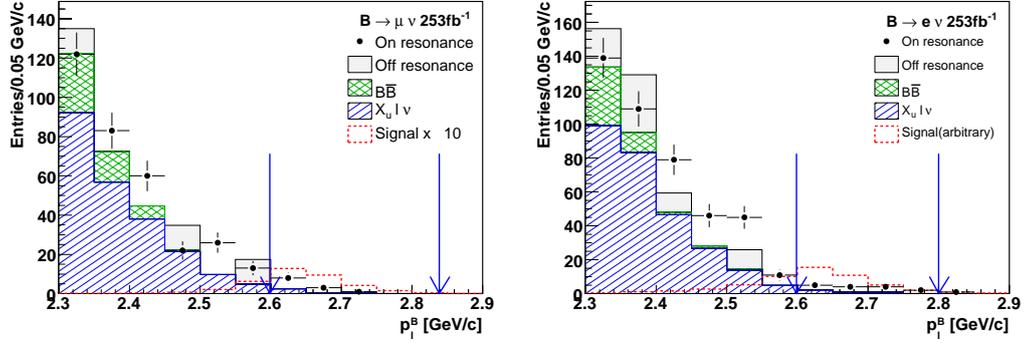

Figure 5.20: $p^B_\ell$ distributions for the signal candidates at Belle. Points show the on-resonance data, and solid histograms show the expected background due to rare $B \to X_u \ell \nu$ decays (hatched, from MC); other $B\bar{B}$ events, principally $B \to X_c \ell \nu$ decays (cross-hatched, also from MC); and continuum events (light shaded, taken from scaled off-resonance data). Dashed histograms are MC $B \to \ell \nu$ signals that are obtained by multiplying the SM expectations by a factor of 10 for the muon mode and by $5 \times 10^6$ for the electron mode. The arrows show the signal regions.

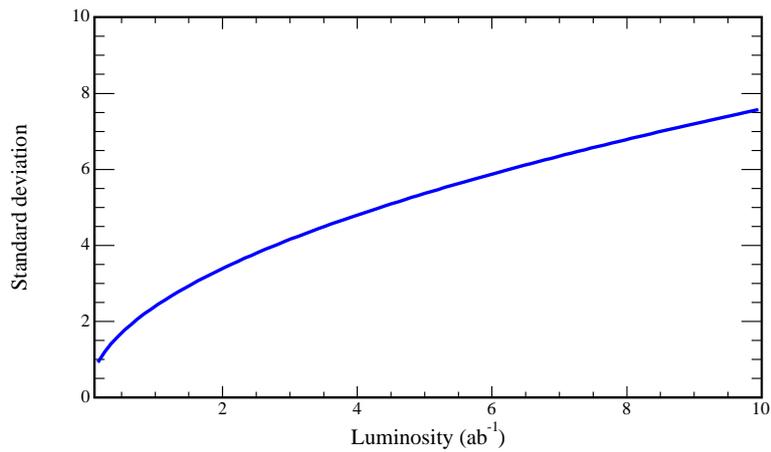

Figure 5.21: The standard deviation for $B^+ \to \mu^+ \nu_\mu$ as a function of luminosity (ab$^{-1}$).

| Decay Mode | Sideband Data | Expected BG | Observed yield | Efficiency [$10^{-5}$] | U.L. |
|---|---|---|---|---|---|
| $B^+ \to K^+ \nu \bar{\nu}$ | 60 | $20 \pm 4$ | 10 | $26.7 \pm 2.9$ | $< 1.4 \times 10^{-5}$ |
| $B^0 \to K^{*0} \nu \bar{\nu}$ | 16 | $4.2 \pm 1.4$ | 7 | $5.1 \pm 0.3$ | $< 3.4 \times 10^{-4}$ |

Table 5.10: Observed yield in the sideband region, expected background, observed yield in the signal region, efficiency (including the $B_{\text{rec}}$ reconstruction efficiency) and the measured 90% C.L. upper limit on the branching fraction for the $B \to K^{(*)} \nu \bar{\nu}$ [94].

We performed a search for the decay $B \to K^{(*)} \nu \bar{\nu}$ in a 492 fb$^{-1}$ data sample [94]. Combinatorial and continuum backgrounds are suppressed by selecting a sample of events with one fully reconstructed $B_{\text{rec}}$. The decay products of the $B$ on the other side of the event are analyzed to search for $B \to K^{(*)} \nu \bar{\nu}$ decay. In the events where a $B_{\text{rec}}$ is reconstructed, we search for decays into a kaon plus two neutrinos. $B^+ \to K^+ \nu \bar{\nu}$ candidate events are required to have one charged track with the charge being opposite to that of the reconstructed $B$. We require the charged particle to be identified as a kaon. $K^{*0}$ mesons are reconstructed from a charged pion and a kaon. The event is required to have zero net charge. We place the requirement on the momentum of the $K^{(*)}$ in the CM, 1.6 GeV/$c < p^* <$ 2.5 GeV/$c$, where the lower bound suppresses the $B\bar{B}$ decays with the $b \to c$ transition and the upper bound the radiative $B \to K^* \gamma$ decays. The direction of the missing momentum is required to be in the range $-0.86 < \cos \theta^*_{\text{miss}} < 0.95$ in the CM frame.

To isolate the signal the remaining energy, calculated by adding the energy of the photons that are not associated with the $B_{\text{rec}}$, $E_{\text{ECL}}$, is used. The $E_{\text{ECL}} < 0.3$ GeV region is defined as the signal region and $0.45 < E_{\text{ECL}} < 1.5$ GeV as the sideband region. The simulated $E_{\text{ECL}}$ shape is used to extrapolate the sideband data to the signal region. The number of MC events in signal region and sideband are counted and their ratio ($r_{MC}$) is obtained. Using the number of data in the sideband and the ratio $r_{MC}$, the number of expected background events in the signal region is estimated. The background estimation from the $E_{\text{ECL}}$ sideband extrapolation is shown in Table 5.10. The numbers of events in sideband region agrees well between MC and data. To obtain the background expected from the MC simulation, $B\bar{B}$ and $e^+ e^- \to u\bar{u}, d\bar{d}, s\bar{s}, c\bar{c}$ events are scaled to equivalent luminosity of the data.

After finalizing the signal selection criteria, the signal region ($E_{\text{ECL}} < 0.3$ GeV) in the on-resonance data is examined. Table 5.10 lists the number of observed events in data in the signal region, together with the expected number of background events in the signal region. Fig. 5.22 shows the $E_{\text{ECL}}$ distributions in the data after all selection requirements except the one on $E_{\text{ECL}}$ have been applied, as well as the expected distribution of background events.

Since no significant excess of events over the expected background is observed, we set upper limits on the branching fractions as given in Tab. 5.10.

Considering the amount of the expected background in $B^0 \to K^{*0} \nu \bar{\nu}$ (Tab. 5.10), the expected signal with the SM branching fraction $Br(B \to K^* \nu \bar{\nu}) = 1.3 \times 10^{-5}$ [95], and no improvements in the measurement method and detector performance, the required luminosity to observe the decay with a 3 $\sigma$ significance is above 100 ab$^{-1}$. However, the value changes drastically if beside the above mentioned hadronic tagging also the semileptonic tagging, mentioned in Sect. 5.1.3 is used. By examination of the results on the $B \to \tau\nu$ decays obtained by hadronic [87] and semileptonic [96] tagging we estimate the efficiency of the latter to be around 4-5 times better than for the hadronic tagging, at the deterioration of the signal to noise ratio by a factor of two. Taking into account also the planned upgrade of the detector, specifically the improved particle identification capabilities and the electromagnetic calorimeter performance, one can expect the improvement in the reconstruction efficiency for $B^0 \to K^{*0}\nu\bar{\nu}$ by around 70%. By these a 3 $\sigma$ significant signal of $B^0 \to K^{*0}\nu\bar{\nu}$ can be observed with around 30 ab$^{-1}$.

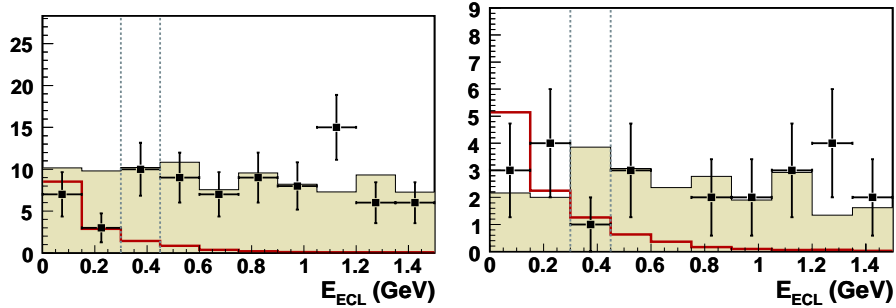

Figure 5.22: $E_{\rm ECL}$ distributions for selected $B^+ \to K^+ \nu\bar{\nu}$ (left) and $B^0 \to K^{*0}\nu\bar{\nu}$ (right) candidates [94]. Shaded (green) histograms show the simulated background distribution normalized to the sideband data. Open (red) histograms represent the SM expected signal multiplied by a factor of 20.

In a more recent paper [97] authors predict a significantly lower branching fraction for these decays, $Br(B \to K^*\nu\bar{\nu}) = 0.68 \times 10^{-5}$. In the paper the decays $B \to K^{(*)}\nu\bar{\nu}$ are analyzed in a model independent way including a possible contribution of the NP right-handed currents parametrized by the Wilson coefficient $C_R^\nu$ (in addition to the SM coefficient $C_L^\nu$). Taking into account the low branching fractions for $B \to K^{(*)}\nu\bar{\nu}$ predicted and the above experimental assumptions one expects to measure both, $Br(B^0 \to K^{*0}\nu\bar{\nu})$ and $Br(B^+ \to K^+\nu\bar{\nu})$ [2] with a relative precision of 30%-35% with an integrated luminosity of 50 ab$^{-1}$. This will allow to overconstrain possible values of $C_R^\nu$ and $C_L^\nu$. The expected sensitivity is illustrated in Fig. 5.23, in terms of

$$\epsilon = \frac{\sqrt{|C_L^\nu|^2 + |C_R^\nu|^2}}{|(C_L^\nu)^{SM}|}$$
$$\eta = \frac{-\Re(C_L^\nu C_R^{\nu*})}{|C_L^\nu|^2 + |C_R^\nu|^2} \ .$$

It can be seen that the sensitivity on $(\epsilon, \eta)$ will be around $\pm 0.2$, which can be compared to the SM value $|(C_L^\nu)^{SM}| \approx 6.4$ [97]. To measure the fraction of longitudinally polarized $K^*$'s in these decays, $F_L$, the direction of the $B$ meson must be known, which is only possible using the hadronic tagging. To reach the necessary sensitivity for additional constraint on the Wilson coefficients from $F_L$ the efficiency of the hadronic tagging must be significantly improved.

---

[2]In [98] the prediction for $Br(B^+ \to K^+\nu\bar{\nu})$ of [97] is further refined to $\sim 15\%$ higher values. Hence the predictions based on this branching fraction are conservative.

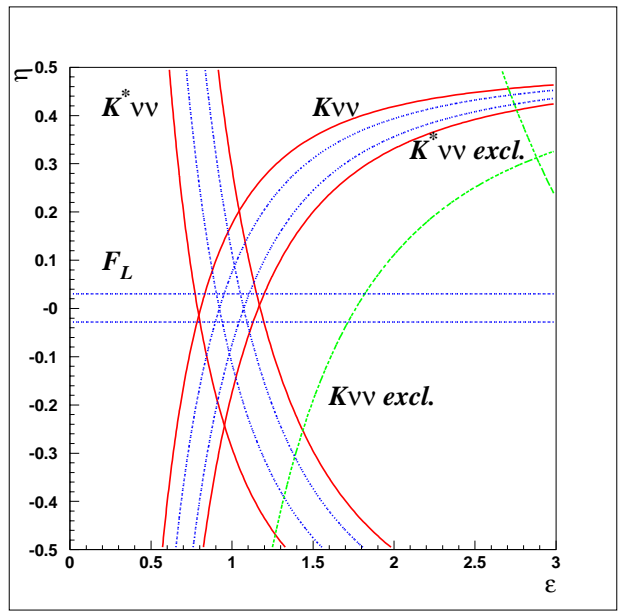

Figure 5.23: Adopted from [97]. Constraints on the $\epsilon(C_L^\nu, C_R^\nu)$ and $\eta(C_L^\nu, C_R^\nu)$ (for definitions see text) arising from the measurement of $Br(B^0 \to K^{*0}\nu\bar{\nu})$ and $Br(B^+ \to K^+\nu\bar{\nu})$ with 50 ab$^{-1}$. Full (red) lines show the expected sensitivity, dotted (blue) lines the predictions including the theoretical uncertainty, and dashed-dotted (green) lines the current 90% C.L. exclusion regions.

## 5.5 $B \to \bar{D}\tau^+\nu_\tau$

The decay $B \to \bar{D}\tau^+\nu_\tau$ is sensitive to the exchange of charged Higgs boson as introduced in the Minimal Supersymmetric Standard Model (MSSM), since the amplitude is roughly proportional to $m_\tau m_b \tan\beta$. The branching fraction of $B \to \bar{D}\tau^+\nu_\tau$ is expected to be large ($\sim 8 \times 10^{-3}$) in the SM, but much more data are required to measure this process because of the presence of two or more neutrinos in the decay final state. This mode consequently requires the tagging of the other side $B$, for which a reconstruction efficiency is estimated to be 0.2%, which has been discussed in Section 5.1.3. Therefore, the study of the MSSM Higgs through this decay mode is only possible with the statistical power of SuperKEKB.

### 5.5.1 Introduction

In the MSSM, the coupling of the charged Higgs bosons, $H^\pm$, to quarks and leptons is given by

$$\mathcal{L}_H = (2\sqrt{2}G_F)^{1/2}\left[\tan\beta\,(\bar{u}_L V_{KM} M_d d_R + \bar{\nu}_L M_\ell \ell_R) + \frac{1}{\tan\beta}\bar{u}_R V_{CKM} M_u d_L\right]H^\pm \\ + h.c., \tag{5.34}$$

where $M_u$ and $M_d$ are the quark mass matrices, $M_\ell$ is the lepton ($\ell = e$, $\mu$, or $\tau$) mass matrix, $V_{CKM}$ is the Cabibbo-Kobayashi-Maskawa matrix, and $\tan\beta = v_2/v_1$ is the ratio of the vacuum expectation values of the Higgs fields. Therefore, the decay amplitude of $B \to \bar{D}\tau^+\nu_\tau$ that is mediated by $\bar{b} \to c\tau^+\nu$ has a term proportional to $M_b M_\tau \tan^2\beta$ [99, 100]. The large $\tau$ mass is an advantage of this decay in measuring the charged Higgs mass compared to other semi-leptonic decays.

Figure 5.24 (top) shows the ratio $B$:

$$B = \frac{\Gamma(B \to \bar{D}\tau^+\nu_\tau)}{\Gamma(B \to \bar{D}\mu^+\nu_\mu)_{\text{SM}}} \tag{5.35}$$

as a function of the charged Higgs mass with several $\tan\beta$ values. The width of each band represents uncertainty in the $B \to D$ semi-leptonic form factor. The form factor is modeled as a function of the momentum transfer $q^2$ using a slope parameter $\rho_1^2$ [101], and the uncertainty is from the error of $\rho_1^2$. The Belle collaboration has measured $\rho_1^2$ as $\rho_1^2 = 1.33 \pm 0.22$ [102]. Figure 5.24 (bottom) shows the $\delta B/B|_{\text{exp}}$ distribution as a function of $\delta\rho_1^2/\rho_1^2|_{\text{exp}}$. In the figure we show several curves with varying $R \equiv M_W \tan\beta/M_H^\pm$.

### 5.5.2 $B \to \bar{D}\tau^+\nu_\tau$ Reconstruction

Final states for the $B \to \bar{D}\tau^+\nu_\tau$ decay include two $\tau$-neutrinos after taking into account the sub-decay of the $\tau$; one additional neutrino exists when $\tau$ decays into a leptonic mode. Because of the presence of two or more invisible particles, the decay of $B \to \bar{D}\tau^+\nu_\tau$ has few kinematic constraints. To reduce the number of combinations of the reconstructed $D$ and $\tau$, we first remove particles originating from the other $B$ ($B_{\text{ful}}$), which does not decay into $B \to \bar{D}\tau^+\nu_\tau$ ($B_{\text{sig}}$). We then reconstruct $D$ and $\tau$ candidates using the remaining particles. Finally, we apply kinematic selection requirements on the reconstructed $D$ and $\tau$ combination to reduce background.

In the following paragraphs we describe the reconstruction of $B \to \bar{D}\tau^+\nu_\tau$. A detailed description of the full reconstruction of $B_{\text{ful}}$ mesons is given in subsection 4.1.2.

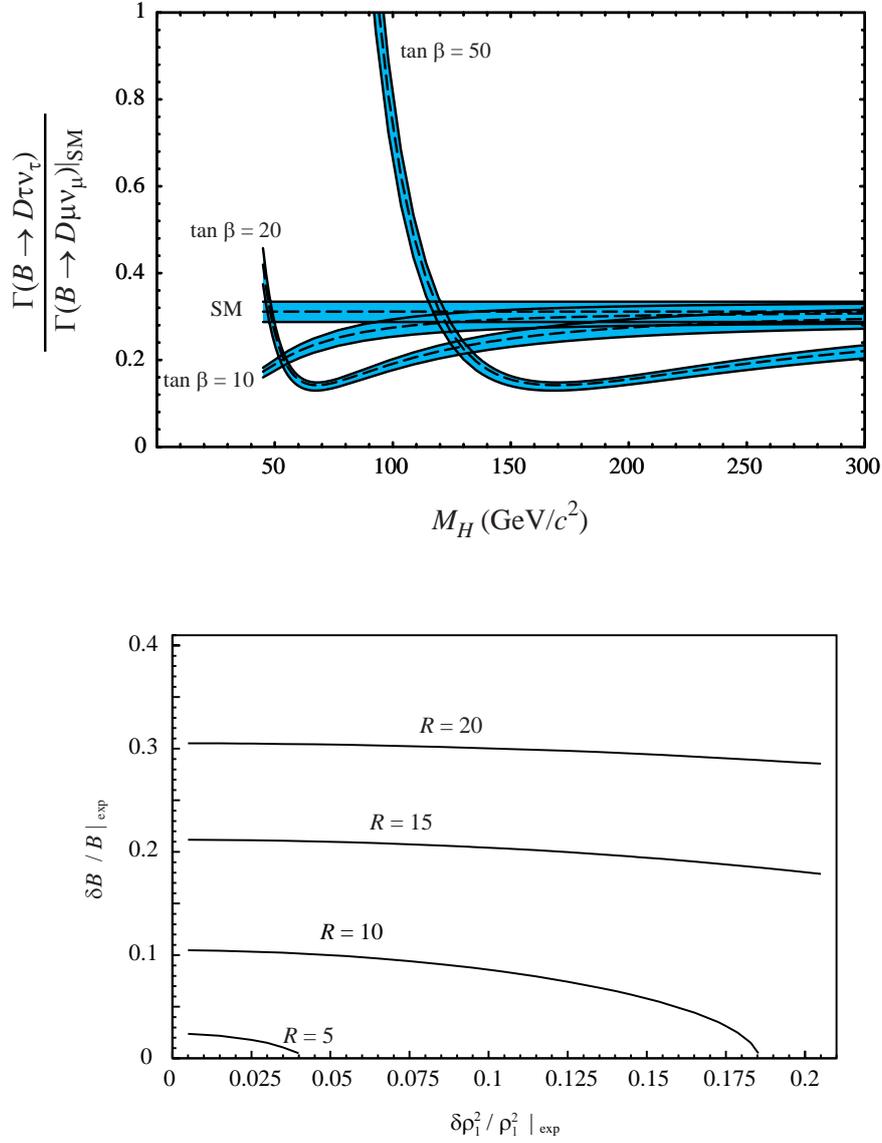

Figure 5.24: The top figure shows the ratio of $\Gamma(B \to \bar{D}\tau^+\nu_\tau)$ to $\Gamma(B \to \bar{D}\mu^+\nu_\mu)$. The flat band is the prediction of the Standard Model; also shown is the charged Higgs contribution, which is a function of the charged Higgs mass with several $\tan\beta$ values. The width of each band is due to uncertainty in the form factor. The bottom figure shows the $\delta B/B|_{\exp}$ distribution as a function of $\delta\rho_1^2/\rho_1^2|_{\exp}$ with varying $R \equiv M_W \tan\beta/M_H^\pm$.

**Light Meson Reconstructions** The $K_S^0$ mesons are reconstructed from a pair of oppositely charged $\pi$ tracks. The invariant mass of the $\pi$ pair should be within $|M_{\pi\pi} - M_{K_S^0}| < 15 \text{GeV}/c^2$. The position of closest approach of the $\pi$ tracks should be displaced from the interaction point in the plane perpendicular to the beam axis.

Neutral pions are reconstructed from a pair of two photons. The $\pi^0$ invariant mass is required to be within $|M_{\gamma\gamma} - M_{\pi^0}| < 16 \text{ MeV}/c^2$. We require the photon energy deposit in the calorimeter to be greater than 50 MeV.

**Charmed Meson Reconstructions** We reconstruct $D$ mesons from $\bar{D}^0 \to K^+\pi^-(\pi^0)$, $K^+\pi^+\pi^-\pi^-(\pi^0)$, $K_S^0\pi^+\pi^-(\pi^0)$, and $K_S^0\pi^0$, where $(\pi^0)$ indicates zero or one neutral pion. Charge conjugate modes are implicitly included throughout this section. These channels cover 35% of the total $\bar{D}^0$ decay width. The charged $D$ meson is reconstructed from $D^+ \to K^-\pi^+\pi^+(\pi^0)$ and $K_S^0\pi^+$, covering 16% of the total $D^+$ decay width.

We construct kaon ($\mathcal{L}(K)$) and pion ($\mathcal{L}(\pi)$) likelihoods to identify the particle species of each charged particle by combining the $dE/dx$ in the drift chamber, the hit in the time of flight counter and the ring imaging Cherenkov counter. The charged particle is identified as a kaon if $\mathcal{L}(K)/[\mathcal{L}(\pi) + \mathcal{L}(K)]$ exceeds 0.4; otherwise it is a pion.

The reconstructed $D^0$ or $D^+$ should have an invariant mass within $|M_{Kn\pi} - M_D| < 30 \text{ MeV}/c^2$.

If more than one $D$ meson is reconstructed, the $D$ meson that has the closest invariant mass to the world average value is taken [4].

**B Meson Reconstructions** The $B \to \bar{D}\tau^+\nu_\tau$ decay is reconstructed from $B^0 \to D^-\tau^+\nu_\tau$ and $B^+ \to \bar{D}^0\tau^+\nu_\tau$, where the $\tau^+$ is identified in one of four following sub-decays: $\tau^+ \to \pi^+\nu_\tau$, $\rho^+(\pi^+\pi^0)\nu_\tau$, $e^+\nu_\tau\bar{\nu}_e$, and $\mu^+\nu_\tau\bar{\nu}_\mu$.

The $B \to \bar{D}\tau^+\nu_\tau$ candidate is reconstructed by adding one charged particle, which is assumed to originate from the $\tau^+$ decay, to the reconstructed $D$ meson. Positively identified protons are rejected.

If the additional charged particle is consistent with the electron or muon hypothesis, the $\tau^+$ decay is treated as a leptonic mode ($\tau^+ \to \ell^+\nu_\tau\bar{\nu}_\ell$)[3]; otherwise the $\tau^+$ decay is considered as a hadronic mode ($\tau^+ \to \pi^+\nu_\tau$).

In the case of $\tau^+ \to \pi^+\nu_\tau$ decay, one neutral pion (if it exists) that is associated to neither $B_{\text{ful}}$ nor $D$ decay is added to the $\pi^+\nu_\tau$ final state to reconstruct $\tau^+ \to \rho^+(\pi^+\pi^0)\nu_\tau$ mode. In this case, the $\pi^+\pi^0$ invariant mass should be within $|M_{\pi^+\pi^0} - M_{\rho^+}| < 300 \text{ MeV}/c^2$.

To reject $B \to \bar{D}^*\tau\nu_\tau$ events, we reconstruct $D^*$'s by adding a $\pi^0$ to the reconstructed $D$. When $|(M_{Kn\pi\pi^0} - M_{Kn\pi}) - \Delta M_{D^*-D}| < 10 \text{ MeV}/c^2$, we discard the event.

If any charged particle and/or $K_L^0$ candidate remains after the reconstructions of $B_{\text{ful}}$ and $B_{\text{sig}}$, the event is rejected.

### 5.5.3 Kinematic Event Selection

We use three kinematic parameters to select $B \to \bar{D}\tau^+\nu_\tau$ signal events from the reconstructed candidates.

The first is the residual cluster energy in the calorimeter ($E_{\text{res}}$). We require $E_{\text{res}} < 100$ MeV.

The second is the missing-mass squared defined by

$$|MM|^2 \equiv |p_{B_{\text{sig}}} - p_D - p_{X^+}|^2, \tag{5.36}$$

---

[3]Throughout this section, the symbol $\ell$ indicates leptons except $\tau$ unless otherwise specified.

where $p_{X^+}$ is the charged particle momentum originating from the $\tau^+$ decay. The momentum $p_{B_{\text{sig}}}$ is given by $p_{B_{\text{sig}}} = p_{\Upsilon(4S)} - p_{B_{\text{ful}}}$.

The last is the cosine of the angle between the momenta of the two $\tau$-neutrinos ($\cos\theta$) in the frame where $\vec{p}_{B_{\text{sig}}} = \vec{p}_D$. This parameter can only be defined for the $\tau^+ \to h^+\bar{\nu}_\tau$ sub-decay. Energy-momentum conservation for the $B_{\text{sig}} \to D\tau^+(h^+\bar{\nu}_\tau)\nu_\tau$ decay is expressed by

$$p_{B_{\text{sig}}} = p_D + p_{h^+} + p_{\bar{\nu}_\tau} + p_{\nu_\tau}. \tag{5.37}$$

The $\tau^+$ and neutrino masses are given by

$$(p_{h^+} + p_{\bar{\nu}_\tau})^2 = m_\tau^2, \tag{5.38}$$

$$p_{\bar{\nu}_\tau}^2 = p_{\nu_\tau}^2 = 0. \tag{5.39}$$

We then boost the system to the frame where

$$\vec{p}_{B_{\text{sig}}} = \vec{p}_D. \tag{5.40}$$

Using Eq. (5.37)-(5.40), we have

$$(E_{B_{\text{sig}}} - E_D)^2 - 2E_{\nu_\tau}(E_{B_{\text{sig}}} - E_D) = m_\tau^2. \tag{5.41}$$

Then, the energies of the two neutrinos can be expressed in terms of measurable parameters as

$$E_{\nu_\tau} = \frac{(E_{B_{\text{sig}}} - E_D)^2 - m_\tau^2}{2(E_{B_{\text{sig}}} - E_D)}, \qquad E_{\bar{\nu}_\tau} = E_{B_{\text{sig}}} - E_D - E_{h^+} - E_{\nu_\tau}. \tag{5.42}$$

Finally, we can measure $\cos\theta$ using the following equation:

$$\begin{aligned}(\vec{p}_{B_{\text{sig}}} - \vec{p}_D - \vec{p}_{h^+})^2 &= (\vec{p}_{\bar{\nu}_\tau} + \vec{p}_{\nu_\tau})^2 \\ &= 2\vec{p}_{\bar{\nu}_\tau} \cdot \vec{p}_{\nu_\tau} \\ &= 2E_{\bar{\nu}_\tau}E_{\nu_\tau}\cos\theta.\end{aligned} \tag{5.43}$$

The $\cos\theta$ of the signal events is limited from $-1$ to $+1$, while that of the background events is unrestricted.

The background contamination is studied by using generic $B$ decay MC samples. We generate samples of $3.2 \times 10^6$ MC events with the fast simulator for generic charged and neutral $B$ decays. In this case, $8 \times 10^{-3}$ is taken as the branching fraction of $B \to \bar{D}\tau^+\nu_\tau$ decay for both charged and neutral $B$. To increase statistics, the $B_{\text{ful}}$ is pseudo-reconstructed using generator information, i.e.: the $B_{\text{ful}}$ is fully reconstructed with perfect efficiency and purity.

Figure 5.25 shows the distribution of the reconstructed $B^+ \to \bar{D}^0\tau^+(h^+\bar{\nu}_\tau)\nu_\tau$ candidates in the $|MM|^2$-$\cos\theta$ plane. The left column figures show the signal events only and the right ones show the background distribution (gray points) with the signal distributions (black points) superimposed. We determine the optimal signal region to be $|MM|^2 > 0.1$ (GeV/$c^2$)$^2$ and $-1.0 < \cos\theta < 0.8$, irrespectively of the $\tau^+ \to \pi^+\bar{\nu}_\tau$ or $\tau^+ \to \rho^+\bar{\nu}_\tau$ by maximizing $S/\sqrt{S+B}$ (FOM), where $S$ and $B$ are the numbers of reconstructed signal and background events, respectively.

Figure 5.26 shows the $|MM|^2$ distribution of the reconstructed candidates of $B^+ \to \bar{D}^0\tau^+(\ell^+\bar{\nu}_\tau\nu_\ell)\nu_\tau$. For that decay irrespectively of the lepton flavor from the $\tau$ decay, we determine the selection criteria $|MM|^2 > 1.2$ (GeV/$c^2$)$^2$ by maximizing the FOM.

We find that $|MM|^2$ and $\cos\theta$ for the neutral $B$ decay have similar distributions in both signal and background as the charged $B$ decay. Therefore, the selection criteria determined by charged $B$ decays are also applied to neutral $B$ decays.

Table 5.11 summarizes the reconstruction efficiency. The numbers of reconstructed signal and the background events in the generic $B$ decay MC samples are also shown.

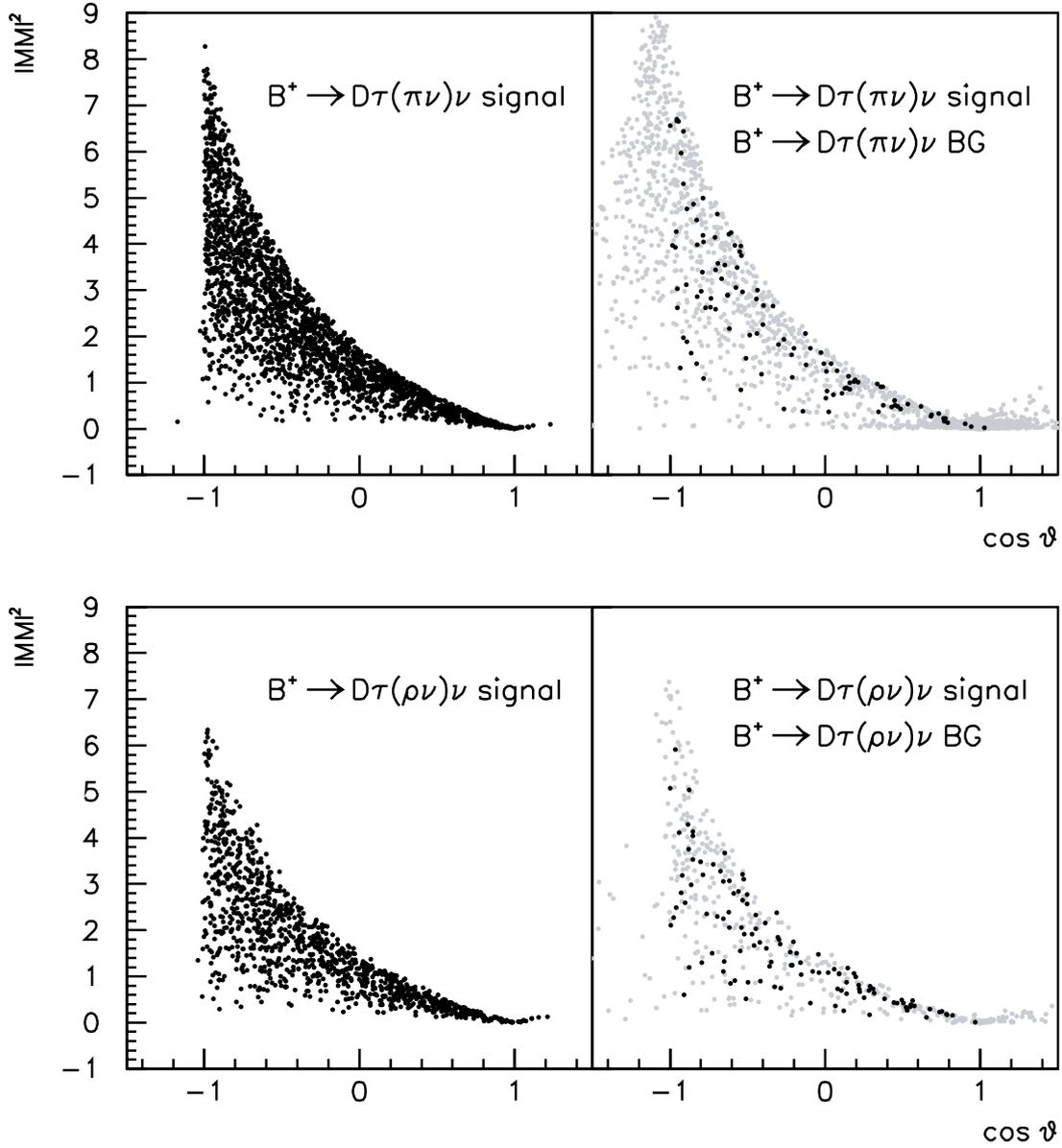

Figure 5.25: A scatter plot of the reconstructed $B^+ \to \bar{D}^0 \tau^+ (h^+ \bar{\nu}_\tau) \nu_\tau$ candidates in the $|MM|^2$-$\cos\theta$ plane. The upper two figures are obtained in the $\tau^+ \to \pi^+ \bar{\nu}_\tau$ decay, and the lower two figures are obtained in the $\tau^+ \to \rho^+(\pi^+ \pi^0) \bar{\nu}_\tau$ decay. The left figures show the distributions obtained from signal MC. In the right figures, the background distributions are shown by gray points and the signal distributions are shown by black points.

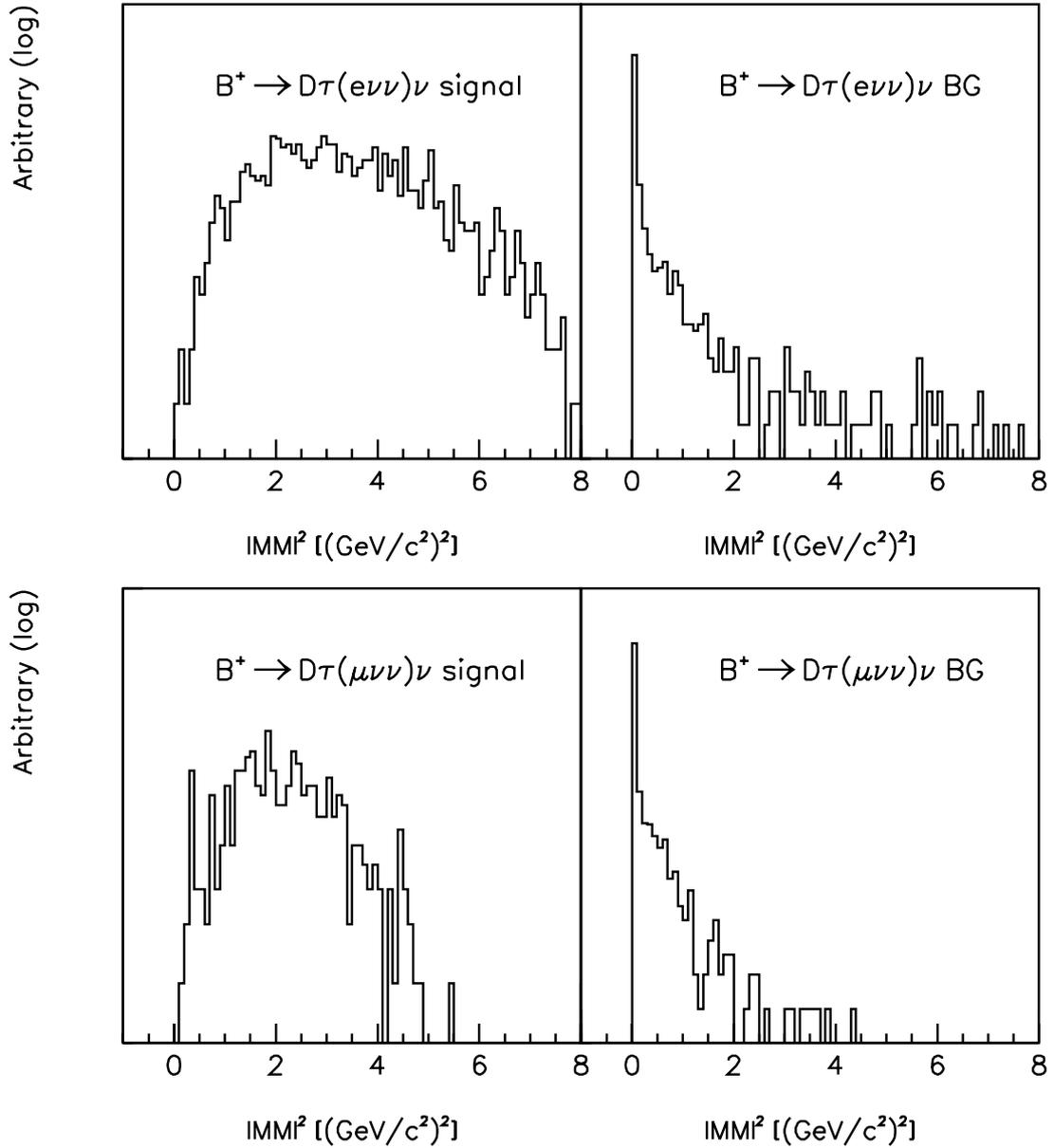

Figure 5.26: The $|MM|^2$ distribution for reconstructed $B^+ \to \bar{D}^0 \tau^+(e^+\bar{\nu}_\tau\nu_e)\nu_\tau$ candidates (upper) and $B^+ \to \bar{D}^0\tau^+(\mu^+\bar{\nu}_\tau\nu_\mu)\nu_\tau$ candidates (lower). The left figures show the distributions obtained from signal MC, and the right figures show the background.

| Decay mode | efficiency (%) | $Br$ | $N_{\text{sig}}$ | $N_{\text{bkg}}$ |
|---|---|---|---|---|
| $\bar{D}^0\tau^+(\ell^+\bar{\nu}_\tau\nu_\ell)\nu_\tau$ | $4.9 \pm 0.3$ | $13.5 \times 10^{-4}$ | 213 | 293 |
| $\bar{D}^0\tau^+(h^+\bar{\nu}_\tau)\nu_\tau$ | $10.9 \pm 0.5$ | $13.6 \times 10^{-4}$ | 476 | 2085 |
| $D^-\tau^+(\ell^+\bar{\nu}_\tau\nu_\ell)\nu_\tau$ | $0.7 \pm 0.2$ | $6.2 \times 10^{-4}$ | 15 | 22 |
| $D^-\tau^+(h^+\bar{\nu}_\tau)\nu_\tau$ | $3.3 \pm 0.4$ | $6.4 \times 10^{-4}$ | 68 | 194 |

Table 5.11: Summary of the reconstruction efficiencies. The numbers of reconstructed signal ($N_{\text{sig}}$) and the background events ($N_{\text{bkg}}$) in generic $B$ decay MC samples are also shown, where $Br(B \to \bar{D}\tau^+\nu_\tau) = 8 \times 10^{-3}$ is assumed.

### 5.5.4 Background Components

The dominant background source in the $B^+ \to \bar{D}^0\tau^+(h^+\bar{\nu}_\tau)\nu_\tau$ signal region is the decay $B^+ \to D^{*-}\ell^+\nu_\ell\pi^+$, due mainly to a missing $\ell^+$ and a slow pion from the $D^{*-}$. The next largest background modes are $B^+ \to \bar{D}^{*0}\ell^+\nu_\ell$ with a missing $\gamma$ or $\pi^0$ from the $\bar{D}^{*0}$ and misidentification of $\ell^+$ as $\pi^+$, and the $B^+ \to \bar{D}^{*0}\tau^+\nu_\tau$ with a missing slow pion. In this study the $B^+ \to \bar{D}^0\tau^+(\ell^+\bar{\nu}_\tau\nu_\ell)\nu_\tau$ decay is considered as a background for the $B^+ \to D^-\tau^+(h^+\bar{\nu}_\tau)\nu_\tau$ analysis; they contribute to the signal region due mostly to misidentification of the $\ell^+$ from the $\tau^+$ decay as $\pi^+$. The sum of the background modes listed above are $\sim 45\%$ of the total background. The contribution from $D_s$ inclusive decays is $\sim 8\%$ of the total.

The largest contribution to the $B^+ \to \bar{D}^-\tau^+(\ell^+\bar{\nu}_\tau\nu_\ell)\nu_\tau$ signal region comes from $B^+ \to \bar{D}^{*-}\ell^+\nu_\ell\pi^+$ where both pions from $B^+$ and $\bar{D}^{*-}$ are missed. The next largest background components are $B^+ \to \bar{D}^{*0}\ell^+\nu_\ell$ and $B^+ \to \bar{D}^{*0}\tau^+\nu_\tau$ when the $\gamma$ or $\pi^0$ from the $\bar{D}^{*0}$ are missed.

The dominant background mode in the $B^0 \to D^-\tau^+(h^+\bar{\nu}_\tau)\nu_\tau$ signal region is $B^0 \to D^-\tau^+\nu_\tau$ with a mis-reconstructed $\tau^+$; the next largest is $B^0 \to D^{*-}\mu^+\nu_\mu$, although it is only $\sim 20\%$ of the total. No other background modes make significant contributions in the signal region.

For the $B^0 \to D^-\tau^+(\ell^+\bar{\nu}_\tau\nu_\ell)\nu_\tau$ mode, because of the small MC statistics, we cannot yet evaluate the background.

### 5.5.5 Statistical Significance

As described in subsection 4.1.2, the full reconstruction efficiencies for charged (neutral) $B$ mesons are estimated to be around 0.2% (0.1%) for a purity of about 80%.

Table 5.12 lists the expected signal yields and backgrounds at integrated luminosities of 5 and 50 ab$^{-1}$. The values include a correction for the purity of $B_{\text{ful}}$ reconstruction. We assume $Br(B \to \bar{D}\tau^+\nu_\tau) = 8 \times 10^{-3}$ in the table. The values listed are obtained by scaling the results in Table 5.11 according to the integrated luminosity. The expected uncertainties in the measured branching fraction ($\delta(Br)/Br$) are also shown.

The branching fraction for $B^+ \to \bar{D}^0\tau^+\nu_\tau$ is expected to be determined with $12\sigma$ statistical significance at an integrated luminosity of 5 ab$^{-1}$. The branching fractions of the neutral $B$ modes can also be measured at 50 ab$^{-1}$ with $11\sigma$ significance.

### 5.5.6 Systematic Uncertainty

Major sources of systematic uncertainty in the branching fraction measurement in $B \to \bar{D}\tau^+\nu_\tau$ decay are expected to be $B_{\text{ful}}$ reconstruction efficiency and purity, particle identification efficiency and purity, and the slow pion detection efficiency.

| Decay mode | 5 ab$^{-1}$ | | | 50 ab$^{-1}$ | | |
|---|---|---|---|---|---|---|
| | $N_{\text{sig}}$ | $N_{\text{bkg}}$ | $\delta(Br)/Br$ | $N_{\text{sig}}$ | $N_{\text{bkg}}$ | $\delta(Br)/Br$ |
| $\bar{D}^0\tau^+(\ell^+\bar{\nu}_\tau\nu_\ell)\nu_\tau$ | $280 \pm 20$ | $550 \pm 20$ | 7.9% | $2800 \pm 50$ | $5500 \pm 70$ | 2.5% |
| $\bar{D}^0\tau^+(h^+\bar{\nu}_\tau)\nu_\tau$ | $620 \pm 20$ | $3600 \pm 60$ | | $6200 \pm 80$ | $36000 \pm 200$ | |
| $D^-\tau^+(\ell^+\bar{\nu}_\tau\nu_\ell)\nu_\tau$ | $10 \pm 3$ | $21 \pm 5$ | 28.5% | $98 \pm 10$ | $210 \pm 10$ | 9.0% |
| $D^-\tau^+(h^+\bar{\nu}_\tau)\nu_\tau$ | $45 \pm 7$ | $170 \pm 10$ | | $450 \pm 20$ | $1700 \pm 40$ | |

Table 5.12: The expected signal yields and backgrounds at integrated luminosities of 5 and 50 ab$^{-1}$, assuming $Br(B \to \bar{D}\tau^+\nu_\tau) = 8 \times 10^{-3}$. The expected uncertainties in the measured branching fraction ($\delta(Br)/Br$) are also shown.

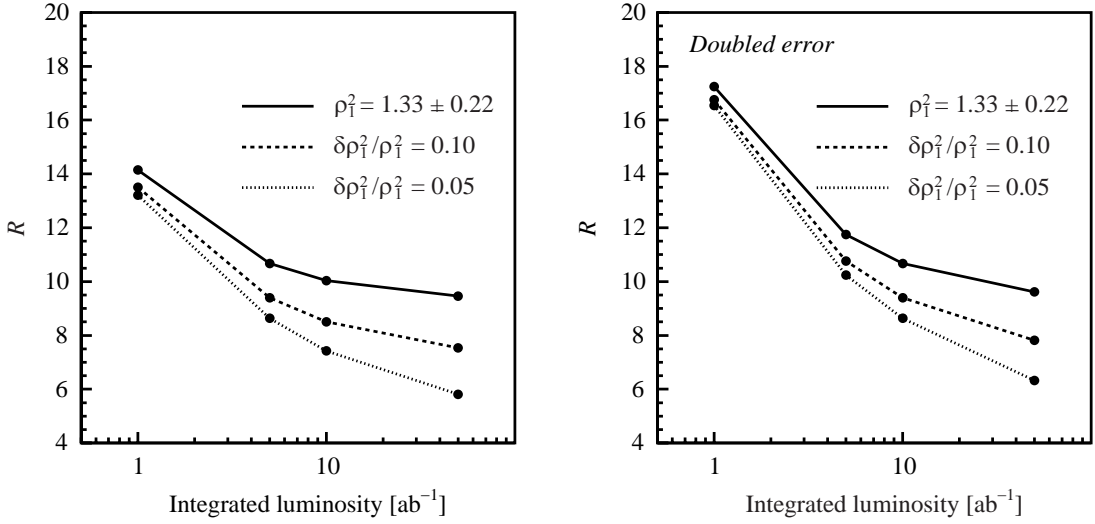

Figure 5.27: The left figure shows the maximum $R$ constrained by the $B^+ \to \bar{D}^0\tau^+\nu_\tau$ branching fraction measurement as a function of the integrated luminosity for several values of $\delta\rho_1^2/\rho_1^2|_{\text{exp}}$. The right figure shows the same plot for the case with double the uncertainty in the measured branching fraction

### 5.5.7 Constraints on the Charged Higgs Mass

Figure 5.27 (left) shows the maximum $R$ value constrained by the branching fraction measurement of $B^+ \to \bar{D}^0\tau^+\nu_\tau$ as a function of the integrated luminosity for several $\delta\rho_1^2/\rho_1^2|_{\text{exp}}$. The right figure shows the case assuming a doubled uncertainty of the measured branching fraction. The systematic uncertainty in the branching fraction measurement is not considered. We ignore the uncertainty in $\delta B/B|_{\text{exp}}$ that comes from the uncertainty in the $B^+ \to \bar{D}^0\mu^+\nu_\mu$ branching fraction because it can be determined much precisely than that of $\bar{D}^0\tau^+\nu_\tau$. We expect

$$M_H > \frac{M_W \tan\beta}{11}$$

at an integrated luminosity of 5 ab$^{-1}$, assuming that the current $\rho_1^2$ precision is unchanged.

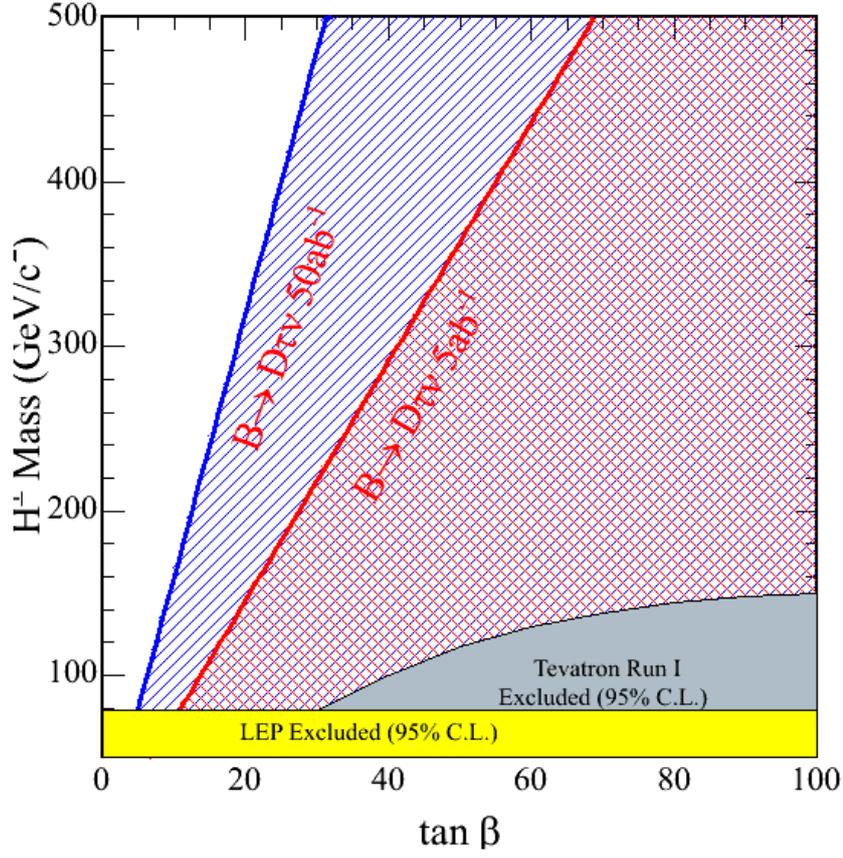

Figure 5.28: Exclusion boundaries in the $[M_{H^+}, \tan\beta]$ plane compared with other experimental searches at LEP and at the Tevatron.

### 5.5.8 Summary

The $B \to \bar{D}\tau^+\nu_\tau$ decay is a sensitive mode to probe the charged Higgs in the MSSM. Using this decay mode we expect that the charged Higgs mass will be constrained by $M_H > M_W \tan\beta/11$ at an integrated luminosity of 5 ab$^{-1}$. Figure 5.28 shows the exclusion boundaries in the $[M_{H^+}, \tan\beta]$ plane at an integrated luminosity of 5 ab$^{-1}$ and 50 ab$^{-1}$.

| Source | Irreducible | Error of $\mathcal{S}$ | Error of $\mathcal{A}$ |
|---|---|---|---|
| Vertex reconstruction | √ | 0.012 | 0.009 |
| Flavor tagging | | 0.004 | 0.003 |
| Resolution function | | 0.006 | 0.001 |
| Physics parameters | | 0.001 | 0.001 |
| Possible fit bias | | 0.007 | 0.004 |
| Background fraction | | 0.006 | 0.002 |
| Background $\Delta t$ shape | | 0.001 | 0.001 |
| Tag-side interference | √ | 0.001 | 0.009 |
| Total | | 0.017 | 0.014 |

Table 5.13: Systematic errors on $\mathcal{A}$ and $\mathcal{S}$ measured with the $J/\psi K^0$ mode at 492 fb$^{-1}$.

## 5.6 $\sin 2\phi_1$

Very precise measurements of $\sin 2\phi_1$, or $\mathcal{S}_{J/\psi K^0_S}$, will remain important at SuperKEKB. There are two major reasons. One is to search for a new $CP$-violating phase in $b \to s$ transitions by testing the SM prediction $\mathcal{S}_{\phi K^0_S} = \mathcal{S}_{J/\psi K^0_S}$ (see Sect. 3.4.4). The other is to check the consistency of the Unitarity Triangle. As explained in Section 2.4.2, $\sin 2\phi_1$ can be determined using the $B^0 \to J/\psi K^0$ mode with very small hadronic uncertainties. It is also insensitive to effects beyond the SM. Thus it serves as a reliable reference point for the SM.

In the present measurements of $\sin 2\phi_1$ $B^0 \to J/\psi K^0_L$, $J/\psi K^{*0}$, $\psi(2S)K^0_S$, $\chi_{c1} K^0_S$ and $\eta_c K^0_S$ decay modes are used in addition to the $B^0 \to J/\psi K^0_S$ decay to reduce the statistical uncertainty. However, with an integrated luminosity of 5 ab$^{-1}$ the systematic uncertainties will become dominant. Therefore, in this study we use only the gold-plated mode $B^0 \to J/\psi(\to \ell^+\ell^-)K^0$, where $K^0$ denotes $K^0_S \to \pi^+\pi^-$ or $K^0_L$, to minimize systematic uncertainties. We perform MC pseudo-experiments assuming that the performance of the detector at SuperKEKB is identical to that of the present Belle detector. The analysis procedure for the measurements of time-dependent $CP$ asymmetries at Belle is described in Section 5.1.2. An example of a fit to a $\Delta t$ distribution for $B^0 \to J/\psi K^0_S$ candidates in a MC pseudo-experiment at 5 ab$^{-1}$ is shown in Fig. 5.29.

Sources of systematic errors include uncertainties in the flavor tagging, in the vertex reconstruction, in the background fractions and $\Delta t$ distributions, in the resolution function, in $\Delta m_d$ and $\tau_{B^0}$, a possible bias in the fit, and the effect of interference [103] in the $f_{\text{tag}}$ final state. Some of these uncertainties are evaluated from control samples, which have large but finite statistics. As the integrated luminosity increases, this part of the systematic error will decrease. In order to estimate the expected systematic error at 5 ab$^{-1}$, we therefore need to separate such *reducible* systematic errors from the other part, which is *irreducible*. In this study, we conservatively assume that uncertainties that do not arise from statistics of control samples are irreducible, and remain the same as the integrated luminosity increases. Further studies on these "*irreducible*" errors will probably find a way to reduce them. Table 5.13 summarizes the sources of systematic errors for $\mathcal{S}$ ($= \sin 2\phi_1$) and $\mathcal{A}$ [104]. All the values are evaluated at 492 fb$^{-1}$.

The total irreducible systematic error for $\mathcal{S}$ is estimated to be 0.012 for $\mathcal{S}_{J/\psi K^0}$. The dominant sources of the irreducible systematic error are uncertainties in the interaction-point profile (0.0072), dependence on the vertex selection-criteria (0.0064), the effect of detector misalignment (0.0056) and possible bias in the $\Delta z$ determination (0.0041). The systematic error for $\mathcal{A}$ is dominated by the tag-side interference and vertex reconstruction. As mentioned above, there will be a possibility to reduce these "*irreducible*" errors from dedicated studies. Table 5.14 lists the expected errors at 492 fb$^{-1}$, 5 ab$^{-1}$ and 50 ab$^{-1}$.

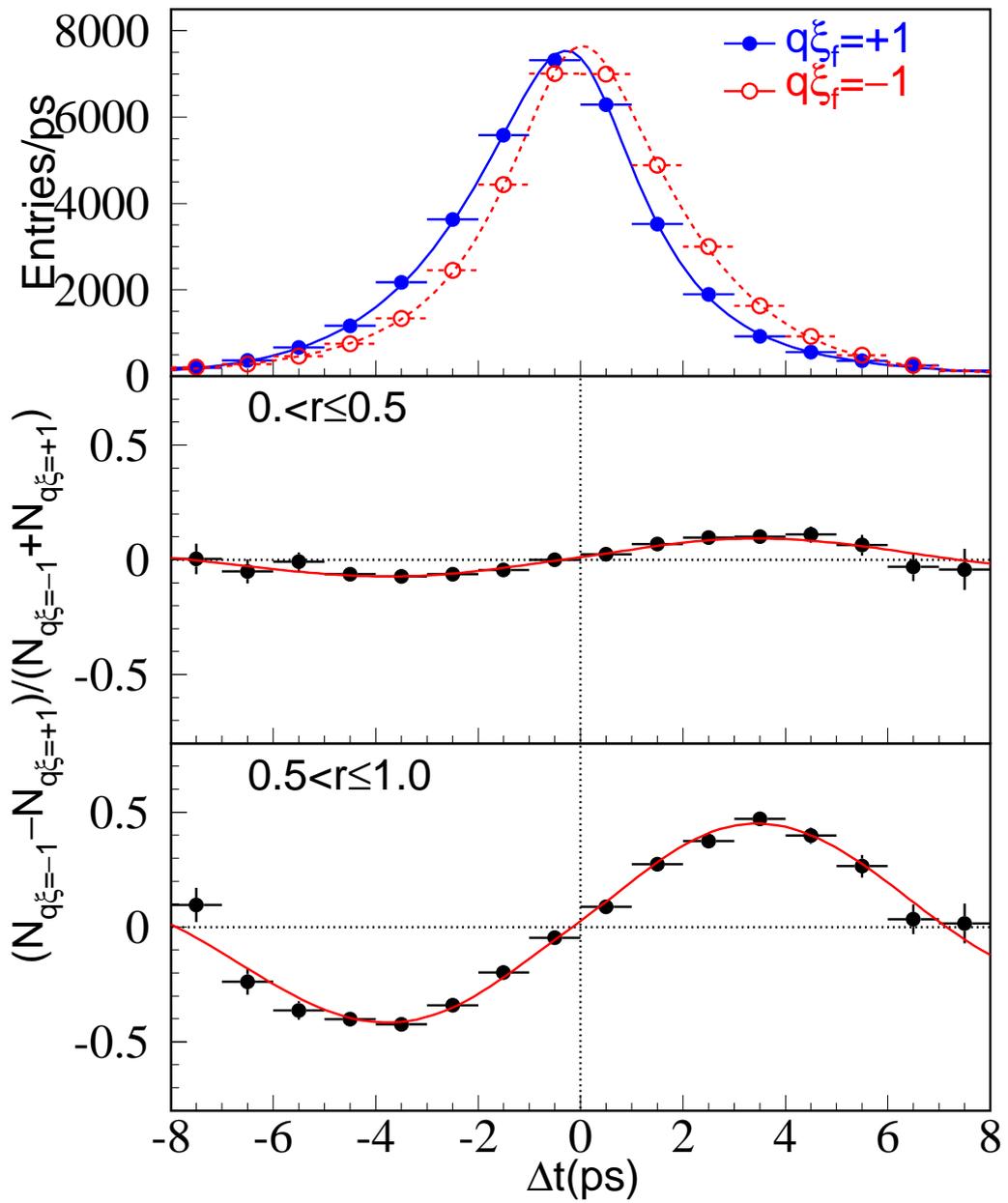

Figure 5.29: An example of a fit to a MC pseudo-experiment at 5 ab$^{-1}$.

|  |  | Statistical | Systematic | | Total |
| --- | --- | --- | --- | --- | --- |
|  |  |  | reducible | irreducible |  |
| $\mathcal{S}_{J/\psi K^0}$ | (492 fb$^{-1}$) | 0.031 | 0.012 |  | 0.036 |
| ($\sin 2\phi_1$) | (5 ab$^{-1}$) | 0.010 | 0.004 | 0.012 | 0.016 |
|  | (50 ab$^{-1}$) | 0.003 | 0.001 |  | 0.012 |
| $\mathcal{A}_{J/\psi K^0}$ | (492 fb$^{-1}$) | 0.021 | 0.006 |  | 0.025 |
|  | (5 ab$^{-1}$) | 0.007 | 0.002 | 0.013 | 0.015 |
|  | (50 ab$^{-1}$) | 0.002 | 0.001 |  | 0.013 |

Table 5.14: Expected errors in time-dependent $CP$ asymmetry measurements with the $B^0 \to J/\psi K^0$ decay at 492 fb$^{-1}$, 5 ab$^{-1}$ and 50 ab$^{-1}$.

The total error for $\mathcal{S}_{J/\psi K^0}$, $\sigma_{\text{tot}}(\mathcal{S}_{J/\psi K^0})$, is obtained from

$$\sigma_{\text{tot}}(\mathcal{S}_{J/\psi K_S^0}) = \sqrt{0.031^2 \times 0.492/\mathcal{L}_{\text{int}} + 0.012^2 \times 0.492/\mathcal{L}_{\text{int}} + 0.012^2}, \quad (5.44)$$

where $\mathcal{L}_{\text{int}}$ is the integrated luminosity in the unit of ab$^{-1}$. We obtain $\sigma_{\text{tot}}(\mathcal{S}_{J/\psi K^0}) = 0.016$ at $\mathcal{L}_{\text{int}} = 5$ ab$^{-1}$ and 0.012 at $\mathcal{L}_{\text{int}} = 50$ ab$^{-1}$. As will be shown in Sect. 6.2.2, the precise measurement of the $\sin 2\phi_1$ with the 50 ab$^{-1}$ of data will remain the most constraining measurement in the combined fit to the unitary triangle.

## 5.7 $\phi_2$

The CKM angle $\phi_2$ is defined as the phase difference between $V_{td}V_{tb}^*$ and $V_{ud}V_{ub}^*$, and can be extracted by measuring the time-dependent $CP$-violating parameter of $b \to u\bar{u}d$ decay transitions [105]. The mixing induced $CP$-violating parameter $\mathcal{S}$ is related with $\phi_2$ via $\mathcal{S} = \sqrt{1-\mathcal{A}^2}\sin(2\phi_2 + \kappa)$, where $\mathcal{A}$ is the direct $CP$-violating parameter and $\kappa$ represents an additional phase induced by $b \to dq\bar{q}$ penguin transitions, where $q = u$ or $d$. The phase $\kappa$ is measurable using isospin relations [106] that provide the relationship among the decay amplitudes and direct $CP$ asymmetries of $B$ and $\overline{B}$ decays. For $B \to \pi\pi$ decays $\kappa$ is expressed in terms of measurable quantities in Eq. (4.6).

The angle $\phi_2$ has been measured using $B \to \pi\pi$, $\rho\pi$ and $\rho\rho$ decays by Belle and BaBar collaborations. All the results yield consistent $\phi_2$ constraints, and the combined $\phi_2$ value $(89.0^{+4.4}_{-4.2})°$ is in agreement with the CKM expectation value $(92.3^{+6.5}_{-6.2})°$ obtained from measurements of other CKM angles and sides [107].

In this section, we give a prospect of the $\phi_2$ measurements by assuming that we accumulate 5 ab$^{-1}$ and 50 ab$^{-1}$ data at a future high luminosity $B$ factory.

### 5.7.1 $\phi_2$ extraction with $B \to \pi\pi$ decays

Both Belle and BaBar collaborations have measured the necessary ingredients of $B \to \pi\pi$ decays for $\phi_2$ extraction: the time-dependent $CP$-violating parameters in $B^0 \to \pi^+\pi^-$ decays, the branching fractions of $B^0 \to \pi^+\pi^-$, $\pi^0\pi^0$ and $B^+ \to \pi^+\pi^0$ decays, and direct $CP$ asymmetry in $B^0 \to \pi^0\pi^0$ decays [108, 109]. Reference [22] compiles these measurements and evaluates the world average (W.A.) values.

To extrapolate the measurement errors to the high luminosity $B$ factory experiment, we categorize them into reducible and irreducible errors. The reducible errors include the statistical errors and the systematic uncertainties of the probability density function (PDF) parameters, particle identification requirements and the possible $CP$ violation effect in the accompanying $B$ meson decays [110]. Our estimation yields the following irreducible systematic errors: 1% (2%) from the $\pi^+$ ($\pi^0$) detection efficiency uncertainty in branching fraction measurements; ±0.01 from the vertex reconstruction uncertainty originating from the SVD mis-alignment in the $\mathcal{S}$ measurements, and ±0.01 from the asymmetry of charged particle detection efficiency in the $\mathcal{A}$ measurements.

All the W.A. measurement values in Ref. [22] are dominated by reducible errors. Hence, to estimate the $\phi_2$ sensitivity with 5 ab$^{-1}$ and 50 ab$^{-1}$ data, we scale the size of the errors by inverse square root of the integrated luminosity, and separately consider the irreducible systematic errors that are unchanged throughout the data accumulation. For the isospin analysis, we use slightly different central values from Ref. [22] to close the isospin triangle. Table 5.15 summarizes the central values and the expected errors of the measurements.

To constrain $\phi_2$, we employ the statistical treatment of Ref. [107]. The isospin relations suggest that the branching fractions and $CP$ asymmetries in $B \to \pi\pi$ decays are described with six parameters, one of which is $\phi_2$. We construct a $\chi^2$ using the six parameters with constraints from the values and errors in Table 5.15. We minimize the $\chi^2(\phi_2)$ for each $\phi_2$ value in the range of $0°$ to $180°$ by varying the remaining five parameters, and calculate the minimum $\chi^2$ value, $\chi^2_{\min}$. The $\chi^2(\phi_2)$ values are translated to a confidence level (C.L.) value using the cumulative distribution of $\Delta\chi^2 = \chi^2(\phi_2) - \chi^2_{\min}$ for one degree of freedom.

It has been pointed out in Ref. [111] that the mixing induced $CP$-violating parameter in $B^0 \to \pi^0\pi^0$ decays, $\mathcal{S}_{\pi^0\pi^0}$, is measurable using photon conversion occurring in the detector material close to the interaction point. The $\pi^0$ Dalitz decay can also be used in the measurement. Electron pairs combined with the information on the interaction region profile enable the determination of the $B$ meson decay vertex which is impossible using neutral particles only. Using a toy MC simulation with the continuum and $B^+ \to \rho^+\pi^0$ backgrounds, the measurement precision of $\mathcal{S}_{\pi^0\pi^0}$ is estimated to be 0.23 with the 50 ab$^{-1}$ data. It should be noted that the upgraded vertex detector would increase the number of photon conversions by around 50% and would thus additionally improve the accuracy of the measurement.

Table 5.15: The measurement central values and expected errors in $B \to \pi\pi$ decays. The branching fractions are given in $10^{-6}$. The errors of W.A. include statistical and systematic uncertainties. The first (second) uncertainty at 5 ab$^{-1}$ and 50 ab$^{-1}$ represents the reducible (irreducible) error.

|  | central values | errors of W.A. [22] | expected errors at 5 ab$^{-1}$ | expected errors at 50 ab$^{-1}$ |
|---|---|---|---|---|
| $\mathcal{B}(\pi^+\pi^-)$ | 5.21 | $\pm 0.20$ | $\pm 0.09 \pm 0.10$ | $\pm 0.03 \pm 0.10$ |
| $\mathcal{B}(\pi^+\pi^0)$ | 5.61 | $\pm 0.40$ | $\pm 0.17 \pm 0.17$ | $\pm 0.05 \pm 0.17$ |
| $\mathcal{B}(\pi^0\pi^0)$ | 1.35 | $\pm 0.21$ | $\pm 0.09 \pm 0.05$ | $\pm 0.03 \pm 0.05$ |
| $\mathcal{S}_{\pi^+\pi^-}$ | $-0.66$ | $\pm 0.08$ | $\pm 0.04 \pm 0.01$ | $\pm 0.01 \pm 0.01$ |
| $\mathcal{A}_{\pi^+\pi^-}$ | $+0.37$ | $\pm 0.07$ | $\pm 0.03 \pm 0.01$ | $\pm 0.01 \pm 0.01$ |
| $\mathcal{S}_{\pi^0\pi^0}$ | $+0.92$ | N.A. | N.A. | $\pm 0.23$ |
| $\mathcal{A}_{\pi^0\pi^0}$ | $+0.16$ | $\pm 0.33$ | $\pm 0.14 \pm 0.01$ | $\pm 0.04 \pm 0.01$ |

Figure 5.30 shows a 1−C.L. plot as a function of $\phi_2$ for three cases: 5 ab$^{-1}$, 50 ab$^{-1}$ without $\mathcal{S}_{\pi^0\pi^0}$ and 50 ab$^{-1}$ with $\mathcal{S}_{\pi^0\pi^0}$. As shown in the figure, the impact of $\mathcal{S}_{\pi^0\pi^0}$ is significant for reducing the ambiguity of the $\phi_2$ solutions. The 1$\sigma$ error for the solution at $\phi_2 = 90°$ is estimated to be $\pm 10°$ and $\pm 3°$ for the 5 ab$^{-1}$ and 50 ab$^{-1}$ data samples, respectively.

We note that the $\phi_2$ measurement precision of $\pm 3°$ is in magnitude similar to the effect of the isospin breaking [112], which represents an inherent uncertainty of the method to extract $\phi_2$, and includes the effect of $\pi^0$-$\eta$-$\eta'$ mixing, $< 1.6°$ [113, 114], contributions of the electroweak penguin transitions, $(1.5 \pm 0.4)°$ [113, 115, 116], and the $\Delta I = 5/2$ contributions caused by electromagnetic rescattering of two pions, $\sim 1\%$ [117].

### 5.7.2 $\phi_2$ extraction with $B \to \rho\pi$ decays

Although the final state of the decay $B^0 \to \rho^+\pi^-$ is not a $CP$ eigenstate, this decay can proceed from both $B^0$ and $\overline{B}^0$. Therefore, the mixing-induced $CP$ violation in $B^0 \to \rho^\pm\pi^\mp$ and $\overline{B}^0 \to \rho^\pm\pi^\mp$ decays are sensitive to $\phi_2$. The $CP$-violation parameters in $B^0 \to \rho^+\pi^-$ were measured by both Belle [118] and BaBar [119] collaborations using the quasi-two-body decay approximation.

Synder and Quinn [120] proposed a time-dependent Dalitz plot analysis in $B^0 \to \pi^+\pi^-\pi^0$ decays for extracting $\phi_2$. The time- and Dalitz plot-dependent differential decay rate is

$$\frac{d\Gamma}{d\Delta t ds_+ ds_-} \propto e^{-|\Delta t|/\tau_{B^0}} \{(|A_{3\pi}|^2 + |\overline{A}_{3\pi}|^2) - q_{\text{tag}}(|A_{3\pi}|^2 - |\overline{A}_{3\pi}|^2)\cos(\Delta m_d \Delta t)$$
$$+ q_{\text{tag}} 2\Im\left[\frac{q}{p} A_{3\pi}^* \overline{A}_{3\pi}\right]\sin(\Delta m_d \Delta t)\}, \qquad (5.45)$$

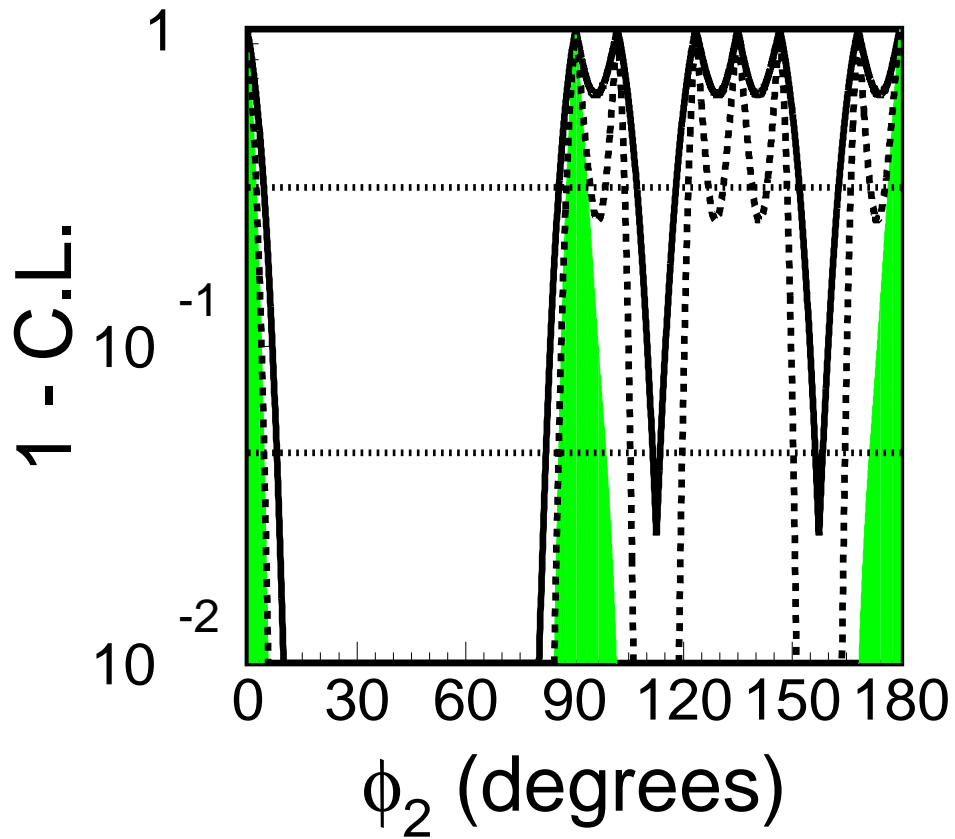

Figure 5.30: Confidence level as a function of $\phi_2$ using $B \to \pi\pi$ decays and isospin relations. The solid and dashed curves show the C.L. with the 5 ab$^{-1}$ and 50 ab$^{-1}$ data without $\mathcal{S}_{\pi^0\pi^0}$, respectively. The filled curve shows the C.L. with the 50 ab$^{-1}$ data and $\mathcal{S}_{\pi^0\pi^0}$. The measurement of $\mathcal{S}_{\pi^0\pi^0}$ is important to reduce the ambiguity among the $\phi_2$ solutions. The two dotted horizontal lines indicate 68.3% ($1\sigma$) and 95.4% ($2\sigma$) C.L.

where
$$s_+ = (p_+ + p_0)^2, \qquad s_- = (p_- + p_0)^2 \qquad (5.46)$$

are the Dalitz plot variables, and $p_+$, $p_-$ and $p_0$ are the four-momenta of $\pi^+$, $\pi^-$ and $\pi^0$, respectively. The amplitudes $A_{3\pi}$ and $\overline{A}_{3\pi}$ are

$$A_{3\pi} = \sum_{\kappa=(+,-,0)} f_\kappa(s_+, s_-) A^\kappa, \qquad \overline{A}_{3\pi} = \sum_{\kappa=(+,-,0)} \overline{f}_\kappa(s_+, s_-) \overline{A}^\kappa, \qquad (5.47)$$

where $A^\kappa$ ($\overline{A}^\kappa$) are complex amplitudes corresponding to $B^0$ ($\overline{B}^0$) $\to \rho^+\pi^-$, $\rho^-\pi^+$, $\rho^0\pi^0$ for $\kappa = +, -, 0$. The function $f_\kappa$ ($\overline{f}_\kappa$) determines the kinematic and dynamical properties of the Dalitz plot of the $B^0$ ($\overline{B}^0$) $\to (\rho\pi)^0$ decays. Using Eq. (5.45) to describe the 3-dimensional distribution one can determine all the magnitudes and relative phases of $A^\kappa$ and $\overline{A}^\kappa$. The six amplitudes are related with $\phi_2$ as

$$e^{+2i\phi_2} = \frac{\overline{A}^+ + \overline{A}^- + \overline{A}^0}{A^+ + A^- + A^0}. \qquad (5.48)$$

Therefore in the limit of high statistics we can determine $\phi_2$ without discrete ambiguities. Belle [121] and BaBar [122] have performed Dalitz plot analyses, and have obtained mild $1\sigma$ constraints for $\phi_2$.

The $\phi_2$ constraints from the $B \to \rho\pi$ decays at present (approximately 1 ab$^{-1}$ of data with Belle and BaBar combined) are not so restrictive compared to those from $B \to \pi\pi$ and $B \to \rho\rho$ decays. With a large data sample at a super $B$ factory, the Dalitz plot analysis is in principle able to pin down a unique $\phi_2$ solution. To this end, we need a better understanding of the contributions from the radially excited $\rho$ states, $\rho(1450)$ and $\rho(1700)$, and from other decay modes, such as $B^0 \to f_0(980)\pi^0$, $\sigma\pi^0$, $\omega\pi^0$ and non-resonant $\pi^+\pi^-\pi^0$, which will be dominant sources of systematic uncertainties. The contributions of the $B^0 \to \pi^+\pi^-\pi^0$ decays other than $(\rho\pi)^0$ are expected to be measured using the Dalitz plot analysis with a large data sample.

For the sensitivity study of the $\phi_2$ extraction using the $B^0 \to \pi^+\pi^-\pi^0$ decay channel, we employ the $CP$-violating parameter values based on the Belle's measurement with a small modification to have a $\phi_2$ solution at 90°. The size of statistical errors is scaled by the inverse square root of the integrated luminosity. We assume that the systematic uncertainties caused by the contributions of $B^0$ decays other than $(\rho\pi)^0$ are reduced to about a third of the values in Ref. [121].

Figure 5.31 shows the obtained C.L. as a function of $\phi_2$ using the combination of the Dalitz plot and the pentagon isospin relations [123]. As shown in the figure, the $\phi_2$ is determined almost uniquely with the 50 ab$^{-1}$ data. The precision of the $\phi_2$ measurement is ±3° (±1.5°) with the 5 ab$^{-1}$ (50 ab$^{-1}$) data.

The electroweak isospin breaking effect in $B \to \rho\pi$ decays is identical to that in $B \to \pi\pi$ and $B \to \rho\rho$ decays. While the isospin breaking in the tree amplitudes has no effect to the $\phi_2$ extraction in the time-dependent Dalitz plot analysis, there is the isospin breaking effect in the penguin amplitudes, mainly caused by $\pi^0$-$\eta$-$\eta'$ mixing. Since the magnitude of the penguin amplitudes are small in the $B \to \rho\pi$ decays, the isospin breaking effect is expected to be small ($< 1°$) [113].

### 5.7.3 $\phi_2$ extraction with $B \to \rho\rho$ decays

The $B \to \rho\rho$ decay contains two vector particles in the final state. Thus, for the measurements of the $CP$-violating parameters in $B^0 \to \rho^+\rho^-$ decays, both $CP$-even and $CP$-odd components have to be taken into account. For each of the $\rho\rho$ helicity amplitudes the same isospin relations are valid as for the $\pi\pi$ final states. The polarization measurements of $B \to \rho\rho$ decays, however, show that the final $\rho\rho$ state is almost purely longitudinally polarized [124–128], indicating that the final state is an almost pure $CP$-even state. This simplifies the isospin analysis by reducing it to the $CP$-even, longitudinally polarized component only. The BaBar collaboration finds evidence for $B^0 \to \rho^0\rho^0$ signal events with a 3.5$\sigma$ significance; the branching fraction is measured to be $\mathcal{B}(B^0 \to \rho^0\rho^0) = (1.07 \pm 0.33 \pm 0.19) \times 10^{-6}$ [128], which is much smaller than those of $B^0 \to \rho^+\rho^-$ and $B^+ \to \rho^+\rho^0$ decays as shown in Table 5.16. Belle uses an even larger data set [129], but observes no signal and puts an upper limit of $\mathcal{B}(B^0 \to \rho^0\rho^0) \leq 1 \times 10^{-6}$ at 90% C.L. Since the tree contribution to $B^0 \to \rho^0\rho^0$ is colour suppressed the decay is sensitive to possible penguin contribution. The low measured rates imply that the penguin pollution in the $B \to \rho\rho$ decays is much smaller than that in the $B \to \pi\pi$ decays. Therefore the $B \to \rho\rho$ decay is a promising channel for measuring $\phi_2$, provided that we measure the $CP$-violating parameters of $B^0 \to \rho^0\rho^0$ decay.

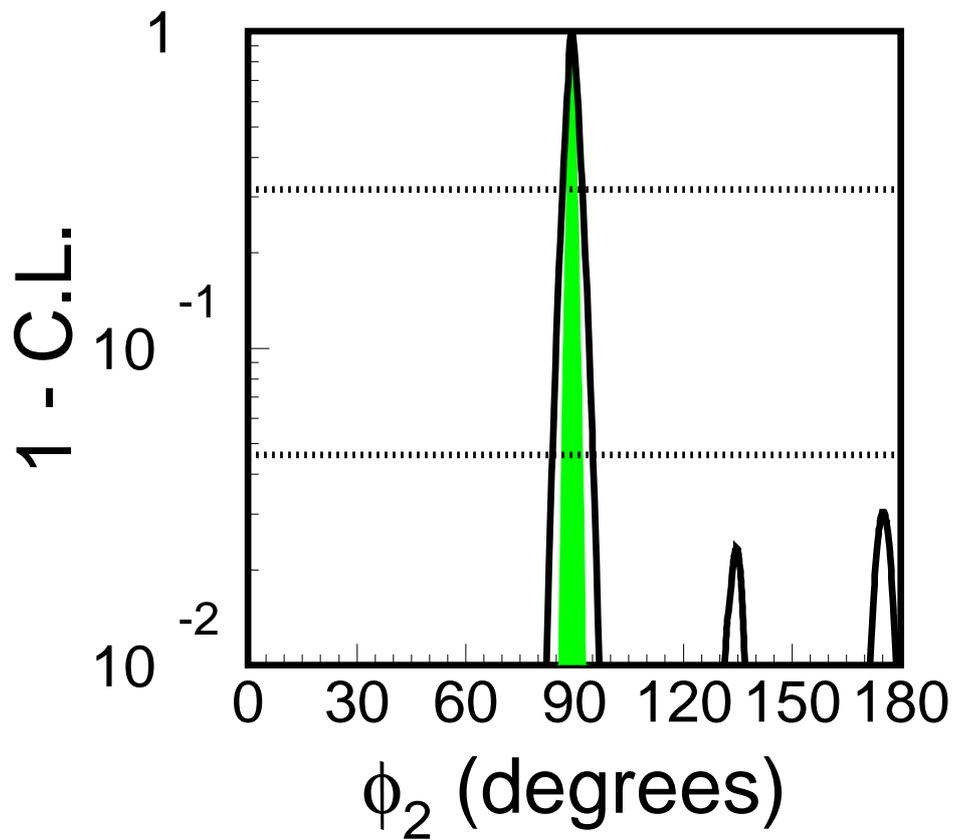

Figure 5.31: Confidence level as a function of $\phi_2$ using the $B \to \rho\pi$ Dalitz plot analysis and isospin relations. The solid and filled curves show the C.L. with 5 ab$^{-1}$ and 50 ab$^{-1}$ data. The horizontal lines indicate 68.3% ($1\sigma$) and 95.4% ($2\sigma$) C.L.

Table 5.16: The measured central values and errors in $B \to \rho\rho$ decays. The branching fractions are in $10^{-6}$. The errors of W.A. are the total errors. The first and second error for 5 ab$^{-1}$ and 50 ab$^{-1}$ data are the reducible and irreducible error, respectively.

|  | central values | errors of W.A. [22] | errors at 5 ab$^{-1}$ | errors at 50 ab$^{-1}$ |
|---|---|---|---|---|
| $\mathcal{B}(\rho^+\rho^-)$ | 23.1 | $\pm 3.3$ | $\pm 1.1 \pm 2.6$ | $\pm 0.4 \pm 2.6$ |
| $\mathcal{B}(\rho^+\rho^0)$ | 18.2 | $\pm 3.0$ | $\pm 0.8 \pm 1.7$ | $\pm 0.3 \pm 1.7$ |
| $\mathcal{B}(\rho^0\rho^0)$ | 1.16 | $\pm 0.46$ | $\pm 0.13 \pm 0.12$ | $\pm 0.04 \pm 0.12$ |
| $\mathcal{S}_{\rho^+\rho^-}$ | 0.00 | $\pm 0.17$ | $\pm 0.08 \pm 0.03$ | $\pm 0.03 \pm 0.03$ |
| $\mathcal{A}_{\rho^+\rho^-}$ | 0.00 | $\pm 0.13$ | $\pm 0.05 \pm 0.03$ | $\pm 0.02 \pm 0.03$ |
| $\mathcal{S}_{\rho^0\rho^0}$ | 0.00 | N.A. | $\pm 0.31 \pm 0.03$ | $\pm 0.10 \pm 0.03$ |
| $\mathcal{A}_{\rho^0\rho^0}$ | 0.00 | N.A. | $\pm 0.26 \pm 0.03$ | $\pm 0.08 \pm 0.03$ |

To estimate the $\phi_2$ sensitivity with the $B \to \rho\rho$ decays, we assume that the longitudinal polarization fraction is 100% for simplicity. We also assume that there is no direct $CP$ violation in $B \to \rho\rho$ decays since the penguin pollution is small. By choosing $\phi_2 = 90°$, all the $CP$-violating parameters are set to zero. Table 5.16 shows the central values and errors of the branching fractions and $CP$-violating parameters. The statistical errors for $\mathcal{S}_{\rho^0\rho^0}$ and $\mathcal{A}_{\rho^0\rho^0}$ are estimated using a toy MC simulation assuming that the signal-to-noise ratio is 0.22, which is inferred from Ref. [128].

For the $B \to \rho\rho$ decays, there are additional systematic error contributions: the self-cross-feed, unknown $B$ decay background contributions, unknown $CP$ violation of the $B$ decay backgrounds and interferences between $B \to \rho\rho$ decays and other $4\pi$ final states such as $B \to a_1\pi$, $\rho\pi\pi$ and $\pi\pi\pi\pi$. We also take into account the uncertainty of the longitudinal fraction; its error is expected to be 1% with a large data sample. We assume that the uncertainties of the interferences are irreducible, and other uncertainties are reducible. We assign 5% for the irreducible error on the branching fraction, and $\pm 0.03$ for the $CP$ violating parameters in $B^0 \to \rho^+\rho^-$ and $\rho^0\rho^0$ decays.

Figure 5.32 shows the obtained confidence level of the $\phi_2$ constraints based on SU(2) isospin relations. The 1$\sigma$ error for the solution at $\phi_2 = 90°$ is $\pm 3°$ ($\pm 1.5°$) for the 5 ab$^{-1}$ (50 ab$^{-1}$) data assuming the isospin relations.

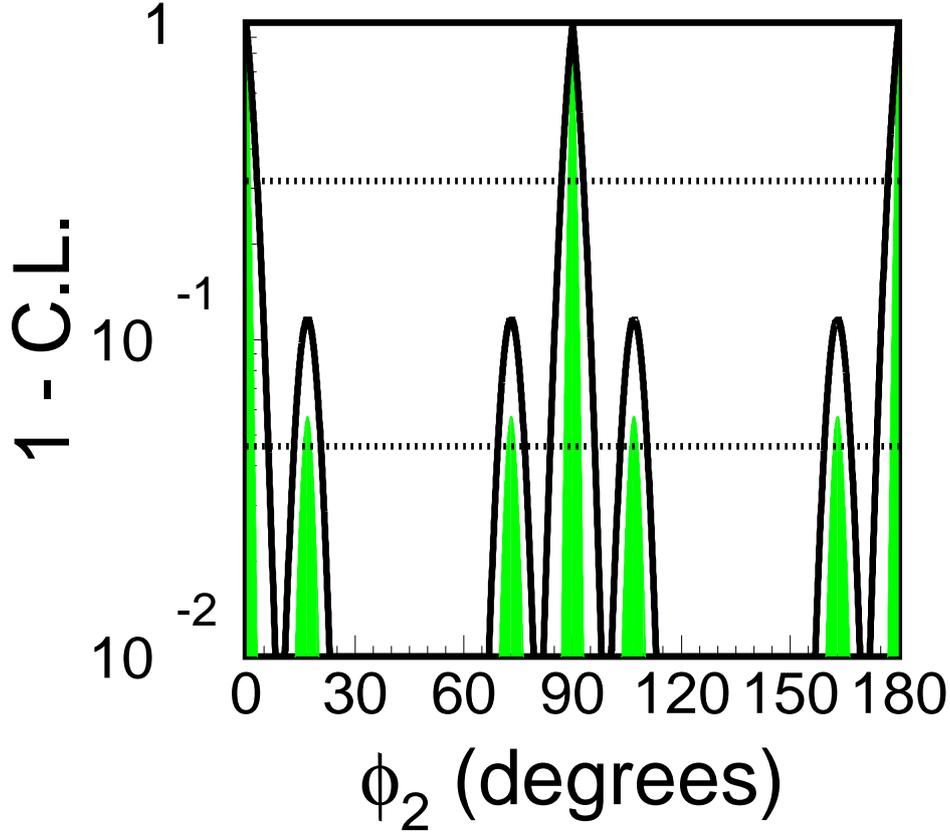

Figure 5.32: Confidence level as a function of $\phi_2$ using $B \to \rho\rho$ decays and isospin relations. The solid and filled curves show the C.L. with the 5 ab$^{-1}$ and 50 ab$^{-1}$ data, respectively. The two dotted horizontal lines indicate C.L.=68.3% ($1\sigma$) and 95.4% ($2\sigma$).

While the electroweak penguin isospin breaking effect in $B \to \rho\rho$ decays is exactly the same as in $B \to \pi\pi$ decays, the $\rho$-$\omega$ mixing of < 2% [112, 113] and the possible existence of an $I = 1$ final state with a fraction $\mathcal{O}(\Gamma_\rho^2/m_\rho^2)$ [130] also contribute to the theoretical uncertainties of the $\phi_2$ extraction. The total uncertainty from those effects is comparable to the $\phi_2$ measurement precision at the high-luminosity $B$ factory.

### 5.7.4 Status of $B \to a_1\pi$ decays

The final state of $B^0 \to a_1^+\pi^-$ decays contains a vector and a pseudoscaler particle, like $B \to \rho\pi$. Hence one would think of the same procedure as for $B \to \rho\pi$ decays to extract the $\phi_2$. However, the $B^0 \to (a_1\pi)^0 \to \pi\pi\pi^0\pi^0$ decays have no overlap region between the $a_1^\pm$ and $a_1^0$ resonance bands, and the interferences with the $B^0 \to \rho^+\rho^-$ decays having the same final state would complicate the analysis. In Ref. [131], the authors propose to measure the time-dependent $CP$-violating parameters with the quasi-two-body decay approximation of $B^0 \to a_1^\pm\pi^\mp$ decays with four charged pions in the final state. The parameters are used to estimate the effective $\phi_2$ ($\phi_2^{\text{eff}}$). The bound of $\Delta\phi_2 = \phi_2^{\text{eff}} - \phi_2$ can be obtained using SU(3) relations between the branching fraction of $B^0 \to a_1^+\pi^-$ and the branching fractions of $B \to a_1 K$, $B \to K_1(1270)\pi$ and $B \to K_1(1400)\pi$ decays.

BaBar collaboration has reported the first measurement of the branching fraction, $\mathcal{B}(B^0 \to$

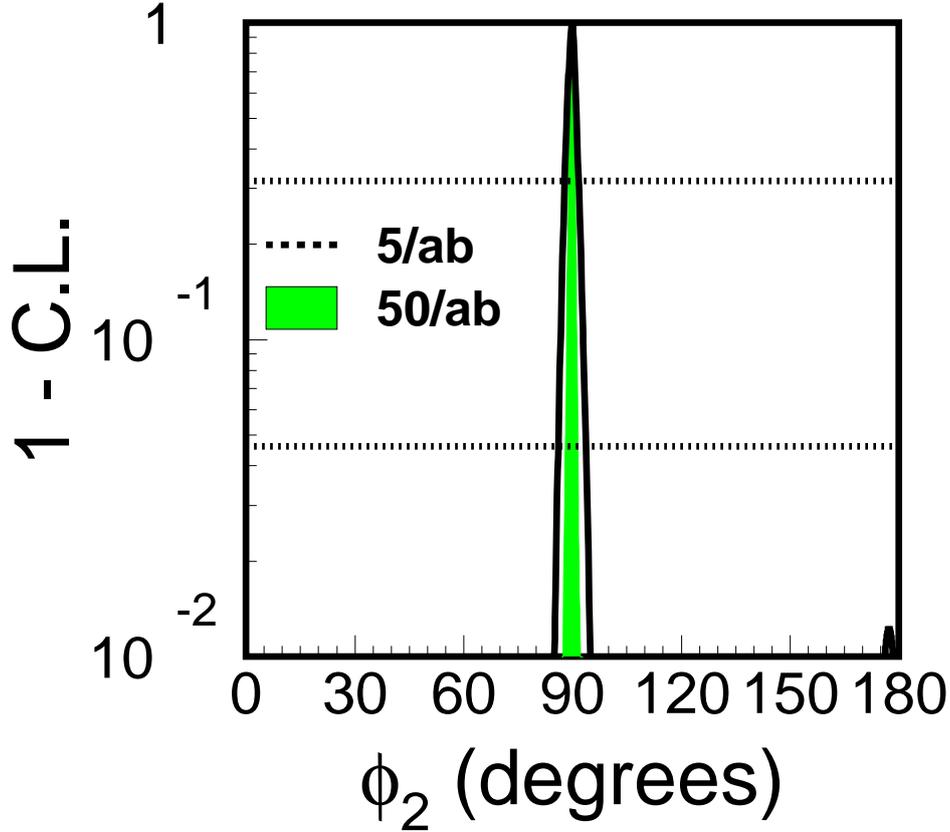

Figure 5.33: Confidence level as a function of $\phi_2$ using $B \to \pi\pi$, $\rho\pi$ and $\rho\rho$ decays. The solid and filled curves show the C.L. with the 5 ab$^{-1}$ and 50 ab$^{-1}$ data, respectively. The two dotted horizontal lines indicate C.L.=68.3% ($1\sigma$) and 95.4% ($2\sigma$).

$a_1^+\pi^-) = (33.2 \pm 3.8 \pm 3.0) \times 10^{-6}$ [132], and time-dependent $CP$-violation parameters with $\phi_2^{\text{eff}} = (78.6 \pm 7.3)°$ [133]. Conservatively assuming that roughly half of the systematic uncertainty scales with the luminosity (mainly uncertainties due to the unknown level of $CP$ violation in the $B\bar{B}$ background and the uncertainty due to the interference between $a_1\pi$ and other $4\pi$ final states; these uncertainties will be reduced with more available data) one can expect a measurement of $\phi_2^{\text{eff}}$ with a precision of around 2°-3° with an integrated luminosity of 5 ab$^{-1}$ using this quasi two-body decay mode.

### 5.7.5 Summary of the $\phi_2$ measurements

We review the prospect of the $\phi_2$ constraints with the $B \to \pi\pi$, $\rho\pi$ and $\rho\rho$ decays at a high luminosity $B$ factory. The $\phi_2$ measurements with the $B \to \pi\pi$ and $\rho\rho$ decays provide an accuracy of a few degrees, comparable with the isospin breaking effect, but a two fold ambiguity remains from the isospin analysis. On the other hand, the $B \to \rho\pi$ decays yield a single solution of $\phi_2$ with a precision of one degree, provided that we reduce the systematic uncertainties caused by the contribution of $B^0$ decays other than $(\rho\pi)^0$ to about a third of the current values. Figure 5.33 shows the combined $\phi_2$ constraints using $B \to \pi\pi$, $\rho\pi$ and $\rho\rho$ decays. The measurement error with the 5 ab$^{-1}$ (50 ab$^{-1}$) data sample is 2° ($\lesssim 1°$).

## 5.8 $\phi_3$

### 5.8.1 Introduction

The CKM weak phase $\phi_3$ is defined as

$$\phi_3 \equiv -\arg\left[\frac{V_{ud}V_{ub}^*}{V_{cd}V_{cb}^*}\right], \quad (5.49)$$

which is independent of the quark phase convention. In the standard phase convention, all elements of the CKM matrix except for $V_{ub}$ and $V_{td}$ are nearly real, and in particular both $V_{ud}$ and $-V_{cd}V_{cb}^*$ are (nearly) real and positive, and we have

$$\phi_3 \equiv \arg V_{ub}^* \quad \text{(standard phase convention)}. \quad (5.50)$$

As pointed out by Gronau and London, one can use the interference between $b \to c\bar{u}s$ and $b \to u\bar{c}s$ transitions to access the angle $\phi_3$. In the case of charged $B \to D^{(*)}K^{(*)}$ decays, where the neutral $D^{(*)}$ meson produced is an admixture of $D^{(*)0}$ (produced by a $b \to c$ transition) and $\bar{D}^{(*)0}$ (produced by a colour-suppressed $b \to u$ transition) states, the final state $f$ is chosen so that both $D^{(*)0}$ and $\bar{D}^{(*)0}$ can contribute. The two amplitudes interfere, and the resulting observables are sensitive to $\phi_3$. There is no penguin contribution, so almost no theoretical uncertainties.

Various methods have been proposed to exploit this interference, the final state $f$ can be a $CP$ eigenstate (e.g. $K^+K^-$, $K_S^0\pi^0$) (GLW method [134, 135]), a suppressed final state (ADS method [136, 137]), or a self-conjugate three-body final state, such as $K_S^0\pi^+\pi^-$ (Dalitz method [138, 139]). All these methods are limited by statistics but have common $B$ parameters, so combining them will bring more than statistics.

The flavor-tagged time-dependent measurement of $D^{(*)-}\pi^+$ and its charge-conjugate mode measure this time $\sin(2\phi_1 + \phi_3)$ and is also affected by a strong phase. One can fully reconstruct the $D^-\pi^+$ and $D^{*-}\pi^+$ final states, or use a partial-reconstruction technique where $\bar{D}^0$ of the decay $D^{*-} \to \bar{D}^0\pi^+$ is not explicitly detected. The latter has more statistics but suffers from a larger background. In both cases, the value of $r$, the ratio of Cabibbo-favored amplitude to the Cabibbo-suppressed amplitude, needs to be input.

More on the theoretical frameworks for each mode as well as the sensitivities are covered in the following subsections. The general strategy of the $\phi_3$ measurement is to perform a simultaneous fit which includes the results from the different methods listed above. The parameters, such as amplitude ratio $r_B$, are common and can be constrained using both ADS and Dalitz methods, and thus improve the $\phi_3$ accuracy.

### 5.8.2 The GLW and ADS methods

In the $B^\pm \to DK^\pm$ decay, the number of $c$ or $\bar{c}$ quark in the final state is one, and as a result the penguin processes $b \to s/d$ cannot contribute. When the neutral $D$ meson is detected in a final state that can come from $D^0$ or $\bar{D}^0$, the two processes $B^- \to D^0K^-$ and $B^- \to \bar{D}^0K^-$ interfere. Diagrams contributing to $B^\pm \to DK^\pm$ are shown in Figure 5.34. They are categorized in terms of the CKM factors involved ($\lambda_c = V_{cb}V_{us}^*$ and $\lambda_u = V_{ub}V_{cs}^*$) and type of processes (T: color-favored tree, C: color-suppressed tree, A: annihilation). $B^- \to D^0K^-$ receives contributions from T and C both with the CKM factor $\lambda_c$, while $B^- \to \bar{D}^0K^-$ can proceed by C and A both with the CKM factor $\lambda_u$. Thus,

$$\arg\frac{A(B^- \to \bar{D}^0K^-)}{A(B^- \to D^0K^-)} = \delta_B - \phi_3, \quad \arg\frac{A(B^+ \to D^0K^-)}{A(B^+ \to \bar{D}^0K^-)} = \delta_B + \phi_3, \quad (5.51)$$

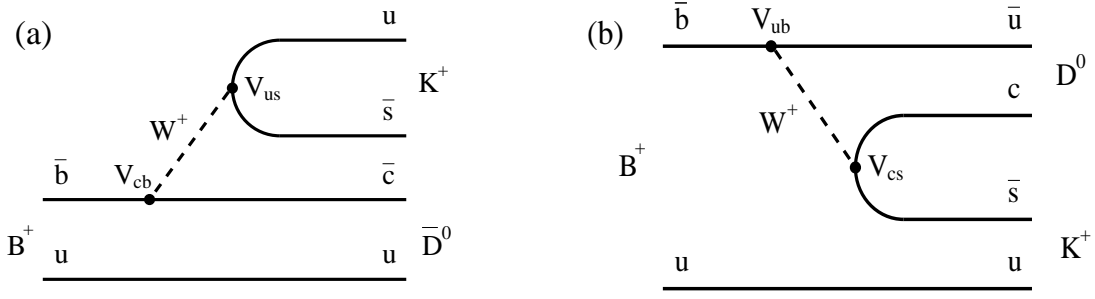

Figure 5.34: Diagrams contributing to the $B^\pm \to DK^\pm$ decays.

where $\delta_B$ is the strong phase difference. We used the notation

$$\arg \frac{\lambda_u}{\lambda_c} = \arg \frac{V_{ub} V_{cs}^*}{V_{cb} V_{us}^*} = -\phi_3, \quad \text{(standard phase convention)} \tag{5.52}$$

and took into account the change of the weak phase sign for charge conjugated decay (strong phase sign is not changed). The absolute size of $A(B^- \to \bar{D}^0 K^-)$ or $A(B^- \to D^0 K^-)$ is the same for charge conjugate decays:

$$|A(B^- \to \bar{D}^0 K^-)| = |A(B^+ \to D^0 K^+)| \equiv B, \tag{5.53}$$
$$|A(B^- \to D^0 K^-)| = |A(B^+ \to \bar{D}^0 K^+)| \equiv A. \tag{5.54}$$

If $D^0$ and $\bar{D}^0$ are detected in a $CP$ eigenstate such as $K^- K^+$ (which is $CP$-even mode; i.e. $D_1 = (D^0 + \bar{D}^0)/\sqrt{2}$), the decay rates of $B^\pm$ are, up to an overall constant of $1/2$,

$$\Gamma(B^- \to D_1 K^-) = \left| A + B e^{i(\delta_B - \phi_3)} \right|^2 = A^2 + B^2 + 2AB \cos(\delta_B - \phi_3), \tag{5.55}$$
$$\Gamma(B^+ \to D_1 K^+) = \left| A + B e^{i(\delta_B + \phi_3)} \right|^2 = A^2 + B^2 + 2AB \cos(\delta_B + \phi_3). \tag{5.56}$$

Then, there can be a decay rate asymmetry [134, 135] between $B^- \to D_1 K^-$ and $B^+ \to D_1 K^+$:

$$A_1 \equiv \frac{\Gamma_{D_1 K^-} - \Gamma_{D_1 K^+}}{\Gamma_{D_1 K^-} + \Gamma_{D_1 K^+}} = \frac{-2 r_B \sin \delta_B \sin \phi_3}{1 + r_B^2 + 2 r_B \cos \delta_B \cos \phi_3}, \tag{5.57}$$

with

$$r_B \equiv \frac{B}{A}. \tag{5.58}$$

The value of $r_B$, the ratio of the magnitudes for the two amplitudes, is expected to be around 0.1, due to color suppression and a numerically slightly smaller combination of the CKM elements. For the $CP$-odd final states (such as $K_S \pi^0$; i.e. $D_2 = (D^0 - \bar{D}^0)/\sqrt{2}$), the asymmetry becomes

$$A_2 \equiv \frac{\Gamma_{D_2 K^-} - \Gamma_{D_2 K^+}}{\Gamma_{D_2 K^-} + \Gamma_{D_2 K^+}} = \frac{2 r_B \sin \delta_B \sin \phi_3}{1 + r_B^2 + 2 r_B \cos \delta_B \cos \phi_3}. \tag{5.59}$$

Belle reported a study with $CP$ modes using a limited sample of $275 \times 10^6$ $B\bar{B}$ events [140], getting clear signals (about 140 signal candidates in $D_1$ and $D_2$ cases) but no significant asymmetry.

When the final state is not $CP$ eigenstate ($f \neq \bar{f}$), the decay rates for $B^{\mp} \to D^{(*)}K^{\mp}$, $D^{(*)} \to f$, can be written as

$$\Gamma_{\mp} \propto r_B^2 + r_D^2 + 2r_B r_D \cos(\delta_B + \delta_D \mp \phi_3), \tag{5.60}$$

where the ratio of $D^{(*)}$ decay amplitudes is defined as $r_D = \frac{|A(D^{(*)0} \to f)|}{|A(\bar{D}^{(*)0} \to f)|}$. The strong phase differences between the $B$ and $D^{(*)}$ decay amplitudes are given by $\delta_B$ and $\delta_D$, respectively. As shown in [141] there is an effective strong phase difference of $\pi$ between the $D^0\pi^0$ and $D^0\gamma$ final states. This has a significant impact to the ADS method. In this approach, we measure the ratio for $B$ decays to suppressed and favoured final states: $\mathcal{R} = \frac{\mathcal{B}(B \to D_{sup}K)}{\mathcal{B}(B \to D_{fav}K)}$. If we consider such ratios separately for $B^-$ and $B^+$, for both cases $D^* \to D\pi^0$ and $D^* \to D\gamma$, we see

$$\mathcal{R}_{\mp}(D^* \to D\pi^0) = r_B^2 + r_D^2 + 2r_B r_D \cos(\delta_B + \delta_D \mp \phi_3), \tag{5.61}$$
$$\mathcal{R}_{\mp}(D^* \to D\gamma) = r_B^2 + r_D^2 - 2r_B r_D \cos(\delta_B + \delta_D \mp \phi_3). \tag{5.62}$$

Since there are four independent equations, which contain only three unknowns (the value of $r_D$ is known from $D$ decays), $\phi_3$ can be extracted, up to the usual four-fold ambiguity. Compare this to the standard ADS analysis; for any particular $B$ decay one finds only two independent equations, which cannot be solved for three unknowns.

The charge averaged rates are given by

$$\mathcal{R}(D^* \to D\pi^0) = r_B^2 + r_D^2 + 2r_B r_D \cos(\delta_B + \delta_D)\cos(\phi_3), \tag{5.63}$$
$$\mathcal{R}(D^* \to D\gamma) = r_B^2 + r_D^2 - 2r_B r_D \cos(\delta_B + \delta_D)\cos(\phi_3), \tag{5.64}$$

so that

$$\left(\mathcal{R}(D^* \to D\pi^0) + \mathcal{R}(D^* \to D\gamma)\right)/2 = r_B^2 + r_D^2, \tag{5.65}$$

does not depend on any phase. Hence the value of $r_B$ in $B^- \to D^*K^-$ decays can be straightforwardly obtained.

Figure 5.35a shows the $\Delta E$ distribution for $B^- \to DK^-$, $D \to K^+\pi^-$ using 605 fb$^{-1}$ of Belle data [142]. The detection efficiency was estimated by MC to be 15.4% and no evidence for signal is seen. The limit on $r_B$ was obtained assuming maximal interference ($\phi_3 = 0°, \delta_B = 180°$ or $\phi_3 = 180°, \delta_B = 0°$) and is found to be $r_B < 0.19$ at 90% confidence level, as shown in Fig. 5.35b.

### 5.8.3 The Dalitz analysis

A Dalitz plot analysis of a three-body final state of the $D$ meson allows one to obtain all the information required for determination of $\phi_3$ in a single decay mode. The use of a Dalitz plot analysis for the extraction of $\phi_3$ was first discussed by D. Atwood, I. Dunietz and A. Soni, in the context of the ADS method [136, 137]. This technique uses the interference of Cabibbo-favored $D^0 \to K^-\pi^+\pi^0$ and doubly Cabibbo-suppressed $\bar{D}^0 \to K^-\pi^+\pi^0$ decays. However, the small rate for the doubly Cabibbo-suppressed decay limits the sensitivity of this technique. Three body final state $K_S^0\pi^+\pi^-$ [138, 139] has been suggested as a promising mode for the extraction of $\phi_3$.

If an amplitude of $D^0$ decaying to a point in the Dalitz plot, $m^2(K_S^0\pi^+) = m_+^2$ and $m^2(K_S^0\pi^-) = m_-^2$, is given by $f(m_+^2, m_-^2)$, then an amplitude of $\bar{D}^0$ decaying to the same final state can be written as $f(m_-^2, m_+^2)$ where the two arguments are simply exchanged. This is due to the approximate $CP$ conservation in the $D$ decay. Explicitly,

$$A(D^0)(m_{K_S^0\pi^+} = m_+, \; m_{K_S^0\pi^-} = m_-) \equiv f(m_+^2, \; m_-^2)$$
$$A(\bar{D}^0)(m_{K_S^0\pi^+} = m_+, \; m_{K_S^0\pi^-} = m_-) \equiv f(m_-^2, \; m_+^2) \tag{5.66}$$

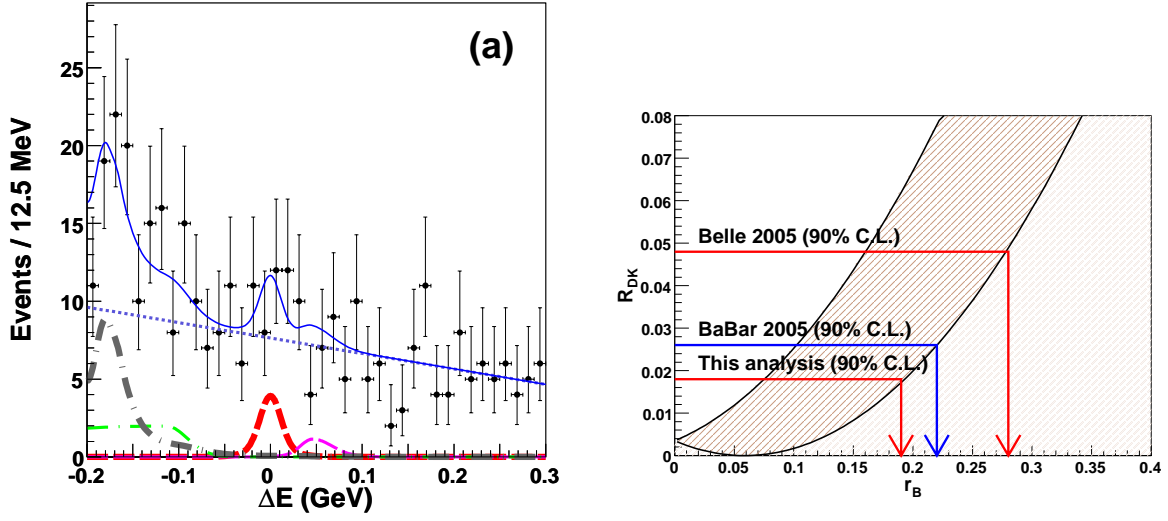

Figure 5.35: (a) $\Delta E$ distribution for $B^- \to D[K^+\pi^-]K^-$ using a Belle data sample of 605 fb$^{-1}$, (b) the corresponding $r_B$ constraint, $r_B < 0.19$ at 90% CL.

The total decay amplitude for $B^- \to \tilde{D}K^-$ followed by $\tilde{D} \to K_S^0\pi^+\pi^-$ is then

$$\begin{aligned} M_- &= A(D^0_{\to K_S^0\pi^+\pi^-}K^-) + A(\bar{D}^0_{\to K_S^0\pi^+\pi^-}K^-) \\ &= A[f(m_+^2, \; m_-^2) + r_B e^{i(\delta_B - \phi_3)} f(m_-^2, \; m_+^2)], \end{aligned} \tag{5.67}$$

where $A$, $r_B$, $\delta_B$ are defined in the previous section.

The total decay amplitude for $B^+ \to \tilde{D}K^+$, $\tilde{D} \to K_S^0\pi^+\pi^-$, where $m^2(K_S^0\pi^+) = m_+^2$ and $m^2(K_S^0\pi^-) = m_-^2$, is similarly

$$\begin{aligned} M_+ &= A(\bar{D}^0_{\to K_S^0\pi^+\pi^-}K^+) + A(D^0_{\to K_S^0\pi^+\pi^-}K^+) \\ &= A[f(m_-^2, \; m_+^2) + r_B e^{i(\delta_B + \phi_3)} f(m_+^2, \; m_-^2)], \end{aligned} \tag{5.68}$$

where the sign of $\phi_3$ has flipped while the strong phase $\delta_B$ remains the same.

The Dalitz plot for $D^0 \to K_S^0\pi^+\pi^-$ based on the Belle measurement [143] is shown in Fig. 5.36. As can be seen from the Dalitz plot, the distribution is highly asymmetric under the exchange of $m_{K_S^0\pi^+}$ and $m_{K_S^0\pi^-}$ which indicates that the interference in the Dalitz plot has a good sensitivity on the phase between the two terms of (5.67) and (5.68). The phases $\delta_B - \phi_3$ and $\delta_B + \phi_3$ are obtained from the separate fits for $B^-$ and $B^+$ decays; $\phi_3$ can be extracted from them.

Figure 5.37 shows the $\Delta E$ and $M_{bc}$ distributions for $B^\pm \to DK^\pm$, $D \to K_S^0\pi^+\pi^-$, based on 605 fb$^{-1}$ of data, and the Dalitz plots of the $D$ decays are shown in Fig. 5.38 separately for $B^+$ and $B^-$ [143]. The results of an unbinned maximum likelihood fit for the parameters $r_B$, $\delta_B$ and $\phi_3$ are shown in Fig. 5.39.

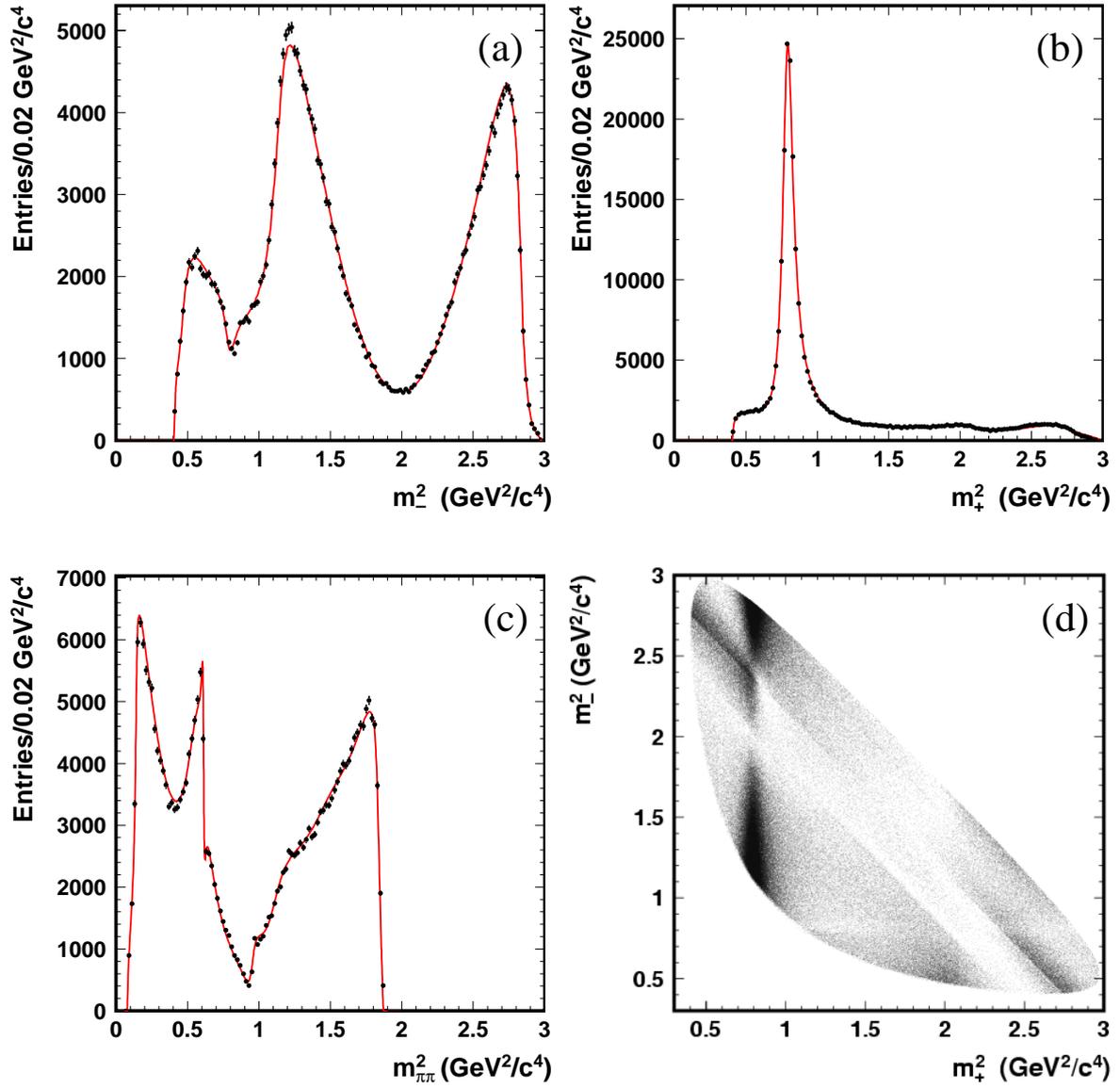

Figure 5.36: (a) the $K^0_S\pi^-$, (b) $K^0_S\pi^+$, (c) $\pi^+\pi^-$ invariant masses and (d) the Dalitz plot for $D^0 \to K^0_S\pi^+\pi^-$ based on the Belle measurement.

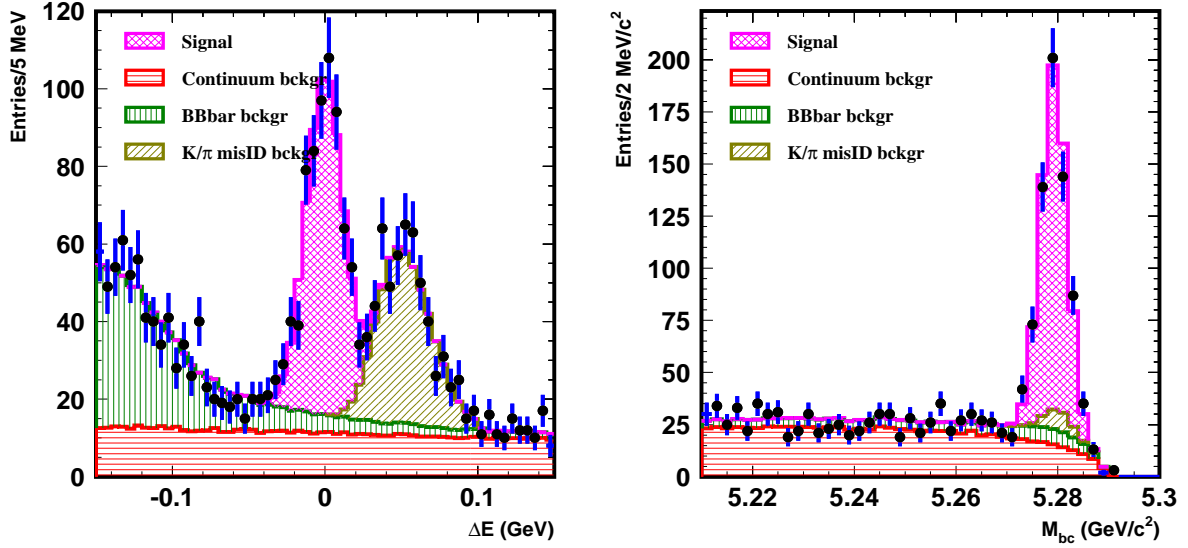

Figure 5.37: The $\Delta E$ and $M_{bc}$ distributions for $B^{\pm} \to DK^{\pm}$, $D \to K_S^0 \pi^+ \pi^-$, based on 605 fb$^{-1}$ of data.

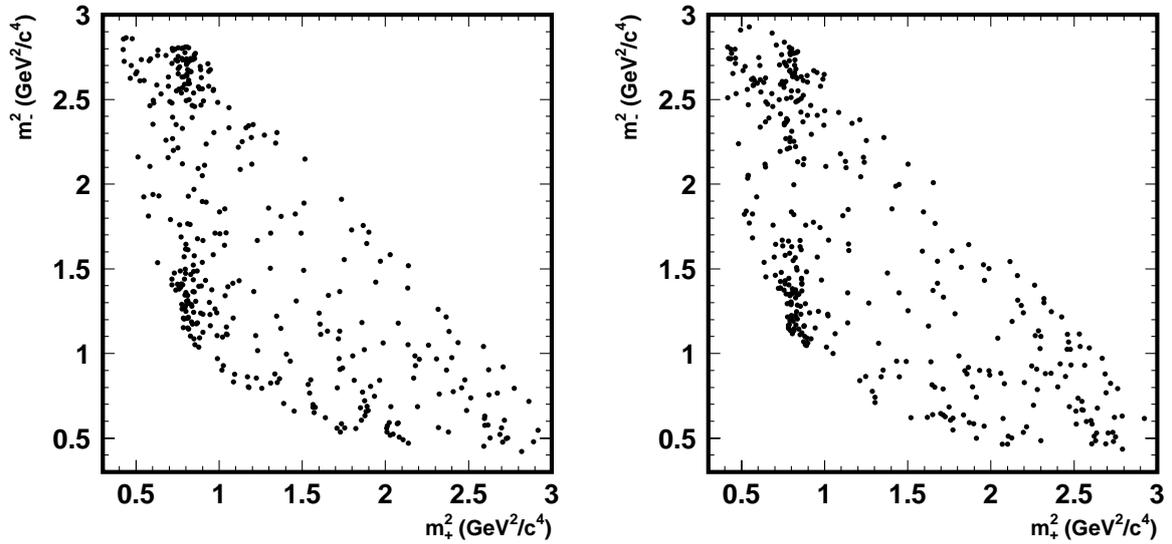

Figure 5.38: Dalitz distributions of $\bar{D}^0 \to K_S^0 \pi^+ \pi^-$ decays from selected $B^+ \to DK^+$ (left) and $B^- \to DK^-$ (right) candidates, based on a data sample of 605 fb$^{-1}$ at Belle.

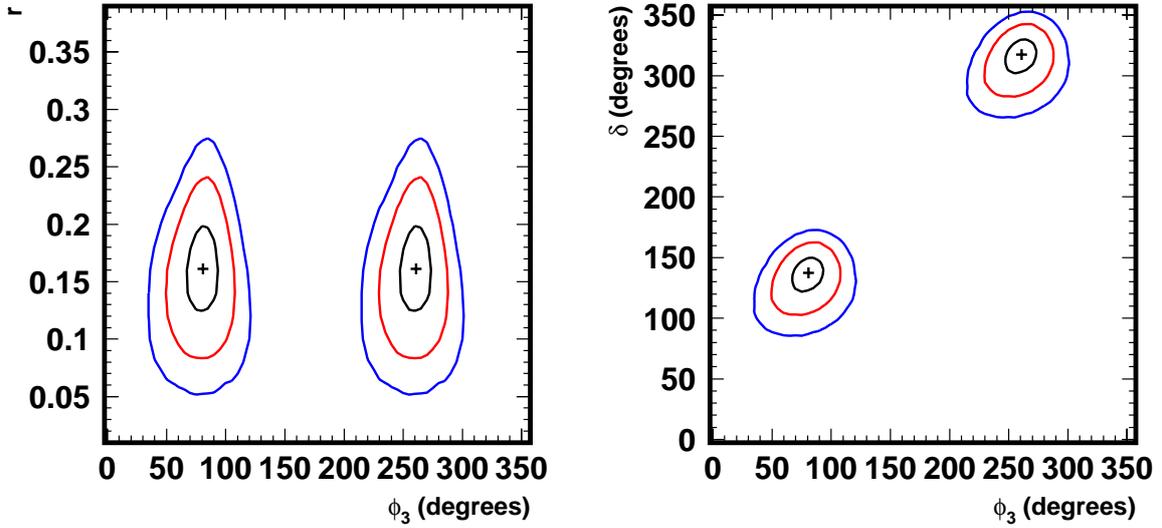

Figure 5.39: Projections of confidence regions for the $B^+ \to DK^+$ mode on the $(r_B, \phi_3)$ and $(\delta_B, \phi_3)$ planes with a data sample of 605 fb$^{-1}$ at Belle.

A simultaneous fit with $DK$, $D^*K$ ($D^* \to D\pi^0$ and $D^* \to D\gamma$) final states yields $\phi_3 = (78.4\ ^{+10.8}_{-11.6}(\text{stat}) \pm 3.6(\text{syst}) \pm 8.9(\text{model}))°$, where the first error is statistical, the second are experimental systematic errors, and the third error is additional systematic errors due to the $D$ decay model dependence. The other parameters, $r_B$ and $\delta_B$, are summarized in the Table 5.17.

|  | $D^0 K^-$ | $D^{*0} K^-$ |
| --- | --- | --- |
| $r_B$ | $0.160^{+0.040}_{-0.038} \pm 0.011^{+0.050}_{-0.010}$ | $0.196^{+0.072}_{-0.069} \pm 0.012^{+0.062}_{-0.012}$ |
| $\delta_B$ (°) | $136.7^{+13.0}_{-15.8} \pm 4.0 \pm 22.9$ | $341.9^{+18.0}_{-19.6} \pm 3.0 \pm 22.9$ |

Table 5.17: $r_B$ and $\delta_B$ parameters obtained for the $DK$, $D^*K$ ($D^* \to D\pi^0$ and $D^* \to D\gamma$) final states.

### 5.8.4 $\phi_3$ sensitivity with 50 ab$^{-1}$

Concerning the Dalitz method, statistical and experimental systematic errors will be reduced significantly with luminosity. However the total uncertainty will be limited by the model error of the order of 10°. A new method [138, 144] was suggested to overcome this problem. It has been shown that the $D$ model uncertainty can be eliminated using the sample of neutral $D$ mesons decaying into $CP$ eigenstates. This sample can be obtained at charm factories, such as CLEO-c [145] that is currently taking data, and the BESIII/BEPCII complex [146], which is an upgrade program at Beijing; these experiments provide copious decays of $\psi(3770)$ mesons into two neutral $D$ mesons.

The density of $D^0 \to K_S^0 \pi^+ \pi^-$ Dalitz plot is given by the absolute value of the amplitude $f_D$ squared: $p_D = p_D(m_+^2, m_-^2) = |f_D(m_+^2, m_-^2)|^2$. In the limit of negligible $CP$-violation in the $D$ system the Dalitz plot for $\bar{D}^0$ is described by $\bar{p}_D(m_+^2, m_-^2) = p_D(m_-^2, m_+^2)$. The density of the $D$ decay Dalitz plot from $B \to DK$ process is expressed as $p_{B\pm} = |f_D + r_B e^{\delta_B \pm \phi_3} \bar{f}_D|^2 = p_D + r_B^2 \bar{p}_D + 2\sqrt{p_D \bar{p}_D}(x_\pm c + y_\pm s)$, where $x_\pm, y_\pm$ include the $\phi_3$, strong phase $\delta_B$ and amplitude ratio $r_B$: $x_\pm = r_B \cos(\delta_B \pm \phi_3)$, $y_\pm = r_B \sin(\delta_B \pm \phi_3)$. The functions $c$ and $s$ are the cosine and sine of the strong phase difference $\Delta \delta_D$ between the symmetric Dalitz plot points: $c = \cos(\delta_D(m_+^2, m_-^2) - \delta_D(m_-^2, m_+^2)) = \cos \Delta \delta_D$, $s = \sin(\delta_D(m_+^2, m_-^2) - \delta_D(m_-^2, m_+^2)) = \sin \Delta \delta_D$. The phase difference $\Delta \delta_D$ can be obtained from the sample of $D$ meson decays to $CP$-eigenstate, for example $K_S \pi^+ \pi^-$. The Dalitz plot density of such a decay is $p_{CP} = |f_D \pm \bar{f}_D|^2 = p_D + \bar{p}_D \pm 2\sqrt{p_D \bar{p}_D} c$ (the normalization is arbitrary). Another possibility is to use a sample, where both $D$ mesons (we denote them as $D$ and $D'$) from $\psi(3770)$ meson decay into $K^0 \pi^+ \pi^-$ state. The four-dimensional density of two correlated Dalitz plots is $p_{\text{corr}}(m_+^2, m_-^2, m_+'^2, m_-'^2) = |f_D f'_D - f'_D \bar{f}_D|^2 = p_D \bar{p}'_D + \bar{p}_D p'_D - 2\sqrt{p_D \bar{p}_D p'_D \bar{p}'_D}(cc' + ss')$, This decay is sensitive to both the $c$ and $s$.

The Dalitz plot is divided into $2\mathcal{N}$ bins placed symmetrically w.r.t. $m_-^2 \leftrightarrow m_+^2$ transformation. Then the expected number of events in the bins of the Dalitz plot from $B \to DK$ decay is $\langle N_i \rangle = h_B[K_i + r_B^2 K_{-i} + 2\sqrt{K_i K_{-i}}(xc_i + ys_i)]$, where $K_i$ is the number of events in the bins in the Dalitz plot of the flavour eigenstate of $D^0$, $h_B$ is the normalization constant. The coefficients $K_i$ are obtained precisely from very large sample of $D^0$ decays in flavor eigenstate, which is accessible at $B$-factories. Coefficients $c_i$ and $s_i$ are determined in charm data analysis. Then the fit is performed, extracting $x$ and $y$. The $\phi_3$, $r_B$ and $\delta_B$ are obtained from $x$ and $y$.

The feasibility study [144, 147] shows that statistical accuracy of the model-independent binned approach is only 10-30% worse compared to the unbinned technique.

Using $\sim 10^4$ tagged $D_{CP}$ decays, corresponding to $\sim 15$ fb$^{-1}$ of integrated luminosity collected at the $\psi(3770)$ peak, the errors on $x, y$ observables would be $\sigma_{x,y} \sim 0.0025$, whereas the irreducible experimental systematic is estimated to be $\sigma_{x,y} \sim 0.003$. The current statistical errors (with 605 fb$^{-1}$) are scaled for the expected sample of 50 ab$^{-1}$: $\sigma_{x,y}(DK) \sim 0.005$, $\sigma_{x,y}(D^*K) \sim 0.013$ and $\sigma_{x,y}(DK^*) \sim 0.016$. Assuming a conservative value of $r_B$ of 0.1 and the central value of $\phi_3$, toy MC samples were generated and for each one the estimation of $\phi_3$ was obtained (as illustrated in Fig. 5.40).

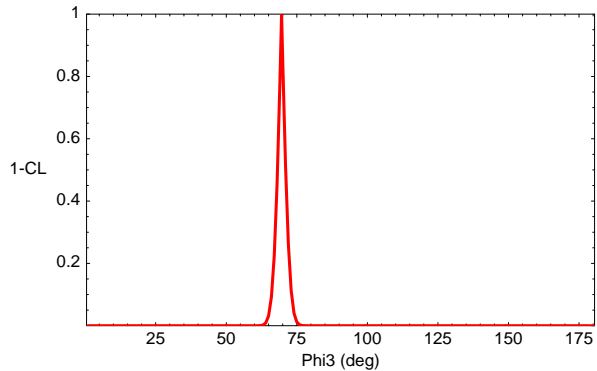

Figure 5.40: $\phi_3$ result obtained for one Dalitz experiment assuming the errors described in the text.

The average $\phi_3$ error obtained is $\sim 2°$.

The statistical error of this measurement will be dominating and can be improved by using

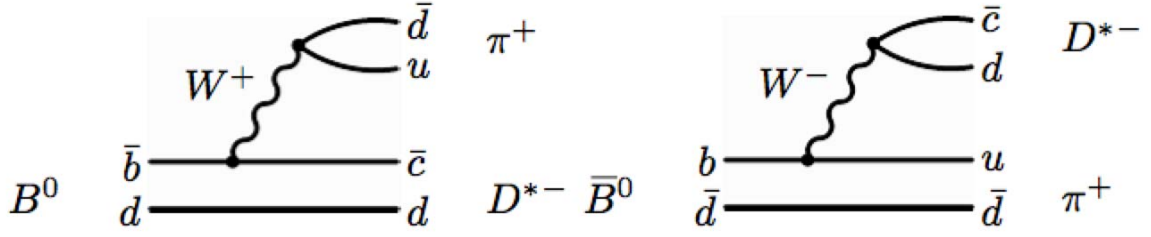

Figure 5.41: Diagrams for (left) $B^0 \to D^{*-}\pi^+$ and (right) $\bar{B}^0 \to D^{*-}\pi^+$. Those for $\bar{B}^0 \to D^{*+}\pi^-$ and $B^0 \to D^{*+}\pi^-$ can be obtained by charge conjugation.

other $D$ decay channels: $K_S^0 K^+ K^-$, $\pi^+\pi^-\pi^0$ and $K_S^0 \pi^+\pi^-\pi^0$.

The above result of the Dalitz method can be combined with ADS and GLW results. For the latter the experimental systematic error is estimated to be $\sigma_{A,R} \sim 0.01$, dominated by the detector charge asymmetry, whereas the statistical errors are scaled from the results of Section 5.8.2. Under these assumptions, the combination of methods yields an error on the angle $\phi_3$ of about 1.5°.

### 5.8.5 $D^{(*)-}\pi^+$ modes

For the final state $D^{(*)-}\pi^+$, the amplitudes of diagrams shown in Fig. 5.41 interfere. The two diagrams contributing to the amplitude $a$ (or $\bar{b}$) have the same CKM factors, and information on $\phi_3$ is contained in

$$\rho \equiv \frac{q\bar{b}}{pa} = -r\exp(\delta - \phi_w), \quad \phi_w \equiv 2\phi_1 - \phi_3, \tag{5.69}$$

where $B_H = pB^0 - q\bar{B}^0$, $r \equiv |\rho|$ and $\delta$ is the relative strong phase between $\bar{b}$ and $a$. The time-dependent distributions are then given by

$$\begin{aligned}
\Gamma_{\ell^-,D^-\pi^+}(\Delta t) &= Ne^{-\gamma|\Delta t|}\left[(1+r^2) + (1-r^2)\cos\Delta m\Delta t - 2r\sin(\phi_w - \delta)\sin\Delta m\Delta t\right], \\
\Gamma_{\ell^+,D^+\pi^-}(\Delta t) &= Ne^{-\gamma|\Delta t|}\left[(1+r^2) + (1-r^2)\cos\Delta m\Delta t + 2r\sin(\phi_w + \delta)\sin\Delta m\Delta t\right], \\
\Gamma_{\ell^-,D^+\pi^-}(\Delta t) &= Ne^{-\gamma|\Delta t|}\left[(1+r^2) - (1-r^2)\cos\Delta m\Delta t - 2r\sin(\phi_w + \delta)\sin\Delta m\Delta t\right], \\
\Gamma_{\ell^+,D^-\pi^+}(\Delta t) &= Ne^{-\gamma|\Delta t|}\left[(1+r^2) - (1-r^2)\cos\Delta m\Delta t + 2r\sin(\phi_w - \delta)\sin\Delta m\Delta t\right],
\end{aligned} \tag{5.70}$$

where $\Gamma_{\ell^-,D^-\pi^+}(\Delta t)$ is the distribution of $\Delta t \equiv t_{D\pi} - t_{\text{tag}}$ when the tag-side is $\bar{B}^0$, etc., and $D\pi$ can also be $D^*\pi$ in which case $r$ and $\delta$ will be replaced by $r^*$ and $\delta^*$, respectively. The two relevant observables are $r\sin(\phi_w + \delta)$ and $r\sin(\phi_w - \delta)$. The value of $r$ cannot be obtained by the fit itself, while the expected value of $r^{(*)}$ is roughly 0.02. One way to obtain $r^*$ experimentally is to use the $SU(3)$-related modes $B^0 \to D_s^{(*)-}\pi^+$ [148]. However, there will be uncertainty associated with the validity of $SU(3)$. The size of the exchange diagram which is missing for $D_s^{(*)-}\pi^+$ can be inferred from the measurement of $D_s^{(*)+}K^-$.

The distributions are plotted in Fig. 5.42 where the value of $r$ is artificially enhanced to 0.1 in order to show the $CP$ violating effects clearly. The top two of Eq. (5.70) can be called unmixed modes and the bottom two mixed modes. As can be noticed in the figure, most information on CP violation is in the mixed modes where CP violation appears as the height asymmetry of $\Delta t > 0$ vs $\Delta t < 0$ and the shift of the minimum from $\Delta t = 0$. Note that $r\sin(\phi_w + \delta)$ can be obtained from the third distribution alone and $r\sin(\phi_w - \delta)$ from the fourth alone. The final state $D^{*-}\pi^+$ can be detected by full reconstruction using the standard technique, or can also be reconstructed by a partial reconstruction method where the $\bar{D}^0$ meson in $D^{*-} \to \bar{D}^0\pi^-$ is not explicitly reconstructed. The $D^{(*)-}\pi^+$ methods work even when there are sizable exchange diagrams while the value of $r^{(*)}$ needs to be supplied externally.

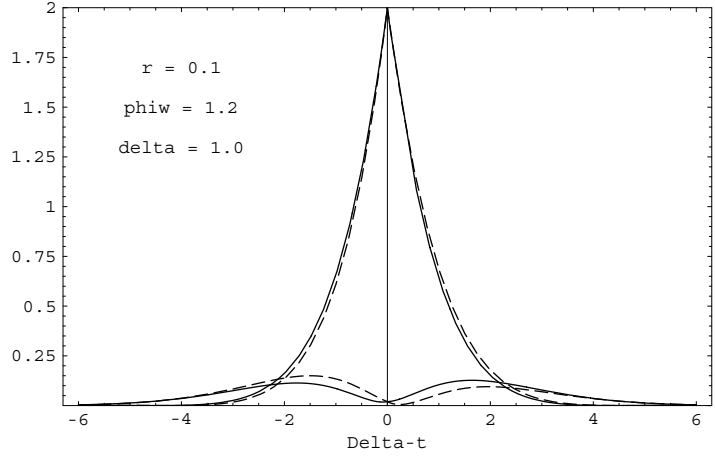

Figure 5.42: The $\Delta t$ distributions of the flavor-tagged $D\pi$ modes for $r = 0.1$, $\phi_w = 1.2$ rad, and $\delta = 1.0$ rad. The solid lines are for $D^-\pi^+$ final state and dashed lines are for $D^+\pi^-$ final states. The mixing parameter $x$ is taken to be 0.71.

The distributions of $\Delta t$ for $\bar{B}^0$-tagged $D^{*+}\pi^-$ and $D^{*+}\pi^-$ are shown in Fig. 5.43 for events with good tagging quality. The analysis was performed on 357 fb$^{-1}$ of data [149]. The full-reconstruction analysis gave the following results:

$$\begin{aligned}
2r^*\sin(\phi_w + \delta^*) &= 0.050 \pm 0.029 \pm 0.013, \\
2r^*\sin(\phi_w - \delta^*) &= 0.028 \pm 0.028 \pm 0.013, \\
2r\sin(\phi_w + \delta) &= 0.031 \pm 0.030 \pm 0.012, \\
2r\sin(\phi_w - \delta) &= 0.068 \pm 0.029 \pm 0.012.
\end{aligned} \quad (5.71)$$

For 5 ab$^{-1}$ and 50 ab$^{-1}$ of data, the statistical errors will be

$$\sigma(2r^*\sin(\phi_w + \delta^*)) \sim \sigma(2r\sin(\phi_w + \delta)) \sim \begin{cases} 0.009 & (5 \text{ ab}^{-1}) \\ 0.003 & (50 \text{ ab}^{-1}) \end{cases} \quad (5.72)$$

We note that the value of $r^{(*)}$ is approximately 0.02; thus, the above errors correspond to the errors on $\sin(\phi_w \pm \delta^{(*)})$ of 0.23 and 0.07, respectively. If the value of $r^{(*)}$ is known, $\phi_w = 2\phi_1 + \phi_3$ (and $\delta^{(*)}$) can be extracted from $\sin(\phi_w \pm \delta^{(*)})$. At this time, determining the value of $r^{(*)}$ to 23% seems reasonably possible while 7% seems quite challenging. On the other hand, our knowledge on $r^{(*)}$ will improve significantly by the time 50 ab$^{-1}$ of data is taken. It is possible that the systematic error on $\phi_3$ due to $r^{(*)}$ is not overwhelming even with 50 ab$^{-1}$ of data.

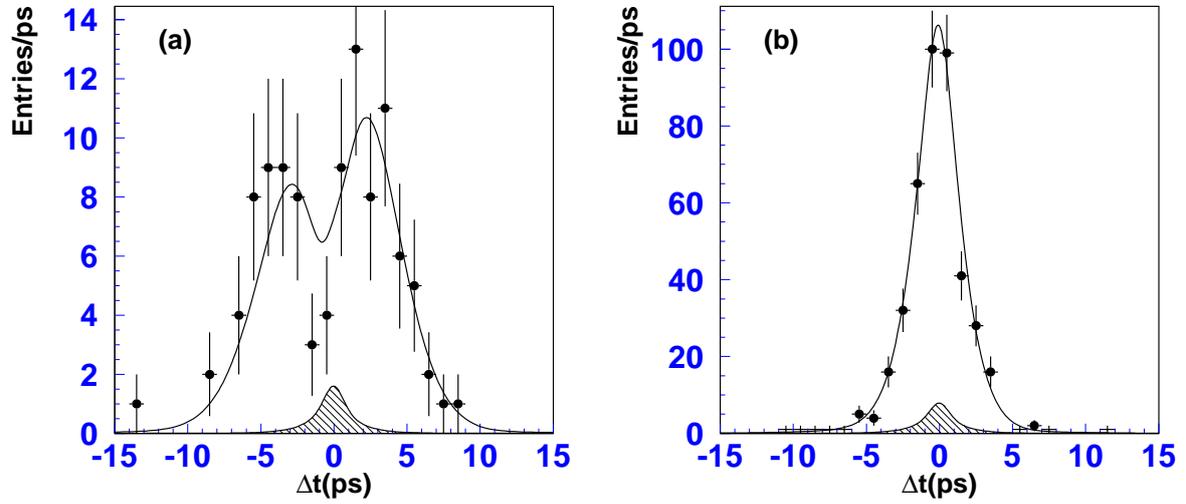

Figure 5.43: The distributions of $\Delta t$ for $\bar{B}^0$-tagged and fully-reconstructed $D^{*+}\pi^-$ (a) and $D^{*+}\pi^-$ (b) candidates at 357 fb$^{-1}$. Only events with good-quality tag are shown.

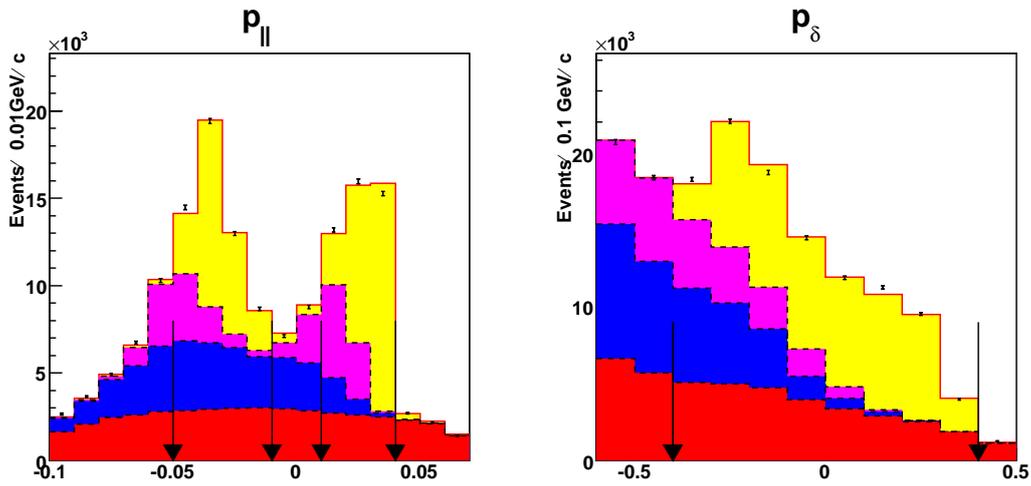

Figure 5.44: Results of the kinematic fits to the $D^*\pi$ candidates based on 605 fb$^{-1}$ of Belle data. The contributions are $D^*\pi$ (yellow), $D^*\rho$ (magenta), correlated background (blue) and uncorrelated background (red).

The statistical situation with the partially-reconstructed $D^{*+}\pi^-$ is substantially better. The helicity of $D^{*+}$ in the $B$ rest frame is 0 due to conservation of angular momentum. Thus, the distribution of the decay angle $\theta^*$ in $D^{*+} \to D^0\pi^+$ decay should have the $\cos^2\theta^*$ dependence. One can reconstruct $\cos\theta^*$ without explicitly detecting $D^0$, and the distributions based on 605 fb$^{-1}$ of data are shown in Fig. 5.44 separately for the two lepton-tagged datasets. There are $50196 \pm 286$ signal events in the same-flavor sample and the opposite-flavor sample. As a result of a $\Delta t$ fit, the statistical and systematical errors on $2r^* \sin(\phi_w \pm \delta^*)$ were found to be 0.019 and 0.012 respectively [150]. The statistics uncertainty can be extrapolated to 5 ab$^{-1}$ and 50 ab$^{-1}$ of data as

$$\sigma(2r^*\sin(\phi_w + \delta^*)) \sim \left\{ \begin{array}{ll} 0.007 & (\text{ 5 ab}^{-1}\text{ }) \\ 0.002 & (50\text{ ab}^{-1}\text{ }) \end{array} \right. \quad (\text{partial rec.}). \tag{5.73}$$

The background fraction is about 30% and comes mainly from $B$ decays such as $D^*\rho$ and $D^{**}\pi$. The accuracy of these background estimations is likely to improve with statistics. The uncertainty of $r^*$, however, may become limiting systematics.

### 5.8.6 Conclusion

We used three independent measurements (in $D^{(*)}K$ modes with the GLW, ADS and Dalitz methods), to estimate a combined $\phi_3$ uncertainty of 1.5° using a data sample of 50 ab$^{-1}$. All the methods listed above are theoretically clean, and can thus provide reliable measurement of the angle $\phi_3$ as a test of the Standard Model and to search for new physics effects.

The flavor-tagged time-dependent measurement of $D^{(*)-}\pi^+$ allows one to obtain a different combination of angles (($2\phi_1 + \phi_3$)) with an unprecedented precision providing another independent check of the Standard Model.

## 5.9 $|V_{ub}|$

### 5.9.1 Introduction

Precise determination of the magnitude of the Cabibbo-Kobayashi-Maskawa matrix element $V_{ub}$ plays a crucial role in over-constraining the unitarity triangle beacuse it directly determines the side directly opposite the best measured angle ($\phi_1$). Measured values of $|V_{ub}|$ are insensitive to new physics effects as they are determined from tree level processes in the Standard Model. A precise measurement, with a relative error of 5% or lower, will be accessible with data from a high luminosity $B$-factory.

Currently $|V_{ub}|$ is best determined from data of semileptonic $B \to X_u \ell \nu$ decays, where $X_u$ denotes a hadronic system containing a $u$-quark. Purely hadronic channels suffer from insurmountable theoretical uncertainties while the purely leptonic $B$ decay has a far smaller rate and an associated unknown of the $B$-meson decay constant. Experimentally $|V_{ub}|$ is measured in inclusive and exclusive channels independently. Each suffer from different experimental and theoretical uncertainties and as such comparison of the results of the two channels provides a powerful consistency check.

The best present error on $|V_{ub}|$ in any one measurement is about 7% depending on the theory method used to extract the value [151]. The statistical error is a significant component contributing 4.5% while the systematic error which includes theoretical uncertainties totals to about 5.5%.

The high luminosity at the SuperKEKB will enable us to perform high statistics measurements of $B \to X_u \ell \nu$ decays with improved "$B$ tagging". This will lead to rate measurements with reduced experimental systematic errors and also to $|V_{ub}|$ extraction much less biased by theoretical ambiguities because almost all of the available phase space will be accessed. The current $B$ factories have performed extremely well with ground breaking measurements that, enhanced, will yield a precision on $|V_{ub}|$ from 2% to 5% - bounds that are respectively ambitious and conservative.

In this section, we discuss the strategy and prospects for determination of $|V_{ub}|$ at the SuperKEKB/Belle experiment using both inclusive and exclusive semileptonic $B \to X_u \ell \nu$ decays.

### 5.9.2 Theoretical formalisms for the semileptonic $B$ decays

**Inclusive decays**

The amplitude for $B \to X_u l \nu$ inclusive decay can be computed in perturbative QCD using the Operator Product Expansion (OPE). Since the $b$ quark inside the $B$ meson has momentum $(m_b v + k)^\mu$, where $k$ is the residual momentum of $O(\Lambda_{QCD})$, the OPE is carried out by expanding the quark propagator as

$$\frac{1}{(m_b v + k - q)^2} = \frac{1}{(m_b v - q)^2} \left[ 1 - \frac{(m_b v - q) \cdot k}{(m_b v - q)^2} - \frac{k^2}{(m_b v - q)^2} + \cdots \right]. \quad (5.74)$$

Denoting the invariant mass and the energy of the hadron state $X_u$ as $m_X$ and $E_X$, the first term is of order $E_X \Lambda_{QCD}/m_X^2$ while the second term is of order $\Lambda_{QCD}^2/m_X^2$.

The phase space can be divided into the following three regions (see Figure 5.45): (i) a generic region where $\Lambda_{QCD}/m_X$, $(E_X \Lambda_{QCD})/m_X^2 \ll 1$. In this region, the differential decay rate can be successfully expanded by the OPE; (ii) a shape function region (or collinear region) where $\Lambda_{QCD}/m_X \ll 1$ and $(E_X \Lambda_{QCD})/m_X^2 \sim 1$. In this region, a class of $\Lambda_{QCD} E/m_X^2$ terms must be resummed, which can be described by the shape function of the $B$ meson [152–154] and its higher twist corrections; (iii) a resonance region where $\Lambda_{QCD}/m_X \sim 1$. In this region, the differential decay rate is dominated by a few exclusive states so that neither the OPE nor the twist expansion work.

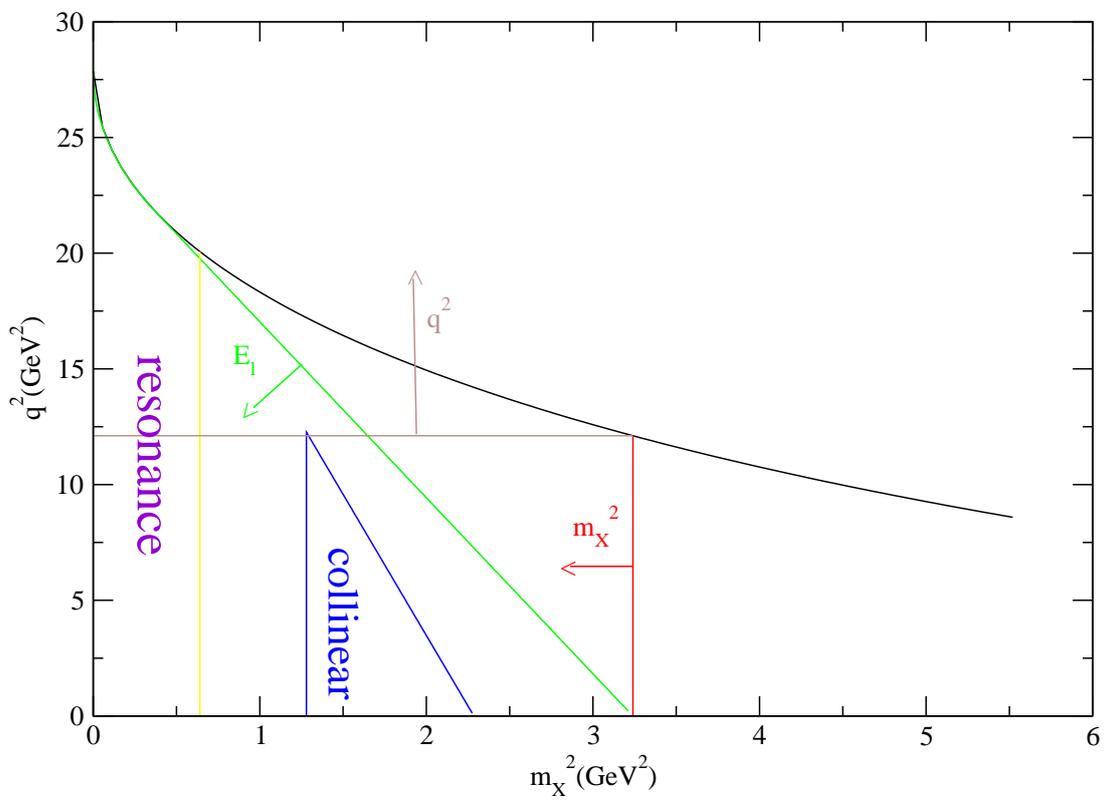

Figure 5.45: Phase space for $B \to X_u l\nu$ decay.

Since the $B \to X_u l \nu$ decay suffers from $B \to X_c l \nu$ decay background one has to introduce the selection criteria of the following kinds:

- lepton energy : $E_l > (m_B^2 - m_D^2)/(2m_B)$,
- hadron invariant mass : $m_X^2 < m_D^2$,
- lepton mass : $q^2 < (m_B - m_D)^2$,
- hadronic light cone : $P_+ \equiv E_H - |\vec{P}_H| < m_D^2/m_B$,
- combined $(q^2, m_X^2)$ cut.

For the $E_l$ selection, only 10% of the rate is available, which is dominated by the shape function region. In this region, the lepton energy spectrum at the leading order is given by

$$\frac{d\Gamma}{dE_l} = \frac{G_F^2 |V_{ub}|^2 m_b^4}{96\pi^3} \int d\omega \theta(m_b - 2E_l - \omega) f(\omega), \tag{5.75}$$

where $f(\omega)$ is the shape function defined by

$$f(\omega) = \frac{1}{2m_B} \langle B | \bar{h} \delta(in \cdot D + \omega) | B \rangle. \tag{5.76}$$

Since this function is a universal quantity of the $B$ meson, it can be measured experimentally through other processes. For example, the photon spectrum of $B \to X_s \gamma$ decay is given by [155]

$$\frac{d\Gamma}{dE_\gamma} = \frac{G_F^2 |V_{tb} V_{ts}^*|^2 \alpha_{QED} m_b^5}{32\pi^4} f(\omega). \tag{5.77}$$

Therefore, the extraction of $|V_{ub}|$ can be done without the theoretical uncertainty of the shape function, if one considers a ratio of weighted integrals over the endpoint regions of $B \to X_u l \nu$ decays and the photon spectrum in $B \to X_s \gamma$ decays as [156, 157]

$$\left| \frac{V_{ub}}{V_{tb} V_{ts}^*} \right|^2 = \frac{3\alpha_{QED}}{\pi} K_{pert}(E_0) \frac{\Gamma_u(E_0)}{\Gamma_s(E_0)} + \mathcal{O}(\Lambda/M_B), \tag{5.78}$$

where

$$\Gamma_u(E_0) = \int_{E_0}^{M_B/2} dE_l \frac{d\Gamma(B \to X_u l \nu)}{dE_l}, \tag{5.79}$$

$$\Gamma_s(E_0) = \frac{2}{M_B} \int_{E_0}^{M_B/2} dE_\gamma (E_\gamma - E_0) \frac{d\Gamma(B \to X_s \gamma)}{dE_\gamma}. \tag{5.80}$$

The coefficient $K_{pert}(E_0)$ is a factor from short-distance effect which can be calculated in perturbative QCD. There are two major sources of uncertainties. The first is the unknown higher twist corrections to the shape function $1/m_b$ [158–161]. Based on model calculations this higher twist correction is expected to be of order 15% for $E_l^{cut} = 2.3$ GeV. The second is the weak annihilation contribution of $O(1/m_b^2)$, which is estimated as 10% for $E_l^{cut} = 2.3$ GeV [162, 163]. These two uncertainties can be reduced below 10% by lowering the lepton energy cut by combining with other cuts.

For the $m_X^2$ cut [164–168], 80% of the kinematic range is available but still the results are sensitive to the shape function. This cut also suffers from the singularity due to the bremsstrahlung diagram when the partonic invariant mass $s = (m_b - v)^2$ is zero.

For the pure $q^2$ cut, 20% of the rate is available and the decay rate is not sensitive to the shape function [169]. However, the kinematic range is sensitive to the resonance region and the convergence of the OPE as well as the convergence in the perturbative expansion in $\alpha_s$ are slower. The largest error comes from the weak annihilation contribution of $O(1/m_b^3)$. The rate with $q^2$ is also sensitive to the uncertainty of $m_b$ [170] as can be seen from the $m_b$ dependence of the partial decay rate parametrized as $\Gamma(q^2 > q_{cut}^2) \propto m_b^{\Delta(q_{cut}^2)}$, where $\Delta(q_{cut}^2) \sim 10 + \dfrac{q_{cut}^2 - (m_B - m_D)^2}{1\text{GeV}^2}$.

For the $P_+$ cut, 80% of the rate is available and the decay rate is sensitive to the shape function, it's similar to the $m_X$ cut but in contrast provides a "buffer zone" against charm background [171]. The event fraction in this case has been calculated according to the prescription of Bosch, Lange, Neubert and Paz (BLNP). In BLNP the differential rates are factorized into perturbatively calculable functions $H_i$ (high scales $\mu_h \sim m_b$) and $J_i$ (final state hadronic jets $\mu_i \sim \sqrt{s_H} \sim \sqrt{m_b \Lambda_{QCD}}$) and universal shape functions $S_i$ (non-perturbative physics).

To summarize, possible sources of errors are (1) perturbative error from unknown two-loop corrections, (2) shape function contributions and bremsstrahlung, (3) uncertainties in $m_b$, and (4) $O(1/m_b^3)$ power corrections. The optimized method would be obtained by combining the $q^2$ and $m_X^2$ cuts [172] or a pure $P_+$ cut [171]. The kinematical constraints $m_X < m_D$ and $q_{cut}^2 > m_B m_b - (m_X^{cut})^2$, and $P_+ < M_D^2/M_B$ reduce the charm background. If we raise $q_{cut}^2$ the errors (3) and (4) gets larger while if we lower $q_{cut}^2$ the errors (2) gets larger and (1) is small in the intermediate $q^2$ region. Thus it is important to find the best cut that minimizes the sum of these errors.

In Table 5.18, we give results of [172] and [171] for the errors of the partial decay rate normalized by the total tree level parton decay rate defined as

$$\frac{G_F^2 |V_{ub}|^2 m_b^5}{192\pi^3} G(q_{\text{cut}}^2, m_{\text{cut}}) \equiv \int_{\hat{q}_{\text{cut}}^2}^{1} d\hat{q}^2 \int_0^{\hat{s}_0} d\hat{s} \frac{d\Gamma}{d\hat{q}^2 d\hat{s}}. \tag{5.81}$$

In order to achieve a $|V_{ub}|$ determination with an accuracy at the level of a few percent, $q_{cut}^2 = 6$ GeV$^2$ (and $m_{Xcut}^2 = m_D^2$) is the optimal choice. The dominant error in this case is the uncertainty in $m_b$. It is therefore important to determine the bottom quark mass to 30 MeV/c$^2$ accuracy.

**Exclusive decays**

The exclusive semileptonic decay $B \to \pi l \nu$ determines the CKM matrix element $|V_{ub}|$ through the following formula,

$$\frac{d\Gamma}{dq^2} = \frac{G_F^2}{24\pi^3} |(v \cdot k_\pi)^2 - m_\pi^2|^{3/2} |V_{ub}|^2 |f^+(q^2)|^2, \tag{5.82}$$

where the form factor $f^+$ is defined as

$$\langle \pi(k) | \bar{q} \gamma^\mu b | B(p) \rangle = f^+(q^2) \left[ (p+k)^\mu - \frac{m_B^2 - m_\pi^2}{q^2} q^\mu \right] + f^0(q^2) \frac{m_B^2 - m_\pi^2}{q^2} q^\mu, \tag{5.83}$$

with $p$ and $k$ the $B$ and $\pi$ meson momenta. $q = p - k$ is the momentum transfer and $q^2 = m_B^2 + m_\pi^2 - 2m_B v \cdot k$, where $v$ is the velocity of the $B$ meson. Since the most promising approach in which systematic improvement based on the first principle calculations is possible is lattice QCD, we focus on the lattice computation of the form factors.

| Cuts on $(q^2, m_X^2)$ | $G(q^2_{\text{cut}}, m_{\text{cut}})$ | $\Delta_{\text{struct}} G$ | $\Delta_{\text{pert}} G$ | $\Delta_{m_b} G$ $\pm 80/30\,\text{MeV}$ | $\Delta_{1/m^3} G$ | $\Delta G$ |
|---|---|---|---|---|---|---|
| Combined cuts | | | | | | |
| $6\,\text{GeV}^2, 1.86\,\text{GeV}$ | 0.38 | $-4\%$ | $4\%$ | $13\%/5\%$ | $6\%$ | $15\%/9\%$ |
| $8\,\text{GeV}^2, 1.7\,\text{GeV}$ | 0.27 | $-6\%$ | $6\%$ | $15\%/6\%$ | $8\%$ | $18\%/12\%$ |
| $11\,\text{GeV}^2, 1.5\,\text{GeV}$ | 0.15 | $-7\%$ | $13\%$ | $18\%/7\%$ | $16\%$ | $27\%/22\%$ |
| Pure $q^2$ cuts | | | | | | |
| $(m_B - m_D)^2, m_D$ | 0.14 | $--$ | $15\%$ | $19\%/7\%$ | $18\%$ | $30\%/24\%$ |
| $(m_B - m_{D^*})^2, m_{D^*}$ | 0.17 | $--$ | $13\%$ | $17\%/7\%$ | $14\%$ | $26\%/20\%$ |
| Pure $P_+$ cuts | $G$ | $\Delta_{\text{struct}} G$ | $\Delta_{\text{pert}} G$ | $\Delta_{m_b} G$ $\pm 70\,\text{MeV}$ | $\Delta_{1/m^3} G$ | $\Delta G$ |
| $0.66\,\text{GeV}^2$ | 0.80 | $9\%$ | $8\%$ | $10\%$ | $10\%$ | $19\%$ |

Table 5.18: $G(q^2_{\text{cut}}, m_{\text{cut}})$ and its errors for different choices of $(q^2_{\text{cut}}, m_{\text{cut}})$. $\Delta_{\text{struct}} G$ gives the fractional effect of the structure function $f(k_+)$ in the simple model which is not included in the error estimate. $\Delta_{\text{pert}} G$ corresponds to pertubative QCD uncertainties. $\Delta_{m_b} G$ is the propagated uncertainty on the $b$-quark mass. $\Delta_{1/m^3} G$ relates to uncertainties on uncalculated power corrections. The total error is obtained by adding each error in quadrature. The two values correspond to $\Delta m_b^{1S} = \pm 80\,\text{MeV}$ and $\pm 30\,\text{MeV}$. Table is from [172] with the exception of the last row, which is from [171]; note that here the uncertainties are within the context of a different methodology to that of [172], therefore a direct comparison should be treated with caution.

Lattice calculations suffer from three major limitations. One is the discretization error from the large energy of the initial and final hadrons. In order to avoid such error, spatial momenta must be much smaller than the cutoff, i.e. $|\vec{p}_B|, |\vec{k}_\pi| < 1$ GeV. This means that the form factors can be computed reliably only in the range of $v \cdot k \equiv E_\pi < 1$ GeV or equivalently $q^2 > 18$ GeV$^2$. Another limitation is the fact that due to limited computer power the light quark mass range for practical simulations is $m_s/3 \leq m_q \leq m_s$ or $m_\pi = 0.4 \sim 0.8$ GeV. In order to obtain physical results chiral extrapolation in the light quark masses is necessary. The last limitation is the large discretization error from the $b$-quark mass. In the present simulations, the lattice cutoff is limited to $a^{-1} = 2 - 3$ GeV, so that the $b$-quark mass in lattice units is larger than unity. This makes the discretization error of $O(am_b)$ completely out of control. In order to avoid this error, one either carries out simulations around a charm quark mass region and extrapolates the result in the inverse heavy quark mass $1/m_Q$ (extrapolation method), or uses heavy quark effective theories, such as NRQCD action or Fermilab action (HQET). In both cases, extrapolation or interpolation of the form factors in $1/m_Q$ may be performed using the HQET-motivated form factors $f_1(v \cdot k)$ and $f_2(v \cdot k)$ [173]

$$\langle \pi(k) | \bar{q} \gamma^\mu b | B(p) \rangle = 2 \left[ f_1(v \cdot k) v^\mu + f_2(v \cdot k) \frac{k^\mu}{v \cdot k} \right]. \quad (5.84)$$

With this choice the heavy quark scaling law is explicit, and the form factors are simply expanded in terms of $1/m_Q$.

Up until 2004 all lattice calculations of the form factors have been performed only in the quenched approximation, in which the sea quark effects are neglected. Since then, two groups, namely Fermilab/MILC and HPQCD have performed unquenched calculations [174], [175]. Both employ "$n_F = 2+1$ flavor" MILC configurations, which use three flavors of improved staggered quarks: two degenerate light quarks and one heavy quark ($\approx m_s$). However the groups use different heavy quark discretizations: Fermilab/MILC use Fermilab quarks and HPQCD use non-relativistic heavy quarks (NRQCD). Errors typically arise from:

- Monte Carlo statistics and fitting,

- Lattice spacing tuning, $a$ and quark masses,

- Matching lattice gauge theory to continuum QCD, which can alse be broken down into relativistic errors, discretization errors and perturbation theory,

- Extrapolation to continuum, and

- Chiral extrapolation to physical up and down quark masses.

The matching and chiral extrapolation errors are dominant. Generic lattice quark actions will have discretization errors and are proportional to $(am_Q)^n$. Here the heavy quark discretization errors in principle can be eliminated order by order by using knowledge of heavy quark and nonrelativistic quark limits of QCD. In practice this is difficult at higher orders and therefore errors due to inexact matching must be estimated. Fermilab combine all errors associated with the discretizing action and estimate errors using knowledge of short-distance coefficients and power counting. HPQCD quote relativistic errors when moving from QCD to NRQCD and perturbation theory errors for NRQCD to Lattice Gauge theory, with the errors estimation relying on power counting. Fermilab derives a lattice gauge theory from continuum QCD using HQET while HPQCD uses non-relativistic QCD.

The last step of chiral extrapolation is necessary to recover the physical values of the up and down quark mass and is performed using staggered chiral perturbation theory. This method takes into account the next-to-leading order light quark mass dependence and light quark discretization effects through to order $\alpha_s^2 a^2 \Lambda_{\text{QCD}}^2$.

The lattice results are shown in Fig. 5.46(top). The total systematic uncertainties are 11% for the form factors in both calculations. The largest quoted error comes from the perturbative matching between the lattice and continuum theory for the HPQCD result, and from the discretization effect for the Fermilab/MILC result. Two unquenched results agree with each other within errors. These are also consistent with previous quenched calculations [176–179].

By integrating $f_+(q^2)$ over $16\tilde{\text{GeV}}^2 \leq q^2 \leq q_{\max}^2$ using the partial branching fraction as provided by HFAG, the CKM matrix element is obtained as

$$|V_{ub}|_{\text{HPQCD}} = 3.33 \pm 0.21^{+0.58}_{-0.38},$$

$$|V_{ub}|_{\text{FNAL/MILC}} = 3.55 \pm 0.20^{+0.61}_{-0.40},$$

where the first error is experimental and the second is from the lattice form factor calculation.

Improvements can be made on general grounds with higher order matching and improved action in the case of Fermilab and lighter quark masses and finer lattice spacings and additional lattice calculations. For $B \to \pi l \nu$, there is a possibility to generate data at additional $q^2$ points with the following methods:

- Moving NRQCD to generate lattice data at low $q^2$ (high pion momentum) while keeping statistical errors under control [180, 181].

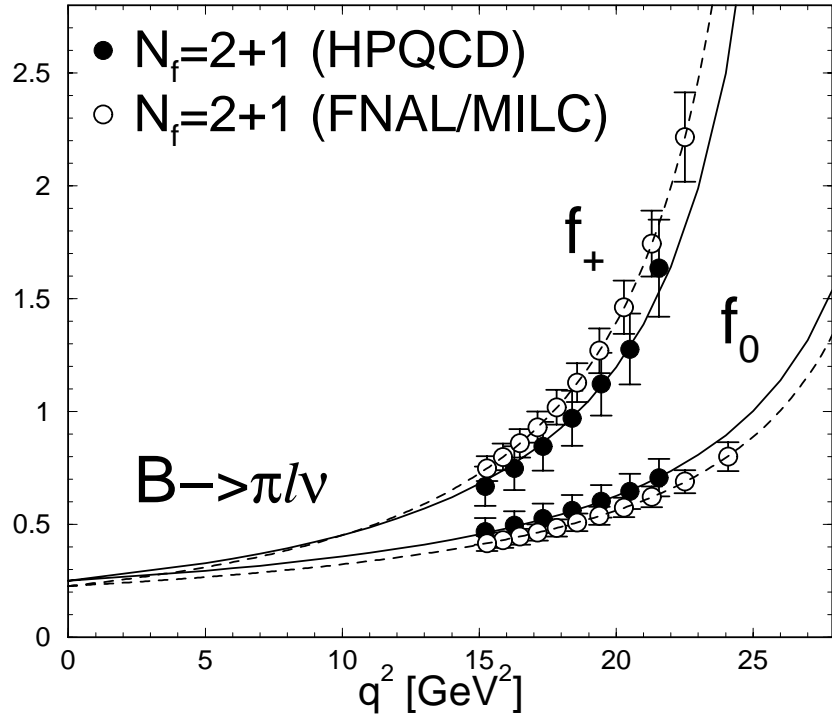

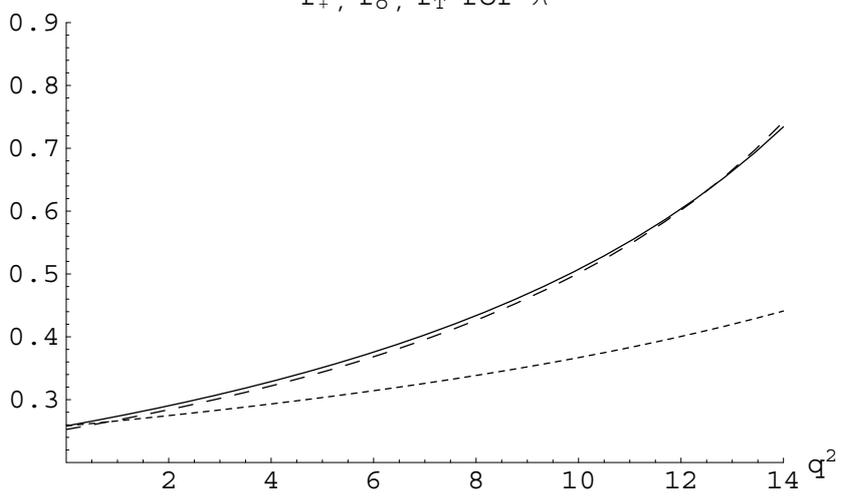

Figure 5.46: (top) Unquenched $B \to \pi l \nu$ form factors by Fermilab and HPQCD lattice groups. (bottom) Light Cone Sum Rule calculation of $f_+(q^2)$ (solid line) and $f_0(q^2)$ (short dashes).

- Twisted Boundary conditions to generate additional high $q^2$ data points with pion momenta at non-integer values of $2\pi/L$ ($L$ = spatial lattice size) [182, 183].

Using the heavy quark symmetry is another way of improvement. Since the CLEO-c experiment can measure form factors for $D \to \pi l \nu$ to a few percent accuracy, their results will be a good approximation for the $B \to \pi l \nu$ form factors. Then the task for lattice QCD is to provide the $1/m_Q$ dependence of the form factors. The $B$ to $D$ ratio $\dfrac{d\Gamma(B \to \pi l\nu)/d(v \cdot k_\pi)}{d\Gamma(D \to \pi l\nu)/d(v \cdot k_\pi)}$ with the same recoil energy $v \cdot k_\pi$ would be a nice quantity to measure on the lattice, since a large part of the statistical error, the perturbative error and the chiral extrapolation errors are expected to cancel in this ratio.

In the large recoil momentum region, where the final state meson has large energy in the rest frame of the decaying $B$ ($E \gg \Lambda_{\rm QCD}$) the light-cone QCD sum rule (LCSR) may be used to calculate the form factors [184–188]. Recall that in lattice calulations $q^2 > 15 \text{ GeV}^2$ owing to $\pi$ energies necessarily being smaller than the inverse lattice spacing. In Figure 5.46(bottom) the latest result [188] is shown, it covers the region $q^2 < 14 \text{ GeV}^2$. The form factor at zero recoil has an irreducible uncertainty of 7%, estimated from the variation of sum rule specific parameters, the overall uncertainty is 12%, in future this could fall to as low as 10%.

Model independent bounds for the whole $q^2$ range can be obtained with dispersion relation, perturbative QCD, and lattice QCD data [189]. Reducing the lattice errors or having other inputs would significantly improve the results. Recently a more elaborate studies was performed [190–194]. For example in [191] the method relies on lattice calculations for the high $q^2$ region, while at $q = 0 \text{ GeV}^2$ a model independent constraint obtained from $B \to \pi\pi$ using the soft-collinear effective theory is imposed, and the shape of $q^2$ is constrained using QCD dispersion relations. The overall uncertainty is dominated by the corresponding errors from the lattice points used. A future direction of research is to attempt to take into account the additional error correlation implied by the dispersion relations when lattice input points are included that are closer together.

The UKQCD collaboration [195] and the SPQcdR collaboration [196] performed studies of $B \to \rho l \nu$ form factors. Both collaborations use an $O(a)$-improved Wilson action for the heavy quark and extrapolate the numerical results of $m_Q \sim m_c$ toward the physical b quark mass. UKQCD obtained the partially integrated decay rate in the region $12.7 \text{ GeV}^2 < q^2 < 18.2 \text{ GeV}^2$ as $\Gamma = (4.9^{+12+\ 0}_{-10-14}) \times 10^{12} s^{-1} |V_{ub}|^2$.

### 5.9.3 Measurement of inclusive $b \to u$ semileptonic decays

In $\Upsilon(4S)$ experiments, a correct measurement of $(q^2, m_X)$ is only possible when the accompanying $B$ decays are fully reconstructed. This technique, referred to as "full reconstruction tagging", allows for isolation of particles from the signal $B$ decay for correct reconstruction of $m_X$, and also to determine the momentum vector of the signal-side $B$ meson. The latter helps to improve reconstruction of the missing neutrino, leading to correct reconstruction of $q^2$ and better discrimination of $B \to X_c \ell \nu$ background leaking into the signal phase space. Full reconstruction tagging also allows us to determine the flavor and charge of the signal-side $B$. It helps to identify the signal lepton using the correlation between the $B$ flavor and lepton charge, and also to measure the decay rate separately for neutral and charged $B$ mesons. However, this method requires a large accumulation of $B\overline{B}$ data because of the relatively small efficiency in the full reconstruction of the accompanying $B$'s (a few times 0.1%). Therefore, the high luminosity of SuperKEKB provides an unique opportunity to perform this measurement with high statistics.

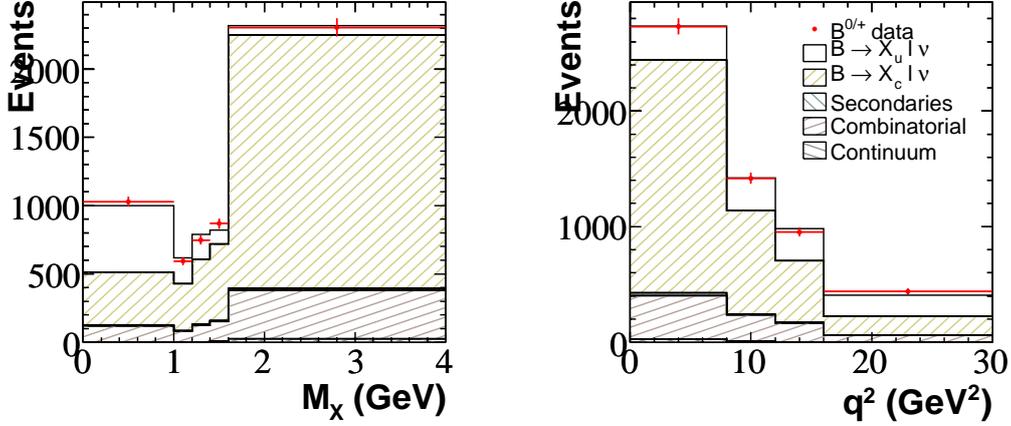

Figure 5.47: Projected distributions of reconstructed $m_X$ and $q^2$ with a full reconstruction analysis using 605 fb$^{-1}$ of data [151].

Belle reported a measurement of $|V_{ub}|$ using full reconstruction tagging which accessed enough phase space to allow a simple and robust theoretical treatment, that is in principle solely OPE based [151]. In a 605 fb$^{-1}$ data sample the signal was extracted in a fit to the $m_X$ vs $q^2$ spectrum, with lepton energy threshold of $E_\ell > 1.0$ GeV. It is the most precise measurement to date and resets the starting point for what can be achieved at a super B factory. The projected distributions of $m_X$ and $q^2$ spectra are shown in Fig 5.47. In the low $m_X$ region, one can see a clear $B \to X_u \ell \nu$ signal. In this data sample, the number of fully reconstructed $B^+$ and $B^0$ candidates in the signal region after background subtraction is $6.9 \times 10^5$ and $4.8 \times 10^5$, respectively. This corresponds to a reconstruction efficiency ($\epsilon_{\text{frec}}$) of about 0.2%.

In the signal region, for example, the observed number of events is 5544 with a fitted background ($B$) of 4512, resulting in a signal-to-background ratio ($S/B$) of 0.23. The efficiency to detect the $B \to X_u \ell \nu$ signal once the accompanying $B$ is fully reconstructed ($\epsilon_{b \to u}$) is about 22%.

Based on extrapolation of the present Belle results and realistic assumptions for the improvement in $\epsilon_{\text{frec}}$ and $S/B$, we have estimated the $|V_{ub}|$ precision at a SuperKEKB/Belle experiment. We consider here the statistical, experimental systematic and theoretical errors, and estimate them as follows.

**Statistical error:** The expected statistical error can be obtained assuming that the uncertainty of the measured branching fraction scales with the signal yield ($S$) and background ($B$) as $\sigma(Br)/Br = \sqrt{S+B}/S$ [4]. Hence the statistical error on $|V_{ub}|$ ($\Delta_{stat}$) scales with the integrated luminosity ($L$) as

$$\Delta_{stat} = \frac{1}{2} \times \sqrt{\frac{1 + (S/B \times f_{S/B})^{-1}}{n_0 \times f_{\text{frec}} \times \epsilon_{b \to u} \times L}} = \frac{0.7}{\sqrt{L}} \quad, \quad (5.85)$$

where $n_0$ is the rate of $B \to X_u \ell \nu$ decays after full reconstruction, estimated as $1052/(0.22 \times 605 \text{ fb}^{-1}) = 8.0/\text{fb}^{-1}$. The factor $f_{frec}$ accounts for the improvement in $\epsilon_{\text{frec}}$, for which expect a 50% gain, $f_{frec} = 1.5$. The factor $f_{S/B}$ accounts for a possible improvement in $S/B$ from the results and is assumed to be $f_{S/B} = 1.2$ (expected to improve by at least 20% with the introduction of more modes and improved detector performance, for example $K_L^0$ rejection and better $K^\pm/\pi^\pm$ separation). The relative statistical errors are estimated to be 1.0% and 0.3% with data samples of 5 ab$^{-1}$ and 50 ab$^{-1}$ respectively.

---

[4] Note that this is conservative since the additional information on the distribution of decays is used in the measurement. The stat. uncertainty on $|V_{ub}|$ follows from $\sigma(|V_{ub}|)/|V_{ub}| = (1/2)\sigma(Br)/Br$.

**Experimental systematic error:** The major source of experimental systematic error ($\Delta_{syst}$) on the Belle measurement is the signal model uncertainty, at 5.6% it is the leading contribution to the systematic error of 7.6% on the branching fraction value. Signal modelling will improve with better measurements of the charmed semileptonic and inclusive radiative $B$-meson decays as well as the signal spectra themselves. Realistically we can expect the model uncertainty to be reduced by half by the time 5 ab$^{-1}$ is accumulated, thus 3%, contributing less than a 2% error to $|V_{ub}|$. The next leading source of uncertainty are detector related, at 4.8%. Increased luminosity can reduce this value as well, as systematic uncertainties heavily rely on calibration samples taken from data, and hence too are limited by statistics. Our estimation gives $\Delta_{syst} \approx 3\%$ in $|V_{ub}|$ for 5 ab$^{-1}$, and with a factor 10 increase in integrated luminosity, realistically one can expect a 50% reduction in the systematic error, therefore $\Delta_{syst} \approx 1.5\%$ at 50 ab$^{-1}$.

**Theoretical error:** With the measurement accessing approximately 90% of the available phase space the explicit theory error is already reduced to a very low level. A good gauge of the uncertainty can be gained from the OPE derived expression for the rate, which to leading order $1/m_b$ and $\alpha_s$ is given by

$$\Gamma(B \to X_u l\nu) = \frac{G_F^2 m_b^5}{192\pi^3}|V_{ub}|^2 \left[1 + \frac{\lambda_1}{2m_b^2} - \frac{9\lambda_2}{2m_b^2}\right], \qquad (5.86)$$

where $m_b$ is the b-quark mass and $\lambda_{1,2}$ parameterise matrix elements of local HQET operators that physically represent, respectively, the $b$-quark's kinetic energy and magnetic interaction with the light degrees of freedom within the $B$-meson. The relative uncertainty on $|V_{ub}|$ scales with that of the b-quark mass multiplied by a factor $\frac{5}{2}$. The dominant error in this case is the uncertainty in the $b$-quark mass. The $b$-quark has been measured to within 55 MeV uncertainty using improved experimental data, where the theoretical and experimental uncertainties are comparable [197]. The theory error contribution is about 30 MeV. Ignoring experimental uncertainties this translates to a uncertainty of about 2% on $|V_{ub}|$. These may be reduced since theorists have used conservative bounds on unknown higher order parameters due to lack of experimental input, namely the higher order moments of inclusive $B$ decay spectra. With increased data, these experimental inputs would become availiable. The relative theoretical errors are estimated to be 4% and 2% with data samples of 5 ab$^{-1}$ and 50 ab$^{-1}$ respectively. This predicted outcome is optimistic and follows without calling into question the current strategy used in assigning theory errors. Global fits which arrive at the mass, so far, assume simplistic estimation of correlations between the theory predictions which, if changed, have recently been demonstrated to have a large impact on the result [5]. A thorough program for the determination of the $b$-quark mass will need to be undertaken at a super $B$ factory, if the modelling and theory error is to be reduced below 5%. It is not automatic for theory errors to fall over time, in fact in past periods, they have increased.

---

[5] See talk given by C. Schwanda at CKM 2008

Adding the above three error sources in quadrature, the overall relative uncertainty are estimated be 5% and 3% with data samples of 5 ab$^{-1}$ and 50 ab$^{-1}$ respectively, and are dominated by theoretical uncertainties. In conclusion, with a precise determination of the $b$-quark mass, a $|V_{ub}|$ error of less than 5% is achievable at $L = 50$ ab$^{-1}$. Figure 5.48 demonstrates the expected improvement of the $|V_{ub}|$ error as a function of the integrated luminosity $L$, within an analysis that applies the loose choice of $E_\ell > 1.0$ GeV.

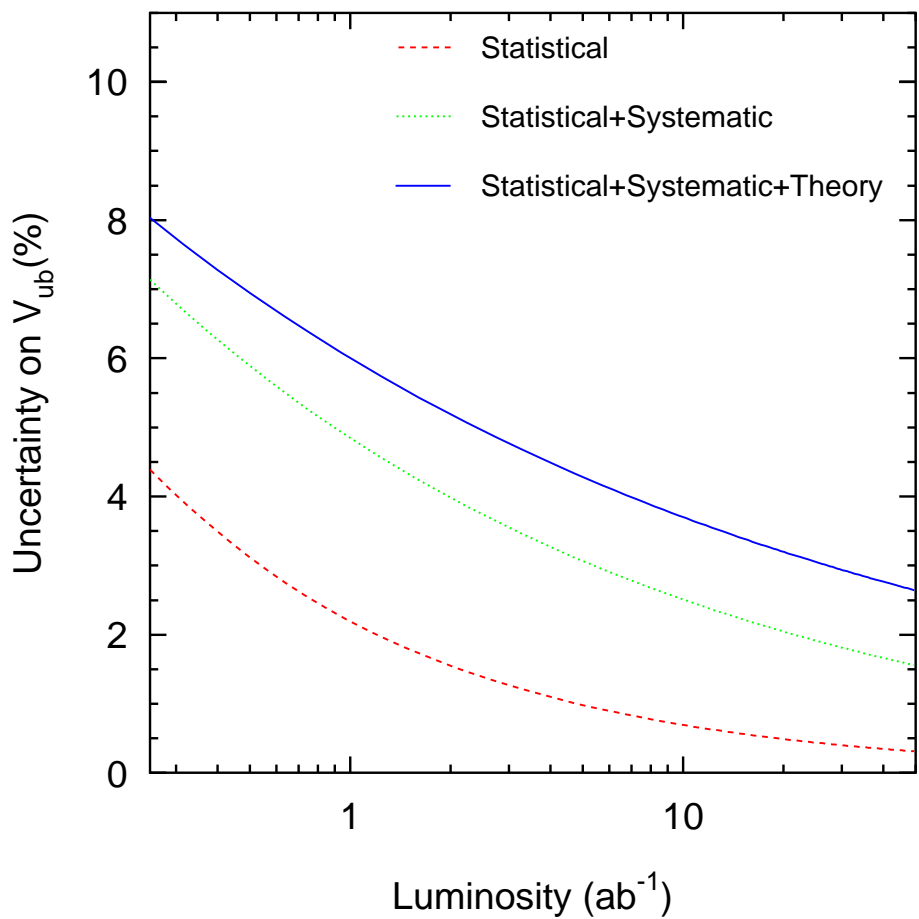

Figure 5.48: Expected improvement of $|V_{ub}|$ error as a function of $L$.

### 5.9.4 Measurement of exclusive $b \to u$ semileptonic decays

The measurement of the inclusive $B \to X_u \ell \nu$ decay is insensitive to theoretical ambiguities, but experimentally challenging because of the large background from $B \to X_c \ell \nu$ decays. Complementary to this, the measurement of exclusive decays, such as $B \to \pi \ell \nu$ and $B \to \rho \ell \nu$, provides experimentally cleaner information, but is subject to large theoretical uncertainties in the form factors. On the experimental side, it is essential to provide precise data for the differential rates, $d\Gamma/dq^2$, of each exclusive channel. This is because $d\Gamma/dq^2$ varies depending on theory models, as such data helps to test the models. Precise data in the high $q^2$ region is especially important, since lattice-QCD calculations, the most promising tool for a reliable model-independent determination of $|V_{ub}|$, are possible only in the region $q^2 > 15$ GeV$^2$, as discussed in Section 5.9.2. A high luminosity SuperKEKB/Belle experiment will enable us to measure the $q^2$ distributions with high statistics and less systematic uncertainties than the best measurement to date [198]. Combined with improvements in lattice QCD with unquenched calculations in the future, this will lead to useful determinations of $|V_{ub}|$.

One of the key experimental issues in measuring the exclusive $B \to \pi\ell^+\nu/\rho\ell^+\nu$ decays is the reconstruction of the undetected neutrino in the final state. Information on missing energy and momentum of the event have been used to infer information about the missing neutrino ("neutrino reconstruction"). This method, originally developed by CLEO, has been applied in existing measurements and exploits the known kinematics of the $e^+e^- \to \Upsilon(4S)$ reaction and near $4\pi$ coverage ("*hermeticity*") of the detector. In reality, however, hermeticity of a detector is never complete. This lack of hermeticity allows background from both $B\overline{B}$ and cross-feed (*e.g.*, $\pi\ell\nu \leftrightarrow \rho\ell\nu$) to contribute, where the latter is serious especially in the high $q^2$ region. For instance, the recent BABAR measurement [198] provides $Br(B \to \pi\ell\nu)$ with a statistical error of 5 (10)% and an experimental systematic error of 6 (8)% for the whole $q^2$ ($q^2 > 16$ GeV$^2$) region. The experimental systematic errors are mainly associated with the neutrino reconstruction; 4% in the whole $q^2$ region and 5% in the $q^2 > 16$ GeV$^2$ region. The BABAR result has been obtained with an event sample of $\sim 206$ fb$^{-1}$, and we can quickly improve the statistical error to a few %. However, the systematic error arising mainly from the neutrino reconstruction will soon limit the experimental uncertainties.

As in the case of the inclusive measurement, which is discussed in the previous section, analyses with full reconstruction tagging improve the situation substantially, and make best use of the high luminosity at SuperKEKB/Belle. Figure 5.49 compares the missing mass resolution for (a) full reconstruction tagging (for $B \to D^0\ell^+\nu$ with 78 fb$^{-1}$), and (b) classical $\nu$ reconstruction (for $B \to \omega\ell^+\nu$ with 78 fb$^{-1}$) in the present Belle analyses. An improvement in the FWHM by almost a factor of 50 is seen for the full reconstruction analysis. We can also consider semileptonic tagging, where one tags more abundant $B \to D^{(*)}\ell^+\nu$ decays in the accompanying $B$ decays. This technique provides about 4 times more statistics by sacrificing purity and $q^2$ resolution. Belle has reported an inclusive analysis result using this method [199].

Figure 5.50 shows the expected improvement of the experimental error in $|V_{ub}|$ as a function of the integrated luminosity $L$, for the whole $q^2$ and the high $q^2$ regions. Here, the errors are estimated by extrapolating the present Belle analysis using semileptonic tagging, and are compared with the classical neutrino reconstruction. One can see that the classical neutrino reconstruction will soon hit the systematic limit. At a few times 100 fb$^{-1}$, the tagged-$B$ analysis will provide more precise results. The experimental precisions in $|V_{ub}|$ expected at 5 ab$^{-1}$ and 50 ab$^{-1}$ are 2.3% and 1.9%, respectively, for the whole $q^2$ region, and 2.9% and 2.1%, respectively, for the $q^2 > 16$ GeV$^2$ region.

Theoretical uncertainties are dominated by lattice QCD, which currently stands at 12% but is anticipated to go as low as 5%. To date, and for the forseeable future, the uncertainty from the lattice will set the lower bound on the uncertainty on $|V_{ub}|$. Nonetheless exclusive measurements, though not as precise as their inclusive counterparts, are very important. Although not yet significant the values of $|V_{ub}|$ as calculated from inclusive measurements are systematically higher than those from exclusive processes (see for example Ref. [22] for a recent survey of results). Whether this discrepancy will persist is an open question and can only be resolved with precise measurements, those which are only attainable from a SuperB factory.

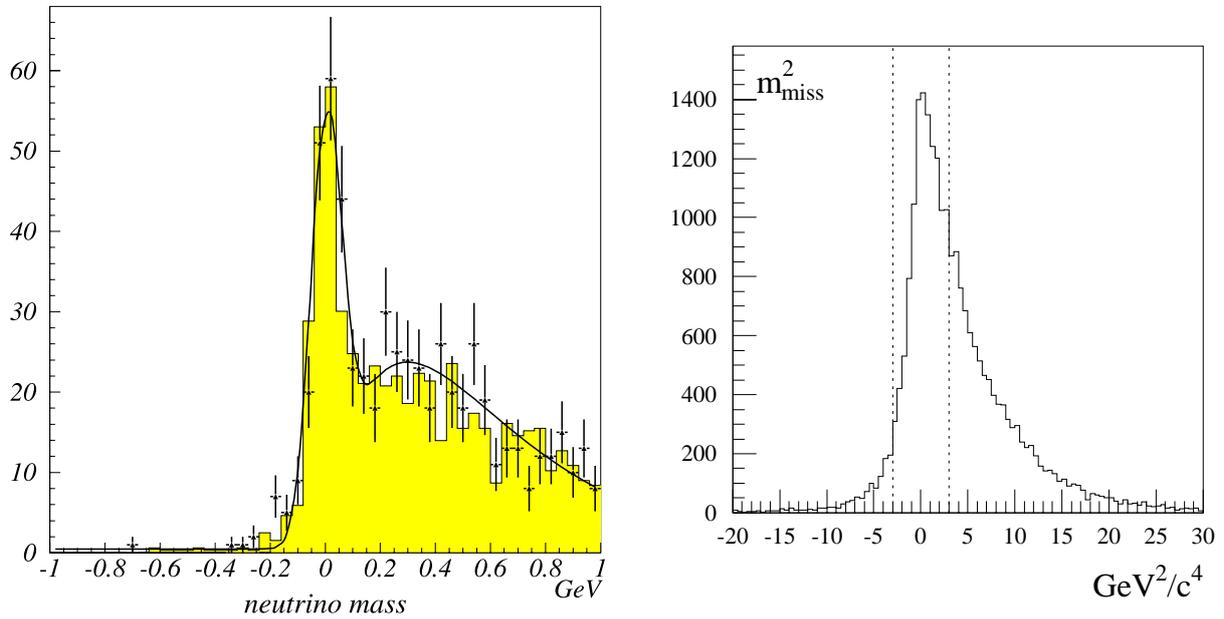

Figure 5.49: Missing mass resolution in (left) a full reconstruction analysis and (right) a classical neutrino reconstruction analysis.

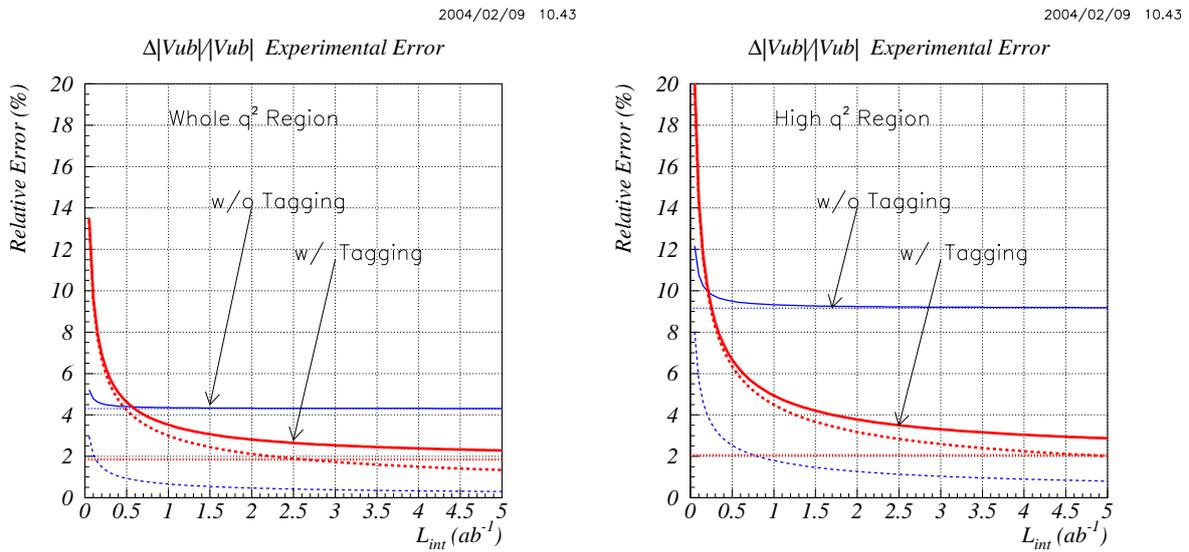

Figure 5.50: Expected improvement of the experimental error in $|V_{ub}|$ (solid lines) as a function of the integrated luminosity $L$, (left) for the whole $q^2$ and (right) the high $q^2$ regions.

## 5.10 Tau decays

### 5.10.1 Introduction

Lepton flavor violation (LFV) in $\tau$ decays is one of the most important physics target in the Super KEKB project. Here we give a brief review of various theoretical predictions available in recent years, and the experimental achievements and prospects at the Super KEKB experiment.

Lepton flavour conservation in the Standard Model (SM) is associated with neutrinos being massless. Observations of neutrino oscillations imply a nonzero mass and hence the mixing of lepton flavours, which is violating the lepton flavor conservation. With finite but tiny masses compared to the weak scale ($m_W \sim 80.4$ GeV), charged lepton flavour violating processes are however strongly suppressed and beyond experimental reach, since

$$\mathcal{B}(\tau \to l\gamma) = \frac{3\alpha}{32\pi} |\sum_i U_{\tau i}^* U_{\mu i} \frac{\triangle_{3i}^2}{m_W^2}|^2 \leq 10^{-53} \sim 10^{-49}, \quad (5.87)$$

where $U$ is the PMNS neutrino mixing matrix and $\triangle_{ij}^2$ is difference of neutrino mass squares, $\triangle_{ij}^2 = m_{\nu_i}^2 - m_{\nu_j}^2$ [200].

The situation is quite different if there are new particles which have masses of the order of the weak scale and couples to leptons. In fact, many extensions of SM, such as supersymmetry (SUSY), little Higgs models and extra dimensions predict enhanced LFV decays. Some of recent theoretical predictions relevant for the LFV $\tau$ decays are summarized in Table 5.19. For the LFV $\tau$ decays, the branching fractions can be as high as in the experimentally obtainable range $10^{-9} \sim 10^{-7}$ and the upper bound is in fact limited by recent B-Factory experiments results.

| model | Ref. |
| --- | --- |
| SUSY + Seesaw | [207], [208], [209], [201], [202] |
| SUSY + GUTs | [203], [204], [205], [206], [210] |
| SUSY(Higgs mediated) | [211], [212], [213], [214], [215], [216], [217] |
| Unconstrained MSSM | [218] |
| SUSY+R-parity violating | [219], [220] |
| Little Higgs | [221], [222] |
| Non-universal Z' | [223] |
| Extra-dimension | [224], [225], [226] |
| Left-Right Symmetric model | [227] |
| $\nu_R$ at O(10TeV) | [228], [229] |
| Others | [230] |

Table 5.19: A compilation of the theoretical predictions of the LFV $\tau$ decays.

The current limit on the $\mu \to e\gamma$ is set at $\mathcal{B}(\mu^+ \to e^+\gamma) < 1.2 \times 10^{-11}$ (90% C.L.) by the MEGA experiment at the Los Alamos Meson Physics Facility [231]. The MEG experiment at PSI is now starting the physics run with a sensitivity with $10^{-13}$ [232]. For the $\mu - e$ conversion experiments, two experiments at sensitivity of $10^{-16}$ are proposed at J-PARK [233] and FNAL [234]. And ambiguous future project at a sensitivity of $10^{-18}$ is designed at J-PARK [235].

There are exciting projects on the LFV searches on the muon sector [236]. However, even if $\mu^+ \to e^+\gamma$ is discovered at some levels, it will not provide sufficient information to determine the underlying LFV mechanism or even identify the correct underlying theory. In addition we do not know which LFV decay mode will first be discovered. For example, even in the presence of the bound of $\mu \to e\gamma$ at $O(10^{-11})$ the rate of $\tau \to \mu\gamma$ can be as high as $O(10^{-7})$ in the MSSM model. This is because different off-diagonal elements of the slepton mass matrix $(m_{\tilde{L}})_{ij}$ contribute to $\mu^- \to e^-\gamma$, and $\tau^- \to \mu^-\gamma$, $\tau^- \to e^-\gamma$. These elements $(m_{\tilde{L}})_{ij}$ are unique source of the LFV in the MSSM model and are free parameters of model as described below.

**Radiative LFV $\tau$ decays**

In SUSY models, LFV $\tau$ decays can receive significant contributions from diagrams containing SUSY particles in the loops, via slepton mixing. In this case, the largest branching fraction is expected in the radiative $\tau$ decays such as $\tau^- \to \mu\gamma$ and $\tau^- \to e\gamma$.

The branching fraction for $\tau \to \mu(e)\gamma$ for large $\tan\beta$ region can be parametrized using three parameters

$$\begin{aligned}
B(\tau \to \mu(e)\gamma)_\gamma &= \frac{12\pi\alpha^3}{G_F^2} \frac{1}{\sin^4\theta_w} |\delta_{3i}|^2 \frac{\tan^2\beta}{m_{\text{SUSY}}^4} \left(\frac{1}{30} + \frac{\tan^2\theta_w}{24}\right)^2 \\
&\simeq (1.5 \times 10^2)|\delta_{3i}|^2 \left(\frac{100\text{GeV}}{m_{\text{SUSY}}}\right)^4 \left(\frac{\tan\beta}{60}\right)^2,
\end{aligned} \quad (5.88)$$

where $m_{\text{SUSY}}$ is a typical SUSY particle mass ($m_{\text{SUSY}} = m_0 = m_{1/2}$) and $\tan\beta$ is the ratio of the vacuum expectation values of the neutral Higgs bosons. $\delta_{3i}$ is the slepton mass matrix divided by $m_{SUSY}^2$,

$$\delta_{3i} = \left(\frac{(m_L^2)_{3i}}{m_{\text{SUSY}}^2}\right),$$

where $i = 2$ for $\tau^- \to \mu\gamma$ and $i = 1$ for $\tau^- \to e\gamma$. Since the off-diagonal components of the slepton mass matrix are not determined within the SUSY frame, one needs to specify models such as SUSY+seesaw and SUSY+GUTS in order to predict the value of $\delta_{3i}$.

Figure 5.51, for example, shows the prediction of $\mathcal{B}(\tau \to \mu\gamma)$ as a function of $m_{\text{SUSY}} = M_{1/2}$ at fixed $\tan\beta$. The plots are obtained by scanning the LHC accessible SUSY-GUT(SO(10)) parameter space [210]. In the same scenario, the expected correlation between $\mathcal{B}(\tau \to \mu\gamma)$ and $\mathcal{B}(\mu \to e\gamma)$ is shown in Fig. 5.52. In some of the parameter space, $\mathcal{B}(\tau \to \mu\gamma)$ can be as high as $\sim 10^{-8}$ even though $\mathcal{B}(\mu \to e\gamma)$ is only at the $10^{-13}$ level.

In the MSSM see-saw and/or GUT scenario the $\delta_{ij} = (m_{\tilde{L}})_{ij}/m_{\text{SUSY}}$ terms are related with the physics parameters at very high (GUT) scales, such as the right-handed neutrino masses $M_{R,j}$ and their mixings, which are unknown. In other words, if supersymmetry particles are discovered at the LHC, the charged lepton flavor violation can become a powerful tool to explore the physics at very high (GUT) energy scale.

**Higgs mediated LFV processes**

The radiative LFV decay rate (Eq.(5.88)) exhibiting a $m_{\text{SUSY}}^{-4}$ dependence becomes small when a typical SUSY mass is larger than $\sim 1$ TeV. In that case the modes mediated by the Higgs boson exchange, such as decays to 3 leptons ($\tau^- \to 3\ell$) or lepton plus scalar/vector ($\tau \to \ell\eta, \ell\eta'$, $\ell f_0(980)$) become relatively important. The predictions for branching fractions of various LFV processes mediated by the Higgs bosons are shown in Fig. 5.53 as a function of Higgs mass. Note that the radiative LFV decays also occur from the Higgs mediated diagrams. Recently calculations taking into account the full diagrams were carried out by M. Herrero el al. in the SUSY-seesaw scenario without mass-insertion approximation [216]. One of the results for

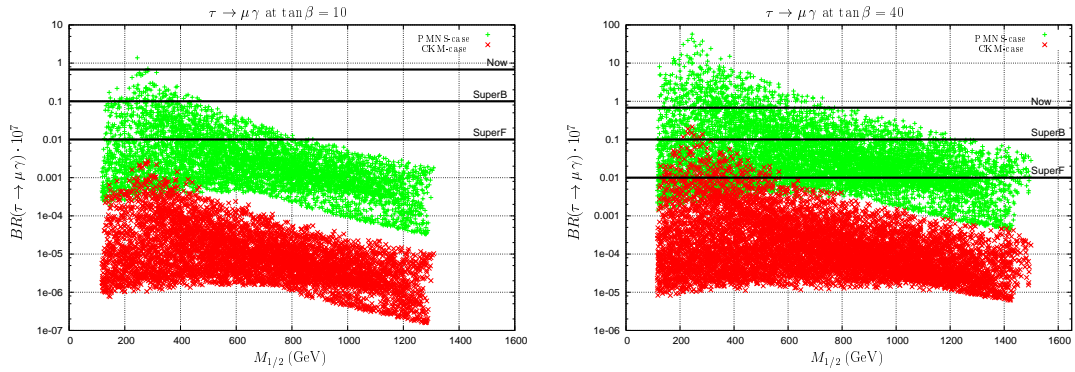

Figure 5.51: Scatter plot of $\mathcal{B}(\tau \to \mu\gamma)$ vs. $M_{1/2}$. The plots are obtained by scanning the LHC accessible SUSY-GUT(SO(10)) parameter space at fixed $\tan\beta$. Green and red points indicate two different assumptions for the neutrino Yukawa coupling ($Y_\nu$) [210].

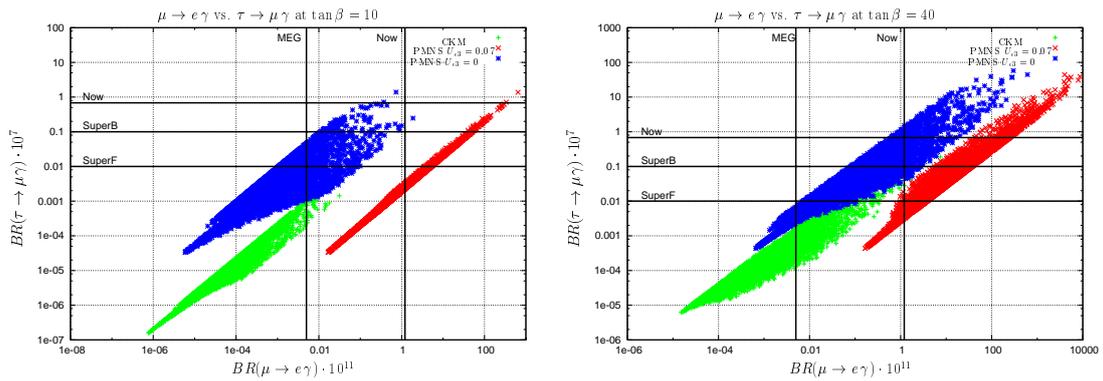

Figure 5.52: Correlation between $\mu \to e\gamma$ and $\tau \to \mu\gamma$ in SUSY-GUTs scenario. The plots are done by scanning the LHC accessible parameter space at fixed $\tan\beta$. Lines denote the present bounds and future sensitivities [210].

$\tau \to \mu\eta$ and $\tau \to \mu f_0(980)$ is given in Fig. 5.54, where the dependence on the right-handed neutrino masses at very high energy (GUT scale) are given. Larger LFV rates are expected for higher mass of right-handed neutrino, $m_{N_3} = O(10^{14})$GeV.

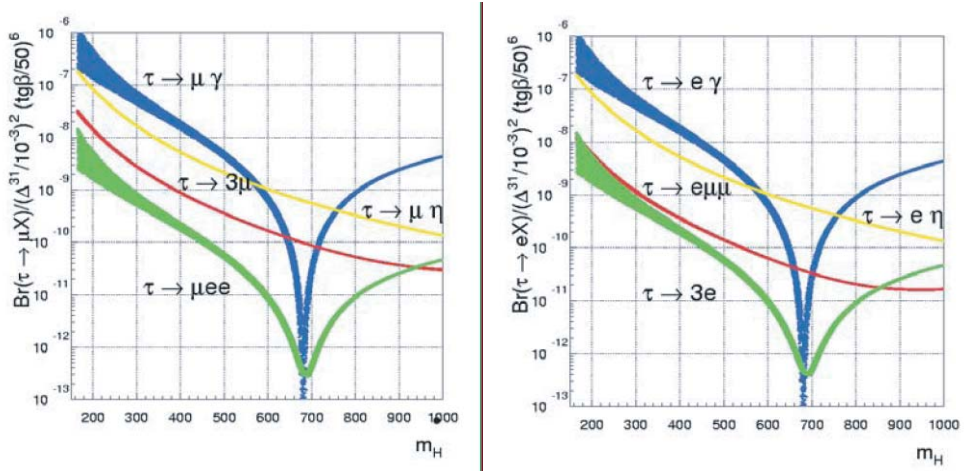

Figure 5.53: Branching fractions of various $\tau \to \mu X$ and $\tau \to eX$ LFV processes vs. the Higgs boson mass $m_H$ in the decoupling limit $m_{SUSY} \gg m_W$ for $X = \gamma, \mu\mu, ee, \eta$ [215].

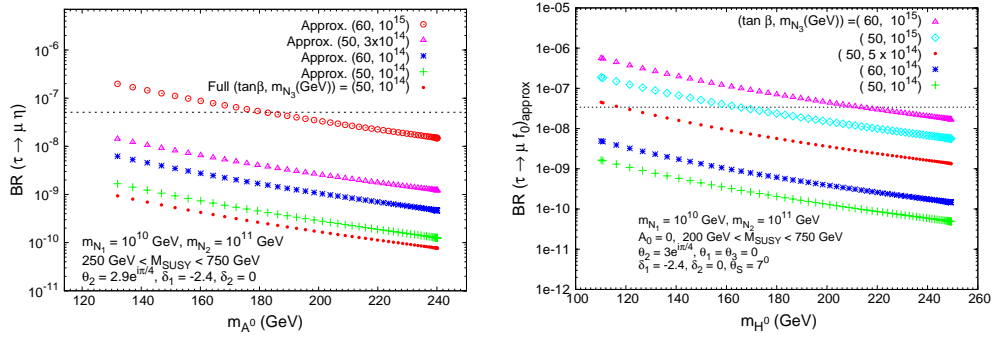

Figure 5.54: Branching fraction of $\tau \to \mu\eta$ and $\tau \to \mu f_0(980)$ as a function of the relevant Higgs masses in the SUSY-seesaw models. The horizontal dashed line in each plot is the present upper bound [217].

Based on the full-diagram calculation, they aprovide approximate formula for $\mathcal{B}(\tau^- \to \mu^- \eta)$ and $\mathcal{B}(\tau^- \to \mu^- f_0(980))$ [202];

$$\mathcal{B}(\tau^- \to \mu^- \eta)_{\text{Happrox}} = 1.2 \times 10^{-7} \times |\delta_{32}|^2 \left(\frac{\tan\beta}{60}\right)^6 \times \left(\frac{100\text{GeV}}{m_A}\right)^4, \quad (5.89)$$

$$\mathcal{B}(\tau^- \to \mu^- f_0(980))_{\text{Happrox}} = (0.42 - 7.3) \times 10^{-8} \times |\delta_{32}|^2 \left(\frac{\tan\beta}{60}\right)^6 \times \left(\frac{100\text{GeV}}{m_H^0}\right)^4, \quad (5.90)$$

where $m_A$ and $m_H^0$ are the pseudo-scalar and scalar Higgs masses, respectively.

**Model dependence**

In general, full set of measurements of $\mu$ and $\tau$ LFV processes is required to identify the underlying physics because in many models there are strong correlations between the expected rates of various channels. For example, the ratios of various branching fractions in the little Higgs model with T-parity and MSSM without and with significant Higgs contributions, shown in Table 5.20, indicate clear difference between the underlying physics scenario. Searches for various LFV $\tau$ decays will be crucial to discriminate the possible underlying models.

| ratio | LHT | MSSM($\gamma$) | MSSM (Higgs) |
|---|---|---|---|
| $\frac{\mathcal{B}(\tau^-\to e^-e^+e^-)}{\mathcal{B}(\tau^-\to e^-\gamma)}$ | 0.4 ... 2.3 | $\sim 1\cdot 10^{-2}$ | $\sim 1\cdot 10^{-2}$ |
| $\frac{\mathcal{B}(\tau^-\to \mu^-\mu^+\mu^-)}{\mathcal{B}(\tau^-\to \mu^-\gamma)}$ | 0.4 ... 2.3 | $\sim 2\cdot 10^{-3}$ | 0.06...0.1 |
| $\frac{\mathcal{B}(\tau^-\to e^-\mu^+\mu^-)}{\mathcal{B}(\tau^-\to e^-\gamma)}$ | 0.3 ... 1.6 | $\sim 2\cdot 10^{-3}$ | 0.02...0.04 |
| $\frac{\mathcal{B}(\tau^-\to \mu^-e^+e^-)}{\mathcal{B}(\tau^-\to \mu^-\gamma)}$ | 0.3 ... 1.6 | $\sim 1\cdot 10^{-2}$ | $\sim 1\cdot 10^{-2}$ |
| $\frac{\mathcal{B}(\tau^-\to e^-e^+e^-)}{\mathcal{B}(\tau^-\to e^-\mu^+\mu^-)}$ | 1.3 ... 1.7 | $\sim 5$ | 0.3...0.5 |
| $\frac{\mathcal{B}(\tau^-\to \mu^-\mu^+\mu^-)}{\mathcal{B}(\tau^-\to \mu^-e^+e^-)}$ | 1.2 ... 1.6 | $\sim 0.2$ | 5...10 |
| $\frac{\mathcal{B}(\mu^-\to e^-e^+e^-)}{\mathcal{B}(\mu^-\to e^-\gamma)}$ | 0.4 ... 2.3 | $\sim 6\cdot 10^{-3}$ | $\sim 6\cdot 10^{-3}$ |
| $\frac{\mathcal{B}(\mu^-\text{ Ti}\to e^-\text{Ti})}{\mathcal{B}(\mu^-\to e^-\gamma)}$ | $10^{-2}...10^2$ | $\sim 5\cdot 10^{-3}$ | 0.08...0.15 |

Table 5.20: Comparison of various ratios of branching fractions in the little Higgs model with T-parity (LHT) and in the MSSM without and with significant Higgs contributions [221].

### 5.10.2 Status of B-factory experiments

B-factory experiments (Belle and Babar) have been collecting data at the $\Upsilon(4S)$ resonance since 1999. Babar has accumulated 575 fb$^{-1}$ data and stopped data taking in 2008, while Belle has accumulated more than 1000 fb$^{-1}$ data until winter 2010. With more than 1.5ab$^{-1}$ of data currently being collected by the two experiments, and the $e^+e^- \to \tau^+\tau^-$ cross section of 0.919 nb, the total sample of $\tau$ pairs at the $e^+e^-$ colliders now exceeds $10^9$ $\tau$, which allows for probing the LFV processes at $O(10^{-8}) - O(10^{-7})$ levels.

The general approach of the LFV $\tau$ decays searches is based on common procedures of selection of the charged and neutral final state particles, removing non-$\tau$ events and keeping the signal efficiency as high as possible. This is accomplished by dividing the candidate event into hemispheres in the center of mass frame (Fig 5.58(a)). Each hemisphere is then considered as a possible candidate for the LFV decay under consideration. The experimental signature is

$$\{\tau^- \to \text{LFV decay, e.g. } \ell^-\gamma, \ell^-h^0\ldots\} \;+\; \{\tau^+ \to (\text{track})^+ + (n_\gamma^{\text{TAG}} \geq 0) + \text{missing}\},$$

or a charge conjugate of the above. To search for exclusive decay modes, we select low-multiplicity (2 or 4 charged tracks) events with zero net charge, and separate the signal side and tag side into two hemispheres using the thrust axis. While these requirements remove $B\bar{B}$ and continuum events, a significant amount of Bhabha, muon-pair and two-photon processes remains. Generic $\tau\tau$ events cannot be exclusively reconstructed since $\tau$ decays include neutrinos. Therefore the background contamination from generic $\tau\tau$ events cannot be totally avoided. Unlike the SM $\tau$ decays which have at least one neutrino, the LFV decay products have no undetectable particles, and one can apply tight conditions on the energy $E_X$ and invariant mass $m_X$ of the decay products. $E_X$ is approximately equal to the beam energy in the center of mass frame and the decay products invariant mass $m_X$ is equal to that of $\tau$. A two-dimensional signal region in the $m_X$ and $\Delta E = E_{\text{beam}} - E_X$ plane is thus used to separate possible signal from the SM $\tau$ decay backgrounds. In order to optimize the signal sensitivity an elliptic shape is used for the signal area as shown in Fig. 5.55. The signal regions are determined by scanning ellipse parameters to minimize the ellipse area and to obtain the highest sensitivity.

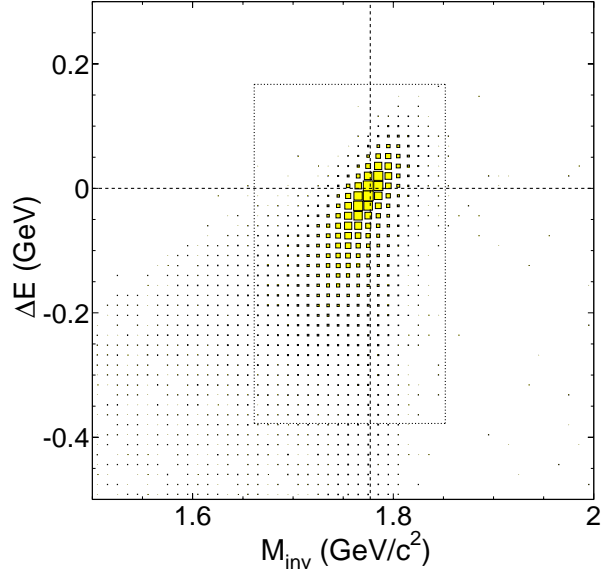

Figure 5.55: $\Delta E$ vs. $M_{\text{inv}}$ distribution for $\tau^- \to \mu^- \gamma$ decay.

Selection criteria are optimized using Monte Carlo simulation of the signal and backgrounds to give the best discovery potential of the signal. For reduction of the background events, we introduce kinematic constraints. For example, we apply a requirement on the relation between the missing momentum ($p_{\text{miss}}$) and the missing mass-squared ($m^2_{\text{miss}}$) for various $\tau$ LFV analyses. This cut reduces the background events quite effectively; 98% of generic $\tau$ pairs and 80-90% of radiative Bhabha, muon-pair and continuum events are removed, while 76% of the signal events is retained for the $\tau^- \to \mu^- \gamma$ analysis (see Figure 5.56).

Typical values of signal efficiency are between 3% to 10%, depending on the mode under study. The efficiency components of the generic LFV decays selection are roughly as follows: trigger (90% ), acceptance/reconstruction (70% ), charged-particle topology selection (70% ), particle identification (50 % ), additional background reduction (50 % ), $\Delta E$ vesus $m_X$ signal region requirement (50 % ).

The evaluation of the upper limit is based on the prescription of Feldman-Cousins and its extensions to include systematic errors.

### 5.10.3 Results on the LFV $\tau$ Decays from B-factory Experiments

Experimentally, LFV $\tau$ decays can be classified as, 1) radiative decays ($\tau \to \ell\gamma$) , 2) three leptons ($\tau \to 3\ell$), 3) lepton plus pseudo-scalar ($\tau \to \ell P^0$), 4) lepton plus vector ($\tau \to lV^0$), 5) lepton plus two hadrons and 6) lepton plus three hadrons. Since the background situation is different for the case 1) and others, we show them separately.

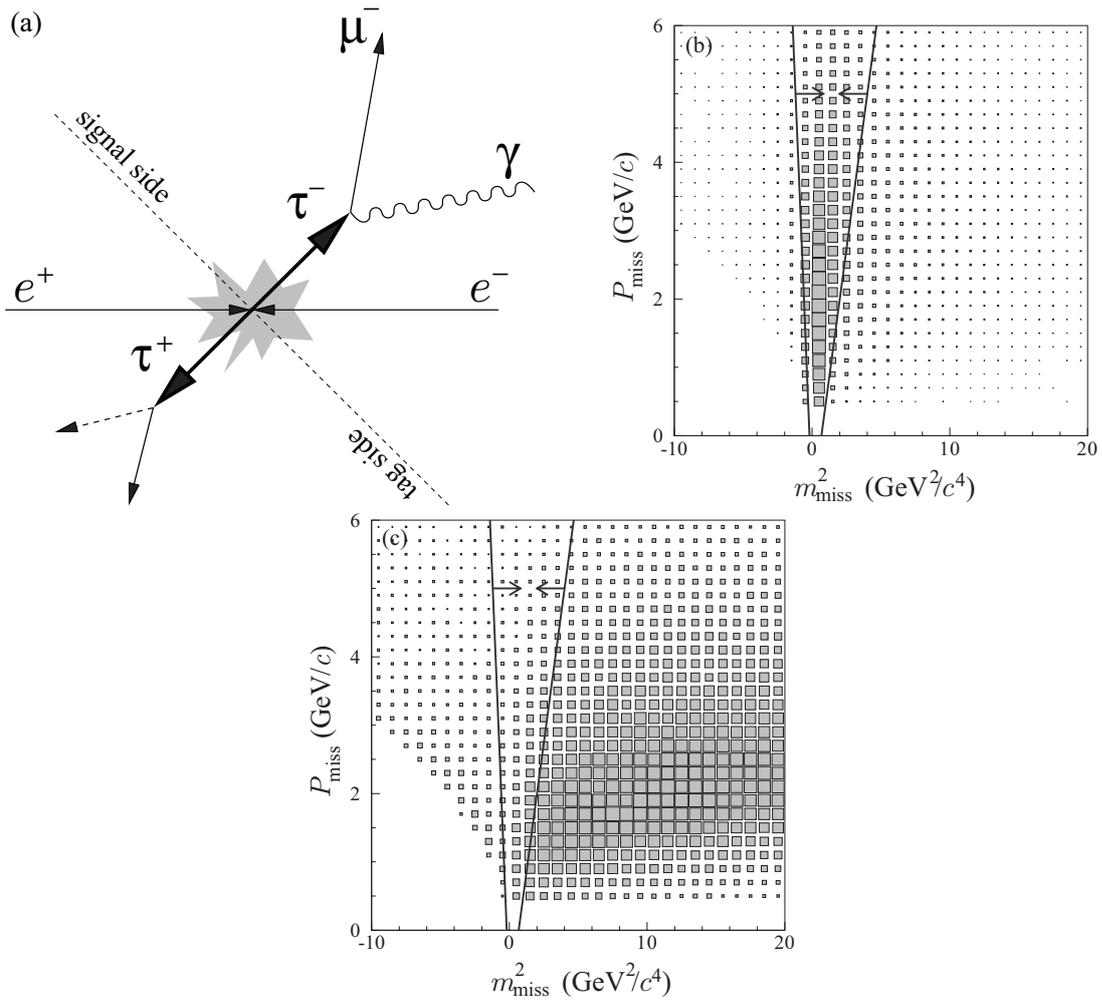

Figure 5.56: (a) Illustration of $e^+e^- \to \tau^+\tau^-$ event topology. Distribution of $p_{\text{miss}}$ vs. $m^2_{\text{miss}}$ for (b) $\tau^- \to \mu^-\gamma$ signal MC events and (c) generic tau pair MC. The area between the two lines is selected.

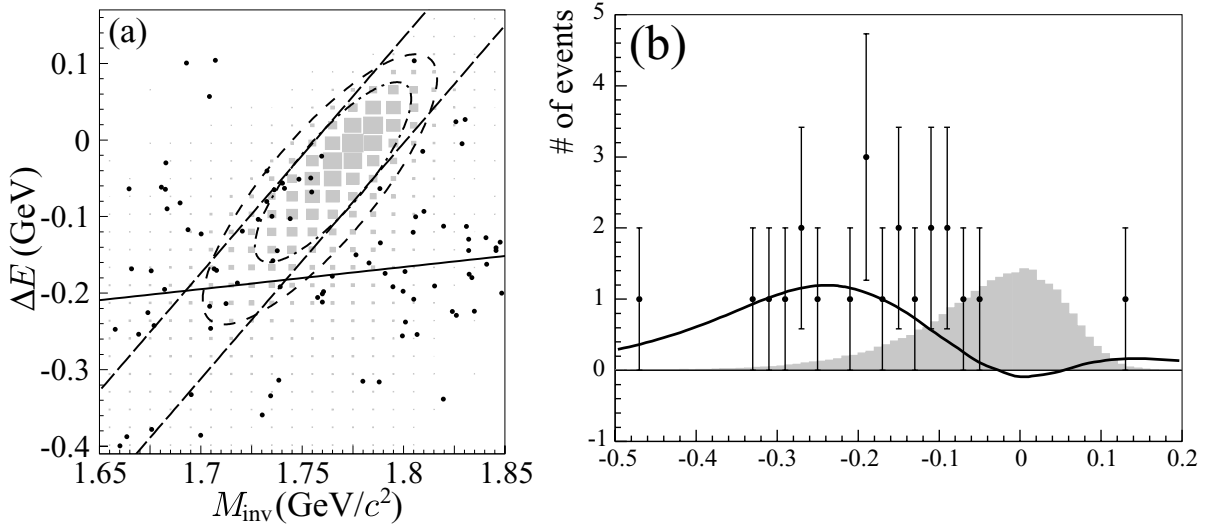

Figure 5.57: (a) $M_{\text{inv}} - \Delta E$ distribution for $\tau^- \to \mu^- \gamma$ decays with 535 fb$^{-1}$ of data. Dots are the data, and shaded boxes indicate the signal MC. The dashed ellipse shows the $3\sigma$ blinded region and the dot-dashed ellipse is the $2\sigma$ signal region. Dashed lines indicate the $\pm 2\sigma$ band of the shorter ellipse axis. The solid line indicates a dense background region. (b) $\Delta E$ data distribution in the $\pm 2\sigma$ $M_{\text{inv}} - \Delta E$ region. Points with error bars are data, and the shaded histogram is the signal MC assuming a branching ratio of $5 \times 10^{-7}$. The solid curve shows the best fit.

|  | BaBar | | Belle | |
|---|---|---|---|---|
|  | $\mathcal{B}(\times 10^{-8})$ | Lum.(fb$^{-1}$) | $\mathcal{B}(\times 10^{-8})$ | Lum.(fb$^{-1}$) |
| $\tau \to \mu + \gamma$ | 4.4 [237] | 470 | 4.5 [238] | 535 |
| $\tau \to e + \gamma$ | 3.3 [237] | 470 | 12 [238] | 535 |
| $\tau \to \ell + \ell' + \ell''$ | 1.8–3.3 [239] | 468 | 1.5–2.7 [240] | 782 |
| $\tau \to \ell + P^0$ | 11–16 [241] | 339 | 7–12 [242] | 401 |
| $\tau \to \mu/e + K_s^0$ | 4.0/3.3 [243] | 469 | 2.3/2.6 [244] | 671 |
| $\tau \to \ell + V^0$ | 2–18 [245] | 451 | 5.9–13 [246] | 543 |
| $\tau \to \mu/e + \omega$ | 10/11 [247] | 384 | 8.0/18 [246] | 543 |
| $\tau \to \ell f_0(980)$ | — | — | 3.2–3.4 [248] | 671 |
| $\tau \to \ell + h' + h''$ | 7.0–48 [249] | 221 | 3.3–16 [250] | 671 |
| $\tau \to \mu/e + K_s^0 K_s^0$ | — | — | 7.1/8.0 [244] | 671 |
| $\tau \to h + \Lambda/\bar{\Lambda}$ | 5.8–15 [251] | 237 | 7.2–14 [252] | 154 |

Table 5.21: Current upper limits for $\tau$ LFV decays with the amount of analyzed data at BaBar and Belle. The meaning of symbols: $\ell = e$ or $\mu$, $P^0 = \eta, \eta'$, or $\pi^0$, $h = \pi$ or $K$ and $V^0 = \rho^0, K^*, \bar{K}^{*0}$, or $\phi$. For $\tau \to \ell f_0(980)$, the upper limits are given for the product $\mathcal{B}(\tau \to \ell f_0(980)) \times \mathcal{B}(f_0 \to \pi\pi)$.

**$\tau \to \ell\gamma$ decays**

The $\tau^- \to \ell^-\gamma$ modes suffer from background as seen in Fig. 5.57. The dominant background remaining after the event selection originates from $\tau^+\tau^-$ events with initial state radiation, $\tau^+\tau^-\gamma$, where one of $\tau$'s decays semi-leptonically and the lepton and the radiated photon compose signal candidates. Multi-photon radiative muon-pairs (and Bhabha events) are the second largest background; one of leptons and a radiated photon form a signal candidate, and the other lepton satisfies the selection for the tag side. These backgrounds cannot be discriminated from the true signal. The background distribution is intensively studied using actual data and MC simulation. A thorough understanding of the origin and properties of the background was obtained. The number of signal events in the final sample were obtained by performing an unbinned extended maximum likelihood fit. The fit results are shown in Fig. 5.57 (b).

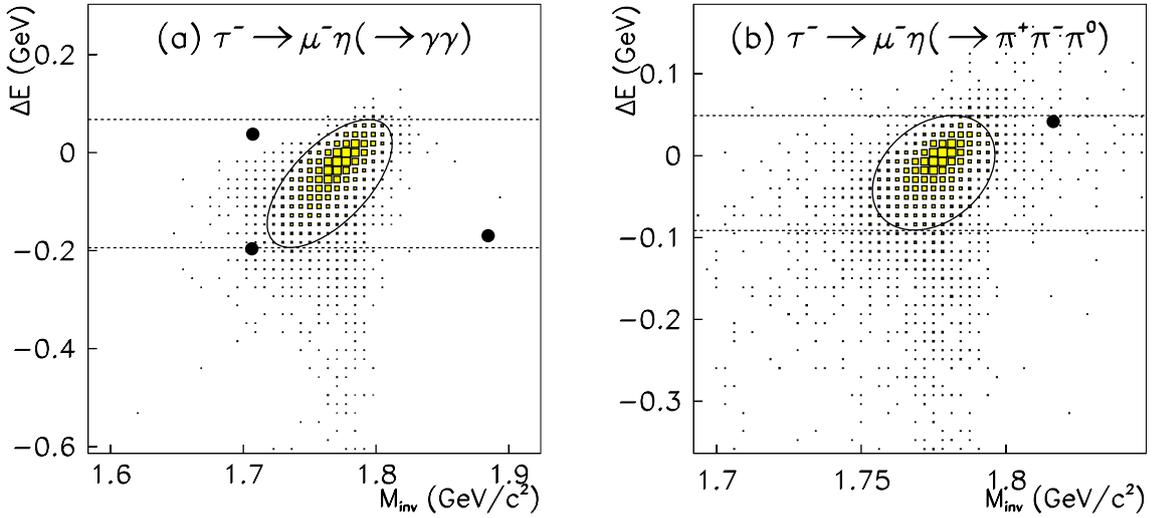

Figure 5.58: Scatter plots of data in the $M_{\rm inv} - \Delta E$ plane corresponding to the $\pm 10\sigma$ area for (a) $\tau^- \to \mu^-\eta(\eta \to \gamma\gamma)$, and (b) $\tau^- \to \mu^-\eta(\eta \to \pi^+\pi^-\pi^0)$ decays, respectively. The data are indicated by filled circles. Filled boxes show the MC signal distribution with arbitrary normalization. The elliptical signal region shown by the solid curve is used for evaluating the signal yield. The region between the dotted horizontal lines excluding the signal region is used to estimate the expected background in the signal region.

**$\tau \to 3\ell$ and $\tau \to \ell + P^0$ decays**

For the modes with three leptons or one lepton plus hadrons, the backgrounds are negligible at the current available luminosity. For $\tau^- \to \ell^- P^0 (P^0 = \eta, \eta', \pi^0, f_0(980))$ and $\tau^- \to \ell^-\ell^+\ell^-$ modes, no signal candidates were found in the signal regions (see Fig. 5.58 and Fig. 5.59). We evaluated the expected backgrounds in the signal region from the sidebands and extracted the number of signal candidates.

The current results from the B-factory experiments for various LFV $\tau$ decays are summarized in Table 5.21 and are shown in Fig. 5.60. We have studied 51 decay modes in total. In all modes, the sensitivity to $\tau$ LFV decays are improved by two orders of magnitude from the previous

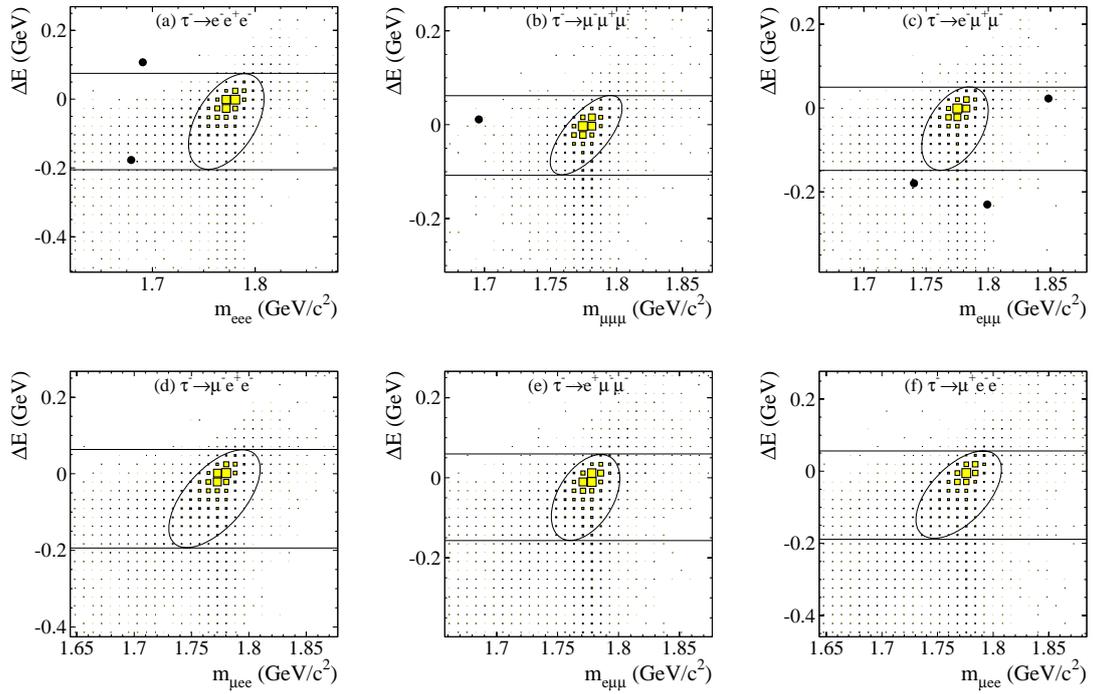

Figure 5.59: Scatter plots in the $M_{3l} - \Delta E$ plane: (a)-(f) correspond to the $\pm 20\sigma$ area for the $\tau^- \to e^-e^+e^-$, $\mu^-\mu^+\mu^-$, $e^-\mu^+\mu^-$, $\mu^-e^+e^-$, $e^+\mu^-\mu^-$ and $\mu^+-e^-e^-$ modes, respectively. The data are indicated by solid circles. Filled boxes show the MC signal distribution with arbitrary normalization. The elliptical signal regions shown by the solid curves are used for evaluating the signal yield. The region between the horizontal solid lines excluding the signal region is used to estimate the background expected in the signal region.

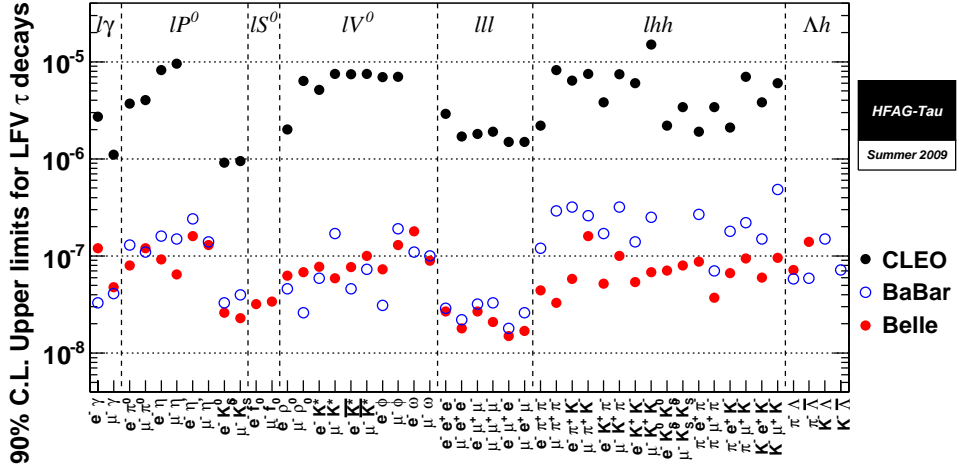

Figure 5.60: Upper limits on the LFV decays of the $\tau$ lepton.

CLEO experiment. Upper limits for branching fractions are at the $10^{-8}$ level and the searches are approaching regions sensitive to new physics.

In all modes, the sensitivity of *tau* LFV decays are improved by 2 order of magnitude from the previous CLEO experiment. Upper limits for branching fractions go down to the $10^{-8}$ level and the searches are approaching regions sensitive to the new physics.

### 5.10.4 Sensitivity to New Physics Model

The B-factory results significantly constrain the parameter space of NP models. As an example, Fig. 5.61 shows the excluded region in the $m_H - \tan\beta$ plane. The excluded region, however, depends on $\delta_{32}$ which is related to the physics at the very high energy (GUT) scale. If the value of $\delta_{32} > 1$, wide region at large $\tau\beta$ has already been excluded, while if $\delta_{32} < 0.1$ most of the shown $\tan\beta - m_{H^0}$ plane is still. If SUSY particles and Higgs bosons are discovered by the LHC experiments, LFV decays can be used to restrict the parameter $\delta_{32}$.

### 5.10.5 Prospects for SuperKEKB

With 50 ab$^{-1}$ of data accummulated at SuperKEKB the sample of $\tau$-pair events available for the measurements would increase to $5 \times 10^{10}$.

Figure 5.62 shows the history of the search for LFV $\tau$ decays and projected upper limits for SuperKEKB as a function of the luminosity for $\tau^- \to \mu^-\gamma$, $\mu^-\eta$ and $\mu^-\mu^+\mu^-$ decays. The expected sensitivities are obtained with an unbinned extended maximum likelihood method assuming the current background conditions. For $\tau^- \to \ell^-\gamma$ modes, the background level is not negligible and the upper limit is proportional to $1/\sqrt{N_{\tau\tau}}$, For the $\tau^- \to \ell^- h^0$ and $\tau \to \ell^-\ell^+\ell^-$ modes, the background is negligible and the upper limit is proportional to $1/N_{\tau\tau}$ up to several ab$^{-1}$, and then gradually changes to $1/\sqrt{N_{\tau\tau}}$ dependence as background candidates appear. Therefore, if the current signal-to-background condition is still maintained, the ultimate goal at 50 ab$^{-1}$ could be a branching fraction sensitivity of $1 \times 10^{-9}$.

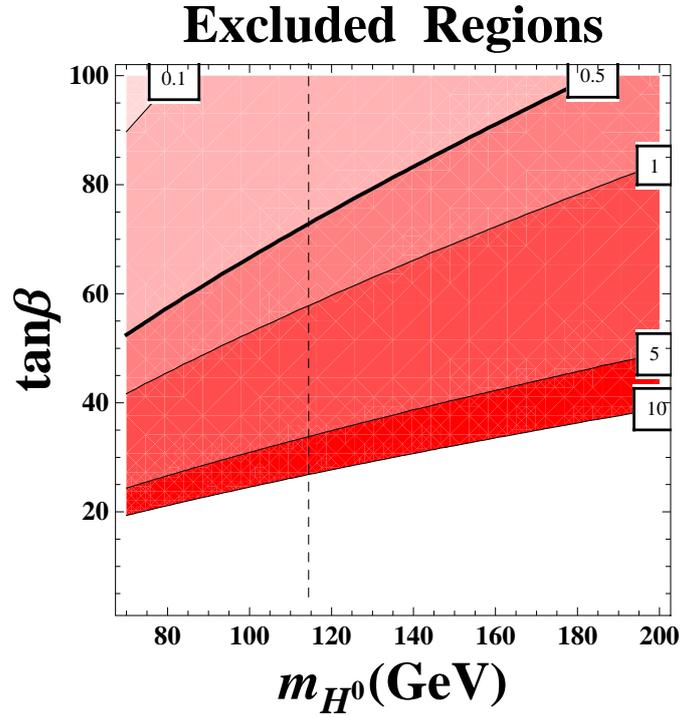

Figure 5.61: Excluded region in the $(m_{H^0}, \tan\beta)$ plane from the Belle upper limit $\mathcal{B}(\tau \to \mu f_0) < 3.4 \times 10^{-8}$. The excluded area are those above the contour lines corresponding to fixed $\delta_{32}| = 0.1, 0.5, 1, 5, 10$ [217].

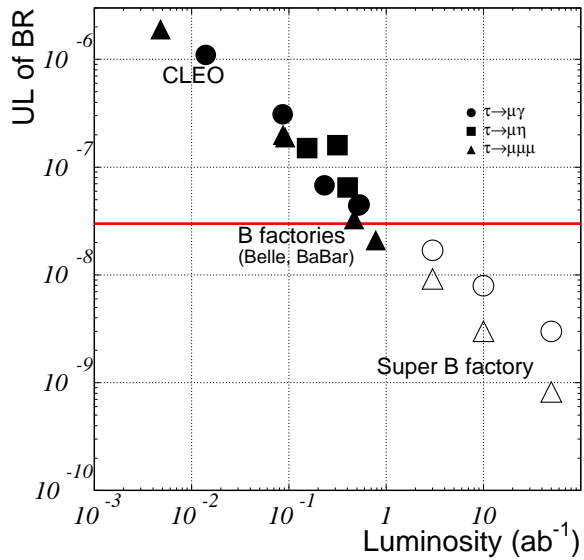

Figure 5.62: History and anticipated upper limits on branching fractions for $\tau^- \to \mu^-\gamma$, $\mu^-\eta$ and $\mu^-\mu^+\mu^-$ decays as a function of the integrated luminosity.

Can we further improve the sensitivity?

Experimental sensitivity is in general determined by three key elements: statistics, resolutions, and a signal-to-background ratio; the achieved sensitivity depends on how well the systematic uncertainty of these aspects is controlled. We discuss here how to improve the sensitivities for LFV $\tau$ decays based on the experience from the $\tau^- \to \ell^- \gamma$ analysis at Belle.

Besides accumulating higher luminosity, it is important to improve the detection efficiency including trigger performance. However, at SuperKEKB, the selection criteria will become tighter and efficiencies will be lower than those at the present $B$ factories.

**Resolution** Improvements of momentum and energy resolution for charged tracks and photons can reduce the size of the signal region in the $M_{\text{inv}}$ - $\Delta E$ plane, which results in a better signal-to-background ratio and improved sensitivity. For the $\tau^- \to \mu^- \gamma$ mode, the energy leakage in the calorimeter and initial state radiation cause a long low-energy tail. We thus need to use a wide signal region for a sufficient efficiency (see Figure 5.55). While the effect of initial state radiation cannot be controlled, the effect of the calorimeter leakage can be improved, for instance by using longer crystals and by improving the reconstruction method to reduce lateral energy leakage.

**Particle identification** Improvements of particle identification are also essential for all the LFV decay modes. The efficiency of $e$-id is already at 98% and further improvements are neither feasible not very helpful. On the other hand, the $\mu$-id efficiency is around 90% and the efficiency increase may reduce the background contamination from events of the type , such as $(\mu\gamma)$+not$-\mu$ events in the $\tau^- \to \mu^- \gamma$ search. A KLM counter with finer segmentation in the radial direction could provide a better muon identification. K/$\pi$ separation is also important in searches for LFV $\tau$ decays that include hadrons in the final state. The Time-Of-Propagation counter is a very good candidate for the improvement.

**Signal-to-background** A requirement on the missing quantities, $p_{\text{miss}} - m^2_{\text{miss}}$, plays an important role for background suppression as seen in Figure 5.56. These criteria provide much better sensitivity than what was achieved by the CLEO experiment. We also apply a new selection for $\tau^- \to e^- \gamma$, using the opening angle between the tagged track and missing particle direction; this criterion effectively removes the radiative Bhabha background events. We need to introduce these kind of selection criteria in all future analyses.

### 5.10.6 Summary

A search for Lepton Flavour Violation is a well motivated part of the SuperKEKB program. Experimental observations of the charged lepton LFV decays would represent an unambiguous evidence of new physics (NP). Even if such processes are not observed, experimental searches and the resulting upper limits on the branching fractions are a powerful tool for constraining the parameter space of NP models.

Many models beyond the SM allow for LFV in $\tau$ decays and a wide range of the parameter space is allowed at present. Some models predict relatively large branching fractions while others give extremely small values. Systematic and extensive investigation of various $\tau$ decay modes can provide a powerful means to select models, restrict their parameter space, and possibly discover new phenomena beyond the Standard Model. Any measured SUSY signal (e.g. at the LHC) would immediately provide constraints on the LFV charged fermion decays.

## 5.11 $B_s^0$ physics at $\Upsilon(5S)$

### 5.11.1 Overview

The physics potential of the Super KEKB collider can be extended significantly by running at the energy of the $\Upsilon(10860)$ resonance, usually denoted as the $\Upsilon(5S)$. The advantage is that at this center-of-mass energy it is possible to produce pairs of $B_s^0/\bar{B}_s^0$ mesons, which are kinematically forbidden when running at the $\Upsilon(4S)$ resonance. This would provide an opportunity to study $B_s^0$ decays in a relatively low background environment as compared to that at a hadron collider. In order to operate the KEKB accelerator in the $\Upsilon(5S)$ mass range, the electron and positron beam energies would need to be increased by 2.7% relative to their energies at the $\Upsilon(4S)$ resonance; this increase would result in the same Lorentz boost factor of $\beta\gamma = 0.425$. A technical capacity to increase the boost would be valuable for $B_s^0$ meson studies, because it could provide an opportunity for a better $B_s^0$ vertex spatial separation and, possibly, measurement of $B_s^0$-mixing.

In order to estimate an expected accuracy of $B_s^0$ measurements, the number of $B_s^0$ mesons in a $\Upsilon(5S)$ data sample has to be determined. To calculate the $B_s^0$ number in a dataset with known integrated luminosity $\mathcal{L}_{\text{int}}$, the two parameters should be measured: the total $b\bar{b}$ production cross section at the $e^+e^-$ center-of-mass energy $\sigma_{b\bar{b}}^{\Upsilon(5S)}$, and the fraction $f_s$ of $B_s^0$ events among all $b\bar{b}$ events. Then the number of $B_s^0$ mesons in a dataset can be calculated as: $N_{B_s^0} = 2 \times \mathcal{L}_{\text{int}} \times \sigma_{b\bar{b}}^{\Upsilon(5S)} \times f_s$. By now CLEO [253] and Belle [254] experiments have measured the $\sigma_{b\bar{b}}^{\Upsilon(5S)}$ with the mean PDG value of $(0.302 \pm 0.014)$ nb and the $f_s$ with the mean PDG value of $f_s = (19.5^{+3.0}_{-2.3})$%. Using these values the number of $B_s^0$ mesons is estimated to be $\sim 5.9 \times 10^8$ in the dataset of $\mathcal{L}_{\text{int}} = 5$ ab$^{-1}$ taken at the $\Upsilon(5S)$.

$B_s^0$ mesons can be produced in the $\Upsilon(5S)$ decays to $B_s^0\bar{B}_s^0$, $B_s^0\bar{B}_s^*$, $B_s^*\bar{B}_s^0$ or $B_s^*\bar{B}_s^*$ intermediate decay channels. The excited state decays to the ground state via $B_s^* \to B_s^0\gamma$. For some $B_s^0$ studies with a high background level it is reasonable to select events only from the $B_s^*\bar{B}_s^*$ channel. In this case the fraction $f_{B_s^*\bar{B}_s^*}$ of $B_s^*\bar{B}_s^*$ events over all $B_s^{(*)}\bar{B}_s^{(*)}$ events should be also determined. This fraction was recently measured by Belle [255] to be $f_{B_s^*\bar{B}_s^*} = (90.1^{+3.8}_{-4.0} \pm 0.2)$%.

A detailed Monte Carlo simulation shows that, using the full reconstruction technique, $B_s^0$ signals are well-identified using the variables $E_B^* - E_{beam}$ and $P_B^*$, where $E_B^*$ and $P_B^*$ are the reconstructed energy and momentum of the $B_s^0$ candidate in the $e^+e^-$ center-of-mass frame, and $E_{beam}$ is the beam energy in this frame. Extracting the total yield in the $B_s^0$ signal regions would allow one to measure absolute (rather than relative) branching fractions, which are very difficult to measure at a hadron machine.

The first evidence for $B_s^0$ production at the $\Upsilon(5S)$ was found by the CLEO collaboration [253, 256] using a dataset of 0.42 fb$^{-1}$ collected in 2003. To test the feasibility of a $B_s^0$ physics program the Belle collaboration collected at the $\Upsilon(5S)$ a dataset of 1.86 fb$^{-1}$ in 2005. After the successful analysis of these data [254, 257], Belle collected a larger sample of 21.7 fb$^{-1}$ in 2006. At the end of 2008 the Belle collaboration was taking data at the $\Upsilon(5S)$ for about two months and an additional dataset of about 30 fb$^{-1}$ was collected.

Many $B_s^0$ decay modes can be observed and studied at an $e^+e^-$ Super B Factory. The strange "partners" of topical $B_d$ decays can be reconstructed, such as Cabibbo-favored $B_s^0 \to D_s^-\pi^+$ and $B_s^0 \to D_s^-D_s^+$ decays, color-suppressed $B_s^0 \to \bar{D}^0 K^0$ and $B_s^0 \to J/\Psi\phi$ decays, Cabibbo-suppressed $B_s^0 \to D_s^- K^+$ and $B_s^0 \to J/\Psi K^0$ decays, the electroweak penguin decay $B_s^0 \to \phi\gamma$, and the $b$ to $u$ transition $B_s^0 \to K^-\pi^+$ as well as many other charmless decays [258]. Final states containing photons, $\pi^0$ and $\eta$ mesons, which are difficult to separate from background in a hadron collider environment, would be well-identified at an $e^+e^-$ machine. In addition, $\Upsilon(5S)$ decays are well-suited for studying large multiplicity $B_s^0$ decays due to the lower particle momenta, the almost 100% trigger efficiency, and the excellent $\pi/K$ discrimination. Inclusive $B_s^0$ branching fractions can also be measured, in particular the inclusive leptonic and inclusive fast photon branching fractions. Theoretical predictions for inclusive channels are usually more reliable and, therefore, well suited for comparison with experimental measurements. Partial-reconstruction techniques can also be used, for example, to measure the exclusive $B_s^0 \to D_s^+ l^- \nu$ and $B_s^0 \to K^+ l^- \nu$ decays. The numbers of $B_s^0$ decays reconstructed in several topical decay modes are listed in Table 5.22, along with the corresponding numbers of events (when available) for the LHCB [259, 260] hadron collider experiment. The event yields listed correspond to one year of running; the corresponding luminosity at the Super $B$ Factory would be $5\,\text{ab}^{-1}$.

| Decay | Branching fraction | Super $B$ ($5\,\text{ab}^{-1}$) | LHCB (per year) |
|---|---|---|---|
| $D_s^- \pi^+$ | $3.7 \cdot 10^{-3}$ | 34 K | 80 K |
| $J/\Psi \phi$ | $1.2 \cdot 10^{-3}$ | 9 K | 120 K |
| $K^- \pi^+$ | $5.0 \cdot 10^{-6}$ | 0.5 K | 5.3 K |
| $\phi \gamma$ | $4.0 \cdot 10^{-5}$ | 4 K | 9.3 K |
| $D_s^{(*)+} D_s^{(*)-}$ | $5 \cdot 10^{-2}$ | 2.5 K | |

Table 5.22: The number of fully-reconstructed $B_s^0$ mesons expected from $\Upsilon(5S)$ decays with $\mathcal{L}_{int} = 5\,\text{ab}^{-1}$, and by the LHCB experiment per year of running.

The time-integrated $CP$ asymmetry $A \equiv (N_{\overline{B}_s \to f} - N_{B_s \to \bar{f}})/(N_{\overline{B}_s \to f} + N_{B_s \to \bar{f}})$ would allow one to measure *direct $CP$* violation in $B_s^0$ decays. A good candidate for this measurement is $B_s^0 \to K^- \pi^+$: assuming a branching fraction of $5 \times 10^{-6}$ and an asymmetry of $\sim 40\%$ [261], one would observe a $\sim 5\sigma$ effect with $5\,\text{ab}^{-1}$ of data. The penguin diagram causing the direct $CP$-violation is very sensitive to nonstandard contributions, and measuring a direct $CP$ asymmetry could probe physics beyond the Standard Model.

### 5.11.2 Measurement of $B_s^0 \to \gamma\gamma$

Many theoretical papers have been devoted to the $B_s^0 \to \gamma\gamma$ decay. This decay is generally expected to be a good candidate to test the Standard Model (SM). First estimates for the decay branching fraction of the order of $10^{-7}$ were obtained within first order QCD calculations [262]. Later, more detailed calculations demonstrated that the next order QCD corrections increase the branching fraction up to a value of $\sim 0.5 \times 10^{-6}$ [263, 264] or even $\sim (1.0 - 1.2) \times 10^{-6}$ [265]. The contributions from long-distance effects, where the two photon final state is produced through intermediate states, in particular $B_s^0 \to \phi\gamma \to \gamma\gamma$ [266], $B_s^0 \to \Psi\phi \to \gamma\gamma$ [265], or $B_s^0 \to D_s^{(*)+} D_s^{(*)-} \to \gamma\gamma$ [267], have also been considered. Generally the relative contributions of these processes have been found to be small, leading to positive or negative corrections of 20% or less. We can conclude that the branching fraction estimates obtained within the SM and calculated using different theoretical approaches are mostly located in the range $\mathcal{B}(B_s^0 \to \gamma\gamma) = (0.4 - 1.0) \times 10^{-6}$. Using the $23.6\,\text{fb}^{-1}$ dataset collected at the $\Upsilon(5S)$, the Belle experiment has obtained the world best upper limit $\mathcal{B}(B_s^0 \to \gamma\gamma) < 8.7 \times 10^{-6}$ [268]. This limit is an order of magnitude larger than the SM prediction.

Within the SM framework the dominant contribution to the $B_s^0 \to \gamma\gamma$ decay is described by a penguin annihilation diagram. Beyond the Standard Model (BSM) physics can also contribute to the loop of the penguin diagram [269–272]. Potentially such BSM contributions should affect the $B \to X\gamma$, $B \to X\gamma\gamma$ and $B_s^0 \to \gamma\gamma$ decays in a similar manner. The inclusive $B \to X\gamma$ decay branching fraction is well measured experimentally and can, therefore, provide a strong constraint on many BSM models [269]. However, some BSM contributions can strongly affect the $B_s^0 \to \gamma\gamma$ decay branching fraction, but their contribution to the $B \to X\gamma$ decay is small relatively to the SM contributions. For such models the $B_s^0 \to \gamma\gamma$ branching fraction measurement could provide the best constraint. In particular, assuming a reasonable agreement within the errors between the experimental measurements and theoretical predictions for the $B \to X\gamma$ branching fraction, the $B_s^0 \to \gamma\gamma$ branching fraction can be enhanced up to $\sim 2 \times 10^{-6}$ within the two Higgs doublet model with charged Higgs [270], up to $\sim 3 \times 10^{-6}$ assuming specific four quark generation matrix elements [271], and even up to $\sim 5 \times 10^{-6}$ within R-parity violation SUSY models with neutralino exchange [272].

It has to be mentioned that the $B_s^0 \to \gamma\gamma$ decay final state can be in the $CP$-odd or $CP$-even eigenstate. The measurement of the $CP$-odd and $CP$-even production ratio could allow us to study $CP$ violating effects. Because this ratio can be predicted with confidence within different BSM models, the comparison of these theoretical predictions with experiment could provide an important test of these BSM models. Unfortunately, the experimental separation of these $CP$ eigenstates, using data collected at the $\Upsilon(5S)$ resonance, requires a huge statistics, which cannot be collected with the luminosity expected at proposed Super $B$ Factories.

With standard selection cuts the reconstruction efficiency for the $B_s^0 \to \gamma\gamma$ decay is expected to be $\sim 20\%$, assuming that photons are reconstructed only in the central region of the Belle electromagnetic calorimeter, $33° < \theta_\gamma < 128°$, where $\theta_\gamma$ is the polar angle in the laboratory system. We expect that a Super Belle detector configuration will allow us to increase the angular interval. Potentially the efficiency can be increased to $\sim 30\%$, if the backward and forward electromagnetic calorimeter regions will be at least partially included in a photon reconstruction procedure. The acceptance will depend on the chosen detector configuration. Unfortunately the amount of material, which the photon traverses before reaching the calorimeter, increases rather rapidly with increasing angle $\theta_\gamma$, leading to the energy resolution degradation. The final reconstruction efficiency of the $B_s^0 \to \gamma\gamma$ decay is expected to be (10-15)%, after the application of strong background suppression requirements.

Finally we can estimate the yield of about 30-70 events for the $B_s^0 \to \gamma\gamma$ mode in a $\Upsilon(5S)$ data sample of 5 ab$^{-1}$ in the $\Upsilon(5S) \to B_s^* \bar{B}_s^*$ channel. Dominant backgrounds are due to event remnants from neighboring bunch-crossings; these backgrounds should have specific behaviors. Significant suppression of this background can be achieved in the current Belle detector using bunch-crossing timing information. This information will be even more critical in a future Super Belle analysis of the $B_s^0 \to \gamma\gamma$ decay. Additional background suppression can be provided by applying cuts on the full event energy or using topological cuts. As obtained from the 1.86 fb$^{-1}$ dataset collected at the $\Upsilon(5S)$ engineering runs, the signal to background ratio of about 1:2 can be achieved using strong background suppression requirements. This implies that 5 ab$^{-1}$ of data collected at the $\Upsilon(5S)$ resonance should be enough to observe the $B_s^0 \to \gamma\gamma$ signal of $\sim (3-5)\sigma$ assuming the SM branching fraction.

### 5.11.3 Test of SM: $\Delta\Gamma_s/\Gamma_s$ and weak phase $\beta_s$

In the Standard Model the normalized width difference $\Delta\Gamma_s/\Gamma_s$ is expected to be about (10-12)% [273–283], and negligibly small $CP$ violation in $B_s^0$ mixing is predicted. The latter results from the fact that the $CP$ violating phase $\beta_s = arg(\mathrm{V}_{ts})$ of the Cabibbo-Kobayashi-Maskawa matrix element $V_{ts}$ is small ($\leq 0.03$), and the $B_s^0$ mass eigenstates coincide with the $CP$ eigenstates. Although $CP$ violation in $B_s^0$ mixing is small within the SM framework, this conclusion does not hold if BSM physics contributions are large. Similar to the SM mixing described by box diagrams, BSM physics can also contribute in box diagrams, resulting in a sizable $CP$ violation. Therefore the observation of sizable $CP$ violation in the $B_s^0$ mixing would be an evidence of BSM physics.

To measure $B_s^0$ mixing, and $CP$ violation in mixing, at $e^+e^-$ $B$ factories using conventional time-dependent technique, the two $B_s^0$ vertexes should be reconstructed with an accuracy of about $\sim 10$ $\mu$m. The time difference of these $B_s^0$ decays can then be obtained using the same technique as is obtained for $B^0$ mesons at the $\Upsilon(4S)$. Unfortunately the current detector upgrade plans do not assume $\sim 10$ $\mu$m resolution to be reached for considered vertex detectors. Potentially the sufficient vertex separation can be obtained by increasing beam energy asymmetry ($\Upsilon(5S)$ boost), however this option is technically limited. Alternatively, the $CP$ violation in the $B_s^0$ decays can be exposed in a manner similar to the $CP$ violation studies in $K^0$ decays, exploiting a lifetime difference between the $CP$-odd and $CP$-even $B_s^0$ states. The current theoretical prediction of $\Delta\Gamma_s/\Gamma_s \approx 12\%$ provides an opportunity to separate statistically the long-lived and short-lived $B_s^0$ components.

Generally three methods can be proposed to search for the $CP$ violation in the $B_s^0$ decays at a super $B$ Factory.

**1.** Comparison of direct and indirect $\Delta\Gamma_s/\Gamma_s$ measurements.

The relation of the $B_s^0$ lifetime difference between the short-lived and long-lived components, $\Delta\Gamma_s$, and the difference between the $CP$-even and $CP$-odd decay rates, $\Delta\Gamma_s^{CP}$, has been discussed in several theoretical papers [284–287]. These values are expected to be equal within the SM framework:

$$\Delta\Gamma_s = \Delta\Gamma_s^{CP} \equiv \Gamma(B_s^{even}) - \Gamma(B_s^{odd}) \qquad (5.91)$$

Potentially, Beyond Standard Model physics processes could provide a large phase, contributing to $B_s^0$ mixing and resulting in a reduction of $\Delta\Gamma_s/\Gamma_s$ [285, 286]:

$$\frac{\Delta\Gamma_s}{\Gamma_s} = \frac{\Delta\Gamma_s^{CP}}{\Gamma_s} \cos\phi \qquad (5.92)$$

where $\phi = arg(-M_{12}/\Gamma_{12})$ and $M_{12}$ and $\Gamma_{12}$ are the $B_s^0$ mass matrix and decay matrix elements, respectively[6]. As is mentioned above, within the SM the phase $\phi$ is close to zero, and a direct $\Delta\Gamma_s/\Gamma_s$ measurement should coincide with $\Delta\Gamma_s^{CP}/\Gamma_s$. Therefore, SM can be tested by comparing the $\Delta\Gamma_s/\Gamma_s$ value obtained from a direct $B_s^0$ lifetime measurement and $\Delta\Gamma_s^{CP}/\Gamma_s$, which can be determined from the difference between summed $CP$-even and $CP$-odd $B_s^0$ decay branching fractions.

The $B_s^0 \to D_s^{(*)+} D_s^{(*)-}$ decays are expected to dominate among $CP$-eigenstate decays, because of large branching fractions, of the order of $(1-2)\%$ for each mode. Moreover, within the heavy quark limit, the $B_s^0 \to D_s^{(*)+} D_s^{(*)-}$ decays are expected to be predominantly $CP$-even states [286]. Other $B_s^0$ decays with fixed $CP$, such as $J/\Psi\eta$ and $J\psi\phi$ modes, have much smaller branching fractions of the order of $10^{-3}$ or less, and would lead to a few percent correction in $\Delta\Gamma_s$ calculations.

Following this approach, an approximate equality was obtained in ref. [286]:

---

[6]Note the similarity of the above equation with Eq. 5.113 used in $D^0$ decays to $CP$ eigenstates, $y_{CP} = y\cos\phi$, valid when $|q/p| = 1$ and discussed in Sect. 5.12.5. A difference on the side of equation at which the $\cos\phi$ appears is due to the fact that in the equation discussed at this place $\Delta\Gamma_{CP}$ represents the difference of decay widths of the $CP$-even and $CP$-odd component of $B_s$ mesons, while in Eq. 5.113 the $y_{CP}$ involves the difference of effective lifetimes as measured in decays to flavour specific and $CP$ eigenstates.

$$2\mathcal{B}(B_s^0 \to D_s^{(*)+}D_s^{(*)-}) \approx \Delta\Gamma_s^{CP}/\Gamma_s \qquad (5.93)$$

Equation 5.93 was obtained under two assumptions: a) that $B_s^0 \to D_s^{*+}D_s^-$ and $B_s^0 \to D_s^{*+}D_s^{*-}$ decays are predominantly $CP$-even states (expected in the heavy quark limit) and b) that $\Delta\Gamma_s^{CP}$ is saturated by the $B_s^0 \to D_s^{(*)+}D_s^{(*)-}$ decays (expected in the Shifman-Voloshin limit [288]). The last assumption means that the contribution of multi-body $CP$-eigenstate decay modes, like $B_s^0 \to D_s^+D_s^-\pi^0$, to $\Gamma(B_s^{even})$ (or maybe to $\Gamma(B_s^{odd})$) is small compared to the dominant contribution of two-body $B_s^0 \to D_s^{(*)+}D_s^{(*)-}$ decays. These two assumptions are expected to be correct within 10% [286] and can be potentially tested by studies of relevant exclusive $B_s^0$ decays.

The lifetime difference $\Delta\Gamma_s/\Gamma_s$ can be obtained from a lifetime distribution for events with one fully reconstructed $B_s^0$ with fixed $CP$ (the best candidates are $B_s^0 \to D_s^{(*)+}D_s^{(*)-}$ and $B_s^0 \to J/\Psi\eta$ decays). Within the SM, the $CP$ eigenstates coincide with the mass eigenstates. Hence the reconstructed $CP$-even $B_s^0$ meson will decay following the exponential distribution $e^{-\Gamma_{short}t}$, whereas to the decay of the second $B_s^0$ only the $CP$-odd component will contribute [7] and will decay according to $e^{-\Gamma_{long}t}$. In practice at the $\Upsilon(5S)$ one measures the time difference $\Delta t = t(B_s^{signal}) - t(B_s^{tag})$. The corresponding $\Delta t$ distribution would have an asymmetric shape, depending on the sign of $\Delta t$; if the signal $B_s^0$ meson is reconstructed in the $CP$-even final state it would be proportional to $e^{-\Gamma_{short}\Delta t}$ for $\Delta t > 0$ and to $e^{+\Gamma_{long}\Delta t}$ for $\Delta t < 0$. A fit to this distribution yields both $\Gamma_{short}$ and $\Gamma_{long}$. The vertex smearing effects, due to detector resolution and event kinematics, should be corrected using MC simulation and have a moderate impact on accuracy of $\Delta\Gamma_s$ measurement. The fixed $CP$ $B_s^0$ lifetime measurement can be complemented by a non-$CP$ eigenstate $B_s^0$ meson lifetime measurement to improve the accuracy.

MC simulation demonstrates, that the accuracy in the direct $\Delta\Gamma_s/\Gamma_s$ measurement of $\sim 1.2\%$ (for a central value of 12.0%) can be obtained with a $\sim 5\text{ab}^{-1}$ dataset at the $\Upsilon(5S)$, where the $B_s^0 \to J/\Psi\phi$ mode beside others mentioned above was also included in analysis [8]. The statistical accuracy of the indirect $\Delta\Gamma_s^{CP}/\Gamma_s$ measurement is expected to be much better, but will be limited to $\sim 5\%$ by model-dependent systematics.

**2.** Test of the shape of lifetime distribution for events with fully reconstructed $B_s^0$ mesons with fixed $CP$.

The $CP$ violation effects of any BSM origin will mix long-lived and short-lived $B_s^0$ states for a fixed $CP$ eigenstate. The study of lifetime distribution for these eigenstates will be important test of $CP$ violation in the $B_s^0$ decays. The lifetime distribution should be fitted by the sum of two exponentials. $CP$ violation would be established if both components are observed in a fixed $CP$ eigenstate $B_s^0$ data sample. A two exponential fixed lifetime fit to a single exponential MC data sample distribution demonstrates that a $(3\text{-}4)\,\sigma$ effect can be observed with a $\sim 5\text{ab}^{-1}$ dataset at the $\Upsilon(5S)$ in the case of a 30% admixture of the long-lived component. Additional systematic uncertainties are expected to be at the level of statistical uncertainty.

**3.** The correlated production of two same $CP$ sign $B_s^0$ mesons in the $\Upsilon(5S)$ decay.

Two same $CP$ sign $B_s^*$ mesons cannot be produced in the $\Upsilon(5S)$ decay. The $CP$ value of the $B_s^*$ and a daughter $B_s^0$ are fully correlated. $CP$ violation in the $B_s^0$ system is negligibly small in the SM framework. Therefore an observation of such $\Upsilon(5S)$ decay would clearly indicate the $CP$ violation in the $B_s^0$ system and, consequently, some BSM contribution.

---

[7]Note that while the $\Upsilon(5S)$ has $CP = 1$, the two $B_s^0$ mesons have opposite $CP$ due to the relative orbital angular momentum $L = 1$.

[8]In the $J/\Psi\phi$ final state the angular distribution must be examined to separate the contributions of $CP$-even and $CP$-odd component.

In the $\Upsilon(5S) \to B_s^* \bar{B}_s^*$ decay, one $B_s^0$ should be fully reconstructed in a $CP$ final state and the second $B_s^0$ decay can be partially reconstructed. Technically, we can fully reconstruct one $B_s^0$ meson in the $B_s^0 \to D_s^{(*)+} D_s^{(*)-}$ or $B_s^0 \to J/\Psi\eta$ modes. The other $B_s^0$ can be partially reconstructed in the $B_s^0 \to D_s^{(*)+} D_s^{(*)-}$ decay mode (with only single reconstructed $D_s$), where energy and momentum balance of the $\Upsilon(5S)$ event can be used to constrain the second $B_s^0$ decay. Assuming the maximal $CP$ violation, about 5 same $CP$ sign events can be observed with a $\sim 5\text{ab}^{-1}$ dataset at the $\Upsilon(5S)$.

### 5.11.4 Potential of $\Lambda_b$ studies at $e^+e^- \to \Lambda_b \bar{\Lambda}_b$ process near mass threshold

Interesting physics can be expected at the higher $e^+e^-$ center-of-mass energy. The energy just above the mass threshold of the $\Lambda_b \bar{\Lambda}_b$ pair (and perhaps above $B_c \bar{B}_c$ mass) is of special interest. At very high energy hadron-hadron machines, the relative rate of $b$ baryon production over all $b$-quark production is $(9.9 \pm 1.7)\%$, and approximately 90% of produced $b$ baryons are $\Lambda_b$ baryons. This rate is comparable with the high energy $B_s^0$ production rate of $(10.4 \pm 1.4)\%$. Potentially, the relative rate of the $\Lambda_b$ baryon production at $e^+e^-$ colliders running at an energy just above the $\Lambda_b \bar{\Lambda}_b$ mass threshold should be about 10% as well. Generally, it is better to record data at an energy of a high lying $\Upsilon$ resonance to increase the $b\bar{b}$ production cross section, however even non-resonant $b\bar{b}$ continuum cross section is expected to be about half of the full $b\bar{b}$ cross section at the $\Upsilon(5S)$. If the $\Lambda_b$ baryon studies can be done at Belle II, this could further extend the proposed physics program.

The CLEO Collaboration investigated the region around the $\Lambda_b \bar{\Lambda}_b$ mass threshold [289]. They scan the CMS energy region between 11227 MeV and 11383 MeV in 3 MeV intervals and collected an integrated luminosity of 14 to 20 pb$^{-1}$ at each point. Although they did not find evidences of resonant or threshold enhancements, a low significance anti-proton and $\bar{\Lambda}^0$ production enhancement is seen in the data at the energy around 11310 MeV. Dedicated runs with the KEKB accelerator and Belle II detector could clarify the situation.

Interesting physics can be done at $e^+e^-$ colliders with the $\Lambda_b$ baryons. Decays of the $\Lambda_b$ baryon are well suited for BSM searches because, in contrast to $B$ and $B_s^0$ mesons, the $\Lambda_b$ has a spin. This provides the possibility to measure directly the parity violation and triple product correlations, which are expected to appear in angular distributions of decay products. Examining the $\Lambda_b$ decays described by penguin diagrams, we obtain a unique situation, where the $CP$ violation is exhibited in a branching fraction charge asymmetry, the $P$- and $T$-violations are exhibited in triple product correlations and New Physics processes could contribute in loop diagrams with an amplitude of the same order of magnitude as SM processes.

The decay $\Lambda_b \to \Lambda \mu^+ \mu^-$ (some theoretical discussions can be found in [290, 291]) is of special interest. This process is described by electromagnetic penguin or box diagrams and, therefore, is very sensitive to different BSM contributions. Possible studies of $\Lambda_b \to \Lambda \mu^+ \mu^-$ decay have been recently discussed in several papers, where the non-resonant muon pair production [290] or resonant $J/\psi \to \mu^+ \mu^-$ production [291] was considered. The strong interest in this decay results from the possibility to study time reversal (T) violation effects, using triple product correlations. These are reflected in the angular distributions of the final state $\Lambda$ and $\mu^+ \mu^-$ pair, and are T-odd under time reversal. The studies can be seen as a complementary test of $CP$ violation by assuming the correctness of the $CPT$ theorem, or might be used to search for $CPT$ violation. Moreover, because the $\Lambda_b \to \Lambda \mu^+ \mu^-$ decay is described by loop diagrams, this decay is natural process to search for BSM effects. The detailed procedure to explore this decay mode is not yet developed. Both experimental and theoretical improvements are required to investigate $\Lambda_b$ decays in detail.

Table 5.23: Expected number of reconstructed $D^{*+} \to D^0\pi^+$, $D^0 \to K^-\pi^+$ decays at different facilities (projected using [292], [293], [294], [295]). †For charm factory the expected yield of $\Psi(3770) \to D^0\bar{D}^0$, $D^0 \to K^-\pi^+$, $\bar{D}^0 \to K^+\pi^-$ is quoted.

| Facility | num. of $D^{*+} \to D^0\pi^+$, $D^0 \to K^-\pi^+$ | int. luminosity [fb$^{-1}$] | Comment |
|---|---|---|---|
| existing B factories | $2.5 \times 10^6$ | 1000 | final data set |
| Super-KEKB | $14 \times 10^6$ | 5000 | purity $\sim 99\%$ |
| Charm factory† | $4 \times 10^4$ | 20 | both $D^0$'s reconstructed; $D^0\bar{D}^0$ in coherent state |
| Tevatron | $0.5 \times 10^6$ | 0.35 | |
| LHCb | $15 \times 10^6$ | 2 | |

## 5.12 Charm Physics

### 5.12.1 Introduction

A large increase of various results in the field of charm quark physics arising in the recent years from the existing $B$-factories reflects two underlying reasons. First, a realization that the $B$ factories represent also a superior source of charmed hadrons, with

$$\sigma(e^+e^- \to B\bar{B}) \approx 1.1 \text{ nb}, \quad \sigma(e^+e^- \to c\bar{c}) \approx 1.3 \text{ nb}. \tag{5.94}$$

Secondly, a dual role of charm physics in the contemporary flavor physics - as a test ground for theoretical calculations (especially lattice QCD), enabling even more precise measurements in the system of $B$ mesons, as well as a stand-alone field of search for phenomena not included in the SM.

To set the stage for some estimates of the expected sensitivities it may be instructive to quote the yield of reconstructed benchmark decays $D^{*+} \to D^0\pi^+$, $D^0 \to K^-\pi^+$ at different facilities. For a modest Super-KEKB integrated luminosity of $L = 5$ ab$^{-1}$ we can expect $14 \times 10^6$ signal decays with a purity of above 99%. With few years of running at the design luminosity one expects an order of magnitude larger data sample.

Charm factories operating at Cornell and in Beijing have somewhat less than two orders of magnitude lower luminosity. The first milestone to be achieved is an integrated luminosity of 2 fb$^{-1}$, to be expected around the end of 2009. With BESIII running four years at the design luminosity, the accumulated data set should correspond to $L \approx 20$ fb$^{-1}$. Pairs of charmed mesons at these facilities possess completely different properties than the ones produced in the fragmentation of $c\bar{c}$ pairs, as discussed below. LHCb at LHC will reconstruct about the same amount of benchmark decays with an integrated luminosity of 2 fb$^{-1}$, roughly corresponding to one year of data taking at the nominal luminosity, as the Super-B factory with $L = 5$ ab$^{-1}$, but in a significantly more demanding environment of hadronic collisions.

### 5.12.2 Spectroscopy of charmed baryons

Important field of confrontation of experimental results to the predictions of QCD is the spectroscopy of charmed hadrons. Due to their specific interest in production of coherent states of $D$ meson pairs, the charm factories cannot contribute to these experimental efforts. Both, baryons and mesons composed of charm quarks offer a rich spectrum of states that can serve as a laboratory to test theoretical predictions based on the quark models, heavy quark symmetry (HQS) and others [296], [297]. Numerous measurement possibilities exist and it is impossible to comprehensively list those. In the following only some hopefully instructive examples are given.

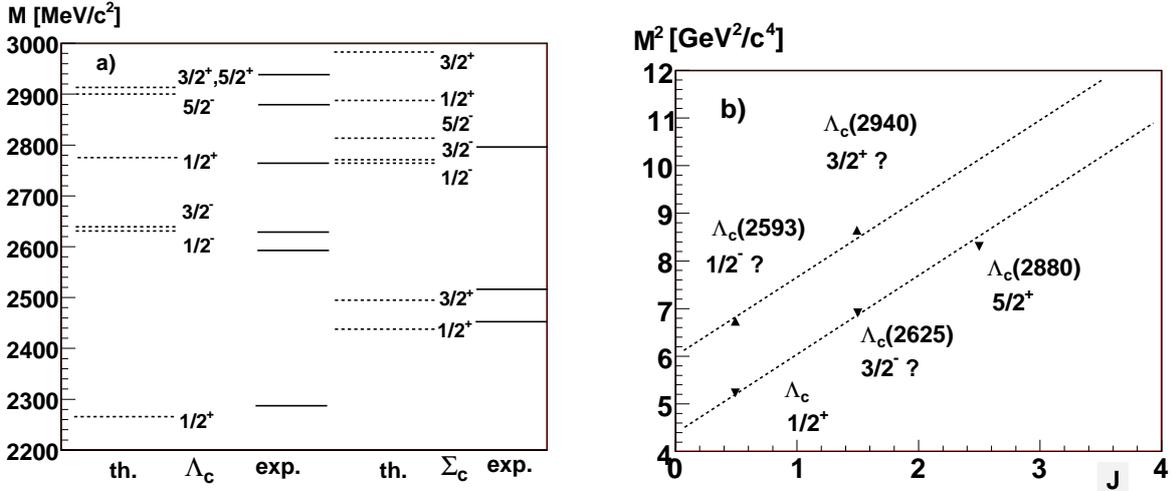

Figure 5.63: (a) Predicted (dashed line) [296] and measured (full line) masses of $\Lambda_c$ and $\Sigma_c$ baryons. Only the lowest predicted radial excitations are shown, and isospin splittings are not included. (b) Square of masses of several known $\Lambda_c$ states as a function of their assumed spins. For the states where the spin assignment is not experimentally tested the question mark is added. The naive linear Regge trajectories with the slope $\sim 1.55$ GeV$^2$ are plotted as dashed lines. The accuracy of the mass determination is better than shown by data points.

Recent experimental results in the field of charmed baryons reveal several new states, as well as (at least in some cases) the determination of their quantum numbers [298], [299], [300], [301]. The latter proves to be crucial for such tests (in order to relate the predicted states to the observed ones). Although the mass of the $c$ quark is large and thus the flavor symmetry of baryons composed of $u, d, s$ and $c$ quarks is broken, the $SU(4)_{\text{flavor}}$ multiplets represent a good visualization of the baryonic states to be expected [24]. In the ground ($L = 0$) 20'-plet ($J^P = 1/2^+$) there are 9 baryons with one $c$ quark, all experimentally observed. In the 20-plet ($J^P = 3/2^+$) all 6 baryons with a single charm quark have been observed. Or at least believed to be observed, since the spin has been experimentally tested for only one of those, $\Lambda_c^+$. For some states the assignment of the spin, based on the quark model, is relatively straightforward (e.g. for $\Sigma_c(2455)$), while for others even the isospin is not known (e.g. $\Lambda_c(2765)$). From the rich spectrum of orbitally excited charmed baryons, 9 candidates have been found so far. Eight of those have been discovered at the $e^+e^-$ colliders operating at the $\Upsilon(4S)$ [24]. For these candidates the determination of the quantum numbers is, due to a large number of predicted states in the relevant mass range, of even larger importance. The experimental situation for $\Lambda_c$ and $\Sigma_c$ states is shown in Fig.5.63(a), where experimentally observed states are shown next to the predictions from the quark model [296]. Note that the isospin splittings are not shown, and for predicted states only the lowest radial excitations are included.

The spectroscopic measurements usually involve two stages: a) reconstruction of badly known or even yet unobserved states, in an appropriate decay mode, and b) study of isospin related decays, angular distributions and/or any other observables yielding information on the quantum numbers of the observed states. For the step a) most commonly the invariant mass of the final state particles is used, although other methods can be applied (see for example section 5.12.3 on double $c\bar{c}$ production). For narrow states where the resolution is comparable to the natural width or even dominates the width of the invariant mass signal, a usual practice is to use the mass difference w.r.t. a well known mass of one of the final state particles. If one is dealing with a cascade decay of the type $R \to AB$, $A \to ab$, the use of $\Delta M \equiv M(abB) - M_A$ reduces the uncertainty on $M(R) = \Delta M + M_A$ caused by the measurement error on $M(ab)$ (masses of particles denoted as $M(ab)$ represent the measured invariant mass, while $M_A$ denotes the nominal mass known from other measurements). An alternative approach is to refit the tracks $a, b$ to originate from a common production point, imposing the nominal mass of $A$ as a constraint in the fit.

The mass constraint fit to $\Lambda_c^+$ was used in the recent studies of the $\Lambda_c(2880)$ spin and parity [300], decaying to $\Lambda_c^+ \pi^- \pi^+$. The study can serve as an example of the possibilities offered by the largely increased statistical power of the samples. In Fig. 5.64 the mass distribution of $\Sigma_c(2455)\pi$ combinations, where $\Sigma_c(2455)$ decays to $\Lambda_c \pi$ is shown. The distribution represents the result of the first stage of the measurement.

The measured angular distribution of $\Lambda_c(2880)^+ \to \Sigma_c(2455)\pi$, $\Sigma_c(2455) \to \Lambda_c \pi$ decays is sensitive to the spin of $\Lambda_c(2880)$, and strongly prefers the value of

$$J^P(\Lambda_c(2880)) = 5/2^+ \quad . \tag{5.95}$$

This represents the second step of studies, mentioned above. The value was obtained from a fit of the predicted helicity angle distributions for $J(\Lambda_c(2880)) \to 1/2(\Sigma_c) + 0(\pi)$ decays, proceeding through $L = 0, 1, 2$ or 3 wave (it should be noted that the parity of the decaying state cannot be determined from the angular distribution alone). The expected angular distributions can be found in [302].

To determine the parity in the framework of HQS, the ratio of branching fractions for decays through intermediate $\Sigma_c$,

$$\frac{Br(\Lambda_c(2880)^+ \to \Sigma_c(2520)\pi)}{Br(\Lambda_c(2880)^+ \to \Sigma_c(2455)\pi)} = 0.225 \pm 0.062 \pm 0.025 \quad , \tag{5.96}$$

is used. The value favors a positive parity assignment, but with the current accuracy cannot distinguish between the two possible angular momenta of the light quarks system in $\Lambda_c(2880)$, $J_\ell^P = 2^+, 3^+$ (prediction for the ratio of branching fractions of 0.23 and 0.36, respectively [303]). The precision is currently limited by the statistical error. With the data available with 5 ab$^{-1}$ this uncertainty would be less than the systematic one (although it should be noted that the latter depends on the knowledge of $\Sigma_c$ parameters and will decrease with larger statistical power of the sample as well), and the quoted ratio would be determined with an accuracy of $\pm 0.03$. As shown in Fig. 5.64, in the same publication Belle observed a significant contribution of $\Lambda_c(2940)^+ \to \Sigma_c(2455)\pi$ decays, comprising 220 signal events ($\Lambda_c(2940)$ was first observed in the decays to $D^0 p$ [299]). While the current statistics does not allow for a similar angular analysis as in the case of $\Lambda_c(2880)$, the increased data sample ($\sim 2 \times 10^3$ signal events with $L$=5 ab$^{-1}$) would enable also $\Lambda_c(2940)$ spin determination.

Not only HQS but also predictions involving Regge trajectories can be tested by studies of these excited states. This is illustrated in Fig. 5.63(b) where $M^2$ of various known $\Lambda_c$ states is plotted versus the assumed spin values. The lines indicate a naive linear Regge trajectories, which provide a good description of data on the light quark hadrons. For charmed baryons deviations from this linear behavior are observed (note that the accuracies of the mass determination are better than the size of the data points). With the increased data set one will be able to test the so far uncertain spin assignments, search for even higher excited states, test the deviations from linearity and the predictions of more sophisticated models [304], and try to understand the underlying QCD foundations.

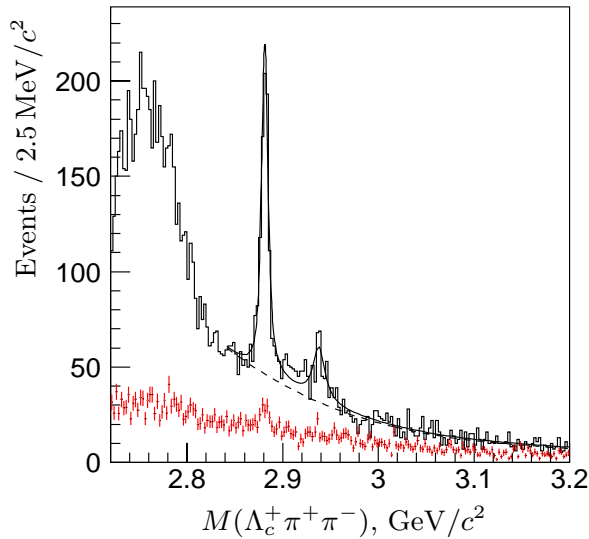

Figure 5.64: Invariant mass of selected $\Sigma_c(2455)\pi$ combinations from [300]. Beside a strong $\Lambda_c(2880)$ signal, also a signal of around 220 events is observed at the mass of around 2940 MeV/$c^2$.

Isospin assignment of recently discovered $\Xi_c(2980, 3077)$ [301] is experimentally verified by the observation of the charged as well as the neutral states. Baryons are reconstructed in decays to $\Lambda_c^+ K^- \pi^+$ and $\Lambda_c^+ K_S \pi^-$. Statistical significance of the $\Xi_c(2980)^0$ signal is only 1.5 standard deviations and would increase roughly by a factor of 3 using the 5 ab$^{-1}$ data set. This would present an evidence that $\Xi_c(2980)$ is indeed an isodoublet, as anticipated from the available data. Needless to say that the analysis of angular distributions, providing an insight into the spins of the observed states, would be possible with the expected data set.

A goal of these and similar studies, possible with a larger data set available at Super-KEKB, would be a well established ensemble of charmed hadrons with determined quantum numbers, to be confronted to theoretical understanding of the strongly bound states.

Separately from the above mentioned hadrons which are understood in terms of a conventional quark model, there exist a growing evidence of more exotic states. Here, "exotic" should be (at least for now) understood in terms of no obvious predicted candidates for the experimentally observed resonances. They are discussed in section 5.13.

Last but not least, up-to-date there is no strong experimental evidence for the existence of doubly charmed baryons. While the double $c\bar{c}$ production is discussed in the next section, a possibility of search for such baryons should be mentioned at this place.

### 5.12.3 Double $c\bar{c}$ production

Important information for the correct quantitative application of QCD to quarkonia production is obtained by the study of double charmonium production in $e^+e^-$ annihilation. Using two-body kinematics, exclusive final states can be identified by studying the spectrum of the mass recoiling against one of two charmonia in the final state, denoted by $(c\bar{c})^{\text{tag}}$, which is reconstructed exclusively. The recoil mass is determined as

$$M_{\text{recoil}}((c\bar{c})^{\text{tag}}) = \sqrt{(E_{\text{CMS}} - E^*_{\text{tag}})^2 - p^{*2}_{\text{tag}}}, \qquad (5.97)$$

with $E_{\text{CMS}}$, $E^*_{\text{tag}}$ and $p^*_{\text{tag}}$ denoting the $e^+e^-$ center-of-mass (CMS) energy, and $(c\bar{c})^{\text{tag}}$ energy and momentum in this frame, respectively. The observed spectra for $(c\bar{c})^{\text{tag}} = J/\psi$, $\psi(2S)$, or $\chi_{c1,2}$ mesons are shown in Fig. 5.65 [305]. The clear peaks are seen in the spectrum of mass recoiling against $J/\psi$ and $\psi(2S)$, a hint for the $J/\psi \chi_{c1}$ production is also seen in the $M_{\text{recoil}}(\chi_{c1})$ spectrum. Measurements of $J/\psi \eta_c$, $J/\psi \chi_{c0}$ and $J/\psi \eta_c(2S)$ production [306, 307] have been in significant disagreement with the non-relativistic leading-order (LO) calculations [308]. Later it has been shown that the non-relativistic approach may not be applicable [309]. Recently the next-to-leading order (NLO) calculations that also include relativistic corrections appear to agree with the measurements, within relatively large uncertainties due to the mass of the charm quark and renormalization scale [310]. The detailed study of double charmonium production with the integrated luminosity of $5\,\text{ab}^{-1}$ allows to measure cross sections for production of numerous exclusive final states with a precision of better than $0.5 - 1\,\text{fb}$ (for comparison, $\sigma(e^+e^- \to J/\psi \eta_c)$, as measured in [306], is around 25 fb), and to study their angular distributions. A large part of the systematic error is expected to scale with the luminosity (reconstruction efficiency dependence on the angular distribution, appropriateness of the fitting function). Moreover, important information can be obtained by studying the dependence of the double charmonium production on $\sqrt{s}$. This can be achieved either by direct measurements at different $E_{CMS}$ energies in the $\Upsilon(1S-5S)$ region, or by using the initial state radiation.

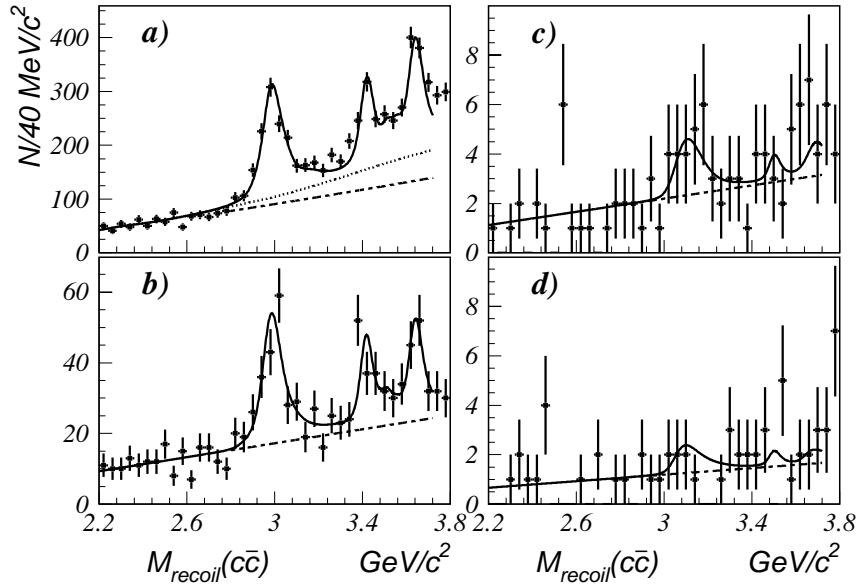

Figure 5.65: The mass of the system recoiling against the reconstructed a) $J/\psi$, b) $\psi(2S)$, c) $\chi_{c1}$ and d) $\chi_{c2}$.

The process $e^+e^- \to J/\psi(c\bar{c})$ can be regarded as a mini-factory of charmonia with a positive charge-conjugation parity. This process can serve for both, searching for new states as well as to study the decays of already known states. For example, with $L = 5\,\text{ab}^{-1}$ of data the expected number of tagged $\eta_c(2S)$ in the $M_{\text{recoil}}(J/\psi)$ spectrum is around 5000. A full reconstruction of the process $e^+e^- \to J/\psi\eta_c(2S) \to (\ell^+\ell^-)(K_S K\pi)$ allows to measure the absolute branching fraction of the decay $\eta_c(2S) \to K_S K\pi$ with an accuracy of $\sim 20\%$. Studies of various double charmonium final states [306] have demonstrated that there is no significant suppression of the production of radially excited states: the cross sections for $J/\psi\eta_c$, $\psi(2S)\eta_c$, $J/\psi\eta_c(2S)$ and $\psi(2S)\eta_c(2S)$ are comparable. Thus observations of $\eta_c(nS)$ and $\chi_{c0}(nP)$ are anticipated with high statistics data. New charmonium-like states, the $X(3940)$ and $X(4160)$, have been observed in the spectrum recoiling against $J/\psi$, and are good candidates for the $\eta_c(3S)$ and $\eta_c(4S)$ mesons. However, to identify these states an angular analysis is required, which is still not possible due to the lack of the statistics. The quantum numbers of the $X(3940)$ and $X(4160)$ can be definitely measured at the super$B$-factory. A search for exclusive decays of the already observed and new states, $(c\bar{c}) \to ab$, can be performed avoiding the usual large efficiency loss in exclusive reconstruction. For the two-body decays only a single final state particle $a$ is reconstructed, while the other can be identified in the recoil mass against $J/\psi a$. The method has been exploited to study $X(3940) \to D^*\bar{D}$ and $X(4160) \to D^*\bar{D}^*$ decays [311].

Another surprising result obtained at Belle is the measured cross section for $e^+e^- \to J/\psi c\bar{c}$, where $c\bar{c}$ pair can fragment either into a charmonium state or charmed hadrons. It has been found to be equal to $0.74 \pm 0.08^{+0.09}_{-0.08}$ pb [305] and is larger than the predictions of perturbative QCD $\sigma(e^+e^- \to c\bar{c}c\bar{c}) = 0.1 - 0.4$ pb [312, 313]. Moreover, an essential fraction of the $e^+e^- \to c\bar{c}c\bar{c}$ should fragment into four charmed hadrons or doubly charmed baryons, rather than $J/\psi c\bar{c}$. An interesting opportunity to test the predictions of perturbative QCD and quark-hadron duality may be available with a high statistics of Super$B$-factory data through the measurement of the production of two $c\bar{c}$ pairs, fragmenting into four charmed hadrons. In order to test the prediction one can search for the production of two equally flavored charmed hadrons, like $D^0 D^0$ or $D^0 D^+$, in the same $e^+e^-$ collision. Summing over all possible pairs of ground state charmed hadrons one can measure $\sigma(e^+e^- \to c\bar{c}c\bar{c})$ in a model-independent way. The data taken at the $\Upsilon(4S)$ resonance is contaminated by the same flavor charmed hadrons from $B\bar{B}$ decays. However, part of the data is usually taken below the $\Upsilon(4S)$ resonance. Assuming that at the Super$B$-factory 10% of the running time will be spent to take the continuum data one could expect to start seeing the signal once the integrated luminosity reaches 5 ab$^{-1}$.

### 5.12.4 Semileptonic and leptonic $D$ meson decays

On the long term the most accurate predictions in the non-perturbative regime of QCD are expected from the lattice calculations (LQCD). To illustrate the way charm measurements with an increased data size can help in validation of these calculations we can turn to the measurement of the least precisely known CKM element, $|V_{ub}|$. While inclusive semileptonic decays of $B \to X_u \ell\nu$ are exploited to determine $|V_{ub}|$ with a relative precision of 8% [9], a complementary method of measuring exclusive decays (most prominently $B \to \pi\ell\nu$) yields a value of $|V_{ub}| = (3.84 \pm ^{0.67}_{0.49}) \times 10^{-3}$. Knowing that $Br(B \to \pi\ell\nu)$ is measured with an experimental accuracy of 8%, it is clear that the main uncertainty arises from the knowledge of the form factor $f_+(q^2)$, which describes the non-perturbative part of the hadronic current in the differential decay rate $d\Gamma/dq^2 = (G_F^2|V_{ub}|^2/24\pi^3)p_\pi^3|f_+(q^2)|^2$. Unquenched LQCD calculations [314] determine the form factor with a precision that propagates to an error of 10%-14% on $|V_{ub}|$. In the near future one can hope the precision of the calculations to be significantly improved, contributing a twice smaller error on $|V_{ub}|$ [24] and hence approaching the current experimental accuracy. It is thus important to be able to test the predictions with an experimental accuracy

---
[9]For details on $|V_{ub}|$ determination see Sect.5.9.

significantly better than ∼5%.

Measurements of semileptonic charm meson decay rates provide an independent test of LQCD calculations of the form factors. Measured differential decay rate of $D^0 \to \pi^- \ell^+ \nu$ by Belle [315] yields the normalization of the form factor at $q^2 = 0$,

$$f_+^\pi(0) = 0.624 \pm 0.020 \pm 0.030 . \tag{5.98}$$

The accuracy of the result obtained using only 282 fb$^{-1}$ of data is comparable to the recent measurement by Cleo [316]. The total error of this measurement is also comparable with the current precision of the LQCD prediction [317],

$$f_+^\pi(0)^{\text{LQCD}} = 0.64 \pm 0.03 \pm 0.06 . \tag{5.99}$$

With the 5 ab$^{-1}$ data set the experimental stat. accuracy would be decreased by a factor of 4. Since background is estimated using the data one can expect also the systematic error to be significantly decreased. Hence the experimental results would be accurate enough to test theoretical predictions of much higher precision than currently available (see Fig. 5.66(a) and Table 5.24).

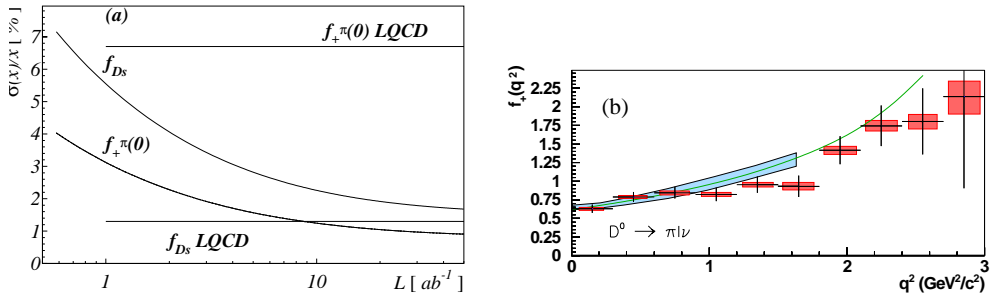

Figure 5.66: (a) The expected uncertainty of measured $f_+^\pi(0)$ and $f_{D_s}$. Horizontal lines represent current accuracy of LQCD calculations. (b) Form factor for $D^0 \to \pi^- \ell^+ \nu$ as measured in [315] (points with error bars), prediction of LQCD [317] (full line) and the corresponding error (shaded band), and the expected uncertainty of experimental points using 5 ab$^{-1}$ of data (shaded rectangles).

To determine the normalization of the form factor, the semileptonic branching fraction of $D^0$ mesons must be measured. Since the total number of charmed mesons produced in $e^+e^-$ collisions has a relatively large uncertainty, the conventional method at the $B$-factories is to determine the rate of decays relatively to another, well measured decay mode. An alternative procedure, resulting in a clean sample of semileptonic decays was developed in [315]. It uses $e^+e^- \to D_{\text{tag}}^{(*)} D_{\text{sig}}^{*\pm} X$ events, where the tagging meson $D_{\text{tag}}^{(*)}$ is fully reconstructed. $X$ represent any combination of primary particles (neutral and charged). Knowing the 4-momentum of the initial $e^+e^-$ system, $D_{\text{tag}}^{(*)}$ and of the selected $X$ combination, the 4-momentum of the signal mesons is obtained as

$$P(D_{\text{sig}}^{*\pm}) = P(e^+e^-) - P(D_{\text{tag}}^{(*)}) - P(X) . \tag{5.100}$$

A fit of the $D_{\text{sig}}^{*\pm}$ momentum to the above equation, with the nominal $D^*$ mass as a constraint, improves the momentum resolution. In the next step a slow pion from $D^{*+} \to D^0 \pi^+$ is searched for and the above missing momentum procedure is repeated to obtain the momentum of the signal $D^0$. By this one obtains an inclusive sample of charmed mesons, serving as a normalization for calculation of the branching fraction to a chosen final state, e.g. $\pi\ell\nu$. The latter is reconstructed by selecting appropriate final state tracks and, if necessary, repeating the missing momentum calculation once again. The yield of inclusively reconstructed $D^0$ mesons with the described method is around 200 events/fb$^{-1}$.

Interpretation of the measured differential decay rate, $d\Gamma(D^0 \to h\ell\nu)/dq^2$, requires parameterization of the form factor dependence on $q^2$. In the past most often a simple pole model was used [318],

$$f_+^h(q^2) = \frac{f_+^h(0)}{1 - q^2/(m_{\text{pole}}^h)^2} \quad , \tag{5.101}$$

gradually replaced by its generalization, the modified pole model [319],

$$f_+^h(q^2) = \frac{f_+^h(0)}{(1 - q^2/(m_{\text{pole}}^h)^2)(1 - \alpha^h q^2/(m_{\text{pole}}^h)^2)} \quad . \tag{5.102}$$

In the simple pole model, the pole mass $m_{\text{pole}}$ is expected to be the mass of the lightest $J^P = 1^-$ meson above the kinematic limit for the decay, $D^*$ and $D_s^*$ for the $\pi\ell\nu$ and $K\ell\nu$ decays, respectively. The modified model incorporates possible contribution of higher mass states. The average of precise measurements of the pole masses by Belle and Cleo [315, 316] (by means of the fit of Eq. 5.101 to the measured differential distribution) is

$$m_{\text{pole}}^\pi = (1.95 \pm 0.04) \text{ GeV}/c^2, \quad m_{\text{pole}}^K = (1.93 \pm 0.03) \text{ GeV}/c^2 \quad , \tag{5.103}$$

significantly deviating from the simple expectations. This points to the fact that at least a simple pole model is not a good parametrization of the form factor. The statistical precision of the current reconstructed samples at $B$- and charm-factory is insufficient to judge whether a fit using the modified pole model, with $m_{\text{pole}}$ fixed to the expected value, offers a significantly better description of the data. Only recently new parameterizations [320] started to be confronted with the measurements [316].

In determining the shape of the form factor, $q^2$ resolution is of crucial importance. It is found to be compatible between $B$- and charm-factory,

$$\sigma(q^2) = 0.015 - 0.025 \text{ GeV}^2 \tag{5.104}$$

(use of the method described above is assumed for the $B$-factory). Illustration of experimental verification of the $f_+$ $q^2$-dependence is shown in Fig. 5.66(b). Points with error bars correspond to the measurement of $f_+(q^2)$ in $D^0 \to \pi^-\ell^+\nu$ decays [315] using 280 fb$^{-1}$ of data, compared to the calculation of [317] (full line). The light dashed area is the current precision of the LQCD and the filled rectangles represent the projected experimental error reduced by a factor of $\sim 4$. A precise measurement of $f_+(q^2)$, especially at high $q^2$ values, where also the theory predictions are most uncertain, is clearly anticipated and wished for.

A pseudoscalar to vector meson transitions, e.g. $D_s^+ \to \phi\ell\nu$, can be described (neglecting the mass of the lepton) by three independent form factors, $A_1(q^2)$, $A_2(q^2)$ and $V(q^2)$. The decay is determined by the lepton invariant mass $q^2$, and three decay angles (for details, see [321]). In order to constrain the values of the form factors a 4-dimensional distribution of the $q^2$ and angular observables must be determined and an appropriate model fit to the data. A large sample of signal events is clearly needed. BaBar collaboration determined the ratios of the form factor normalizations, $r_V = V(0)/A_1(0)$ and $r_2 = A_2(0)/A_1(0)$, using 78 fb$^{-1}$ of data [322]. The uncertainty of the result, $r_V = 1.636 \pm 0.067 \pm 0.038$, $r_2 = 0.705 \pm 0.056 \pm 0.029$, is dominated by the available statistics of the analyzed sample. Even more so, as the major part of the systematic uncertainty (background modeling, limited statistics of the simulated sample) would decrease significantly in the future. Taking this into account one can expect a precision of ±0.02 for both, $r_V$ and $r_2$, using 5 ab$^{-1}$ of data. Unquenched LQCD calculations of the form factors for the pseudoscalar to vector transition are not yet available.

An area where the dual role of charm physics is also exposed is the study of purely leptonic decays of charmed mesons. Within the SM the decay width is expressed as $\Gamma(D \to \ell^+\nu) = (G_F^2/8\pi)f_D^2 m_\ell^2 m_D(1 - m_\ell^2/m_D^2)^2|V_{cq}|^2$, where $|V_{cq}|$ is the appropriate CKM element and $f_D$ is the meson decay constant. The analogous decay constants of $B$ mesons play an important role in constraining the apex of the unitarity triangle using the ratio of $B_d^0$ and $B_s^0$ meson oscillation frequencies. The ratio is related to the CKM elements through $\Delta m_s/\Delta m_d = f_{B_s}^2 B_{B_s}|V_{ts}|^2/f_{B_d}^2 B_{B_d}|V_{td}|^2$. Both oscillation frequencies are measured with a relative precision of less than 1% [24] [323]. On the other hand, the theoretical knowledge of $f_{B_s}^2 B_{B_s}/f_{B_d}^2 B_{B_d}$ is limited to around 5%. Once again a precise calculation of this factor would be needed and the leptonic decays of $D$ mesons can provide a test of such calculations.

In a recent measurement by Belle [324] a sample of around 60 inclusive $D_s$ meson decays is reconstructed in each fb$^{-1}$ of data using the recoil technique analogous to the one described above for the $D \to \pi\ell\nu$ measurement. Such a method enables an absolute measurement of the $D_s^+ \to \ell^+\nu$ branching fraction which is crucial for the accurate determination of the decay constant. The uncertainty of the branching fraction using the $B$-factory data set is dominated by the errors which will be reduced by increasing the available data set. The relative uncertainty using 548 fb$^{-1}$ of data amounts to around 14% [324], only around a factor of two larger than in the recent measurement by Cleo-c [325]. Using the equation for the decay width given above the result of Cleo collaboration is converted into the value of the $D_s$ meson decay constant [325]

$$f_{D_s} = (259.5 \pm 6.6 \pm 3.1) \text{ MeV}. \tag{5.105}$$

This can be compared to the LQCD prediction [326],

$$f_{D_s}^{\text{LQCD}} = (241 \pm 3) \text{ MeV}. \tag{5.106}$$

Having 5 ab$^{-1}$ data at hand the branching fraction can be measured with an accuracy of around ∼4% corresponding to a ∼2% error on $f_{D_s}$. This is a sufficient accuracy to test the LQCD calculations. It should be noted that measuring the ratio of $Br(D_s^+ \to \ell^+\nu)/Br(D^+ \to \ell^+\nu)$ has an advantage of cancellation of at least part of the systematic error, which holds also for the lattice calculation.

As a rough illustration of $f_{D_s}$ and $f_+^\pi(0)$ determination accuracy the projected errors are plotted as a function of the integrated luminosity in Fig.5.66(c). It should be noted that the prediction of the systematic errors is uncertain and hence the total projected error is a rough estimate. Nevertheless the figure can give a feeling on the possibility of LQCD prediction testing using the increased data-sets at the Super-KEKB. Table 5.24 shows some expectations for the accuracies of the parameters, following from the studies of the (semi)leptonic charmed meson decays.

### 5.12.5 $D^0 - \bar{D}^0$ mixing and $CP$ violation

Until recently the only system of neutral mesons in which the phenomena of mixing between the particle and anti-particle has not yet been observed was the system of neutral $D^0$ mesons. The reason for this lies in the fact that this is the only system where oscillations between particle and anti-particle are governed by the exchange of the down-type $(d, s, b)$ quarks in the box diagram loops. Time evolution of an initially produced flavor eigenstate $D^0$ or $\bar{D}^0$ is governed by parameters

$$x = \frac{m_1 - m_2}{\Gamma}, \quad y = \frac{\Gamma_1 - \Gamma_2}{2\Gamma} \quad , \tag{5.107}$$

where $m_{1,2}$ and $\Gamma_{1,2}$ are the masses and the widths of the mass eigenstates, respectively. $\Gamma$ is their average width, $\Gamma = (\Gamma_1 + \Gamma_2)/2$. Since the majority of $D$ meson decays is Cabibbo allowed (CA), while the $D^0 - \bar{D}^0$ transitions are twice Cabibbo suppressed (DCS), it follows that the oscillations are slow compared to the decay rate

$$\frac{m_1 - m_2}{\Gamma}, \quad \frac{\Gamma_1 - \Gamma_2}{2\Gamma} \propto \mathcal{O}(\sin^2 \theta_C) << 1 \quad . \tag{5.108}$$

Table 5.24: Expected relative errors on some of the parameters, determined in the studies of (semi)leptonic $D_{(s)}$ decays.

| Decay mode | parameter | $L = 5$ ab$^{-1}$ | | $L = 50$ ab$^{-1}$ | |
|---|---|---|---|---|---|
| | | $\sigma_{\text{stat}}$ [%] | $\sigma_{\text{syst}}$ [%] | $\sigma_{\text{stat}}$ [%] | $\sigma_{\text{syst}}$ [%] |
| $D^0 \to \pi \ell \nu$ | $f_+^\pi(0)$ | 1 | 1.5 | 0.2 | 1 |
| $D_s \to \mu \nu$ | $f_{D_s}$ | 2 | 2 | 0.6 | 1.5 |
| $D_s \to \phi \ell \nu$ | $r_V$ | 0.6 | 1.0 | 0.2 | 1.0 |
| $D_s \to \phi \ell \nu$ | $r_2$ | 1.0 | 2.4 | 0.3 | 2.4 |

Beside this, in the limit of an exact flavor symmetry the box diagram process yields zero contribution due to the GIM mechanism. Non-zero values of $x$ and $y$ arise due to the $SU(3)_{\text{flavor}}$ breaking, i.e. $m_d^2 \neq m_s^2$ ($b$ quark contribution is negligible due to the small value of $|V_{ub}|$). Two approaches are used to estimate the resulting values of the mixing parameters within the SM. The inclusive one is based on the Operator Product Expansion [327] and finds typical values $|x|, |y| \sim \mathcal{O}(10^{-3})$. The exclusive method uses a series of exclusive states accessible to both $D^0$ and $\bar{D}^0$, thus coupling the two states [328]. The method shows that

$$|x|, |y| \sim \mathcal{O}(10^{-2}) \tag{5.109}$$

can be expected within the SM. Due to the small expected mixing rate, as-yet-unobserved particles contributing to the box diagram (NP) could significantly affect the measured values [329]. As most of the processes with low expected rates, the $D^0$ mixing is thus an interesting area to search for possible contribution of NP.

The SM $CP$ violation ($CPV$) effects in the system of $D^0$ mesons are expected to be experimentally even more difficult to observe than the mixing itself. In the mixing and $CP$ violation processes involving $D^0$ mesons, to a good approximation only the first two generations of quarks are involved. Using the Wolfenstein parameterization of the CKM matrix the complex phase in the corresponding elements arises in the $V_{cs}$ [330] as $\eta A^2 \lambda^4 \sim \mathcal{O}(10^{-3})$. This value can serve as a rough estimate of the magnitude of the $CPV$ effects one could expect. A small $CPV$ within the SM opens a possibility of the search for the unknown processes that might significantly increase the expected values.

The mixing parameters $x$ and $y$ affect the time dependent as well as the time integrated rates of individual $D^0$ decay modes and are hence experimentally observable. In all the measurements described below, $D^{*+} \to D^0 \pi_s^+$ decays, in which the charge of the characteristic low momentum pion determines the flavor of the produced $D^0$, are used. Selecting $D^*$ mesons with high enough momentum, typically around 2.5 GeV/$c$, eliminates the contribution of $D$ mesons produced in $B$ decays. These have a different decay time distribution (due to a finite decay time of $B$ mesons). $D^0$ daughter tracks are refitted to a common vertex. The production point of the $D^0$ is obtained as an intersection of its trajectory, slow $\pi$ track and the $e^+e^-$ interaction region, and together with the decay vertex determines the meson decay length $L$. Typical resolution achieved on the decay time,

$$t = \frac{m(D^0)L}{p} \tag{5.110}$$

is

$$\sigma(t) \sim \frac{\tau(D^0)}{2} \tag{5.111}$$

with $\tau(D^0) = (410.1 \pm 1.5)$ fs [24].

In 2007 the first experimental evidence for the $D^0 - \bar{D}^0$ mixing emerged [331] [332]. Belle has measured the difference in the apparent lifetime (assuming the time distributions to be exponential) when a $D^0$ meson decays to the $CP$ eigenstates $f_{CP}$, $f_{CP} = K^+K^-, \pi^+\pi^-$, and when it decays to the final state $K^-\pi^+$ [331]. The relative difference is defined as

$$y_{CP} = \frac{\tau(K^-\pi^+)}{\tau(f_{CP})} - 1, \tag{5.112}$$

which can in terms of the mixing parameters be expressed as [333]

$$y_{CP} = y\cos\phi - \frac{1}{2}A_M x\sin\phi. \tag{5.113}$$

Here, $A_M$ and $\phi$ parameterize the $CPV$ in the mixing and interference between mixing and decays, respectively. If $CPV$ can be neglected, $A_M = \phi = 0$ and $y_{CP} = y$. The result of the measurement using 540 fb$^{-1}$ of data is $y_{CP} = (1.31 \pm 0.32 \pm 0.25)\%$, significantly deviating from zero. While this evidence represents a large step forward in experimental sensitivity to the mixing parameters, with 5 ab$^{-1}$ of data at hand one can determine $y_{CP}$ with a total uncertainty reduced to only $\pm 0.15\%$. The contribution of systematic error to the latter value (when sources expected to scale with the luminosity are taken into account) is around $\pm 0.10\%$ and hence the statistical error is still slightly larger than the systematic. With an order of magnitude larger data set the total error would clearly depend only on a precise understanding of the detector and sources of systematic errors which do not completely cancel in the measured lifetime ratio. Main sources of the systematic uncertainty will arise from a possible small bias in the decay time determination[10] and possible differences in the decay time distributions of background events in the signal and sideband intervals. Such a precise measurement would be of large importance to discriminate between different theoretical predictions (SM ones, once they become more accurate, as well as predictions for NP contributions).

With the measured $y_{CP}$ being at the upper limit of the values expected within the SM, it is even more important to search for the possible $CPV$ effects which, if observed, would at the current level of sensitivity represent a sign of NP processes. In the same measurement search for the $CPV$ is performed by measuring the quantity

$$A_\Gamma = \frac{\tau(\bar{D}^0 \to K^-K^+) - \tau(D^0 \to K^+K^-)}{\tau(\bar{D}^0 \to K^-K^+) + \tau(D^0 \to K^+K^-)}. \tag{5.114}$$

---

[10]The effect is a consequence of the correlation between the opening angle among final state tracks and the reconstructed decay time.

This asymmetry is related to the mixing and $CPV$ parameters through [333]

$$A_\Gamma = \frac{1}{2} A_M y \cos\phi - x \sin\phi. \quad (5.115)$$

No $CPV$ is observed, the result of the measurement being $A_\Gamma = (0.01 \pm 0.30 \pm 0.15)\%$. The uncertainty with 5 ab$^{-1}$ of data would be $\pm 0.12\%$, to which the systematic error not scaling with the luminosity contributes around $0.06\%$. The latter would dominate the uncertainty with 50 ab$^{-1}$ and would arise from similar sources as for the $y_{CP}$ measurement.

Another sensitive measurement became available recently [334] by performing a time-dependent Dalitz analysis of $D^0 \to K_S \pi^+ \pi^-$ decays. Different intermediate states ($CP$ even or odd, like $K_S f_0$ or $K_S \rho^0$, or flavour eigenstates, like $K^*(890)^+ \pi^-$), populate different regions of the Dalitz space, and exhibit a different decay-time distributions, depending directly on $x$ and $y$. A simultaneous fit of the decay-time and Dalitz distribution thus allows one to determine the two parameters. Results of this measurement, performed on a 540 fb$^{-1}$ data set, are $x = 0.80 \pm 0.29 ^{+0.09}_{-0.07} {}^{+0.15}_{-0.14}$ and $y = 0.33 \pm 0.24 ^{+0.07}_{-0.12} {}^{+0.08}_{-0.09}$, where the first error is statistical, the second experimental systematic and the last systematic due to the model used to describe the Dalitz plot. The study of self-conjugated multi-body final states in $D^0$ decays is the only method which directly determines the parameter $x$. Using the extrapolation of the errors from the above measurement and assuming the experimental and part of the model systematic error to be reduced, one can hope for an accuracy of around $\pm 0.15\%$ ($\pm 0.10\%$) and $\pm 0.10\%$ ($\pm 0.08\%$) for $x$ and $y$, respectively, with 5 ab$^{-1}$ (50 ab$^{-1}$). The main non-scaling systematic uncertainty will be due to the model dependence in the description of the Dalitz distribution. Since the expected accuracy on $y$ is similar to the one from the measurement of $D^0$ decays into the $CP$ eigenstates one can gain further in the precision by averaging the two uncorrelated results.

$CPV$ allowed fits to $D^0 \to K_S \pi^+ \pi^-$ decays reveal a similar sensitivity to the $A_M$ parameter as obtained from decays into the $CP$ eigenstates, and additional sensitivity to the $CP$ violating phase $\phi$. Using 540 fb$^{-1}$ of data the resulting values are $|q/p|(\approx 1 + A_M/2) = 0.86 \pm 0.30 \pm 0.10$ and $\phi = (-0.24 \pm 0.30 \pm 0.10)$ $rad$. The expected accuracies of $|q/p|$ and $\phi$ determination from the $D^0 \to K_S \pi^+ \pi^-$ study with 5 ab$^{-1}$ (50 ab$^{-1}$) are 0.13 (0.10) and 0.12 $rad$ (0.07 $rad$), respectively, where the main non-scaling systematic uncertainty is again expected to arise from the limitations related to the Dalitz model.

Additional decay modes enabling an analogous measurement as described above include $D^0 \to K_S K^+ K^-$ and $D^0 \to \pi^0 \pi^+ \pi^-$. These will improve the overall accuracy of $x$ and $y$ parameters determination, but the prediction of the improvement is difficult. The sensitivity of individual decay channel to the mixing parameters is determined by the strong phase variation over the Dalitz plane which is apriori unknown.

Hadronic decays to non-$CP$ eigenstates provide a similar sensitivity to the mixing as the methods described above. Study of the decay-time dependence of $D^0 \to K^+ \pi^-$ allows one to separate the DCS contribution from the mixing one, $D^0 \to \bar{D}^0 \to K^+ \pi^-$. The time distribution can be written as

$$\Gamma[D^0 \to K^+ \pi^-] = e^{-\Gamma t} |A_{K^- \pi^+}|^2 \left[ R_D + \sqrt{R_D}(y' \cos\phi - x' \sin\phi)\Gamma t + \frac{x'^2 + y'^2}{4} (\Gamma t)^2 \right], (5.116)$$

where $R_D$ is the ratio of DCS and CA decay rates. The study of the time dependence allows only for the determination of parameters $x'^2$ and $y'$, related to $x$ and $y$ by an unknown strong phase difference between the DCS and CA decays, $\delta_{K\pi}$; $x' = x \cos\delta_{K\pi} + y \sin\delta_{K\pi}$, $y' = y \cos\delta_{K\pi} - x \sin\delta_{K\pi}$. This phase can be measured independently at the charm factory [335], although at this time the value is not well constrained yet: $\cos\delta_{K\pi} = 1.09 \pm 0.66$ (once the integrated luminosity of 20 fb$^{-1}$ is collected at the charm factory, the error on $\cos\delta_{K\pi}$ is expected to decrease to $\pm 0.04$ [336]). The most stringent constraints on $x'^2$ and $y'$ parameters were obtained in [337], using 400 fb$^{-1}$ of data. The fit to the $t$ distribution of $D^0 \to K^+\pi^-$ decays, assuming negligible $CPV$, results in $x'^2 = (0.18 \pm^{0.21}_{0.23}) \times 10^{-3}$, $y' = (0.6 \pm^{4.0}_{3.9}) \times 10^{-3}$ and $R_D = (3.64 \pm 0.17) \times 10^{-3}$, including the systematic error (the latter increases the stat. error by 12%). To properly model the $t$ distributions, this method requires a good understanding of the detector time resolution. With 5 ab$^{-1}$ one expects an accuracy of $\pm 0.10 \times 10^{-3}$ and 0.2% on $x'^2$ and $y'$, respectively. While at this integrated luminosity the statistical error is still comparable to the systematic one, with 50 ab$^{-1}$ it becomes negligible, resulting in rough expectations of $\pm 0.08 \times 10^{-3}$ and 0.15% for the corresponding total uncertainties. The remaining systematics is expected to be dominated by imperfect modeling of the decay time resolution function.

The fit to the decay-time distribution which allows for the $CPV$ is sensitive to the parameters $A_M$ and $A_D$. The result for the latter, which parameterizes the $CPV$ in the DCS decays, is $A_D = (23 \pm 47) \times 10^{-3}$, where the uncertainty is statistical only. Assuming approximately the same fraction of scaling and non-scaling uncertainty as for the $x'^2$ measurement and propagating the error to 5 ab$^{-1}$ and 50 ab$^{-1}$ one finds the anticipated values of 2.1% and 1.7%.

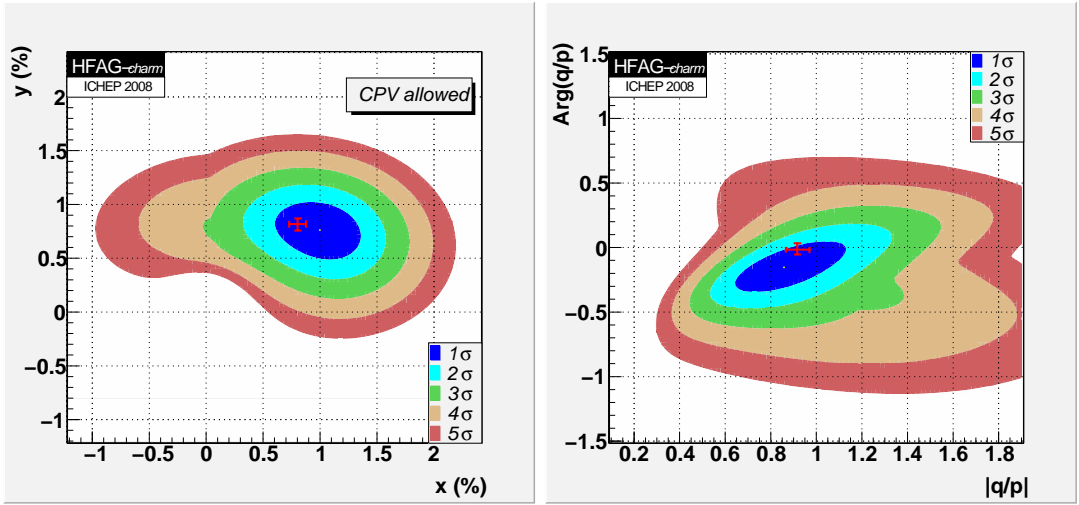

Figure 5.67: Left: The probability contours for the current world average values of the mixing parameters $x$ and $y$ [338]. The error bars denote the expected accuracy on the parameters with 50 ab$^{-1}$ following from the extrapolation of results using $K^+K^-$, $\pi^+\pi^-$, $K^+\pi^-$ and $K_S\pi^+\pi^-$ final states. Right: The probability contours for the current world average values of the $CPV$ parameters $|q/p|$ and $\phi$ [338]. The error bars denote the expected accuracy with the same assumptions as above.

Summary of the expected accuracies achievable with the Super-KEKB data set is given in Table 5.25. To get a general impression on how the combination of different measurements reflects in the precision of the parameters related to the $D^0 - \bar{D}^0$ mixing, one can do a simple $\chi^2$ minimization of the projected results described above ($D^0 \to K^+K^-/\pi^+\pi^-$, $D^0 \to K_S\pi^+\pi^-$, $D^0 \to K^+\pi^-$) in terms of the mixing parameters $x$ and $y$, $CPV$ parameters $|q/p|$ and $\phi$, and $\delta_{K\pi}$. We assume Gaussian p.d.f's, with the observed correlations among variables in a single measurement [338] taken into account. The assumed central values in this averaging procedure are $x = y = 0.8\%$, $|q/p| = 1$, $A_D = \phi = 0$ and $\delta_{K\pi} = 0.38$ rad, roughly corresponding to the current world average values. The probability contours of the current average [338] are shown in Fig. 5.67. Results of the fits are presented in Table 5.25 in the column denoted "Combined". The uncertainties following from the fit are also illustrated with error bars in Fig. 5.67. Note that the expected accuracies shown are derived using only the existing results and will be further improved using the so far unused methods, for example the decay time dependent Dalitz analyses of $D^0 \to K_S K^+ K^-$ and $D^0 \to \pi^+\pi^-\pi^0$ decays.

Table 5.25: Approximate expected accuracies of different parameters describing the $D^0$ mixing and $CPV$. The first uncertainty is the statistical error combined with part of the systematic error expected to scale with the luminosity, and the second is the rest of the systematic uncertainty.

| Parameter | Current | $L =5$ ab$^{-1}$ | | $L =50$ ab$^{-1}$ | |
|---|---|---|---|---|---|
| | | | Combined | | Combined |
| $y_{CP}$ [%] | $\pm 0.39 \pm 0.10$ | $\pm 0.12 \pm 0.10$ | | $\pm 0.05 \pm 0.10$ | |
| $A_\Gamma$ [%] | $\pm 0.33 \pm 0.06$ | $\pm 0.10 \pm 0.06$ | | $\pm 0.04 \pm 0.06$ | |
| $x$ [%] | $\pm 0.31 \pm 0.10$ | $\pm 0.10 \pm 0.10$ | $\pm 0.12$ | $\pm 0.03 \pm 0.10$ | $\pm 0.09$ |
| $y$ [%] | $\pm 0.26 \pm 0.07$ | $\pm 0.08 \pm 0.07$ | $\pm 0.09$ | $\pm 0.03 \pm 0.07$ | $\pm 0.06$ |
| $|q/p|$ | $\pm 0.30 \pm 0.08$ | $\pm 0.10 \pm 0.16$ | $\pm 0.08$ | $\pm 0.03 \pm 0.16$ | $\pm 0.06$ |
| $\phi$ [rad] | $\pm 0.30 \pm 0.06$ | $\pm 0.10 \pm 0.06$ | $\pm 0.08$ | $\pm 0.03 \pm 0.05$ | $\pm 0.05$ |
| $x'^2$ [$10^{-3}$] | $\pm 0.23 \pm 0.09$ | $\pm 0.07 \pm 0.09$ | | $\pm 0.02 \pm 0.09$ | |
| $y'$ [%] | $\pm 0.42 \pm 0.16$ | $\pm 0.12 \pm 0.16$ | | $\pm 0.04 \pm 0.16$ | |
| $\delta_{K\pi}$ [rad] | | | $\pm 0.11$ | | $\pm 0.08$ |

It is thus fair to conclude that already with a modest Super-KEKB integrated luminosity the $D^0$ mixing parameters will be measured with a $\mathcal{O}(10^{-3})$ accuracy. Probably more importantly for the search of NP, the $CPV$ parameters can be constrained to around $5 \times 10^{-3}$ with 50 ab$^{-1}$ of data. This propagates into the sensitivity on various $CPV$ asymmetries of around $5 \times 10^{-4}$ [11] The expected accuracy of $|q/p|$ and $\phi$, for example, propagates into an uncertainty of around $\pm 0.07\%$ on the asymmetry $A_\Gamma$. The expected 65% C.L. region in $(\phi, A_M)$ plane is sketched in Fig. 5.67, right.

---

[11]In terms of the underlying parameters the factors determining the observable asymmetries are typically of the form $x \sin\phi$, $y(|q/p| - 1) \cos\phi$.

## 5.13 Charmonia and new particles

In the history of physics, many new and important insights have derived from detailed studies of "well understood" systems: precise measurements of the motion of planets in the solar system led to the discovery of an anomalous precession of the perihelion of Mercury's orbit, which provided an important impetus for general relativity; high resolution measurements of atomic spectra were the key to the discovery of fermion spin.

In hadronic physics, the best understood quark-antiquark systems are heavy quarkonia, *i.e*, $c\bar{c}$ or $b\bar{b}$ mesons. Here, because the quarks are massive, they are nearly non-relativistic and ordinary quantum mechanics is applicable. Moreover, lattice calculations are particularly well suited to heavy quark systems. As these improve we can expect reliable first-principle calculations of quarkonium properties with good precision.

The discovery of the missing $c\bar{c}$ states and the precise measurements of properties of the already observed are important for a number of reasons. First of all, measurements of the masses of these states will pin down unknown parameters of charmonium models, such as the strength of the fine and hyperfine terms in the inter-quark potential. Second, some properties of these states are highly constrained by theory. Measured variances from theoretical predictions could indicate new and unexpected phenomena.

The QCD-motivated models predict an existence of hadrons of more complex structure than conventional mesons or baryons, such as hybrids [341] and multiquark states of either molecular [342], tetraquark [343] or hadrocharmonium [344] configuration. As the conventional charmonium spectrum is much cleaner with respect to dense spectrum of light states, therefore exotic states containing $c\bar{c}$ are expected to be identified easier than the ones predicted in the light spectrum. Any resonance observed in addition to predicted multiplets might give a hint of such an exotic spectroscopy.

Some of the recently observed charmonium-like $XYZ$ states could be candidates for the exotic hadrons mentioned. However most of them still await confirmation or need their properties to be further studied before any decisive interpretation is made.

A Super-KEKB factory would provide *superb* opportunities for precision, high sensitivity measurements of the charmonium system. A reason for that is not only large statistics expected but also number of processes suitable for studying the $c\bar{c}$ system which allow ones to perform comprehensive studies. In $B$ meson decays, the $b \to c\bar{c}s$ subprocess is CKM-favored and, thus, final states containing charmonium particles are common. The simplest $B$ meson decays yielding charmonium state $(X_{c\bar{c}})$ are $B \to KX_{c\bar{c}}$; their large branching fractions $(\mathcal{O}(10^{-3}))$ assure large number of detected charmonia. Even right at the $\Upsilon(4S)$ peak, the cross-section for the continuum production of $c\bar{c}$ quark pairs is higher than that for $b\bar{b}$ pairs. One of the continuum processes useful for charmonium studies is double $c\bar{c}$ production elaborated before. Valuable spectroscopic technique occurs to be also an initial state radiation (ISR) process which allows one to access states bearing quantum numbers $J^{PC} = 1^{--}$. In environment of Super-KEKB factory charmonia are also formed in $\gamma\gamma$ fusion and can be further reconstructed exclusively in numerous final states. Thus, a "Super-$B$" factory is also a "super-charm" factory that will support a variety of interesting studies of charmed particles and charmonium physics.

### 5.13.1 Charmonia below open-charm threshold

All the states that are predicted to have masses below the thresholds for open charm production have been already discovered. Observations made within last few years, such as discovery of the $\eta_c(2S)$ [345] as well as of the most elusive state $h_c$ [346], have completed the list of low lying charmonia.

The $\eta_c(2S)$ had evaded detection for nearly thirty years and was finally discovered in 2002 by Belle via its $K_S K\pi$ decay mode in exclusive $B^- \to K^- K_S K\pi$ decays [345] and subsequently confirmed by other Belle measurements [347, 348] as well as by BaBar [349] and CLEO [350]. The measured $M[\psi(2S)] - M[\eta_c(2S)]$ hyperfine splitting ($49 \pm 4$ MeV) was at the lower end of the theoretically preferred range ($43 \sim 75$ MeV) [351, 352]. The agreement between theory and experiment is substantially improved when the coupling of the $\psi(2S)$ to $D\bar{D}$ channels is taken into account [353].

Although being studied systematically in numerous processes by several experiments, the $\eta_c(2S)$ still requires further explorations. Its only decay mode observed so far is $K_s K^+\pi^-$, which is surprising, given many $\eta_c(2S)$ decay channels observed.

The $\eta_c(2S)$ has not been observed in $\psi(2S) \to \gamma\eta_c(2S)$ M1 transition [354], typically used by CLEO to study properties of charmonia, for example the properties of the $\eta_c$ [355]. Moreover the M1 transition occurs to distort line-shape of the charmonium produced and thus bias its mass and width measurement [24]. Therefore high precision study of the $\eta_c(2S)$ properties and search for its exclusive decay modes shall be a task for Super-KEKB factory where precision measurements can be performed for the $\eta_c(2S)$ produced in various processes. A discovery process, $B \to \eta_c(2S)K$ with $\eta_c(2S) \to K_s K^+\pi^-$, looks very promising, as one expects here about 35000 signal events for 5 ab$^{-1}$ data sample and an accuracy of the $\eta_c(2S)$ mass and width of about 1 MeV and 3 MeV, respectively. These errors as well as an additional one coming from model describing an interference between the $\eta_c(2S)$ signal with complicated features of the $K_s K^+\pi^-$ background can be significantly improved with increased statistics.

The $\eta_c(2S)$ produced in $\gamma\gamma$ fusion have been also studied with Belle data sample of about 500 fb$^{-1}$ [356]. For the $\eta_c(2S)$ reconstructed in $K_s K^+\pi^-$ decay mode mass and width were measured with precision higher than other existing measurements. Also in this case the interference effects between the $\eta_c(2S)$ signal and continuum background, ignored in preceding studies, were found to affect the measured parameters. Hints of $\eta_c(2S) \to 4K$, $4\pi$, $2K2\pi$ signals seen assure that additional decay modes of $\eta_c(2S)$ can be observed with Super-KEKB factory data sample.

As mentioned in Sect.5.12.3, with very high statistics one can use the double charmonium production, especially $e^+e^- \to J/\psi X_{c\bar{c}}$ process, to measure absolute branching fractions of $\eta_c(1, 2S)$ (or other $C$-even charmonia) decaying to various final states. But this method, being based on rather low resolution recoil-mass technique, can hardly be used for precise mass or width measurements.

The $h_c$ has proven to be the most elusive of the low-lying charmonium states. The signal of the $h_c$ candidates found by the E760 experiment at Fermilab [357] has not been confirmed by a subsequent experiment E835 [358]. The $h_c$ was finally observed by CLEO in 2005 via $\psi(2S) \to \pi^0 h_c$ with $h_c \to \gamma\eta_c$ in both, inclusive analysis (without $\eta_c$ reconstruction) and exclusive reconstruction of $\eta_c$ [346]. The recent CLEO study yields $M(h_c) = 3525.2 \pm 0.18 \pm 0.12$ MeV and P-wave hyperfine splitting $M(h_c) - M(1^3 P_{c\ cog}) = 0.08 \pm 0.18 \pm 0.12$ MeV [359]. This is an important test of perturbative QCD and validation of relativistic potential calculations. Indeed one has expected $M(h_c) \equiv M(1^1 P_{c1}) \approx M(1^3 P_{c\ cog})$, as P-wave splitting describing spatial behavior of $c\bar{c}$ potential is proportional to $\delta^3(\vec{r})$ for a Coulomb-like $c\bar{c}$ interaction and thus tiny.

Confirmation of the $h_c$ by Belle will be a challenge even with Super-KEKB factory data. Searches in $B$ meson decays are hampered by the fact that the exclusive decay process $B \to K h_c$, which violates factorization, is expected to be small. Using a technique based on a suggestion by Suzuki [360], Belle looked for the $h_c$ via the $B \to K h_c$ and $h_c \to \gamma\eta_c$ decay chain. The hope was that the large expected decay branching ratio for $h_c \to \gamma\eta_c$ ($\sim 60\%$) might compensate for the small, factorization-suppressed rate for $B \to K h_c$. So far, no signal has been seen, but the branching ratio upper limit ($\sim 4 \times 10^{-5}$ with 250 $fb^{-1}$ of data) is not very restrictive [361]. This is because of the low efficiency for reconstructing the $\eta_c$ in low-background channels. Future searches at higher luminosity will improve this limit, but only by a factor that goes as the square-root of the increase in luminosity.

The process $e^+e^- \to \eta_c h_c$ is allowed by $C$-parity conservation. Thus, another approach for Belle $h_c$ searches would be to use experimentally distinct $\eta_c$ decay final states, such as $p\bar{p}$ and $4K$ modes, to study states recoiling against continuum $\eta_c$ in $e^+e^- \to c\bar{c}c\bar{c}$ processes. Since the useful $\eta_c$ modes have branching fractions ($\sim 10^{-3}$) that are $\sim 10^{-2}$ smaller than those for the $J/\psi$ this method will require a large data sample. If we assume that the $\sigma(e^+e^- \to \eta_c h_c) \simeq \sigma(e^+e^- \to J/\psi\chi_{c0})$, about 10 ab$^{-1}$ of data would be needed to confirm the $h_c$ by this method.

### 5.13.2 Charmonia above open-charm threshold

As for the states above the $D\bar{D}$ mass threshold, despite recent experimental progress, situation is not well established. Of the predicted $n = 1$ $D$-wave charmonia only $\psi(3770)$ ($1^3D_{c1}$) has been observed. Remaining ones, $1^3D_{c2,c3}$, $1^1D_{c2}$, with masses predicted to be close to the $\psi(3770)$ mass, are supposed to be narrow and decay to lower-lying charmonia as their decays to $D\bar{D}^{(*)}$ are forbidden due to either parity conservation or suppressed by angular momentum barrier. Therefore observations of all the $D$-waves states is feasible. Especially discovery of $1^3D_{c2}$ is important for a possible identification with $X(3872)$ (see Sect.5.13.3), which - even though unlikely - has not been ruled out. Since the exclusive decay $B \to (1^3D_{c2})K$ is expected to be strongly suppressed by factorization, here also searches in exclusive decays are not promising. Since the $\gamma\chi_{c1}$ and $\pi^+\pi^-J/\psi$ modes are expected to be strong and are experimentally distinct, inclusive searches for $B \to (1^3D_{c2})X$ are feasible.

The next unobserved multiplet, $2^3P_{c1,c2,c3}$ and $2^1P_{c1}$ ($\chi'_{c1,c2,c3}$ and $h'_c$) should lie in mass range of 3800-3980 MeV and have widths of 30-150 MeV. The $Z(3930)$ observed as a peak in the spectrum of $D\bar{D}$ produced in $\gamma\gamma$ and having quantum numbers determined to be $2^{++}$, is a good candidate for $\chi'_{c2}$ [362]. This interpretation could be confirmed by observation of the $D\bar{D}^{(*)}$ final state predicted to have a large, about 25%, branching fraction for $\chi'_{c2}$. Given the signal yield observed for $Z(3930) \to D\bar{D}$ with data sample of about 400 fb$^{-1}$, at least 5 ab$^{-1}$ sample would be necessary to observe significant $Z(3930) \to D\bar{D}^{(*)}$ signal, if one accounts for the $D^*$ reconstruction efficiency and lower partial width predicted.

As mentioned in section 5.12.3, the $X(3940)$ and $X(4160)$ states observed in $e^+e^- \to J/\psi X$ could be candidates for $\eta_c(3S)$ and $\eta_c(4S)$, although the X masses are below potential model estimations being 4050 MeV for $\eta_c(3S)$ and 4400 MeV for $\eta_c(4S)$. Once a data sample of 5 ab$^{-1}$ is available, the assignments suggested could be verified by observing the $X(3940)$ and $X(4160)$ signals, respectively, in $\gamma\gamma \to D\bar{D}^*$ and $\gamma\gamma \to D^*\bar{D}^*$ followed by angular analyses.

The known charmonia above the $D\bar{D}$ mass threshold: $\psi(3770)$, $\psi(4040)$, $\psi(4160)$ and $\psi(4415)$ are $1^{--}$ states corresponding respectively to $1^3D_1$, $3^3S_1$, $4^3S_1$ and $2^3D_1$. Their parameters have been determined from the fit to the $R$ spectrum measured in energy scans performed by BESII experiment [363]. However fitting such inclusive data is complicated and yields the resonant parameters being strongly model-dependent. To reduce this effect, one could fit exclusive cross-sections for $e^+e^- \to D^{(*)}\bar{D}^{(*)}$ accessed through ISR process, once they are measured with Super-KEKB factory statistics. Also a line-shape of the $\psi(3770)$ which was found to be anomalous and indicating evidence for a new structure additional to the single $\psi(3770)$ resonance [364], could be further investigated. The latter may help to solve the puzzle related to the large inclusive non-$D\bar{D}$ branching fraction of the $\psi(3770)$, that is not confirmed by searches for exclusive non-$D\bar{D}$ decays.

### 5.13.3 Charmonium-like states

An important task for the Super-KEKB factory will be further studies of the $XYZ$ states. The recent results do not seem to bring us any closer to solve the puzzle which started in 2003, when Belle discovered a narrow charmonium-like $\pi^+\pi^- J/\psi$ state with a mass of 3872 MeV in exclusive $B \to K\pi\pi J/\psi$ decays [365]. This state, called the $X(3872)$, has been confirmed by CDF and $D\emptyset$ to be produced in $p\bar{p}$ collisions, as well as by BaBar [366]. Recently the CDF experiment has precisely measured mass of the $X(3872)$ to be $m_{X(3872)} = 3871.61 \pm 0.16 \pm 0.19$ MeV/$c^2$. In addition to $J/\psi\pi^+\pi^-$ where dipion mass spectrum is consistent with originating from $\rho \to \pi\pi$ [367], also an evidence of radiative decays to $J/\psi\gamma$ [368] and recently to $\psi(2S)\gamma$ [369] indicates a positive $C$-parity for the $X(3872)$. The mentioned properties along with results of the CDF angular analysis [370] strongly favor $J^{PC} = 1^{++}$ and $2^{-+}$ for the $X(3872)$. The $X(3872)$ was originally considered to be the $1^3D_{c2}$, however a closer examination of its properties indicated problems with this interpretation: the 3872 MeV mass is substantially above model expectations of $\sim 3815$ MeV; the decay rate to $\gamma\chi_{c1}$ is too small; the shape of the $M(\pi\pi)$ distribution is too strongly peaked at high $\pi\pi$ masses; and the inferred exclusive branching ratio for $B \to (1^3D_{c2})K$ is too large. As a result, a number of theorists have speculated that this particle may not be a $c\bar{c}$ charmonium state, but, instead, a new type of four-quark meson.

The $X(3872)$ mass being in close vicinity of the sum of the $D^0$ and $D^{*0}$ masses has triggered speculations that $X(3872)$ is a molecular bound state of $D^0$ and $\bar{D}^{*0}$ [371]. A narrow near-threshold enhancement which could originate from the $X(3872)$, have been observed in the mass distribution of $D^0\bar{D}^{*0}$ system produced in the $B \to KD^0\bar{D}^{*0}$ decays (Fig. 5.68) [372]. Its measured mass ($m = 3872.6^{+0.5}_{-0.4} \pm 0.4$ MeV/$c^2$) and large branching fraction for the $D^0\bar{D}^{*0}$ decay mode with respect to $J/\psi\pi^+\pi^-$ decays, support the molecular interpretation of the $X(3872)$. Although the resonance line-shape can be important for investigating the nature of the $X(3872)$ [373], studying the $X(3872)$ shape is not feasible with the current statistics. Also precise measurement of mass and width of the enhancement can be only performed with Super-KEKB factory data where having 5 ab$^{-1}$ would allow one to reconstruct about 400 signal events. As the large branching fraction of radiative $X(3872)$ decays, especially of $X(3872) \to \psi(2S)\gamma$, are problematic for molecular scenario [374], confirmation of existing evidence for $X(3872) \to J/\psi\gamma$ and $\psi(2S)\gamma$ radiative decays as well as search for less distinct ones such as $\chi_{c1}\gamma$, will be an important task on the way to understanding the nature of $X(3872)$.

Tetraquark explanation of $X(3872)$ [343] [375] predicts that the doublet should exist corresponding to $[cu][\bar{c}\bar{u}]$ and $[cd][\bar{c}\bar{d}]$ states produced respectively in charged $B^+ \to K^+ J/\psi\pi^+\pi^-$ and neutral $B^0 \to K^0 J/\psi\pi^+\pi^-$ decays with masses different by a few MeV. However recent studies have not revealed any significant mass difference between the $X$ produced in charged versus neutral $B$ decays [376] (Fig. 5.69). The signal yield for $X(3872) \to J/\psi\pi^+\pi^-$ (about 150 events for 600 fb$^{-1}$ data) once enlarged by a factor of a few shall be sufficient for determination of width of the $X(3872)$ and a shape of the resonance. Remaining action items related to hadronic $X(3872)$ transitions and feasible with increased data sample, would be: search for $J/\psi\pi^0\pi^0$ decay mode which for any conventional resonance should obey the isospin relation, confirmation of isospin violating mode $J/\psi\pi^+\pi^-\pi^0$ and search for the charged partner of the $X(3872)$ predicted by the tetraquark models in the $J/\psi\pi^+\pi^0$ and not observed so far [377].

Other states which await understanding are $Y(4008)$, $Y(4260)$, $Y(4360)$ and $Y(4660)$ discussed below in more details. These are $1^{--}$ resonances observed in the $J/\psi\pi^+\pi^-$ and $\psi(2S)\pi^+\pi^-$ systems produced in the ISR process. Estimation of their partial decay widths for numerous open-charm final states, which are expected to be large for conventional charmonia in this mass region, will give an insight into the nature of the $Y$-family.

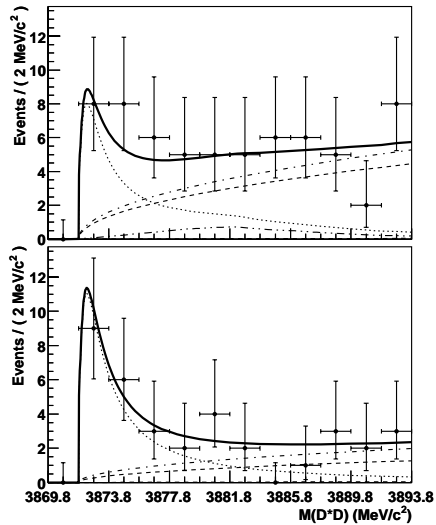

Figure 5.68: $M(D^0\bar{D}^{*0})$ distributions in $B \to KD^0\bar{D}^{*0}$ decays for $D^{*0} \to D^0\gamma$ (left) and $D^{*0} \to D^0\pi^0$ (right). Solid line is the result of the fit, dotted line the contribution of the signal, other lines denote backgrounds.

$Z^+(4430)$, the first charmonium-like state of non-zero electric charge has been observed in the $\pi^+\psi(2S)$ decay channel in a study of $B \to K\pi^+\psi(2S)$ decays performed by Belle [378]. This observation, based on a simple fit to the $\pi^+\psi(2S)$ mass distribution, has been confirmed through the Dalitz-plot analysis of $B \to K\pi^+\psi(2S)$ reaction [379]. The Dalitz plot studied and fit results are shown in Fig. 5.70. Spin hypotheses of $J = 0$ and 1 have resulted in comparable fit qualities and thus cannot be distinguished with current data sample. Therefore much larger statistics is necessary to determine the spin-parity of the $Z^+(4430)$ as well as to further study a structure at $M^2(\psi(2S)\pi^+) \sim 18.5\,\text{GeV}^2/c^4$, the significance of which remains below $4\sigma$ with the current luminosity.

Being a charged state the $Z^+(4430)$ has minimum quark content ($c\bar{c}u\bar{d}$), and thus must be exotic. Theoretical explanations have suggested that since the mass of the $Z^+(4430)$ is close to the $D^*\bar{D}_1(2420)$ threshold it could be either an $S$-wave threshold effect [380] or a $D^*\bar{D}_1(2420)$ molecule [381], whereas tetraquark hypothesis considers the $Z^+(4430)$ to be a diquark-antidiquark state with the $[cu][\bar{c}\bar{d}]$ configuration [382] and predicts an existence of its neutral partner decaying to $\psi(2S)\pi^0$ or $\psi(2S)\eta$. In the molecular scenario the dominating decay modes should be $D^*\bar{D}^*\pi$ whereas in the tetraquark one: $D^{(*)}\bar{D}^*$ and $J\psi\pi$ decay channels in addition to $\psi(2S)\pi$. Obviously the mentioned decay modes are worth studying, however most of them require significantly larger statistics because of either photons or slow pions present in the final state and causing efficiency loss.

Recently BaBar in a search for the $Z^+(4430)$ in the $\pi^+\psi(2S)$ and $\pi^+J/\psi$ decays modes has not found significant $Z^+(4430)$ signal in any of the systems studied [383] and claimed that Belle and BaBar data remain statistically consistent. This calls for further, high statistics studies of the $Z^+(4430)$.

Two other charged resonance-like structures have been observed by Belle [384] in the $\pi^+\chi_{c1}$ mass distribution near $4.1\,\text{GeV}/c^2$ in the $\bar{B}^0 \to K^-\pi^+\chi_{c1}$ decays through full analysis of their Dalitz plot (Fig. 5.71). Also in this case Super-KEKB statistics is necessary to determine quantum numbers and precisely measure masses and widths of the resonances observed.

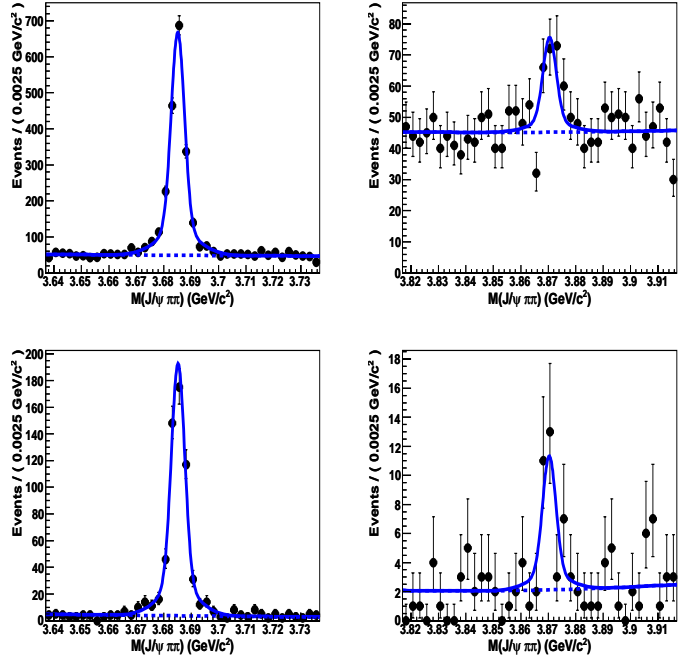

Figure 5.69: $M(J/\psi\pi^+\pi^-)$ distribution for the $\psi(2S)$ (used as a reference) and $X(3872)$ regions for charged $B^+ \to K^+ J/\psi\pi^+\pi^-$ decays (respectively left and right upper plots) and for neutral $B^0 \to K_s^0 J/\psi\pi^+\pi^-$ decays (left and right bottom plots).

### 5.13.4 Spectroscopy and cross-sections through ISR

Among unexpected charmonium-like states the $Y(4260)$, $Y(4008)$ [385–387], $Y(4360)$ and $Y(4660)$ [388, 389] are produced via $e^+e^-$ annihilation have quantum numbers $J^{PC} = 1^{--}$ (measured parameters of $Y$ states are presented in Tab.5.26). The absence of open charm production associated with these new states is inconsistent with their interpretation as conventional charmonium. Partial widths of $Y$'s decay channels to charmonium plus light hadrons are found to be much larger than those usual for conventional charmonium states. The $Y(4260)$ and $Y(4360)$ signals observed in $e^+e^- \to Y(4260)(\to J/\psi\pi^+\pi^-)\gamma_{ISR}$ and $e^+e^- \to Y(4360)(\to \psi(2S)\pi^+\pi^-)\gamma_{ISR}$, respectively, are shown in Fig. 5.72.

On the other hand, the parameters of the conventional charmonium $1^{--}$ states obtained from fits to the inclusive cross section [390] remain model dependent. Since interference between different resonant structures depends upon the specific final states, measurements of exclusive cross sections for charmed meson and baryon pairs in the 4 to 5 GeV energy range are needed to clarify the situation.

Initial-state radiation (ISR) provides a powerful tool for measuring exclusive $e^+e^-$ cross sections at $\sqrt{s}$ smaller than the initial $e^+e^-$ center-of-mass energy at $B$-factories. ISR allows one to obtain cross sections over a broad energy range, while the high luminosity of the $B$-factories compensates for the suppression associated with the emission of a hard photon.

During the past three years numerous measurements of exclusive $e^+e^-$ cross sections for charmed hadron pairs have been reported. Most of these measurements were performed at $B$-factories using initial-state radiation (ISR). Belle presented the first results on the $e^+e^-$ cross sections to the $D\overline{D}$, $D^+D^{*-}$, $D^{*+}D^{*-}$, $D^0D^-\pi^+$, $D^0D^{*-}\pi^+$ and $\Lambda_c^+\Lambda_c^-$ final states [391–395] shown in Fig. 5.73 a), b), c), d), e), f), respectively. BaBar measured $e^+e^-$ cross sections to $D\overline{D}$ and recently to the $D\overline{D}^*$, $D^*\overline{D}^*$ final states [396, 397]. CLEO-c performed a scan over the energy range from 3.97 to 4.26GeV and measured exclusive cross sections for the $D_s\overline{D}_s$, $D_s\overline{D}_s{}^*$ and $D_s^*\overline{D}_s{}^*$ final states at thirteen points with high accuracy [398]. The measured by Belle open charm final states nearly saturate the total cross section for charm hadron production in $e^+e^-$ annihilation in the $\sqrt{s}$ region up to $\sim 4.5$GeV. The exclusive cross sections for charm strange meson pairs have been measured to be an order of magnitude smaller than charm meson production [398]. Charm baryon-antibaryon pair production occurs at energies above 4.5GeV where experimental data are limited or unavailable.

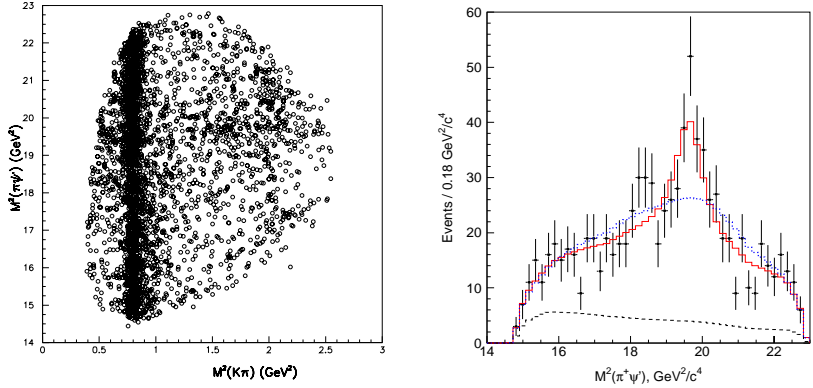

Figure 5.70: $M^2(\pi^+\psi(2S))$ vs. $M^2(K\pi^+)$ Dalitz plot distribution for $B \to K\pi^+\psi(2S)$ decays and its $M^2(\pi^+\psi(2S))$ projection with the $K^* \to K\pi$ mass regions vetoed. Dot-dashed (blue) histogram is the Dalitz-plot fit result for the model with all known $K^*$ mesons only; full (red) histogram represents the fit result with $K^*$'s and $Z \to \psi(2S)\pi^+$ resonance.

No clear evidence for open charm production associated with any of the $Y$ states has been observed. In fact the $Y(4260)$ peak position appears to be close to a local minimum of both the total hadronic cross section [399] and of the exclusive cross section for $e^+e^- \to D^*\overline{D}^*$ [392, 397]. The $X(4630)$, recently found in the $e^+e^- \to \Lambda_c^+\Lambda_c^-$ cross section as a near-threshold enhancement [394], has a mass and width (assuming the $X(4630)$ to be a resonance) consistent within errors with those of the $Y(4660)$ (Tab.5.26). However, this coincidence does not exclude other interpretations of the $X(4630)$, for example, as a conventional charmonium state [400, 401] or as a baryon-antibaryon threshold effect [402]. Finally, this year Belle reported no evidence for $Y(4260) \to D^0 D^{*-}\pi^+$ decays as predicted by hybrid models [403].

Although main contributions to inclusive cross section for $e^+e^-$ annihilation to hadrons have been measured, many other open charm final states remain unknown. Among them $D^{(*)}\overline{D}^{(*)}(n)\pi$ and $\Lambda_c^+\Lambda_c^-(n)\pi$ which are seen in the mass spectra recoiling against $D^{(*)}\gamma_{\text{ISR}}$ and $\Lambda_c^+\gamma_{\text{ISR}}$, respectively. The charm strange final states $D_s^{(*)}\overline{D}_s^{(*)}$ have been measured by CLEO-c with high accuracy, but in the limited energy range below 4.26GeV only [398].

The main problem to extract the cross sections of the mentioned above charm final states is the lack of statistics. The selection of signal events using full reconstruction of the final states plus the $\gamma_{\text{ISR}}$ photon, suffers from low efficiency due to the small $D_{(s)}^{(*)}$ reconstruction efficiencies, small branching fractions and the low geometrical acceptance for the $\gamma_{\text{ISR}}$, which tends to be emitted along the beam line. Using a partial reconstruction leads to serious background problems. Moreover small statistics does not allow to perform angular analysis of the already measured

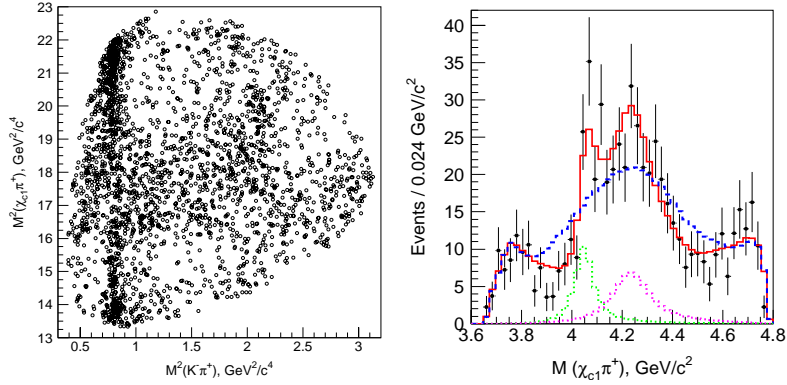

Figure 5.71: Left: $M^2(\pi^+\chi_{c1})$ vs. $M^2(K\pi^+)$ Dalitz plot distribution for $\bar{B}^0 \to K^-\pi^+\chi_{c1}$ decays. Right: $M(\pi^+\chi_{c1})$ distribution for $1.0 < M^2(K^-\pi^+) < 1.75$ GeV$^2/c^4$. Dashed (blue) histogram is the Dalitz-plot fit result for the model with all known $K^*$'s; full (red) histogram represents the fit result with all known $K^*$'s and two $\pi^+\chi_{c1}$ resonances; dotted (green and magenta) histograms represent the contributions of the $\pi^+\chi_{c1}$ resonances.

cross sections, which could shed light on the origin of the cross section structures.

Table 5.26: Measured parameters of the $Y$ states

| State | $M$, MeV/$c^2$ | $\Gamma_{\rm tot}$, MeV | $J^{PC}$ | Decay Modes | Production | Collaboration |
|---|---|---|---|---|---|---|
| $Y(4008)$ | $4008 \pm 40^{+114}_{-28}$ | $226 \pm 44 \pm 87$ | $1^{--}$ | $\pi^+\pi^- J/\psi$ | $e^+e^-$(ISR) | Belle 07 [389] |
| $Y(4260)$ | $4259 \pm 8^{+2}_{-6}$ | $88 \pm 23^{+6}_{-4}$ | $1^{--}$ | $\pi^+\pi^- J/\psi$ | $e^+e^-$(ISR) | BaBar 05 [385] |
| $Y(4260)$ | $4252 \pm 6^{+2}_{-3}$ | $105 \pm 18^{+4}_{-6}$ | $1^{--}$ | $\pi^+\pi^- J/\psi$ | $e^+e^-$(ISR) | BaBar 08 [387] |
| $Y(4260)$ | $4247 \pm 12^{+17}_{-32}$ | $108 \pm 19 \pm 10$ | $1^{--}$ | $\pi^+\pi^- J/\psi$ | $e^+e^-$(ISR) | Belle 07 [386] |
| $Y(4325)$ | $4324 \pm 24$ | $172 \pm 33$ | $1^{--}$ | $\pi^+\pi^- \psi(2S)$ | $e^+e^-$(ISR) | BaBar 06 [388] |
| $Y(4325)$ | $4361 \pm 9 \pm 9$ | $74 \pm 15 \pm 10$ | $1^{--}$ | $\pi^+\pi^- \psi(2S)$ | $e^+e^-$(ISR) | Belle 07 [389] |
| $Y(4660)$ | $4664 \pm 11 \pm 54$ | $48 \pm 15 \pm 3$ | $1^{--}$ | $\pi^+\pi^- \psi(2S)$ | $e^+e^-$(ISR) | Belle 07 [389] |
| $X(4630)$ | $4634^{+8+5}_{-7-8}$ | $92^{+40+10}_{-24-21}$ | $1^{--}$ | $\Lambda_c^+\Lambda_c^-$ | $e^+e^-$(ISR) | Belle 08 [394] |

### 5.13.5 Summary

Super-KEKB factory will provide opportunities for unprecedented studies of the properties of charmonium systems and to search for the missing charmonium states. These measurements will provide stringent tests of hadronic models in the filed where they are supposed to be most reliable.

Taking history as our guide, we are confident that new and surprising discoveries will be made.

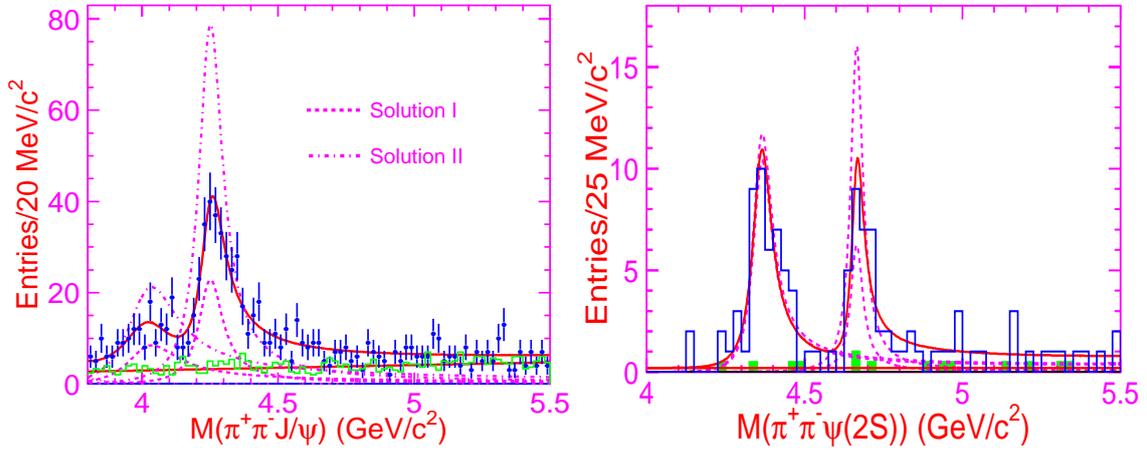

Figure 5.72: $M(J/\psi\pi^+\pi^-)$ (left) and $M(\psi(2S)\pi^+\pi^-)$ (right) distributions from Belle [386,389]. The dashed curves show the $Y$ state contributions for the two fit solutions corresponding to the destructive and constructive interferences between the resonances.

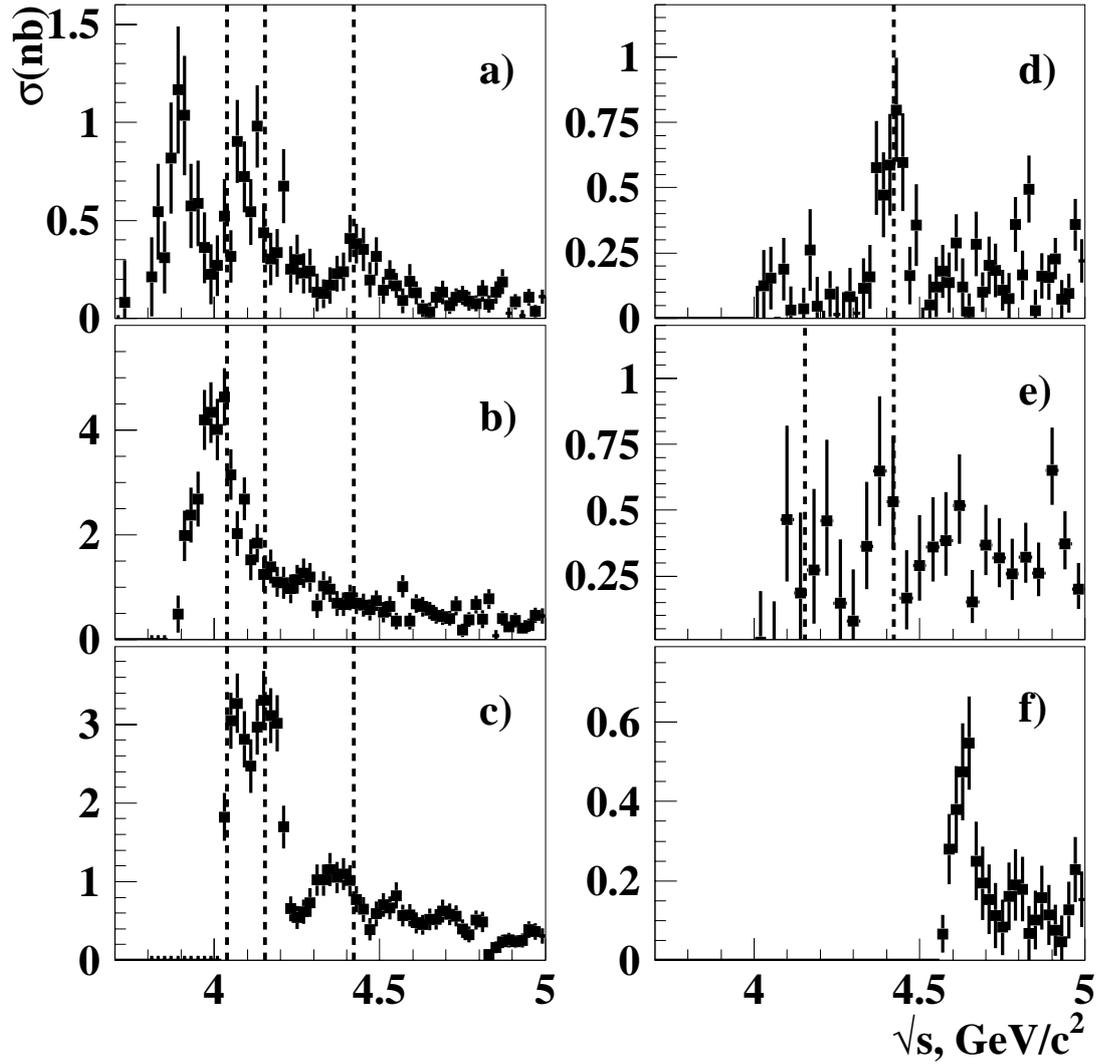

Figure 5.73: The exclusive cross sections for a) $e^+e^- \to D\overline{D}$; b) $e^+e^- \to D^+D^{*-}$; c) $e^+e^- \to D^{*+}D^{*-}$; d) $e^+e^- \to D^0 D^- \pi^+$; e) $e^+e^- \to D^0 D^{*-}\pi^+$ and f) $e^+e^- \to \Lambda_c^+\Lambda_c^-$. The error bars include only the statistical uncertainties. The dotted lines correspond to the $\psi(4040)$, $\psi(4160)$ and $\psi(4415)$ masses [24].

## 5.14 Electroweak physics

The standard-electro weak model has two fundamental parameters, which are directly related to measurable quantities: the parameter $\rho = M_W^2/(M_Z^2 \sin^2\Theta_W)$, which is unity in the standard model and the Weinberg angle $\Theta_W$, which determines the relative contributions of electromagnetic and weak forces. Both parameters have been measured at the $e^+e^-$ colliders PETRA [404] and TRISTAN [405] with a typical precision of about 6% before the high precision measurements at the Z-pole became available from LEP and SLD. In addition experiments employing neutrino proton scattering have measured $\sin^2\Theta_W$ with an accuracy of typically 10% [406]. The errors on these older measurements are too large for the present investigation and will not be considered any further.

Recently a growing interest is being observed to revisit this type of physics and repeat $\sin^2\Theta_W$ and $\rho$ measurements with high precision. There are two aspects to this renewed interest in a precision determination of the fundamental electro-weak parameters. One is related to a measurement of the NuTeV collaboration at Fermilab [407], which observes values of $\rho$ and $\sin^2\Theta_W$ not in agreement with the Standard model. The second motivation is that the scale dependence of gauge couplings has been observed for the strong coupling constant $\alpha_s$ and the electromagnetic coupling $\alpha$, however, not yet for the coupling constant of the weak isospin group SU(2).

The quantity $\sin^2\Theta_W$ is related to the coupling parameters of two gauge groups, U(1) for the electromagnetic part and SU(2) for the weak isospin part of the standard electro weak model. Both couplings have a different scale dependence resulting in a scale dependence of a more complicated nature for $\sin^2\Theta_W$ [408] as shown in Fig. 5.74. To test the scale dependence, data below the Z-pole with an accuracy of a few $10^{-4}$ will serve the purpose.

Two experiments are being proposed to measure $\sin^2\Theta_W$ at center of mass energies below 1 GeV, QWEAK [409] at the Jefferson Lab in Virginia and one at SLAC, E-158 [410]. The QWEAK experiment will deduce $\sin^2\Theta_W$ from elastic scattering of polarized electrons off protons, while the SLAC experiment measures Moller scattering of polarized electrons.

The NuTeV detector at Fermilab consists of an 18 m long, 690 ton active steel-scintillator target with drift chambers as tracking devices followed by an iron-toroid spectrometer. High purity $\nu_\mu$ and $\overline{\nu}_\mu$ beams resulting from interactions of 800 GeV protons in a BeO target can be directed onto the detector.

In principle $\sin^2\Theta_W$ can be derived from a measurement of the ratio between neutral current (NC) and charged current interactions (CC) in a nuclear target for just one neutrino species. This approach is, however, subject to large QCD corrections. The corrections can be minimized by a determination of the ratio $R^-$ with

$$R^- = \frac{\sigma(\nu_\mu N \to \nu_\mu X) - \sigma(\overline{\nu}_\mu N \to \overline{\nu}_\mu X)}{\sigma(\nu_\mu N \to \mu^- X) - \sigma(\overline{\nu}_\mu N \to \mu^+ X)}$$

NuTeV obtains with this method $\sin^2\Theta_W = 0.2277 \pm 0.0013 \pm 0.0009$ [407], a value which is $3\sigma$ above the value expected for the standard electro weak model. From a two parameter fit to $\rho$ and $\sin^2\Theta_W$ they conclude that one of the quantities, but not both of them can be made to agree with the Standard model value. Their result is sensitive to new physics in the W and Z sector and suggests a smaller left-handed NC coupling to light quarks than expected. But this result also depends on hadronic corrections and nuclear structure functions, which is the main criticism with respect to an interpretation in terms of deviations from the standard model.

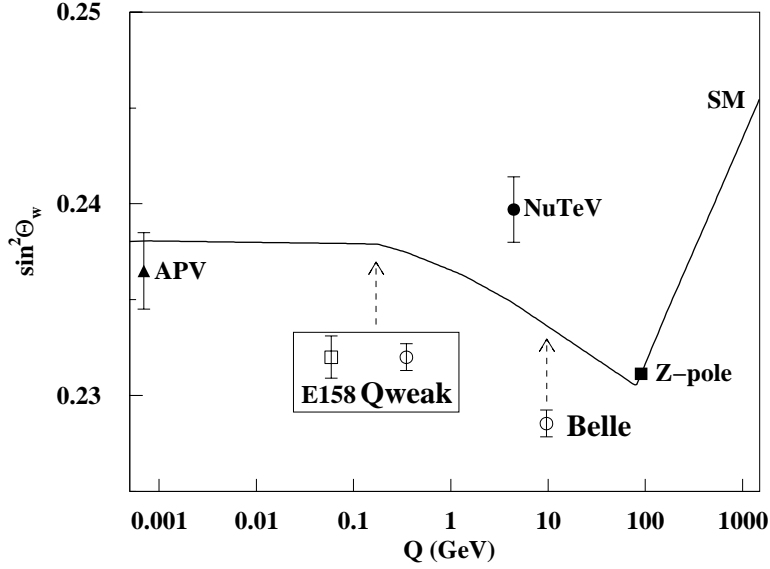

Figure 5.74: Scale dependence of $\sin^2\Theta_W$. The full symbols show the current situation, while the open symbols with error bars for the proposed experiments QWEAK [409], SLAC E-158 [410] and Belle are placed at the correct cm energy with arbitrarily chosen vertical positions. The previous measurements are determinations from atomic parity violation (APV) [411], deep inelastic neutrino scattering (NuTeV) [407], and from Z-pole asymmetries (LEP [412] and SLD [413]).

In the electro-weak process $e^+e^- \to \mu^+\mu^-$ the values for $\sin^2\Theta_W$ and $\rho$ are derived from a fit to the distribution of the centre of mass production angle ($\Theta^*$) of $\mu$ pairs with respect to the axis of the incoming positron in the $e^+e^-$ center of mass system [404].

$$\frac{d\sigma}{d\Omega} = \frac{\alpha^2}{4s}(C_1(1 + \cos^2\Theta^*) + C_2 \cos\Theta^*)$$

with the following definitions:

$$C_1 = 1 + 2v_e v_\mu \chi + (v_e^2 + a_e^2)(v_\mu^2 + a_\mu^2)\chi^2$$

$$C_2 = -4a_e a_\mu \chi + 8v_e a_e a_\mu \chi^2$$

$$v_{e,\mu} = -1 + 4\sin^2\Theta_W \qquad a_{e,\mu} = -1$$

The quantity $\chi$ may be written in two different ways, as a function of $\sin^2\Theta_W$ or as a function of $\rho$

$$\chi = \frac{1}{16 \sin^2\Theta_W \cos^2\Theta_W} \frac{s}{(s - M_Z^2)}$$

or

$$\chi = \frac{\rho G_F M_Z^2}{8\pi\alpha\sqrt{2}} \frac{s}{(s - M_Z^2)}$$

where s is the square of the center of mass energy and $G_F = 1.17 \times 10^{-5}$ GeV$^{-2}$ is the Fermi coupling constant.

For the purpose of estimating statistical errors and significance levels it is sufficient to consider the forward-backward charge asymmetry integrated over all angles, which in the approximation $C_1 \approx 1$, $C_2 \approx -4a_e a_\mu \chi$ equals

$$A_{FB} = \frac{3C_2}{8C_1} = -\frac{3}{2}\chi \ .$$

In an actual experiment one would fit the angular distribution and hence increase the sensitivity.

A guideline for estimating the significance of a measurement at $\sqrt{s} = 10$ GeV would be the capability of distinguishing the value of $\sin^2\Theta_W$ at the Z-pole from a value it would assume if it were running according to the predictions of the Standard model. With the Z-pole value of $\sin^2\Theta_W = 0.232$ the forward-backward asymmetry assumes a value of $A_{FB}(\sqrt{s} = 10 \ GeV) = 6.34 \times 10^{-3}$. For $\sin^2\Theta_W$ running according to the standard model, the asymmetry is $A_{FB}(\sqrt{s} = 10 \ GeV) = 6.38 \times 10^{-3}$. The charge asymmetries for a running and a constant coupling coupling constant thus differ by $\delta(A_{FB}) = 4 \times 10^{-5}$.

From the above relation between $A_{FB}$ and $\chi$ one can estimate the sensitivity of $A_{FB}$ on $\sin^2\theta_W$. At 10 GeV center of mass energy, the statistical error on $\sin^2\Theta_W$ $\sigma(\sin^2\Theta_W)$ is around 50 times larger than the error on the forward-backward asymmetry:

$$\sigma(\sin^2\Theta_W) \approx 50 \ \sigma(A_{FB})$$

With a statistical error of $\sigma(A_{FB}) = \pm 1 \times 10^{-5}$ on the charge asymmetry the corresponding error on $\sin^2\Theta_W$ is $\sigma(\sin^2\Theta_W) = 5 \times 10^{-4}$.

The number of events required to achieve this accuracy is certainly smaller than $10^{10}$ events[12], because when fitting an angular distribution much more efficient use is being made of the experimental data. Thus for an accumulated luminosity of 10 ab$^{-1}$ with around $10^{10}$ produced $e^+e^- \to \mu^+\mu^-$ events a statistically very significant measurement can be made, which is compatible with that of the two dedicated experiments as may be inferred from Fig. 5.74.

The $\mu$-pairs from the decay of the $\Upsilon(4S)$ will have a different asymmetry than those of the continuum, as with $\Upsilon$ as an intermediate state the Z-boson couples to b-quarks and the relative weight between $\gamma$ and Z-exchange is altered with respect to the continuum. The branching fraction of BR($\Upsilon(4S) \to \mu^+\mu^-) = 3 \times 10^{-5}$ is, however, so small that corrections to the angular dependence will become less important and they can be calculated with sufficient accuracy, as all couplings and weak charges are known. The experiment is thus a continuum experiment.

Radiative corrections are small and can be calculated with sufficient accuracy [414]. These corrections come in two categories, the emission of real photons from the incoming electrons and positrons resulting in a downward shift of the center of mass energy and higher order loop corrections. The initial state radiation is well understood and corresponding MC programs are at hand. The loop corrections affect $\Theta_W$ and $\rho$.

Loop corrections are absorbed into the definition of an effective Weinberg angle $\sin^2\Theta_{eff} = \kappa(Q^2)\sin^2\Theta_W$ with $\kappa(M_Z^2) = 1.040$ at the Z-pole. At 10 GeV center of mass energy we expect $\kappa$ to be of the order of 1.02. The value of $\kappa$ is mainly due to the running of the electromagnetic coupling constant $\alpha$. The contribution from electroweak loops depending on the Higgs and the top masses is one order of magnitude smaller. Loop corrections to $\rho$ are of the order of 0.5% and depend on the running of $\alpha$ as well as on the Higgs and top masses. The uncertainty introduced by radiative corrections is estimated to be at the level of $10^{-6}$.

---

[12]Estimating the stat. uncertainty of the asymmetry by $\sigma(A_{FB}) = 1/\sqrt{N}$.

A potential source of systematic errors are radiative returns of the $\Upsilon(1S)$ and higher resonances. These events can be removed by a cut on the invariant mass of the two leptons.

The angular distribution of electron pairs is known with high accuracy and therefore electron pairs can be used to check the charge symmetry of the detector and other systematic errors like those, which may be related to the fact that the muon momenta in the forward and backward hemispheres are different due to the asymmetric energies of the colliding electron and positron beams. Systematic errors due to a charge asymmetry of the detector will, thus, decrease with increasing statistics. They are estimated at $10^{-4}$ for an accumulated luminosity of 1 ab$^{-1}$.

Muon identification need not be very restrictive, because there are only very few reactions, which could fake muon pairs, if collinearity of the two tracks is requested. Tau pairs will be rejected by collinearity cuts and invariant mass cuts, the cross section for pion pairs is too small to present a serious background and electron pairs are easily identified by their unique signature in the CsI crystals. As a first guess, a loose identification requirement for a single muon track would be sufficient, while the second collinear particle need not be strictly identified as a muon, as long it is incompatible with the electron hypothesis.

The presence of SU(2) breaking forces could result in a different phenomenological pattern for reactions, where quarks are present or only leptons are involved. Leptoquarks as an example could alter the result of a $\sin^2\Theta_W$ measurement in neutrino-nucleon scattering without a measurable impact on the charge asymmetry of muon pairs in e$^+$e$^-$ annihilation. Therefore one could imagine to face a situation, where purely leptonic experiments would agree with the Standard model, while reactions involving nucleons wouldn't. In this case one could extend the program at SuperKEKB to a determination of the charge asymmetry of jets. This is a very ambitious measurement, which needs extensive MC studies to investigate the feasibility of such a measurement.

In summary, SuperKEKB will be capable of performing a statistically significant measurement of the Weinberg angle $\sin^2\Theta_W$ in order to prove the running of the U(1)/SU(2) couplings of the standard electroweak model and set limits on new weak isospin breaking interactions. In order to achieve this, data corresponding to about 10 ab$^{-1}$ are needed to add an independent measurement point of competitive accuracy to other low energy experiments. Higher statistics will of course lead to even more stringent tests of the Standard model.

## 5.15 Summary

Tables 5.27 and 5.28 summarize the sensitivities for some of key observables described in the previous sections. As a comparison, we also list expected sensitivities at LHCb whenever available. It is seen that most of key observables are accessible only at the $e^+e^-$ $B$ factories. The advantage of the clean environment at SuperKEKB is thus clear. Note that the $B$ physics program at hadron colliders has its own unique measurements that are not accessible at $e^+e^-$ $B$ factories. Examples include rare $B_s$ decays such as $B_s \to \mu^+\mu^-$. Thus $B$ physics programs at hadron colliders also help scrutinize the rich phenomenologies of $B$ meson decays.

| Observable | Belle 2006 ($\sim$0.5 ab$^{-1}$) | SuperKEKB (5 ab$^{-1}$) | (50 ab$^{-1}$) | $^\dagger$LHCb (2 fb$^{-1}$) | (10 fb$^{-1}$) |
|---|---|---|---|---|---|
| Hadronic $b \to s$ transitions | | | | | |
| $\Delta \mathcal{S}_{\phi K^0}$ | 0.22 | 0.073 | 0.029 | | 0.14 |
| $\Delta \mathcal{S}_{\eta' K^0}$ | 0.11 | 0.038 | 0.020 | | |
| $\Delta \mathcal{S}_{K^0_S K^0_S K^0_S}$ | 0.33 | 0.105 | 0.037 | - | - |
| $\Delta \mathcal{A}_{\pi^0 K^0_S}$ | 0.15 | 0.072 | 0.042 | - | - |
| $\mathcal{A}_{\phi\phi K^+}$ | 0.17 | 0.05 | 0.014 | | |
| $\phi_1^{eff}(\phi K_S)$ Dalitz | | 3.3° | 1.5° | | |
| Radiative/electroweak $b \to s$ transitions | | | | | |
| $\mathcal{S}_{K^0_S \pi^0 \gamma}$ | 0.32 | 0.10 | 0.03 | - | - |
| $\mathcal{B}(B \to X_s \gamma)$ | 13% | 7% | 6% | - | - |
| $A_{CP}(B \to X_s \gamma)$ | 0.058 | 0.01 | 0.005 | - | - |
| $C_9$ from $\overline{A}_{\text{FB}}(B \to K^*\ell^+\ell^-)$ | - | 11% | 4% | | |
| $C_{10}$ from $\overline{A}_{\text{FB}}(B \to K^*\ell^+\ell^-)$ | - | 13% | 4% | | |
| $C_7/C_9$ from $\overline{A}_{\text{FB}}(B \to K^*\ell^+\ell^-)$ | - | | 5% | | 7% |
| $R_K$ | | 0.07 | 0.02 | | 0.043 |
| $\mathcal{B}(B^+ \to K^+ \nu\nu)$ | $^{\dagger\dagger} < 3\ \mathcal{B}_{\text{SM}}$ | | 30% | - | - |
| $\mathcal{B}(B^0 \to K^{*0} \nu\bar{\nu})$ | $^{\dagger\dagger} < 40\ \mathcal{B}_{\text{SM}}$ | | 35% | - | - |
| Radiative/electroweak $b \to d$ transitions | | | | | |
| $\mathcal{S}_{\rho\gamma}$ | - | 0.3 | 0.15 | | |
| $\mathcal{B}(B \to X_d \gamma)$ | - | 24% (syst.) | | - | - |
| Leptonic/semileptonic $B$ decays | | | | | |
| $\mathcal{B}(B^+ \to \tau^+ \nu)$ | 3.5$\sigma$ | 10% | 3% | | |
| $\mathcal{B}(B^+ \to \mu^+ \nu)$ | $^{\dagger\dagger} < 2.4 \mathcal{B}_{\text{SM}}$ | 4.3 ab$^{-1}$ for 5$\sigma$ discovery | | - | - |
| $\mathcal{B}(B^+ \to D\tau\nu)$ | - | 8% | 3% | - | - |
| $\mathcal{B}(B^0 \to D\tau\nu)$ | - | 30% | 10% | - | - |
| LFV in $\tau$ decays (U.L. at 90% C.L.) | | | | | |
| $\mathcal{B}(\tau \to \mu\gamma)\ [10^{-9}]$ | 45 | 10 | 5 | - | - |
| $\mathcal{B}(\tau \to \mu\eta)\ [10^{-9}]$ | 65 | 5 | 2 | - | - |
| $\mathcal{B}(\tau \to \mu\mu\mu)\ [10^{-9}]$ | 21 | 3 | 1 | - | - |
| Unitarity triangle parameters | | | | | |
| $\sin 2\phi_1$ | 0.026 | 0.016 | 0.012 | $\sim$0.02 | $\sim$0.01 |
| $\phi_2\ (\pi\pi)$ | 11° | 10° | 3° | - | - |
| $\phi_2\ (\rho\pi)$ | 68° < $\phi_2$ < 95° | 3° | 1.5° | 10° | 4.5° |
| $\phi_2\ (\rho\rho)$ | 62° < $\phi_2$ < 107° | 3° | 1.5° | - | - |
| $\phi_2$ (combined) | | 2° | $\lesssim$ 1° | 10° | 4.5° |
| $\phi_3\ (D^{(*)} K^{(*)})$ (Dalitz mod. ind.) | 20° | 7° | 2° | 8° | |
| $\phi_3\ (DK^{(*)})$ (ADS+GLW) | - | 16° | 5° | 5-15° | |
| $\phi_3\ (D^{(*)}\pi)$ | - | 18° | 6° | | |
| $\phi_3$ (combined) | | 6° | 1.5° | 4.2° | 2.4° |
| $|V_{ub}|$ (inclusive) | 6% | 5% | 3% | - | - |
| $|V_{ub}|$ (exclusive) | 15% | 12% (LQCD) | 5% (LQCD) | - | - |
| $^{\dagger\dagger\dagger} \bar{\rho}$ | 20.0% | | 3.4% | | |
| $^{\dagger\dagger\dagger} \bar{\eta}$ | 15.7% | | 1.7% | | |

Table 5.27: Summary of sensitivity (I). Branching fraction limits in the table are at the 90% confidence level. $^\dagger$Values for LHCb are statistical only and are taken from [415] unless otherwise stated. $^{\dagger\dagger}$ $\mathcal{B}_{\text{SM}}$ represents the expected branching fraction in the SM; $\mathcal{B}_{\text{SM}} = 5 \times 10^{-6}$ for $\mathcal{B}(B^+ \to K^+ \nu\nu)$, $7 \times 10^{-6}$ for $B^0 \to K^{*0} \nu\bar{\nu}$ and $7.07 \times 10^{-7}$ for $\mathcal{B}(B^+ \to \mu\nu)$ are used in this table. $^{\dagger\dagger\dagger}$See the next chapter for details.

| Observable | Belle | Belle/SuperKEKB | LHCb[†] | |
|---|---|---|---|---|
| | | | (2 fb$^{-1}$) | (10 fb$^{-1}$) |
| $B_s$ physics | (25 fb$^{-1}$) | (5 ab$^{-1}$) | | |
| $\mathcal{B}(B_s \to \gamma\gamma)$ | $< 8.7 \times 10^{-6}$ | $0.25 \times 10^{-6}$ | - | - |
| $\Delta\Gamma_s^{CP}/\Gamma_s$ ($Br(B_s \to D_s^{(*)}D_s^{(*)})$) | 3% | 1% (model dependency) | - | - |
| $\Delta\Gamma_s/\Gamma_s$ ($B_s \to f_{CP}$ t-dependent) | - | 1.2% | - | - |
| $\phi_s$ (with $B_s \to J/\psi\phi$ etc.) | - | - | 0.02 | 0.01 |
| $\mathcal{B}(B_s \to \mu^+\mu^-)$ | - | | 6 fb$^{-1}$ for $5\sigma$ discovery | |
| $\phi_3$ ($B_s \to KK$) | - | | 7-10° | |
| $\phi_3$ ($B_s \to D_s K$) | - | | 13° | |
| $\Upsilon$ decays | (3 fb$^{-1}$) | (500 fb$^{-1}$) | | |
| $\mathcal{B}(\Upsilon(1S) \to$ invisible) | $< 2.5 \times 10^{-3}$ | $< 2 \times 10^{-4}$ | | |
| | ($\sim$0.5 ab$^{-1}$)[‡] | (5 ab$^{-1}$) | (50 ab$^{-1}$) | |
| Charm physics | | | | |
| $D$ mixing parameters | | | | |
| $x$ | 0.25% | 0.12% | 0.09% | 0.25%[††] |
| $y$ | 0.16% | 0.10% | 0.05% | 0.05%[††] |
| $\delta_{K\pi}$ | 10° | 6° | 4° | |
| $|q/p|$ | 0.16 | 0.1 | 0.05 | |
| $\phi$ | 0.13 rad | 0.08 rad | 0.05 rad | |
| $A_D$ | 2.4% | 1% | 0.3% | |
| New particles[ℵ] | | | | |
| $\gamma\gamma \to Z(3930) \to D\bar{D}^*$ | | $> 3\sigma$ | | |
| $B \to K X(3872)(\to D^0\bar{D}^{*0})$ | | 400 events | | |
| $B \to K X(3872)(\to J/\psi\pi^+\pi^-)$ | | 1250 events | | |
| $B \to K Z^+(4430)(\to \psi'\pi^+)$ | | 1000 events | | |
| $e^+e^- \to \gamma_{\text{ISR}} Y(4260)(\to J/\psi\pi^+\pi^-)$ | | 3000 events | | |
| Electroweak parameters | | ($\sim$10 ab$^{-1}$) | | |
| $\sin^2\Theta_W$ | - | $3 \times 10^{-4}$ | | |

Table 5.28: Summary of sensitivity studies (II). [†]Values for LHCb are taken from [415] unless otherwise stated. [‡]For $D$ mixing parameters the world average of results from [22] is quoted. [††]LHCb sensitivities on $x$ and $y$ are estimated from sensitivities on $x'^2$, $y'$ and $y_{CP}$ [295] assuming $x = y = 0.80\%$ and $\delta_{K\pi} = 25° \pm 6°$. [ℵ]Due to a large number of various possible measurements we only list expected signal yields for a few interesting processes.

# References

[1] A. G. Akeroyd *et al.* [SuperKEKB Physics Working Group], "Physics at super B factory," arXiv:hep-ex/0406071.

[2] A. B. Carter and A. I. Sanda, "CP Violation In Cascade Decays Of B Mesons," Phys. Rev. Lett. **45**, 952 (1980).

[3] A. B. Carter and A. I. Sanda, "CP Violation In B Meson Decays," Phys. Rev. D **23**, 1567 (1981).

[4] I. I. Bigi and A. I. Sanda, "Notes On The Observability Of CP Violations In B Decays," Nucl. Phys. B **193**, 85 (1981).

[5] K. Abe *et al.* [Belle Collaboration], "An improved measurement of mixing-induced CP violation in the neutral B meson system. ((B))," Phys. Rev. D **66**, 071102 (2002).

[6] H. Tajima *et al.*, "Proper-time resolution function for measurement of time evolution of $B$ mesons at the KEK B-factory," Nucl. Instrum. Meth. A **533**, 370 (2004).

[7] T. Matsumoto, "Tagging with fully reconstructed $B$'s in $B$-factory experiments," ACAT03 conference, Dec. 1-5, 2003, KEK, Japan.

[8] B. Aubert *et al.* [BABAR Collaboration], "Measurement of the inclusive charmless semileptonic branching ratio of $B$ mesons and determination of $|V_{ub}|$," Phys. Rev. Lett. **92**, 071802 (2004).

[9] del Re, Daniele, "Measurement of $V_{ub}$ studying inclusive semileptonic decays on the recoil of fully reconstructed $B$'s with the BaBar experiment," Ph.D. thesis, Dec. 5, 2002.

[10] A. Sugiyama, "$B \to X_u l \nu$ measurement using the pseudo full reconstruction of $\Upsilon(4s)$ at BELLE," Belle Note No. 509.

[11] B. Aubert *et al.* [BABAR Collaboration], "A search for $B^+ \to K^+ \nu \bar{\nu}$. ((B))," International Conference on High Energy Physics 2002, arXiv:hep-ex/0207069.

[12] B. Aubert *et al.* [BABAR Collaboration], "A search for $B^+ \to^t au^+ \nu_\tau$ recoiling against $B^- \to D^0 l^- \nu_l X$," arXiv:hep-ex/0303034.

[13] For a recent review, see for example, Y. Nir, "CP violation: The CKM matrix and new physics," Nucl. Phys. Proc. Suppl. **117**, 111 (2003).

[14] T. Moroi, "CP violation in $B_d \to \phi K_S$ in SUSY GUT with right-handed neutrinos," Phys. Lett. B **493**, 366 (2000).

[15] S. Baek, T. Goto, Y. Okada and K. i. Okumura, "Neutrino oscillation, SUSY GUT and B decay," Phys. Rev. D **63**, 051701 (2001).

[16] D. Chang, A. Masiero and H. Murayama, "Neutrino mixing and large CP violation in B physics," Phys. Rev. D **67**, 075013 (2003).

[17] For a recent review, see for example, W. Bernreuther, "CP violation and baryogenesis," Lect. Notes Phys. **591**, 237 (2002).

[18] Y. Grossman and M. P. Worah, "CP asymmetries in $B$ decays with new physics in decay amplitudes," Phys. Lett. B **395**, 241 (1997).

[19] K.-F. Chen et al. [Belle Collaboration], "Observation of Time-Dependent $CP$ Violation in $B^0 \to \eta' K^0$ Decays and Improved Measurements of $CP$ Asymmetries in $B^0 \to \phi K^0$, $K_S^0 K_S^0 K_S^0$ and $B^0 \to J/\psi K^0$ Decays," Phys. Rev. Lett. **98**, 031802 (2007).

[20] B. Aubert it et al. [BABAR Collaboration], "Measurement of $CP$-violating asymmetries in the $B^0 \to K^+ K^- K^0$ Dalitz plot," arXiv:hep-ex/0607112.

[21] B. Aubert it et al. [BABAR Collaboration], "Observation of CP violation in $B^0 \to \eta' K^0$ decays," Phys. Rev. Lett. **98**, 031801 (2007).

[22] E. Barberio et al. [Heavy Flavor Averaging Group (HFAG)], "Averages of b-hadron and c-hadron Properties at the End of 2007," arXiv:0808.1297; updates at http://www.slac.stanford.edu/xorg/hfag/charm/index.html.

[23] J. Blatt and V. E. Weisskopf, *Theoretical Nuclear Physics*, J. Wiley & Sons, New York (1952).

[24] C. Amsler et al. [Particle Data Group], "Review of Particle Physics," Phys. Lett. B **667** 1, (2008).

[25] S. M. Flatté, "Coupled - Channel Analysis of the $\pi - \eta$ and $K - \bar{K}$ Systems Near $K - \bar{K}$ Threshold," Phys. Lett. B **63**, 224 (1976).

[26] A. Garmash et al. [Belle Collaboration], "Dalitz analysis of the three-body charmless decays $B^+ \to K^+ \pi^+ \pi^-$ and $B^+ \to K^+ K^+ K^-$," Phys. Rev. D **71**, 092003 (2005).

[27] B. Aubert et al. [BaBar Collaboration], "Measurement of $CP$-Violating Asymmetries in the $B^0 \to K^+ K^- K_S^0$ Dalitz Plot," arXiv:0808.0700 [hep-ex].

[28] M. Ablikim et al. [BES Collaboration], "Resonances in $J/\psi \to \phi \pi^+ \pi^-$ and $\phi K^+ K^-$," Phys. Lett. B **607**, 243 (2005).

[29] J. Dalseno, presentation at 34th International Conference on High Energy Physics (ICHEP2008), http://www.hep.upenn.edu/ichep08/talks/misc/download_slides?Talk_id=163

[30] M. Hazumi, "Large direct CP violation in $B \to \phi \phi X_s$ decays," Phys. Lett. B **583**, 285 (2004).

[31] F. Fang et al. [Belle Collaboration], "Measurement of branching fractions for $B \to \eta_c K^*$ decays," Phys. Rev. Lett. **90**, 071801 (2003).

[32] H. C. Huang *et al.* [Belle Collaboration], "Evidence For $B \to \Phi\Phi K$," Phys. Rev. Lett. **91**, 241802 (2003).

[33] N. G. Deshpande and J. Trampetic, "Exclusive and semiinclusive $B$ decays based on $b \to s\eta_c$ transition," Phys. Lett. B **339**, 270 (1994).

[34] M. Gourdin, Y. Y. Keum and X. Y. Pham, "Testing factorization in color suppressed beauty decays with the $B \to \eta_c + K(K^*)$ modes," Phys. Rev. D **51**, 3510 (1995).

[35] M. R. Ahmady and E. Kou, "Combined $B \to X_s\psi$ and $B \to X_s\eta_c$ decays as a test of factorization," Eur. Phys. J. C **1**, 243 (1998).

[36] KEK, Tsukuba Progress Report 2000, "The Belle detector," Nucl. Instrum. Meth. A **479**, 117 (2002).

[37] J. Kaneko *et al.* [Belle Collaboration], "Measurement of the electroweak penguin process $B \to X_s l^+ l^-$," Phys. Rev. Lett. **90**, 021801 (2003).

[38] Y. Grossman, G. Isidori and M. P. Worah, "CP asymmetry in $B_d \to \phi K_S$: Standard model pollution," Phys. Rev. D **58**, 057504 (1998).

[39] D. London and A. Soni, "Measuring the CP angle $\beta$ in hadronic $b \to s$ penguin decays," Phys. Lett. B **407**, 61 (1997).

[40] M. Beneke and M. Neubert, "QCD factorization for $B \to PP$ and $B \to PV$ decays," Nucl. Phys. B **675**, 333 (2003).

[41] H.-Y. Cheng, C.-K. Chua and A. Soni, "$CP$-violating asymmetries in $B^0$ decays to $K^+ K^- K^0_{S(L)}$ and $K^0_S K^0_S K^0_{S(L)}$," Phys. Rev. D **72**, 094003 (2005).

[42] A. R. Williamson and J. Zupan, "Two body $B$ decays with isosinglet final states in SCET," Phys. Rev. D **74**, 014003 (2006).

[43] Y. Grossman, Z. Ligeti, Y. Nir and H. Quinn, "SU(3) relations and the CP asymmetries in B decays to eta' K(S), Phi K(S) and K+ K- K(S)," Phys. Rev. D **68**, 015004 (2003).

[44] H.-Y. Cheng, C.-K. Chua and A. Soni, "Effects of final-state interactions on mixing-induced CP violation in penguin-dominated B decays," Phys. Rev. D **72**, 014006 (2005).

[45] M. Gronau, J. L. Rosner and J. Zupan, "Updated bounds on CP asymmetries in $B^0 \to \eta' K^0_S and B^0 \to \pi^0 K^0_S$," Phys. Rev. D **74**, 093003 (2006).

[46] M. Misiak *et al.*, "Estimate of $B(-\bar{B} \to X_s \gamma) at O(\alpha_s^2)$," Phys. Rev. Lett. **98**, 022002 (2007).

[47] A. Ali *et al.*, "Improved model independent analysis of semileptonic and radiative rare B decays," Phys. Rev. D**66**, 034002 (2002);

T. Hurth, "Present status of inclusive rare B decays," Rev. Mod. Phys. **75**, 1159 (2003).

[48] R. Ammar *et al.* [Cleo Collaboration], "Evidence for penguins: First observation of $B \to K^*(892)\gamma$," Phys. Rev. Lett. **71**, 674 (1993).

[49] D. Mohapatra *et al.* [Belle Collaboration], "Observation of $b \to d\gamma$ and determination of $|V(t_d)/V(t_s)|$," Phys. Rev. Lett. **96**, 221601 (2006).

[50] B. Aubert *et al.* [BaBar Collaboration], "Branching fraction measurements of $B^+ \to \rho^+\gamma$, $B^0 \to \rho^0\gamma$, and $B^0 \to \omega\gamma$," Phys. Rev. Lett. **98**, 151802 (2007).

[51] K. Abe *et al.* [Belle Collaboration], "Observation of the decay $B \to Kl^+l^-$," Phys. Rev. Lett. **88**, 021801 (2002).

[52] J. Kaneko *et al.* [Belle Collaboration], "Measurement of the electroweak penguin process $B \to X_s l+l-$," Phys. Rev. Lett. **90**, 021801 (2003).

[53] A. Ishikawa *et al.* [Belle Collaboration], "Observation of $B \to K^*l^+l^-$," Phys. Rev. Lett. **91**, 261601 (2003).

[54] A. Ishikawa *et al.* [Belle Collaboration], "Measurement of Forward-Backward Asymmetry and Wilson Coefficients in $B \to K^*l^+l^-$," Phys. Rev. Lett. **96**, 251801 (2006).

[55] P. Gambino, U. Haisch, M. Misiak, "Determining the sign of the $b \to s\gamma$ amplitude," Phys. Rev. Lett. **94**, 061803 (2005), and references therein.

[56] S. Fukae, C. S. Kim, T. Morozumi and T. Yoshikawa, "A model independent analysis of the rare $B$ decay $B \to X_s l^+l^-$," Phys. Rev. D **59**, 074013 (1999).

[57] G. Hiller and F. Kruger, "More model-independent analysis of $b \to s$ processes," Phys. Rev. D**69**, 074020 (2004).

[58] O. L. Buchmüller, H.U. Flächer, "Fit to moment from $B \to X_s l \bar{n} u$ and $B \to X_s \gamma$ decays using heavy quark expansions in the kinetic scheme," Phys. Rev. D**73**, 073008 (2006).

[59] C. W. Bauer *et al.*, "Global analysis of inclusive B decays," Phys. Rev. D**70**, 094017 (2004).

[60] K. Abe *et al.* [Belle Collaboration], "Determination of $|V_{cb}|$ and $m_b$ from Inclusive $B \to X_c l\nu$ and $B-> X_s\gamma$ Decays at Belle," arXiv:hep-ex/0611047;

C. Schwanda *et al.* [Belle Collaboration], "Measurement of the Moments of the Photon Energy Spectrum in $B \to X_s\gamma$ Decays and Determination of $|V(_{cb})|$ and $m(b)$ at Belle," Phys. Rev. D**78**, 032016 (2008).

[61] C. S. Lim, T. Morozumi and A. I. Sanda, "A Prediction For $d\Gamma(B \to sl\bar{l})/dQ^2$ Including The Long Distance Effects," Phys. Lett. B **218**, 343 (1989).

[62] M. Misiak *et al.*, "Estimate of $B(-\bar{B} \to X_s\gamma) at O(\alpha_s^2)$," Phys. Rev. Lett. **98**, 022002 (2007).

[63] J. R. Andersen, E. Gardi, "Radiative B decay spectrum: DGE at NNLO," JHEP **0701**, 029 (2007).

[64] K. Abe *et al.* [Belle Collaboration], "Improved Measurement of Inclusive Radiative B-meson decays," AIP Conf. Proc. **1078**, 342 (2009). arXiv:0804.1580.

[65] A. Arhrib, C. K. Chua and W. S. Hou, "Supersymmetric model contributions to $B_d^0 - \bar{B}_d^0$ mixing and $B \to \pi\pi$, $\rho\gamma$ decays," Eur. Phys. J. C **21**, 567 (2001).

[66] A. G. Akeroyd and S. Recksiegel, "Direct CP asymmetry of $B \to X_{d,s}\gamma$ in a model with vector quarks," Phys. Lett. B **525**, 81 (2002).

[67] N. Taniguchi *et al.* [Belle Collaboration], "Measurement of branching fractions, isospin and CP-violating asymmetries for exclusive $b \to d\gamma$ modes," Phys. Rev. Lett. **101**, 111801 (2008).

[68] T. Hurth, E. Lunghi and W. Porod, "Untagged $B \to X_s\gamma$ CP asymmetry as a probe for new physics," Nucl. Phys. B**704**, 56 (2005).

[69] S. Nishida *et al.* [Belle Collaboration], "Measurement of the CP asymmetry in $B \to X_s\gamma$," Phys. Rev. Lett. **93**, 031803 (2004).

[70] J. M. Soares, "CP violation in radiative b decays," Nucl. Phys. B **367**, 575 (1991).

[71] A. G. Akeroyd, Y. Y. Keum and S. Recksiegel, "Effect of supersymmetric phases on the direct CP asymmetry of $B \to X_d\gamma$," Phys. Lett. B **507**, 252 (2001).

[72] B. Dutta, Y. Mimura, "Large Phase of $B_s - \bar{B}_s$ Mixing in Supersymmetric Grand Unified Theories," Phys. Rev. D**78**, 071702 (2008).

[73] D. Atwood, M. Gronau, A. Soni, "Mixing induced CP asymmetries in radiative B decays in and beyond the standard model," Phys. Rev. Lett. **79**, 185 (1997).

[74] Y. Ushiroda *et al.* [Belle Collaboration], "Time-Dependent CP Asymmetries in $B^0 \to K_s^0\pi^0\gamma$ transitions," Phys. Rev. D**74**, 111104 (2006).

[75] C. S. Kim, Y. G. Kim, C. D. Lu and T. Morozumi, "Azimuthal angle distribution in $B \to K^*(\to K\pi)l^+l^-$ at low invariant $m_{l^+l^-}$ region," Phys. Rev. D **62**, 034013 (2000).

[76] A. Ali *et al.*, "A Comparative study of the decays $B \to (K, K^*)l^+l^-$ in standard model and supersymmetric theories," Phys. Rev. D**61**, 074024 (2000).

[77] P. Gambino, U. Haisch, M. Misiak, "Determining the sign of the $b \to s\gamma$ amplitude," Phys. Rev. Lett. **94**, 061803 (2005).

[78] A. Ishikawa *et al.* [Belle Collaboration], "Measurement of Forward-Backward Asymmetry and Wilson Coefficients in $B \to K^*\ell^+\ell^-$," hep-ex/0508009.

[79] I. Adachi *et al.* [Belle Collaboration], "Measurement of the Differential Branching Fraction and Forward-Backward Asymmetry for $B \to K(*)l^+l^-$," arXiv:0810.0335.

[80] Y. Wang and D. Atwood, "Rate difference between $b \to s\mu^+\mu^-$ and $b \to se^+e^-$ in SUSY with large $\tan\beta$," Phys. Rev. D **68**, 094016 (2003).

[81] T. Aaltonen *et al.* [CDF Collaboration], "Search for $B_s^0 \to \mu^+\mu^-$ and $B_d^0 \to \mu^+\mu^-$ decays with $2fb^{-1}$ of $p - \bar{p}$ collisions," Phys. Rev. Lett. **100**, 101802 (2008).

[82] W. S. Hou, "Enhanced charged Higgs boson effects in $B^- \to \tau\bar{\nu}$, $\mu\bar{\nu}$ and $b \to \tau\bar{\mu} + X$," Phys. Rev. D **48**, 2342 (1993).

[83] A. G. Akeroyd and S. Recksiegel, "The effect of $H^\pm$ on $B^\pm \to \tau^\pm\nu/\tau$ and $B^\pm \to \mu^\pm\nu/\mu$," J. Phys. G **29**, 2311 (2003).

[84] D. Guetta and E. Nardi, "Searching for new physics in rare $B \to \tau$ decays," Phys. Rev. D **58**, 012001 (1998).

[85] S. Baek and Y. G. Kim, "Constraints on the R-parity violating couplings from $B^{\pm} \to l^{\pm}\nu$ decays," Phys. Rev. D **60**, 077701 (1999).

[86] A. G. Akeroyd and S. Recksiegel, "R parity violating enhancement of $B/u^+ \to l^+\nu$ and $B/c^+ \to l^+\nu$," Phys. Lett. B **541**, 121 (2002).

[87] K. Ikado *et al.* [Belle Collaboration], "A search for $B \to \tau\nu$," Phys. Rev. Lett. **97**, 251802 (2006).

[88] G. Buchalla and A. J. Buras, "QCD corrections to rare K and B decays for arbitrary top quark mass," Nucl. Phys. B **400**, 225 (1993).

[89] Y. Grossman, Z. Ligeti and E. Nardi, "First limit on inclusive $B \to X_s \nu\bar{\nu}$ decay and constraints on new physics," Nucl. Phys. B **465**, 369 (1996) [Erratum-ibid. B **480**, 753 (1996)].

[90] P. Colangelo, F. De Fazio, P. Santorelli and E. Scrimieri, "Rare $B \to K^{(*)}\nu\bar{\nu}$ decays at B factories," Phys. Lett. B **395**, 339 (1997).

[91] R. Barate *et al.* [ALEPH Collaboration], "Measurements of $BR(b \to \tau^- \bar{\nu}_\tau X)$ and $BR(b \to \tau - \bar{\nu}_\tau D^{*\pm} X)$ and upper limits on $BR(B^- \to \tau^- \bar{\nu}_\tau)$ and $BR(b \to s\nu\bar{n}u)$," Eur. Phys. J. C **19**, 213 (2001).

[92] T. E. Browder *et al.* [CLEO Collaboration], "A search for $B \to \tau\nu$," Phys. Rev. Lett. **86**, 2950 (2001).

[93] B. Aubert *et al.* [BABAR Collaboration], "A Search for the decay $B^- \to K^- \nu\bar{\nu}$," arXiv:hep-ex/0304020.

[94] K.-F. Chen [Belle Collaboration], "A search for $B \to \tau\nu$," Phys. Rev. Lett. **99**, 221802 (2007).

[95] G. Buchalla *et al.*, "Phenomenology of nonstandard Z couplings in exclusive semileptonic $b \to s$ transitions," Phys. Rev. D **63**, 014015 (2001).

[96] I. Adeachi *et al.* [Belle Collaboration], "Measurement of $B^- \to \tau^- \nu_{\bar{\tau}}$ Decay With a Semileptonic Tagging Method," arXiv:0809.3834.

[97] W. Altmannshofer *et al.*, "New strategies for New Physics search in $B \to K^*\nu\bar{\nu}$, $B \to K\bar{\nu}\nu$ and $B \to X_s\nu\bar{\nu}$ decays," JHEP **0904**, 022 (2009).

[98] J. F. Kamenik, C. Smith, "Tree-level contributions to the rare decays $B^+ \to \pi^+\nu\bar{\nu}$, $B^+ \to K^+\nu\bar{\nu}$, and $B^+ \to K^{*+}\nu\bar{\nu}$ in the Standard Model," Phys. Lett. B **680**, 471 (2009).

[99] M. Tanaka, "Charged Higgs effects on exclusive semitauonic B decays," Z. Phys. C **67**, 321 (1995).

[100] T. Miki, T. Miura and M. Tanaka, "Effects of charged Higgs boson and QCD corrections in $\bar{B} \to D\tau\bar{\nu}_\tau$," arXiv:hep-ph/0210051.

[101] I. Caprini, L. Lellouch and M. Neubert, "Dispersive bounds on the shape of $\bar{B} \to D^{(*)}l\nu$ form factors," Nucl. Phys. B **530**, 153 (1998).

[102] K. Abe *et al.* [Belle Collaboration], "Determination of $|V_{cb}|$ from the semileptonic decay $\bar{B}^0 \to D^{*+}l^-\bar{\nu}$," KEK-PREPRINT-2001-82, *Prepared for 20th International Symposium on Lepton and Photon Interactions at High Energies (LP 01), Rome, Italy, 23-28 Jul 2001.*

[103] O. Long, M. Baak, R. N. Cahn and D. Kirkby, "Impact of tag-side interference on time dependent CP asymmetry measurements using coherent $B^0\bar{B}^0$ pairs," Phys. Rev. D **68**, 034010 (2003).

[104] K.-F. Chen *et al.* [Belle Collaboration], "Observation of time-dependent CP violation in $B^0 \to \eta' K^0$ decays and improved measurements of CP asymmetries in $B^0 \to \phi K^0, K_s^0 K_s^0 K_s^0$ and $B^0 \to J/\psi K^0$ decays," Phys. Rev. Lett. **98**, 031802 (2007).

[105] A. B. Carter and A. I. Sanda, "CP Violation in Cascade Decays of B Mesons," Phys. Rev. Lett. **45**, 952 (1980);

A. B. Carter and A. I. Sanda, "CP Violation in B Meson Decays," Phys. Rev. D **23**, 1567 (1981);

I. I. Bigi and A. I. Sanda, "Notes on the Observability of CP Violations in B Decays," Nucl. Phys. B **193**, 85 (1981);

M. Gronau, "CP Violation in Neutral B Decays to CP Eigenstates," Phys. Rev. Lett. **63**, 1451 (1989).

[106] M. Gronau and D. London, "Isospin analysis of CP asymmetries in B decays," Phys. Rev. Lett. **65**, 3381 (1990).

[107] J. Charles *et al.*, "CP violation and the CKM matrix: Assessing the impact of the asymmetric B factories," Eur. Phys. J. C **41**, 1 (2005); `http://ckmfitter.in2p3.fr`.

[108] H. Ishino *et al.* [Belle Collaboration], "Observation of Direct CP-Violation in $B^0 \to \pi^+\pi^-$ Decays and Model-Independent Constraints on $\phi_2$," Phys. Rev. Lett. **98**, 211801 (2007);

S.-W. Lin *et al.* [Belle Collaboration], "Measurements of Branching Fractions for $B \to K\pi$ and $B \to \pi\pi$ Decays with 449 million $B\bar{B}$ Pairs," Phys. Rev. Lett. **99**, 121601 (2007);

K. Abe *et al.* [Belle Collaboration], "Improved measurement of $B^0 \to \pi^0\pi^0$," arXiv:hep-ex/0610065.

[109] B. Aubert *et al.* [BaBar Collaboration], "Measurement of CP Asymmetries and Branching Fractions in $B \to \pi\pi$ and $B \to K\pi$ decays," arXiv:hep-ex/0607106;

B. Aubert *et al.* [BaBar Collaboration], "Improved Measurements of the Branching Fractions for $B^0 \to \pi^+\pi^-$ and $B^0 \to K^+\pi^-$, and a Search for $B^0 \to K^+K^-$," Phys. Rev. D **75**, 012008 (2007);

B. Aubert *et al.* [BaBar Collaboration], "Observation of CP violation in $B^0 \to K^+\pi^-$ and $B^0 \to \pi^+\pi^-$," Phys. Rev. Lett. **99**, 021603 (2007).

[110] O. Long, M. Baak, R.N. Cahn, and D. Kirkby, "Impact of tag side interference on time dependent CP asymmetry measurements using coherent $B^0\bar{B}^0$ pairs," Phys. Rev. D **68**, 034010 (2003).

[111] H. Ishino, M. Hazumi, M. Nakao and T. Yoshikawa, "New Measurements Using External Photon Conversion at a High Luminosity B Factory," arXiv:hep-ex/0703039, submitted to Phys. Rev. Lett.

[112] J. Zupan, "Penguin pollution estimates relevant for $phi_2/\alpha$ extraction," Nucl. Phys. Proc. Supp. **170**, 33 (2007).

[113] M. Gronau and J. Zupan, "Isospin-breaking effects on $\alpha$ extracted in $B \to \pi\pi$, $\rho\rho$, $\rho\pi$," Phys. Rev. D **71**, 074017 (2005).

[114] S. Gardner, "How isospin violation mocks 'new' physics: $\pi^0 - \eta$, $\eta'$ mixing in $B \to \pi\pi$ decays," Phys. Rev. D **59**, 077502 (1999);

S. Gardner, "Towards a precision determination of $\alpha$ in $B \to \pi\pi$ decays," Phys. Rev. D **72**, 034015 (2005).

[115] M. Gronau, D. Pirjol and T.M. Yan, "Model independent electroweak penguins in B decays to two pseudoscalars," Phys. Rev. D **60**, 034021 (1999); [Erratum-ibid. D **69**, 119901 (2004)].

[116] M. Neubert and J.L. Rosner, "New bound on $\gamma$ from $B^{\pm} \to \pi K$ decays," Phys. Lett. B **441**, 403 (1998);

A. J. Buras and R. Fleischer, "A General analysis of gamma determinations from $B \to \pi K$ decays," Eur. Phys. J. C **11**, 93 (1999).

[117] F. J. Botella, D. London and J. P. Silva, "Looking for $\Delta I = 5/2$ amplitude components in $B \to \pi\pi$ and $B \to \rho\rho$ experiments," Phys. Rev. D **73**, 071501 (2006).

[118] C. C. Wang et al. [Belle Collaboration], "Study of $B^0 \to \rho^{\pm}\pi^{\pm}$ time-dependent CP violation at Belle," Phys. Rev. Lett. **94**, 121801 (2005).

[119] B. Aubert et al. [BaBar Collaboration], "Measurements of branching fractions and CP violating asymmetries in $B^0 \to \rho^{\pm}h^{\pm}$ decays," Phys. Rev. Lett. **91**, 201802 (2003);

B. Aubert et al. [BaBar Collaboration], "Measurement of CP-Violating Asymmetries in $B^0 \to (\rho\pi)^0$ Using a Time-Dependent Dalitz Plot Analysis," arXiv:hep-ex/0408099.

[120] A. E. Snyder and H. R. Quinn, "Measuring CP asymmetry in $B \to \rho\pi$ decays without ambiguities," Phys. Rev. D **48**, 2139 (1993).

[121] A. Kusaka et al. [Belle Collaboration], "Measurement of CP Asymmetry in a Time-Dependent Dalitz Analysis of $B^0 \to (\rho\pi)^0$ and a Constraint on the CKM Angle $\phi_2$," Phys. Rev. Lett. **98**, 221602 (2007);

A. Kusaka et al. [Belle Collaboration], "Measurement of CP asymmetries and branching fractions in a time-dependent Dalitz analysis of $B^0 \to (\rho\pi)^0$ and a constraint on the quark mixing angle $\phi_2$," Phys. Rev. D **77**, 072001 (2008).

[122] B. Aubert et al. [BaBar Collaboration], "Measurement of CP-violating asymmetries in $B^0 \to (\rho\pi)^0$ using a time-dependent Dalitz plot analysis," Phys. Rev. D **76**, 012004 (2007).

[123] H. J. Lipkin, Y. Nir, H. R. Quinn and A. E. Snyder, "Penguin trapping with isospin analysis and CP asymmetries in B decays," Phys. Rev. D **44**, 1454 (1991).

[124] A. Somov et al. [Belle Collaboration], "Measurement of the branching fraction, polarization, and CP asymmetry for $B^0 \to \rho^+\rho^-$ decays, and determination of the CKM phase $\phi_2$," Phys. Rev. Lett. **96**, 171801 (2006);

A. Somov et al. [Belle Collaboration], "Improved measurement of CP-violating parameters in $B^0 \to \rho^+\rho^-$ decays," Phys. Rev. D **76**, 011104 (2007).

[125] B. Aubert *et al.* [BaBar Collaboration], "Updated Measurement of the CKM Angle $\alpha$ Using $B^0 \to \rho^+\rho^-$ Decays," arXiv:hep-ex/0607098.

[126] J. Zhang *et al.* [Belle Collaboration], "Observation of $B^+ \to \rho^+\rho^0$," Phys. Rev. Lett. **91**, 221801 (2003).

[127] B. Aubert *et al.* [BaBar Collaboration], "Measurements of branching fraction, polarization, and charge asymmetry of $B^\pm \to \rho^\pm\rho^0$ and a search for $B^\pm \to \rho^\pm f^0(980)$," Phys. Rev. Lett. **97**, 261801 (2006).

[128] B. Aubert *et al.* [BaBar Collaboration], "Evidence for $B^0 \to \rho^0\rho^0$ decay and implications for the CKM angle $\alpha$," Phys. Rev. Lett. **98**, 111801 (2007).

[129] C. C. Chiang *et al.* [Belle Collaboration], "Measurement of $B^0 \to \pi^+\pi^-\pi^+\pi^-$ Decays and Search for $B^0 \to \rho^0\rho^0$," Phys. Rev. D **78**, 111102 (2008).

[130] A. F. Falk, Z. Ligeti, Y. Nir and H. Quinn, "Comment on extracting alpha from $B \to \rho\rho$," Phys. Rev. D **69**, 011502 (2004).

[131] M. Gronau and J. Zupan, "Weak phase $\alpha$ from $B^0 \to a_1^\pm(1260)\pi^\pm$," Phys. Rev. D **73**, 057502 (2006).

[132] B. Aubert *et al.* [BaBar Collaboration], "Observation of $B^0$ Meson Decay to $a_1^\pm(1260)\pi^\pm$," Phys. Rev. Lett. **97**, 051802 (2006)

[133] B. Aubert *et al.* [BaBar Collaboration], "Measurements of CP-Violating Asymmetries in $B^0 \to a_1^\pm(1260)\pi^\pm$ Decays," Phys. Rev. Lett. **98**, 181803 (2007).

[134] M. Gronau and D. London., "How To Determine All The Angles Of The Unitarity Triangle From $B_d^0 \to DK_S$ And $B_s^0 \to D^0$," Phys. Lett. B **253**, 483 (1991).

[135] M. Gronau and D. Wyler, "On determining a weak phase from CP asymmetries in charged $B$ decays," Phys. Lett. B **265**, 172 (1991).

[136] D. Atwood, I. Dunietz and A. Soni, "Enhanced CP violation with $B \to KD^0(\bar{D}^0)$ modes and extraction of the CKM angle $\gamma$," Phys. Rev. Lett. **78**, 3257 (1997).

[137] D. Atwood, I. Dunietz and A. Soni, "Improved methods for observing CP violation in $B^\pm \to KD$ and measuring the CKM phase $\gamma$," Phys. Rev. D **63**, 036005 (2001).

[138] A. Giri, Y. Grossman, A. Soffer and J. Zupan, "Determining $\gamma$ using $B^\pm \to DK^\pm$ with multibody $D$ decays," Phys. Rev. D **68**, 054018 (2003).

[139] A. Bondar, Proceedings of BINP Special Analysis Meeting on Dalitz Analysis, 24-26 Sep. 2002, unpublished.

[140] K. Abe *et al.* [Belle Collaboration], "Study of $B^\pm \to D_{CP}K^\pm$ and $D_{CP}^*K^\pm$ Decays", Phys. Rev. D **73**, 051106 (2006).

[141] A. Bondar and T. Gershon, "On $\phi_3$ Measurements Using $B^- \to D^*K^-$ Decays", Phys. Rev. D **70**, 091503 (2004).

[142] Y. Horii, K. Trabelsi, H. Yamamoto, *et al.* [Belle Collaboration], "Study of the Suppressed $B$ meson Decay $B^- \to DK^-$, $D \to K^+\pi^-$", Phys. Rev. D **78**, 071901 (2008).

[143] A. Poluektov *et al.* [Belle Collaboration], "Updated Measurement of $\phi_3$ with a Dalitz Plot Analysis of $B^+ \to D^{(*)}K^+$ Decay", arXiv:0803.3375v1 [hep-ex].

[144] A. Bondar and A. Poluektov, "Feasibility study of model-independent approach to $\phi_3$ measurement using Dalitz plot analysis," Eur. Phys. J. C **47**, 347 (2006).

[145] Y. Kubota *et al.* [CLEO Collaboration], "The CLEO-II detector," Nucl. Instrum. Meth. A **320**, 66 (1992);

D. Peterson *et al.*, "The CLEO III drift chamber," Nucl. Instrum. Meth. A **478**, 142 (2002);

M. Artuso *et al.*, "Construction, pattern recognition and performance of the CLEO-III LiF-TEA RICH detector," Nucl. Instrum. Meth. A **502**, 91 (2003);

R. A. Briere *et al.*, Cornell University Report CLNS-01-1742.

[146] The BES Detector, preliminary design report, IHEP-BEPCII-SB-13 (2004).

[147] A. Bondar and A. Poluektov, "Model-independent measurement of the angle $\phi_3$ using Dalitz plot analysis," Proceedings of the CKM2006 workshop, Nagoya, Japan, 2006.

[148] N.J. Joshi, K. Trabelsi, T. Aziz *et al.* [Belle Collaboration], "Measurement of the Branching Fractions for $B^0 \to D_s*+\pi^-$ and $B^0 \to D_s^{*-}K^+$ Decays", Phys. Rev. D **81**, 031101 (2010).

[149] F.J. Ronga, T. R. Sarangi, *et al.* [Belle Collaboration], "Measurements of CP Violation in $B^0 \to D^{*-}\pi^+$ and $B^0 \to D^-\pi^+$ Decays", Phys. Rev. D **73**, 092003 (2006).

[150] S.Bahinipati *et. al*, [Belle Collaboration], "Measurements of time-dependent $CP$ Asymmetries in $B \to D^{*\mp}\pi^{\pm}$ decays using a partial reconstruction technique", arXiv:0809.3203 [hep-ex].

[151] P. Urquijo *et al.* [Belle Collaboration], "Measurement of $|V_{ub}|$ from Inclusive Charmless Semileptonic $B$ Decays," arXiv:0907.0379.

[152] M. Neubert, "QCD based interpretation of the lepton spectrum in inclusive $\bar{B} \to X_u l \bar{\nu}$ decays," Phys. Rev. D **49**, 3392 (1994).

[153] T. Mannel and M. Neubert, "Resummation of nonperturbative corrections to the lepton spectrum in inclusive $B \to X l \bar{\nu}$ decays," Phys. Rev. D **50**, 2037 (1994).

[154] I. I. Bigi, M. A. Shifman, N. G. Uraltsev and A. I. Vainshtein, "On the motion of heavy quarks inside hadrons: Universal distributions and inclusive decays," Int. J. Mod. Phys. A **9**, 2467 (1994).

[155] M. Neubert, "Analysis of the photon spectrum in inclusive $B \to X_s \gamma$ decays," Phys. Rev. D **49**, 4623 (1994).

[156] A. K. Leibovich, I. Low and I. Z. Rothstein, "Extracting $V_{ub}$ without recourse to structure functions," Phys. Rev. D **61**, 053006 (2000).

[157] M. Neubert, "Note on the extraction of $|V_{ub}|$ using radiative B decays," Phys. Lett. B **513**, 88 (2001).

[158] C. W. Bauer, M. E. Luke and T. Mannel, "Light-cone distribution functions for $B$ decays at subleading order in $1/m_b$," Phys. Rev. D **68**, 094001 (2003).

[159] C. W. Bauer, M. Luke and T. Mannel, "Subleading shape functions in $B \to X_u l \bar{\nu}$ and the determination of $|V_{ub}|$," Phys. Lett. B **543**, 261 (2002).

[160] A. K. Leibovich, Z. Ligeti and M. B. Wise, "Enhanced subleading structure functions in semileptonic $B$ decay," Phys. Lett. B **539**, 242 (2002).

[161] M. Neubert, "Subleading shape functions and the determination of $|V_{ub}|$," Phys. Lett. B **543**, 269 (2002).

[162] I. I. Bigi and N. G. Uraltsev, "Weak annihilation and the endpoint spectrum in semileptonic $B$ decays," Nucl. Phys. B **423**, 33 (1994).

[163] M. B. Voloshin, "Nonfactorization effects in heavy mesons and determination of $|V_{ub}|$ from inclusive semileptonic $B$ decays," Phys. Lett. B **515**, 74 (2001).

[164] A. F. Falk, Z. Ligeti and M. B. Wise, "$V_{ub}$ from the hadronic invariant mass spectrum in semileptonic $B$ decay," Phys. Lett. B **406**, 225 (1997).

[165] R. D. Dikeman and N. G. Uraltsev, "Key distributions for charmless semileptonic $B$ decay," Nucl. Phys. B **509**, 378 (1998).

[166] I. I. Bigi, R. D. Dikeman and N. Uraltsev, "The hadronic recoil mass spectrum in semileptonic $B$ decays and extracting $|V_{ub}|$ in a model-insensitive way," Eur. Phys. J. C **4**, 453 (1998).

[167] A. K. Leibovich, I. Low and I. Z. Rothstein, "On the resummed hadronic spectra of inclusive $B$ decays," Phys. Rev. D **62**, 014010 (2000).

[168] A. K. Leibovich, I. Low and I. Z. Rothstein, "Extracting $|V_{ub}|$ from the hadronic mass spectrum of inclusive $B$ decays," Phys. Lett. B **486**, 86 (2000).

[169] C. W. Bauer, Z. Ligeti and M. E. Luke, "A model independent determination of $|V_{ub}|$," Phys. Lett. B **479**, 395 (2000).

[170] M. Neubert, "On the inclusive determination of $|V_{ub}|$ from the lepton invariant mass spectrum," JHEP **0007**, 022 (2000).

[171] S. W. Bosch, B. O. Lange, M. Neubert and G. Paz, "Proposal for a precision measurement of $|V_{ub}|$," Phys. Rev. Lett. **93**, 221801 (2004).

[172] C. W. Bauer, Z. Ligeti and M. E. Luke, "Precision determination of $|V_{ub}|$ from inclusive decays," Phys. Rev. D **64**, 113004 (2001).

[173] G. Burdman, Z. Ligeti, M. Neubert and Y. Nir, "The Decay $B \to \pi l \nu$ in heavy quark effective theory," Phys. Rev. D **49**, 2331 (1994).

[174] M. Okamoto *et al.*, "Semileptonic $D \to \pi/K$ and $B \to \pi/D$ decays in 2+1 flavor lattice QCD," Nucl. Phys. Proc. Suppl. **140**, 461 (2005).

[175] E. Dalgic, A. Gray, M. Wingate, C. T. H. Davies, G. P. Lepage and J. Shigemitsu, "B Meson Semileptonic Form Factors from Unquenched Lattice QCD," Phys. Rev. D **73**, 074502 (2006); [Erratum-ibid. D **75**, 119906 (2007)].

[176] K. C. Bowler *et al.* [UKQCD Collaboration], "Improved $B \to \pi l \nu_l$ form factors from the lattice," Phys. Lett. B **486**, 111 (2000).

[177] A. Abada, D. Becirevic, P. Boucaud, J. P. Leroy, V. Lubicz and F. Mescia, "Heavy → light semileptonic decays of pseudoscalar mesons from lattice QCD," Nucl. Phys. B **619**, 565 (2001).

[178] A. X. El-Khadra, A. S. Kronfeld, P. B. Mackenzie, S. M. Ryan and J. N. Simone, "The semileptonic decays $B \to \pi l\nu$ and $D \to \pi l\nu$ from lattice QCD," Phys. Rev. D **64**, 014502 (2001).

[179] S. Aoki *et al.* [JLQCD Collaboration], "Differential decay rate of $B \to \pi l\nu$ semileptonic decay with lattice NRQCD," Phys. Rev. D **64**, 114505 (2001).

[180] K. M. Foley and G. P. Lepage, "Moving NRQCD for B form factors at high recoil," Nucl. Phys. Proc. Suppl. **119**, 635 (2003).

[181] C. T. H. Davies, K. Y. Wong and G. P. Lepage, "B meson decays from moving-NRQCD on fine MILC lattices," PoSLAT2006 **099** (2006) and arXiv:hep-lat/0611009.

[182] P. F. Bedaque, "Aharonov-Bohm effect and nucleon nucleon phase shifts on the lattice," Phys. Lett. B **593**, 82 (2004).

[183] C. T. Sachrajda and G. Villadoro, "Twisted boundary conditions in lattice simulations," Phys. Lett. B **609**, 73 (2005).

[184] A. Khodjamirian and R. Ruckl, "QCD sum rules for exclusive decays of heavy mesons," Adv. Ser. Direct. High Energy Phys. **15**, 345 (1998).

[185] P. Ball, "$B \to \pi$ and $B \to K$ transitions from QCD sum rules on the light-cone," JHEP **9809**, 005 (1998).

[186] A. Khodjamirian, R. Ruckl, S. Weinzierl, C. W. Winhart and O. I. Yakovlev, "Predictions on $B \to \pi l\nu_l$, $D \to \pi l\nu_l$ and $D \to K l\nu_l$ from QCD light-cone sum rules," Phys. Rev. D **62**, 114002 (2000).

[187] P. Ball and R. Zwicky, "Improved analysis of $B \to \pi e \nu$ from QCD sum rules on the light-cone," JHEP **0110**, 019 (2001).

[188] P. Ball and R. Zwicky, "New results on $B \to \pi$, $K$, $\eta$ decay formfactors from light-cone sum rules," Phys. Rev. D **71**, 014015 (2005).

[189] L. Lellouch, "Lattice-Constrained Unitarity Bounds for $\bar{B}^0 \to \pi^+ \ell^- \bar{\nu}_\ell$ Decays," Nucl. Phys. B **479**, 353 (1996).

[190] M. Fukunaga and T. Onogi, "A model independent determination of $|V_{ub}|$ using the global $q^2$ dependence of the dispersive bounds on the $B \to \pi l\nu$ form factors," Phys. Rev. D **71**, 034506 (2005).

[191] M. C. Arnesen, B. Grinstein, I. Z. Rothstein and I. W. Stewart, "A precision model independent determination of $|V_{ub}|$ from $B \to \pi e\nu$," Phys. Rev. Lett. **95**, 071802 (2005).

[192] T. Becher and R. J. Hill, "Comment on form factor shape and extraction of $|V_{ub}|$ from $B \to \pi l \nu$," Phys. Lett. B **633**, 61 (2006).

[193] R. J. Hill, "The modern description of semileptonic meson form factors," eConf **C060409**, 027 (2006).

[194] J. M. Flynn and J. Nieves, "$|V_{ub}|$ from Exclusive Semileptonic $B \to \pi$ Decays," Phys. Lett. B **649**, 269 (2007).

[195] K. C. Bowler, J. F. Gill, C. M. Maynard and J. M. Flynn, "$B \to \rho l \nu$ form factors in lattice QCD," JHEP **0405**, 035 (2004).

[196] A. Abada, D. Becirevic, P. Boucaud, M. Flynn, .P. Leroy, V. Lubicz and F. Mescia [SPQcdR Collaboration], "Heavy to light vector meson semileptonic decays," Nucl. Phys. Proc. Suppl. **119**, 625 (2003).

[197] C. Schwanda et al. [Belle Collaboration], "Measurement of the Moments of the Photon Energy Spectrum in $B \to X_s \gamma$ Decays and Determination of $|V_{cb}|$ and $m_b$ at Belle," Phys. Rev. D **78**, 032016 (2008).

[198] B. Aubert et al. [BABAR Collaboration], "Measurement of the $B^0 \to \pi^- l^+ \nu$ form-factor shape and branching fraction, and determination of $|V_{ub}|$ with a loose neutrino reconstruction technique," Phys. Rev. Lett. **98**, 091801 (2007).

[199] T. Hokuue et al. [Belle Collaboration], "Measurements of branching fractions and $q^2$ distributions for $B \to \pi l \nu$ and $B \to \rho l \nu$ decays with $B \to D(*) l \nu$ decay tagging," Phys. Lett. B **648**, 139 (2007).

[200] W.J. Marciano and A.I. Sanda, Phys. Lett. B 67, 303 (1977); B.W. Lee and R.E. Shrock, Phys. Rev. D 16, 1444 (1977); T.P. Cheng and L.F. Li, Phys. Rev. D 16, 1425 (1977).

[201] S. Antusch, E. Arganda, M. J. Herrero and A. M. Teixeira, "Impact of theta(13) on lepton flavour violating processes within SUSY seesaw," JHEP **0611** (2006) 090 [arXiv:hep-ph/0607263].

[202] E. Arganda, M. J. Herrero and J. Portoles, "Lepton flavour violating semileptonic tau decays in constrained MSSM-seesaw scenarios," JHEP **0806** (2008) 079 [arXiv:0803.2039 [hep-ph]].

[203] J. Hisano, T. Moroi, K. Tobe and M. Yamaguchi, "Exact event rates of lepton flavor violating processes in supersymmetric SU(5) model," Phys. Lett. B **391**, 341 (1997) [Erratum-ibid. B **397**, 357 (1997)] [arXiv:hep-ph/9605296].

[204] N. Arkani-Hamed, H. C. Cheng and L. J. Hall, "Flavor mixing signals for realistic supersymmetric unification," Phys. Rev. D **53**, 413 (1996) [arXiv:hep-ph/9508288].

[205] J. Hisano, D. Nomura, Y. Okada, Y. Shimizu and M. Tanaka, "Enhancement of $\mu \to e\gamma$ in the supersymmetric SU(5) GUT at large $\tan\beta$," Phys. Rev. D **58**, 116010 (1998) [arXiv:hep-ph/9805367].

[206] T. Fukuyama, T. Kikuchi and N. Okada, "Lepton flavor violating processes and muon g-2 in minimal supersymmetric SO(10) model," Phys. Rev. D **68** (2003) 033012 [arXiv:hep-ph/0304190].

[207] J. Hisano, T. Moroi, K. Tobe, M. Yamaguchi and T. Yanagida, "Lepton flavor violation in the supersymmetric standard model with seesaw induced neutrino masses," Phys. Lett. B **357**, 579 (1995) [arXiv:hep-ph/9501407].

[208] J. Hisano, T. Moroi, K. Tobe and M. Yamaguchi, "Lepton-Flavor Violation via Right-Handed Neutrino Yukawa Couplings in Supersymmetric Standard Model," Phys. Rev. D **53**, 2442 (1996) [arXiv:hep-ph/9510309].

[209] J. R. Ellis, J. Hisano, M. Raidal and Y. Shimizu, "A new parametrization of the seesaw mechanism and applications in supersymmetric models," Phys. Rev. D **66**, 115013 (2002) [arXiv:hep-ph/0206110].

[210] L. Calibbi, A. Faccia, A. Masiero and S. K. Vempati, "Lepton flavour violation from SUSY-GUTs: Where do we stand for MEG, PRISM PRIME and a super flavour factory," Phys. Rev. D **74** (2006) 116002 [arXiv:hep-ph/0605139].

[211] K. S. Babu and C. Kolda, "Higgs-mediated $\tau \to 3\mu$ in the supersymmetric seesaw model," Phys. Rev. Lett. **89**, 241802 (2002) [arXiv:hep-ph/0206310].

[212] A. Dedes, J. R. Ellis and M. Raidal, "Higgs mediated $B^0_{s,d} \to \mu\tau$, $e\tau$ and $\tau \to 3\mu$, $e\mu\mu$ decays in supersymmetric seesaw models," Phys. Lett. B **549** (2002) 159 [arXiv:hep-ph/0209207].

[213] M. Sher, "$\tau \to \mu\eta$ in supersymmetric models," Phys. Rev. D **66**, 057301 (2002) [arXiv:hep-ph/0207136].

[214] A. Brignole and A. Rossi, "Lepton flavour violating decays of supersymmetric Higgs bosons," Phys. Lett. B **566** (2003) 217 [arXiv:hep-ph/0304081].

[215] P. Paradisi, "Higgs-mediated $\tau \to \mu$ and $\tau \to e$ transitions in II Higgs doublet model and supersymmetry," JHEP **0602** (2006) 050 [arXiv:hep-ph/0508054].

[216] M. Herrero, J. Portoles and A. Rodriguez-Sanchez, "Sensitivity to the Higgs sector of SUSY-seesaw models via LFV tau decays," arXiv:0909.0724 [hep-ph].

[217] M. J. Herrero, J. Portoles and A. M. Rodriguez-Sanchez, "Sensitivity to the Higgs Sector of SUSY-Seesaw Models in the Lepton Flavour Violating $\tau \to \mu f_0(980)$ decay," Phys. Rev. D **80** (2009) 015023 [arXiv:0903.5151 [hep-ph]].

[218] A. Brignole and A. Rossi, "Anatomy and phenomenology of mu tau lepton flavour violation in the MSSM," Nucl. Phys. B **701** (2004) 3 [arXiv:hep-ph/0404211].

[219] J. P. Saha and A. Kundu, "Constraints on R-parity violating supersymmetry from leptonic and semileptonic $\tau$, $B_d$ and $B_s$ decays," Phys. Rev. D **66**, 054021 (2002) [arXiv:hep-ph/0205046].

[220] R. Barbier *et al.*, "R-parity violating supersymmetry," Phys. Rept. **420** (2005) 1 [arXiv:hep-ph/0406039].

[221] M. Blanke, A. J. Buras, B. Duling, A. Poschenrieder and C. Tarantino, "Charged Lepton Flavour Violation and $(g-2)_\mu$ in the Littlest Higgs Model with $T^-$ Parity: A Clear Distinction from Supersymmetry," JHEP **0705** (2007) 013 [arXiv:hep-ph/0702136].

[222] C. X. Yue and S. Zhao, "Lepton flavor violating signals of a little Higgs model at the high energy linear e+ e- colliders," Eur. Phys. J. C **50** (2007) 897 [arXiv:hep-ph/0701017].

[223] C. x. Yue, Y. m. Zhang and L. j. Liu, "Non-universal gauge bosons Z' and lepton flavor-violation tau decays," Phys. Lett. B **547** (2002) 252 [arXiv:hep-ph/0209291].

[224] N. Arkani-Hamed, S. Dimopoulos, G. R. Dvali and J. March-Russell, "Neutrino masses from large extra dimensions," Phys. Rev. D **65**, 024032 (2002) [arXiv:hep-ph/9811448].

[225] K. R. Dienes, E. Dudas and T. Gherghetta, "Light neutrinos without heavy mass scales: A higher-dimensional seesaw mechanism," Nucl. Phys. B **557**, 25 (1999) [arXiv:hep-ph/9811428].

[226] A. De Gouvea, G. F. Giudice, A. Strumia and K. Tobe, "Phenomenological implications of neutrinos in extra dimensions," Nucl. Phys. B **623**, 395 (2002) [arXiv:hep-ph/0107156].

[227] A. G. Akeroyd, M. Aoki and Y. Okada, "Lepton Flavour Violating tau Decays in the Left-Right Symmetric Model," Phys. Rev. D **76** (2007) 013004 [arXiv:hep-ph/0610344].

[228] A. Ilakovac and A. Pilaftsis, "Flavor violating charged lepton decays in a GUT and superstring inspired standard model," Nucl. Phys. B **437** (1995) 491 [arXiv:hep-ph/9403398].

[229] A. Ilakovac, "Lepton flavor violation in the standard model extended by heavy singlet Dirac neutrinos," Phys. Rev. D **62** (2000) 036010 [arXiv:hep-ph/9910213].

[230] G. Cvetic, C. Dib, C. S. Kim and J. D. Kim, "On lepton flavor violation in tau decays," Phys. Rev. D **66**, 034008 (2002) [Erratum-ibid. D **68**, 059901 (2003)] [arXiv:hep-ph/0202212].

[231] M. Ahmed et al. [MEGA Collaboration], Phys. Rev. D **65** (2002) 112002.

[232] J. Adam et al. [MEG collaboration], "A limit for the $\mu \to e\gamma$ decay from the MEG experiment," arXiv:0908.2594 [hep-ex].

[233] D. Bryman et al. [COMET Collaboration], J-PARK Proposal P21 (2007).

[234] R.M. Carey, et al. [Mu2e Callaboration] Fermilab Letter of Intent (2007).

[235] Y. Kuno et al. [PRISM Collaboration], J-PARK Letter of Intent (2006).

[236] W. J. Marciano, T. Mori and J. M. Roney, "Charged Lepton Flavor Violation Experiments," Ann. Rev. Nucl. Part. Sci. **58** (2008) 315.

[237] B. Aubert et al. [BABAR Collaboration], "Searches for Lepton Flavor Violation in the Decays $\tau \to e\gamma$ and $\tau \to \mu\gamma$," arXiv:0908.2381 [hep-ex].

[238] K. Hayasaka et al. [Belle Collaboration], "New search for $\tau \to mu\gamma$ and $\tau \to e\gamma$ decays at Belle," Phys. Lett. B **666** (2008) 16 [arXiv:0705.0650 [hep-ex]].

[239] G. Marchiori, [BABAR Collaboration], "Search for Lepton Flavor Violating Decays $\tau^\pm \to \ell\ell\ell$", Presented at CIPANP09. [arXiv:0909.3870].

[240] Y. Miyazaki et al. [Belle Collaboration], "Search for Lepton Flavor Violating Decays $\tau^\pm \to \ell ll$", [arxiv:1001.3221].

[241] B. Aubert et al. [BABAR Collaboration], "Search for Lepton Flavor Violating Decays $\tau^\pm \to \ell^\pm \pi^0, \ell^\pm \eta, \ell^\pm \eta'$," Phys. Rev. Lett. **98** (2007) 061803 [arXiv:hep-ex/0610067].

[242] Y. Miyazaki et al. [Belle Collaboration], "Search for lepton flavor violating $\tau^-$ decays into $\ell\eta, \ell\eta'$ and $\ell\pi^0$," Phys. Lett. B **648** (2007) 341 [arXiv:hep-ex/0703009].

[243] B. Aubert *et al.* [BaBar Collaboration], "Search for Lepton Flavour Violating Decays $\tau \to \ell K_S^0$ with the BaBar Experiment," Phys. Rev. D **79** (2009) 012004 [arXiv:0812.3804 [hep-ex]].

[244] Y. Miyazaki *et al.* [Belle Collaboration], "Search for lepton flavor violating $\tau^-$ decays into $\ell$- Ks", submitted to PLB. See http://www.slac.stanford.edu/xorg/hfag/tau/HFAG-TAU-LFV.htm

[245] B. Aubert *et al.* [BABAR Collaboration], "Improved limits on lepton flavor violating $\tau$ decays to $\ell\phi$, $\ell\rho$, $\ell K^*$ and $\ell \bar{K}^*$," Phys. Rev. Lett. **103** (2009) 021801 [arXiv:0904.0339 [hep-ex]].

[246] Y. Nishio *et al.* [Belle Collaboration], "Search for lepton-flavor-violating $\tau \to \ell V^0$ decays at Belle," Phys. Lett. B **664** (2008) 35 [arXiv:0801.2475 [hep-ex]].

[247] B. Aubert *et al.* [BABAR Collaboration], "Search for Lepton Flavor Violating Decays $\tau^\pm \to \ell^\pm \omega$ ($\ell = e, \mu$)," Phys. Rev. Lett. **100** (2008) 071802 [arXiv:0711.0980 [hep-ex]].

[248] Y. Miyazaki *et al.* [Belle Collaboration], "Search for Lepton-Flavor-Violating tau Decays into Lepton and $f_0(980)$ Meson," Phys. Lett. B **672** (2009) 317 [arXiv:0810.3519 [hep-ex]].

[249] B. Aubert *et al.* [BaBar Collaboration], "Search for lepton-flavor and lepton-number violation in the decay $\tau^- \to \ell^\mp h^\pm h'^-$," Phys. Rev. Lett. **95** (2005) 191801 [arXiv:hep-ex/0506066].

[250] Y. Miyazaki *et al.* [Belle Collaboration], Phys. Lett. B **682** (2010) 355 "Search for Lepton Flavor and Lepton Number Violating tau Decays into a Lepton and Two Charged Mesons," [arXiv:0908.3156].

[251] G. D. Lafferty [BABAR Collaboration], "Lepton flavour violation and baryon number non-conservation in $\tau \to \Lambda + h$," Nucl. Phys. Proc. Suppl. **169** (2007) 186.

[252] Y. Miyazaki *et al.* [Belle Collaboration], "Search for Lepton and Baryon Number Violating $\tau^-$ Decays into $\bar{\Lambda}\pi^-$ and $\Lambda\pi^-$," Phys. Lett. B **632** (2006) 51 [arXiv:hep-ex/0508044].

[253] G. S. Huang *et al.* [CLEO Collaboration], "Measurement of B($\Upsilon(5S) \to B_s^{(*)}\bar{B}_s^{(*)}$) using $\phi$ mesons," Phys. Rev. D **75**, 012002 (2007).

[254] A. Drutskoy *et al.* [Belle Collaboration], "Measurement of inclusive $D_s$, $D^0$ and $J/\psi$ rates and determination of the $B_s^{(*)}\bar{B}_s^{(*)}$ production fraction in $b\bar{b}$ events at the $\Upsilon(5S)$ resonance," Phys. Rev. Lett. **98**, 052001 (2007).

[255] R. Louvot *et al.* [Belle Collaboration], "Measurement of the Decay $B_s^0 \to D_s^- pi^+$ and Evidence for $B_s^0 \to D_s^\pm K^\pm$ in $e^+e^-$ Annihilation at $\sqrt{s} \sim$10.87 GeV," Phys. Rev. Lett. **102**, 021801 (2009).

[256] G. Bonvicini *et al.* [CLEO Collaboration], "Observation of $B_s^*\bar{B}_s^*$ production at the $\Upsilon(5S)$ resonance," Phys. Rev. Lett. **96**, 022002 (2006).

[257] A. Drutskoy *et al.* [Belle Collaboration], "Measurements of exclusive $B_s^0$ decays at the $\Upsilon(5S)$," Phys. Rev. D **76**, 012002 (2007).

[258] J. f. Sun, G. h. Zhu and D. s. Du, "Phenomenological analysis of charmless decays $B_s \to PP$, $PV$ with QCD factorization. ((U))," Phys. Rev. D **68**, 054003 (2003).

[259] P. Ball *et al.*, "B decays at the LHC," arXiv:hep-ph/0003238.

[260] N. Harnew [LHCB Collaboration], "B-Physics Prospects with the LHCb Experiment," Phys. Atom. Nucl. **71**, 588 (2008).

[261] H. J. Lipkin, "Is observed direct $CP$ violation in $B_d \to K^+\pi^-$ due to new physics? Check standard model prediction of equal violation in $B_s \to K^-\pi^+$," Phys. Lett. B **621**, 126 (2005).

[262] G.-L. Lin, J. Liu and Y.-P. Yao, "Flavor Changing Two Photon Decay," Phys. Rev. Lett. **64**, 1498 (1990).

[263] C.-H. V. Chang, G.-L. Lin, and Y.-P. Yao, "QCD corrections to $b \to s\gamma\gamma$ and exclusive $B_s \to \gamma\gamma$ decay," Phys. Lett. B **415**, 395 (1997).

[264] L. Reina, G. Ricciardi, and A. Soni, "QCD corrections to $b \to s\gamma\gamma$ induced decays: $B \to X_s\gamma\gamma$ and $B_s \to \gamma\gamma$," Phys. Rev. D **56**, 5805 (1997).

[265] S. W. Bosch, G. Buchalla, "The Double radiative decays $B \to \gamma\gamma$ in the heavy quark limit," JHEP **0208**, 054 (2002).

[266] G. Hiller and E. O. Iltan, "Leading logarithmic QCD corrections to the $B_s \to \gamma\gamma$ decay rate including long distance effects through $B_s \to \phi\gamma \to \gamma\gamma$," Phys. Lett. B **409**, 425 (1997).

[267] D. Choudhury, J. Ellis, "Estimates of long distance contributions to the $B_s \to \gamma\gamma$ decay," Phys. Lett. B **433**, 102 (1998).

[268] J. Wicht *et al.* [Belle Collaboration], "Observation of $B_s \to \phi\gamma$ and Search for $B_s \to \gamma\gamma$ Decays at Belle," Phys. Rev. Lett. **100**, 121801 (2008)

[269] S. Bertolini and J. Matias, "The $b \to s\gamma\gamma$ transition in softly broken supersymmetry," Phys. Rev. D **57**, 4197 (1998).

[270] T. M. Aliev, G. Hiller, and E. O. Iltan, "Leading logarithmic QCD corrections to the $B_s \to \gamma\gamma$ decays in the two Higgs doublet model," Nucl. Phys. B **515**, 321 (1998).

[271] W. J. Huo, C. D. Lu, and Z. J. Xiao, "$B_{s,d} \to \gamma\gamma$ Decay with the Fourth Generation," hep-ph/0302177.

[272] A. Gemintern, S. Bar-Shalom, and G. Eilam, "$B \to X_s\gamma\gamma$ and $B_s \to \gamma\gamma$ in supersymmetry with broken R-parity," Phys. Rev. D **70**, 035008 (2004).

[273] M. Beneke, G. Buchalla and I. Dunietz, "Width Difference in the $B_s - \bar{B}_s$ System," Phys. Rev. D **54**, 4419 (1996).

[274] K. Hartkorn and H. G. Moser, "A New Method Of Measuring $\Delta\Gamma)/\Gamma$ In The $B_S^0 - \bar{B}_S^0$ System," Eur. Phys. J. C **8**, 381 (1999).

[275] R. Barate *et al.* [ALEPH Collaboration], "A study of the decay width difference in the $B_s^0 - \bar{B}_s^0$ system using $\phi\phi$ correlations," Phys. Lett. B **486**, 286 (2000).

[276] M. Beneke, G. Buchalla, C. Greub, A. Lenz and U. Nierste, "Next-to-leading order QCD corrections to the lifetime difference of $B_s$ mesons," Phys. Lett. B **459**, 631 (1999).

[277] M. Ciuchini, E. Franco, V. Lubicz, F. Mescia and C. Tarantino, "Lifetime differences and CP violation parameters of neutral B mesons at the next-to-leading order in QCD," JHEP **0308**, 031 (2003).

[278] S. Hashimoto, K. I. Ishikawa, T. Onogi and N. Yamada, "Width difference in the $B_s - \bar{B}_s$ system with lattice NRQCD," Phys. Rev. D **62**, 034504 (2000).

[279] S. Hashimoto, K. I. Ishikawa, T. Onogi, M. Sakamoto, N. Tsutsui and N. Yamada, "Renormalization of the $\Delta B = 2$ four-quark operators in lattice NRQCD," Phys. Rev. D **62**, 114502 (2000).

[280] D. Becirevic, D. Meloni, A. Retico, V. Gimenez, V. Lubicz and G. Martinelli, "A theoretical prediction of the $B_s$ meson lifetime difference," Eur. Phys. J. C **18**, 157 (2000).

[281] J. Flynn and C. J. D. Lin, "$B_s^0 \bar{B}_s^0$ mixing and $b$ hadron lifetimes from lattice QCD," J. Phys. G **27**, 1245 (2001).

[282] S. Aoki *et al.* [JLQCD Collaboration], "$B^0 - \bar{B}^0$ mixing in quenched lattice QCD. ((U)) ((W))," Phys. Rev. D **67**, 014506 (2003).

[283] M. Beneke and A. Lenz, "Lifetime difference of $B_s$ mesons: Theory status," J. Phys. G **27**, 1219 (2001).

[284] R. Aleksan *et al.*, "Estimation of $\Delta\gamma$ for the $B_s - \bar{B}_s$ system: Exclusive decays and the parton model," Phys. Lett. B **316**, 567 (1993).

[285] Yu. Grossman, "The $B_s$ width difference beyond the standard model," Phys. Lett. B **380**, 99 (1996).

[286] I. Dunietz, R. Fleischer, U. Nierste, "In pursuit of new physics with $B_s$ decays," Phys. Rev. D **63**, 114015 (2001).

[287] K. Anikeev *et al.*, "B Physics at the Tevatron: Run II and Beyond," Workshop Report, hep-ph/0201071.

[288] M. A. Shifman and M. B. Voloshin, "On Production of $D$ and $D^*$ Mesons in B Meson Decays," Yad. Fiz. **47**, 801 (1988); Sov. J. Nucl. Phys. **47**, 511 (1988).

[289] D. Besson et al. [CLEO Collaboration], "Search for $e^+e^- \to \Lambda_b^0 \bar{\Lambda}_b^0$ near threshold," Phys. Rev. D **71**, 012004 (2005).

[290] C. H. Chen, C. Q. Geng, J. N. Ng, "T violation in $\Lambda_b \to \Lambda l^+ l^-$ decays," Nucl. Phys. Proc. Suppl. **115**, 263 (2003);

T. M. Aliev, V. Bashiry, M. Savci, "Double-lepton polarization asymmetries in $\Lambda_b \to \Lambda l^+ l^-$ decay," Eur. Phys. J. C **38**, 283 (2004);

A. K. Giri, R. Mohanta, "Effect of R-parity violation on the rare decay $Lambda_b \to \Lambda \mu^+ \mu^-$," J. Phys. G **31**, 1559 (2005).

[291] Z. J. Ajaltouni, E. Conte, "Angular Analysis of $\Lambda_b$ Decays into $\Lambda$: Applications to Time-Odd Observables and CP violation in $\Lambda_b$ Decays," hep-ph/0409262;

O. Leitner, Z. J. Ajaltouni, E. Conte, "Testing Fundamental Symmetries with $\Lambda_b \to \Lambda$ -Vector Decay," hep-ph/0602043.

[292] L. Zhang *et al.* [Belle Collaboration], "Improved constraints on $D^0 - \bar{D}^0$ mixing in $D^0 \to K^+\pi^-$ decays at BELLE," Phys. Rev. Lett. **96**, 151801 (2006).

[293] D. M. Asner *et al.* [Cleo Collaboration], "Determination of the $D^0 \to K^+\pi^-$ Relative Strong Phase Using Quantum-Correlated Measurements in $e^+e^- \to D^0\bar{D}^0$ at CLEO," Phys. Rev. D **78**, 012001 (2008).

[294] A. Abulencia *et al.* [CDF Collaboration], "Measurement of the ratio of branching fractions $B(D^0 \to K^+\pi^-)/B(D^0 \to K^-\pi^+)$ using the CDF II Detector," Phys. Rev. D **74**, 031109 (2006).

[295] P. Spradlin, G. Wilkinson, F. Xing, "A study of $D^0 \to hh'$ decays for $D^0 - \bar{D}^0$ mixing measurements," LHCb public note LHCb-2007-049.

[296] S. Capstick, N. Isgur, "Baryons in a Relativized Quark Model with Chromodynamics," Phys. Rev. D **34**, 2809 (1986).

[297] N. Isgur, M. B. Wise, "Spectroscopy with heavy quark symmetry," Phys. Rev. Lett. **66**, 1130 (1991);

Y. Oh, B. Park, "Excited states of heavy baryons in the Skyrme model," Phys. Rev. D **53**, 1605 (1996);

S. Migura *et al.*, "Charmed baryons in a relativistic quark model," Eur. Phys. J. A **28**, 41 (2006).

[298] R. Mizuk *et al.* [Belle Collaboration], "Observation of an isotriplet of excited charmed baryons decaying to $Lambda_c^+\pi$," Phys. Rev. Lett. **94**, 122002 (2005).

[299] B. Aubert *et al.* [BaBar Collaboration], "Observation of a charmed baryon decaying to $D^0 p$ at a mass near $2.94 GeV/c^2$," Phys. Rev. Lett. **98**, 012001 (2007).

[300] R. Mizuk *et al.* [Belle Collaboration], "Experimental Constraints on the Spin and Parity of the $Lambda_c(2880)$," Phys. Rev. Lett. **98**, 262001 (2007).

[301] R. Chistov *et al.* [Belle Collaboration], "Observation of new states decaying into $Lambda_c^+ K^-\pi^+$ and $Lambda_c^+ K_S^0\pi^-$," Phys. Rev. Lett. **97**, 162001 (2006).

[302] H. M. Pilkuhn, *The interactions of hadrons*, North-Holland Pub. (1967).

[303] H. Y. Cheng, C. K. Chua, "Strong Decays of Charmed Baryons in Heavy Hadron Chiral Perturbation Theory," Phys. Rev. D **75**, 014006 (2007).

[304] A. Selem, F. Wilczek, "Hadron Systematics and Emergent Diquarks," arXiv:hep-ph/0602128.

[305] P. Pakhlov *et al.* [Belle Collaboration], to be submitted soon.

[306] K. Abe *et al.* [Belle Collaboration], "Study of double charmonium production in $e^+e^-$ annihilation at $s^{1/2}$ 10.6 GeV," Phys. Rev. D **70**, 071102 (2004).

[307] B. Aubert *et al.* [BaBar Collaboration], "Measurement of double charmonium production in $e^+e^-$ annihilations at $s^{1/2} = 10.6$ GeV," Phys. Rev. D **72**, 031101 (2005).

[308] E. Braaten, J. Lee, "Exclusive double charmonium production from $e^+e^-$ annihilation into a virtual photon," Phys. Rev. D **67**, 054007 (2003).

[309] A. E. Bondar, V. L. Chernyak, "Is the BELLE result for the cross section $\sigma(e^+e^- \to J/\psi + \eta_c)$ a real difficulty for QCD?," Phys. Lett. B **612**, 215 (2005).

[310] Y. Zhang *et al.*, "Next-to-leading order QCD correction to $e^+e^- \to J/\psi + \eta_c$ at $s^{1/2} = 10.6$ GeV," Phys. Rev. Lett. **96**, 092001 (2006).

[311] K. Abe *et al.* [Belle Collaboration], "Observation of a new charmonium state in double charmonium production in $e^+e^-$ annihilation at $s^{1/2}$ 10.6 GeV," Phys. Rev. Lett. **98**, 082001 (2007).

[312] D. Kang *et al.*, "Inclusive production of four charm hadrons in $e^+e^-$ annihilation at B factories," Phys. Rev. D **71**, 071501 (2005).

[313] V. V. Kiselev, A. K. Likhoded, and M. V. Shevlyagin, "Double charmed baryon production at B factory," Phys. Lett. B **332**, 411 (1994);

A. V. Berezhnoy and A. K. Likhoded, "Quark-hadron duality and production of charmonia and doubly charmed baryons in $e^+e^-$ annihilation," Phys. Atom. Nucl. **70**, 478 (2007).

[314] M. Okamoto *et al.* [Fermilab/MILC Collaboration], "Semileptonic $D \to \pi/K$ and $B \to \pi/D$ decays in 2+1 flavor lattice QCD," Nucl. Phys. Proc. Suppl. **140**, 461 (2005);

J. Shigemitsu *et al.* [HPQCD Collaboration], "Semileptonic B decays with $N(f) = 2 + 1$ dynamical quarks," Nucl. Phys. Proc. Suppl. **140**, 464 (2005).

[315] L. Widhalm *et al.* [Belle Collaboration], "Measurement of $D^0 \to \pi l \nu (K l \nu)$ Form Factors and Absolute Branching Fractions," Phys. Rev. Lett. **97**, 061804 (2006).

[316] J. Y. Ge *et al.* [Cleo Collaboration], "Study of $D^0 \to \pi^- e^+ \nu_e$, $D^+ \to \pi^0 e^+ \nu_e$, $D^0 \to K^- e^+ \nu_e$, and $D^+ \to \bar{K}^0 e^+ \nu_e$ in Tagged Decays of the $\psi(3770)$ Resonance," Phys. Rev. D **79**, 052010 (2009).

[317] C. Aubin *et al.*, "Semileptonic decays of D mesons in three-flavor lattice QCD," Phys. Rev. Lett. **94**, 011601 (2005).

[318] G. Amoros, S. Noguera, J. Portoles, "Semileptonic decays of charmed mesons in the effective action of QCD," Eur. Phys. J. C **27**, 243 (2003).

[319] M. Bečirevič, A. B. Kaidalov, "Comment on the heavy —¿ light form-factors," Phys. Lett. B **478**, 417 (2000).

[320] T. Becher, R. Hill, "Comment on form-factor shape and extraction of $|V_{ub}|$ from $B \to \pi l \nu$," Phys. Lett. B **633**, 61 (2006).

[321] G. Köpp *et al.*, "Angular correlations for semileptonic D meson decays," Z. Phys. C **48**, 327 (1990).

[322] B. Aubert *et al.* [BaBar Collaboration], "Measurement of the Hadronic Form Factors in $D_s \to \phi e \nu$ Decays," arXiv:hep-ex/0607085.

[323] A. Abulencia *et al.* [CDF Collaboration], "Observation of $B_s^0 - \bar{B}_s^0$ Oscillations," Phys. Rev. Lett. **97**, 242003 (2006);

V. M. Abazov *et al.* [D0 Collaboration], "First direct two-sided bound on the $B_s^0$ oscillation frequency," Phys. Rev. Lett. **97**, 021802 (2006).

[324] L. Widhalm *et al.* [Belle Collaboration], "Measurement of $B(D_s^+ \to \mu\nu)$," Phys. Rev. Lett. **100**, 241801 (2008).

[325] J. P. Alexander *et al/* [Cleo Collaboration], "Measurement of $B(D_s^+ \to l^+\nu)$ and the Decay Constant $f_{D_s^+}$ From $600pb^{-1}$ of $e^+e^-$ Annihilation Data Near 4170 MeV," Phys. Rev. D **79**, 052001 (2009).

[326] E. Follana *et al.* [HPQCD and UKQCD Collaboration], "High Precision determination of the $\pi$, K, D and $D_s$ decay constants from lattice QCD," Phys. Rev. Lett. **100**, 062002 (2008).

[327] I. I. Bigi, N. Uraltsev, "$D^0 - \bar{D}^0$ oscillations as a probe of quark hadron duality," Nucl. Phys. B **592**, 92 (2001).

[328] A. F. Falk *et al.*, "SU(3) breaking and $D^0 - \bar{D}^0$ mixing," Phys. Rev. D**65**, 054034 (2002);

A. F. Falk *et al.*, "The $D^0 - \bar{D}^0$ mass difference from a dispersion relation," Phys. Rev. D**69**, 114021 (2004).

[329] A. A. Petrov, "Charm mixing in the Standard Model and beyond," Int. J. Mod. Phys. A **21**, 5686 (2006);

E. Golowich, S. Pakvasa, A. A. Petrov, "New Physics contributions to the lifetime difference in $D^0 - \bar{D}^0$ mixing," Phys. Rev. Lett. **98**, 181801 (2007).

[330] S. Bianco, F. Fabbri, D. Benson, I. I. Bigi, "A Cicerone for the Physics of Charm," Riv. Nuov. Cim. **26**, 7 (2003).

[331] M. Starič *et al.* [Belle Collaboration], "Evidence for $D^0 - \bar{D}^0$ Mixing," Phys. Rev. Lett. **98**, 211803 (2007).

[332] B. Aubert *et al.* [BABAR Collaboration], "Evidence for $D^0 - \bar{D}^0$ Mixing," Phys. Rev. Lett. **98**, 211802 (2007).

[333] S. Bergmann *et al.*, "Lessons from CLEO and FOCUS measurements of $D^0 - \bar{D}^0$ mixing parameters," Phys. Lett. B **486**, 418 (2000).

[334] L. M. Zhang *et al.* [Belle Collaboration], "Measurement of $D^0 - \bar{D}^0$ mixing in $D^0 \to K_s^0\pi^+\pi^-$ decays," Phys. Rev. Lett. **99**, 131803 (2007).

[335] D. Asner *et al.* [Cleo Collaboration], "$D^0\bar{D}^0$ Quantum Correlations, Mixing, and Strong Phases," Int. J. Mod. Phys. A **21**, 5456 (2006).

[336] X. D. Cheng *et al.*, "Strong Phase and $D^0 - D^0bar$ mixing at BES-III," Phys. Rev. D **75**, 094019 (2007).

[337] L. Zhang *et al.* [Belle Collaboration], "Improved constraints on $D^0 - \bar{D}^0$ mixing in $D^0 \to K^+\pi^-$ decays at BELLE," Phys. Rev. Lett. **96**, 151801 (2006).

[338] E. Barberio *et al.* [HFAG Group], "Averages of b-hadron and c-hadron Properties at the End of 2007," arXiv:0808.1297; updates at http://www.slac.stanford.edu/xorg/hfag/charm/index.html.

[339] G. Burdman *et al.*, "Rare charm decays in the standard model and beyond," Phys. Rev. D **66**, 014009 (2002).

[340] G. Burdman, I. Shipsey, "$D^0 - \bar{D}^0$ mixing and rare charm decays," Ann. Rev. Nucl. Part. Sci. **53**, 431 (2003).

[341] T. Barnes, F. E. Close et al., "Hybrid and conventional mesons in the flux tube model: Numerical studies and their phenomenological implications," Phys. Rev. D **52**, 5242 (1995).

[342] M. B. Voloshin, L. B. Okun "Hadron Molecules and Charmonium Atom," JETP Lett. **23**, 333 (1976);

A. De Rujula, H. Georgi et al., "Molecular Charmonium: A New Spectroscopy?" Phys. Rev. Lett. **38**, 317 (1977);

N. A. Tornqvist, "From the deuteron to deusons, an analysis of deuteron - like meson meson bound states," Z. Phys. C **61**, 525 (1994).

[343] L. Maiani, F. Piccinini et al., "Diquark-antidiquarks with hidden or open charm and the nature of X(3872)," Phys. Rev. D **71**, 014028 (2005).

[344] S. Dubynskiy, M. B. Voloshin, "Hadro-Charmonium," Phys. Lett. B **666**, 344 (2008).

[345] S. K. Choi et al. [Belle Collaboration], "Observation of the $\eta_c(2S)$ in exclusive $B \to KK_S K^- \pi^+$ decays," Phys. Rev. Lett. **89**, 102001 (2002).

[346] J. L. Rosner et al. [Cleo Collaboration], "Observation of $h_c(P(1)-1)$ state of charmonium," Phys. Rev. Lett. **95**, 102003 (2005);

P. Rubin et al. [Cleo Collaboration], "Improved measurement of the electroweak penguin process $B \to X_s l^+ l^-$," Phys. Rev. D **72**, 092005 (2005).

[347] K. Abe et al. [Belle Collaboration], "Observation of double $c\bar{c}$ production in $e^+e^-$ annihilation at $s^{1/2}$ 10.6 GeV," Phys. Rev. Lett. **89**, 142001 (2002).

[348] K. Abe et al. [Belle Collaboration], "Comment on "$e^+e^-$ annhilation into $J/\psi J/\psi$"," arXiv:hep-ex/0306015.

[349] B. Aubert et al. [BaBaR Collaboration], "Measurements of the mass and width of the $eta_c$ meson and of an $eta_c(2S)$ candidate," Phys. Rev. Lett. **92**, 142002 (2004).

[350] D. M. Asner et al. [Cleo Collaboration], "Observation of $eta'_c$ Production in $\gamma\gamma$ Fusion at CLEO," Phys. Rev. Lett. **92**, 142001 (2004).

[351] W. Buchmuller and S. H. H. Tye, "Quarkonia and Quantum Chromodynamics," Phys. Rev. D **24**, 132 (1981).

[352] T. A. Lahde and D. O. Riska, "Pion rescattering in two pion decay of heavy quarkonia," Nucl. Phys. A **707**, 425 (2002).

[353] E. Eichten, K. Lane and C. Quigg, "Charmonium levels near threshold and the narrow state $X(3872) \to \pi^+\pi^- J/\psi$," Phys. Rev. D **69**, 094019 (2004).

[354] Preliminary results shown at QWG2008.

[355] R. E. Mitchel et al. [Cleo Collaboration], "$J/\psi$ and $\psi(2S)$ Radiative Decays to $\eta_c$," Phys. Rev. Lett. **102**, 011801 (2009).

[356] Nakazawa - talk at Photon2007 and Uehara - talk at QWG2008.

[357] T. A. Armstrong et al., "Observation of the p wave singlet state of charmonium," Phys. Rev. Lett. **69**, 2337 (1992).

[358] A. Tomaradze, Talk presented at Hadron 2001, Prodvino, Russia, Aug. 25 - Sept. 1, 2001.

[359] S. Dobbs et al. [Cleo Collaboration], "Precision Measurement of the Mass of the $h_c(P-1(1))$ State of Charmonium," Phys. Rev. Lett. **101**, 182003 (2008).

[360] M. Suzuki, "Search of 1P1 charmonium in B decay," Phys. Rev. D **66**, 037503 (2002).

[361] F. Fang et al. [Belle Collaboration], "Search for the $h_c$ meson in $B^\pm \to h_c K^\pm$," Phys. Rev. D **74**, 012007 (2006).

[362] S. Uehara et al. [Belle Collaboration], "Observation of a $\chi'(c2)$ candidate in $\gamma\gamma \to D\bar{D}$ production at BELLE," Phys. Rev. Lett. **96**, 082003 (2006).

[363] M. Ablikim et al. [BES Collaboration], "Determination of the $\psi(3770)$, $\psi(4040)$, $\psi(4160)$ and $\psi(4415)$ resonance parameters," Phys. Lett. B **660**, 315 (2008).

[364] M. Ablikim et al. [BES Collaboration], "Anomalous Line Shape of the Cross Section for $e^+e^- \to$ Hadrons in the Center-of-Mass Energy Region between 3.650 and 3.872 GeV," Phys. Rev. Lett. **101**, 102004 (2008).

[365] S. K. Choi et al. [Belle Collaboration], "Observation of a narrow charmonium - like state in exclusive $B^\pm \to K^\pm \pi^+ \pi^- J/\psi$ decays," Phys. Rev. Lett. **91**, 262001 (2003).

[366] D. Acosta et al. [CDF Collaboration], "Observation of the narrow state $X(3872) \to J/\psi \pi^+ \pi^-$ in $\bar{p}p$ collisions at $s^{1/2} = 1.96$-TeV," Phys. Rev. Lett. **93**, 072001 (2004);

V. M. Abazov et al. [D0 Collaboration], "Observation and properties of the X(3872) decaying to $J/\psi \pi^+ \pi^-$ in $p\bar{p}$ collisions at $s^{1/2} = 1.96$-TeV," Phys. Rev. Lett. **93**, 162002 (2004);

B. Aubert et al. [BaBar Collaboration], "Study of the $B^- \to J/\psi K^- \pi^+ \pi^-$ decay and measurement of the $B^- \to X(3872) K^-$ branching fraction," Phys. Rev. D **71**, 071103 (2005).

[367] A. Abulencia et al. [CDF Collaboration], "Measurement of the dipion mass spectrum in $X(3872) \to J/\psi \pi^+ \pi^-$ decays," Phys. Rev. Lett. **96**, 102002 (2006).

[368] K. Abe et al. [Belle Collaboration], "Evidence for $X(3872) \to \gamma J/\psi$ and the sub-threshold decay $X(3872) \to \omega J/\psi$," arXiv:hep-ex/0505037.

[369] B. Aubert et al. [BaBar Collaboration], "Evidence for $X(3872) \to \psi(2S)\gamma$ in $B^\pm \to X(3872) K^\pm$ decays, and a study of $B \to c\bar{c}\gamma K$," Phys. Rev. Lett. **102**, 132001 (2009).

[370] A. Abulencia et al. [CDF Collaboration], "Analysis of the quantum numbers $J^{PC}$ of the X(3872)," Phys. Rev. Lett. **98**, 132002 (2007).

[371] E. S. Swanson, "Short range structure in the X(3872)," Phys. Lett. B **588**, 189 (2004);

F. E. Close, P. R. Page, "The $D^0 \bar{D}^0$ threshold resonance," Phys. Lett. B **578**, 119 (2004);

M. B. Voloshin, "Interference and binding effects in decays of possible molecular component of X(3872)," Phys. Lett. B **579**, 316 (2004).

[372] B. Aubert *et al.* [BaBar Collaboration], "Study of Resonances in Exclusive B Decays to $\bar{D}^*D^*K$," Phys. Rev. D **77**, 011102 (2008);

I. Adachi *et al.* [Belle Collaboration], "Study of the $B \to X(3872)(\to D^{*0}\bar{D}^0)K$ decay," arXiv:0810.0358.

[373] C. Hanhart *et al.*, "Reconciling the X(3872) with the near-threshold enhancement in the $D^0\bar{D}^{*0}$ final state," Phys. Rev. D **76**, 034007 (2007);

M. B. Voloshin, "Isospin properties of the X state near the $D\bar{D}^*$ threshold," Phys. Rev. D **76**, 014007 (2007);

E. Braaten, M. Lu, "The Effects of charged charm mesons on the line shapes of the X(3872)," Phys. Rev. D **77**, 014029 (2008).

[374] E. S. Swanson, "Diagnostic decays of the X(3872)," Phys. Lett. B **598**, 197 (2004).

[375] D. Ebert, R. N. Faustov, V. O. Galkin, "Masses of heavy tetraquarks in the relativistic quark model," Phys. Lett. B **634**, 214 (2006).

[376] B. Aubert *et al.* [BaBar Collaboration], "A Study of $B \to X(3872)K$, with $X(3872) \to J/\psi\pi^+\pi^-$," Phys. Rev. D **77**, 111101 (2008);

I. Adachi *et al.* [Belle Collaboration], "Study of X(3872) in $B$ meson decays," arXiv:0809.1224

[377] B. Aubert *et al.* [BaBar Collaboration], "Search for a charged partner of the X(3872) in the B meson decay $B \to X^-K$, $X^- \to J/\psi\pi^-\pi^0$," Phys. Rev. D **71**, 031501 (2007).

[378] S. K. Choi *et al.* [Belle Collaboration], "Observation of a resonance-like structure in the $\pi^\pm\psi'$ mass distribution in exclusive $B \to K\pi^\pm\psi'$ decays," Phys. Rev. Lett. **100**, 142001 (2008).

[379] R. Mizuk *et al.* [Belle Collaboration], "Dalitz analysis of $B \to K\pi^+\psi'$ decays and the $Z(4430)^+$," Phys. Rev. D **80**, 031104 (2009).

[380] J. L. Rosner, "Threshold effect and $\pi^\pm\psi(2S)$ peak," Phys. Rev. D **76**, 114002 (2007).

[381] C. Meng, K. T. Chao, "$Z^+(4430)$ as a resonance in the $D_1(D_1')D^*$ channel," arXiv:0708.4222;

X. Liu, Y. R. Liu *et al.*, "Is $Z^+(4430)$ a loosely bound molecular state?" Phys. Rev. D **77**, 034003 (2008).

[382] L. Maiani, A. D. Polosa, "The Charged Z(4433): Towards a New Spectroscopy," arXiv:0708.3997.

[383] A. G. Mokhtar [BaBar Collaboration], "Searches for exotic X,Y, and Z states with BABAR," arXiv:0810.1073.

[384] R. Mizuk *et al.* [Belle Collaboration], "Observation of two resonance-like structures in the $\pi^+\chi_{c1}$ mass distribution in exclusive $\bar{B}^0 \to K^-\pi^+\chi_{c1}$ decays," Phys. Rev. D **78**, 072004 (2008).

[385] B. Aubert *et al.* [BaBar Collaboration], "Observation of a broad structure in the $\pi^+\pi^-J/\psi$ mass spectrum around $4.26 Gev/c^2$," Phys. Rev. Lett. **95**, 142001 (2005).

[386] C. Z. Yuan *et al.* [Belle Collaboration], "Measurement of $e^+e^- \to \pi^+\pi^- J/\psi$ cross-section via initial state radiation at Belle," Phys. Rev. Lett. **99**, 182004 (2007).

[387] B. Aubert *et al.* [BaBar Collaboration], "Study of the $\pi^+\pi^- J/\psi$ Mass Spectrum via Initial-State Radiation at BABAR," arXiv:0808.1543.

[388] B. Aubert *et al.* [BaBar Collaboration], "Evidence of a broad structure at an invariant mass of $4.32 GeV/c^2$ in the reaction $e^+e^- \to \pi^+\pi^-\psi(2S)$ measured at BaBar," Phys. Rev. Lett. **98**, 212001 (2007).

[389] X. L. Wang *et al.* [Belle Collaboration], "Observation of Two Resonant Structures in $e^+e^- \to \pi^+\pi^-\psi(2S)$ via Initial State Radiation at Belle," Phys. Rev. Lett. **99**, 142002 (2007).

[390] M. Ablikim *et al.* [BES Collaboration], "Determination of the $\psi(3770)$, $\psi(4040)$, $\psi(4160)$ and $\psi(4415)$ resonance parameters," Phys. Lett. B **660**, 315 (2008).

[391] G. Pakhlova *et al.* [Belle Collaboration], "Measurement of the near-threshold $e^+e^- \to D\bar{D}$ cross section using initial-state radiation," Phys. Rev. D **77**, 011103 (2008).

[392] G. Pakhlova *et al.* [Belle Collaboration], "Measurement of the near-threshold $e^+e^- \to D^{*\pm}D^{*\mp}$ cross section using initial-state radiation," Phys. Rev. Lett. **98**, 092001 (2007).

[393] G. Pakhlova *et al.* [Belle Collaboration], "Observation of $\psi(4415) \to D\bar{D}^{*2}(2460)$ decay using initial-state radiation," Phys. Rev. Lett. **100**, 062001 (2008).

[394] G. Pakhlova *et al.* [Belle Collaboration], "Observation of a near-threshold enhancement in the $e^+e^- \to \Lambda_c^+\Lambda_c^-$ cross section using initial-state radiation," Phys. Rev. Lett. **101**, 172001 (2008).

[395] G. Pakhlova *et al.* [Belle Collaboration], "Measurement of the $e^+e^- \to D^0 D^{*-}\pi^+$ cross section using initial-state radiation," Phys. Rev. D **80**, 091101 (2009).

[396] B. Aubert *et al.* [BaBar Collaboration], "Study of the Exclusive Initial-State Radiation Production of the $D\bar{D}$ System," Phys. Rev. D **76**, 111105 (2007).

[397] B. Aubert *et al.* [BaBar Collaboration], "Exclusive Initial-State-Radiation Production of the $D\bar{D}$, $D^*\bar{D}^*$, and $D^*\bar{D}^*$ Systems," Phys. Rev. D **79**, 092001 (2009).

[398] D. Cronin-Hennessy *et al.* [CLEO Collaboration], "Measurement of Charm Production Cross Sections in $e^+e^-$ Annihilation at Energies between 3.97 and 4.26 GeV," Phys. Rev. D **80**, 072001 (2009).

[399] J. Z. Bai *et al.* [BES Collaboration], "Measurements of the cross-section for $e^+e^- \to$ hadrons at center-of-mass energies from 2 GeV to 5 GeV," Phys. Rev. Lett. **88**, 101802 (2002).

[400] J. Segovia, D. R. Entem, F. Fernandez, "Charm spectroscopy beyond the constituent quark model," arXiv:0810.2875;

B. Q. Li and K. T. Chao, "Higher Charmonia and X,Y,Z states with Screened Potential," Phys. Rev. D **79**, 094004 (2009).

[401] G. J. Ding *et al.*, "Canonical Charmonium Interpretation for Y(4360) and Y(4660)," Phys. Rev. D **77**, 014033 (2008);

A. M. Badalian, B. L. G. Bakker, I. V. Danilkin, "The S - D mixing and di-electron widths of higher charmonium $1^{--}$ states," Phys. Atom. Nucl. **72**, 638 (2009).

[402] For example:

S. Anderson *et al.* [CLEO Collaboration], "First observation of the decays $B^0 \to D^{*-}p\bar{p}\pi^+$ and $B^0 \to D^{*-}p\bar{n}$," Phys. Rev. Lett. **86**, 2732 (2001);

K. Abe *et al.* [Belle Collaboration], "Observation of $\bar{B}^0 \to D^{0*}p\bar{p}$," Phys. Rev. Lett. **89**, 151802 (2002);

K. Abe *et al.* [Belle Collaboration], "Observation of $B^{\pm} \to p\bar{p}K^{\pm}$," Phys. Rev. Lett. **88**, 181803 (2002);

M. Z. Wang *et al.* [Belle Collaboration], "Observation of $B^0 \to p\bar{\Lambda}\pi^-$," Phys. Rev. Lett. **90**, 201802 (2003);

N. Gabyshev *et al.* [Belle Collaboration], "Study of decay mechanisms in $B^- \to \Lambda_c^+ \bar{p}\pi^-$ decays and observation of low-mass structure in the $Lambda_c^+\bar{p}$ system," Phys. Rev. Lett. **97**, 242001 (2006);

J. L. Rosner, "Low mass baryon anti-baryon enhancements in B decays," Phys. Rev. D **68**, 014004 (2003);

Eef van Beveren, X. Liu, R. Coimbra, G. Rupp, "Possible $\psi(5S)$, $\psi(4D)$, $\psi(6S)$ and $\psi(5D)$ signals in $\Lambda_c\bar{\Lambda}_c$," Europhys. Lett. **85**, 61002 (2009);

Eef van Beveren, G. Rupp, "Production of hadron pairs in electron-positron annihilation near the $K^+K^-$, $D\bar{D}$, $B\bar{B}$, and $\Lambda_c^+\Lambda_c^-$ thresholds," Phys. Rev. D **80**, 074001 (2009).

[403] For example:

S. L. Zhu, "The Possible interpretations of Y(4260)," Phys. Lett. B **625**, 212 (2005);

F. E. Close, P. R. Page, "Gluonic charmonium resonances at BaBar and BELLE?" Phys. Lett. B **628**, 215 (2005);

E. Kou, O. Pene, "Suppressed decay into open charm for the Y(4260) being an hybrid," Phys. Lett. B **631**, 164 (2005).

[404] W. Bartel *et.al.*, "Tests Of The Standard Model In Leptonic Reactions At Petra Energies," Z. Phys. C **30**, 371 (1986);

R. Marshall, "A Determination of $\sin^2(\theta_W)$ and the Lepton and Quark Electroweak Couplings from $e^+e^-$ Data," Z. Phys. C **43**, 607 (1989) (Review article).

[405] M. Miura *et al.*, "Precise measurement of the $e^+e^- \to \mu^+\mu^-$ reaction at $s^{1/2} = 57.77$ GeV," Phys. Rev. D **57**, 5345 (1998).

[406] L. A. Ahrens *et al.*, "Precise Determination of $\sin^2(\theta_W)$ from Measurements of the Differential Cross-sections for $\nu_\mu p \to \nu_\mu p$ and $\bar{\nu}_\mu p \to \bar{\nu}_\mu p$," Phys. Rev. Lett. **56**, 1107 (1986), Erratum-ibid. **56**, 1883, 1986.

[407] G. P. Zeller *et al.* [NuTev Collaboration], "A Precise determination of electroweak parameters in neutrino nucleon scattering," Phys. Rev. Lett. **88**, 091802 (2002).

[408] M. J. Ramsey-Musof, "Low-energy parity violation and new physics," Phys. Rev. C **60** 015501 (1999).

[409] D. S. Armstrong *et al.*, "Qweak: A Precision measurement of the proton's weak charge," AIP Conf. Proc. **698**, 172 (2004).

[410] P. L. Anthony, "Precision measurement of the weak mixing angle in Moller scattering," Phys. Rev. Lett. **95**, 081601 (2005).

[411] C. S. Wood *et al.*, "Measurement of Parity Nonconservation and an Anapole Moment in Cesium," Science **275**, 1759 (1997).

[412] J. Alcarez *et al.* [Electroweak Working Group], "A Combination of Preliminary Electroweak Measurements and Constraints on the Standard Model," hep-ex/0511027.

[413] K. Abe *et al.* [SLD Collaboration], "A High precision measurement of the left-right Z boson cross-section asymmetry," Phys. Rev. Lett. **84**, 5945 (2000).

[414] M. Boehm, W. Hollik, "Model Dependence Of The Electromagnetic Corrections To Lepton Pair Production In Electron - Positron Collisions," Z. Phys. C**23**, 31 (1984);

M. Boehm, A. Denner, W. Hollik, "Radiative Corrections to Bhabha Scattering at High-Energies. Virtual and Soft Photon Corrections," Nucl. Phys. B **304**, 687 (1988);

[415] O. Schneider, talk given at the 1st LHCb Collaboration Upgrade Workshop, Jan. 11-12, 2007, http://www.nesc.ac.uk/action/esi/contribution.cfm?Title=729.

## Chapter 6

# Study of New Physics Scenarios at SuperKEKB

## 6.1 SUSY

### 6.1.1 Comparison of different SUSY breaking scenarios

As discussed in the previous sections, there are a number of processes in which one may find observable effects of physics beyond the Standard Model (BSM). In particular, processes induced by loop diagrams can be sensitive to new particles at the TeV energy scale, as the Standard Model contribution is relatively suppressed. Such processes include $\mathcal{S}_{\phi K_S^0}$, $A_{CP}^{\rm dir}(b \to s\gamma)$, $A_{CP}^{\rm mix}(b \to s\gamma)$, $Br(b \to s\ell^+\ell^-)$, $A_{FB}(b \to s\ell^+\ell^-)$, and $Br(\tau \to \mu\gamma)$. In this section we consider some specific new physics models in the context of supersymmetry (SUSY) and investigate how their effects could be seen in these processes.

The hierarchy problem of the Higgs mass suggests that the physics beyond the Standard Model most likely exists at the TeV energy scale, and hence the LHC has a good chance of discovering new particles. However, the flavor structure of the BSM will still remain to be investigated, as hadron collider experiments are largely insensitive to flavor-violating processes. In SUSY models, for example, flavor physics is the key to investigating the mechanism of SUSY breaking, which is expected to exist at a higher energy scale.

As discussed in Section 3.3, even in the minimal supersymmetric extension of the SM (MSSM), most of the parameter space is already excluded by the present FCNC constraints, and hence one has to consider some structure in the soft SUSY breaking terms such that the FCNC processes are naturally suppressed. In this study we consider the following three models according to refs. [1, 2].

- *Minimal supergravity model (mSUGRA)*
  In this model one assumes that supersymmetry is broken in some invisible (hidden) sector and its effect is mediated to the visible (observable) world only through the gravitational interaction. Since gravity is insensitive to flavor, the induced soft breaking terms are flavor blind at the scale where they are generated. Namely, all sfermions have degenerate mass $m_0^2$; trilinear couplings of scalars are proportional to Yukawa couplings; all gauginos are degenerate with mass $M_{1/2}$. Even though the soft breaking terms are flavor blind, they could induce flavor violation as they evolve from the SUSY breaking scale (assumed here to be $M_X \simeq 2 \times 10^{16}$ GeV for simplicity) to the electroweak scale by the renormalization group equations, since the Yukawa couplings are flavor dependent. In general, FCNC effects are small for this model.

- *SU(5) SUSY GUT with right-handed neutrinos*
  Grand Unification (GUT) is one of the motivations to consider supersymmetric models, as they naturally lead to the unification of the gauge couplings. In GUT models quarks and leptons belong to the same multiplet at the GUT scale. Therefore the flavor mixings of the quark and lepton sectors are related to each other. In particular, large neutrino mixing can induce large squark mixing for right-handed down-type squarks [3–8]. In this class of models we consider two cases where the masses of the right-handed neutrinos are *degenerate* or *non-degenerate*. Because of the large neutrino mixing, the bulk of the parameter space is excluded in the degenerate case due to the $\mathcal{B}(\mu \to e\gamma)$ constraint, and as a result the effect on other FCNC processes is limited. On the other hand, for the non-degenerate case one can expect larger FCNC effects.

- *U(2) flavor symmetry*
  The family structure of the quarks and leptons could be explained by some flavor symmetry among generations. Although $U(3)$ is a natural candidate for such a symmetry, it is badly broken by the top quark Yukawa coupling. Therefore, here we consider a $U(2)$ flavor symmetry among the first two generations in the context of SUSY. We assume some breaking structure of the $U(2)$ symmetry for the Yukawa couplings, squark mass matrices, and the scalar trilinear couplings.

These models still contain many parameters. We randomly choose a number of points from the multi-dimensional parameter space, while fixing the gluino mass at 600 GeV/c$^2$ and $\tan\beta = 30$. Among these points, those that are already excluded from existing mass bounds and FCNC constraints, such as $\epsilon_K$, $b \to s\gamma$ branching fraction, and $\mu \to e\gamma$ branching ratio, are eliminated. The results for several quantities are then shown in the following scatter plots.

First of all, the expectations from these models for two interesting FCNC processes $\mathcal{S}_{\phi K_S^0}$ and $\mathcal{S}_{b \to s\gamma}$ are shown in Fig. 6.1, together with the expected sensitivity at SuperKEKB, expected at the modest integrated luminosity of 5 ab$^{-1}$. Each dot in the plots shows a randomly chosen parameter point of the model. One can observe a significant deviation from the SM for the $SU(5)$ SUSY GUT with a right-handed neutrino (non-degenerate case) while the mSUGRA model gives comparatively small deviations from the SM for these FCNC processes. We emphasize that these different models could have almost an identical mass spectrum of SUSY particles as shown on the bottom panels in the figure. This means that the flavor structure of BSM models is barely probed by measurements at high energy collider experiments.

Correlations for other interesting FCNC processes are shown in Figs. 6.2–6.5 for the above four SUSY models. Also shown in the figures are expected errors at 50 ab$^{-1}$. With these additional correlations, we can further narrow down possible scenarios; for instance, large direct $CP$ violation in $b \to s\gamma$ would imply that the U(2) model was favored. On the other hand, observation of $\tau \to \mu\gamma$ decays would strongly suggest the non-degenerate SU(5) model. Even if the deviation from the mSUGRA in $\mathcal{S}_{\phi K_S^0}$ and $\mathcal{S}_{b \to s\gamma}$ turns out to be too small to distinguish among different scenarios, there are many other possibilities that can be pursued at SuperKEKB.

In order to illustrate the constraints on the parameter space of various NP models, we show the values of $S(\phi K_S)$ and $S(K^*\gamma)$ as a function of the gluino mass (for a fixed value of $\tan\beta = 30$) [10] in Fig. 6.6. The expected experimental sensitivity at 50 ab$^{-1}$ is also shown. The actual constraints of course depend on the measured central values, but it is clear that the measurements at Super KEKB can severely constrain possible parameters of the models.

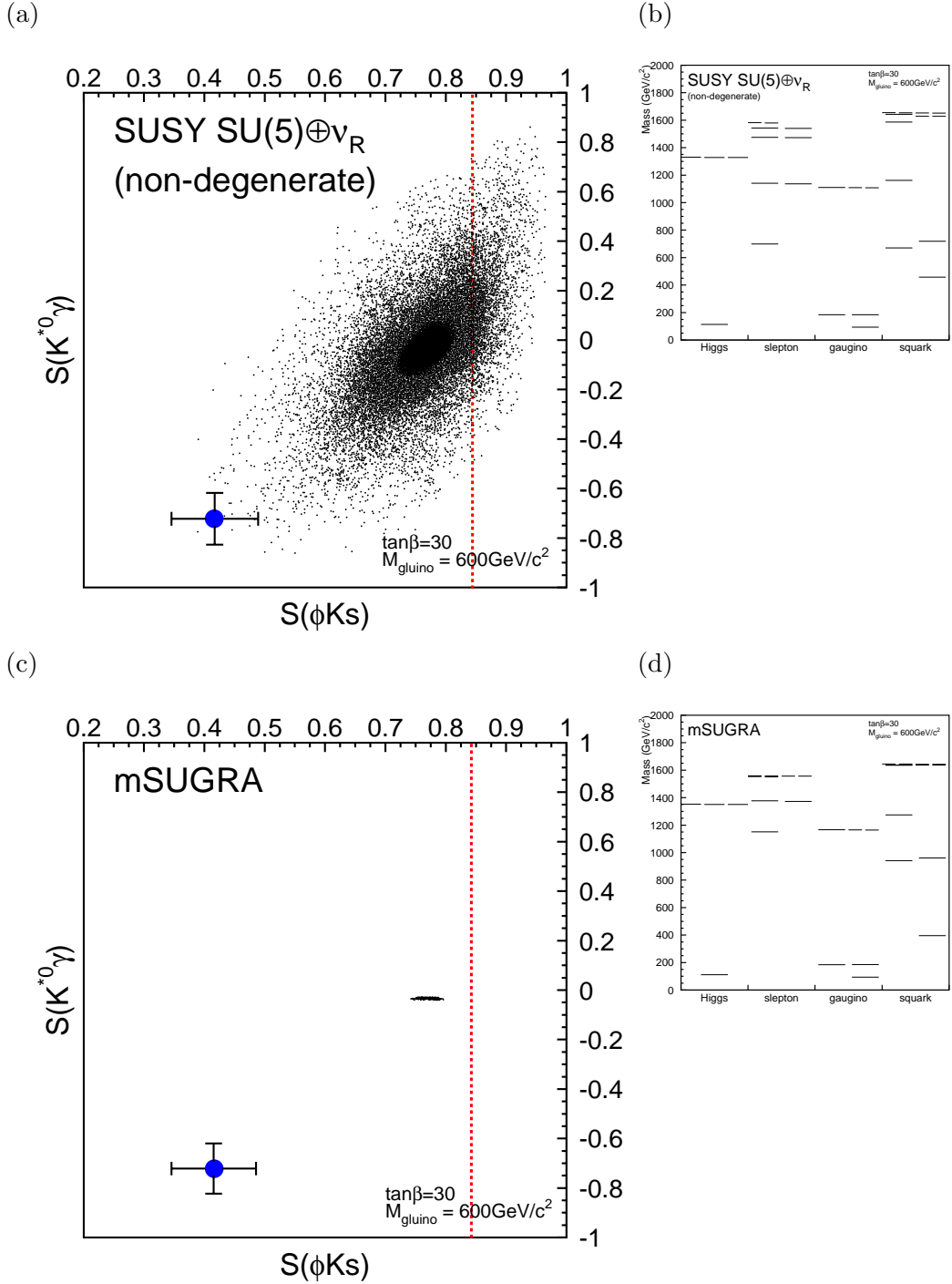

Figure 6.1: $\mathcal{S}_{K^{*0}\gamma}$ and $\mathcal{S}_{\phi K_S^0}$ for various parameters in (a) SUSY $SU(5)$ with right-handed neutrinos (the non-degenerate case), and (c) mSUGRA. Circles with error bars indicate the expected result from a certain parameter set in the $SU(5)$ SUSY GUT, where the errors are obtained at 5 ab$^{-1}$. The present experimental upper limit on $S(\phi K_S)$ at 99% C.L. is also shown by the dashed vertical line [9]. Associated small figures show examples of mass spectra of SUSY particles for (b) the SUSY $SU(5)$ and (d) mSUGRA.

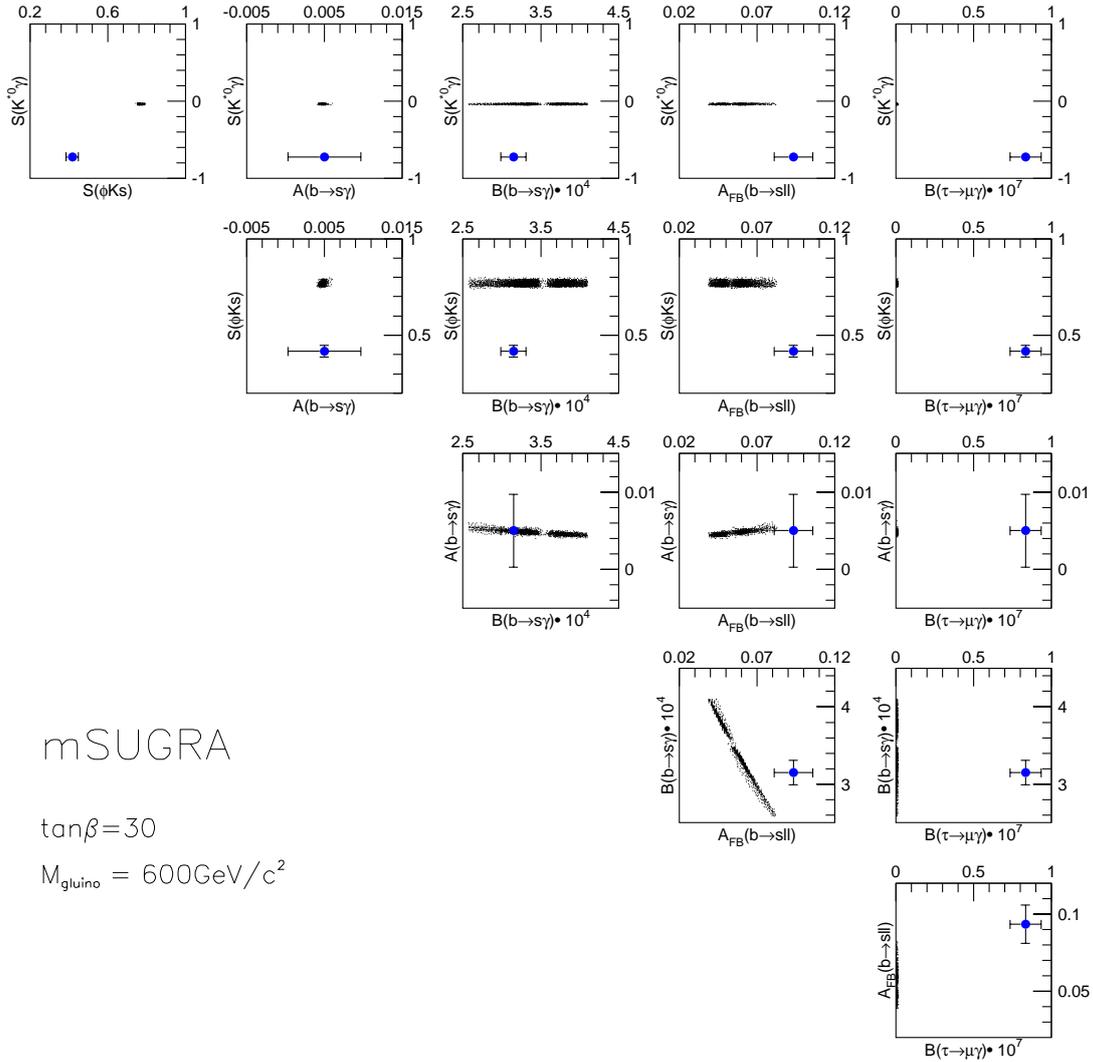

Figure 6.2: Correlations between key observables in the case of mSUGRA. Dots show the possible range in mSUGRA. The circles correspond to a certain parameter set of non-degenerate SUSY SU(5) GUT model with $\nu_R$. Expected errors with an integrated luminosity of 50 ab$^{-1}$ are shown by bars associated with the circles.

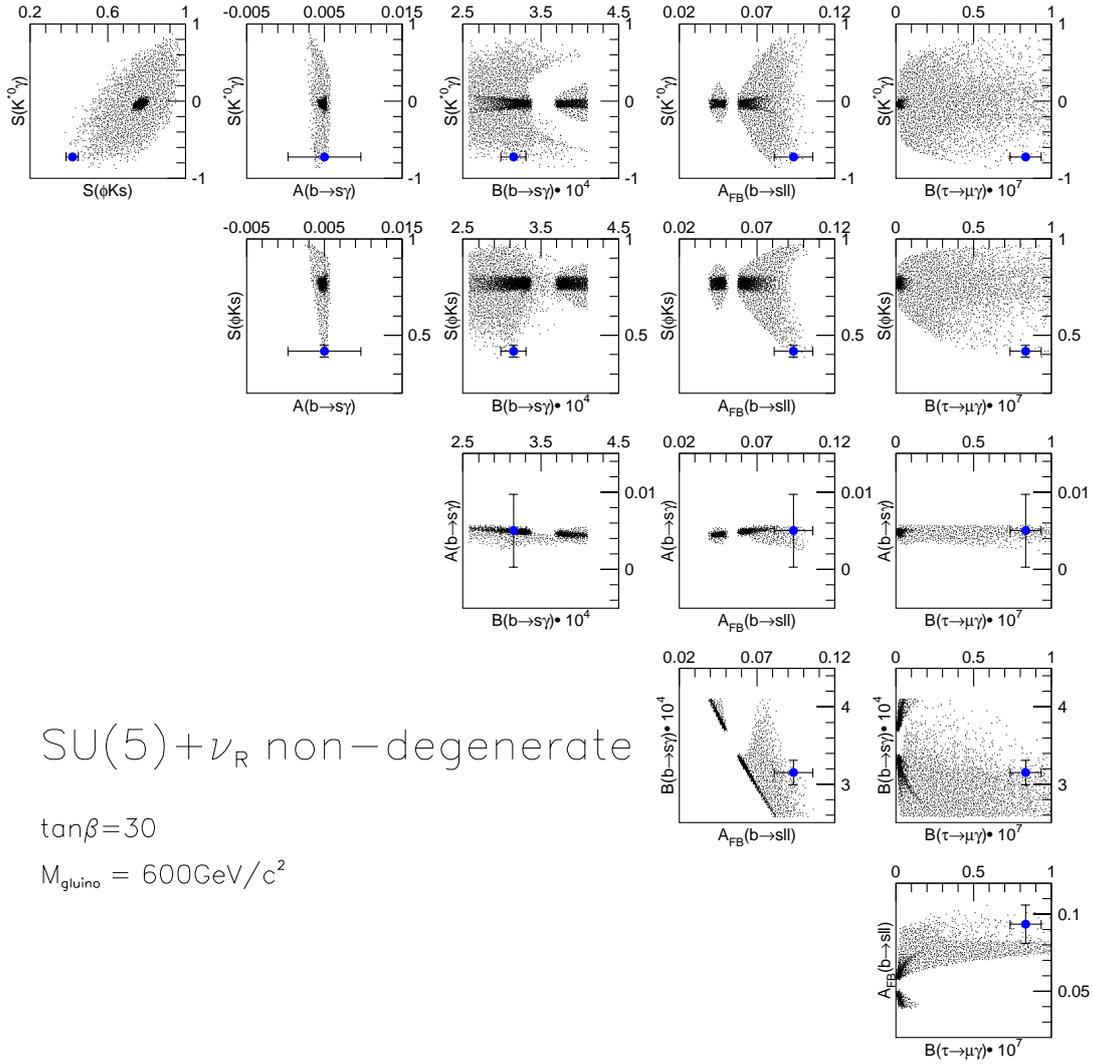

Figure 6.3: Correlations between key observables in the case of the non-degenerate SUSY SU(5) GUT model with $\nu_R$. Dots show the possible range in the model. The circles correspond to a certain parameter set in this model. Expected errors with an integrated luminosity of 50 ab$^{-1}$ are shown by bars associated with the circles.

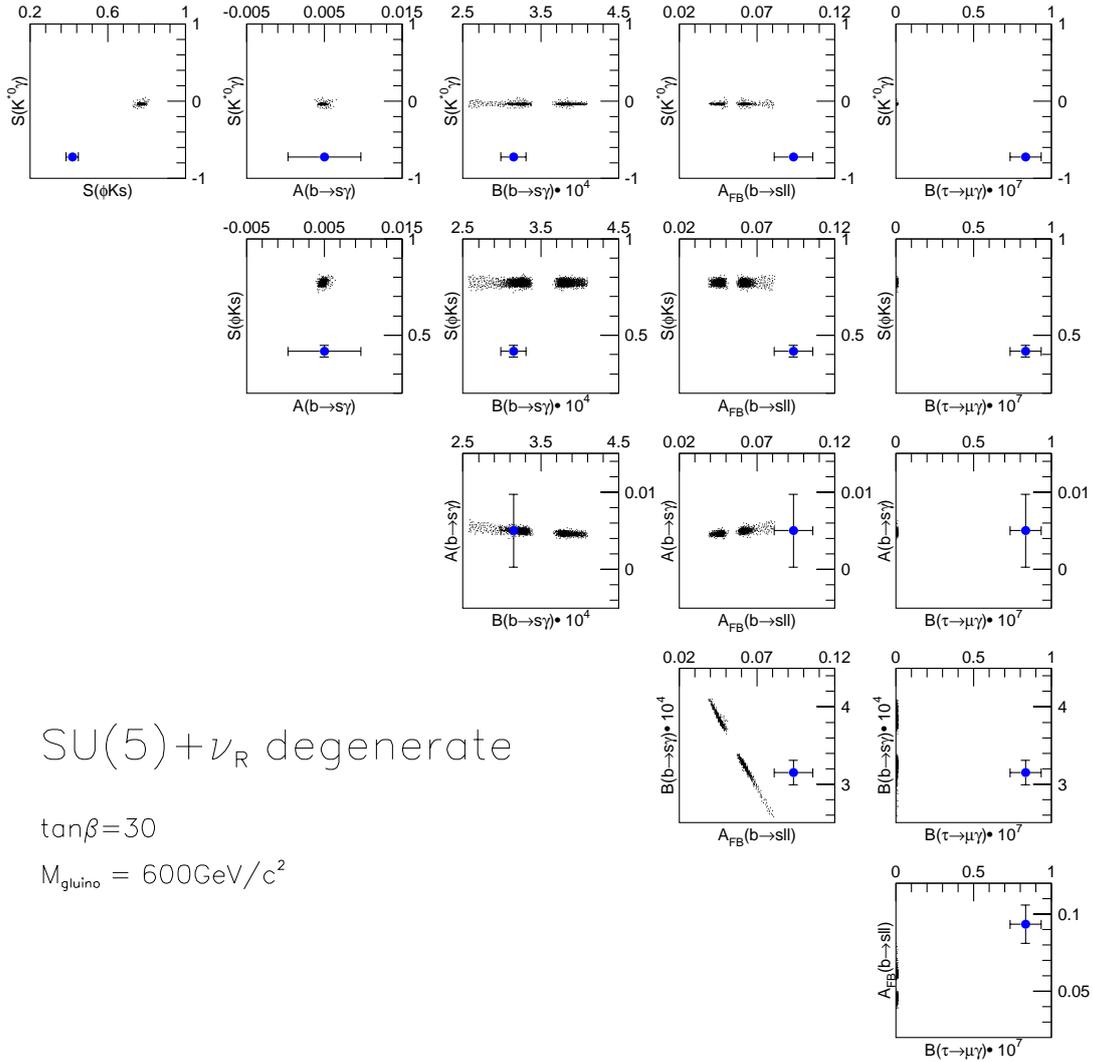

Figure 6.4: Correlations between key observables in the case of the degenerate SUSY SU(5) GUT model. Dots show the possible range in the model. The circles correspond to a certain parameter set of the non-degenerate SUSY SU(5) GUT model with $\nu_R$. Expected errors with an integrated luminosity of 50 ab$^{-1}$ are shown by bars associated with the circles.

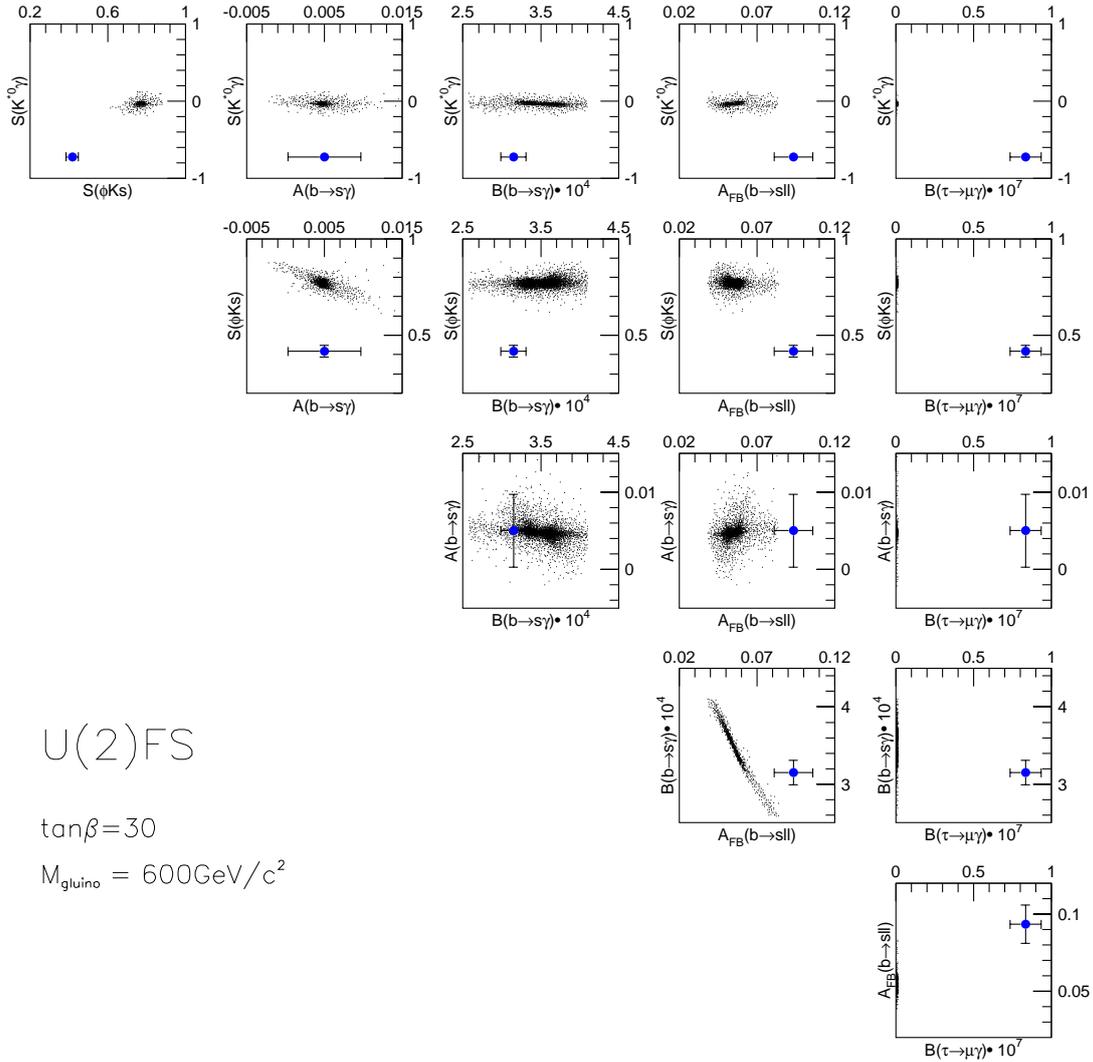

Figure 6.5: Correlations between key observables in the case of the U(2) flavor symmetry model. Dots show the possible range in the model. The circles correspond to a certain parameter set of the non-degenerate SUSY SU(5) GUT model with $\nu_R$. Expected errors with an integrated luminosity of 50 ab$^{-1}$ are shown by bars associated with the circles.

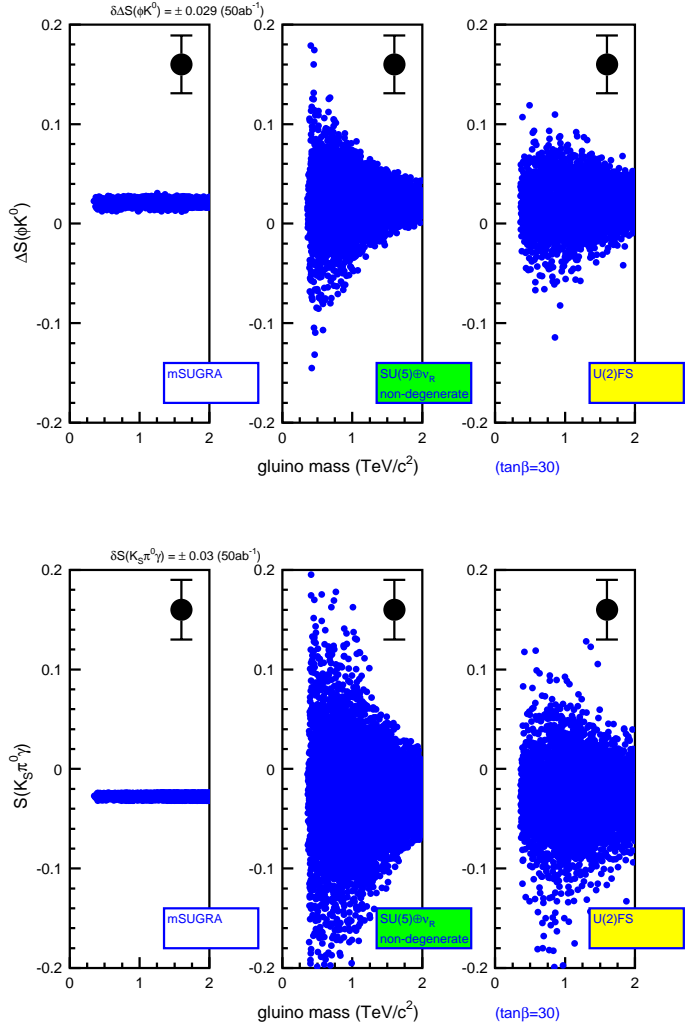

Figure 6.6: From [10]. Top: Difference of time-dependent asymmetries in $B \to \phi K_S$ and $B \to J/\psi K_S$, $\Delta S(\phi K^0)$, in three different models as a function of the gluino mass. Dots with error bars denote the expected accuracy of the measurement with 50 ab$^{-1}$. Bottom: Time-dependent asymmetry in $B \to K_S \pi^0 \gamma$, $S(K_S \pi^0 \gamma)$, as a function of the gluino mass, with the same notation as above.

## 6.2 The CKM fit as a model independent approach

### 6.2.1 Strategy

At SuperKEKB, there are a number of processes from which we can measure the fundamental parameters of the Standard Model (SM), *i.e.* quark mixing angles and phases. In this section we discuss how these measurements at SuperKEKB may be used to gain insight into the flavor structure of new physics without assuming some particular model.

Fitting the CKM parameters using the available data is a common practice. One draws the constraints obtained from several measurements in the parameter space of $(\bar\rho, \bar\eta)$, which are the least known parameters in the usual parametrization of the CKM matrix. If one found an inconsistency among the different input quantities, it would indicate the existence of a new physics effect for some quantities. We investigate how SuperKEKB can contribute to narrowing down the allowed region from each measurement.

Once deviations from the SM are established, the next question is which quantities are affected by the new physics and how. To consider such questions, it is natural to assume that tree-level processes are not affected much by the new physics. If the Standard Model is a low energy effective theory of some more fundamental theory, the new physics effects will manifest themselves as non-renormalizable higher dimensional operators. In tree-level processes such operators are relatively suppressed, while in loop-induced FCNC processes the higher dimensional operators could compete with the loop effects of the Standard Model.

Therefore, we may proceed in three steps. Firstly, we obtain a constraint on $(\bar\rho, \bar\eta)$ using only tree-level processes, *i.e.* $|V_{ub}|$ from the semileptonic decay $b \to u\ell\bar\nu$ and $\phi_3$ from the $DK$ asymmetry. The result can be considered as the SM value of $(\bar\rho, \bar\eta)$ even in the presence of the new physics. Secondly, the allowed region for $(\bar\rho, \bar\eta)$ is compared with the determination through $B^0 - \bar B^0$ mixing, *i.e.* $\Delta m_d$ and $\sin 2\phi_1$ from $B \to J/\psi K_S^0$. If we assume that the decay amplitude for $B \to J/\psi K_S^0$ is dominated by the tree-level contribution, this region of $(\bar\rho, \bar\eta)$ contains loop effects from $\Delta B = 2$ processes. Finally, we consider a parametrization of new physics contributions to $B^0 - \bar B^0$ mixing with the amplitude $M_{12}$ expressed as $M_{12} = M_{12}^{\rm SM} + M_{12}^{\rm new}$, and obtain a constraint on $M_{12}^{\rm new}$ [11–13]. Such a study of the CKM parameter fit is described in the next section.

The separation of the Standard Model and new physics contributions can also be formulated for radiative and semileptonic $B$ decays: $b \to s\gamma$ and $b \to s\ell^+\ell^-$. As discussed in Section 3.5, in the language of the effective Hamiltonian the new physics contribution can be expressed in terms of Wilson coefficients.

A similar strategy can be applied to hadronic decay amplitudes. For instance, the difference between the values of $\sin 2\phi_1$ measured in $J/\psi K_S^0$ and $\phi K_S^0$ decays implies some new physics effect in the $b \to s\bar s s$ penguin process. One can parametrize the $b \to s\bar s s$ amplitude in terms of SM and new physics amplitudes. In this case, however, the calculation of the decay amplitudes from the fundamental theory is much more difficult and the utility of the parametrization is limited.

### 6.2.2 The CKM fit

We perform a global fit to the CKM parameters $(\bar\rho, \bar\eta)$ using the CKMfitter package [14][1]. If no new physics effect exists, all the measurements should agree with each other in the $(\bar\rho, \bar\eta)$ plane and can be combined to obtain the constraint on them. On the other hand, the effect of new physics may cause discrepancies among these measurements.

---

[1] We use the CKMFitter package with the $R$-fit option. The parameter values are the default values used in the package if not mentioned explicitly.

| measurements | Belle (~0.5ab$^{-1}$) | | SuperKEKB (50ab$^{-1}$) | |
|---|---|---|---|---|
| | central value | error | central value | error |
| $\Delta m_d$ | 0.507 ps$^{-1}$ | 0.8% | 0.507ps$^{-1}$ | 0.8% |
| $\sin 2\phi_1 (b \to c)$ | 0.642 | 0.026 | 0.734 | 0.012 |
| $\phi_2$ | 93.0$^o$ | 11$^o$ | 94.6$^o$ | 2$^o$ |
| $\phi_3$ | 53.0$^o$ | 19$^o$ | 61.6$^o$ | 3$^o$ |
| $V_{ub}$ (inclusive) | $4.09 \times 10^{-3}$ | 6.3% | $3.94 \times 10^{-3}$ | 3.0% |
| $Br(B \to \tau\nu)$ | $1.79 \times 10^{-4}$ | 38% | $1.13 \times 10^{-4}$ | 4% |
| $\frac{Br(B\to(\rho,\omega)\gamma)}{Br(B\to K^*\gamma)}$ | 0.034 | 26% | 0.032 | 5% |
| $V_{cb}$ | $0.417 \times 10^{-3}$ | 0.16% | $0.417 \times 10^{-3}$ | 0.16% |
| $\delta m_s$ | 17.77 ps$^{-1}$ | 0.7% | 17.77 ps$^{-1}$ | 0.06 % |
| $\epsilon_K$ | 0.002221 | 3.6% | 0.002221 | 3.6% |

Table 6.1: Summary of measurements used in the CKM fit. Quoted errors in $V_{ub}$ are experimental errors only. Central values of SuperKEKB measurements correspond to $\bar{\rho} = 0.193$ and $\bar{\eta} = 0.356$.

| parameters | central values | errors | |
|---|---|---|---|
| | | Belle (~0.5ab$^{-1}$) | SuperKEKB (50ab$^{-1}$) |
| $f_{Bs}$ | 0.237 | 14% | 4% |
| $f_{Bs}\sqrt{B_s}$ | 0.277 | 13% | 4% |
| $\xi$ | 1.24 | 5% | 2% |
| $V_{ub}$ theory | - | 7% | 4% |
| $B \to (\rho,\omega)\gamma$ theory | - | 11% | 8% |

Table 6.2: Summary of theoretical parameters used in the CKM-fit.

The strategy is to look for the new physics effects from only one source at a time, in order to quantitatively identify the new physics contribution. Assuming that the tree-level processes are insensitive to new physics effects, we determine the SM value of $(\bar{\rho}, \bar{\eta})$ using $|V_{ub}|$ from $B \to X_u \ell \nu$ decays and $\phi_3$ from $B^\pm \to D^0[\to K^0_S \pi^+ \pi^-] K^\pm$ [2]. We then examine the $b \to d$ box diagram, i.e. $B^0 - \bar{B}^0$ oscillations, for which the available measurements are $\sin 2\phi_1$ from $B^0 \to c\bar{c} K^0$ and $\Delta m_d$.

The experimental measurements used in the fit are listed in Table 6.1, where the errors are based on the sensitivity studies in the previous sections. In addition to the measurements at SuperKEKB, other measurements of $\Delta m_s$ and $\epsilon_K$ are included in the fits. In particular, a very high precision measurement of $\Delta m_s$ is expected by the LHCb experiment in near future [16] which gives a strong constraint in $(\bar{\rho}, \bar{\eta})$ combined with the $\Delta m_d$ measurement at SuperKEKB. The errors in the table include both statistical and experimental systematic errors. They are combined and symmetrized.

---

[2] There is a possible effect of new physics on the Dalitz distributions through $D^0 - \bar{D}^0$ mixing or $CP$ violation in $D^0$. However, we can check and measure the effect from the data using soft-pion tagged $D^0$ decays ($D^{*+} \to D^0 \pi^+_s$) [15], and verify whether its effect to the $\phi_3$ measurement is small enough.

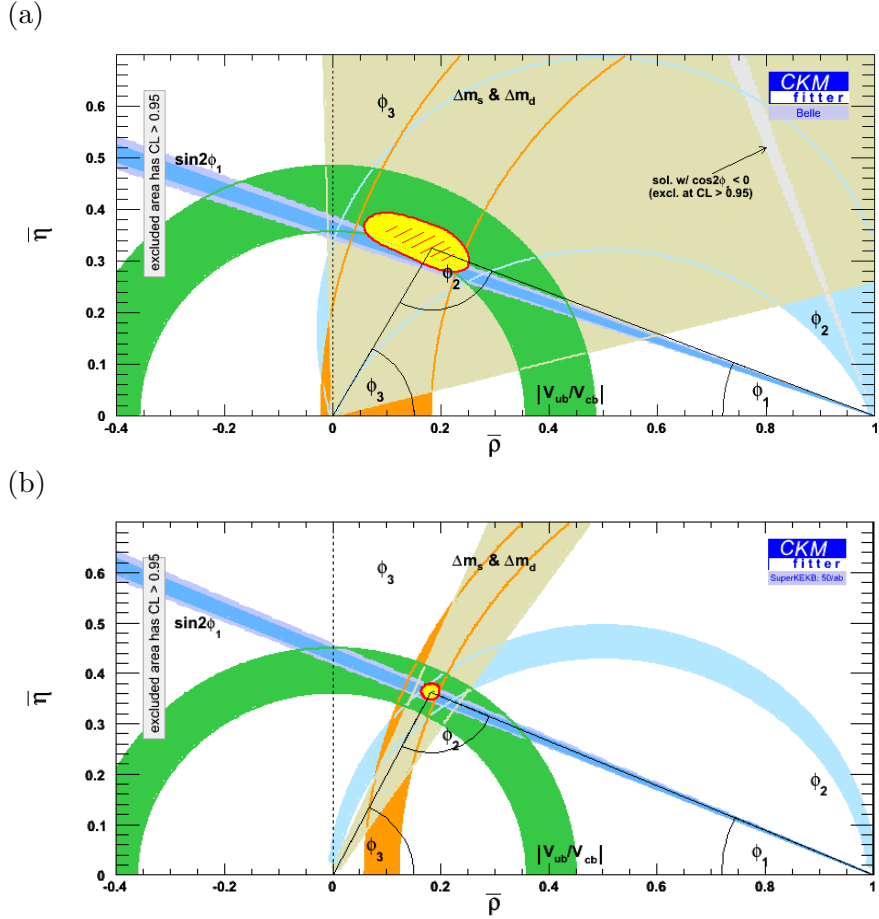

Figure 6.7: Constraints on the CKM unitarity triangle at (a) Belle ($\sim 0.5$ ab$^{-1}$) and (b) SuperKEKB (50 ab$^{-1}$).

Table 6.2 summarizes the theoretical parameters used in the fit. The parameters are mainly given by the lattice QCD calculations. The current largest uncertainty is in the calculations of $f_B$ and $B_B$ for the $|V_{tb}|/|V_{ts}|$ derivation from the measured $\Delta m_d$ and $\Delta m_s$. To reduce the uncertainty, the calculations for $B_s$ mesons are adopted and they are related to be $B_d$ parameters through a SU(3) correction factor $\xi$ which is better known. We adopted S.Sharpe's predictions [17] for the uncertainties in these parameters for the fit.

The constraint on $(\bar{\rho}, \bar{\eta})$ using all these measurements for the SM case is shown in Figure 6.7. The plots are shown for (a) current Belle ($\sim 500$ fb$^{-1}$) and (b) 50 ab$^{-1}$. Drastic improvements are expected with the statistical power of the SuperKEKB. Table 6.3 shows the errors in determined $(\bar{\rho}, \bar{\eta})$ for the cases of the current Belle ($\sim 0.5$ ab$^{-1}$) and the SuperKEKB(50 ab$^{-1}$). An O(1%) determination of $(\bar{\rho}, \bar{\eta})$ apex becomes possible at SuperKEKB.

The effect of New Physics(NP) is expected to appear in the measurements where the loop diagrams contribute. In the SuperKEKB measurements, possible NP effects are expected to contribute, for example, to the $B_d$ box diagram and influence the measurements of $\sin 2\phi_1$ ($b \to c$) and $\Delta m_d$. On the other hand, the measurements of $\phi_3$ by $B \to D^{(*)}K^{(*)}$ and $|V_{ub}|$ by $B \to X_u l\nu$ are governed by pure tree-level diagrams only and do not include the NP effect. The comparison of the $(\bar{\rho}, \bar{\eta})$ apex determined by these two measurements separately could be a sensitive probe to NP. Fig. 6.8 a) shows the constraint in $(\bar{\rho}, \bar{\eta})$ obtained by the measurements of $|V_{ub}|$ and $\phi_3$ only, while b) is that of $\sin 2\phi_1$ and $\Delta m_d$ only.

|                          | $\sigma(\bar{\rho})$ | $\sigma(\bar{\eta})$ |
|--------------------------|----------|----------|
| Belle ($\sim$0.5 ab$^{-1}$) | 20.0%    | 15.7%    |
| SuperKEKB (50 ab$^{-1}$)   | 3.4%     | 1.7%     |

Table 6.3: Errors in the $(\bar{\rho}, \bar{\eta})$ constraints at Belle and SuperKEKB.

The effect of new physics on the $b \to d$ box diagram can be parameterized by modifying the mixing amplitude $M$ as

$$M = r_d^2 M_{\rm SM} \exp(-i2\theta_d), \qquad (6.1)$$

where $M_{\rm SM}$ is the contribution from the SM, while $r_d$ and $\theta_d$ are is the magnitude and phase of the NP contribution. The CKM fit including the NP parameterization is performed for the measurements of $\sin 2\phi_1$ and $\Delta m_d$, together with the tree level measurements $\phi_3$ and $|V_{ub}|$, and the obtained constraints in $r_d^2$ and $2\theta_d$ are shown in Fig. 6.2.2. The accuracies of the determination of the magnitude $r_d^2$ and the phase $2\theta_d$ are summarized in Table 6.4.

| Parameter | Belle ($\sim$0.5 ab$^{-1}$) | SuperKEKB (50 ab$^{-1}$) |
|-----------|-----------|-----------|
| $r_d^2$   | $\pm 0.7$ | $\pm 0.15$ |
| $2\theta_d$ | $\pm 11^o$ | $\pm 3^o$ |

Table 6.4: Errors in the NP parameters at Belle and SuperKEKB.

(a)

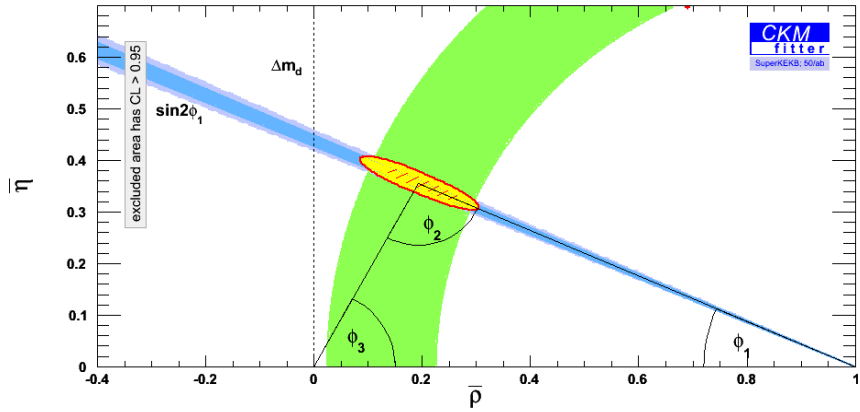

(b)

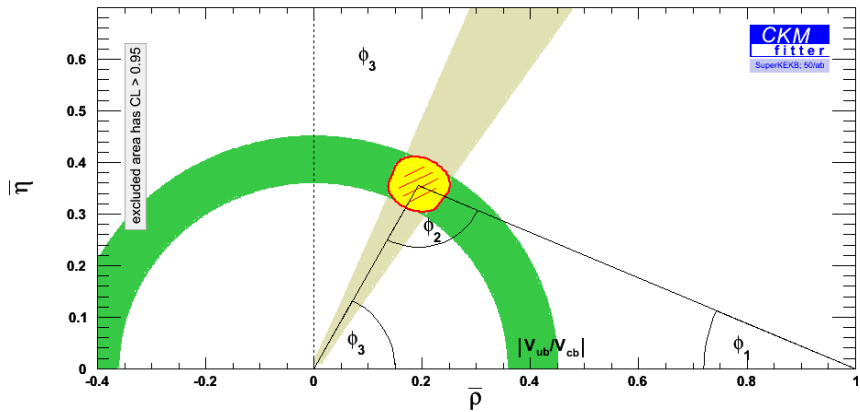

Figure 6.8: Constraints on the CKM unitarity triangle by $\phi_3$ and $|V_{ub}|$ measurements only (a), and by $\sin 2\phi_1$ and $\Delta m_d$ only.

(a) 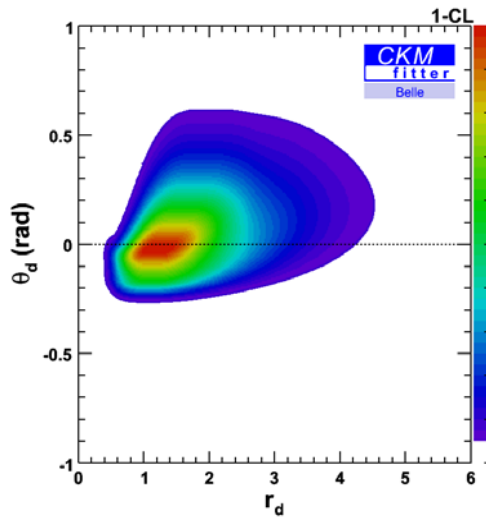 (b) 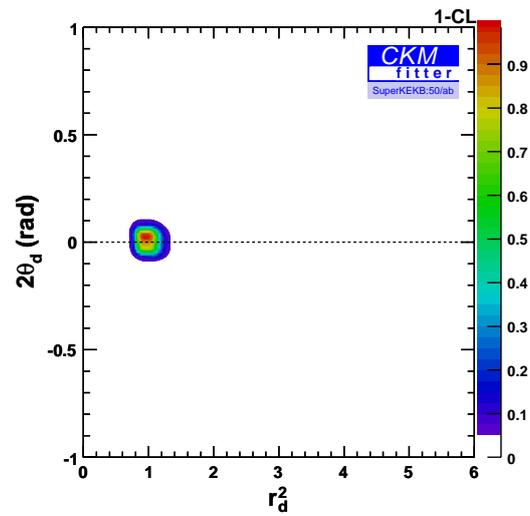

Figure 6.9: Constraints on the NP parameter plane $(r_d^2, 2\theta_d)$ at Belle (a) and SuperKEKB (b).

# References

[1] T. Goto, Y. Okada, Y. Shimizu, T. Shindou and M. Tanaka, "Exploring flavor structure of supersymmetry breaking at B factories," Phys. Rev. D **66**, 035009 (2002).

[2] T. Goto, Y. Okada, Y. Shimizu, T. Shindou and M. Tanaka, "Exploring flavor structure of supersymmetry breaking from rare B decays and unitarity triangle," Phys. Rev. D**70**, 035012 (2004).

[3] S. Baek, T. Goto, Y. Okada and K. Okumura, "Neutrino oscillation, SUSY GUT and B decay," Phys. Rev. D **63**, 051701 (2001).

[4] T. Moroi, "Effects of the right-handed neutrinos on $\Delta S = 2$ and $\Delta B = 2$ processes in supersymmetric SU(5) model," JHEP **0003**, 019 (2000).

[5] T. Moroi, "CP violation in $B_d \to \phi K_S$ in SUSY GUT with right-handed neutrinos," Phys. Lett. B **493**, 366 (2000).

[6] N. Akama, Y. Kiyo, S. Komine and T. Moroi, "CP violation in kaon system in supersymmetric SU(5) model with seesaw-induced neutrino masses," Phys. Rev. D **64**, 095012 (2001).

[7] S. Baek, T. Goto, Y. Okada and K. i. Okumura, "Muon anomalous magnetic moment, lepton flavor violation, and flavor changing neutral current processes in SUSY GUT with right-handed neutrino," Phys. Rev. D **64**, 095001 (2001).

[8] D. Chang, A. Masiero and H. Murayama, "Neutrino mixing and large CP violation in B physics," Phys. Rev. D **67**, 075013 (2003).

[9] E. Barberio *et al.* [Heavy Flavor Averaging Group (HFAG)], "Averages of b-hadron and c-hadron Properties at the End of 2007," arXiv:0808.1297; updates at `http://www.slac.stanford.edu/xorg/hfag/charm/index.html`.

[10] T. Browder *et al.*, "On the Physics Case of a Super Flavour Factory," JHEP **0802**, 110 (2008).

[11] J. M. Soares and L. Wolfenstein, "CP violation in the decays $B^0 \to \psi K_S$ and $B^0 \to \pi^+\pi^-$: A Probe for new physics," Phys. Rev. D **47**, 1021 (1993).

[12] T. Goto, N. Kitazawa, Y. Okada and M. Tanaka, "Model independent analysis of $B - \bar{B}$ mixing and CP violation in $B$ decays," Phys. Rev. D **53**, 6662 (1996).

[13] R. N. Cahn and M. P. Worah, "Constraining the CKM parameters using CP violation in semi-leptonic B decays," Phys. Rev. D **60**, 076006 (1999).

[14] A. Hocker, H. Lacker, S. Laplace and F. Le Diberder, "A new approach to a global fit of the CKM matrix," Eur. Phys. J. C **21**, 225 (2001).

[15] L. M. Zhang *et al.*, [Belle Collaboration], "Measurement of $D^0 - \bar{D}^0$ mixing in $D^0 \to K_s^0 \pi^+ \pi^-$ decays," Phys. Rev. Lett. **99**, 131803 (2007).

[16] O. Schneider, talk given at the 1st LHCb Collaboration Upgrade Workshop, Edinburgh, Jan. 11-12, 2007, `http://www.nesc.ac.uk/action/esi/contribution.cmf?Title=729`.

[17] S. Sharpe, talk given at "Lattice QCD: Presend and Future", Orsay, April 14-16 2004, `http://events.lal.in2p3.fr/conferences/lqcd/`.

# Chapter 7

# Summary

The grand challenge of elementary particle physics is to identify the fundamental elements of Nature and uncover the ultimate theory of their creation, interactions and annihilation. To this end, all elementary particles and the forces between them should be described in a unified picture. Such unification naturally requires a deep understanding of physical laws at a very high-energy scale; for instance the unification of electroweak and strong forces is expected to occur at around $10^{16}$ GeV, which is often called the GUT scale.

It is unlikely that the GUT scale will be realized at any accelerator-based experiment even in the distant future. However, there are a few very promising ways to promote our grand challenge. One such approach is to elucidate the nature of quantum loop effects by producing as many particles as possible. This provides the rationale to pursue the luminosity frontier.

There is no doubt that past experiments at the luminosity (or intensity) frontier of the age have yielded epoch-making results. This good tradition has been followed by Belle and BaBar, experiments at energy-asymmetric $e^+e^-$ $B$ factories KEKB and PEPII, which have observed $CP$ violation in the neutral $B$ meson system. The result is in good agreement with the predictions of the Kobayashi-Maskawa model of $CP$ violation described by the Cabibbo-Kobayashi-Maskawa (CKM) matrix. We are now confident that the CKM phase is the dominant source of $CP$ violation. The measurements performed by the two $B$ factories experimentally verified theoretical predictions to the extent sufficient for the Nobel prize in physics awarded to M. Kobayashi and T. Maskawa in 2008. In 2003 the KEKB collider achieved its design luminosity of $1 \times 10^{34}$ cm$^{-2}$s$^{-1}$. By 2010 the Belle experiment will accumulate an integrated luminosity of 1 ab$^{-1}$. The Unitarity Triangle has been determined with a precision of $\mathcal{O}(10)\%$. Various other quantities in $B$ meson decays have also become accessible in the era of the $B$ factories. In particular, the first observation of direct $CP$ violation in charmless $B$ decays has been obtained.

Over the past thirty years, the success of the Standard Model, which incorporates the KM mechanism, has become increasingly firm. This strongly indicates that the Standard Model is *the* effective low-energy description of Nature. Yet there are several reasons to believe that physics beyond the Standard Model should exist. One of the most outstanding problems is the quadratically divergent radiative correction to the Higgs mass, which requires a fine tuning of the bare Higgs mass unless the cutoff scale is $\mathcal{O}(1)$ TeV. This suggests that the new physics lies at the energy scale of $\mathcal{O}(1)$ TeV. There is a good chance that LHC will discover new elementary particles such as supersymmetric (SUSY) particles. With this vision in mind, we raise an important question "what should be a role of the luminosity frontier in the LHC era ?"

To answer the question, we note that the flavor sector of the Standard Model is quite successful in spite of the problem in the Higgs sector. This is connected to the observation that Flavor-Changing-Neutral-Currents (FCNCs) are highly suppressed. Indeed, if one considers a general new physics model without any mechanism to suppress FCNC processes, present experimental results on $B$ physics imply that the new physics energy scale should be larger than $\mathcal{O}(10^3)$ TeV. This apparent mismatch is called *the new physics flavor problem*. To overcome the problem, any new physics at the TeV scale should have a mechanism to suppress FCNC processes, which often results in a distinctive flavor structure at low energy. Therefore, the indispensable roles of the luminosity frontier are to observe deviations from the Standard Model in flavor physics, and more importantly, to distinguish between different new physics models by a close examination of the flavor structure. Comprehensive studies of $B$ meson decays in the clean $e^+e^-$ environment provide the ideal solution for this purpose, which is not possible at LHC nor even at the future linear collider.

These provide the primary motivation for SuperKEKB, a major upgrade of KEKB. Its design luminosity is $8 \times 10^{35}$ cm$^{-2}$s$^{-1}$, which is 40 times as large as the peak luminosity achieved by KEKB. Various FCNC processes, such as the radiative decay $b \to s\gamma$, the semileptonic decay $b \to sl^+l^-$, and the hadronic decays $b \to s\bar{q}q$ and $b \to d\bar{q}q$, can be studied with unprecedented precision. All of these processes are suppressed in the Standard Model by the GIM mechanism, and therefore the effect of new physics is relatively enhanced. New observables that are currently out of reach will also become accessible. In addition to $B$ meson decays, FCNC processes in $\tau$ and charm decays will also be studied at SuperKEKB.

The Belle detector will be upgraded to take full advantage of the high luminosity of SuperKEKB. In spite of harsh beam backgrounds, the detector performance will be at least as good as the present Belle detector and improvements in several aspects are envisaged. Tables 5.27 and 5.28 summarize the physics reach at SuperKEKB. As a reference, measurements expected at LHCb are also listed. One of the big advantages of SuperKEKB is the capability to reconstruct rare decays that have $\gamma$'s, $\pi^0$'s, $K_L^0$'s or neutrinos in the final states. There are several key observables in the above tables that require this capability. Also important are time-dependent $CP$ asymmetry measurements using only a $K_S^0$ and a constraint from the interaction point to determine the $B$ decay vertices. Examples include $B^0 \to K^{*0}(\to K_S^0\pi^0)\gamma$, $\pi^0 K_S^0$ and $K_S^0 K_S^0 K_S^0$. These fundamental measurements cannot be carried out at hadron colliders.

Figure 5.4(right) shows a comparison between time-dependent $CP$ asymmetries in $B^0 \to J/\psi K_S^0$, which is dominated by the $b \to c\bar{c}s$ tree process, and $B^0 \to \phi K_S^0$, which is governed by the $b \to s\bar{s}s$ FCNC (penguin) process. It demonstrates how well a possible new $CP$-violating phase can be measured. Such a new source of $CP$ violation may revolutionize the understanding of the origin of the matter-dominated Universe, which is one of the major unresolved issues in cosmology. Figure 6.1 shows correlations between time-dependent $CP$ asymmetries in $B^0 \to K^{*0}\gamma$ and $B^0 \to \phi K_S^0$ decays in two representative new physics models with different SUSY breaking scenarios; the SU(5) SUSY GUT with right-handed neutrinos and the minimal supergravity model. The two can be clearly distinguished. This demonstrates that SuperKEKB is sensitive to a quantum phase even at the GUT scale. Note that these two models may have rather similar mass spectra. It will therefore be very difficult to distinguish one from the other at LHC. If SUSY particles are discovered at LHC, the origin of SUSY breaking will become one of the primary themes in elementary particle physics. SuperKEKB will play a leading role in such studies.

We emphasize that the example above is just one of several useful correlations that can be measured only at SuperKEKB. The true value of SuperKEKB is a capability to observe the pattern as a whole, which allows us to differentiate a variety of new physics scenarios. It is so to speak "*DNA identification of new physics*", in that each measurement does not yield a basic physical parameter of the new physics but provides an essential piece of the overall flavor structure. This strategy works better when we accumulate more data. Thus the target annual integrated luminosity of more than 5 ab$^{-1}$ is not a luxury but necessity, and stable long-term operation of SuperKEKB is necessary to meet the requirements.

Determination of the Unitarity Triangle will also be pushed forward and will be incorporated in the global pattern mentioned above. This can be done at SuperKEKB using redundant measurements of all three angles and all three sides of the Unitarity Triangle. In particular, $\phi_2$ measurements and $V_{ub}$ measurements require the reconstruction of $\pi^0$ mesons and neutrinos and are thus unique to a Super $B$-Factory. An inconsistency among these measurements implies new physics. Figure 6.7 shows the constraints from the Belle experiment and the projection for the case with 50 ab$^{-1}$, respectively. The clear difference between two figures demonstrates the power of the ultimate precision of $\mathcal{O}(1)\%$, which will be obtained at SuperKEKB. This precision will allow us to test many new physics scenarios that employ new flavor symmetries. Such tests will serve as a necessary step toward our grand challenge of resolving the origin of hierarchy in the flavor structure.

We thus conclude that the physics case at SuperKEKB is compelling. It will be the place to elucidate *the new physics flavor problem* in the LHC era. The physics program at SuperKEKB is not only complementary to the next-generation energy frontier programme, but is an essential element of the grand challenge in elementary particle physics.